\documentclass[12pt]{article}
\pdfoutput=1

\usepackage{amsmath,amssymb,amscd}
\usepackage{listings}
\usepackage{caption}
\usepackage{dsfont}
\usepackage{slashed}
\usepackage{color}
\usepackage{ulem}
\usepackage[pdftex]{graphicx}
\usepackage{epstopdf}
\usepackage{subfigure}
\usepackage{epsfig}
\usepackage{listings}
\usepackage{caption}
\usepackage{cite}
\usepackage{multirow}
\usepackage{booktabs}
\usepackage{hyperref}
\usepackage{float}

\newcommand{\beq}{\begin{equation}}
\newcommand{\eeq}{\end{equation}}
\newcommand{\nbea}{\begin{align*}}
\newcommand{\neea}{\end{align*}}
\newcommand{\nbeq}{\begin{equation*}}
\newcommand{\neeq}{\end{equation*}}

\usepackage{multirow}
\usepackage{array}
\newcolumntype{M}[1]{>{\centering\arraybackslash}m{#1}}
\newcolumntype{N}{@{}m{0pt}@{}}

\newcommand{\EEG}{\rm e^+ e^-\rightarrow \gamma\gamma}
\newcommand{\EEGG}{\rm e^+ e^-\rightarrow \gamma\gamma(\gamma)}
\newcommand{\EEGGG}{\rm e^+ e^-\rightarrow \gamma\gamma\gamma}
\newcommand{\EEGGGG}{\rm e^+ e^-\rightarrow \gamma\gamma\gamma\gamma}
\newcommand{\EEllgg}{\rm e^+ e^-\rightarrow \ell^{+}\ell^{-}\gamma\gamma}
\newcommand{\EELLGG}{\rm e^+ e^-\rightarrow l^+l^-\gamma\gamma}
\newcommand{\EEEEG}{\rm e^+ e^-\rightarrow e^+e^-(\gamma)}
\newcommand{\EEEE}{\rm e^+ e^-\rightarrow e^+e^-}

\newcommand{\Lagr}{\mathcal{L}}

\begin{document}

\pagestyle{plain}

\baselineskip=21pt

\begin{center}

{\large {\bf Utilizing Theory and Experiment to Search for new structures of Fundamental Particles}}

\vskip 0.3in

\bf  J\"{u}rgen Ulbricht \textsuperscript{a},~
\bf  Minghui Liu \textsuperscript{b},~

\vskip 0.3in

{\small {\it

\vspace{0.25cm}
\vspace{0.25cm}
\vspace{0.25cm}
\textsuperscript{a} Swiss Institute of Technology ETH Z\"{u}rich, CH-8093 Z\"{u}rich, Switzerland \\
\vspace{0.25cm}
\textsuperscript{b} University of Science and Technology of China, USTC, Hefei, Anhui 230026, China \\
}
}

\vskip 0.3in


\end{center}


\baselineskip=18pt \noindent

\vskip 3mm

\vskip 5mm



\begin{abstract}
Our current understanding of the Big Bang tells us that the universe was not a dimensionless point at the 
time ZERO, but already equipped with four dimensions. The fundamental idea behind this work, which 
explores physics beyond the Standard Model, is to abandon the POINT-LIKE nature of the FUNDAMENTAL PARTICLE.
The investigation is conducted in a theoretical and an experimental part. In the theoretical part, 
we present the `` Empirical Toy Ansats about a Microstructure of Fundamental Particles (ETAMFP) ''.
The numerical coincidence of the three spatial interactions ``Strong, Weak, and electromagnetic '', together 
with the corresponding three charges ``Color, Weak, and electric'', is used to develop the ground 
state of ALL FUNDAMENTAL particle in ONE SCHEMA.
The time coordinate serves to interpret the higher states of the fundamental particle as vibrational states 
of the coordinates x, y, and z (Monopole, Dipole, Quadrupole). In addition to this ETAMPF model, we are 
interested in the size of a geometrically extended object. We consider the electron, the most fundamental 
particle with a lifetime of $ 6.6\times 10^{28} $ years, which is longer than the lifetime of the universe, as our object 
of study. We investigated a Lorentz-contracted GYROSCOPE, a Soliton structure in the context of General 
Relativity, and finally the Gross-Pitaevskii equation to develop an  ELECTRON wave function. 
This wave function allowed for the determination of an upper limit for the size of the electron, and the ETAMPF 
model makes it possible to determine a lower limit for the size of the electron. 
Experimental testing of the finite size of the electron revealed that the $ \EEG $  reaction is sensitive 
to the long-range QED interaction, and the $\EEEEG$ BHABHA reaction is sensitive to the short-range 
Weak interaction. According to the theoretical investigation, the outer size of the electron is $ r_{theory}= 3.2\times 10^{-14} $[m]  
and the experimental size is $ r_{exp}= 3.18\times 10^{-14} $[m]. The theoretical size of the inner kernel is 
$ r_{theory-kernel}= 3.3\times 10^{-15} $[m] and the experimental size is $ r_{exp-kernel}= 3.9\times 10^{-15} $[m].
The agreement between our investigations in the theoretical and experimental domains of our work 
generally shows that it is possible to combine all fundamental particles into a common scheme ``The 
ETAMPF model''. Especially the electron as a geometrically extended object with two fundamental 
boundaries. These findings open a window to the fusion of General Relativity and Quantum Mechanics 
and an explanation of the wave-particle duality.
\end{abstract}

\tableofcontents

\section{Theoretical and Experimental Investigations into Particle Models beyond Standard Theory.}
\label{Theory111}

The High Energy Physics group at ETH Z\"{u}rich and USTC Hefei has been investigating the ``Substructure of FUNDAMENTAL PARTICLES ( FP's )~\footnote{``FUNDAMENTAL PARTICLES'' is equal to ``ELEMENTARY PARTICLES'' in the Standard Model}''
for the past 33 years in a collaboration between USTC and ETHZ. Following the temporal structure shown in Fig.{\ref{Invest-timing} of the 
study, the group introduced an ``Empirical Toy Model Ansatz about a Microstructure of Fundamental Particles'' (the ETAMFP 
model), in a theoretical chain. 
Two NEW particles describe ALL FP's in a ``Classical theoretical approach to calculating the size of FP's'', using a  ``Spinning 
superconducting electrovacuum solitons'' \cite{Irina111} and performed a calculation of  
wave function using the substructure of FP via the E-field of a non-point electron with the Gross-Pitaevskii equation.
The wave function allows the calculation of two geometric extensions of an electron, an outer and a kernel 
limit of the electron size, in accordance with the ETAMFP model.
In parallel, the group analysed the size of elementary particles in an experimental chain 
using  ``Experiments to test the size of FP's via direct contact term''. 
In particular, for the electron, a ``Comparison of theoretical data with 
experiment'' find agreement between experimental measurements and theory.
The significance $  \sigma $ in the experiment for the outer size of the electron 
was 5 $  \sigma $ and for the kernel of the electron 2 $  \sigma $ \cite{Firstpaper}.

The entire study of ``Theoretical and Experimental Investigations into Particle Models beyond Standard Theory'' 
presented in Fig. {\ref{Invest-timing} allow the following conclusion. If the theoretical condition
of a point-particle as a fundamental particle in the quantised Standard Model is replaced to a geometrically 
extended particle and the fundamental particle is treated like a superconductor in Einstein's Theory 
of Relativity, a very important connection between the quantised Standard Model and Einstein's Theory 
of Relativity appears.  The theoretical investigation find two boundaries for an Electron, in agreement 
by the experimental findings of two geometric boundaries of the  Electron in $ e^{+}e^{-}\rightarrow \gamma \gamma $ 
scattering, which is sensitive to electromagnetic interactions, and by the weak 
interaction of $ e^{+}e^{-}\rightarrow e^{+}e^{-} $ BHABHA scattering.

\newpage

\begin{figure}[htbp]
\begin{center}
\rotatebox{00}{
 \includegraphics[width=18.0cm,height=10.0cm]{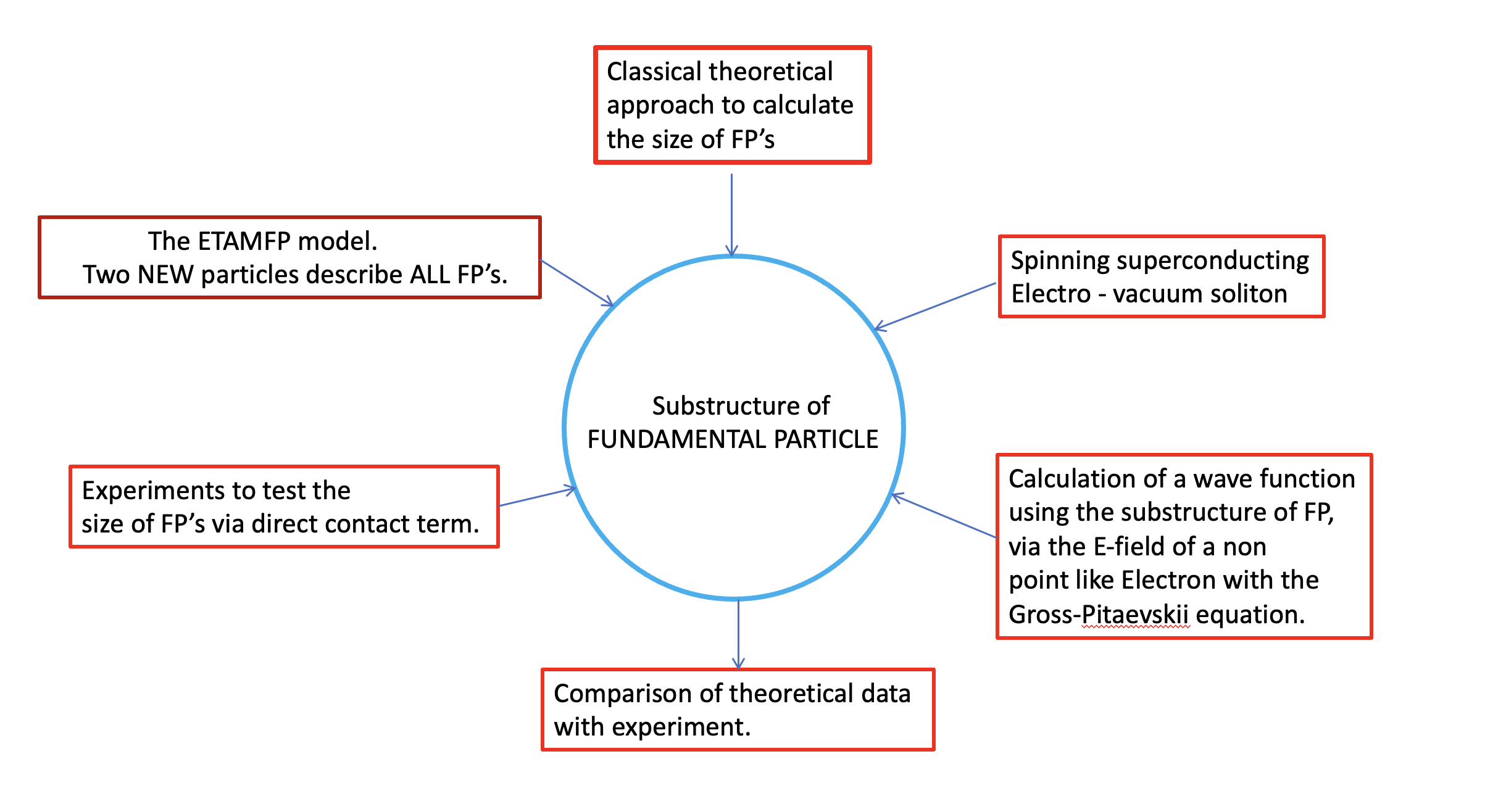}}
\end{center}
\caption{Timing of the investigation of finit size of FP's, performed by ETHZ and USTC. }
\label{Invest-timing}
\end{figure}

\subsection{The history of the ETAMFP model about a new Substructure of Fundamental Particles.}
\label{History ETAMPF}

The investigation of physics beyond the Standard Model (SM) and the Big Bang Model (BBM) 
has been carried out with enormous effort in the past.

In 1783, John Michell introduced the black hole, a body so massive that even light cannot escape \cite{Michell}.
The black hole has a one-way surface, the event horizon, into which objects can fall, but nothing can escape.
Black Holes are still the subject of intense investigations like non static black holes, Einstein-Yang-Mills sphalerons and black 
holes, vacuum black holes, soliton and black holes, five-dimensional Black Hole, Skyrme black hole, tiny black holes, 
hairy black holes, cosmic coloured black holes and non-abelian black holes \cite{Brodbeck}.

In 1931, Paul Dirac introduced the magnetic monopole, a hypothetical particle, a magnet with only one pole.
He showed that the existence of magnetic monopoles would explain the quantisation of electric charge in the universe.
\cite{Dirac1}. Recently, calculations of non-Abilian monopolies, monopolies in the BHI-Higgs theory, 
gravitational lumps, self-gravitating monopolies, and monopoly-antimonopoly pairs 
have been published \cite{Ali}.

Since the introduction of the Yang-Mills theory in 1954 \cite{Jang1}, the theoretical effort to 
search for fundamental particle-like structures has increased.

In 1958, T. Skyrme introduced Skyrmions, a mathematical model for the formation of baryons, 
subatomic particles \cite{Lipkin}. Their topological charge was identified with the baryon number.
They are not individual baryons, but rather coherent states of known baryons and higher 
resonances on a compact manifold associated with the spin and flavour symmetry groups. 
Extensive theoretical studies on the various parameters of Skyrmions have been conducted in recent years.
Skyrmions and their stability \cite{Tokuno}, vibrational energies and excitation \cite{Davies}, 
calculating density, mass, and size \cite{Byung}, Barion nucleus production \cite{Manton}, 
spin and isospin \cite{Souza}, \cite{Ioannidou}, Skyrmions - Black holes \cite{Brihaye}, Gravitational Skyrmions \cite{Shiiki},
Skyriom stars \cite{Skyriom1} and monopole - Skyrioms \cite{Skyriom2}. 

Various theoretical models are investigated, such as the Landau-Lifshits ansatz, the Yang-Mills ansatz, 
the Einstein-Skyrmion model, as well as condensates, lattices, and fluids. Skyrmion scattering 
has also been investigated \cite{Skyriom3}.

In 1961, Yoichiro Nambu and Giovanni Jona-Lasinio introduced the Nambu-Jona-Lasinio model \cite{Lasinio}.
The model is a theory of nucleons and mesons composed of interacting Dirac fermions with chiral 
symmetry. It is a dynamical model of elementary particles based on an analogy to superconductivity.
The dynamic creation of a condensate from fermion interactions inspired many theories of electroweak 
symmetry breaking, such as the Technicolor theory \cite{Lasinio1}, the top quark condensate \cite{Lasinio2}, 
the quark condensate \cite{Lasinio3}, and the nonlocal quark model \cite{Lasinio4}.

In 1968, Rosen introduced a configuration of a charged scalar field that is classically stable under 
small perturbations \cite{Rosen}. These stable configurations of multiple scalar fields were 
investigated by Friedberg, Lee, and Sirlin in 1976 \cite{Rosen1}.
The name ``Q-ball'' and the proof of quantum mechanical stability were introduced by Sidney 
Coleman \cite{Rosen2}. Recently, studies on rotating Q-balls have been published \cite{Rosen3}.

In 1984, F. Klinkhamer and N. Manton first time described a Sphaleron  \cite{Sphaleron} , a static solution to the electroweak field 
equations of the Standard Model of particle physics. A sphaleron is a saddle point of the electroweak 
potential energy of the surface $  z=x^{2}-y^{2} $ in three-dimensional analytic geometry and is involved in 
processes that violate the baryon and lepton numbers.
In some theories, at the higher temperatures of the early universe, Sphalerons convert an imbalance in the number 
of leptons and antileptons formed by the first leptogenesis into an imbalance in the number of baryons 
and antibaryons \cite{Sphaleron1}.
The most important theoretical investigations concern electroweak Sphalerons with spin and charge \cite{Sphaleron2}, 
Sphalerons and Strings \cite{Sphaleron3}, Sphalerons and Black Holes \cite{Sphaleron4}, Sphalerons and vortex rings \cite{Sphaleron5}, 
Sphalerons and Higgs \cite{Sphaleron6}, and QCD Sphalerons \cite{Sphaleron7}.

In 1997, articles appeared on the Dilaton, a hypothetical particle in string theory. The Dilaton is a particle 
of a scalar field associated with gravity. Investigations were conducted on the mass of the Dilaton \cite{Dilaton}, Dilatonic 
Inflation  \cite{Dilaton1}, the Dilaton and Quantum Cosmology \cite{Dilaton2}, the Dilaton and Dark Energy \cite{Dilaton3}, 
and the Dilaton in Einstein-Yang-Mills theory \cite{Dilaton4}.

Recent research on Solitons was published between 1997 and 2008. The soliton phenomenon was 
first described by John Scott Russell (1808 - 1882) \cite{Soliton}.
A soliton is a self-amplifying solitary wave that maintains its shape at constant speed. Solitons arise as solutions to a 
class of weakly nonlinear dispersive partial differential equations that describe physical systems.
They play an important role in particle physics. In recent years, investigations have been conducted on non-Abelian Solitons, 
gravitating and spinning Solitons, Solitons, Quark - Solitons, Sine-Gordon Solitons, anomalous Abelian Solitons, 
and Quark-Solitons \cite{Soliton1}.
 
In 2000, a semiclassical instability of de Sitter space and the formation of black hole pairs in de Sitter space 
were published \cite{De Sitter}.  De Sitter space is the maximally symmetric vacuum solution of Einstein's field equation with 
a repulsive cosmological constant $ \Lambda $. De Sitter space was discovered by Willem de Sitter in 1917 \cite{De Sitter1}.
  
In 2004, investigations were conducted on the excitation of the physical vacuum \cite{vacuum}. 
The vacuum state \cite{vacuum1} is the quantum state with the lowest possible energy.
The vacuum state contains fleeting electromagnetic waves and particles that pop into and out of existence. 
In perturbation theory, the properties of the vacuum are analogous to the properties of the ground state of a harmonic oscillator.
In the Standard Model, the non-zero vacuum expectation value of the Higgs field is the mechanism by which the 
other fields in the theory acquire mass. The uncertainty principle, in the form $ \Delta E \Delta t \geq \hbar $, 
states that in a vacuum, for a short time $ \Delta t $, one or more particles with an energy $ \Delta E $ 
above the vacuum value can be created.

In 2007, theoretical solutions for Einstein-Yang-Mills strings and superconducting electroweak strings were presented \cite{Yang-Mills}.

After the LEP was commissioned in 1989, all four LEP detectors launched a program to explore physical 
phenomena beyond the Standard Model, in particular the reaction $ \EEGG $ response to the non-point like behaviour 
of FP's. As shown in  Fig.{\ref{Invest-timing} ``Timeline of the investigation of the finite size of FP's'', carried out by ETHZ and USTC, 
the interest of  ETHZ and USTC is focused on investigating the possible consequences of the non-point like behaviour of FP's.

In the quantised Standard Model, it is entirely sufficient to describe our experimental knowledge in physics with 
19 parameters. It is not necessary to use a microstructure of the FP's.
However, one must accept that an FP, like an electron, is geometrically a point with finite mass, charge, spin, 
magnetic moment, and electric dipole moment. Microscopically, the density of the FP is infinite; the charge 
must be point-like. The idea of a point with a zero diameter that can generate a magnetic moment and an 
electric dipole moment contradicts a three-dimensional, logical space structure.
If we remove the condition that the FP must be a point and allow for an extended microscopic structure, we can avoid 
these objections in understanding. Such an approach points to the unquantised BBM.
In such a model, an object with rest mass must be described by a density mass accumulation resulting from a stress-energy 
tensor $ T_{k}^{i}\neq 0 $ in a finite space. The particle would already be a geometrical extended object.
It should be emphasised that both theories, SM and BBM, describe the same physical object very successfully.
Therefore, it seems reasonable to search for an ansatz to a geometrically extended object in the transition between SM and BBM.
    
In the following sections, we introduce an ``Empirical Toy Ansatz about a Microstructure of Fundamental Particles''.
We first follow current knowledge of the running coupling constants of the Standard Model and the temporal evolution of the very
early Big Bang theory to develop a microstructure of FP's. Both theories are coupled by a common energy scale $ \Lambda $ which 
defines according the Uncertainty Principle a size. We assume the time-scale-size development of both theories is stored in the size
of a fundamental particle. This implements that the fraction of elementary particles close to a diameter of zero is historically compared to the highest 
possible scales of the Standard Model and the Big Bang theory, the Planck scale. This logic forms the expansion 
of elementary particles from size zero to the present size. In this sense, the size of Fundamental Particles is the direct consequence of the temporal evolution of 
 the Standard Model and the Big Bang theory from time t = 0 to the time of nucleosynthesis $t=t_{nucleosyntesis} $.
   
To test the validity of this ansats, we compare this model step by step with the experimentally measured 
parameters describing the FP's and examine whether the ansatz can reproduce these parameters. 
First, we examine whether the approach can organise all known fermions and bosons into a common scheme. 
We then investigate the effects of the magnetic and electric dipole moment of the FP's on their size.     

\subsection{Empirical Toy Ansats about a Microstructure of Fundamental Particles (ETAMFP).}
\label{Empirical toy ansatz}

As discussed in the chapter on the SM model, two fundamental conditions must be met to test small distances in a 
scattering experiment, such as the $\EEGG$ reaction. First, it is necessary to perform the test at the correct test energy Q.
As shown in the chapter on the Standard Model (Equ.3), the test energy Q should be equal to the center-of-mass 
energy ECM Q = $E_{CM}$ at the interaction point of the $ e^{+}e^{-} $ beam, since the test size $\lambda $ is 
directly inversely proportional to the test energy $ \lambda \sim f(1/E_{CM} $. Second, it is necessary to record a large 
number of events during the reaction under study.
By taking into account radiation corrections of the SM it is possible to significantly reduce the test size,
 as demonstrated, for example, by measuring the electric dipole moment of the electron \cite{dipol}. Since FP's in 
 the SM are point particles, experimental proof of this fact would require an infinitely high 
 test ECM energy or infinitely high statistics.
Both conditions would pass  the Grand Unification and the Planck scale, since the experiment would have to 
be performed at such ECM energies or radiative corrections would have to be calculated up to these scales. 
The predictions of the running coupling constant in SM and SUSY models are summarised in Fig. {\ref{Running-coupling}.
Fig.{\ref{Timing-BP} demonstrate that the physics conditions have changed at these scale energies dramatically.

\begin{figure}[htbp]
\begin{center}
\rotatebox{00}{
 \includegraphics[width=12.0cm,height=16.0cm]{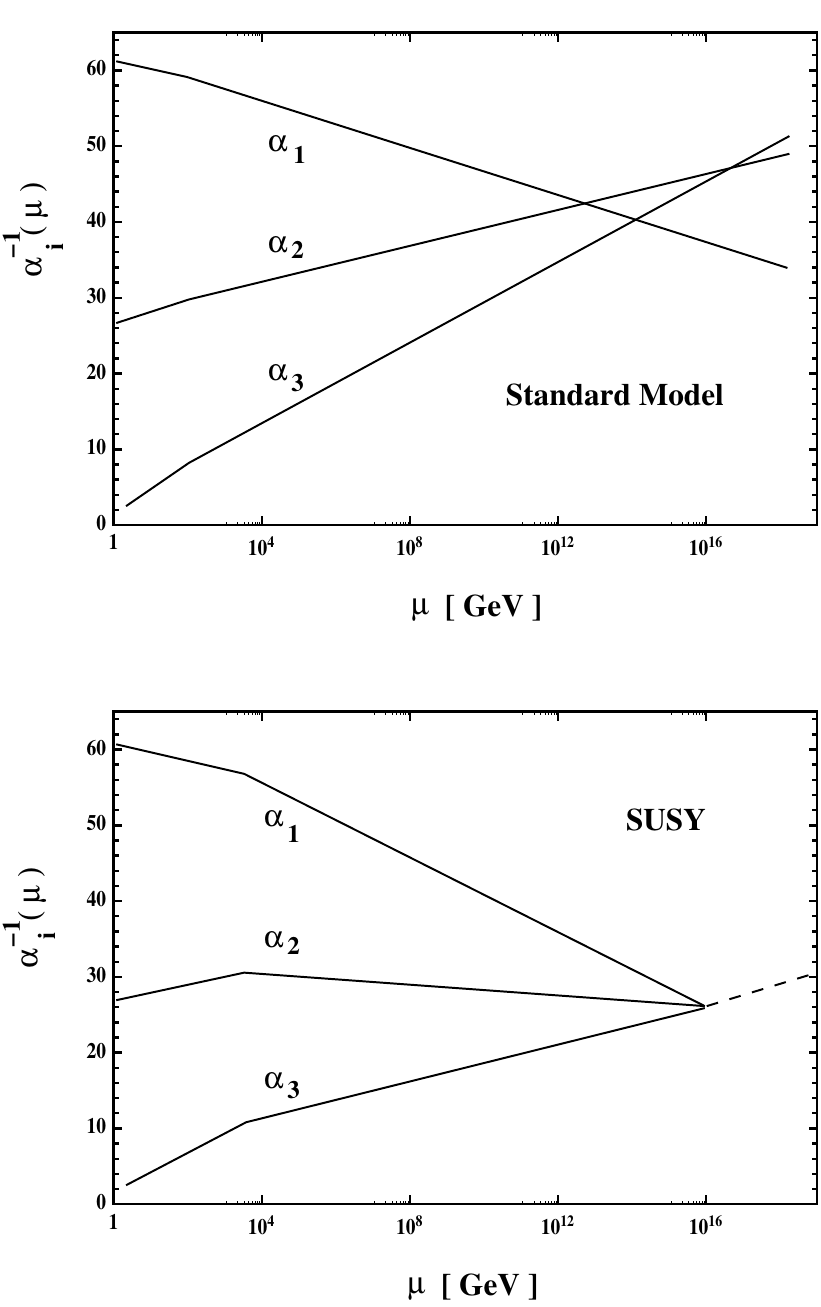}}
\end{center}
\caption{The variation of $  \alpha _{i} $  with the scale $ \mu (GeV) $ for SM and SUSY model \cite{Runnig111}.}
\label{Running-coupling} 
\end{figure}

\newpage

\begin{figure}[htbp]
\begin{center}
\rotatebox{00}{
  \includegraphics[width=16.0cm,height=14.0cm]{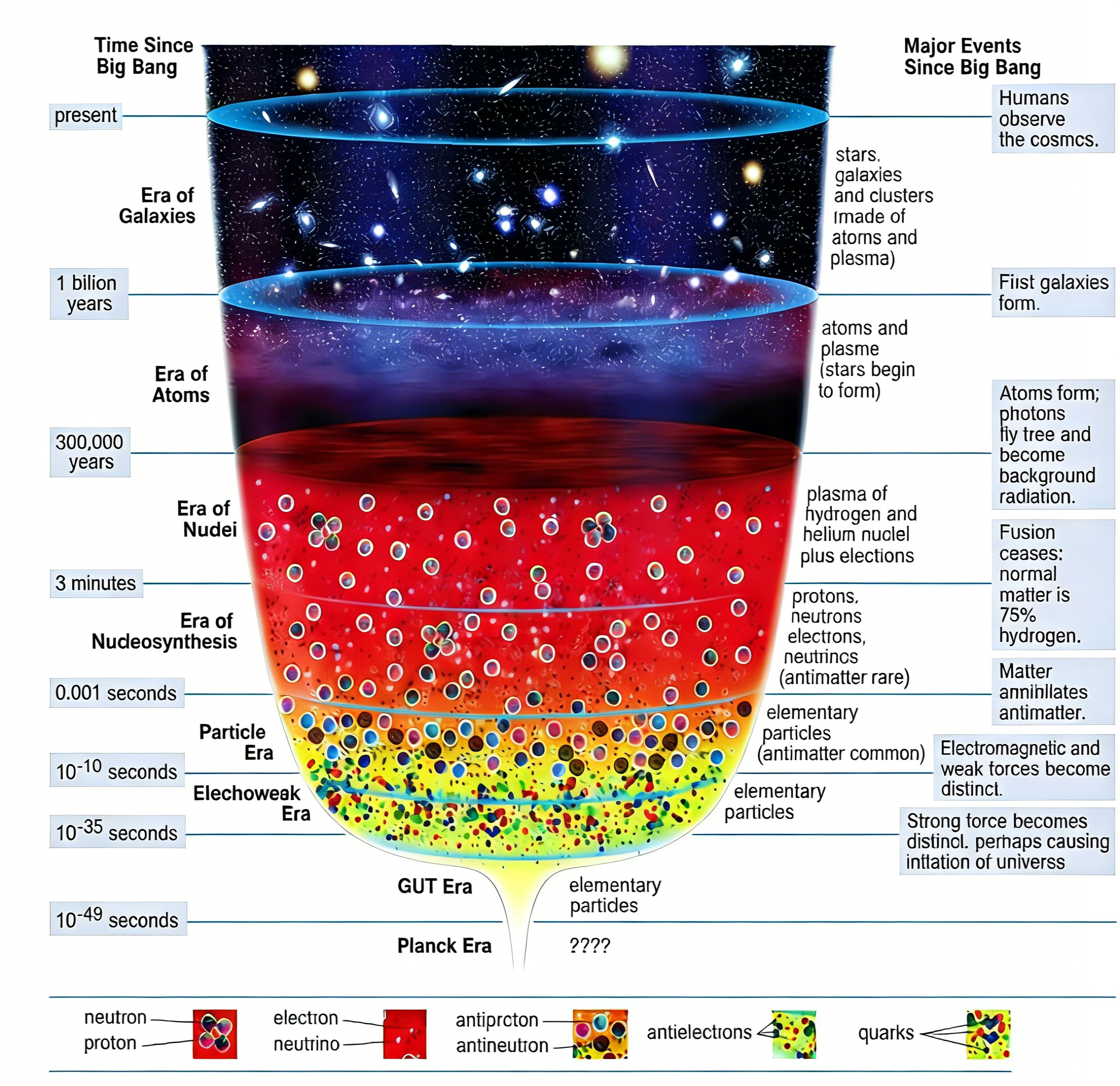}}
\end{center}
\caption{Big Bang Model timing \cite{Timing222}.}
\label{Timing-BP}
\end{figure}

It is not feasible to test these extreme energy conditions experimentally or to perform them with an infinite number of events.
The experimentally available energies of known accelerators, both planned and in operation, 
are shown in Fig.{\ref{Time-accelerator}. They reach approximately 15 TeV 
and simulate a time after the Big Bang of approximately  $  10^{-20} $ s.
The leading parameters defining the physical conditions in the SM model are the test energy Q and the scale 
$ \mu $, which are clearly illustrated in Fig. {\ref{Running-coupling}.
Essential in the BBM model are the time after the Big Bang ``t'' (Chapter  ``Primordial nucleosynthesis'' Equ.3 ),
the temperature ``T'' (Chapter ``Primordial nucleosynthesis'' Equ.3 ) , and the scale $ \mu $, which are clearly 
illustrated in Fig.{\ref{Timing-BP} and Fig.{\ref{Time-accelerator}.

To reconstruct the same physical conditions for a given parameter set in the SM and BBM models, it is 
essential that both models satisfy reverse time invariance. If an experiment is conducted on a 
defined scale $ \mu $ according Fig.{\ref{Time-accelerator}, we assume always to measure the same 
physical properties depending on this scale, for example, at the LEP energies $ W_{\pm } $ and $ Z^{0} $.
This result is independent of the date of the experiment. We always measure $W_{\pm }$ and $Z^{0}$ 
on the LEP scale in spring or autumn. The logical consequence of this experimental fact is that 
the universe remembers its origin.

\begin{figure}[htbp]
\begin{center}
\rotatebox{00}{
 \includegraphics[width=17.0cm,height=13.0cm]{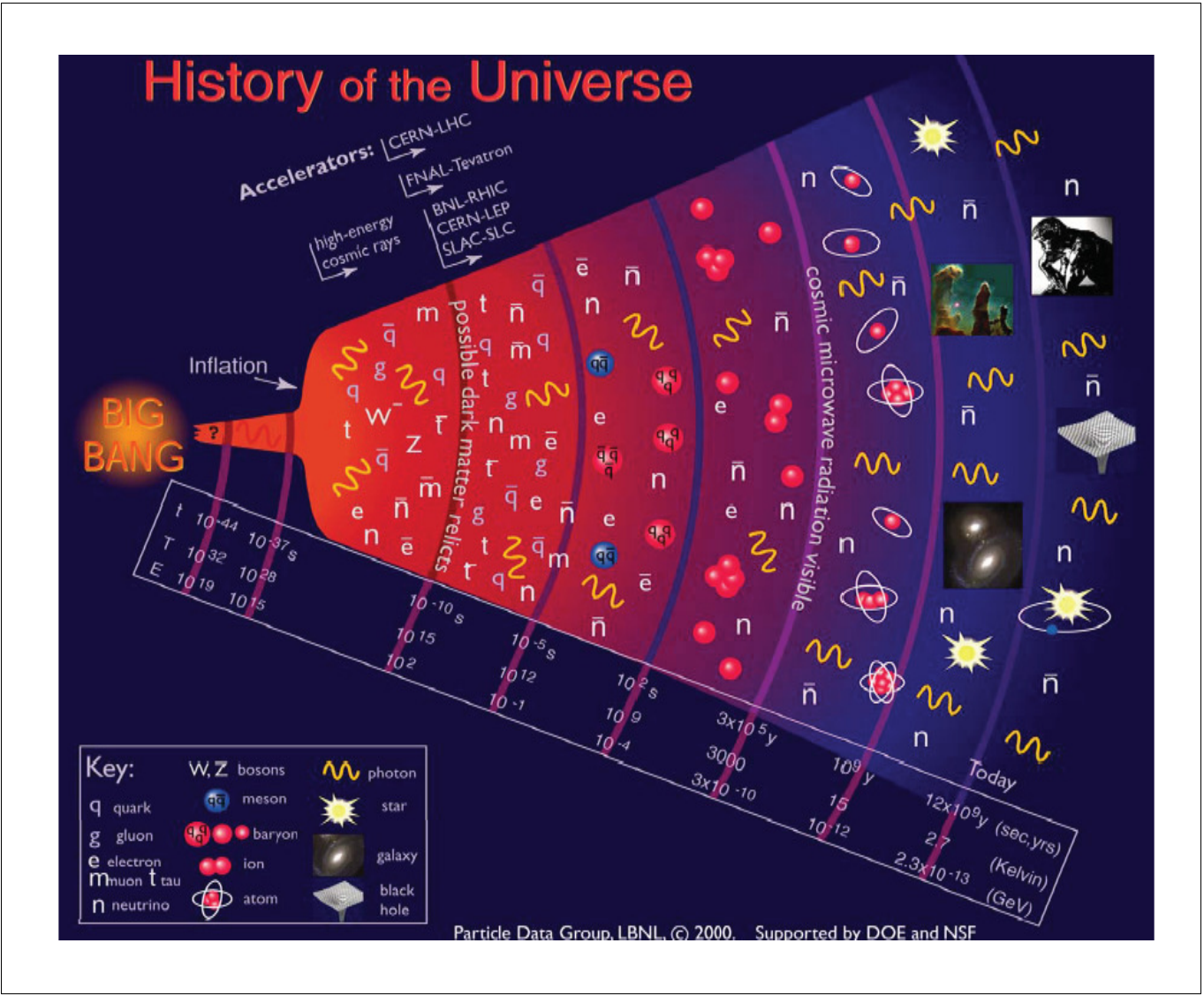}}
\end{center}
\caption{Big Bang Model timing including accelerator energies for $ E=\mu $ \cite{Timing555}. }
\label{Time-accelerator}
\end{figure}

The real experiment was conducted by nature itself during the creation of the universe. The entire scale and time 
dependence must be stored in the physical laws of the SM and BBM models. The elementary particles are created in this process.
For this reason, we assume that the same thing happens with the formation of FP's. 
Following this philosophy, our empirical toy ansatz to a geometric microstructure of FP's 
assumes that the evolution of the size of FP's from zero to the present 
size is a result of the early history of the universe, which is stored in the geometric expansion 
of the FP's, or that the geometric size of FP's is a result of the 
temporal development of the history of the very young universe.

\subsubsection{The experimental paradox to measure small distances}
 
 The experimental paradox of measuring small distances is well documented in the $\EEGG$ reaction. 
 It is not enough to simply increase the test energy $ Q=E_{CM} $  of the experiment to measure 
 smaller distances. In parallel, the physical parameters that depend on the energy 
 scale $ \mu $ change the conditions for the entire experiment.
 In this sense, the experiment is a race between the test size and the scale $ \mu $, 
 which defines the conditions for the entire experiment.
 
 We would like to discuss this paradox in more detail in a ``Gedanken'' experiment. 
 Starting with the $\EEGG$ reaction at the LEP, the $E_{CM}$ energy is approximately 200 GeV.
 This is above the production threshold for W boson pairs of $m_{W}^{\pm }=91.1876\pm 0.0021$ GeV.
As discussed in the chapter on Standard Model (SM) Eq. 9, the ECM energy according to 
Fig. {\ref{Running-coupling} and SM Eq. 9 is on a scale of approximately $ \mu = 200$ GeV.
Considering the high statistics of the $ \EEGG $ reaction, this results in a test size of approximately $ 10^{-18} $ cm, 
measured in reference \cite{POINT-PAPER}. As analysed in chapter SM, this scale yields three FP 
families (see Table 1 and Fig. 2) and three interactions (see Fig. 4 and Fig. 5). 
It is possible to create $Z^{0}$ and $W^{\pm }$ bosons. According to Eq.9 in the chapter SM model, 
the parameter set does not change with increasing test energy up to the Grand Unification Scale of about 1015 GeV.
At this scale, the three interactions unify at different energies or at a single energy, depending 
on the Standard Model or SUSY. Beyond this scale, up to the Planck scale, only one interaction remains. 
In this Era, massive superheavy particles are being discussed. These particles would play an 
important role in proton decay \cite{PROTON-decay}.

Beyond the Planck scale of $10^{19}$ GeV or $10^{-13}$ cm test size, quantised gravitation must be taken into 
account. In 1955, John Wheeler \cite{Weeler} devised a spacetime foam beyond this scale.
The foam is supposedly the foundations of the fabric of the universe \cite{Weeler1}. At this scale, the uncertainty principle 
allows for the creation and annihilation of particles without violating conservation laws. 
Space shrinks, and the energy of the virtual particles increases.
According Einstein's theory of general relativity curves energy spacetime,
This suggests that the energy of the fluctuations would be large enough to cause significant 
deviations from the smooth spacetime observed at larger scales.
These fluctuations could serve as a candidate for the seeds of the primordial perturbations in the cosmic inflation 
because the Large-Scale Structure of the universe is assumed to form a foam as a result of the growth 
of initially small density perturbations due to gravitational instability \cite{Instability}. 
The character of this foam is still being debated. \cite{Condenstate}. One possibility that the foam is a condensate 
was discussed by B.L.Hu \cite{Condenstate1}. If this is true, the gedanken experiment will 
detect a condensate between $  10^{-33} $ cm and zero cm.

Up to this condition of the universe in the experiment, we used exclusively the SM model, which defines the test 
size only implicitly via the uncertainty principle. The test size of our experiment is 
explicitly represented in the BBM model, especially at these very high scales.
In this model, the horizon distance or size of the universe in the vicinity the Planck scale clearly defines 
an upper limit for the test size of our gedanken experiment, provided that the test size is to be causally linked.
If we use the common scale link between SM and BBM, as discussed in SM Table 3, 
it is possible to repeat the gedanken experiment in the BBM model.
For the discussion, it is useful to repeat the experiment, starting in  the vicinity of the Planck Scale down to the 
scale $\mu$= 200 GeV, and to use the time sequences of the BBM model  discussed 
in the chapter ``Cosmology of the Early Universe'', in particular ``Fig. 2: The eight intervals of the Big Bang model''. 
If time-reversal invariance holds from the Planck size down to test size zero, we would like to
 emphasise that any experimental physicist will unequivocally measure a superfluid condensate 
 in this Era. Between the Planck Era and the GUT Era, the horizon distance, or rather the size of 
 the universe, increases, the experiment is dominated by only one interaction, and mass already exists.
The gedanken experiment will therefore measure one interaction and mass of superheavy 
particles. On the GUT scale, the size of the universe is sufficient large that three interactions 
with three types of charge get suddenly generated.
In the following Electroweak Era, the size of the universe will be large enough to accommodate FP's. 
The gedanken experiment will measure three FP families, including three interactions discussed in the SM model.
We will end our gedanken experiment at this Scale.

If time reversal invariance holds for the SM and BBM models from the vicinity of the Planck Era to 
the present Era, the universe can remember how it was created.
We would like to discuss two scenarios. First, we assume that the Planck-Era condensate exists 
only in the Planck-Era. Second, we assume that the condensate exists beyond the Planck-Era 
and focus particularly on  geometrical extent of the FP's.

\begin{itemize}
\item In the first scenario, we assume that the Planck-Era condensate exists only in the Planck-scale Era 
 and disappears once the universe exceeds this scale. In the GUT Era, it is replaced by superheavy 
 particles of the size s of the universe $ 0\leq s\leq s_{GUT} $.  
 After the GUT Era, the regime of three interactions dominates the size ``s'' of the universe $  0\leq s\leq s_{today} $. 
 Condensates or superheavy particles, as well as phase transition energies from the different Era's, 
 no longer exist on this scale.The parameters that define the different Eras act precisely at the energy in which the universe 
 passes through the adjacent scale $\mu$ of the respective phase transition.
 
 \item  In the second scenario, we assume that the condensates of the gravitationally 
 dominated Planck Era still exist after the universe has undergone the phase 
 transition from the Planck to the GUT Era. 
  Primordial inhomogeneities caused by quantum fluctuations in the Planck Era can propagate 
  into the GUT Era. As the universe expands to the GUT scale, the horizon distance of the 
  condensate increases, but remains within the GUT size.
  In the GUT Era, the primordial inhomogeneities form onion-like aggregates with an inner condensate core and an outer mass shell.
 Hypothetical super-heave neutrals are discussed in this Era.  
  These aggregates would be a candidate for the seeds of inflation to generate scalar density 
  fluctuations, as discussed in the chapter ``Key Events in the Life of the Universe''.
  The conditions discussed so far, including superheavy neutral particles, 
  persist even after passing the GUT scale within the size of the universe on that scale.
  According to the GUT scale, these aggregates could serve as kernels, adding three types of 
  charge (color, electromagnetic, and weak). The three neighbouring interactions - strong, 
  electromagnetic, and weak - become evident. Especially when the strong force becomes evident, 
  this appears to be the cause of inflation.  
  After the inflation of the universe, the horizon distance is significantly larger than any possible 
  elementary particle created and any relic size of the Planck and GUT scales. With increasing 
  time t after the Big Bang and decreasing temperature T, a quark-gluon plasma is formed.
  Protons, neutrons, electrons, and neutrinos are generated in a high-temperature plasma. Ultimately, the 
  universe follows the regime of Eq. 3 to 11 described in the chapter ``Primordial Nucleosynthesis''. 
  From a high-temperature plasma, FP's and ultimately atoms are produced and stabilised.
  In this scenario, the geometric structure of a Fundamental Particle is the direct consequence of the 
  evolution of primordial inhomogeneities from the Planck Era to the present Era. A relic of the 
  Planck-Era condensate, the mass shell of the GUT Era, and subsequent charges, as well as 
  remnants of the Planck and GUT scales, should still exist today. According to our scenario, 
  this structure should be present in every Fundamental Particle .
\end{itemize}

In the first scenario, no information about the evolution of the universe would be available today. 
The conditions would be directly linked to the scale energy and would occur in time t after the Big Bang. 

In the second scenario, the Planck scale at $ 10^{19} $ GeV or the Planck length $  l_{PL}=10^{-33} $ cm, 
is compared with the experimental limits of the FP size in Reference \cite{POINT-PAPER} 
of about $ 10^{-18} $ cm. It is obvious that FP's of this size are not created at this scale.
The FP's are 15 orders of magnitude larger. However, this scenario traces a possibility 
to describe the internal structure of an FP from the Planck scale to the present scale. 
Therefore, we will use the second scenario for the further discussion.

\subsubsection{The geometrical approach of the ETAMFP model.}

Following the second scenario, the geometric microstructure of an fundamental particle is defined 
by the historical development of the SM and BBM models.  Each FP particle stores every energy 
state in size and coupling that the Universe has passed through during its evolution, 
as shown in chapter SM Fig. 5: ``The variation of $ \alpha_{i}$ with the scale 
$ \mu (GeV)$ for SM and SUSY models''. In such a case, this microstructure can 
be directly derived from the chapter SM model Fig. 5 and the BBM model just 
under discussion ``Big Bang Model timing'' Fig.{\ref{Timing-BP}. 
Based on these two figures, Fig. {\ref{Timing-FP} shows the first approach of a schematic 
time development compared to SM and BBM theory of a geometrically extended FP in the 
ETAMFP model.

\begin{figure}[htbp]
\begin{center}
\rotatebox{00}{
 \includegraphics[width=16.0cm,height=12.0cm]{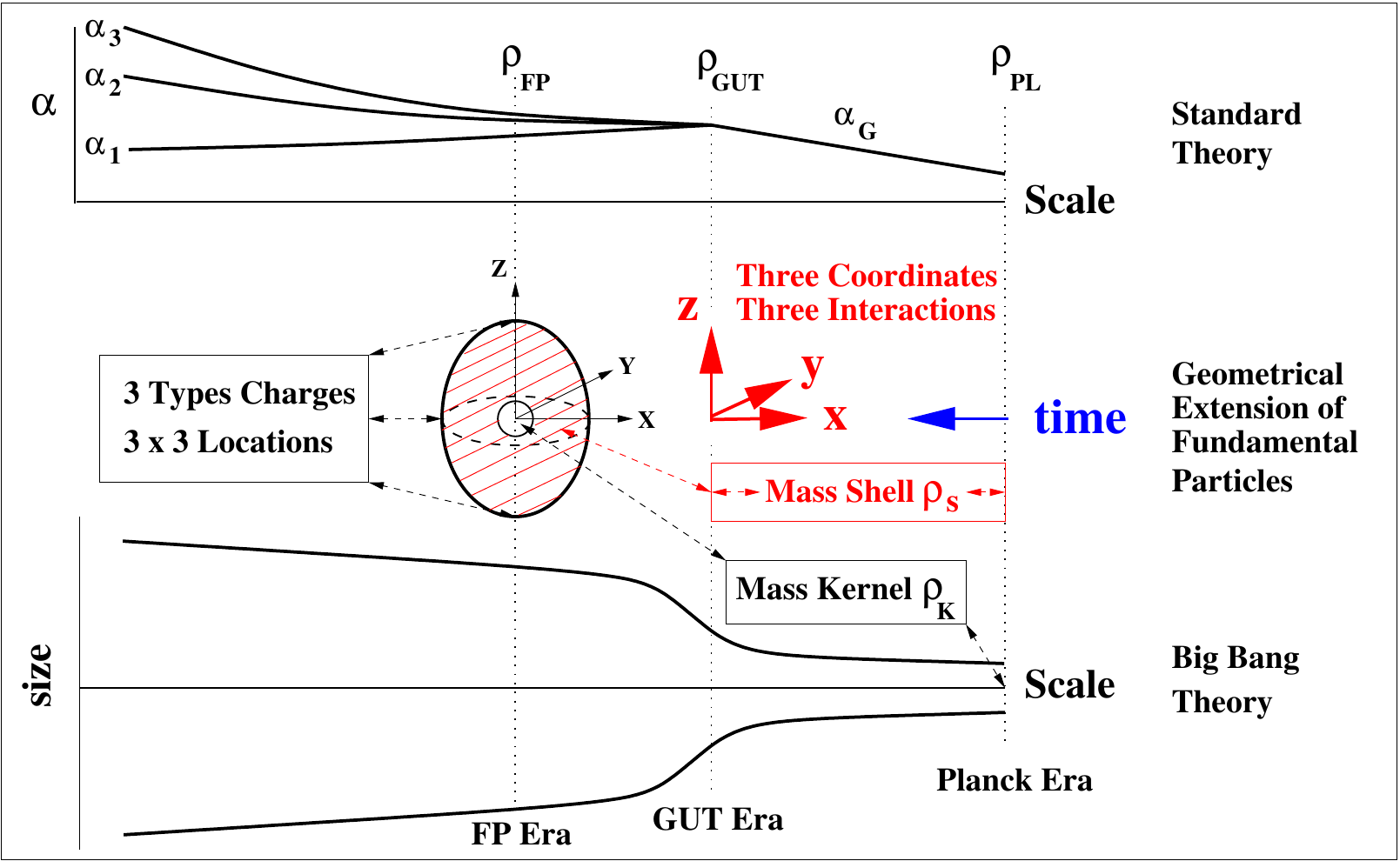}}
\end{center}
\caption{Schematic development of geometrical extended FP in the ETAMPF model. }
\label{Timing-FP}
\end{figure}

The currently known mass of the universe consists of baryonic matter $ \Omega _{B}=0.044\pm 0.004$, 
dark matter $ \Omega _{DM}=0.22\pm 0.04 $, and dark energy $ \Omega_{DE} =0.7 $. 
Since all these contributions originate from a single source, the discussed procedure 
should be generally valid for all three contributions. In the following discussion, we have 
focused only on baryonic matter, since this is where the FP's are located and describe 
Fig. {\ref{Timing-FP}}  from the middle part, upper part and lower part.

In the middle part of Figure {\ref{Timing-FP}}, we describe the historical process from time 
$t_{Planck}=5.39\times 10^{-44}$ s of the Planck Era to the present day in 
terms of the mass density, size, and charge distribution that a point-like observer would 
measure after the creation of an elementary particle. 
The status of the FP's in the FP Era of approximately 200 GeV, the CM energy of the CERN collider, is shown.
It is an object with a non-rotating mass kernel with a density of $ \rho _{k} $, which is
the heritage of the strongly attenuated Planck density $\rho _{PL} $.
The kernel carries mass and the three charges color C, electric Q and weak $ T_{3} $. 
 The size of the kernel radius $ r_{k} $ is the relic of the Planck radius in 
the course of the temporal evolution of the universe. 
The kernel ends at a domain wall that separates it from the rotating mass shell (Spin) of the GUT Era. 
The domain wall has a density of $\varrho_{s}$ and an outer radius of $r_{0}$. 
 The mass density $ \varrho_{s} $ is the heritage of the strongly attenuated GUT density 
 $ \varrho_{GUT} $, and the quantity $ r_{0} $ is the radius on the GUT scale expanded by 
 the temporal evolution of the universe. The positioning of the three different charges at nine locations 
 are possible. We could place three electric charges Q on the x-coordinate, 
three weak charges $ T_{3} $ on the y-coordinate, and three 
colours C on the z-coordinate.

The upper part schematically shows the scale dependence of the Running Coupling Constant, of the coupling 
$ \alpha _{3} $, $ \alpha _{2}$ and $ \alpha _{1}$ as a function of the scale of the SM 
Fig. {\ref{Running-coupling}. See also SM model chapter Fig.3. 

In the lower part, we relate the temporal evolution of the Big Bang theory from 
Fig.{\ref{Timing-BP} to the SM model with the corresponding energy scales 
of the BBM model Planck Era, the GUT Era, and the FP Era.

To discuss Fig. {\ref{Timing-FP} in more detail, we introduce a point-like observer 
that moves step by step from the Big Bang time t = 0 to the current data. 

In the first step, the observer begins to examine the details of the universe from the 
Big Bang time t = 0 to the Planck scale time $ t_{Planck}\approx 5.4^{-44} $ sec. 
He enters an Era about which the actual knowledge is limited and all assumptions 
must be considered speculative. If a four-dimensional universe 
already exists at time t = 0, a volume explosion is the starting point.
All different geometric points expand homogeneously \cite{WMAP222} in four-dimensional space at time t = 0.
The time t = 0 is transferred to these geometric points through something like ``Quantum Entanglement'', since 
a transition through electromagnetic interaction is not possible, as it is unavailable at that time.
 A completely homogeneous universe would contain NO galaxies, stars, or Fundamental Particles.
The completely homogeneous universe must be overlaid by a small ``Quantum Fluctuation'' which are 
responsible for our existing baryon universe. 
As discussed, the currently known mass of the universe consists of baryonic matter $ \Omega _{B}=0.044\pm 0.004$, 
dark matter $ \Omega _{DM}=0.22\pm 0.04 $, and dark energy $ \Omega_{DE} =0.7 $ \cite{WMAP222} .
The observer focuses his interest on the universe that 
emerged from these quantum fluctuations: galaxies, stars, and elementary particles.
In this case, the observer measures a quantised energy-mass condensate 
resulting from the volumetric explosion with a density of $\varrho_{PL}$. 
In this mass- and time-dominated condensate, the charges color, electricity, 
and weak charge are already present. The components of this condensate are in direct contact with each other, at a 
distance of less than $ r\leq 10^{-33} $ cm, and cannot interact, like in the field theory of the SM model.
The statistically distributed volume explosion does not lead to a total spin other than zero. 
The consequences of this assumption are discussed in the following chapter, 
``Relationships between Continuous Symmetries, Gravity, and Spin''.

Next, the observer crosses the Planck-scale domain boundary at the time $ t_{Planck}\approx 5.4^{-44} $ sec,  
energy $ Q_{Planck}=1.2\times 10^{19} $ GeV and Plank Length of  $ r_{Planck}\sim 1.6\times 10^{-33} $ cm 
and enters the GUT-scale Era at time $ t_{GUT}\approx 1.0^{-36} $ sec, energy $ Q_{GUT}=1.0\times 10^{16} $ GeV.
The universe passes through the inflationary Era approximately $ t_{Inflation}\approx 1.0^{-32} $ sec after the Big Bang.
This makes it difficult to calculate the size of the universe because it is expanding rapidly.
An approximation is $ r_{GUT}\sim 10 $ cm. In nature, transitions from one Era to the next are typically smooth.
Therefore, we expect the gravitational, mass-dominated regime to smoothly shift to the GUT scale.
A decreasing mass density is likely to emerge, forming a mass shell around the 
Planck-era mass core. This era is dominated by only one interaction.
If it were possible to describe this Era using a high-group field theory \cite{PROTON-decay},
a geometric structure consisting of two particles interacting via a gauge 
particle would be possible. If the total spin of the universe were zero, these
particles, which describe the mass shell, could possess a spin that leads to a total spin of zero.
All known spins, 1/2, 1, and zero, which add vectorially to the total spin zero, would be possible.
As a further consequence, the mass shell could rotate around the Planck kernel, with a mass density
of $\varrho_{s}$ in a geometric microscopic image.
 
Next, on its time-space journey, the observer crosses the domain boundary of the GUT Era at
$ Q=1.2\times 10^{19} $ GeV at a mass density of $ \varrho _{GUT}$. On the scale of the Grand 
Unification in the temporal evolution of the universe, the three interactions - Strong, 
Electromagnetic, and Weak - appear, accompanied by the three neighbouring types 
of charge: strong, electromagnetic, and weak. In the SM model they unify at three 
energies in the SUSY model as shown in Fig.{\ref{Timing-FP}} at one energy.
This is illustrated for the SUSY model in the upper part of the Fig.{\ref{Timing-FP}}. The tree interactions are 
significantly stronger as gravity in this Era and the size of the universe is large 
enough to allow three independent interactions in a three-dimensional space.
For the observer focusing on the spin-axis particle-like aggregates, which have 
been scaled up to this scale by Planck-Era quantum fluctuations, there are 
three possible geometric positions for the placement of charges on a coordinate 
axis: one position at the center and two on each surface of the object.
A total of $ 3\times 3=9 $ positions are possible.

According to time reversal invariance, the conclusion of the last step would be.
In the ETAMPF model, all of these particles would be the result 
of primordial quantum fluctuations already in the Planck Era.
Scalar density fluctuations, which trigger large-scale structures, 
would be particularly important. The geometric size of the 
universe will increase according to the comoving scale, but
the size of the FP's, stabilised by the four fundamental forces, will remain the same.
According to the ETAMPF model, the remnants of the Planck Era, including the Planck size, 
will also remain stable, and the GUT Era, including the GUT size, will remain stable and 
remain within the FP's.

In order to develop a ``Basic scheme of the extended fundamental particles in the ETAMFP 
model'', it is useful to transfer the details of Fig. {\ref{Timing-FP} into a new scheme 
of the fundamental  particles, which is shown in detail in Fig. {\ref{Basic-ETAMFP} 
containing the same information as Fig. {\ref{Timing-FP}.


\begin{figure}[htbp]
\begin{center}
\rotatebox{00}{
 \includegraphics[width=14.0cm,height=9.0cm]{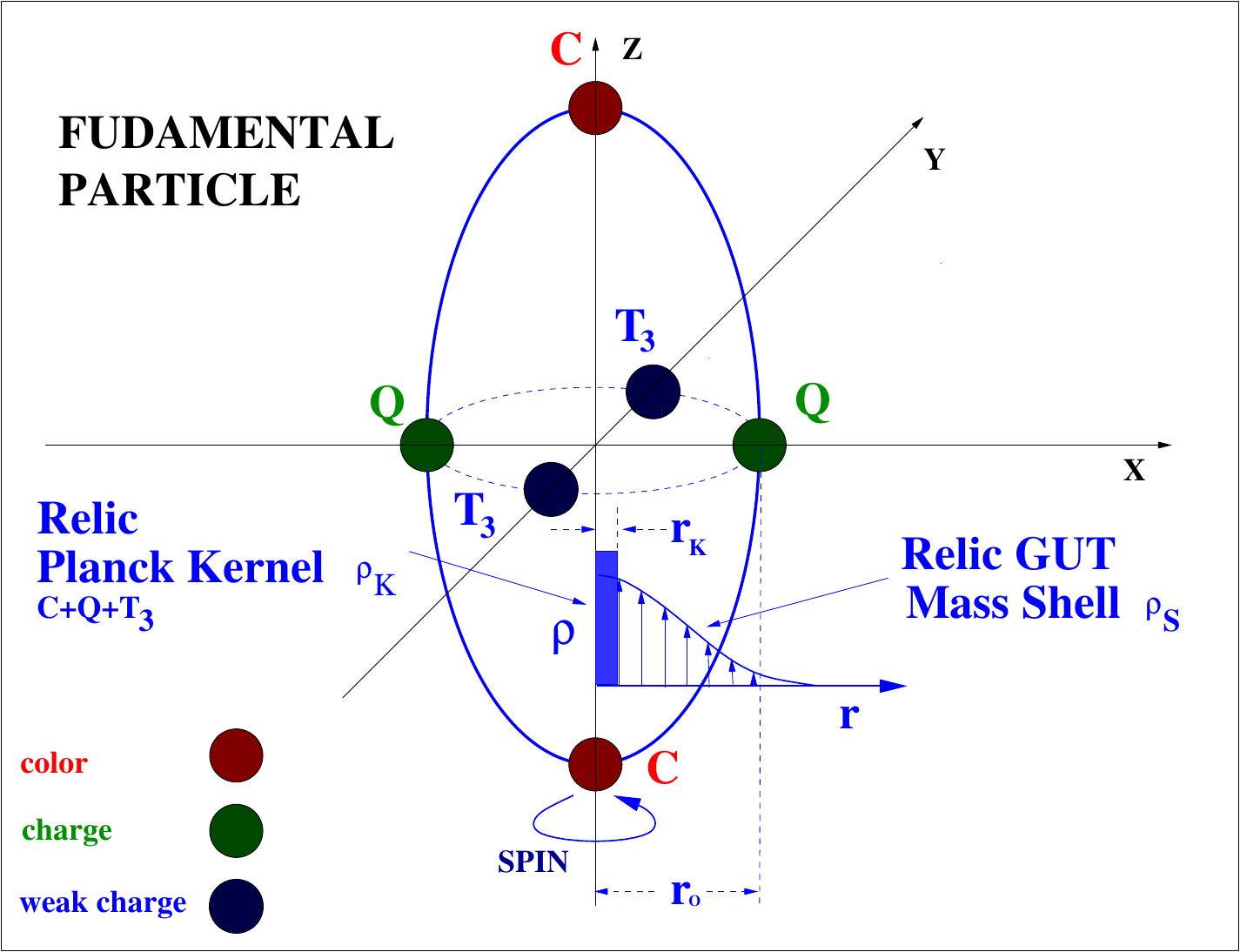}}
\end{center}
\caption{Basic scheme of extended fundamental particle in the ETAMFP model. }
\label{Basic-ETAMFP}
\end{figure}

The scheme includes three coordinates. According to Fig. {\ref{Timing-FP} 
in Fig. {\ref{Basic-ETAMFP}, three electric charges Q 
could be placed on the x-coordinate, three weak charges $ T_{3} $ on the y-coordinate, 
and three colours C on the z-coordinate.
A ``Relic Planck Kernel'' with density $ \varrho _{K} $ and radius $ r _{K} $ 
at the center of the scheme is introduced. 
This Planck Kernel is surrounded by a  ``Relic GUT Mass Shell'' with density
 $ \varrho _{S} $ and radius $ r _{0} $ and SPIN. 

The full picture of quantum field theory with FP's and gauge bosons would be formed up. 
In the direct contact regime at the GUT scale the displacement 
between the charges is tiny, repulsive forces could be superimposed by attractive forces. 
A superposition of similarly strong repulsive and attractive forces opens up the possibility 
that a stable geometric arrangement could exist at a certain distance from the center of the 
FP in which the different forces balance each other out. 
The force of the negative charge changes from repulsive to attractive depending on the 
distance of the same charge. Such an arrangement could form a highly stable energy minimum. 
For example, the extremely long lifetime of the electron.

\subsubsection{Links between continuous symmetries, gravitation and spin}

The discussion on ``Links between continuous symmetries, gravitation 
and spin'' investigate the question: ``Is the numerical link between 
four dimensions and four interactions a coincidence, or does this 
link arise from physical laws ? And is the spin in the SM merely 
a quantum number or the result of a geometrically extended object of FP's ?''
In the following discussion, we take a more classical ansatz to answering these questions.
 
To link the charge and mass distributions of the FP's in Fig.{\ref{Basic-ETAMFP} with the 
SM and BBM models, we used the running coupling constant of the SM in 
Fig.{\ref{Running-coupling} and the space-time evolution of the BBM in Fig.{\ref{Timing-BP}.
Embedded in the continuous symmetries of four-dimensional space-time, further 
information is required to link gravity and spin to the discussed microstructure of FP's.
We would like to consider four possible connections of the gravitational interaction 
with the time of four-dimensional spacetime and later discuss the consequences 
for the FP's with regard to the local isotropy of space with respect to rotation ( Spin ).

First at scales below the Grand Unification, the three interactions, strong, electromagnetic, 
and weak,  initially appear to be clearly separated, since the geometric size of 
the universe is sufficient to distinguish them. These three interactions
are embedded in space, which is described, for example, by the Cartesian coordinates x, y, and z.
However, there are four coordinates and four interactions in total. The last possible coordinate is time.
In line with our discussion above, it would be useful to link time to gravitational interaction. 
This would be particularly important from GUT scale to Planck scale, where gravitational interaction 
dominates, but should also hold at the present scale.

Second we like to stress that an experiment which investigates the center of a SM model 
point like FP, will meet unambiguously the conditions at the Grand unification scale where 
the gravitation is not negligible any more and next the the Planck Scale where the gravitation 
get dominant.  At the Planck Scale the Planck conditions of mass, length, time and density of
 Eq.\ref{PlMass}, Eq.\ref{Pllenght}, Eq.\ref{Pltime} and Eq.\ref{Pltensity} of the gravitation 
 are absolute crucial.
 
\begin{equation}
m_{Pl}=\sqrt{\hbar c / G } \simeq 2 \times 10^{-5} g
\label{PlMass}
\end{equation}

\begin{equation}
l_{Pl}=\sqrt{\hbar G / c^{3} } \simeq 1.6 \times 10^{-33} cm
\label{Pllenght}
\end{equation}

\begin{equation}
t_{Pl}=l_{Pl}/c\simeq 5 \times 10^{-44} s
\label{Pltime}
\end{equation}

\begin{equation}
\rho_{Pl}=c^{5}/(\hbar G^{2}) \simeq 5 \times 10^{93} g/cm^{3}
\label{Pltensity}
\end{equation}

Under these conditions, a complete geometrically extended three-dimensional FP is not 
possible because the scale is too high or the test distance is too small.
However, if our assumption that the FP has accumulated the history of the SM and the BBM
in its geometric structure is correct, this implements that the inner part of the FP is dominated 
by gravity and the structure is predominantly one-dimensional.
Since the three-dimensional test distance is extremely small $ 10^{-33} $ cm
time would again be a good candidate for this dimension.

Third, the mass-energy equivalence $ E = m \times c^{2} $ proposed and interpreted 
by Einstein in 1905 as a general principle resulting from the relativistic symmetries of 
space and time also points to time as a connection with gravitation..

Fourth, the uncertainty principle $ \Delta E \times \Delta t \sim \hbar $ again links time to energy. 
The energy-mass equivalents again indicate a connection between time and gravity.
We use these four indications to link time to the gravitational interaction in the further discussion of this paper.

In the SM, the local isotropy of spacetime leads, via Noether's Theorem, to the conservation 
of angular momentum and ultimately to the spin of the FP's currently existing on the scale. 
In the SM, spin is a quantum number describing a property of the particle.
For the point particle of the SM, spin has no microscopic significance, since a rotating point 
with zero diameter is meaningless. If we drop the point particle condition and allow a 
geometrical extension of the FP's as in Fig.{\ref{Basic-ETAMFP}, rotation of the object is possible, 
and spin, magnetic moment, and electric dipole moment are naturally included  within the object. 
We know from experiments that, for example, the spin axis of an electron is antiparallel
to the axis of the electron magnetic moment.
For our extended object in Fig.{\ref{Basic-ETAMFP}, this implies that the charge rotates together 
with the center of mass in the same plane of rotation. The geometrical extension of the mass 
$ \varrho (r) $ at the center and the position of the charges at $ r_{0} $ generate a classical 
angular momentum, a magnetic moment, and an electric dipole moment.
As we discussed in the geometric approach, the question remains at what time  in 
the evolution of the universe spin appears? Spin is linked to the emergence of the mass 
of the FP, since in the microscopic picture, only a rotating mass like a gyroscope leads 
to a stable spin axis. In our ETAMFP model, the mass of the FP is a relic of the time between t = 0 and Planck 
time as the Planck kernel, and between Planck time and GUT time as the mass shell. 
It seems unlikely that the volume explosion generates spin before Planck time, since this 
process is statistically uniformly distributed over the entire explosion volume. 
This would mean that the universe has a total spin of zero. However, according to the 
Planck scale, the universe could already be geometrically sufficiently extended to 
form fine-sized particles that interact directly with each other like a  condensate, as in 
high-group field theory \cite{PROTON-decay} .
If a quantised spin arises in this case, two particles with spin interact with each other via a 
particle field. The spin of the particles involved would be well-defined.
If we assume, as in the SM model, that the particles have spin 1/2, the field particles 
must have spin 1 or zero, since only $ 2\times \frac{\vec{1}}{2}+\vec{1}=\vec{0} $ or 
$ 2\times \frac{\vec{1}}{2}+\vec{0}=\vec{0} $ result in zero.
In a self-interacting mode with two field particles, higher spins such as 2 are also possible, 
since $\vec{2}+\vec{2}=\vec{0}$. Following this discussion, it seems reasonable to assume 
that spin in the universe occurs in the time window between the Planck and GUT scales.

\subsubsection{Forces and stability}
\label{Forces and stability}

As discussed in the geometric approach, the dependence of the coupling constants
$ \alpha _{3}$, $ \alpha _{2}$, and $ \alpha _{1}$ as a function of the scale Q(GeV ) in
the case shown in Fig.{\ref{Running-coupling}  and the time-size dependence of the same scale in the BBM
in the case shown in Fig. {\ref{Timing-BP} contain information about which forces act on which scale and
how these forces depend on the particle size scale under investigation.

At the LEP scale, around 200 GeV, a clear concept of forces and FP's exists.
The SM contains the three strong, electromagnetic, and weak forces depicted in Fig.{\ref{Running-coupling} , 
which interact with three families of FP's, as shown in Chapter SM, Fig. 2 and Fig. 28.
Up to the grand unification scale, the forces are sorted by their strength: $\alpha_{3}>\alpha_{2}>\alpha_{1}$. 
At this scale, at approximately $10^{15}$ GeV, the three forces get unified into one force. 
The corresponding test distance in our discussed gedanken experiment is so small at this scale 
that the Standard Model's concept of two geometrically separated FP's interacting via a similarly 
geometrically separated gauge boson is questionable.
In this case, it seems reasonable to assume that the interaction is similar a direct contact term interaction,
 such as that used in the Lagrange model \cite{Direct-Contact}. Between the grand unification scale and the Planck scale, 
 only one force exists; above the Planck scale, gravity dominates.
From the experiment the known FP's are in particular in the case of the electron with a mean 
life time $  \tau >4.6\times 10^{26} $ yr extreme stable. This experimental fact requires that the distance 
dependence of the forces acting in the FP's must lead to highly stable conditions.

The aim of the following discussion is to develop a general scenario based on the 
experimental facts and the scale-time-radius dependence of the SM and BBM 
models to imagine a possible radius dependence of the four known forces that 
could lead to such highly stable conditions.
Of particular interest is the Grand Unification Scale ``GUT Era''
in Fig. {\ref{Timing-FP}, since all four interacting forces are in direct contact.
In total, four different forces exist: the strong color force ``C'', the weak force, the electromagnetic force ``EM''
and the gravitational force ``$ \mathrm{G_{DS}} $''.
Classically, the electromagnetic force ``EM'' is plus infinity at r = 0, and the gravitational force
 ``$\mathrm{G_{DS}}$'' is negative infinity. Forces that are infinite at r = 0 cannot form a 
 stable state like the electron.
 
Only forces that change from a positive to a negative gradient with increasing distance from r = 0 can enable stable states.
One of such force is the de Sitter force $ \mathrm{G_{DS}}$ \cite{Desiter1}. This force is negative 
at large distances, passes through ZERO at a distance of $ \mathrm{r_{P}} $, and is positive at r = 0.

In Fig.{\ref{Forces1}, the relationship between the distance behaviour of the de Sitter force 
$ \mathrm{G_{DS}}$ (Red line), the electromagnetic force $ \mathrm{EM}$ (Green line), the 
classical gravitational force $ \mathrm{G_{SS}}$ (dashed blue line), the strong color force
$ \mathrm{C}$ (Magenta), and possible superpositions of the forces $ \mathrm{G_{DS}+EM} $ 
and $ \mathrm{G_{DS}+C} $ (Black line) are presented in an artistic overall view.

The De Sitter  force $ \mathrm{G_{DS}}$ is well documented, 
electromagnetic force $ \mathrm{EM}$  and strong color force 
$ \mathrm{C}$ distance behaviour is speculative connected to the De Sitter  force.

In detail, if we start with a radius r = 0 at the center of the FP, it is necessary to first discuss the 
radius dependence of gravity at distances close to the Planck scale $ \mathrm{r_{P}} $ 
and the Grand Unification scale $ \mathrm{r_{GU}}$.
If we trace the classical gravitational force $\mathrm{G_{SS}}$ as a function of the 
distance between two masses in a Schwarzschild vacuum \cite{Scharzschild}, as shown in Fig. {\ref{Forces1} 
(Blue dashed line), the force f1 is infinite at r = 0. Such behaviour does not lead to a stable 
mass particle, which, according to our experimental findings, is required for a FP, for example, an electron.

\begin{figure}[htbp]
\begin{center}
\rotatebox{00}{
 \includegraphics[width=14.0cm,height=9.0cm]{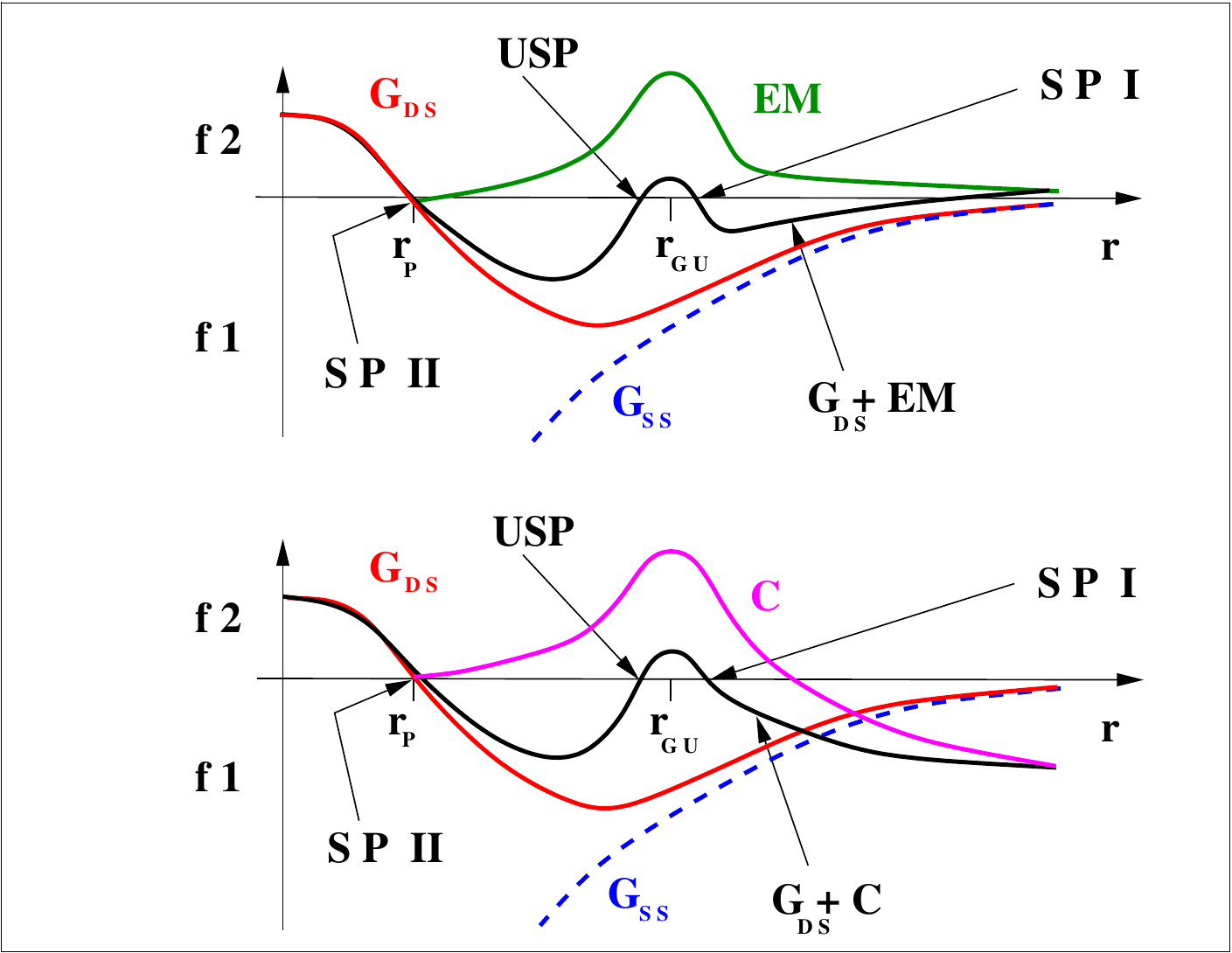}}
\end{center}
\caption{ Artistic scenario of the dependence off the forces of SM and BBM model close to
              the Planck and Grand unification scale. Upper part unification of electromagnetic ( EM ) 
              force with gravitational force  $\mathrm{G_{DS}}$ and lower part unification of color force ( C ) with $\mathrm{G_{DS}}$ }
\label{Forces1}
\end{figure} 
  
 The mass core is only stable if the attractive Schwarzschild force f1 transforms into a repulsive 
 force f2 in the vicinity of $\mathrm{r_{P}}$, similar to a de Sitter \cite{Desiter1} behaviour 
 $\mathrm{G_{DS}}$ (red solid line in Fig. {\ref{Forces1}).
 In such a case, in our direct contact scenario, a repulsive force leads 
 to a repulsive pressure, which is balanced by an attractive pressure generated by the 
 Schwarzschild force. A possible superpositions of the forces $ \mathrm{G_{DS}+EM} $ 
and $ \mathrm{G_{DS}+C} $ (Black line)  presented in Fig.{\ref{Forces1} open such a possibility.
 Two stable positions Point SPII and SPII appear.
 
In such a case, in our direct contact scenario, a repulsive force leads to a repulsive 
pressure, which is balanced by an attractive pressure generated by the Schwarzschild 
force. A possible superposition of the forces $ \mathrm{G_{DS}+EM} $ and 
$ \mathrm{G_{DS}+C} $ (black line), depicted in Fig. {\ref{Forces1}, opens up such a possibility.
Two stable positions emerge, points SPI and SPII but also unstable positions USP.
 
 
 \begin{figure}[htbp]
\begin{center}
\rotatebox{00}{
 \includegraphics[width=12.0cm,height=9.0cm]{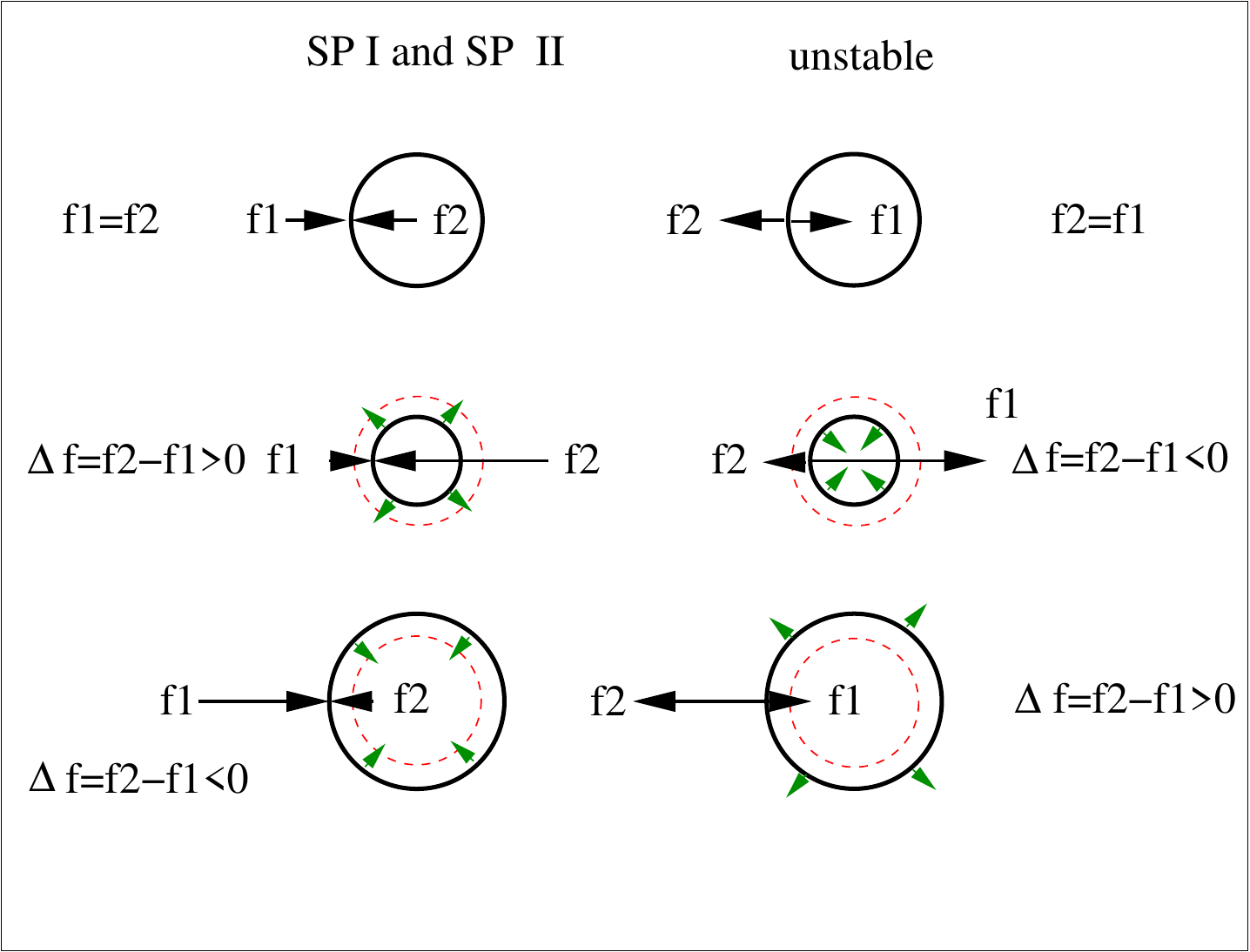}}
\end{center}
\caption{Stability conditions for a mass core of FP's. Stable conditions left side and unstable conditions right side. }
\label{Forces22}
\end{figure}

In Fig. {\ref{Forces22} three possibilities for creating stable and unstable positions are discussed in detail.
In the first row $f1=f2$ or $f2=f1$ in Fig. {\ref{Forces22}, the direction of the forces against each other leads to the stable state
SP I and SP II; conversely, the direction of the forces opposite to each other leads to the unstable state USP.
The second and third row describe the dynamics of the forces.
In row 2, when the stable radius r (black circle) decreases, a repulsive force $ \Delta f=f2-f1>0 $ pushes the mass particle
 back to the stable position (red circle, dashed line). The reverse situation, shown on the right, $ \Delta f=f2-f1< 0 $, 
 leads to an unstable situation.
In contrast, the stable radius r (black circle) increases, a repulsive force $ \Delta f=f2-f1<0$ pushes the mass 
core back to the stable position ( Red circle, dashed line). The reverse situation shown on the right
$ \Delta f=f2-f1 > 0 $ leads to an unstable situation.
In summary, only a vacuum such as the de Sitter vacuum would lead to a stable mass core of the FP's. 

Based on the temporal behaviour of the coupling constant $ \alpha $ in the Standard Theory, the function of the universe size
in the Big Bang Theory, and the evolution of the various forces in radius/time of these theories, shown in Fig. {\ref{Timing-FP}, 
Fig. {\ref{Forces1}, and Fig. {\ref{Forces22}, it is possible to discuss in more detail the unification of the electromagnetic force EM,
the color force C, and the weak force at the radius of the grand unification $ \mathrm{r_{GU} } $, as well as the involvement of gravity
between the Planck and grand unification scales.

Between the radius $ r_{P}<r<r_{GU} $ the SM and BBM predicts one force including one coupling constant $  \alpha _{G} $ and for $ r>r_{GU} $ 
the three discussed forces appear including the three charges color, electromagnetic and weak. The consequence of 
this prediction is that the charges should be geometrical located at the vicinity of the radius $ r_{GU} $. For simplicity we 
assume in the further discussion the charge themselves have a point-like geometrical character.

The electromagnetic interaction EM is repulsive at distances $ r>>r_{GUT} $, as shown in Fig. {\ref{Forces1} (green solid line).
We assume that the force decreases to zero between $ r_{P}<r<r_{GU} $ or is very small at $ r_{P} $. 
Combining the EM force with gravity $ G_{DS}+EM $, as shown in the black solid line at the top of 
Fig. {\ref{Forces1}, a stable point SPI is formed, which appears like gravity's SPII. The repulsive internal 
nuclear forces balance the attractive forces $ r>r_{GU} $, as explained in Fig. {\ref{Forces1} and Fig. {\ref{Forces22}.

As discussed, an electromagnetic charge located at SPI would be stabilised geometrical very similar as the gravitational 
mass core at $ r<r_{P} $ shown in Fig.{\ref{Forces22}. If the stable radius r shown as black circle in Fig.{\ref{Forces22} get smaller a 
repulsive force $ \Delta f=f2-f1>0 $  will push the EM charge back in the stable position ( Middle left red circle broken line in Fig.{\ref{Forces22} ).
In the contrary if the stable radius increases the radius get pushed back from an attractive force $ \Delta f=f2-f1<0$ 
( Lower broken red circle in Fig.{\ref{Forces22} ). In conclusion a De Sitter like 
behaviour of the EM force would lead to a stable geometrical position of an EM charge at point  SPI .

If we choose Cartesian coordinates for the FP's, the three coordinates x, y, and z exist. In Fig. {\ref{Timing-FP}, 
each axis has three stable position center $ x < r_{P} $ SP II , for the EM force, at the (+x-axis) $ r_{GU} $ SP1 
and at the (-x-axis) $ r_{GU} $ SP1.These charges at a distance greater than $ 2\times r_{GU} $ are stabilised by the repulsive force $ GDS + EM $ $ r>>r_{GU} $. At approximately $r_{P}$, the SM and BBM predict that gravity dominates all other forces.
Even a very weak attractive coupling between mass and the three types of charge would
establish a stable position for these charges at the geometric position of maximum mass concentration.
According to Fig. {\ref{Timing-FP}} and Fig. {\ref{Forces1}}, this would be directly at the center of the FP's..

The color force C and the weak force act attractively at $r>>r_{GU}$. If we again assume a change in their character
to a repulsive force near $r_{GU}$ and set it to zero or very small at $r=r_{P}$, the unification with gravity
$G_{DS}+C$, shown as a black line in the lower part of Fig. {\ref{Forces1}, yields a stable point SPI.
If we place the color charge C on the (+ z-axis), an opposite stable state would prevail on the (- z-axis), 
since the color force for $ r>r_{GU} $ would repulsive over a certain distance to stabilise the second 
color charge, which is opposite to the first charge. The situation at $ r=r_{P} $ would be the same as in the EM case.
The scenario for the weak force would be similar on the last free y-axis and is not discussed further.

The crossing point USP in Fig. {\ref{Forces1} leads to a geometrically unstable state, as shown in Fig. {\ref{Forces22} on the right.
The force $ \Delta f=f2-f1<0 $ in the middle right area leads to a collapse to the SPII position, while the force $ \Delta f=f2-f1>0 $
expands the USP to the $ SPI $ position in the lower right area.

Concluding the discussion of the behaviour of the four forces, it can be seen that the above scenario generates
six stable geometric positions for charges at approximately $r=r_{GU}$, plus one position 
for each charge at the center of the FP's, including a stable mass core. This adds three position for 
the fundamental charges. All together nine position. One more position is hidden in the coupling 
at the ``GUT Era'' between Gravity $ \alpha _{G} $ and $ \alpha $ .
The three Cartesian coordinate axes x, y, and z, the three interactions $ \alpha _{3} $, $ \alpha _{2} $, 
and $ \alpha _{1} $, as well as the three possible fundamental charges, can be 
explained by simple geometric reasons. But interaction number four the gravity is also used in the 
overall scheme in Fig. {\ref{Timing-FP}}. This suggests that after the line element 
$ \left | d_{s}\right |^{2} $, another coordinate exists near x, y, z. The K. Schwarzschild metric 
\cite{lambda1} and a neutral, spherically symmetric black hole with a regular de Sitter 
interior \cite{lambda} time t would be good candidate.

\subsubsection{The flashing vacuum}
\label{The flashing vacuum}

The ETAMFP model in Fig. {\ref{Basic-ETAMFP} involves a charge orbiting a kernel.. 
An accelerated charge emits electromagnetic radiation.
Such an object loses energy and is unstable because it has a limited lifetime.
For example, if we consider an electron with a lifetime of $ \tau > 4.6 \times 10^{26} $ 
year, it is obvious that a radiation-free path, as postulated in the Bohr model \cite{Bohr1} , must exist. 
A well-known example of a stable, geometrically extended object is the hydrogen 
atom in Fig. {\ref{Hydrogen1}}.

\newpage

\begin{figure}[htbp]
\begin{center}
\rotatebox{00}{
 \includegraphics[width=14.0cm,height=14.0cm]{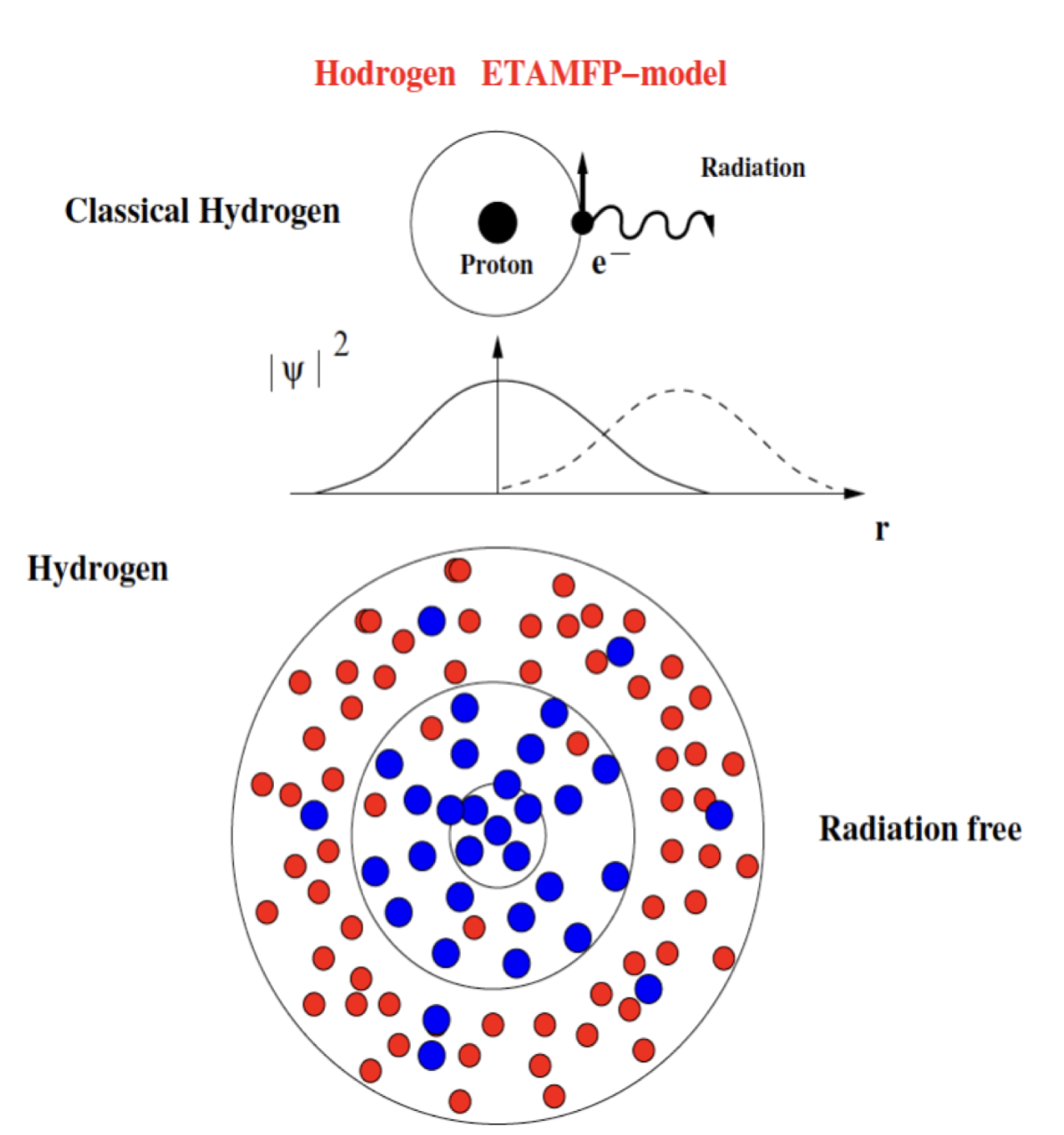}}
\end{center}
\caption{The Hydrogen atom. }
\label{Hydrogen1}
\end{figure}
 
 The upper part of Fig. {\ref{Hydrogen1}} shows that the probability distribution of the SM 
 model $ \left| \psi \right|^{2}=f(r) $ of an electron in the s-state peaks at the center of the 
 proton, but not in a delta function at r = 0, similar to how the semi-stable p-state peaks 
 at a distance $ r>0 $ outside the center of the proton.
 The lower part of the figure {\ref{Hydrogen1}} shows the electron (Red) orbiting the center of the hydrogen 
 atom (Black) in a radiation-free trajectory. It appears and disappears statistically at a certain distance 
 $ r $ from the center of the hydrogen atom, simultaneously generating, with these flashes, a free-orbiting 
 radiation around the center of the hydrogen atom.
 
 To interpret this statistical appearance and disappearance, a second state would be necessary that 
 allows the electron to disappear. A suitable candidate for this state would be the vacuum..
 The vacuum state \cite{Vacuum1} is the quantum state with the lowest possible energy. This state is 
 characterised by special properties. In perturbation theory, for example, the differential cross 
 section of QED \cite{POINT-PAPER} ( Equation 21 in this reference.)
is calculated in the $\EEGG$ reaction, taking into account radiation effects up to $ O(\alpha ^{3}) $. 
It contains fleeting electromagnetic waves and particles that suddenly appear and disappear again.  
The uncertainty principle in the form $ \Delta E\times \Delta t\geq \hbar $ implies that in a vacuum 
one or more particles with energy $ \Delta E $ can be created for a time $ t<\Delta $.
   
In a first scenario. In the vicinity of a black hole, the vacuum becomes unstable and loses energy through 
Hawking radiation \cite{Hawking1}. Vacuum fluctuations cause a particle-antiparticle pair 
to appear near the event horizon of a black hole. One particle of the pair falls 
into the black hole, while the other escapes. In this quantum tunnelling 
effect, one particle-antiparticle pair is screened one tunnels outside the event horizon.
 If the total energy is conserved and this particle escapes the black hole, this results in a loss of energy or mass. 
Hawking radiation occurs near a black hole with all the properties of a black hole, in particular a Schwarzschild 
horizon \cite{Scharzschild} and a singular gravitational potential. In this case, the particle pair is screened.
Finally, it is possible that the black hole evaporates.
   
 Another scenario would be that no particle-antiparticle pair is created, but rather a single particle 
 oscillates statistically back and forth between the real state and the vacuum state. In all of the 
 scenarios discussed, the vacuum has the potential to flash statistically from time to time at 
 any geometric location. This will also be possible in the vicinity of an elementary particle. An FP is not, as far as we know, 
 a black hole, but there are discussions about how an FP could be described by a de Sitter-Schwarzschild 
 geometry, a black hole whose singularity is replaced by a de Sitter core  of a fundamental 
 scale \cite{Irina66},\cite{ Dymnikova22} and \cite{ Dymnikova33 }.   
 If the gravitational potential is no longer singular, the possibility arises that
following the stability discussion in Fig. {\ref{Forces1}
one more vacuum property is possible. 
The vacuum fluctuates near an FP, similar to Hawking radiation, with the difference 
that one pair partner does not exit the FP, but only appears statistically at a certain 
location, while its counterpart remains in the vacuum, similar to a mirror particle in 
a Dirac sea \cite{Dirac22}, or oscillates between real and vacuum states.
 Since a vacuum exists in all models discussed so far  ( See chapter SM model, 
 Table 3 ``Possible connections between the Standard and Big Bang models''), 
 we introduce the flashing vacuum \cite{Gothe1}, as a working hypothesis 
 shown in Fig. {\ref{Hydrogen1}.

Applied to our ETAMFP model shown in Fig.{\ref{Basic-ETAMFP}}, the previous discussion
for the classical charge rotating at radius $r_{0}$ would implement an aggregation 
of statistically blinking charges located in the vicinity of $r_{0}$ with a specific distribution function.
The number of flashes per unit time, together with the distribution function, would define the radius
$ r_{0} $, assuming that the charge is geometrically much smaller than the FP. The integral summed 
over all flashes would define the total charge of the FP. If the flashes are collectively 
superimposed on a rotational velocity, this would be the underlying cause of the 
magnetic moment and the electric dipole moment. The discussion so far has focused
 on electric charge, but especially radiation effects in perturbation theory, also including strong, 
 electromagnetic, and weak interactions. These also implements the attached charges color, 
 electrical, and weak in their fluctuations. The vacuum properties are also of great 
 importance for the Higgs mechanism for generating mass for the FP's.
For this reason, it is very likely that the vacuum is also capable of flashing in a quantum matter mode. 
In our ETAMFP model, the statistical quantum flashes per unit time, together with a distribution function, 
would define the density distribution $\varrho(r)$ as a function of radius r.
The total integral would allow the calculation of the total mass of the FP, and similarly to how the charges 
would generate a superimposed rotational velocity of the flashes, an angular momentum would be 
generated, and the integral would generate the total spin. Here, too, the geometric dimension of such 
a mass quantum must be smaller than the size of an FP.
As an ad hoc hypothesis, we assume that the frequency of these quantum of flashes is high enough to 
be calculated using a classical ansatz. We will follow this approach in the next sections of this paper.

\subsubsection{Conclusion of the empirical toy ansatz  (ETAMFP) to a microstructure of a fundamental particle.}
\label{Conclusion ETAMPF}

We first discussed the experimental paradox of measuring very small distances. In the case of the 
point-like FP's of the Standard Model, it would be necessary to measure distances down to zero.
This unambiguously request infinite high test energies, the CM energy of the experiment will cross the GUT scale and Planck 
 scale. The logic consequence is, the whole physics conditions according to the energy scale of the 
 SM- or BBM-model agrees with  physics laws existing at this scales. 
 The experiment must be conducted below the Planck time of $ t_{Planck}=10^{-44} $ s and the 
 Planck length of $ l_{Planck}=10^{-33} $ cm. To reconstruct the same physical conditions for 
 a given scale $ \mu $ in the SM and BBM models, it is important that both models satisfy inverse time invariance.
 This is confirmed from experiment up to the LEP energies.  For our discussion we assume this is 
 correct up to Planck Era. The logical consequence of this fact is, the universe remembers how it was created.
 In our empirical toy ansatz ETAMFP on the microstructure of an fundamental  particle, we propose 
 that the FP's contain every energy state across its radius that the universe has passed through during its evolution.
 In this scenario, the geometric structure of a Fundamental Particle is the direct consequence of
 the evolution of primordial inhomogeneities from the Planck Era to the present day. 
 In every elementary particle, relics of the condensate from the Planck Era, the mass shell from 
 the GUT Era and subsequent charges, as well as information about the size of the Planck and GUT 
 scales, are still present today. The important new approach is that the ETAMFP model transforms the point-like FP's of the 
 SM into geometrically extended objects. It tracks the evolution of the universe in the SM and BBM models. 
 In particular, in the BBM model, it tracks primordial nucleosynthesis. The FP's are no longer points.
 
 Second, we have developed an approach for a geometrically extended fundamental particle. 
 We use the obvious proposal that the SM and the BBM describe the 
 same nature and use the common connection of the scale $ \mu $ between both models.
This links the scale dependence of the SM model couplings $ \alpha _{1} $, 
$ \alpha _{2} $, and $ \alpha _{3} $ to the time $ t $ after the Big Bang as shown in Fig.{\ref{Timing-FP}.
In this case, it is possible to construct the microstructure of the FP's, taking into 
account the conditions of the temporal evolution of the early Universe, directly 
from the SM and BBM models in three main steps.

In the time window $0<t<t_{Planck}$ Planck, there exists a condensate of mass and the three charges
color, electromagnetic, and weak, which are in direct contact with each other, including the geometric size of the
phase transition from the Planck to the GUT Era.

Between $ t_{Planck}<t<t_{GUT} $ , condensate formation by the force $ \alpha _{G} $ dominates the 
geometric size of the phase transition from the Planck Era to the GUT Era.

Finally, in the window $ t_{GUT}<t<t_{FP} $ the size of the universe is large enough,
the primordial inhomogeneities originating from the Planck Era are endowed with the three independent charges of color C,
electromagnetic Q, and weak $ T_{3} $, which corresponding to the three corresponding interactions $ \alpha $:
strong, electromagnetic, and weak. It is possible to construct a basic scheme for FUNDAMENTAL PARTICLES, 
which is shown in Fig. {\ref{Basic-ETAMFP}. The whole structure follows the inflation and ends at the FP Era 
as fundamental particles.

Third, we discussed links between continuous symmetries gravitation and spin.
After discussing four possible links of gravitational interaction with time in four-dimensional 
spacetime, we conclude that the three interactions, strong, electromagnetic and weak 
dominate in three-dimensional space and the fourth interaction gravity is linked to time.
Following the time course of the SM or BBM model, we assume that spin in the universe 
occurs in the time window between the Planck and GUT scales.

Fourth, we search for a relationship between the forces acting within a FP and its stability. 
Since the electron, for example, is a highly stable particle, the radius dependence of 
the forces in the electron must lead to a very stable state.
It turns out that stable conditions are only possible if the known forces, strong, electromagnetic, 
weak and gravitational forces change from repulsive to attractive or vice versa near or within the 
FP, with the strength of the gravitational force being similar or even dominant.
The discussion of the radius behaviour of the four forces shows that the above scenario 
generates six stable geometric positions for placing charges at approximately $ r\geq r_{GUT}$ 
plus one position for each charge at the center of the FP's, including a stable mass core. 
The geometric positions of the various charges can be explained by simple geometric 
reasons and an energy minimum of the FP mass.

Finally fifth, we investigate why the rotating, highly accelerated charge of the ETAMFP model 
does not radiate electromagnetic energy. From the electron's lifetime of $\tau >4.6\times 10^{26}$ 
year, we conclude that a similar free path length must exist as in the Bohr model.
We use the stable hydrogen Atom to obtain a microscopic picture of how an electron in the 
p-state can orbit the center of the Atom radiationlessly. We introduce the flashing vacuum 
as a candidate to allow the electron to appear and disappear randomly, forming the 
well-known probability distribution $ \left|\psi \right|^{2}=f(r) $.
We extend the vacuum properties defined in the SM and BBM models by the possibility that 
the vacuum fluctuates near an FP, similar to Hawking radiation, with the difference that one 
pair partner does not escape the FP but only statistically appears at a certain location, while 
its counterpart remains in the vacuum, similar to a mirror particle in a Dirac sea, or 
oscillates between real and vacuum states. As a working hypothesis, 
we assume that the frequency of the quantum flashes is large 
enough to be used for a classical ansatz of a radiation-free path length of the charges 
orbiting around the center of the FP.

All together, the geometric structure in our ETAMFP model is generic. Any extended stable or 
semistable structure arising from the primordial inhomogeneities or quantum 
fluctuations in the Planck Era is fundamental. We have summarised the most 
important structures discussed today. All particle-like structures discussed should 
be referred to as fundamental particles.

To simplify the discussion in the following chapters, we will follow the common 
practice of restricting the names of fundamental particles to fermions, bosons, and Higgs.

\subsection{Scheme of geometrical extended fundamental particles and anti-particles}
\label{Scheme of geometrical extended FB's } 

To confront  the ETAMFP model of a geometrically extended FP described in 
Fig. {\ref{Basic-ETAMFP}, we first test whether the model is able to sort all 
known fundamental  particles (Fermions), fundamental  particles (Anti-fermions) 
and interaction particles (Bosons) into a common scheme.
 
\subsubsection{The charges of fundamental particles}

Following the experiment of R. A. Millikan (1911) \cite{Millikan33}, we learned experimentally 
that most fundamental particles carry charges. The charges of the FP's of the 
four interactions, geometrically localise the location of the source of the interaction and the strength of the 
strong, electromagnetic, weak, and gravitational interactions. Details in are discussed in 
Chapter ``Physic beyond Standard Model Chapter 5'' in Fig. 4 (Interaction between fundamental  particles), 

\textbf { A) The charge of the strong, electromagnetic and weak interaction.}

To test the scheme, it is first necessary to summarise the charges of the Standard 
Theory SM at a low scale in the FP Era, as shown in Fig. {\ref{Timing-FP} 
and Fig. {\ref{Basic-ETAMFP}. The interactions are described by the 
charges of the strong interaction (red) (C) on the z-axis, the most fundamental charge
of the electromagnetic interaction (green) ($ Q \pm 1/3 $) on the x-axis, and the charge of the weak interaction
of the weak isospin (blue) ( T and $ T_{3} $) on the y-axis.
The positioning of the electromagnetic charge (green) ($ Q \pm 1/3 $) on the x- or y-axis 
is important to generate a magnetic moment when the spin axis is parallel to the z-axis.
The hyper-charge quantum numbers $ Y $ is a function of the 
electric charge and the weak isospin $ T_{3}$  according Eq.\ref{Charge1}.

\begin{equation}
Q=T_{3}+\frac{Y}{2}
\label{Charge1}
\end{equation}

Besides the colours all the charges of the fermions are summarised in Tab.~\ref{Hypercharge} 
of ref. \cite{PROTON-decay}.


\vspace{10.0mm}
\begin{table}
\caption{Hypercharge quantum numbers of Lepton and Quarks}
\label{Hypercharge}
\begin{center}
\begin{tabular}{||l|l|l|l|l|l|l|l|l|l||}                         \hline
$Lepton$   &$T$&$T_{3}$&$Q$&$Y$&$Quark$&$T$&$T_{3}$&$Q$&$Y$
                                                            \\  \hline
 $\nu_{e}$ &$1/2$&$1/2$&$0$&$-1$&$u_{L}$&$1/2$&$1/2$&$2/3$&$1/3$
                                                            \\  \hline
$e^{-}_{L}$&$1/2$&$-1/2$&$-1$&$-1$&$d_{L}$&$1/2$&$-1/2$&$-1/3$&$1/3$
                                                            \\  \hline
           &     &      &    &    &$u_{R}$&$0  $&$0   $&$2/3 $&$4/3$
                                                            \\  \hline
$e^{-}_{R}$&$0  $&$0   $&$-1$&$-2$&$d_{R}$&$0  $&$0   $&$-1/3$&$-2/3$
                                                            \\  \hline
\end{tabular}
\end{center}
\end{table}

\textbf { B) The pseudo charge of the gravitational interaction.} \\
\label{pseudo charge}

The model in Fig.{\ref{Basic-ETAMFP} describes in three dimensions the 
electromagnetic charge $ Q $ is placed on the x-axis, the weak charge $ T_{3} $ on 
the y-axis and the color charge $ C $ on the z-axis.
We know that the SM and BBM models use four dimensions, 
it is important to examine what type of charge belongs to the fourth dimension, the time.
After we discussed in the chapter ``Links between continuous symmetries gravitation and spin'', we 
reached the conclusion to link in this paper the time to the gravitational interaction. 
Two indications point to the mass as pseudo charge of the gravitational interaction.

First we consider the definition of a charge as the source of a force field in field theory.
The field surrounding a charge contains energy. If we examine the Einstein equations 
Eq.\ref{eq.7a} to Eq.\ref{eq.7d} and ignoring the cosmological term $ \Lambda $, 
energy will implement energy-stress tensors $ T_{\mu \nu}\neq 0 $. 
This also means that the curvature will not be zero.
As a consequence, according to the Einstein equation, there must be 
mass in the vicinity  of the charge. If we assume that 
a charge is the source of a force field and the Einstein equations 
are valid for our case, a charge cannot exist without mass. This fact is unique to all
known charges - coloured, electric, and weak. Consequently, elementary particles 
(fermions and bosons) of this type must also possess mass.
A special case is mass itself in the Einstein equations. If we extend the definition 
of charge as the source of a gravitational field to mass, mass takes on a dual 
significance. In our usual definition, mass behaves like mass in (gr), but is 
simultaneously a charge like color, electric, and weak charges.
This self-interacting phenomenon will equip every spatial 
energy accumulation of a field with mass, in particular the gamma and Higgs. 
The gamma is a hybrid because it is a spatial energy accumulation of electromagnetic 
character, it carries via $ E=m\times c^{2} $ induced mass and is electromagnetic. 
It is sensitive to gravitational interactions via its induced mass, and the gauge particle 
is sensitive to electromagnetic interactions. The Higgs boson, as a boson without spin
 or charge, carries only mass and is responsible for generating the mass of all 
 FB's  Under these circumstances, the Higgs boson would be an ideal candidate for 
 placement on the time axis. Detail information in chapter ``Physic beyond Standard Model Chapter 5''
Lagrangian Eq. 14 and Figure 7.
These considerations combine the four known interactions  strong, electromagnetic, 
weak and gravitational with the four charges colour, electric, weak and mass into 
one scheme and at the same time link the four interactions and charges with the 
four coordinates as discussed in the chapter ``Geometric Approach''.
,

Second is the mathematical structure of the forces $ F_{Q} $ between two charges 
$ Q_{1} $ and $ Q_{2} $ in the Coulomb law Eq.\ref{ChargeE},

\begin{equation}
F_{Q}=\frac{1}{4\pi \varepsilon }\frac{Q_{1}Q_{2}}{r^{2}}
\label{ChargeE}
\end{equation}

the force $ F_{M} $ between two magnetic poles $ p_{1} $ and $ p_{2} $
Eq.\ref{ChargeM}

\begin{equation}
F_{M}=\frac{1}{4\pi \mu }\frac{p_{1}p_{2}}{r^{2}}
\label{ChargeM}
\end{equation}

and the the gravitational force $ F_{G} $ between two masses $ m_{1} $ and $ m_{2} $
Eq.\ref{ChargeG}

\begin{equation}
F_{G}=G\frac{m_{1}m_{2}}{r^{2}}
\label{ChargeG}
\end{equation}

absolute analog. The analogy applies to the electromagnetic force and to the force between masses.
A problem is, the amount of electric charge $Q$ is independent of the speed of the FP's.
In contrast, the rest mass of the particle is subject to the Lorentz transformation, 
which drives the mass to infinity as the particle's speed approaches the speed of light.
Furthermore, the question arises whether there is an analogy between the color charge C 
( The force of Gluons), the electromagnetic charge Q ( The force generated by light), and the 
weak charge T3 (The weak force generated by Gauge Bosons). It is interesting to notice, that 
all of these interactions are a function of the distance between the different charges.

As a working hypothesis, in the following discussion of this paper, we use mass as the 
pseudo-charge of the gravitational force. The geometric size of the FP can be derived from the 
size of the mass core and/or the size of the coloured, electric, and weak charges. 
Without limiting the general discussion, we assume that the coloured, 
electric, and weak charges can be placed at the center of the FP's and 
outside the center. The integral over the three possible axes from outside 
to the center ultimately yields the total charge of the FP's. Likewise, 
the integral over the time axis for $ r_{0} $ from $ r = 0 $ to $ r\geq r_{0} $ 
yields the mass of the FP's. \\

\textbf { C) Effect of charges on a geometrically extended object.} \\

An electric charge at the edge of an extended object with spin creates a 
magnetic moment and, under certain conditions, an electric dipole moment.
Considering the basic scheme of an FP in Fig. {\ref{Basic-ETAMFP}, 
one can assume that the principle for electric charge also applies to color, 
weak, and mass charges. This assumption would result in the four charges 
generating four moments and four dipole moments.
A color charge creates a color moment, an electric charge a magnetic moment, 
a weak charge a weak moment, and a mass moment. The same charges can 
create the equivalent color, electric, weak, and mass dipole moments.
The FP's usually carry more than just one charge, for example, in the case of 
quarks, they carry color charge, electric charge, weak charge, and mass charge.
This would lead to the possibility of mixed moments and dipole moments.

So far, experimental evidence for the magnetic moment of the electron and muon 
\cite{Particle1} and a hint for the electric dipole moment of the electron and muon 
\cite{Particle1} are available. Furthermore, experimental limits and discussions on 
quark color magnetic moments and weak magnetic moments \cite{Particle2} are in preparation.
The scheme presented in Fig.{\ref{Basic-ETAMFP} takes into account current 
experimental knowledge. The color charges on the spin axis (z-axis) would 
produce small  or zero color moment, the electric charges on the x-axis would 
produce a magnetic moment, and the weak charges on the y-axis would 
produce a weak moment. According to Fig. {\ref{Basic-ETAMFP}, 
we assume a geometric distribution of the 
mass density $ \varrho (t,r) $ for the mass. Together with the spin rotation, 
this generates an angular momentum, the spin.
No mass-dipole moment is generated unless the mass distribution is geometrically 
clustering in some way. The Einstein equations prohibit a mass-dipole moment, 
which implements the assumption that mass density distributions are non-clustering.

\subsubsection{Extension of the geometrical approach to four dimensions}
\label{Extension33}

To extend the three-dimensional scheme of Fig.{\ref{Basic-ETAMFP} to 
four dimensions, we introduce the scheme for an left handed electron 
$ e_{L}^{-} $ in Fig.{\ref{Electron11} in an overview and detail.

\newpage

\begin{figure}[htbp]
\begin{center}
\rotatebox{00}{
 \includegraphics[width=14.0cm,height=12.0cm]{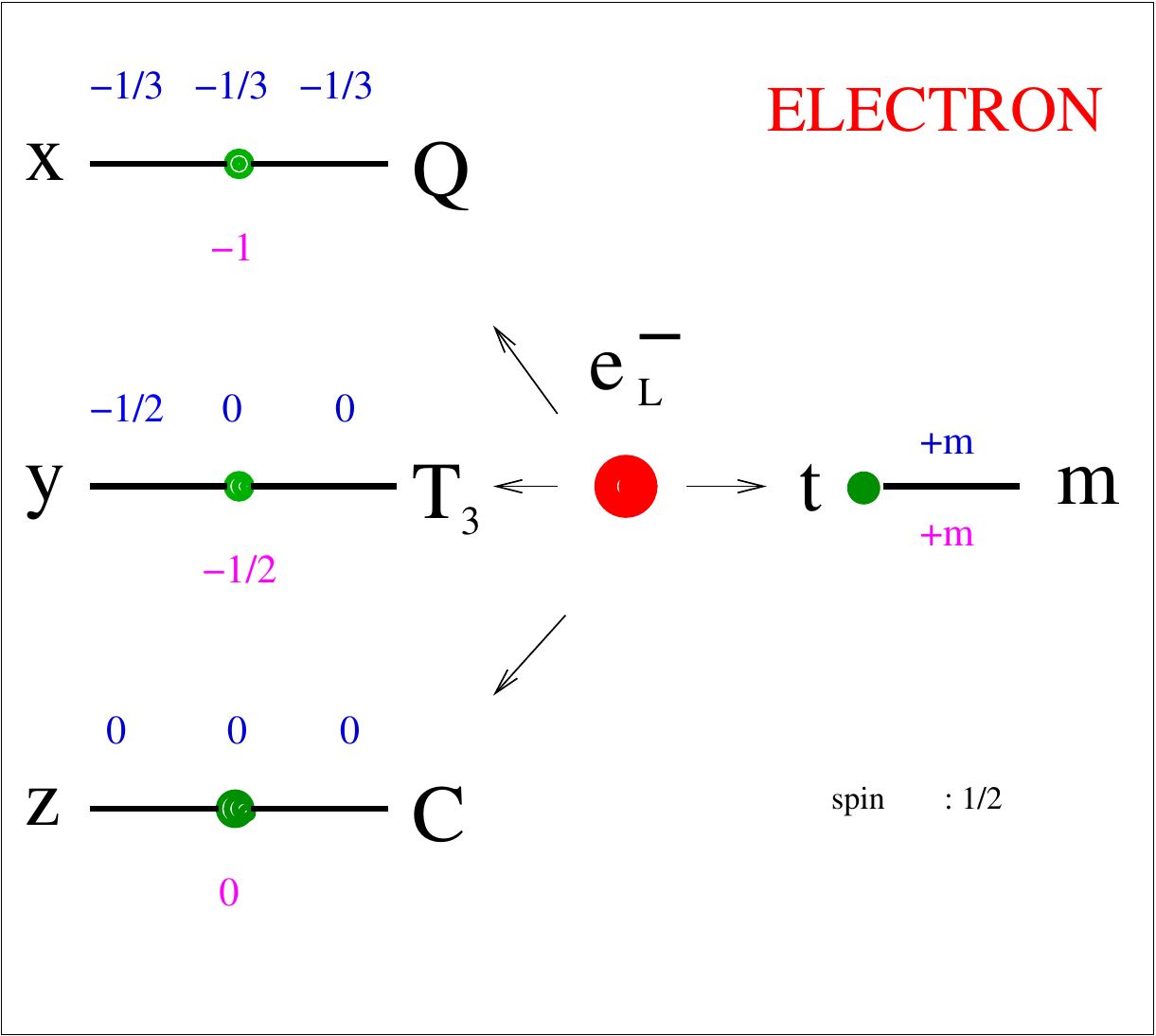}}
\end{center}
\caption{Scheme of the left handed electron $ e_{L}^{-} $.}
\label{Electron11}
\end{figure}

The three Cartesian coordinates are drawn by solid black lines on the left side of the figure.
The three coordinates x, y, and z are shown to the left of the line, and the corresponding charges
 $ Q $, $ T_{3} $, and $ C $ are shown to the right.
The fourth coordinate is shown on the right of the figure. On this side, time ``t'' is on the left, and the 
pseudo charge mass ( m , + m ) of is on the right. The geometric center of the FP is a green dot.
On the left side of the figure, the geometric position $r_{0}$ of the charges $Q$, $T_{3}$ , and $C$ is symbolised by the 
length of the black line. Above the black line, the amount of charge at the left, 
center, and right positions for a coordinate is shown in blue numbers. 
The integral of the charge, measured far from the FP, is shown below the black line in red. 
The commonly used symbol for the  electron FB  $ e_{L}^{-} $  ( In black  )  is 
shown in the center of the figure above a large red dot.
The time coordinate on the right side of the diagram has only one center (green dot)
and a black line to the right, symbolising that there is no negative absolute time.
Above the black line, a mass is assigned in blue if the particle is stable. Below, a mass 
is assigned if the particle has a rest mass ( Red ). The assigned mass $ +m $ means matter and 
$ -m $ means antimatter. The length of the black time line symbolises that the particle has a 
mass. We do not distinguish between rest mass and energy.

Before applying the scheme to all FP's, we show in detail how to use the scheme to 
describe, for example, an electron, as shown in Fig. {\ref{Electron11}.
At the $M_{W}$ scale, four independent interactions exist, and four orthogonal 
coordinates are available. According to the discussion in the section 
``Empirical Toy Ansatz on a Microstructure of Elementary Particles (ETAMFP)'', 
the mass core of the FP is time like, the strong, electromagnetic and weak 
interactions are space like. The spatially extended mass kernel, together with the spin of the FP, assigns three 
different positions for the placement of a charge for each axis, as shown in 
Fig. {\ref{Basic-ETAMFP}. Two at the border of the mass kernel and one in the kernel.  
The spin axis of the FP coincides with the z-direction as a distinguished axis. As discussed, 
we keep the z-axis free for the color charge, place the weak charge on the y-axis, 
and the electric charge on the x-axis. The electron carries no color charge, which 
sets the charge $C$ on the z-axis to zero. The weak hypercharge 
$Y$ ranges from 0 to 4/3 and, according to Eq.\ref{Charge1}, is a function
of the charge $Q$ and the weak isospin $T_{3}$, which ranges from 0 to $\pm 1$ in units of 1/2.
This number is more fundamental and is therefore used as the quantum number
for the scheme under discussion. We have the option of placing the weak charges $T_{3}$,
which could be 0, $\pm 1/2$, and $\pm 1$, on the y-axis.

In the case of the electron $T_{3}=-1/2$, there are three possible positions: left, center and right open.
If the electron had a weak moment, the charge $T_{3}=-1/2$ would have to be placed 
to the left or right of the y-axis. Since this has not been ruled out so far, we use the position
y-axis left. The total charge, measured from the far outside of the electron, would also be
$T_{3}=-1/2$. The electric charge 1/3 ( Blue ) must be placed on the x- or y-axis, since only in this position 
can a magnetic moment and an electric dipole moment be generated together with the spin.
We choose the x-axis. On the x-axis, all three open positions are occupied by $Q = -1/3$. 
The total charge, measured by an observer far outside the mass nucleus, 
would be the integral of all three charges, $Q = -1$ ( Red ).

The real geometric position of the charges is usually not exactly on the x- or y-axis, 
since the forces between the charges bring them into a position of energetic minimum, 
as explained in the chapter ``Forces and Stability'' (Fig. {\ref{Forces1} 
and Fig. {\ref{Forces22}). For example, in the case of the electron alone, a second charge $T_{3}$ on the free 
position of the y-axis would lead to an overall balanced symmetric system, placing 
the various charges precisely on the different axes. Our interest in the following 
chapters lies in a schematic representation. We therefore ignore this detail.
The electron is a very stable particle. For this reason, the mass on the time axis is 
denoted by $ +m $ in blue at the top and $ +m $ in red at the bottom, since this mass
is also the rest mass of the electron. The electron is a fermion with spin 1/2, as shown 
in Fig. {\ref{Electron11}}, bottom right.

\subsubsection{Scheme of the lightest left-handed fundamental particles.}
\label{Scheme lightest left-handed FB}

In Fig.{\ref{FM-left-handed} are displayed the scheme of all the four left handed lightest FP's. 

\begin{figure}[htbp]
\begin{center}
\rotatebox{00}{
 \includegraphics[width=16.0cm,height=12.0cm]{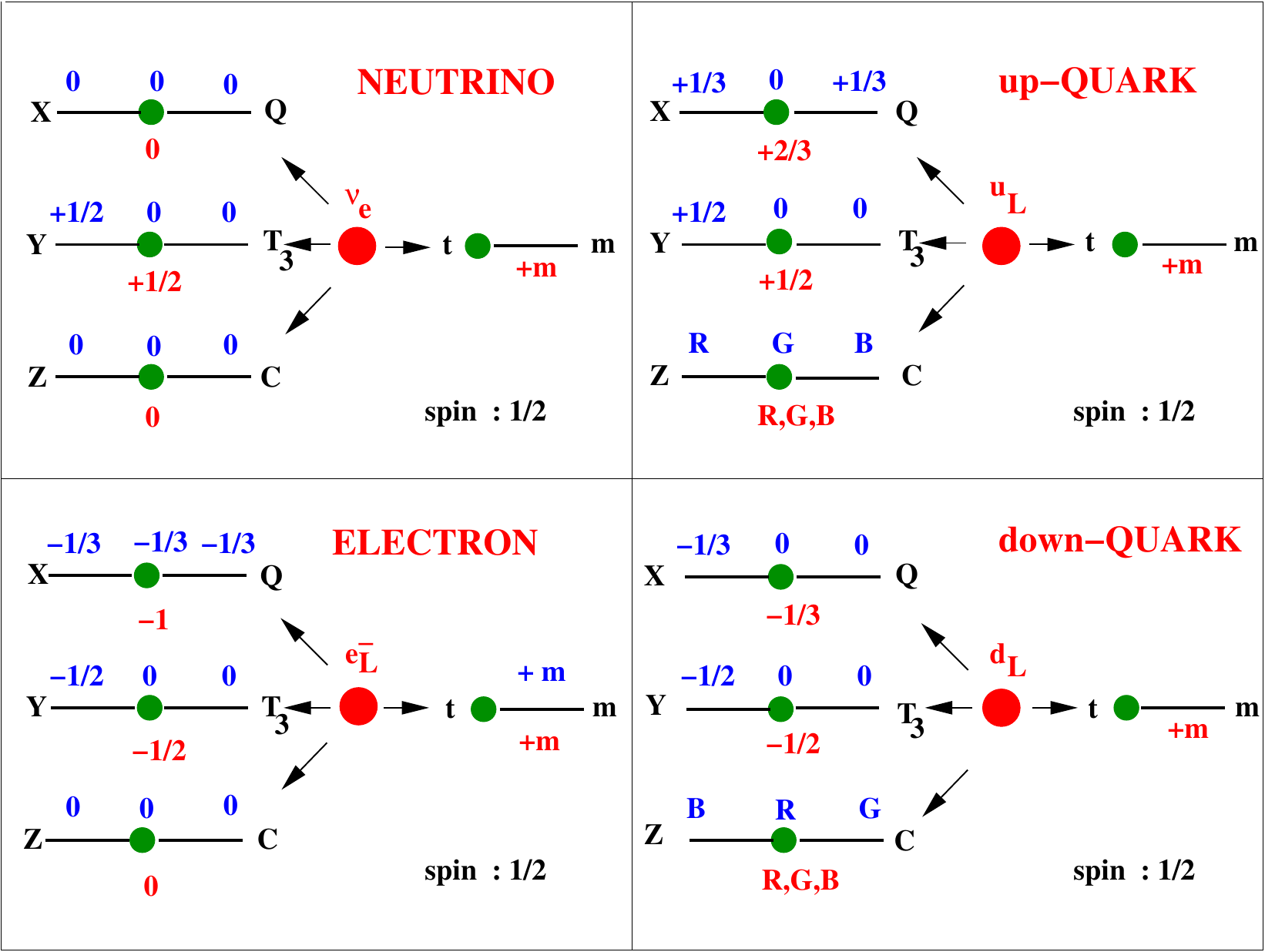}}
\end{center}
\caption{Scheme of the lightest left-handed fundamental particles.}
\label{FM-left-handed}
\end{figure}

For the left handed electron neutrino $  \nu _{e} $  are $ Q = C = 0 $, the mass is $ m< 3 $ eV \cite{Neutrino1}, 
$ T_{3}=+1/2$ and the spin is 1/2. The limitation of the electron neutrino mass $ m < 3 $ eV
is problematic because, after the discovery of neutrino oscillation, a neutrino rest mass must 
exist (see the chapter ``Neutrino Search'').
Below the t-axis, we have therefore assigned the neutrino rest mass $+m$ in red colour. We choose
all electric charges Q to be zero in all possible positions above the x-axis and
correspondingly, the integral charge below the x-axis to be zero.
For the electron neutrino $ \nu _{e} $, there exists a limit for the magnetic moment of 
$ \mu < 1.0\times 10^{-10} \left [ \mu _{B} \right ] $ \cite{Neutrino1}. This opens up the 
possibility that $ \nu _{e} $ carries electric charges. In this case, $ \nu _{e} $ could 
consist of a charge $ Q = - 1/3 $ and an ant-icharge $ Q = +1/3 $ placed above 
it at two of the three free positions on the x-axis.
The scheme would still be consistent with experiment, since the integrated charge
 below the x-axis Q would be zero. For the weak charge $T_{3}=+1/2$, three 
 possible positions above the y-axis are open: left, center, and right.
If $ \nu _{e} $ carried a weak moment, the charge $ T_{3}=+1/2 $ would have to be placed to 
the left or right of the y-axis. Since this has not been ruled out so far, we use the y-axis position at the top left.
If $ \nu _{e} $ did not carry a weak moment, the charge $ T_{3}=+1/2 $ would have to be placed centrally above the y-axis.
The total weak charge of $ \nu _{e} $, measured from the far outside, is also $ T_{3}=+1/2 $ 
and is shown below the y-axis. The $ \nu _{e} $ carries no color, which determines the color charge
$ C = 0 $ above and below the z-axis.

We have discussed the left-handed electron $ e_{L}^{-} $ in great detail in the chapter \ref{Extension33}, 
we only repeat the most important points of this discussion here.
The particle carries three electric charges, $Q = -1/3$, which we have placed in the three 
free positions above the x-axis ( Blue color ) . The integrated charge is located at $Q = -1$ below
the x-axis ( Red color ), $T_{3}=-1/2$ to the left above the y-axis ( Blue color ), and the integrated charge is 
also located at $T_{3}=-1/2$ below the y-axis ( Red color ). 
The $ e_{L}^{-} $  carries no color, which determines the color charge
$ C = 0 $ above and below the z-axis. The particle has spin $1/2$, a 
mass of $m = 0.51$ MeV, and is very stable \cite{Particle1}.
For this reason, the mass $+m$ is assigned above (blue color) and below (red color) the t-axis.
In agreement with the experiment \cite{Particle1}, the scheme assumes that the electron
generates a magnetic moment. Electric dipole moments and weak moments are possible.

The left-handed up quark $u_{L}$ consists of two anti charges $Q = +1/3$, which can be placed 
on two of the three free positions above the x-axis ( Blue color ). The integrated charge is $Q = +2/3$ 
and is shown below the x-axis ( Red color ).
All the possible positions of this two charges would to lead a magnetic moment of the 
$ u_{L} $. Similar like the for the $  \nu _{e} $, for the weak charge of the $ u_{L} $  $ T_{3}=+1/2 $
three possible positions above the y-axis left, center and right are open. If the $ u_{L} $ 
would carry a weak moment it would 
be necessary to put the charge $ T_{3}=+1/2 $ left or right above the y-axis. 
Since this is not excluded so far, we have used the position of the y-axis to the left above ( Blue color )
\cite{Particle2}. If $u_{L}$ did not carry a weak moment, the charge $T_{3}=+1/2$ 
would have to be placed centrally above the y-axis. 
The total weak charge of $u_{L}$, measured from the far outside, is also $T_{3}=+1/2$ 
and is shown below the y-axis ( Red color ). The color can be placed
in a total of 18 combinations (RRR, GGG, BBB, RBG, RGB, BRG, BGR,
GRB, GBR, RGG, GRG, GGR, RBB, BRB, BBR, BRR, RBR, and RRB) above 
the z-axis ( Blue color ). The color is assigned
symbolically to R, G, and B above the z-axis. The total of 18 possible charge combinations are
symbolically represented (R, G, and B) below the z-axis ( Red color ).    
 The color charges could create color moments. The particle has spin 1/2, a rest mass 
 between m = 1.5 and 4 MeV \cite{massUL}, and is stable only in a nucleus. We have 
 classified it as unstable with a rest mass of +m below the t-axis ( Red color ).

  The left-handed down-Quark $ d_{L} $  is equipped with one charges $ Q = -1/3 $ which could 
  be placed at one of the three free positions above the x-axis ( Blue color ). The choice of the position will
  generate on the mass kernel no magnetic moment and at the border a magnetic moment 
  with opposite sign to the up-Quark $ u_{L} $.  
  If the position of the two charges of the up-Quark 
  $ u_{L} $ is like in Fig.{\ref{FM-left-handed} the magnetic moment of the up-Quark $ u_{L} $ 
  must be $ \left| \mu _{u}\right|> \left| \mu _{d}\right| $. 
  The weak charge of the $ d_{L} $ is $ T_{3}=-1/2 $ . 
  Three possible positions (left, center, and right) above the y-axis are open ( Blue color ). 
 If $ d_{L} $ carried a weak moment, it would be necessary to place the 
 charge $ T_{3}=-1/2 $ to the left or right above the y-axis.   
 Depending on further experiments \cite{Particle2}, it might be possible to detect or exclude a 
 weak moment. If it is excluded, the position of the weak charge would have to shift accordingly toward the center.
 The total weak charge of $d_{L}$, measured from the far outside,
is also $T_{3}=-1/2$ and is shown below the y-axis ( Red color ). As with the $u_{L}$ quark, the color
can be placed above the z-axis in a total of 18 combinations (RRR, GGG, BBB, RBG, RGB,
BRG, BGR, GRB, GBR, RGG, GRG, GGR, RBB, BRB, BBR, BRR, RBR, and RRB).
The color is symbolically assigned to R, G, and B above the z-axis ( Blue color ).
 The integrated total possible 18 charge combinations are shown symbolic 
 (R, G, B) below the z-axis ( Red color ). The color charges 
 could generate color moments. The particle has spin 1/2, with a rest mass between 
 m = 4 to 8 MeV \cite{massUL} and is stable only in a nucleus. We assigned it non stable with rest 
 mass $ +m $ below the t-axis in red colour.

\subsubsection{Scheme of the lightest right-handed fundamental particles}

Fig. {\ref{FM-right-handed} shows the scheme of all right-handed lightest FP's. The 
transition from left-handed FP's to right-handed FP's occurs by changing the weak 
charge $ T_{3} $ from $ T_{3}=\left| 1/2\right| $ to $ T_{3}=0 $. The remaining 
quantum numbers remain unchanged. For this reason, we only briefly 
discuss the scheme of right-handed FP's.

\newpage

\begin{figure}[htbp]
\begin{center}
\rotatebox{00}{
 \includegraphics[width=16.0cm,height=12.0cm]{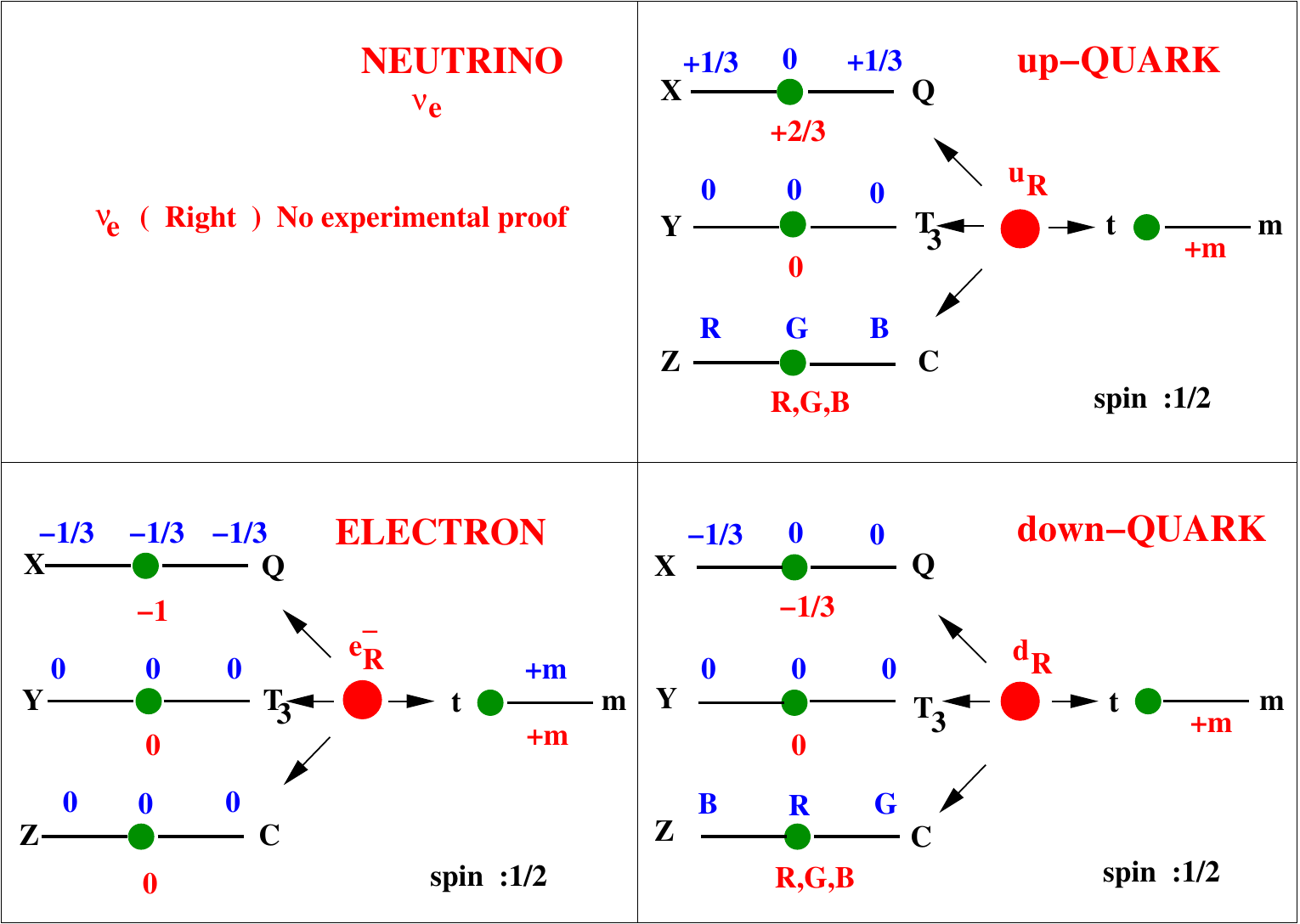}}
\end{center}
\caption{Scheme of the lightest right-handed fundamental particles.}
\label{FM-right-handed}
\end{figure}

All experiments have proofed  that no right-handed neutrino exists. Therefore, 
the scheme for $ \nu _{e}(right) $ is empty.

The right-handed electron $ e_{R}^{-} $ carries three electric charges $ Q = - 1/3 $, 
which we have placed in the three free positions above the x-axis ( Blue color ). 
The integrated charge is $ Q = -1 $ and is shown below the x-axis ( Red color ).
The weak charge $T_{3}=0$ and, accordingly, the integrated charge $T_{3}=0$ and the color are zero.
For this reason, all weak and coloured charges above and below the y- and z-axes are zero.
The particle has spin 1/2, a mass of $m = 0.51 $ MeV, and is very stable \cite{Particle1}. 
For this reason, we assign the mass +m ( Blue color ) above and below the t-axis ( Red color ). 
The scheme request, in agreement with experiment (\cite{Particle1}), that the electron generates a magnetic 
moment. Electric dipoles are possible.

The right-handed up quark $u_{R}$ consists of two anti-charges $ Q = +1/3 $, 
which can be placed on two of the three free positions above the x-axis ( Blue color ).
The integrated charge is $Q = +2/3$ and is shown below the x-axis ( Red color ).
All possible positions
of these two charges would result in a magnetic moment of $u_{R}$. The weak
charge of $u_{R}$ is $T_{3}=0$. Accordingly, we set all weak charges above and
the integrated weak charge below the y-axis to zero.
 The color can be placed in a total of 18 combinations above the Z axis (RRR, GGG, BBB, RBG,
RGB, BRG, BGR, GRB, GBR, RGG, GRG, GGR, RBB, BRB, BBR, BRR,
RBR, and RRB). The color is symbolically assigned to R, G, and B above the Z axis ( Blue color ).
The integrated total possible 18 charge combinations are shown symbolic 
(R, G, B) below the z-axis ( Red color ). The color charges could generate color moments. 
The particle has spin 1/2, with a rest mass between $ m = 1.5 - 4 $ MeV \cite{massUL} 
and is stable only in a nucleus. We assigned it non stable with rest mass $ +m $ below the t-axis
 n red colour.
 
The right-handed down quark $ d_{R} $ consists of one charge $ Q = -1/3 $, 
which can be placed at one of the three free positions above the x-axis ( Blue color ).
The choice of the position will generate on the mass kernel no magnetic 
moment and at the border a magnetic moment with opposite sign to the 
up-Quark $ u_{R} $. If the position of the two charges of the up-Quark $ $ is 
like in Fig.{\ref{FM-right-handed} the magnetic moment of the up-Quark $ u_{R} $ must be 
$\left| \mu _{u}\right|> \left| \mu _{d}\right| $. The weak charge of the $ d_{R} $ is $ T_{3}=0 $. Accordingly we 
set all weak charges above and the integrated weak charge below
the y-axis to zero. Like for the $ u_{R} $ quark the color could be places above 
the z-axis in total 18 combinations ( RRR, GGG, BBB, RBG, RGB, BRG, 
BGR, GRB, GBR, RGG, GRG, GGR, RBB, BRB, BBR, BRR, RBR and RRB ). 
The color is assigned symbolic R, G, B above the z-axis ( Blue color ).
The integrated total of 18 possible charge combinations are symbolically represented (R, G, B) 
below the z-axis ( Red color ). The color charges could generate color moments. The particle has 
spin 1/2, a rest mass between m = 4 and 8 MeV \cite{massUL}, and is stable only in 
the nucleus. We have marked it in red as unstable with a rest mass of $ +m $ below the t-axis. 

\subsubsection{Scheme of the lightest right-handed anti fundamental particles.}

Fig. {\ref{FM-right-handed-anti} displays the scheme of all right-handed lightest anti-FP's.  
  
\newpage
  
 \begin{figure}[htbp]
\begin{center}
\rotatebox{00}{
 \includegraphics[width=16.0cm,height=12.0cm]{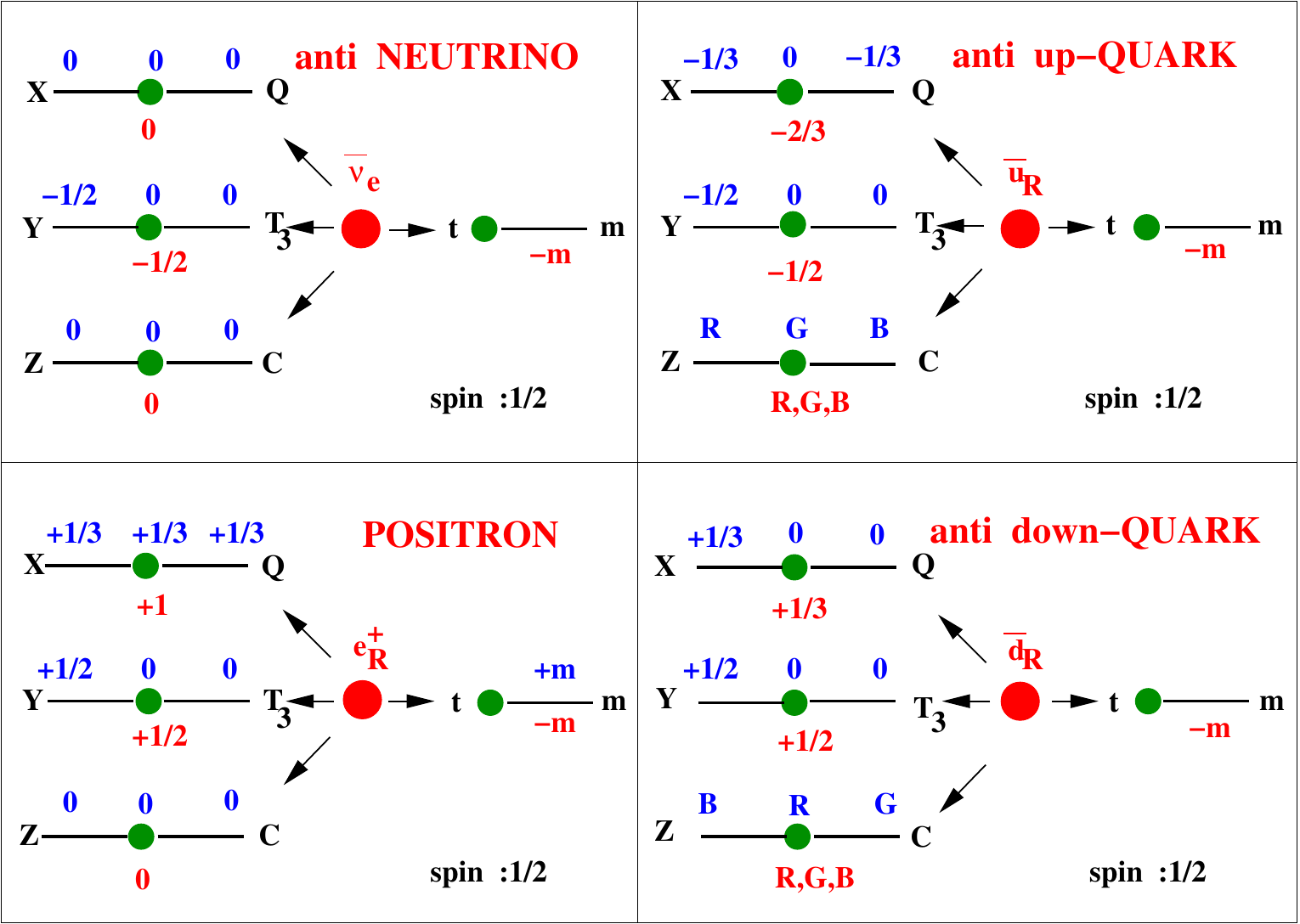}}
\end{center}
\caption{Scheme of the lightest right-handed anti fundamental particles.}
\label{FM-right-handed-anti}
\end{figure}

The experimental results for the right handed anti electron neutrino $  \bar{\nu _{e}} $
 are $ Q = C = 0 $, the mass is small and $ T_{3}=-1/2 $.  We place the electric  
 charge $Q$ above the x-axis at $Q = 0$ ( Blue color ) and the integral charge
below it at $Q = 0$ ( Red color ). It is possible that the particle carries electric charges, as in the left-handed neutrino
$\nu_{e}$. In this case, $\bar{\nu_{e}}$ could be composed of a charge $Q = -1/3$ and
an anti-charge $Q = +1/3$ located at two of the three free positions on the x-axis.
The scheme would still be consistent with experiment, since the integrated charge $Q$ would be zero.
We place the weak charge $T_{3}=-1/2$ (Blue color) above the left side of the y-axis.
Since we have no experimental evidence for a weak moment, this is a random decision.
In this position, $ \bar{\nu _{e}} $ could generate a weak moment,
the same would happen in the right position. The middle position would generate a weak moment of zero.
The integrated weak charge below the y-axis is $T_{3}=-1/2$ ( Red color ).
The color charge $C$ above the z-axis is zero ( Blue color ), as is the integrated color charge
below the z-axis ( Red color ). The stability and mass of the particle have not been experimentally settled..
We assume an unstable particle with finite anti-rest mass and place the anti-mass 
$ -m $ below the t-axis ( Red color ). The intrinsic properties of 
neutrinos and antineutrinos are not yet been experimentally clarified.
It is not known whether each neutrino mass eigenstate
$ \nu _{i} $ is identical to or different from its antiparticle $ \bar{\nu _{i}} $.
For $ \nu _{i}=\bar{\nu _{i}} $ we would have Majorana particles, and
for $ \nu _{i}\neq \bar{\nu _{i}} $ we would call them Dirac particles \cite{Dirac22}.

The right-handed positron $ e_{R}^{+} $ consists of three anti-charges $ +1/3 $ located 
above the x-axis ( Blue color ) and the integrated charge $ Q = +1$ below the x-axis ( Red color ). Similar to the 
left-handed electron $ e_{L}^{-} $, it carries a weak charge $\left|T_{3} \right|=1/2 $, 
but with a positive sign $ T_{3}=+1/2 $. Three possible positions (left, center, and right) are open. If the positron 
carried a weak moment, the charge $T_{3}=+1/2$ would have to be placed to the left or right above the y-axis.
Since this hasn't been ruled out yet, we use the position of the y-axis at above left 
(Blue color). The total weak charge, measured far outside the positron, would 
also be $T_{3}=+1/2$, shown below the y-axis (Red color).
The positron carries no color what sets the color charge $ C = 0 $ above and below 
the z-axis ( Blue - Red color ) The particle has spin 1/2, an anti-mass of $m = 0.51$ MeV, and is very stable. For 
this reason, the mass $+m$ is assigned above (Blue color) and the anti-mass $-m$ 
below the t-axis (Red color). The scheme request that the positron generates a 
magnetic moment, similar to the electron $e_{L}^{-}$. An electric dipole moment is possible.

The right-handed anti-up quark $ \bar{u}_{R} $ is composed of two charges $ Q = - 1/3 $, which
could be placed in two of the three free positions above the x-axis ( Blue color ). The integrated charge is 
$ Q = - 2/3 $ and is shown below the x-axis ( Red color ).
All possible positions of these two charges would result in a magnetic moment 
of $ \bar{u}_{R} $. The weak charge of $ \bar{u}_{R} $ is $T_{3}=-1/2$. 
Three possible positions above the y-axis (left, center, and right) are 
open ( Blue color). If $ \bar{u}_{R} $ carried a weak moment, it would be necessary 
to place the charge $T_{3}=-1/2 $ to the left or right above the y-axis.
Since this is not excluded yet \cite{Particle2}, we use the position of the y-axis to the left 
above ( Blue color ). If $ \bar{u}_{R} $ carried no weak moment, the charge 
$ T_{3}=-1/2 $ would have to be placed in the center above the y-axis. 
The total weak charge of $ \bar{u}_{R} $, measured from the far outside, 
is also $ T_{3}=-1/2 $ and is shown below the y-axis (Red color).
The anti-color can be placed in a total of 18 combinations above the Z-axis 
(RRR, GGG, BBB, RBG, RGB, BRG, BGR, GRB, GBR, RGG, GRG, GGR, RBB, BRB, BBR, BRR, RBR and RRB).
The anti-color is assigned symbolic R, G, B above the z-axis ( Blue color ). The integrated 
total possible 18 charge combinations are shown symbolic (R, G, B) below 
the z-axis ( Red color ). For simplicity we use in Fig.{\ref{FM-right-handed-anti} for color and anti-color the same 
symbol. The anti-color charges could generate anti-color moments. The particle has spin 
1/2 and an anti-rest mass between m = 1.5 and 4 MeV \cite{massUL}.
We classified it as unstable with an anti-rest mass $ -m $ below the t-axis (Red color).

The right-handed anti-down-Quark $ \bar{d}_{R} $ is composed out of one anti-charge 
$ Q = +1/3 $ which could be placed at one of the three free positions above 
the x-axis. The choice of the position will generate on the mass kernel no 
magnetic moment and at the border a magnetic moment with opposite sign 
to the anti-up Quark $ \bar{u}_{R} $ . 
If the position of the two charges of the anti-up-Quark 
$ \bar{u}_{R} $ is like in Fig.{\ref{FM-right-handed-anti} the 
magnetic moment of the anti-up Quark $ \bar{u}_{R} $ must be
 $ \left| \mu _{\bar{u}}\right|>\left| \mu _{\bar{d}}\right| $. Similar like for the $ \nu _{e} $, 
 the weak charge of the $ \bar{d}_{R} $ is $ T_{3}=+1/2 $ .
 Three possible positions above the y-axis (left, center, and right) ( Blue color ) are open.
If $ \bar{d}_{R} $ were a carrier of a weak moment, the charge $ T_{3}=+1/2 $ 
would have to be placed to the left or right above the y-axis. Depending on further 
experiments \cite{Particle2} , it might be possible to detect or rule out a weak moment.
 In the case of exclusion, the position of the weak charge must shift accordingly 
 toward the center. The total weak charge of the $ \bar{d}_{R} $, measured from 
 the far outside, is also $ T_{3}=+1/2 $ and is shown below the y-axis ( Red color ). As with the 
 $ \bar{u}_{R} $ quark, the anti-color could be placed above the z-axis in a total 
 of 18 combinations (RRR, GGG, BBB, RBG, RGB, BRG, BGR, GRB, GBR, 
 RGG, GRG, GGR, RBB, BRB, BBR, BRR, RBR, and RRB).
The anti-color is symbolically assigned B,R,G above the z-axis ( Blue color ).
The integrated total of 18 possible charge combinations are symbolically represented (R,G,B) 
below the z-axis ( Red color ). For simplicity, we use the same symbol for color and 
anti-color in Fig. {\ref{FM-right-handed-anti}. The anti-color charges could generate anti-color moments.
The particle has spin 1/2 and an anti-rest mass between m = 4 and 8 MeV \cite{massUL}.
We have assigned it as unstable with an anti-rest mass $ - m $ below the t-axis ( Red color).

\subsubsection{Scheme of the lightest left-handed anti fundamental particles.}
\label{Scheme lightest anti left-handed FB}

In Fig.{\ref{FM-left-handed-anti} are displayed the scheme of all the left handed lightest anti-FP's. 

\newpage

\begin{figure}[htbp]
\begin{center}
\rotatebox{00}{
 \includegraphics[width=16.0cm,height=12.0cm]{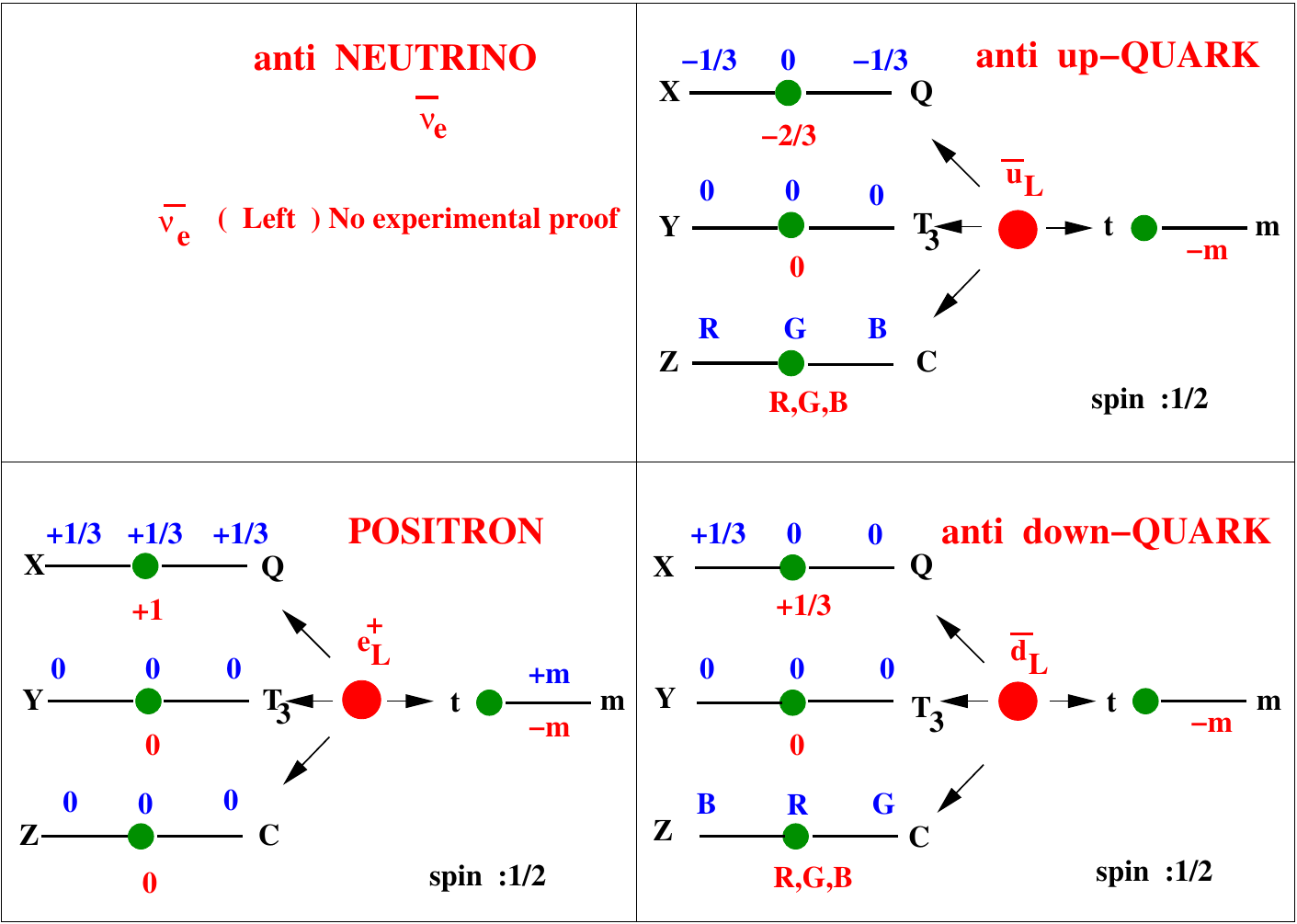}}
\end{center}
\caption{Scheme of the lightest left handed fundamental anti-particles.}
\label{FM-left-handed-anti}
\end{figure}

The transition from right-handed anti-FP's to left-handed anti-FP's occurs by 
changing the weak charge $ T_{3} $ from $ T_{3}=\left| 1/2\right| $ to $ T_{3}=0 $. 
The remaining quantum numbers remain unchanged. For this reason, we only 
briefly discuss the scheme of right-handed anti-FP's.

There is no experimental evidence for a left-handed anti-neutrino. Therefore, 
the scheme for $ \bar{\nu _{e}} $ (left) is empty.

The left-handed positron $ e_{L}^{+} $ carries three electric anti-charges $ Q = +1/3 $, 
which we have placed in the three free positions above the x-axis ( Blue color ). The integrated 
charge is $ Q = +1 $, which is shown below the x-axis ( Red color ). The weak charge $ T_{3}=0 $ 
and, accordingly, the integrated charge $ T_{3}=0 $, and the color is zero.
For this reason, all weak and coloured charges above and below the y- and z-axes 
are zero. The particle has spin 1/2, a mass of m = 0.51 MeV, and is very stable 
\cite{Particle1}. For this reason, we assigned the mass $ +m $ above the t-axis 
(Blue color) and a finite anti-rest mass $ -m $ below the t-axis (Red color).
The scheme request that the $ e_{L}^{+} $ 
generates a magnetic moment an electric dipole. 

The left-handed anti-up quark $ \bar{u_{L}} $ is composed out of two charges $ Q = -1/3 $, 
which could be placed in two of the three free positions above the x-axis ( Blue color ).
The integrated charge is $ Q = -2/3 $ is shown below the x-axis ( Red color ).
All possible positions of these two charges would result in a 
magnetic moment of $ \bar{u_{L}} $. The weak charge of $ \bar{u_{L}} $ is $ T_{3}=0 $. 
Accordingly, we set all weak charges above and the integrated weak charge below the y-axis to zero.
The anti-color can be placed in a total of 18 combinations above the z-axis ( RRR, GGG, BBB,
RBG, RGB, BRG, BGR, GRB, GBR, RGG, GRG, GGR, RBB, BRB, BBR, BRR,
RBR, and RRB ). The anti-color is symbolically assigned to R, G, and B above the Z-axis ( Blue color ).
The integrated total of 18 possible charge combinations are represented symbolically (R, G, B) 
below the z-axis ( Red color ). For simplicity, we use the same symbol for color and anti-color in 
Fig. {\ref{FM-left-handed-anti}. The anti-color charges could generate anti-color moments.
The particle has spin 1/2 and an anti-rest mass between m = 1.5 and 4 MeV \cite{massUL}.
We have classified it as unstable with an anti-rest mass $ -m $ below the t-axis ( Red color ).

The left-handed anti-down-Quark $ \bar{d}_{L} $ is composed out of one anti-charges $ Q = +1/3 $
which could be placed at one of the three free positions above the x-axis ( Blue color ). The choice of the
position will generate on the mass kernel no magnetic moment and at the border a magnetic
moment with opposite sign to the anti-up-Quark $ \bar{u}_{L} $.
If the position of the two charges of the anti-up quark $ \bar{u}_{L} $ is as in 
Fig. {\ref{FM-left-handed-anti} , the magnetic moment of the anti-up quark $ \bar{u}_{L} $ must be
$\left| \mu _{u}\right|>\left| \mu _{d}\right|$. The weak charge of $ \bar{d}_{L} $ is $T_{3}=0$.
Accordingly, we set all weak charges above and the integrated weak charge below the y-axis to zero.
Like for the $ \bar{u}_{L} $ quark the anti-color 
could be places above the z-axis in total 18 combinations ( RRR, GGG, BBB, RBG, RGB, 
BRG, BGR, GRB, GBR, RGG, GRG, GGR, RBB, BRB, BBR, BRR, RBR and RRB ). 
The anti-color is assigned symbolic B, R, G above the z-axis ( Blue color ). The integrated total 
possible 18 charge combinations are shown symbolic ( R, G, B ) below the z-axis ( Red color ). 
For simplicity, we use the same symbol for color and anti-color in Fig. {\ref{FM-left-handed-anti}. 
The anti-color charges could generate anti-color moments. The particle has spin 1/2 and an 
anti-rest mass between m = 4 and 8 MeV \cite{massUL}. We have designated it as unstable 
with an anti-rest mass $-m$ below the t-axis ( Red color ).

\subsubsection{Scheme of the Gauge Bosons.}
\label{Scheme Gauge Bosons}

In the simplest structure of the SM, two fundamental spin 1/2 particles interacting 
with each other through spin 1 point particles for the strong, electro magnetic and 
weak interaction as shown in Fig.{\ref{Interaction1}. 

\newpage

\begin{figure}[htbp]
\vspace{5.0mm}
\begin{center}
\rotatebox{00}{
 \includegraphics[width=16.0cm,height=12.0cm]{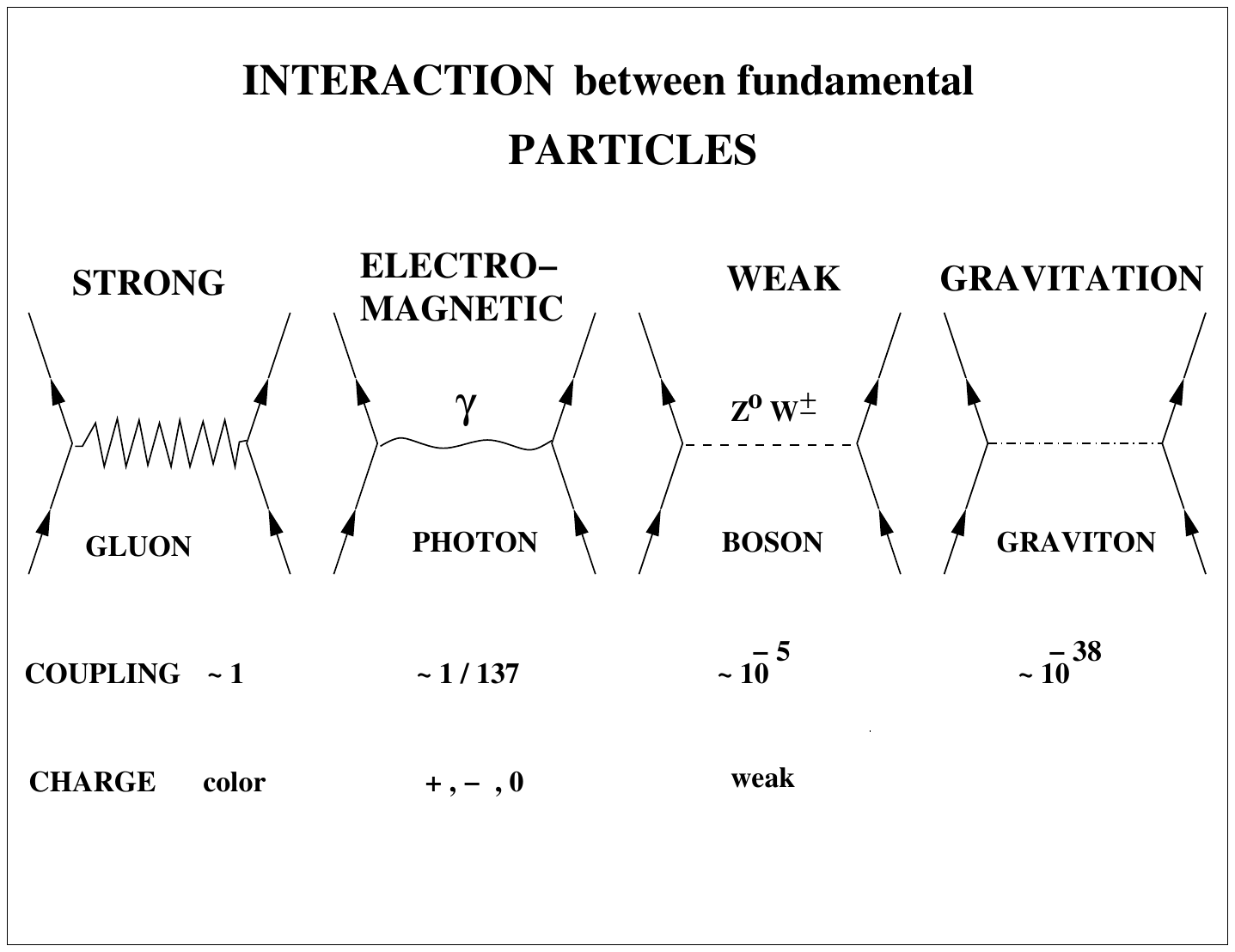}}
\end{center}
\caption{Interactions between fundamental particles}
\label{Interaction1}
\end{figure}

If the scheme for geometrically extended elementary particles, presented for example 
in Fig. {\ref{Basic-ETAMFP}, is generally valid for all FP's, it is important to test whether 
the scheme can also describe the gluons, $ \gamma $ and $ Z^{0} $ , $ W^{\pm } $. \\

\textbf {A) Scheme of the eight Gluons of the strong interaction.}\\

Scheme of the eight Gluons of the strong interaction are shown in Fig.{\ref{8Gluons}.

\newpage

\begin{figure}[htbp]
\vspace{5.0mm}
\begin{center}
\rotatebox{00}{
 \includegraphics[width=16.0cm,height=12.0cm]{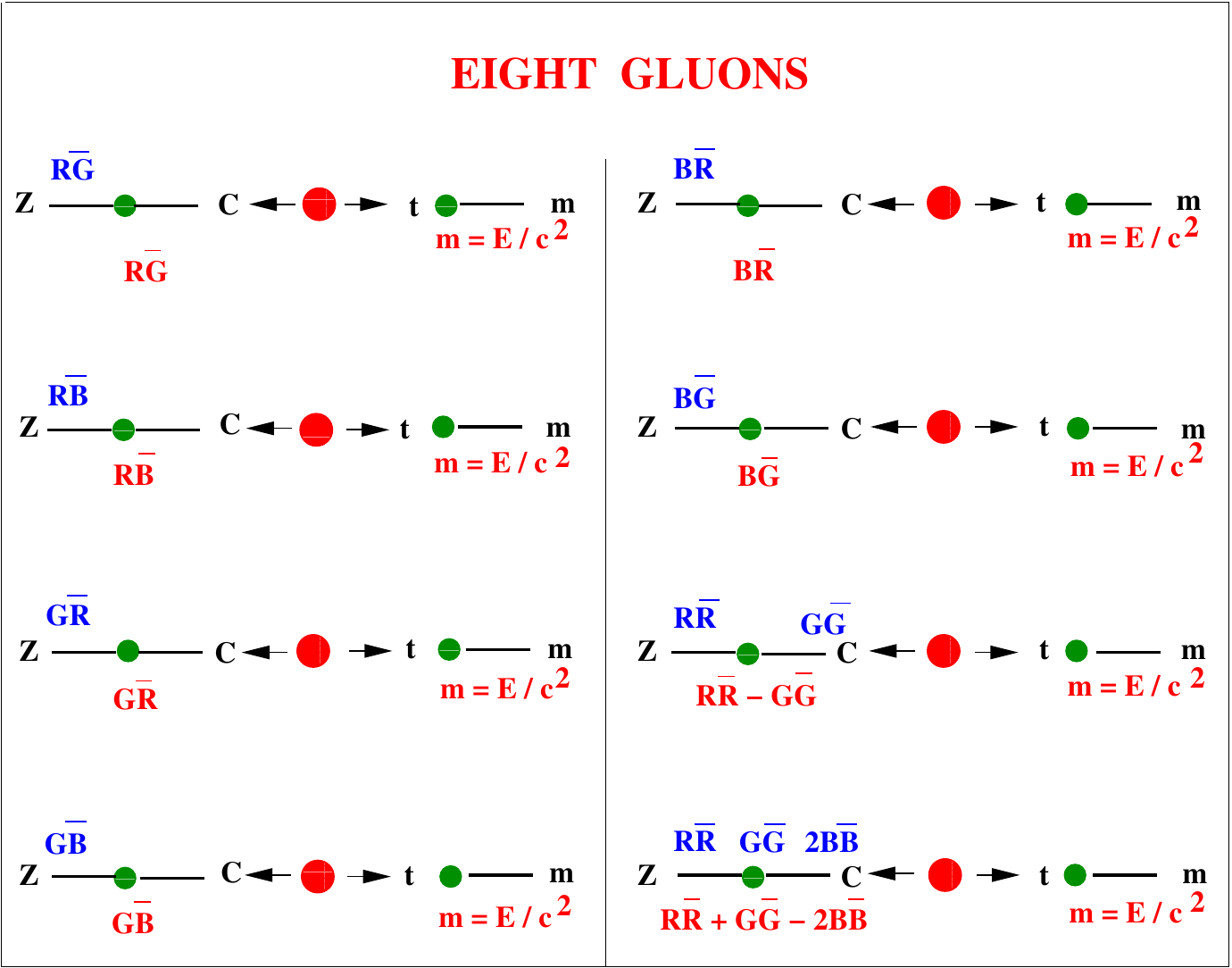}}
\end{center}
\caption{Scheme of the eight Gluons of the strong interaction.}
\label{8Gluons}
\end{figure}

In the SM are eight Gluons necessary for the strong interaction between 
the six quarks. All the Gluons are combinations of colours and anti colours 
located at one single point. The Gluons carry no electromagnetic or weak charge $Q=T_{3}=0$. Their 
rest mass is also zero, but their induced mass is $m=E/c^{2}$. In a scheme 
where the gluon is a geometrically extended object, this results in the 
x- and y-axes being unoccupied. The object looks like a string from 
the perspective of the three charges of the SM. Only the 
induced mass in the kernel will lead to a geometrical extension. 
To respect these experimental findings we populate only the z-axis in agreement 
with Fig.{\ref{Basic-ETAMFP}  and do not show the x-axis and 
y-axis in the scheme. We keep the t-axis as before. We follow this pattern in Fig. {\ref{8Gluons} and locate 
the nine possible color-anti-color combinations ($ R\bar{G},R\bar{B},G\bar{R},G\bar{B},
B\bar{R},B\bar{G},R\bar{R},G\bar{G},B\bar{B}$), as with the fermions, at three positions 
above the z-axis ( Blue color ). The choice of position-left, center, or right-depends on 
a color moment of the Gluons. Since there is no experimental evidence for such a moment, we made a random selection. 
The eight integrated gluon combinations, viewed from the far outside, the gluon
( $R\bar{G},R\bar{B},G\bar{R},G\bar{B},B\bar{R},B\bar{G},R\bar{R}-G\bar{G},R\bar{R}+G\bar{G}-2B\bar{B}$)
are placed below the z-axis ( Red color ). The gluons carry only an induced mass of $m=E/c^{2}$ 
and were therefore placed below the t-axis ( Red color ). The spin of the Gluons assigns on specific direction 
 of the object, we choose in accordance with Fig.{\ref{Basic-ETAMFP} the z-axis. To locate 
 the carrier of the color force, color and anti-color in one location of the three 
 geometrical axis is not the only possibility in the scheme under discussion. It would be
 possible to separate the color and anti-color at two geometrical separated positions. 
 For the combination $ R\bar{R}+G\bar{G}-2B\bar{B} $ maximal six free positions would be necessary. 
 These position would exist at the three axis x, y and z surrounding the porter of the 
 mass kernel as shown for example in Fig.{\ref{Basic-ETAMFP}. Such a configuration would generate
 a spherical closed structure but would need three coordinates. 
 In Fig. {\ref{8Gluons}} we follow the assumption that each free space coordinate acts as 
 a free parameter for the three interactions; strong, electromagnetic and weak and 
 use only one axis, the z-axis. \\
  
  \textbf {B) Scheme of the Gauge bosons $ \gamma $ , $ Z^{0} $  and $ W^{\pm } $}.\\
  
  The scheme of the interaction of the Gauge bosons is shown in Fig. {\ref{Gauge-bosons}.
  
 \newpage 
  
\begin{figure}[htbp]
\vspace{5.0mm}
\begin{center}
\rotatebox{00}{
 \includegraphics[width=16.0cm,height=12.0cm]{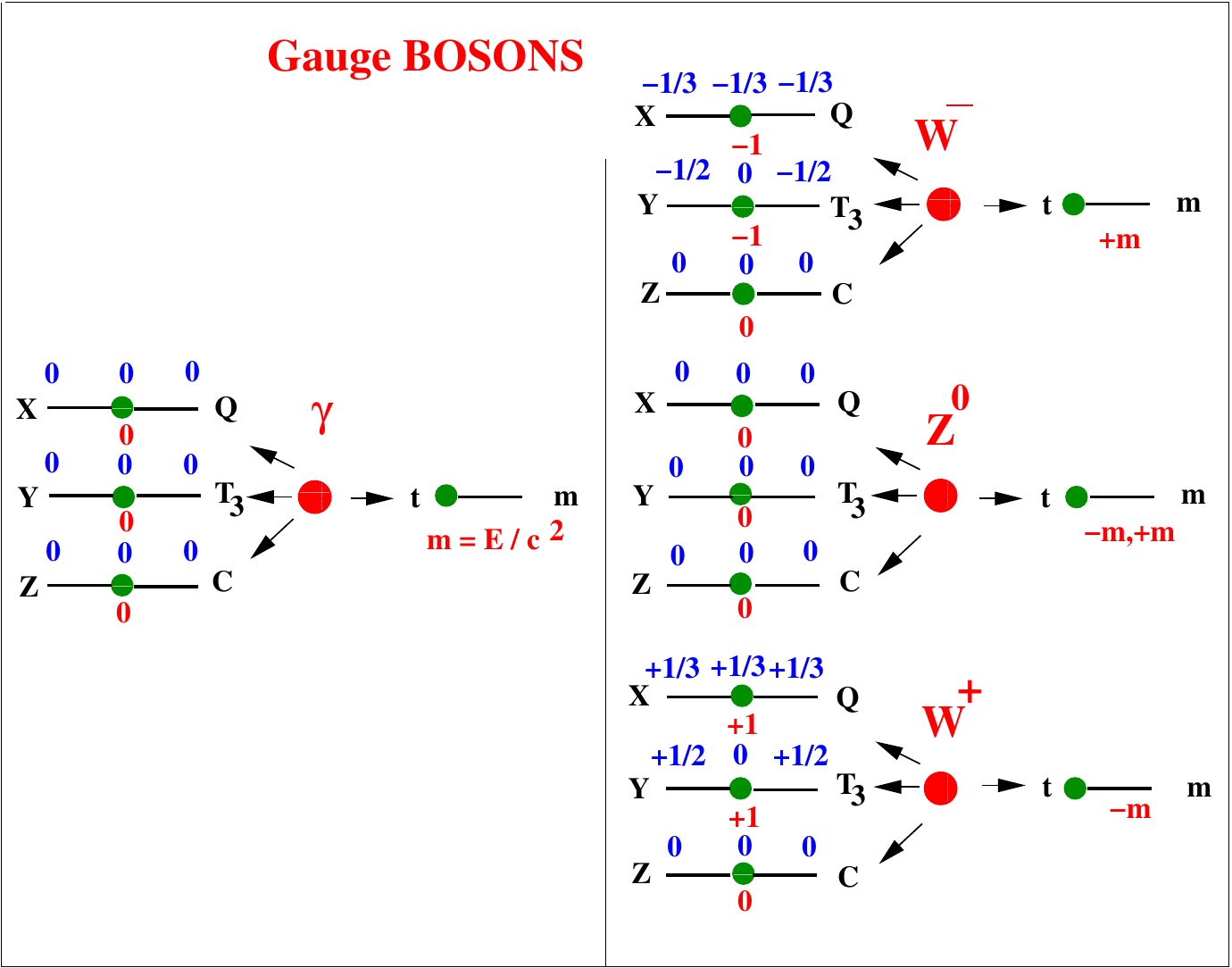}}
\end{center}
\caption{Scheme of the Gauge bosons $ \gamma $, $ Z^{0} $ and $ W^{\pm } $.}
\label{Gauge-bosons}
\end{figure}

The most interesting particle is the $ \gamma $ particle. It carries no charge $ Q=T_{3}=C=0 $, 
no rest mass, only induced mass or energy, and spin = 1. The spin exhibits a unique axis 
and polarisation characteristics such as longitudinal or circular polarisation.  
We know from experiments that it is a wave or a particle, depending on the type of the experimental 
test. It is its own anti-particle. A geometric extension could come from a kernel with 
induced mass surrounded by a charge $ Q = -1/3 $ and an anti-charge $ Q = +1/3 $, or 
from a self-interacting electromagnetic field.    
In the scheme shown in Fig.{\ref{Gauge-bosons}, we set all charges above and below the three 
geometric axes to zero ( Blue - Red color ). We place the induced mass or energy $ E=m/c^{2} $ below the t-axis 
( Red color ). Our scheme would define $ \gamma $ as a poor time like object with an induced mass kernel.
  
The carriers of the weak interaction in the SM model are the weak bosons $ Z^{0} $ and $ W^{\pm}$.
The $ Z^{0} $ is the partner of the $ \gamma $ with charges $ Q=T_{3}=C=0 $ , with spin=1, but with a rest mass of
m = 91.1876 $\pm$ 0.0021 GeV \cite{W-Z} . 
It is its own anti-particle. A geometric extension could be very similar to that of the $  \gamma $
particle, but the mass kernel would be generated by a high rest mass, as shown in 
Fig.{\ref{Gauge-bosons} below the t-axis in $ -m $ and $ +m $ ( Red color ).
As for the $ Z^{0} $ are the charges $ Q=T_{3}=C=0 $, we set in Fig.{\ref{Gauge-bosons}
all the charges above and below the three geometrical axis to zero ( Blue-Red color ).
  
The weak Boson $W^{-}$ carries three electric charges $Q = -1/3$, two weak charges 
$T_{3}=-1/2$, no color charge $C = 0$, and a rest mass of m = 80.425 $\pm$ 0.038 GeV \cite{W-Z}.
In Fig.{\ref{Gauge-bosons}, we place the three electric charges $Q = -1/3$
above the x-axis in the three free positions ( Blue color ).
The scheme request in accordance with the SM, that the $ W^{-} $  carries a magnetic moment \cite{W-Z22}.
The integrated charge is $ Q = -1 $ shown below the x-axis ( Red color ). The two weak charges $ T_{3}=-1/2 $
we put above the y-axis left and right. The integrated weak charge is $T_{3}=-1$ 
and is plotted below the y-axis ( Red color ). The weak bosons 
$W^{\pm }$ are the FP's with an integrated weak charge $T_{3}=\left|1\right|$. Together with 
the fermions with an integrated weak charge $T_{3}=0$ and $T_{3}=\left|1\right|$, this fact 
supports the assumption to cluster the weak charge in units of $ 1/2 $.
This leads to the weak charge sequence of $T_{3}=0,\pm 1/2,\pm 1$ already discussed in 
Fig. {\ref{Electron11}}. The geometric position of the charge $T_{3}$ (left, right) generates 
a weak moment that is larger than the moment generated by the position (center, right) or
(center, left), such that $\mu _{weak}^{left,right}>\mu _{weak}^{centre,right}$.
Since there is no weak experimental evidence for a weak moment of $ W^{-} $, we chose the position
randomly. We set the color charge above and below the z-axis to zero, since $ W^{-} $ carries no 
color charge ( Blue - Red color ). The particle has spin = 1 with a rest mass of $ +m $, which we assigned below the t-axis ( Red color).

The weak boson $W^{+}$ carries three electric anti-charges $Q = +1/3$, two weak charges $T_{3}=+1/2$,
no color charge $C = 0$, and a rest mass of m = 80.425$\pm$0.038 GeV \cite{W-Z}.
In Fig.{\ref{Gauge-bosons} we place the three electric anti-charges $Q = +1/3$
above the x-axis in the three free positions ( Blue color ).
The scheme request in accordance with the SM, that the $ W^{+} $ carries a magnetic moment with opposite sign 
to the $ W^{-} $ moment  \cite{W-Z22}. The integrated anti-charge is $ Q = +1 $
shown below the x-axis ( Red color ).  We place the two weak charges $T_{3}=+1/2$ 
above the y-axis, left and right ( Blue color ). The integrated weak 
charge $T_{3}=+1$ is located below the y-axis ( Red color ). Similar to $W^{-}$, the geometric 
position of charge $T_{3}$ (left, right) generates a weak moment
that is greater than the moment generated by the position (center, right) or (center, left), such that
$\mu_{weak}^{left,right}>\mu_{weak}^{center,right}$.
Since there is no experimental evidence for a weak moment of $ W^{+} $, we chose 
the position randomly. We set the color charge above and below the z-axis to zero, 
since $ W^{+} $ carries no color charge ( Blue - Red color ). The particle has spin = 1 and a very similar 
rest mass to $ W^{-} $: $ (m_{W^{+}}-m_{W^{-}})=-0.2 \pm 0.6 $ GeV \cite{W-Z}.
We set the anti-rest mass of $ W^{+} $ as $ -m $ below the t-axis ( Red color ). \\

\textbf {C) Scheme of the Higgs.}\\

The Higgs - boson is shown Fig.{\ref{Higgs-boson}. 

\begin{figure}[htbp]
\vspace{5.0mm}
\begin{center}
\rotatebox{00}{
 \includegraphics[width=16.0cm,height=10.0cm]{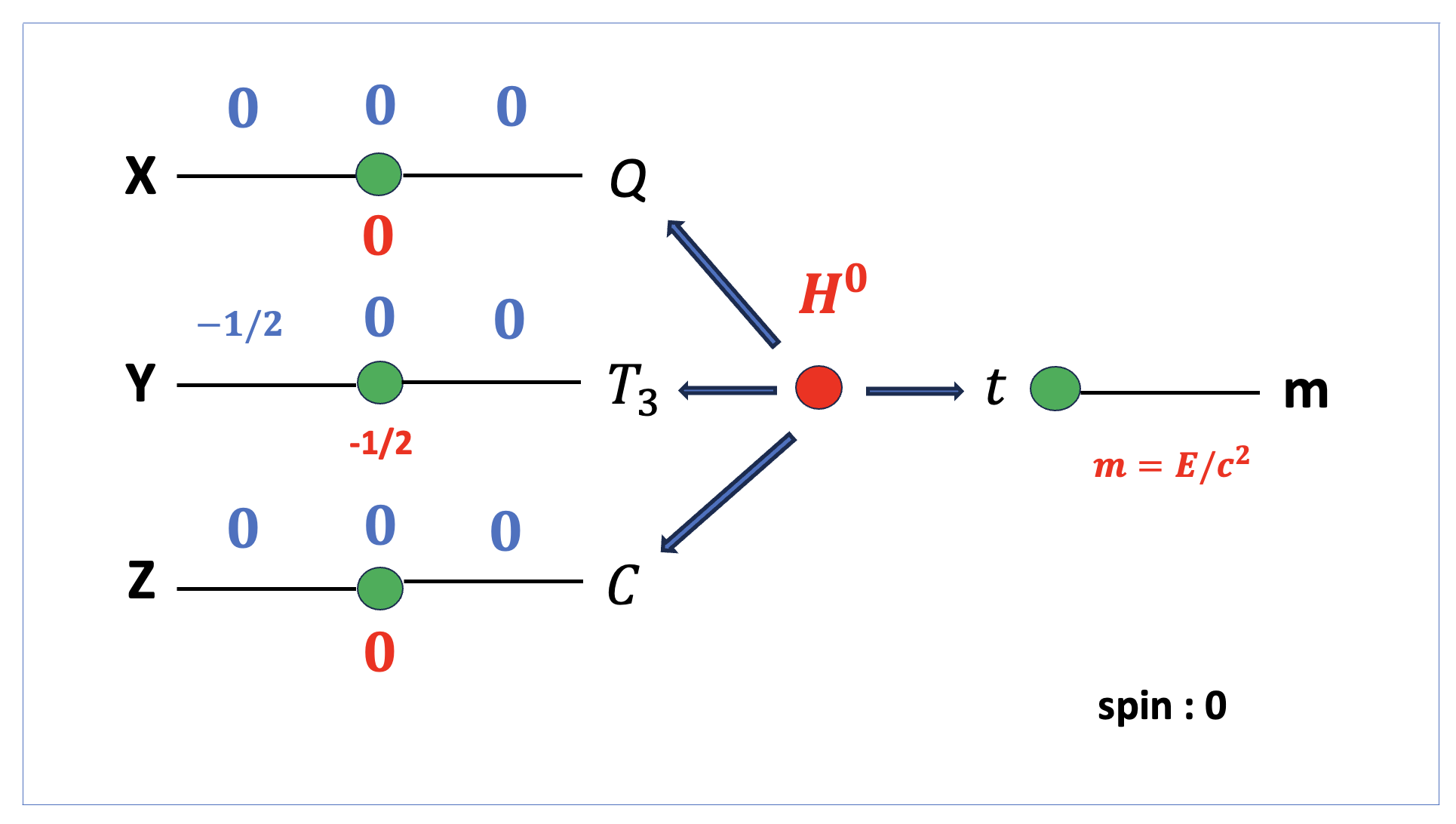}}
\end{center}
\caption{Scheme of the Higgs boson.}
\label{Higgs-boson}
\end{figure}

The Higgs boson is responsible for the mass of all FP's and the Gauge particles. 
It is very similar to the $  \gamma $ particle, with zero electric charge and zero color 
charge. Its spin is zero, its weak isospin $ T_{3}=-1/2 $, its weak 
hypercharge $ Y_{W}=+1$, parity $ P=+1 $, the mass $ m = 125.11\pm 0.11\ GeV/c^{2} $ and a mean lifetime of 
$ t\sim 1.56\times 10^{-22}s $. The important parameter is $T_{3}=-1/2$. 
Since the spin is zero, there are no assigned x, y, or z axes. 
The placement of $T_{3}=-1/2$ is entirely artificial. We choose the y-axis ( Blue color ). 
The integrated $T_{3}=-1/2$ from outside is placed below the  y - axis ( Red color ). Depending on whether 
a weak moment $\mu$ exist, $T_{3}=-1/2$ is placed to the left or right of the y-axis, or, 
if no $\mu$ is exist, in the middle of the y-axis. Since the mass is unstable, we place it below the t-axis ( Red color ).

It is important to notice that the details of the Higgs parameters are still under investigation and it will take
new accelerators and particular time to settle the parameters.

\subsubsection{Excited states of Fermions.}
\label{Exited-Fermions}

The discussed scheme for geometrical extended fundamental particles includes the currently known
lightest left and right handed FP's, lightest left and right handed anti - FP's of the first particle 
family and the gauge bosons. The majority of the FP's belong to the second and third family. 
To test the reliability of the scheme under discussion it is essential to proof is the scheme 
able to describe also the second and third family.\\

\textbf {A) Coincidence between three space coordinates and three families. }\\

It would be possible to include all these particles in the described scheme, but all 
subsequent scheme structures would be very similar, as in Fig. {\ref{FM-left-handed},
Fig. {\ref{FM-right-handed} and Fig. {\ref{FM-right-handed-anti},
Fig. {\ref{FM-left-handed-anti}.
It is noteworthy that the charge, charge position, and left-/right-handed structure of 
the heavier FP's are consistent within the families. The most important parameter for 
experimentally characterising the difference between the three families is the rest mass of the fermions.
The rest mass of families one, two, and three increases exponentially with the family number.
$ m_{\mu }/m_{e}\sim 207 $ and $ m_{\tau }/m_{e}\sim 3477 $ \cite{Particle1}.
In the ETAMFP model in Fig. {\ref{Basic-ETAMFP}}, is the mass of the 
geometrically extended FP's located in the center of the particle.
For this reason, it makes sense to look for a way for the mass sphere to increase
its mass or energy. Since the spin of all three particle families is the same, oscillation 
of the fermion mass sphere would be one way to increase its rest mass or energy.
According to the three Cartesian coordinates, the mass sphere has three 
fundamental degrees of freedom for oscillation. This fact coincide with the 
three families of fermions. For example, $ \mu $ and $ \tau $ would be the excited states of the charge
$ e $ on the x, y, and z axes. A harmonic oscillator with vibrational energy
$ E = \hbar \omega (n_{x} + n_{y} + n_{z}) $ (neglecting zero energy) would allow us to use the
three spatial coordinates x, y, and z of our scheme to generate vibrational states on the x-axis
$ n_{x} = 0,1,2... $, on the y-axis $ n_{y} = 0,1,2... $, and on the z-axis $ n_{z} = 0,1,2... $.
The ground state would be $ n_{x}=1 $ and $ n_{y}=n_{z}=0 $, the first excited 
state $ n_{x}=n_{y}=1 $ and $ n_{z}=0 $, and the third excited state 
$ n_{x}=n_{y}=n_{z}=1 $. The mass sphere of the FP's would perform a 
rotational motion ( Spin )superimposed by an oscillation in three axes, as 
shown in Fig.{\ref{Excited-states1}.

\newpage

\begin{figure}[htbp]
\vspace{5.0mm}
\begin{center}
\rotatebox{00}{
 \includegraphics[width=16.0cm,height=12.0cm]{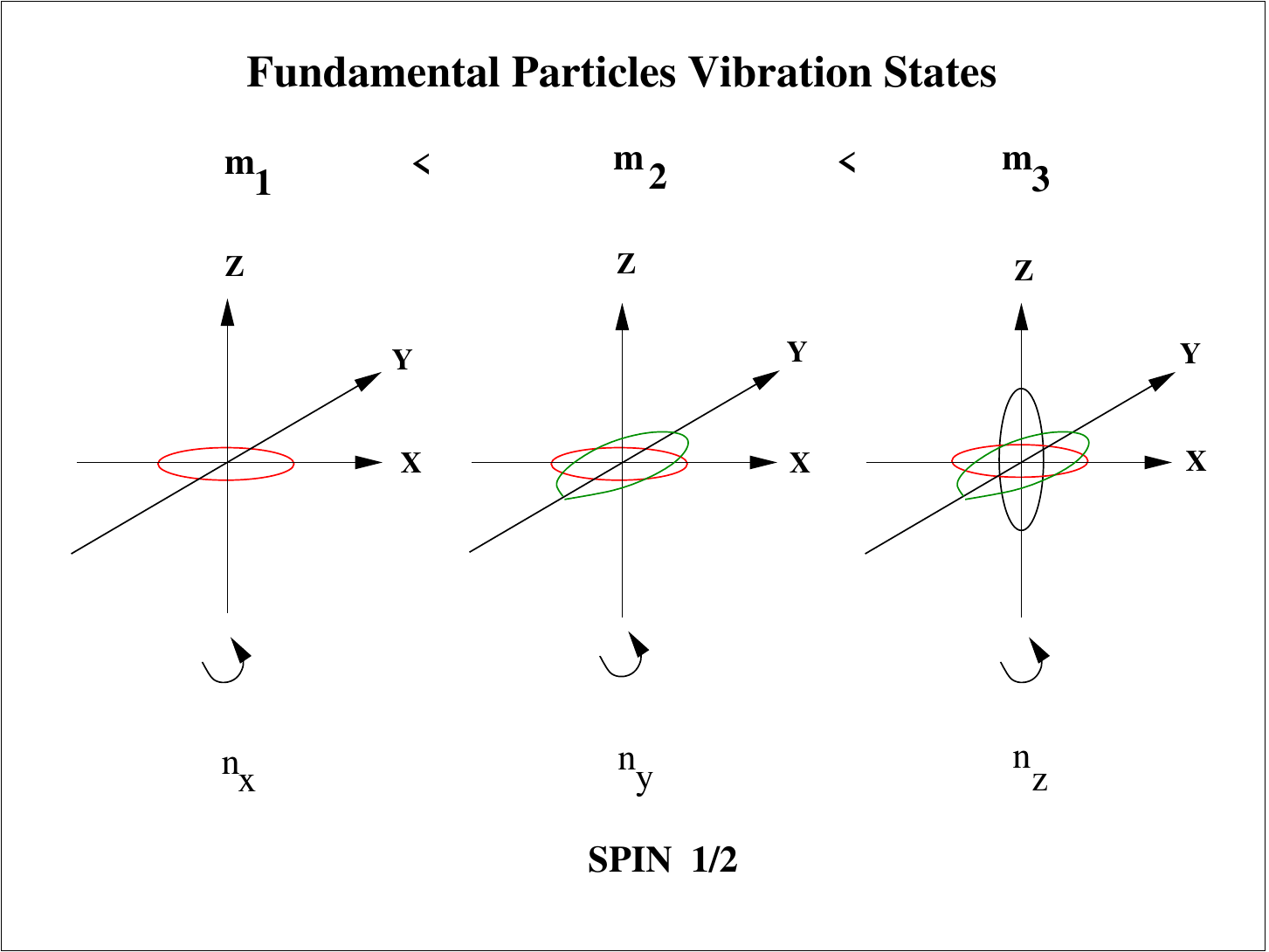}}
\end{center}
\caption{Excited states of fundamental particles.}
\label{Excited-states1}
\end{figure}

The oscillation is limited by the essential condition that the spin of the FP's does not change.
In the center of Fig. {\ref{Excited-states1}}, the three oscillation states are symbolised by three coordinate systems.
The mass of the three families $  m_{1}<m_{2}<m_{3} $ is displayed on top of z-axis of these coordinates. 
The spine 1/2 is symbolised by a twisting arrow below the z-axis. The spin
axis is chosen to be parallel the z-axis. 

The ground state on the x-axis $  n_{x} $ is shown left, 
the first excited state on the y-axis $  n_{y} $ in the middle and the second exited state on the 
z-axis $ n_{z} $  on the right in Fig.{\ref{Excited-states1}. 
The ground state on the x-axis would be a rotating 
spherical mass sphere with a distinct rotating axis parallel the spin symbolised 
as red circle in the x-y plane. 
The first excited state on the y-axis would be a rotating ellipsoidal sphere of mass 
around the z-axis, simultaneously oscillating with its central axis in the x- and 
y-directions, symbolised by a red and green ellipse in the x-y plane.
The third excited state on the z-axis would be an ellipsoidal mass sphere rotating 
around the z-axis and simultaneously oscillating with its central axis in the x-, y-, 
and z-directions, symbolised by a red, green, and black ellipse in the x-y and x-z planes.\\

\textbf {B) Empirical ansats to sort all three particle families into one scheme. }\\

The conditions discussed open up the possibility of sorting all masses of fermions 
in an empirical ansats of a rotator superposed with a strongly degenerate oscillator.
With 2 parameters and 10 constants for 12 fermion masses, it is possible to 
calculate all fermion and anti-fermion masses, including a prediction of the 
mass limits for the $ \nu _{\mu } $ and $ \nu _{\tau } $ neutrinos in 
Eq.\ref{FPmass}.

\begin{eqnarray}
\label{FPmass}
E(k_{i};Q) & = & (A+B|Q|+CQ^{2}+D|Q|^{3})(k_{i})^{f(Q,k_{i})}  \\
f(Q,k_{i}) & = & (R+|Q|V(k_{i}-1)+|Q|(|Q|-1)        \nonumber  \\
           &   & (S(|Q|-1/3)+W(|Q|-1/3)(k_{i}-1)    \nonumber  \\
           &   & +T(|Q|-2/3)+Z(|Q|-2/3)(k_{i}-1)))  \nonumber
\end{eqnarray}

 Eq.\ref{FPmass} contains the two parameters family number $ k_{i} $ and the electric charge $ Q $
of the fermions. For simplicity, the weak charge and the color charge are not used,
but are implicitly hidden in the 10 constants. 
 The first parameter is the family number $ k_i = (n_x + n_y + n_z) $ according to Fig.{\ref{Excited-states1}
with $ n_{x,y,z} $ vibrational state for the x-, y-, and z-axes. The first family is $ n_x = 1 $ and $ n_y = n_z = 0 $,
 which leads to $ k_1 = (n_x + n_y + n_z) = 1 + 0 + 0 = 1 $. The second family is $ n_x = n_y = 1 $ and 
 $ n_z = 0 $, which leads to $ k_2 = (n_x + n_y + n_z) = 1 + 1 + 0 = 2 $. 
 The third family is $ n_{x}=n_{y}=n_{z}=1 $, which leads to $ k_{3}=(n_{x}+n_{y}+n_{z})=1+1+1=3 $.
  More exited states are mathematical possible.  
 The second parameter is the charge $ Q=0,\pm 1,\pm 2/3,\pm 1/3 $ for the leptons/anti-leptons 
 and quarks/anti-quarks. Included in Eq.\ref{FPmass} are the 10 constant factors 
 $ A<3\times 10^{-6} $ MeV, $ B = 42.1358 $  MeV, $ C = - 87.7995 $ MeV and $ D = 46.1747 $MeV 
 to define the masses of the first family. The constant factors for the leptons/anti-leptons 
 are $ R = 7.9617 $ and $ V = - 0.2698 $ ,for the quark/anti-quark family $ Q\pm 2/3 $ the 
 factors $ S = 5.2528 $ and $ W = - 19.3806 $ and for the for the quark/anti-quark family 
 $ Q\pm 1/3 $ the factors $ T = - 77.3429 $  and $ Z = 26.8191 $.  
 These 10 constant factors are fittet to the experimental measured mass of the 
 lightest fermion $  m_{\nu _{e}};m_{e};m_{u};m_{d} $ and the heaviest 
 fermions $  m_{\tau };m_{b};m_{t} $. After this simple approach it is possible
 to predict the masses of fermion family two $  m_{\nu _{\mu}};m_{\mu};m_{s};m_{c} $. \\

\textbf {C) Comparison of theoretical calculations and experimental data. }\\

The comparison of the theoretical calculations according to Eq.\ref{FPmass} and 
the experimental data is shown in Fig. {\ref{Excited-states2}. The diagram shows the family number
$ k_{i} = 1;2,3 $ on the x-axis and the rest mass of the fermions in MeV on the y-axis.

\newpage

\begin{figure}[htbp]
\begin{center}
\rotatebox{00}{
 \includegraphics[width=16.0cm,height=12.0cm]{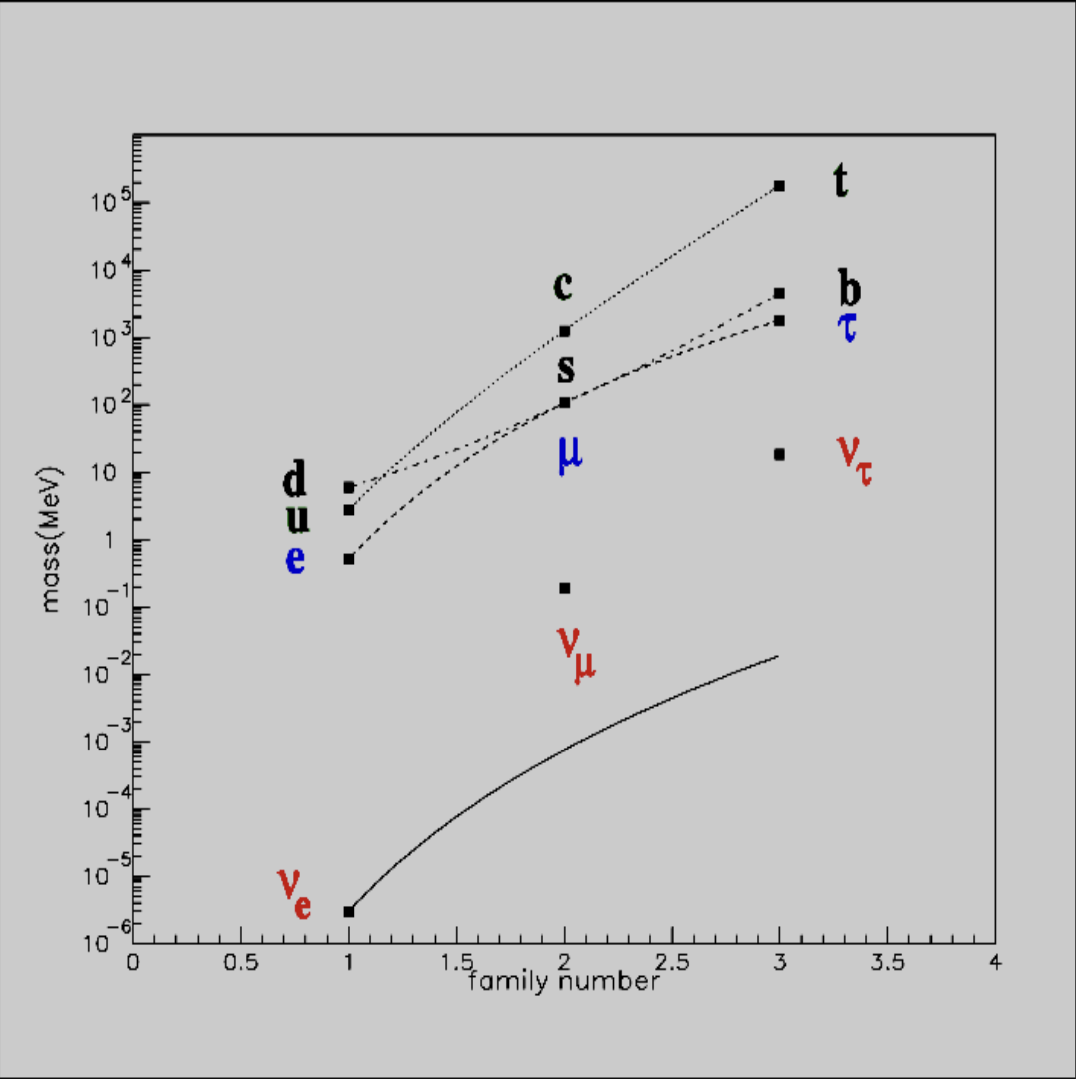}}
\end{center}
\caption{Excited states of fermions..}
\label{Excited-states2}
\end{figure}

The experimental data are shown in black rectangles. The calculation for the rest mass of the 
$ u, c, t $ quark is displayed as black dotted line, for the $ d, s, b $ quark displayed as black dot 
dashed line, for the $  e,\mu ,\tau $ leptons displayed as black dashed line and for the 
$\nu _{e},\nu _{\mu },\nu _{\tau } $ neutrinos displayed as black solid line.

The experimental data are taken from ref. \cite{Ellis1}. For the quarks with the charge Q = 2/3,
 $ u, c $ and $ t $, we used as data the geometrical middle value of the experimental mass 
 limits from ref. \cite{Ellis1}. The same middle values for the experimental rest mass we used 
 for the quarks $ d, s $  and $ b $ with the charge $ Q = -1/3 $. For the charge leptons $ e,\mu $  and $ \tau $
 we used the well measured rest masses from ref. \cite{Ellis1}. For the neutrinos exist mass 
 limits only. For this reason, we use the mass limit for $ \nu _{e} $ as a starting value and calculate 
the mass limits for $ \nu _{\mu } $ and $ \nu _{\tau } $ (black solid line). We 
assume the same exponent for the calculation as for the charged leptons. The 
experimental limits (black rectangles) are significantly higher than the calculation. 
Further experiments are necessary to prove whether this is due to the accuracy of 
the experiment or a calculation effect. The perfect agreement
between the experimentally measured rest mass and the calculation of Eq. \ref{FPmass} for
$ \nu _{e};e;u;d $ and $ \tau ;b,t $ is not surprising, since the parameters in Eq. \ref{FPmass}
are fitted to these parameters. Importantly, Eq. \ref{FPmass} perfectly predicts the masses of $ \mu ;s;c $.
With 12 parameters, it is possible to predict three masses $ \mu ;s;c $
and demonstrate a way to calculate all masses of FP's as a function of mass and family number.  
A moderate agreement between calculation and experimental data is discussed in \cite{Hasan1}.      
          
The mass increase for the first family of $ m_{\nu _{e}}<<m_{e}<m_{u}<m_{d} $ roughly follows
the strength of the weak, electromagnetic, and strong coupling constants of the
three interactions. In Fig. {\ref{Excited-states2}}, it can be seen that the gradient of mass increase per 
 family is very similar for the charged leptons and the quarks with charge $ Q = 2/3 $.
The quarks with charge $ Q = -1/3 $, in particular the mass of the $ d $ quark, do not follow this trend.  
The upper experimental mass limit for the $u$ quark of 4 MeV
just touches the lower mass limit of the $d$ quark, also at 4 MeV. With improved 
experimental data in the coming years, this discrepancy could disappear. 
The trend of the masses of $ \nu _{\mu };\nu _{\tau } $ shows a clear increase 
 in agreement with neutrino oscillations, but further data are needed to draw any conclusion.

  

  
\subsubsection{Conclusion on the scheme of geometrically extended fundamental particles and antiparticles.}
\label{Conclusion ETAMPF}

In conclusion it was possible to demonstrate the hypothesis that the temporal and size evolution 
of the history of the very young Universe forms the geometric extension of the FP's and makes 
it possible to classify all FP's into a common scheme. The mass increase for the first 
family roughly follows the strength of the weak, electromagnetic, 
and strong coupling constants of the three interactions. The scheme predicts a generally similar 
trend in the gradient of mass increase per family for the leptons and quarks. 
The scheme shows that the link between geometric space and time is of essential importance. 
Four interactions correspond to the four coordinates of our universe. In this philosophy, 
the three families represent the ability of the mass kernel to oscillate (link time) in the three 
spatial coordinates. The scheme also shows that the simplest charged, 
highly symmetric geometric object is the electron. It serves as the ground state for all 
leptons with extremely high stability.  
  
The scheme opens a new interpretation of the ``Wave-particle duality'' \cite{WAVE-PARICLE}.
The scheme is not introducing a substructure of two new particles in the FP's like ``PREONS'' \cite{PREON1}.
The FP's are ONE single particle with a KERNEL pus an OUTSIDE HOMOGEN STRUCTURE.
The FP's are NOT points any more. In other words TWO distinct features are intergraded 
in ONE particle. This interpretation would look like a PARTICLE the KERNEL and a wave 
like the outside structure. Consequences of these facts will be discussed in
chapter ``Particle-like structure related to gravity''. 
  
 It is also important that the different charges of the FP's are measured ONLY for the entire particle.
The detailed distribution of these charges in the scheme has not yet been clarified. For example for the
Electron $ e $ the integrated electric charge is $ Q = -1 $. As three charge $ Q=0,\pm 1/3, \pm 2/3 $ are
allowed. Many positions on the x-, y-, and z-axes are possible. It is possible to set 
x-axis $ \ charge\ Q=left \ (-1/3)+middle \ (-1/3)+right \ (-1/3)=-1 $ , 
$ \ charge\ Q=left \ -(+2/3) +middle \ (0) +right +\ (-1/3)=-1 $ or 
$ \ charge\ Q=left \ +(-1/3) +middle \ (0) +right +\ (-2/3)=-1 $ and so on. This details are not known so fare.
In the scheme we use the 3 $\times ( Q= -1/3 ) $ version for the electron.

\subsection{Application of special relativity in a classical approach to study the electron as a Lorentz contracted GYROSCOPE.}
\label{GYROSCOPE}

In the discussed scheme in Fig.{\ref{Basic-ETAMFP},  ``Basic scheme of extended fundamental particle in the ETAMFP model'',
Fig.{\ref{Electron11} ``Scheme of the left handed electron $ e_{L}^{-} $'' and Fig.{\ref{FM-left-handed}
 ``Scheme of the lightest left-handed fundamental particles: Is the electron the most fundamental particle?'' 
It has a spherical geometric structure, is the ground state of leptons, and is very stable, with a 
lifetime of $\tau_{e}>4.6\times 10^{26}$ year. For this reason, it is a very important object to study
the consequences of rest mass, charge, spin, magnetic dipole, and possible electric dipole in accordance 
with our particle scheme of a geometrically extended object.
Inspecting our basic scheme of an geometrical  extended FP in Fig.{\ref{Basic-ETAMFP}, the charges 
$ Q $ and $  T_{3} $ will fulfil a rotation orbiting the coordinate of the mass center $ x = y = z = 0 $. 
Three questions immediately arise from a classical approach. Is the size of the electron larger 
than the Planck size $  10^{-33} $ cm ? At the Planck scale at the latest, gravity must be quantised, 
and a classical picture would become meaningless. Is there a scenario for the forces in the 
electron to form such a stable object? Why don't the accelerated charges Q and $  T_{3} $ 
release energy in the form of radiation, e.g., light?

The experimental limits on the size of FP's are investigated using a direct contact term regime
\cite{CONTACT1} (Fig. \ref{Feynman2}). The limits are in the range of $ 10^{-18} $ cm. Of 
interest is the limit of the electron's electric dipole moment of $ d_{e} < (0.07 \pm 0.07) \times 10^{-26} $ e cm. 
These limits lie far beyond the Planck scale. The difference, at 15 and 7 orders of magnitude, 
is so far from the Planck scale that we believe a classical approach is still valid.

In the chapter Forces and Stability \ref{Forces and stability} we discussed in Fig.{\ref{Forces1} and Fig.{\ref{Forces22}
a possible scenario of how two stable positions in one object  SPI and SPII could schematically exist.

In the case of the electron point SPII in Fig., {\ref{Forces1} would be the non-rotating 
inner mass kernel. The center of mass would extend to the rotating point SPI, where the charges
$ Q $ and $ T_{3} $ would be placed shown in  Fig.{\ref{Forces1} and Fig.{\ref{Basic-ETAMFP}.
This scenario would assume that the mass density $ \varrho _{s} $ as a function of $ r_{m} $
goes to zero at $ r_{m}\sim r_{0}\sim r_{SPI} $, where the two charges $ Q = - 1/3 $ are placed.
The same stability scenario also explains that two electric charges $ Q = - 1/3 $ are placed 
opposite each other on the x-axis shown in Fig. {\ref{FM-left-handed} and one charge 
$ Q = -1/3 $ is placed at $ x = y = z = 0$. Consequently, the magnetic moment and the 
possible electric dipole moment of the electron are generated only from a total charge 
of $ Q = -2/3 $ and not from $ Q = -1 $. For simplicity, we neglect the weak charge $ T_{3} $.

The extremely high experimentally proven stability of the electron requires a radiation-free 
trajectory of the charges around the electron's center. This experimental fact requires that 
the charges do not experience any acceleration, since charges rotating in a circle are accelerated.
In the chapter on a flashing vacuum \ref{The flashing vacuum}, we discussed a microscopic 
scenario of such a radiation-free path. In this picture, the vacuum has the important property 
of moving a charge geometrically along a specific curved path without acceleration.
In this way, the curved path of a charge becomes radiation-free. An important consequence of 
such a path is that the geometric length will be contracted due to the Lorentz transformation 
as the velocity of the charges approaches the speed of light.

For the following calculations, we need numerical information on the parameters mass, 
charge, spin, magnetic and possible electric dipole moment of the electron, as well as 
the Planck Constant, the speed of light, and the Bohr Magneton. These parameters 
are summarised in Table ~\ref{Parameterelectron} and taken from the 
ref. \cite{Electron-Parameter}.

\newpage

\begin{table}
\caption{ Parameters of electron. }
\label{Parameterelectron}
\begin{center}
\begin{tabular}{||l|l|l||}                                      \hline
  parameter           & Symbol        & value
                                                               \\ \hline
  Total Mass         & $ E_{tot} $   & $ 0.51099892 \pm 0.00000004 $ MeV
                                                               \\ \hline
  Charge
                      & $ e  $        & $ 1.60217653(14) \times 10^{-19} $ A s
                                                               \\ \hline
  Weak Charge
                      & $ T_{3}  $    & $ - 1/2 $
                                                               \\ \hline
  Spin                & $ S=\frac{1}{2}\cdot\hbar $ &
                        $ \frac{1}{2} \cdot 6.58211915(56)\times 10^{-22} $ MeV s
                                                               \\ \hline
  Magnetic Moment
                      & $ \mu_{e} $   &
                        $ 1.001159652187 \pm 0.000000000004 \cdot \mu_{B} $
                                                               \\ \hline
  Electric Dipole Moment
                      & $ d_{e} $     & $( 0.07 \pm 0.07 ) 10^{-26} $ e cm
                                                               \\ \hline
  Planck Constant     & $\hbar$       & $ 6.58211915(56) \times 10^{-22} $ MeV s
                                                               \\ \hline
  Speed of Light
                      & $ c $         & $ 299792458 $ m/s
                                                               \\ \hline
  Bohr Magneton
                      & $ \mu_{B}=\frac{e\hbar}{2m_{e}} $ &
                        $ 5.788381804(39) \times 10^{-11} $ MeV/T
                                                               \\ \hline
\end{tabular}
\end{center}
\end{table}

In Tab.~\ref{Parameterelectron}, the electron charges are the source of an electric, weak, 
and gravitational force and should be capable of polarising and curving the vacuum at the 
finiteness  of the mass core. According to the Einstein equation (See, for example, Chapter SM Table 3, 
``Possible links between Standard and Big Bang model''), the electron's mass would arise from these curvatures.
The spin of a geometrically extended electron is classically described by a gyroscope with a defined 
spin axis and moment of inertia. The spin axis is perpendicular to a surface of rotation. The same 
applies to the classical magnetic moment of a charge. In our scheme, the charge rotates 
like a small object around the object's mass core. Since we know from experiments 
that the spin axis is antiparallel to the axis of the magnetic moment, 
it is likely that the charge rotates in the same plane of rotation as the mass core.
Considering the charge in Fig. {\ref{Basic-ETAMFP} as a small object, it must cause an electric 
dipole moment if an observer measures the dipole moment in the x-y plane and the electron is 
polarised parallel to the z-axis. The degree of polarisation and the conditions of the experimental 
setup influence the measured dipole moment. The experiment is extremely difficult and not yet completed.

In the following sections in this chapter we study with an increasing refinement of the classical 
electron model the impact of the different parameters of the electron on the size of the electron.
It is important to realise that these considerations are fare beyond the Standard Model, because
so fare no unified theory for all physics exist.

\subsubsection{Mass and charge radius of an electron as a gyroscope.}

To investigate the order of magnitude in a first simple approach, we describe the microscopic 
structure of the electron as a rotating sphere like a gyroscope filled with mass as shown 
in Fig.{\ref{Basic-ETAMFP}. The charges, spin and spin axis are in accordance also with Fig.{\ref{Basic-ETAMFP}. 
In our scheme, the electron, as the ground state of the leptons, carries no vibrational energy. 
The total energy $E_{tot}$ of a gyroscope under these conditions is the sum of the electron 
energy $m_{0}c^{2}$ and the gyroscope's rotational energy $E_{rot} = \frac {1}{2} \theta \omega^{2}_{m}$, as shown in
Eq.\ref{Etot}.

\begin{equation}
E_{tot} = m_{0} c^{2} + \frac {1}{2} \theta \omega^{2}_{m}
\label{Etot}
\end{equation}

The angular momentum L is shown in Eq.\ref{Ltot}

\begin{equation}
L = \theta \omega_{m} = \frac{1}{2} \hbar
\label{Ltot}
\end{equation}

and the magnetic moment of the electron $ \mu _{e} $ in Eq.\ref{MagMoment}.

\begin{equation}
\mu_{e} = \frac{1}{2} e \omega_{e} r_{e}^{2}
\label{MagMoment}
\end{equation}

$ \theta $ is the moment of inertia of the mass core, $ \omega _{m} $ and $ \omega _{e} $ 
are the angular velocities of the mass core and the charges, and $ r_{e} $ is the radius of a 
small object charge from the center. 

The intention of this first approach is to investigate only the order of magnitude of the electron size. 
Therefore, we calculate possible limits for the angular velocity $\omega_{m}$ from the experimental values.
A lower imit for $ \omega _{m}(min) $ can be derived from Eq.\ref{Ltot}, which 
allows a small but non-zero $ \omega _{m}(min)>0$ for $ \theta \to \infty $.
An upper limit for $ \omega _{m} $ can be calculated if we set in Eq.\ref{Etot} the rest mass $ mc^{2}=0 $. 
Inserting Eq.\ref{Etot} in Eq.\ref{Ltot} and set $ mc^{2}=0 $ it is possible to calculate with values from 
Tab.~\ref{Parameterelectron}  the angular velocity $ \omega _{m}(max) $  in Eq.\ref{AgularVelocite}.

\begin{equation}
\omega_{m}(max) = 4 \frac{E_{tot}}{\hbar} = 3.1 \cdot 10^{21}~~1/s
\label{AgularVelocite}
\end{equation}
 
As already mentioned, it is known from experiment that the rotation plane of the 
charge and mass of the electron is identical. There is no experimental information
available on the angular velocities $\omega_{m}$ and $\omega_{e}$ of the mass and charge.
It is also unknown at what distance from the center $ r_{m} $ the mass density approaches 
zero and at what radius $ r_{e} $ the charge of the small object is located. Following our 
discussion of forces and stability \ref{Forces and stability} in the introduction to this chapter, it is likely to assume 
hat some functional interaction, such as $ \omega_{m} = f ( \omega_{e}) $, exists 
between $ \omega _{m} $ and $ \omega _{e} $.
 
For a working hypothesis, we follow this philosophy by first using the simplest possible 
 assumptions, namely $ r_{m} = r_{e} = r $ and $\omega_{m} = \omega_{e} = \omega $.
Both hypotheses are supported by the discussion in the chapter on the pseudo-charge \ref{pseudo charge}
of gravitational interaction, since the electron's mass is most likely generated by the electron's electric charge. 
According to our ETAMFP model, this charge is mostly located at radius $ r_{e} $ 
and rotates around the center, which supports the assumption that the angular 
velocity of charge and mass is equal.  Using Eq.\ref{MagMoment}, the angular velocity $ \omega $ and, for simplicity, 
the charge $ Q = | 1 | $, the radius $ r(max) $ in Eq.\ref{RaduisAmax} can be 
calculated with $ \omega _{min}>0 $ and $ r(min) $ in Eq.\ref{RaduisAmin} 
with Eq.\ref{AgularVelocite}.

\begin{equation}
r (max) = \sqrt \frac {2 \mu_{e}}{ e \omega(min) } < \infty
\label{RaduisAmax}
\end{equation}

\begin{equation}
r (min) = \sqrt \frac {2 \mu_{e}}{ e \omega(max) } > 1.93 \cdot 10^{-13}~~m
\label{RaduisAmin}
\end{equation}

The limits on the radius estimated in this simple approach are very large.
In particular, the radius $r(max)$ is in complete contradiction to our experimental 
direct contact term regime \cite{CONTACT1} (Fig. \ref{Feynman2}).
We will therefore ignore this bound for further discussion. For the guidance of 
the next approach, the bound on $r(min)$ is more interesting.

The radius $ r(min) $ is far from the Planck or Grand Unification scale 
$ r(min) >> r_{Planck} $, which supports our assumption of using a 
classical approach for these calculations. The velocity of the outer mass radius
and the charge $ v = r(min) \times \omega = 6.0 \cdot 10^{8} \ m/s \ 
(\beta = 2.0) $ is greater than the speed of light $ v > c $.
 
To account for this fact in the next approach, one could use special or general 
relativity. We have already discussed the charge of electron orbits on a radiation-free trajectory.
This is the same condition as if the charge were moving at constant speed 
along a straight line. For this reason, in the following section, we use special 
relativity in the form of Lorentz transformations to avoid velocities $ \beta > 1 $. \\

\subsubsection{Charge radius of electron as a Lorentz contracted gyroscope.} 
\label{Lorentz contracted gyroscope} 

To investigate the next order of magnitude in a second approach, we 
describe the microscopic structure of the electron as before as a rotating 
sphere, similar to a mass-filled gyroscope, as shown in Fig. {\ref{Basic-ETAMFP}.
The charges, spin, and spin axis also correspond to the representation in Fig. {\ref{Basic-ETAMFP}.
We retain the radius condition $ r_{m}=r_{e}=r $. However, in this approach, we consider 
the result of the last calculation and consider the velocity of the outer mass radius
and the charge, is greater than the speed of light, and we consider the mass $ m_{0}c^{2} $.
We just discussed the radiation free path length $ p=2\pi r_{m}=2\pi r_{e}=2\pi r $ of the 
mass radius $ r_{m} $ and charge $ r_{e} $  must be corrected be the special relativity. It is 
necessary for this correction to distinguish between the radius $ r_{0} $ an observer 
placed on the rotating system will measure and the radius $ r_{e} $ an observer far 
away in rest will measure. This led to the conclusion that an external 
observer at rest measures a Lorentz-contracted value p as the charge's 
velocity $ v=\omega_{e}r_{0} $  approaches $ c $. The charge 
does not respond to special relativity, and the angular velocity $ \omega_{e} $ is only 
visible to an outside stationary observer. The moving charge does not behave 
like a clock in the plane of rotation. For this reason, only the angular velocity 
$ \omega _{e} $ exists. The measured magnetic moment comes from a charge 
moving along the trajectory $ p=2\pi r $. What is actually contracted in this 
rotating system is the radius $ r_{0} $, shown in Eq.\ref{Rcontracted}.

\begin{equation}
r_{0}=\frac{r_{e}}{\sqrt {1-\beta^{2}}}
\label{Rcontracted}
\end{equation}

In particular, it is important to examine the effects of the observer's position.
Is the observer outside or inside the electron? For example, is the radius
$r_{0}$ the radius inside the electron and $r_{e}$ the radius that an observer outside
the electron would measure? \\

\textbf {The Lorentz contraction affects magnetic and electric dipole moment of electron.}\\

Based on the data from Tab.~\ref{Parameterelectron} for the magnetic and 
electric dipole moment of the electron, it is noteworthy that $ \mu _{e} $ is a
finite, well-measured number and $ d_{e} $ is very small within the error
still zero. Our simple approach just discussed suggests that the velocity of the charge is close 
to the speed of light. Regarding the Lorentz contraction of the radius $r_{0}$ in 
Eq. \ref{Rcontracted} to $r_{e}$, the question arises as to which radius $r_{0}$
or $r_{e}$, together with the charge, generates the magnetic and electric dipole 
moment of the electron, measured by a stationary observer in rest in the laboratory.
In other words, to which radius $ r_{0} $ or $ r_{e} $ is the experiment sensitive to measure 
$ \mu _{e} $ or $ d_{e} $. For this reason, we investigate  the limits for  $ \mu _{e} $ and 
$ d_{e} $ given from the experimental values and our ETAMFP model to use $ r_{0} $ or $ r_{e} $

We first discuss the limit for the magnetic moment $ \mu _{e} $ .
To set a limit on $ \mu _{e} $ for the case that $ r_{0} $ generates the electron's magnetic 
moment, measured by an external observer at rest, we assume, following the above discussion, the most likely case,
$ \beta \to 1 $. The radius $ r_{0} $ is contracted to $ r_{e} $ by Eq.\ref{Rcontracted}.
The magnetic moment under these circumstances reads as in Eq.\ref{Magmomentr0}.

\begin{equation}
\mu_{e}=\frac{1}{2} e \omega_{e} r_{0}^{2}
       =\frac{1}{2} e \omega_{e} \frac {r_{e}^{2}}{(1-\beta^{2})}
\label{Magmomentr0}
\end{equation}

The velocity of the charge $ v $ is displayed in Eq.(\ref{Qvelocityr0}). 

\begin{equation}
\beta=\frac{v}{c}=\frac{\omega_{e} r_{0}}{c}
                 =\frac{\omega_{e}}{c}\frac{r_{e}}{\sqrt{1-\beta^{2}}}
\label{Qvelocityr0}
\end{equation}

Replacing $ r_{e} $ in Eq.(\ref{Magmomentr0}) with Eq.(\ref{Qvelocityr0})
allows to calculate the discussed limit in Eq.(\ref{Limitmur0}) if we
insert in Eq.(\ref{Limitmur0}) the numerical values from $ e $, $ c $ from
Tab. \ref{Parameterelectron} and $ \omega_{e} $ from
Eq.(\ref{AgularVelocite}).

\begin{equation}
\lim_{\beta\to~ 1}\mu_{e} =
\lim_{\beta\to~ 1} \frac{1}{2} e \frac{\beta^{2} c^{2}}{\omega_{e}}
= e \frac{1}{2} \frac{c^{2}}{\omega_{e}} = 2.3\times 10^{-24 }~~Am^{2}
\label{Limitmur0}
\end{equation}

This limit is close to the experimentally measured value shown in
Eq.(\ref{Comparisomuexpmulim}).

\begin{equation}
\mu_{e} ( experiment ) = 9.3 \times 10^{-24} ~~Am^{2} \sim
\mu_{e} ( limit ) = 2.3 \times 10^{-24}~~Am^{2}
\label{Comparisomuexpmulim}
\end{equation}

For the second case, to set a limit for $ \mu _{e} $ for the case that $ r_{e} $ generates the magnetic 
moment of the electron, measured by an internal observer at rest, we again 
assume the most likely case $ \beta \to 1 $ following the above discussion.
The radius in the rotating system $ r_{0} $ is finite and contracts to $ r_{e}\to 0 $ 
according to Eq.\ref{Rcontracted}. The angular velocity $ \omega _{e} $ is finite. The limit is
zero, as shown in Eq.\ref{MagMomentlowere}.

\begin{equation}
\lim_{r_{e}\to~ 0}\mu_{e} =
\lim_{r_{e}\to~ 0} \frac{1}{2} e \omega_{e} r_{e}^{2} = 0
\label{MagMomentlowere}
\end{equation}

This limit is in absolute contradiction to the experimentally measured value 
in Eq.(\ref{Comparisomuexpmulimre}). 

\begin{equation}
\mu_{e} ( experiment ) = 9.3 \times 10^{-24} ~~Am^{2} >>
\mu_{e} ( limit ) = 0 ~~Am^{2}
\label{Comparisomuexpmulimre}
\end{equation}

To conclude whether $ r_{0} $ or $ r_{e} $ generates the magnetic moment
$ \mu _{e} $ of the electron, the ETAMFP model points to $ r_{0} $ as the radius 
for generating the magnetic moment.

We second discuss the limit for an electric dipole moment $d_{e}$.
Classically, one requires for an electric dipole moment $d_{e}$, a charge $+e$, and $-e$ at a distance $r$.
It is not clear whether our ETAMFP model follows this rule, since very high stable points of
electromagnetic force would exist, as discussed in Fig. {\ref{Forces1}} at points SP I and SP II.
The charge doesn't change from negative to positive, but the force changes from repulsive to attractive.
For this reason, we ignore this fact for the following calculations and check whether $d_{e}\sim 0$.
  
To establish a limit for $ d_{e} $ for the case where $ r_{0} $ generates the electron's electric dipole moment,
measured by an external observer at rest, we again assume the most likely case $ \beta \to 1 $.
The radius $ r_{0} $ is contracted to $ r_{e} $ by Eq.\ref{Rcontracted}. The electric dipole moment under
these circumstances is as in Eq.\ref{Elecmomentr0}, if we replace $ r_{0} $
by $ r_{e} $ using Eq.\ref{Rcontracted}.  
  
\begin{equation}
d_{e} = e r_{0} = e \frac{ r_{e} }{\sqrt{1-\beta^{2}}}
\label{Elecmomentr0}
\end{equation}

Replacing $  r_{0} $ in Eq.\ref{Elecmomentr0} with Eq.\ref{Qvelocityr0} leads 
to the limit for $ d_{e} $ in Eq.\ref{Limitder0} if we use again $ c $  from Tab.~\ref{Parameterelectron}
and $ \omega _{e} $  from Eq.\ref{AgularVelocite}.

\begin{equation}
\lim_{\beta\to~ 1} d_{e} =
\lim_{\beta\to~ 1} e \frac{c}{\omega_{e}}\beta = e \frac{c}{\omega_{e}}
= 9.6 \times 10^{-14}~em
\label{Limitder0}
\end{equation}

This limit is about 17 orders of magnitude different from the experimentally 
measured value in Eq.\ref{Comparisondeexpr0}.

\begin{equation}
d_{e} ( experiment ) = 0.07\times 10^{-28} ~~e m <<
d_{e} ( limit ) = 9.6 \times 10^{-14}~~e m
\label{Comparisondeexpr0}
\end{equation}

To set a limit on $ d_{e} $ for the case that $ r_{e} $ generates the 
electric dipole moment of the electron, measured by an internal observer at 
rest, we again assume the most likely case $ \beta \to 1 $.
The radius in the rotating system $  r_{0} $ will be finite and contracted by equation 
Eq.\ref{Rcontracted} to $ r_{e}=0 $. The angular velocity $ \omega _{e} $
 will by finite. The limit is zero as shown in Eq.(\ref{LimitDipolre}).

\begin{equation}
\lim_{r_{e}\to~ 0}d_{e} =
\lim_{r_{e}\to~ 0} e r_{e} = 0
\label{LimitDipolre}
\end{equation}

This limit roughly agrees with the experimentally measured limit shown 
in Eq. (\ref{Comparisondeexpre}).

\begin{equation}
d_{e} ( experiment ) = ( 0.07 \pm 0.07 ) \times 10^{-28} ~~e m \sim
d_{e} ( limit ) = 0 ~~e m
\label{Comparisondeexpre}
\end{equation}

Taking into account the limits just discussed, the answer from the comparison between 
the experimental values and the calculated limits for the magnetic and electric
dipole moment of the electron concerning which radius $ r_{0} $ or $ r_{e} $ has to be used 
from the Lorentz contraction is rather clear. 

The measurement of the electron's magnetic moment is sensitive to the radius
$r_{0}$. This is the radius that an observer at the center of the electron would measure
in the CM - frame from the center to the charge rotating at $\omega_{e}$.   
  
The measurement of the electric dipole moment $d_{e}$ is sensitive to the 
Lorentz-contracted radius $r_{e}$, which an observer measures when located 
far outside the resting electron in the LAB - frame. The calculation yields a value of $d_{e}\sim 0$. 
The converse assumption, placing the observer on the rotating disk of the electron in the CM - frame 
also leads to a very small value of $d_{e}\sim 0$, but not so close to ZERO as we use $ r_{0} $.
     
It would be possible to use the magnetic dipole moment $ \mu _{e} $ or the electric 
dipole moment $ d_{e} $ to estimate a lower bound on the electron radius. It is not 
clear which bound is more realistic for determining an electron radius. To use
$ \mu _{e} $ would be in good agreement with the ETAMFP - model, a magnetic 
moment $ \mu _{e} $ exists, since an electric current ( cluster ore evenly distributed ) circulating around 
a center generates a $ \mu _{e} $. The calculations discussed point to such an explanation.

In contrast, the use of $d_{e}$ presents some significant problems.
In our model Fig. {\ref{Basic-ETAMFP}, we assumed that the charge
rotating around the electron's center resembles a cluster.
As mentioned above, classical theory requires an electric dipole 
moment, a charge $+e$, and a negative charge $-e$ at a distance of $r$.
Our ETAMFP model Fig. {\ref{Basic-ETAMFP}
consists ONLY of negative charges and NOT a combination of $ +e $ and $ -e $.
Such a distribution would not generate an electric dipole moment, and our 
calculations are consistent with this assumption.    

Considering Fig. {\ref{Forces1}, two stable points exist in the ETAMFP model
for the forces at SP II at $ r_{p} $ and SP I at radius $ r > r_{GU} $. It is logical
that in the ETAMFP model, the point SP I corresponds to the outer radius
of the electron and SP II to the radius of the nucleus.
As mentioned above, the classical dipole moment of the electron is $d_{e}=0$, 
since there is no charge or anti charge in the electron. However, at a short 
distance, the $G_{DS}+EM$ force changes from repulsive to attractive at 
points SP II and SP I. This could lead to charge-anti charge behaviour.     
In particular, the point SP I would be a good candidate for a similar electric 
dipole moment. In other words, $d_{e}$ is not definitively excluded from 
determining an electron radius. These facts also suggest us using the electric 
dipole limit $d_{e}$ to determine the outer radius of the electron radius.
  
\subsubsection {Using $ \mu _{e} $ to set the parameter limits of the electron after the Lorentz contraction.} 
 
Our approach for a geometrically extended electron is so advanced at this stage that 
we can estimate the parameters $ \omega $, rest mass $ m_{0}c^{2} $, rotational energy
$E_{rot}$, $ r_{0} $ and $ r_{e} $ under the condition that the speed of the rotating 
charge is close to the speed of light $ c $. In the further course of this step,
the magnetic moment of the electron $ \mu _{e} $  is described by the Eq.\ref{Magmomentr0} 
and the angular velocity $ \omega _{e} $ is described by the Eq.\ref{Qvelocityr0}, 
where $ \mu _{e} $ and $ \omega _{e} $ depend on $ r _{0} $.
  
To calculate the angular velocity $ \omega _{e} $, we insert Eq.\ref{Qvelocityr0}
into Eq.\ref{Magmomentr0} and get Eq.\ref{AngularVelocityr0}.  
  
 \begin{equation}
\omega_{e}=\frac{1}{2} \frac{e(c\beta)^{2}}{\mu_{e}}
\label{AngularVelocityr0}
\end{equation}

To calculate the rest mass $ m_{0}c^{2} $, we slightly modify Eq.\ref{Etot} to Eq.\ref{EtotErot}.  
  
\begin{equation}
E_{tot} = m_{0} c^{2} + E_{rot}
\label{EtotErot}
\end{equation}

As in the first approach, we assumed that the angular velocity of the center of mass
$ \omega_{m} $ and the charge are equal $ \omega_{m} = \omega_{e} = \omega $. 
This, together with Eq. \ref{Ltot}, allows us to represent the rotational energy 
$ E_{rot} $ in the form of Eq. \ref{ErotOmegaLh}.

\begin{equation}
E_{rot} = \frac{1}{2} \theta \omega^{2} = \frac{1}{2} L \omega
        = \frac{1}{4}\hbar \omega
\label{ErotOmegaLh}
\end{equation}

By substituting Eq. \ref{ErotOmegaLh} into Eq. \ref{EtotErot}, the rest mass $ m_{0} c^{2} $
can be calculated in Eq.\ref{Restmass}.

\begin{equation}
m_{0}c^{2} = E_{tot} - \frac{1}{4} \hbar \omega
\label{Restmass}
\end{equation}

The last missing parameter, the radius $ r_{0} $, can be easily calculated by rewriting Eq.\ref{Qvelocityr0} as
Eq.\ref{Radiusr0}.

\begin{equation}
r_{0} = \frac{\beta c }{\omega}
\label{Radiusr0}
\end{equation}

If we set $ \beta = 1 $ and take for the charge $ e $ ,
the speed of light $ c $ and the magnetic moment
$ \mu_{e} $ the numerical values from
Tab.\ref{Parameterelectron} is is possible to
calculate $ \omega $. We calculated $ \omega $
for the discussed option $ \mu_{e} $ is generated from
a total charge $ 2/3 \times e $ as discussed for our ETAMFP model
in  Fig.{\ref{Basic-ETAMFP} and for the possibility
this charge is $ 1 \times e $. The numerical values are
calculated with Eq.\ref{AngularVelocityr0} and shown
together with the assumption $ \beta \leq 1 $
in Tab.\ref{modelbeta1} first three lines. 
If the angular velocity $ \omega $ is known for the case the charge responsible
for $ \mu_{e} $ is $ 2/3 \times e $ or $ 1 \times e $ it is possible to
calculate with Eq.\ref{Restmass} the rest mass for both
options $ m_{0} c^{2} (2/3~e ) $ and $ m_{0} c^{2} (1~e ) $
displayed in Tab.\ref{modelbeta1} in the fourth and
fifth line. For the same both options it is possible to calculate
with Eq.\ref{ErotOmegaLh} the rotational energy
$ E_{rot}~(2/3~e) $ and $ E_{rot}~(1~e) $ taking into
account $ \omega (2/3~e ) $ and $ \omega (1~e ) $ and
the total energy $ E_{tot} $ with $ \hbar $ from
Tab.\ref{Parameterelectron}. The numerical values are shown
in Tab.\ref{modelbeta1} in the sixth and seventh line.
For both options, it is also possible to calculate the radii $ r_{0} (2/3~e)$
and $ r_{0}(1~e)$ using Eq. \ref{Radiusr0}, taking into account 
$ \omega (2/3~e ) $ and $ \omega (1~e ) $ and
the $ c $ from Table \ref{Parameterelectron} including
$ \beta = 1 $. The numerical values are
given in Table \ref{modelbeta1} in the eighth and ninth rows.
The last row of Table \ref{modelbeta1} gives the numerical
value for the Lorentz-contracted radius
$ r_{e} = r_{0} \sqrt{( 1 - \beta^{2})} \ge 0 $.
This value must, of course, be close to zero.

\newpage

\begin{table}
\caption{Size of electron mass core and charge radius for $\beta = 1$}
\label{modelbeta1}
\begin{center}
\begin{tabular}{||l|l|l||}                                      \hline
  parameter           & Symbol        & value
                                                               \\ \hline
  velocity charge     & $ \beta  $    & $ \le 1 $
                                                               \\ \hline
  angular velocity $ 2/3~e $
                      & $ \omega (2/3~e ) $
                                      & $ \le 5.1696325 \times 10^{20} $ 1/s
                                                               \\ \hline
  angular velocity $ 1~e $
                      & $ \omega (1~e ) $
                                      & $ \le 7.7544488  \times 10^{20} $ 1/s
                                                               \\ \hline
  rest mass $ 2/3~e $ & $ m_{0} c^{2} (2/3~e ) $
                      & $ \ge 0.425931077 $ MeV $ ( 83.3526374 \% ) $
                                                               \\ \hline
  rest mass $ 1~e $   & $ m_{0} c^{2} (1~e ) $
                      & $ \ge 0.383397155 $ MeV $ ( 75.028956 \% ) $
                                                               \\ \hline
  rotational energy $ 2/3~e $
                      & $ E_{rot}~(2/3~e) $
                      & $ \le 0.085067843 $ MeV $  ( 16.6473626  \% ) $
                                                               \\ \hline
  rotational energy $ 1~e $
                      & $ E_{rot}~(1~e) $
                      & $ \le 0.127601765 $ MeV $  ( 24.971044   \% ) $
                                                               \\ \hline
  charge radius $ 2/3~e $
                      & $ r_{0} (2/3~e)$ & $ \ge 5.7991058 \times 10^{-13} $ m
                                                               \\ \hline
  charge radius $ 1~e $
                      & $ r_{0}(1~e)$    & $ \ge 3.8660705 \times 10^{-13} $ m
                                                               \\ \hline
  charge radius rest system
                      & $ r_{e} = r_{0} \sqrt{( 1- \beta^{2})} $
                      & $ \ge 0 $ m
                                                               \\ \hline
\end{tabular}
\end{center}
\end{table}

The current estimate of this approach, just discussed, is that the speed 
of the rotating charge is close to the speed of light. Taking into account 
the Lorentz contraction of the charge rotating around the center of the 
electron, this explains why the magnetic moment is relatively large and 
the electric dipole moment very small. After discussing the scale 
dependence of SM and BBM, the radius $ r_{e} $ will certainly not be zero. 


\subsubsection{Mass and charge radius of electron using spezial relativity as Lorentz contracted gyroscope.}

Our model discussed so far did not contain any specific assumption about 
the electron's mass core. In the discussion on the geometric approach to 
FP's, summarised in Fig. {\ref{Basic-ETAMFP}, we assumed that the 
mass core of the FP's consists of two parts.
An inner, non-rotating mass kernel, the seeds of a quantised Planck-Era energy-mass 
condensate, and a outer rotating mass density distribution, the seeds of mass between 
the Planck and GUT Era. In the chapter on excited states of fermions \ref{Exited-Fermions}, we discussed the 
possibility of describing the electron as the vibrational ground state of leptons in 
Fig. {\ref{Excited-states2}. An energetically favourable object with such a massive 
inner core would be a classical outer rotating sphere, such as a gyroscope, 
with a rotation axis as its spin axis.\\

\textbf {Mass core as Lorentz-contracted rigid sphere.}\\

Following the philosophy of using the simplest possible approach, the most easily
accessible assumption is the use of a rigid sphere with constant mass density,
rotating with angular velocity $ \omega_{m} $ around the rotation axis and having 
a boundary at radius $ r_{0} $. The moment of inertia $ \theta $ in 
Eq. \ref{Imoment} is a function of the mass $ m_{0} $ and the radius $ r_{0} $

\begin{equation}
\theta = \frac{2}{5} m_{0} r_{0}^{2}
\label{Imoment}
\end{equation}

The corresponding rotational energy $ E_{rot} $ is shown in the Eq.\ref{IErot}.

\begin{equation}
E_{rot} = \frac{1}{2} \theta \omega_{m}^{2} = \frac{1}{2} L \omega_{m}
        = \frac{1}{4}\hbar \omega_{m}
\label{IErot}
\end{equation}

In Eq.\ref{Imoment}, we assumed that the radius $ r_{0}(mass)$ 
at which the mass density ends, as shown in Fig.{\ref{Basic-ETAMFP}
corresponds to the radius $ r_{0}(charge)$ at which the charge is located:
$ r_{0}(mass) = r_{0}(charge) = r_{0} $.
If we again assume $\omega_{m}=\omega_{e}=\omega $, as in the
previous chapter on the charge radius, the velocity of the
surface of the mass sphere at the position $r_{0}$ perpendicular
to the spin axis will exceed the speed of light. The conditions
are very similar to those in the case of the charge.
In accordance with our discussion in Chap.\ref{The flashing vacuum} ( The Flashing Vacuum ), 
we assume that the mass $ m_{0} $ and the radius $ r_{0} $ in Eq. \ref{Imoment} must be corrected 
using the Lorentz transformations. In our simple model of a rigid sphere, only the ring of mass with a 
crescent-shaped cross-section in the vicinity $ r_{0} $ is sensitive to $ \beta $.
The mass content of this ring rapidly decreases to zero as its inner radius $ r $ 
approaches $ r_{0} $. This mass decrease dampens the increasing sensitivity 
to $ \beta $ as $ \beta $ approaches $ \beta=1 $ at $ r=r_{0} $.
In contrast, the massless surface ring of the sphere near $ r_{0} $ is fully exposed 
to sensitivity to $ \beta $ and is significantly contracted since $ \beta $ is very close 
to $ \beta=1 $. In the limit of $ \beta=1 $, the surface ring in the 
vicinity $ r_{0} $ is contracted to a point at the center of the sphere.
 
To apply the Lorentz transformations on Eq.\ref{Imoment} we first estimate the magnitude 
of the sensitivity of $ m_{0} $ and $ r_{0} $  to $ \beta $.

We start with the $ \beta $-sensitivity of $ m_{0} $. The mass distribution of a rotating sphere 
is shown in Fig. \ref{Spheremassbeta}. 

\newpage

\begin{figure}
\begin{center}
 \epsfig{file=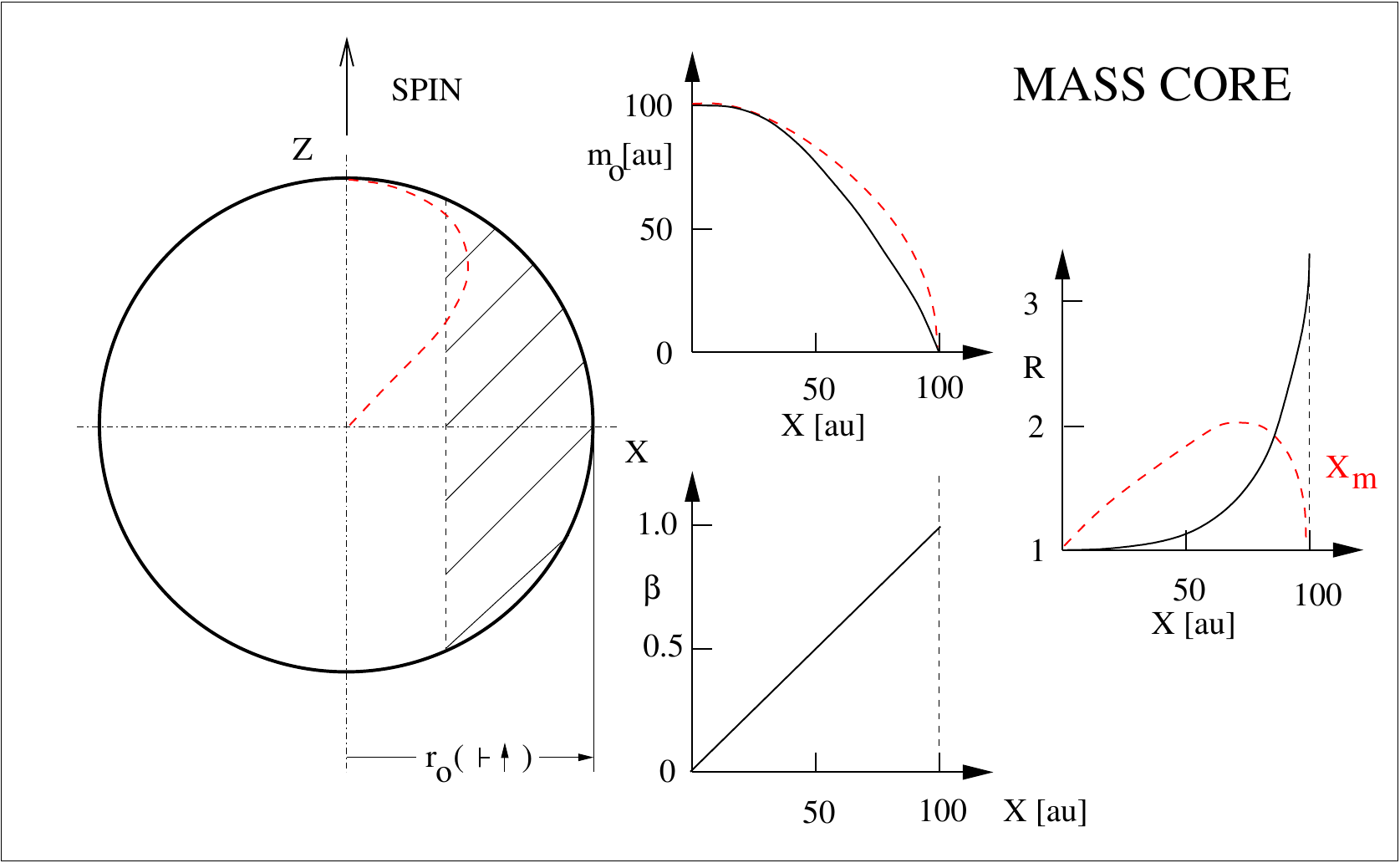,width=14.0cm,height=9.0cm}
\end{center}
\caption{ The spherical electron mass rotating $ \beta \rightarrow 1 $ }
\label{Spheremassbeta}
\end{figure}

On the left side of Fig. \ref{Spheremassbeta}, 
the mass sphere is drawn in two dimensions (z-axis and x-axis), rotating parallel to the z-axis.
Using the Lorentz transformations on Eq.\ref{Imoment},
 in the upper middle section, the mass distribution of the sphere is shown in arbitrary units 
 $ m_{0} $[au] as a function of $ x=X $ [au] at constant mass density. The mass distribution $ m_{0} $[au] (
 solid black line) decreases from $ 100 $ [au] at $ x={0} $ to zero at $ x=r_{0} ( \vdash~\Uparrow ) $. 
 The graph at the bottom middle  of Fig. \ref{Spheremassbeta} shows the linear increase
of $ \beta $ as a function of x according to Eq.\ref{Qvelocityr0}.
The graph at the far right of Fig.\ref{Spheremassbeta} shows the Lorentz factor
$ R = 1 / \sqrt { ( 1-\beta^{2} ) } $ as a function of x as a solid black line.
If we convolve the dependencies $ \beta $, $ R = 1 / \sqrt { ( 1-\beta^{2} ) } $ and 
$ m_{0} $ [au] on $ X $ [au], we can calculate the correction for the mass 
$ \Delta m ( x,\Delta x ) = \Delta m_{0} ( x,\Delta x ) /\sqrt { (1 - \beta^{2}) } $.

The correction $ \Delta m(x,\Delta x)=0 $ at $ X\left [ au \right ]=r=0 $
reaches a maximum at approximately $ X\left [ au \right ]=r=80 $ and approaches
$\Delta m(x,\Delta x)=0 $ at $ X\left [ au \right ]=r=100 $ or $ r=r_{0} $. This is
illustrated in the mass diagram (red dashed line) in Fig. \ref{Spheremassbeta}.
 Comparing the black line with the red dashed line in this diagram, one sees that the Lorentz 
 correction is only active for $ X\left [ au \right ] =r > 40 $, also shown in the black hatched 
 crescent-shaped region on the left side of Fig.\ref{Spheremassbeta}.
 The total mass increase originated by the Lorentz transformation is 
 approximately 11\% if $ \beta =1 $ at $ x = r_{0} ( \vdash~\Uparrow ) $. 
In this simple model, we have so far used a constant mass density distribution from $ r=0 $ to $ r=r_{0} $.
Following the discussion in Chap.\ref{The flashing vacuum} and
Fig. {\ref{Basic-ETAMFP}, we expect a mass distribution that decreases steeply 
from $ r=0 $ to $ r=r_{0} $, to $ \varrho =0 $ in the vicinity  $ r=r_{0} $.
This will suppress the above discussed edge effect of the rotating sphere to be 
negligible at $ x = r_{0} ( \vdash~\Uparrow ) $. For this reason, we ignore the Lorentz correction of 
$ m_{0} $ ( 11\% ) in the following calculations.

Second, we estimate the $ \beta $-dependence on $ r_{0} $. The boundary of the rotating 
sphere is a massless shell rotating at $ x = r_{0} ( \vdash~\Uparrow ) $ with maximum $ \beta =1 $.
The surface velocity
$ \beta $ decreases as the distance $ x $ from the spin axis decreases
$ x < r_{0} ( \vdash~\Uparrow ) $ and finaly becomes $ \beta =0 $ at the spin
axis, as shown in the lower panel on the middle side (black solid line) in Fig.\ref{Spheremassbeta}.
The distance from the spin axis $ x $ in the rotating system is subject to a Lorentz contraction
from Eq. (\ref{shellcontaction1}) to the distance $ x_{m} $ measured in an experiment outside the electron.

\begin{equation}
X_{m}=x_{m} = x \sqrt{1-\beta^{2}}
\label{shellcontaction1}
\end{equation}

The application of the Eq.\ref{shellcontaction1} to $ x $ in arbitrary units is 
shown in the right panel of Fig.\ref{Spheremassbeta} (red dashed line).
The contraction sets $ x_{m} $ to zero for $ \beta =1 $ at $ x = r_{0} ( \vdash~\Uparrow ) $, 
and the radius at $ x = 0 $ remains unchanged. A quarter of this contracted infinitely thin 
shell is shown in the red dashed line on the left side of Fig.\ref{Spheremassbeta}. After inspecting of
Fig.\ref{Spheremassbeta}, it is obvious that the $ \beta $-dependence 
on $ r_{0} $ must be taken into account to calculate the moment of inertia in Eq.\ref{Imoment}.\\

\textbf {Model factor respecting not rotating inner mass kernel.}\\

Following the discussion of the basic scheme in Fig. {\ref{Basic-ETAMFP}},
we expect that not all of the electron's mass is rotating.
This would imply that the inner Planck nucleus does not contribute to the rotational energy.
To account for this possible mass kernel , we introduce a general model factor $B$ for 
non-rotating and rotating inner mass kernels. The following calculation includes the 
possibility of non-rotating and rotating kernels.
We expect this factor to be $ 0 < B < 1 $, since the decreasing influence of the inner, 
non-rotating mass core dominates the maximum possible increase in mass at the edge 
of the mass sphere, which we investigated in the previous paragraph. For this reason, 
we modify the rest mass $ m_{0} $ to the model rest mass $m_{0}^{model}=Bm_{0}$. 

The radius limit, derived from the electron's electric dipole moment, depends only on the 
radius of the charge. We assume that the electron's charge is located at 
radius $ x = r_{0} ( \vdash~\Uparrow ) $. We discussed the limits for this radius 
in Eq.\ref{LimitDipolre} and Eq.\ref{Comparisondeexpre}. Following this discussion 
and Eq.\ref{shellcontaction1}, we set the minimum possible radius $x(min)=r_{e}$ at $z = 0$.
Combining our assumptions about the various angular velocities, the radius, and the model 
factor, we obtain the following results: $\omega_{m}=\omega_{e}=\omega $, 
$ x_{m}(min)=r_{m}=r_{e}=r_{L}$, and $ m_{0}^{model}=Bm_{0} $.
Taking these assumptions into account, we replace the radius $ r_{0} $ in 
Eq.\ref{Imoment} with Eq.\ref{r0mass1}.

\begin{equation}
r_{0} = \frac{r_{m}}{\sqrt{1-\beta^{2}}}
\label{r0mass1}
\end{equation}

Inserting the model factor and Eq.\ref{r0mass1} in Eq.\ref{Imoment} we get the Lorentz corrected mass moment
of inertia $ \theta $  including all our assumptions in in Eq.\ref{LImoment1} .

\begin{equation}
\theta = \frac {2}{5} B m_{0} \frac{r_{m}^{2}}{(1-\beta^{2})}
\label{LImoment1}
\end{equation}

In this phase of the development of our ETAMFP model, we investigate the
 influence of previously introduced assumptions on the mass and charge 
 radius of an electron, considered as a Lorentz-contracted gyroscope, on the model.
 For this reason, in a first step, we develop equations for the velocity
$ \beta $ , the angular velocity $ \omega $ , the energy of $ m_{0}c^{2} $ 
and $ E_{rot} $ , the radius $ r_{0} $ , the
Lorentz-contracted radius $ r_{L} $ as a function of the model factor $ B $.
 
In a second step we study the from the numerical measured parameters 
 in Tab.\ref{Parameterelectron} of the electron given numerical limits for the model factor $ B $ in five plots 
$ \beta ^{2}=f(B),\omega =f(B),E_{rot}=f(B),r_{0}=f(B) \ and \ r_{L}=f(B)$. \\

\textbf {Equations including model factor for rotating mass core.}\\ 

If we substitute Eq.\ref{Qvelocityr0} into Eq.\ref{LImoment1} and use 
Eq.\ref{IErot}, it is possible to calculate $ \beta $ as a function of 
$ \omega $ and $ B m_{0} $ in Eq.\ref{betaI}.

\begin{equation}
\beta^{2}=\frac{5}{4}\frac{\hbar \omega }{ B m_{0} c^{2}}
\label{betaI}
\end{equation}

Next, we replace $ \mu_{e} $ in Eq.\ref{Magmomentr0} by $ \mu_{e} = A \mu_{B} $ 
(With $ A $ as the anomalous magnetic moment), use $ r_{L} $ from
Eq.\ref{Qvelocityr0}, and determine the angular velocity $ \omega_{e} $ as a function of
$ \beta^{2} $ in Eq.\ref{AngularVelocityI}.

\begin{equation}
\omega_{e}=\frac{1}{2} \frac{e(c\beta)^{2}}{A\mu_{B}}
\label{AngularVelocityI}
\end{equation}

The experimental measured parameters $ e $, $ c $,
$ A=1.001159652187 $,
and $ \mu_{B} $ are known and shown in table \ref{Parameterelectron}. 
By substituting Eq.\ref{AngularVelocityI} into Eq.\ref{betaI} and
using Eq.\ref{IErot} with the definition of $ \mu_{B}=(e\hbar)/(2m_{e}) $, we obtain
the angular velocity $ \omega $ as a function of $ B $ in
Eq.\ref{omegaMODEL}.

\begin{equation}
\omega = \frac{E_{tot}}{\hbar}(4-\frac{5~\tilde{e}}{B A })
\label{omegaMODEL}
\end{equation}

According to our chapter on schemes of fermions  \ref{Scheme of geometrical extended FB's }, 
we use the model-dependent charge $ \tilde{e} $ in Eq.\ref{omegaMODEL},
where $ \tilde{e} $ could be $ \tilde{e} = 1 $ or $ 2/3 $.

To find a similar expression for $ \beta $, we insert
$ m_{0}c^{2} $ from Eq.\ref{Restmass}
into Eq.\ref{betaI} and replace $ \omega $
with Eq.\ref{omegaMODEL}. This allows us to calculate
the velocity $ \beta^{2} $ in Eq.\ref{betaMODEL}.

\begin{equation}
\beta^{2} = \frac {4~A~B-5~\tilde{e}}{B~\tilde{e}}
\label{betaMODEL}
\end{equation}

This equation depends only from the anomal magnetic moment $ A $,
the reduced charge $ \tilde{e} $ and the model factor $ B $.

A specific value of the model parameter $ B $ defines
via Eq.\ref{betaMODEL} and Eq.\ref{omegaMODEL}
a velocity $ \beta $ and the angular velocity $ \omega $.
By Inserting $ \omega $ for the calculation of
the energy $ m_{0} c^{2} = f( B ) $
and $ E_{rot} = f( B ) $ we use Eq.\ref{Restmass}
and Eq.\ref{IErot}. 
For the calculation of the
radius $ r_{0} $ we insert $ \beta $ and $ \omega $
in Eq.\ref{Radiusr0}.

For the last missing parameter $ r_{L} = f(B) $
we use the Eq.\ref{Qvelocityr0} solved for $ r_{L} $ and
substitute the angular velocity $ \omega $ from
Eq.\ref{omegaMODEL} and the velocity $ \beta $ from Eq.\ref{betaMODEL}
to calculate Eq.\ref{RlLorentz}.

\begin{equation}
r_{L} = \frac {c}{\omega}\beta\sqrt{1-\beta^{2}} =
        \frac {c}{\frac{E_{tot}}{\hbar}(4 - \frac{5~\tilde{e}}{B A})}
        \sqrt{\frac{4 A B - 5 ~ \tilde{e}}{B ~\tilde{e}}}
        \sqrt{1-\frac{4 A B - 5 ~ \tilde{e}}{B ~\tilde{e}}}
\label{RlLorentz}
\end{equation}
 
 The current state of our calculations now allows us to calculate
the velocity $ \beta $, the angular velocity $ \omega $,
the energy of $ m_{0} c^{2} $ and $ E_{rot} $, the radius $ r_{0} $, and
the Lorentz-contracted radius $ r_{L} $ as a function of the
model factor $ B $.\\
 
\textbf {Numerical limits of the model factor determined by experiments.}\\ 

As announced, in a second step, we investigate the numerical limits for the 
model factor $ B $ dictated by the numerically measured parameters in 
table \ref{Parameterelectron} of the electron in five diagrams: $ \beta^{2} = f ( B ) $,
$ \omega = f ( B ) $, $ E_{rot} = f ( B ) $,
$ r_{0} = f ( B ) $, and $ r_{L} = f ( B ) $.

Important for the following discussions is the test to determine which 
charge $ \tilde{e} = 1 $ or $ \tilde{e} = 2/3 $ is consistent with our 
previously discussed model. In table \ref{modelbeta1}, 
it was not possible to distinguish between the two possibilities.

Substituting $B=1$ and $\tilde{e} = 1$ into Eq.\ref{betaMODEL},
we calculate a negative value of $\beta^{2}=-0.995361392$, which leads to 
an imaginary value of $\beta$. Using the experimental permissible limits for
$0\le\beta^{2}\le1$, we obtain a corresponding limit for
the model factor $B$ in the range of $1.24\le B\le1.66$.
The imaginary value $ \beta $ for $ B = 1 $ is in disagreement the Lorentz 
transformation. The calculated model factor $ B $ in the range $ 1.24 \le B \le 1.66 $  
contradicts our expectation of $ B \le 1 $.

If we substitute $ B = 1 $ and $ \tilde{e} = 2/3 $ into Eq.\ref{betaMODEL}, 
we calculate a $ \beta $ that is slightly larger than one of $ \beta^{2} = 1.006957912 $. 
By substituting the physically permissible limits of $ 0 \le \beta^{2} \le 1 $ into Eq.\ref{betaMODEL}, 
we calculate a model factor in the range of $ 0.832368076 \le B \le 0.998610351 $ shown Fig.\ref{BETAfunctionB}.

\newpage

\begin{figure}
\begin{center}
 \epsfig{file=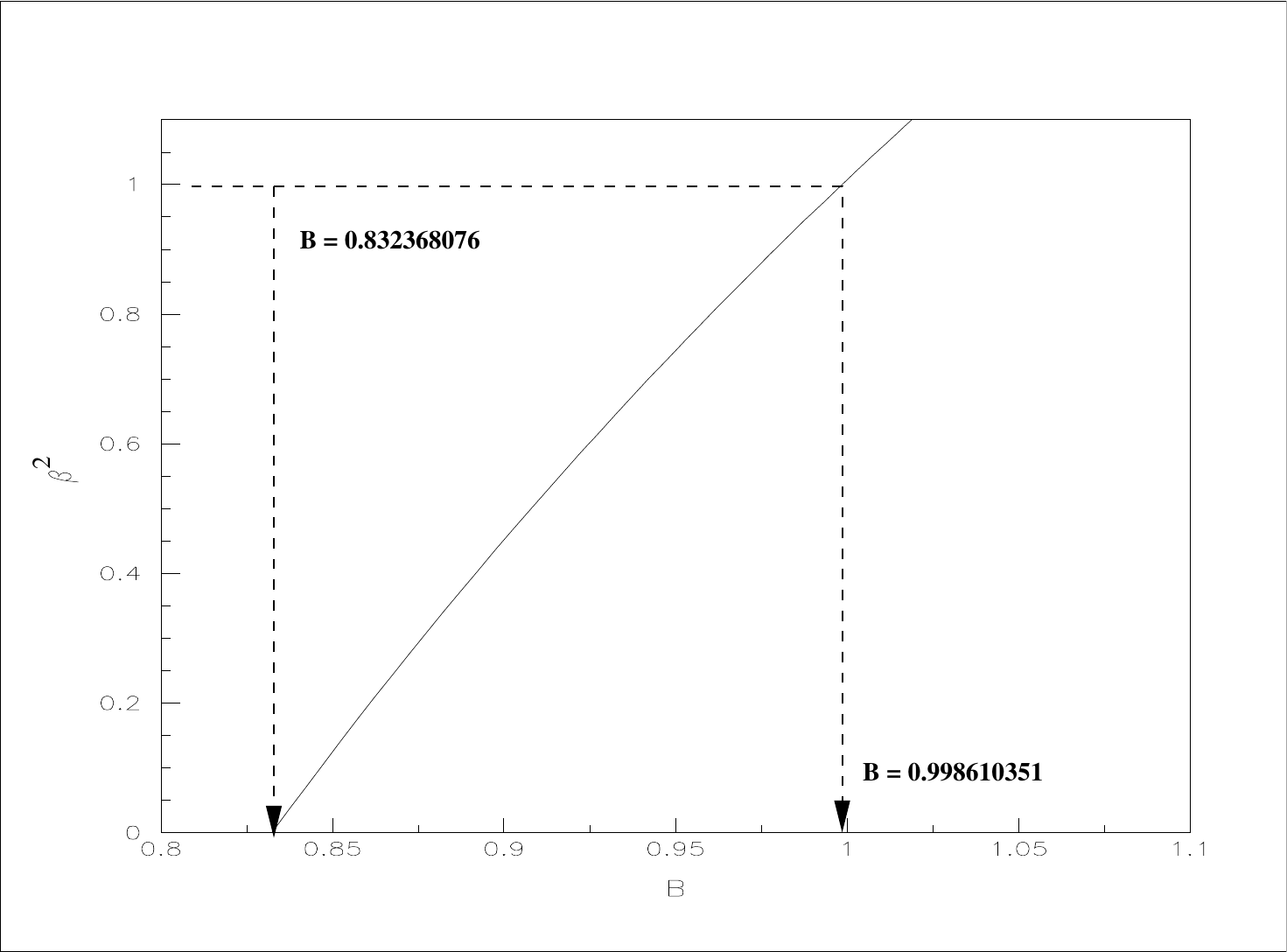,width=16.0cm,height=12.0cm}
\end{center}
\caption{ Possible range of $ B $ as function from $ \beta^{2} $}
\label{BETAfunctionB}
\end{figure}

Interestingly, the anomalous magnetic moment A and $ \tilde{e} = 2/3 $ are the leading 
parameters in Eq.\ref{betaMODEL} that determine the numerical value of $ \beta^{2} $.
According to our ETAMFP model, we expect a real value of $ \beta \le 1 $ and
a model factor of $ B \le 1 $. This is consistent with
our geometric approach in Fig.{\ref{Basic-ETAMFP}}
and the overall fermion scheme, in which we assume that
the electron's charge is placed in units of $ 1/3 $ in
three possible locations. For this reason, we consider only 
the case of $ \tilde{e} = 2/3 $ in the following discussion.

The solid black line in Fig.\ref{BETAfunctionB} represents the 
function discussed. For $ \beta^{2} = 1 $, we find the
upper limit for $ B = 0.99861035 $ and, for a non-rotating sphere, 
correspondingly for $ \beta^{2} = 0 $, a lower limit for
$ B = 0.832368076 $, shown in dashed solid lines with an arrow 
in Fig.\ref{BETAfunctionB}. The limits of $ B $ are used in the following diagrams:
$ \omega = f ( B ) $ , $ E_{rot} = f ( B ) $, $ r_{0} = f ( B ) $ and $ r_{L} = f ( B ) $, 
to investigate the behaviour of $ \omega $, $ E_{rot} $ and
$ r_{L} $ in the context of our ETAMFP model.

To calculate the graph $ \omega = f ( B ) $ we use
Eq.\ref{omegaMODEL} and vary $ B $ from $ B = 0.7 $
to $ B = \infty $ to show the entire possible range of $ \omega $ in Fig.\ref{OMEGAfunctionB}.
The solid black line in Fig.\ref{OMEGAfunctionB} represents the 
function discussed.

\begin{figure}
\begin{center}
 \epsfig{file=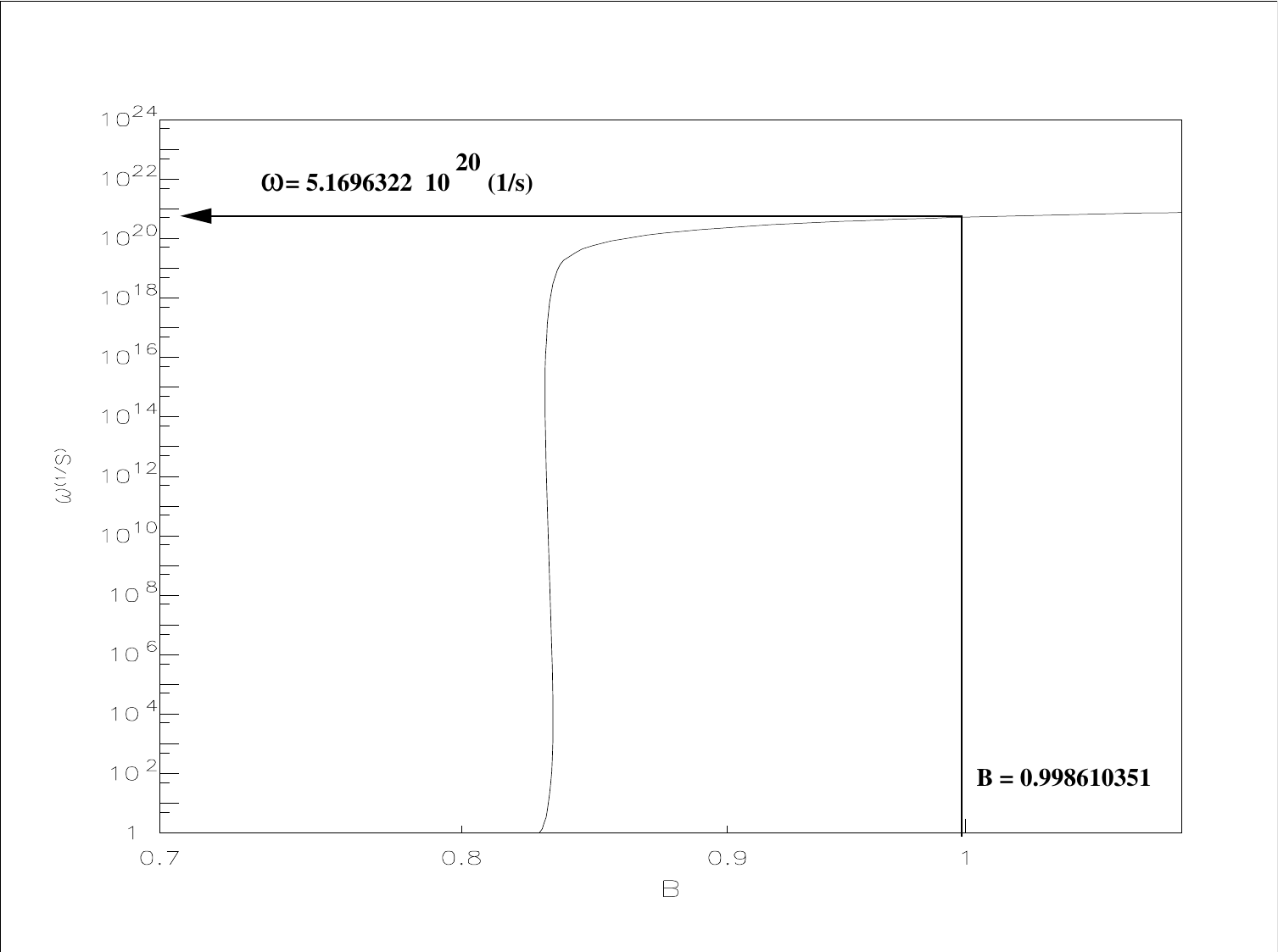,width=16.0cm,height=12.0cm}
\end{center}
\caption{ Possible range of $ \omega $ as function from $ B $ }
\label{OMEGAfunctionB}
\end{figure}

If we substitute $B=0.998610351$ into Eq.\ref{omegaMODEL}, 
which corresponds to $\beta=1$, we calculate a
maximum value of the angular velocity
$\omega_{max} = 5.1696322\times10^{20}~(1/s)$. This value
corresponds to the value in Table \ref{modelbeta1} for
$\omega ( 2/3 e)$ and is shown in Fig.\ref{OMEGAfunctionB}
as a solid straight black line with an arrow. 
If we substitute $B=0.832368076$ into Eq.\ref{omegaMODEL}, which
corresponds to $\beta=0$, we naturally calculate
a minimum value of the angular velocity of
$\omega_{min} = 0.0~(1/s)$, which is only important for the
completeness of the representation.
We obtain a similar result if we substitute $ B = \infty $ into Eq.\ref{omegaMODEL}. We obtain the
maximum possible value of $ \omega_{max}=3.1053763 \times 10^{21} (1/s) $.
This value is only mathematically interesting, since it would exceed $ \beta = 1 $.

To calculate the representation $ E_{rot} = f ( B ) $
and $ m_{0} c^{2} = f ( B ) $ shown in Fig.\ref{ROTmassfunctionB}, we vary
the model parameter $ B $ from $ B = 0.8 $ to $ B = 1.0 $,
and calculate the corresponding angular velocity
$ \omega $ using Eq.\ref{omegaMODEL}. We then substitute $ \omega $ into
Eq.\ref{Restmass} to obtain $ m_{0} c^{2} = f ( B ) $, and
Eq.\ref{IErot} to determine $ E_{rot} = f ( B ) $.

\newpage

\begin{figure}
\begin{center}
 \epsfig{file=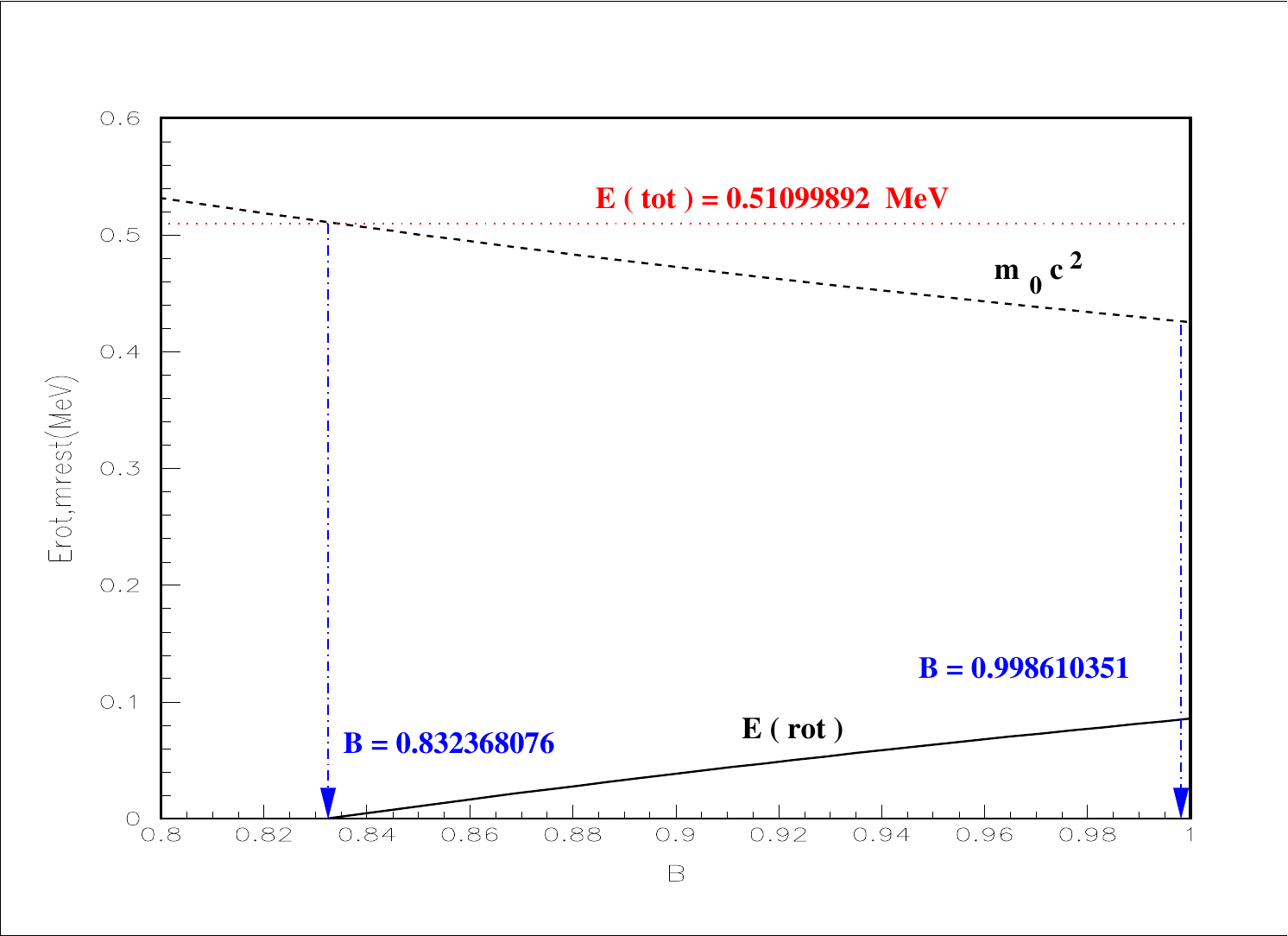,width=16.0cm,height=12.0cm}
\end{center}
\caption{ Possible range of $ m_{0} c^{2}$ and $ E_{rot} $ as function from $ B $ }
\label{ROTmassfunctionB}
\end{figure}

The black dashed line in Fig.\ref{ROTmassfunctionB} shows $ m_{0} c^{2} = f ( B ) $ and the black
solid line $ E_{rot} = f ( B ) $.
The maximum possible fraction for $ B = 0.998610351 $ between
$ m_{0} c^{2} $ and $ E_{rot} $ is $ m_{0} c^{2} = 0.425931082~MeV $
or $ (83.35 \% )$ and $ E_{rot} = 0.085067838~MeV $ or
$ (16.65 \% )$. These values are in agreement with the
numbers for $ m_{0}c^{2}( 2/3 e ) $ and $ E_{rot} ( 2/3 e ) $
in table \ref{modelbeta1} and marked by a blue dot dashed line
with arrow on the right side in Fig.\ref{ROTmassfunctionB}.
The lower limit for $ B = 0.832368076 $ results from
$ E_{rot} = 0.00~MeV $ and $ m_{0} c^{2} = E_{tot} $ and is
marked in Fig. \ref{ROTmassfunctionB} on the left as a blue 
dashed line with an arrow, along with the numerical value
for $ E_{tot} $ as a dotted red line.

To calculate the representation $ r_{0} = f ( B ) $
shown in Fig.\ref{ROfunctionB}, we vary the model parameter
$ B $ from $ B = 0.8 $ to $ B = 1.0 $, calculate the velocity $ \beta^{2} $ using
Eq.\ref{betaMODEL} and the angular velocity $ \omega $ using Eq.\ref{omegaMODEL}
and insert these values in Eq.\ref{Radiusr0} to find $ r_{0} = f ( B ) $
shown as black solid line in Fig.\ref{ROfunctionB}.

\newpage

\begin{figure}
\begin{center}
 \epsfig{file=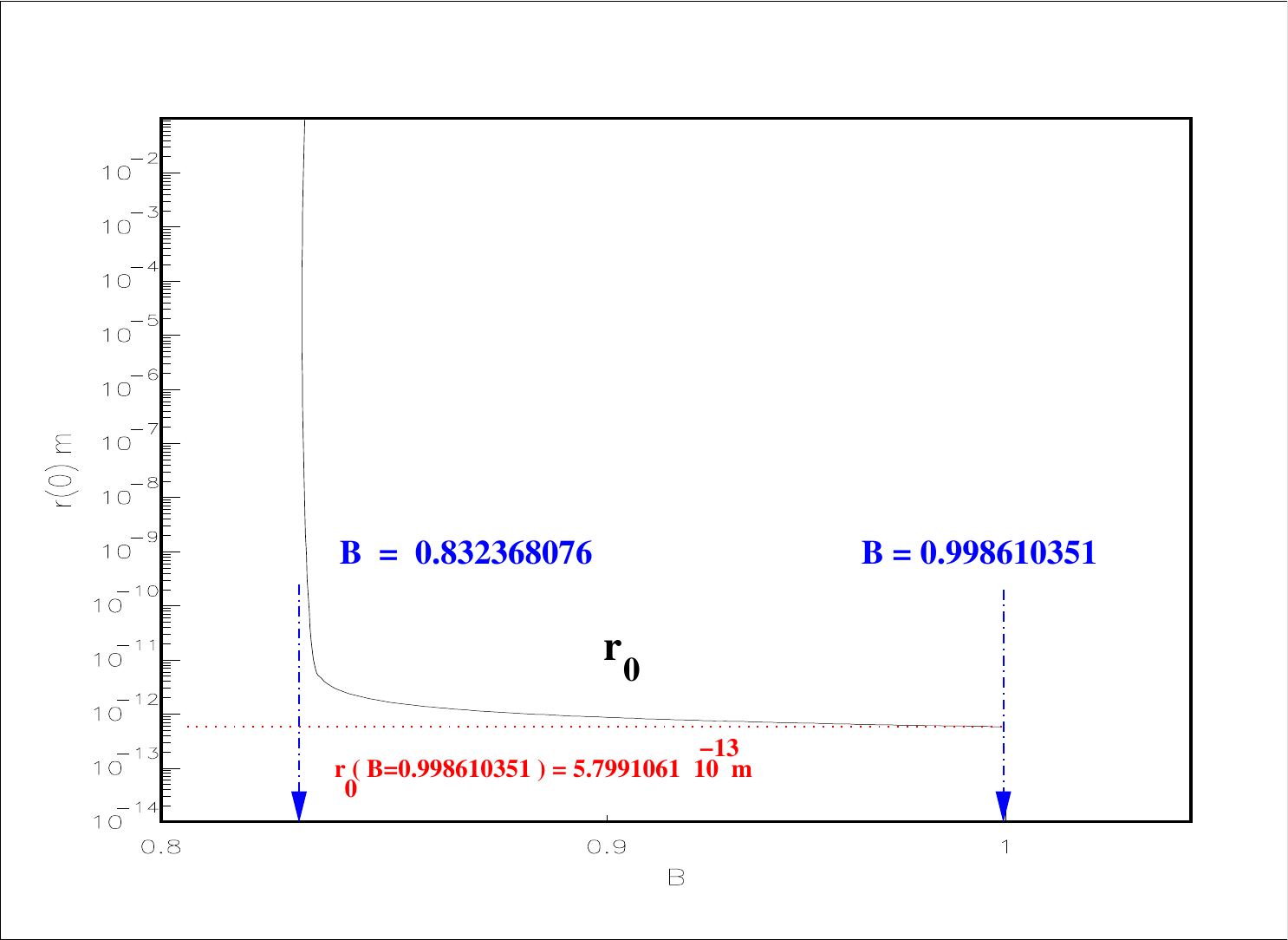,width=16.0cm,height=12.0cm}
\end{center}
\caption{ Possible range of $ r_{0} $ as function from $ B $ }
\label{ROfunctionB}
\end{figure}

An observer located in the center of the electron
rotating with $ \omega $ would measure a minimum possible
$ r_{0} $ allowed by the experimental date of
$ r_{0}=5.7991061 \times 10^{-13}~m $ for the condition
$ \beta = 1 $ and $ B=0.998610351 $. 
This value agrees with the value for $ r_{0} ( 2/3 e ) $ in table \ref{modelbeta1} 
and is marked by the crossing of the red dotted line with the blue dashed
line on the right side in Fig. \ref{ROfunctionB}.
The maximum of this radius increases slowly with decreasing 
$ B $ to $ r_{0} \rightarrow \infty $ at $ B = 0.832368076 $, represented by 
the blue dashed line on the left in Fig. \ref{ROfunctionB}. This 
singularity is certainly not supported by the data, since the 
electron has a finite spin and is merely a mathematical possibility.

To calculate the representation of the Lorentz-contracted radius
$ r_{L} = r_{exp} = f ( B ) $ of the electron, measured by an
observer external to the electron, we vary the model parameter from
$ B = 0.832368076 $ to $ B = 1.0 $ in Eq.\ref{RlLorentz}
and draw the black solid line in Fig. \ref{RexpfunctionB}.

\newpage

\begin{figure}
\begin{center}
 \epsfig{file=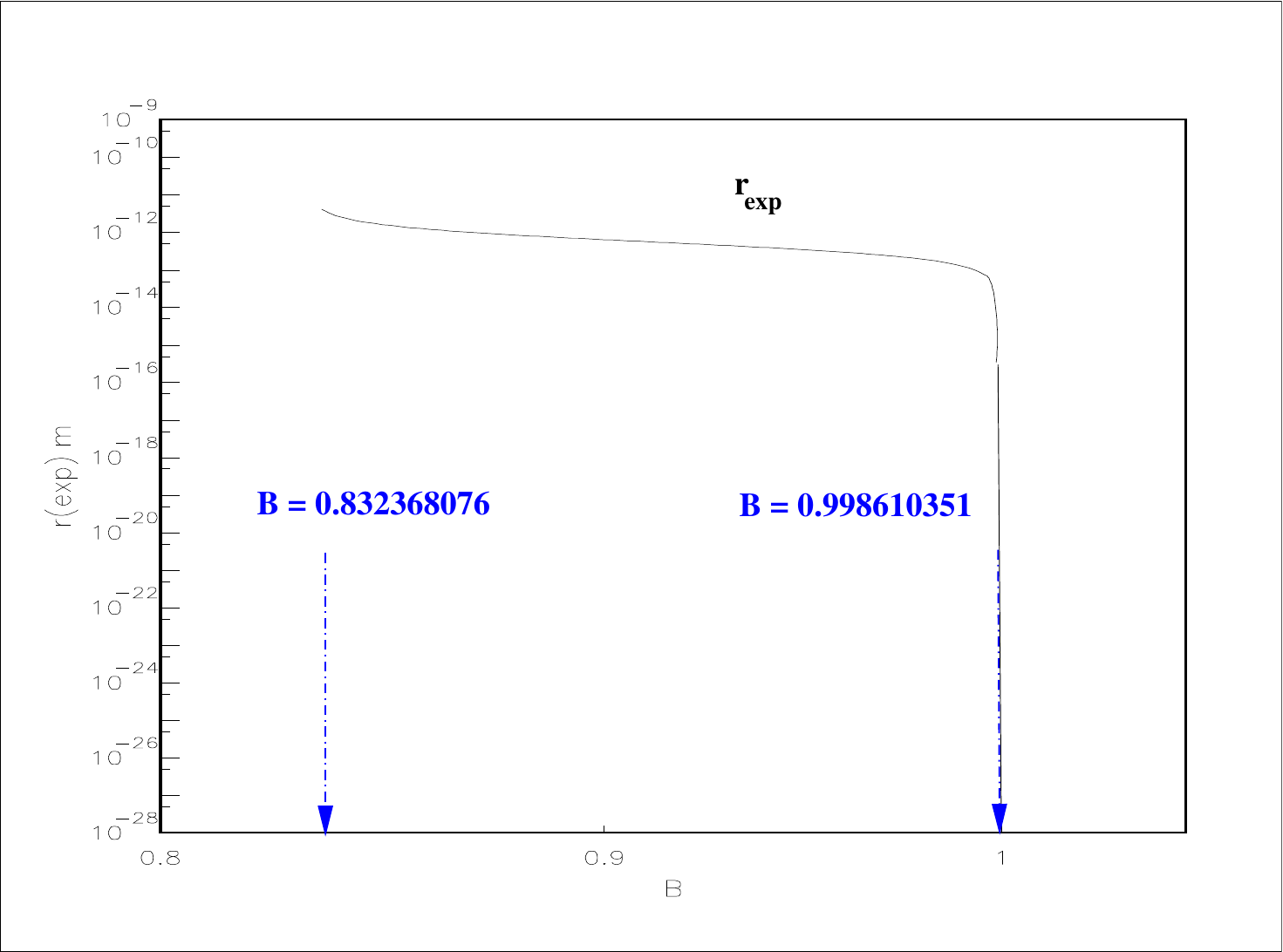,width=16.0cm,height=12.0cm}
\end{center}
\caption{ Possible range of $ r_{L} $ as function from $ B $ }
\label{RexpfunctionB}
\end{figure}

The square root structure of the Eq.\ref{RlLorentz} leads 
to a steep gradient singularity when $ B \rightarrow 0.998610351 $, 
which also includes $ \beta \rightarrow 1 $. This is depicted by a
 blue dot dashed line with an arrow on the right side of Fig. \ref{RexpfunctionB}.
For $ B \rightarrow 0.832 368 076 $ $ r_{L} $ reaches a maximum
marked by a blue dashed dotted line with arrow left side in
Fig.\ref{RexpfunctionB}. The possible
from the experiment and model allowed values ranging from
is $ 8.6 \times 10^{-12}~(m) \ge r_{L} \ge 0.0 ~ (m) $.

One more approximation for $ r_{L} $ would be to use the measurement of the 
magnetic moment $ \mu _{e} $, the charge of the electron $ e $ from 
Table \ref{Parameterelectron}, and the frequency $ \nu =\omega /2\pi $ from Table \ref{modelbeta1}.
to calculate $ r_{L} $. The classical magnetic moment $ \mu _{e} $ is given in Eq.\ref{Mag-Moment},
including the mass of the electron $ m_{e} $ and the angular momentum $ L $.

\begin{equation}
\mu _{e} =\frac{-e}{2m_{e}}L
\label{Mag-Moment}
\end{equation}

The angular momentum $ L $ is a function of $ m_{e} $ the frequency $  \nu  $ and 
$ r_{L} $ shown in Eq.\ref{Angular}.

\begin{equation}
L=2\pi m_{e}\nu r_{L}^{2}
\label{Angular}
\end{equation}

By inserting Eq.\ref{Mag-Moment} into Eq.\ref{Angular}, $ r_{L} $ Eq.\ref{RadiusL} can be calculated.

\begin{equation}
r_{L}=\sqrt{\frac{\mu _{e}}{-e\pi \nu }}=8.39035\times 10^{-13}\left [ m \right ]
\label{RadiusL}
\end{equation}

This radius $ r_{L} $ would be the un-contracted radius $ r_{L}=r_{0} $ of the electron as a function 
of the magnetic moment $ \mu _{e} $, since $ \mu _{e}=f(r_{0}) $.

\subsubsection { Implication of the electron's electric dipole moment on its geometric expansion. }

In all previous calculations, we have not used the electron's electric dipole moment 
$d_{e}$ because the experimental value is very small. The excellent, very important, 
and demanding experiments \cite{Dipol11} have so far measured a value for $d_{e}$ 
shown in Table \ref{Parameterelectron}, which is comparable to zero, which must be 
considered an important limit for the ETAMPF model.

In all previous considerations of the classical approach, the charges of
 the electron for example in Fig. {\ref{Basic-ETAMFP} are separated by a distance 
 $ r_{0} $. A classical dipole consists of two charges of opposite signs 
 separated by a distance $ x $. Our ETAMFP model contains two electric charges $ - 1/3 \times e $
of the same sign ( Not charge/anti-charge ), separated by a distance $ r_{0} $. In this sense,
our model is not a classical dipole, but it is quite similar.
Our approach of the sensitivity of the experiment to measure the electric dipole moment
in Eq.\ref{Comparisondeexpr0} to Eq.\ref{Comparisondeexpre}
leads to the conclusion that the experiments measure the
distance between the charges in the LAB system. Considering the value for
$ r_{e} $ in Table \ref{modelbeta1} and the function of $ r_{e}=f(B) $
in Fig. \ref{RexpfunctionB}, it is obvious that as the velocity
of the charges approaches $ c $, the Lorentz contraction pushes the radius $ r_{e} $
towards zero. This is consistent with the tiny measurement of $d_{e}$.
For these reasons, we identify $r_{e}$ with the measured electric dipole 
moment $d_{e}$. In this sense, $d_{e}$ is a direct measurement of the 
distance between the two charges $-1/3$ orbiting the electron's mass 
kernel, measured by an observer external to the electron in the LAB frame.

Respecting the experimental challenge that a discovery potential of about $ 5 \times \sigma $ 
for measuring the electron's electric dipole moment has not been achieved so far, and 
considering the conditions just discussed, we introduce the dipole moment into our 
ETAMFP model and use the radius $ r_{e} $ derived from the dipole moment.
In a first step, we use the radius $ r_{e} $ measured by the observer in the 
laboratory frame to calculate $ \beta $, the angular velocity $ \omega $, the rest mass 
$ m_{0}c^{2} $, the rotational energy $ E_{rot} $, and the radius $ r_{0}$ of the 
electron without making any specific assumptions about the electron's
 nuclear mass. This means that we ignore the model factor $ B $ 
 discussed above. In a second step, we consider this model factor $ B $.

In the first step, we calculate the radius $ r_{e} $ from the electric dipole moment. 
As shown in Fig. {\ref{Basic-ETAMFP} and Fig. {\ref{Electron11}}, the three 
charges $ -1/3 $ in our scheme generate the electric dipole moment.
It would be possible to group the two outer charges $ -1/3 \times e \times 2 r_{e} $
or the inner charge twice with the outer charges $ 2 \times ( -1/3 \times e \times r_{e} ) $, 
which would lead to the same result of the dipole moment in Eq.\ref{ElectronDipol}.

\begin{equation}
d_{e} = e~l = (\frac{1}{3} e ) ( 2 r_{e} )
            =  2((\frac{1}{3} e ) ( r_{e} ))
            =  ( 0.07 \pm 0.07 ) \times 10^{-26} e~cm
\label{ElectronDipol}
\end{equation}

The numerical value is taken from table \ref{Parameterelectron}.
The according experimental measured radius $ r_{e} $
is shown in Eq.\ref{ElectronDipolRadius}.

\begin{equation}
      r_{e} = \frac{3}{2}~l = 0.105 \times 10^{-28}~m
\label{ElectronDipolRadius}
\end{equation}

To calculate an equation for $ \beta^{2} $,
we substitute $ \omega _{e} $ from Eq.\ref{Qvelocityr0} into
Eq.\ref{Magmomentr0} and obtain Eq.\ref{BETAsquer}.

\begin{equation}
      \beta^{2} = \frac{(\frac{2\mu_{e}}{c~e~r_{e}})^{2}}
                  {1 + {(\frac{2\mu_{e}}{c~e~r_{e}})^{2}}}
\label{BETAsquer}
\end{equation}

To simplify Eq.\ref{BETAsquer}, we introduce the 
constant factor $ K $ of Eq.\ref{Kfactor}.

\begin{equation}
      K = (\frac{2\mu_{e}}{c~e~r_{e}})^{2}
\label{Kfactor}
\end{equation}

We express the equation for the velocity of the charge $ 2/3 \times e $
of the electron $ v = \beta \times c $ in the very compact form of $ \beta^{2} $ in
Eq.\ref{BETAKfactor}.


\begin{equation}
      \beta^{2} = \frac{K}{1+K}
\label{BETAKfactor}
\end{equation}

Since $ K \ge 0 $ is in quadratic form, the value of $ \beta $ lies between
$ 0 \le \beta \le 1 $. To calculate the angular velocity $ \omega_{e} $ of the charges,
we solve Eq.\ref{Qvelocityr0} for $ \omega_{e} $ and insert the K-factor from 
Eq.\ref{Kfactor} into this equation, which is shown in Eq.\ref{omegare}.

\begin{equation}
      \omega_{e} = \frac{c}{r_{e}} \frac{1}{1+K} \sqrt{K}
\label{omegare}
\end{equation}

By inserting Eq.\ref{omegare} into Eq.\ref{Restmass}, the rest mass
$ m_{0} c^{2} $ is defined in Eq.\ref{restmassK}.

\begin{equation}
       m_{0} c^{2} = E_{tot} -\frac{1}{4} \hbar \frac{c}{r_{e}}
                     \frac{1}{1+K}\sqrt{K}
\label{restmassK}
\end{equation}

Inserting Eq.\ref{omegare} into Eq.\ref{IErot} allows the calculation
of the rotational energy $ E_{rot} $ in Eq.\ref{Erotre}.

\begin{equation}
       E_{rot} = \frac{1}{4} \hbar \frac{c}{r_{e}} \frac{1}{1+K}\sqrt{K}
\label{Erotre}
\end{equation}

By inserting Eq.\ref{BETAKfactor} into Eq.\ref{r0mass1},
the charge radius of the electron in the rotating system 
( CM - frame ) can be calculated in Eq.\ref{r0fuctionK}.

\begin{equation}
       r_{0} = r_{e} \sqrt { 1 + K }
\label{r0fuctionK}
\end{equation}

We omit the recalculation of the various parameters from Table \ref{modelbeta1}
using Eq.\ref{ElectronDipol} to Eq.\ref{r0fuctionK},
since the change in the parameters is negligible if we assume $\beta \approx 1$.

As previously announced, in a second step, we incorporate the model factor 
$ B $ into our calculations. First, we calculate the factor $ B $ itself as a function
 of the magnetic moment $ \mu_{e} $, the speed of light $ c $, the charge $ \tilde{e} $,
 and the radius $ r_{e} $, summarised in the factor $ K $ of Eq.\ref{Kfactor}.
Using Eq.\ref{BETAKfactor} and Eq.\ref{betaMODEL}, the model factor $ B $ 
can be calculated in Eq.\ref{BfactorfunctionK}.

\begin{equation}
       B = \frac{(1+K)5\tilde{e}}{4 A ( 1+K ) - K \tilde{e} }
\label{BfactorfunctionK}
\end{equation} 

The factor $B$ is represented in simple form as a function of $K$,
the anomalous magnetic moment $A$, and the reduced charge $\tilde{e}$.
By inserting the factor $B$ into the equations from Eq.\ref{betaI} to
Eq.\ref{RlLorentz}, the influence of the factor $B$ can be investigated, 
which corresponds to the influence of the electric dipole moment of 
the electron on our Lorentz-contracted mass core.

For the further discussion of our ETAMFP model, it is important to investigate 
the numerical influence of the measured value of the electric dipole moment 
on our ETAMFP model. We inserting for this reason the electric dipole moment
from Tab. \ref{Parameterelectron} in Eq.\ref{ElectronDipol} to
Eq.\ref{r0fuctionK} to calculate the numerical values of the parameters
$ \beta $, the angular velocity $ \omega $, rest mass $ m_{0}c^{2} $,
rotational energy $ E_{rot} $ and the radius of the $ r_{0} $.
The numerical values are shown in Tab. \ref{electronPARAMETERSdipol}.

\newpage

\begin{table}
\caption{Size of electron mass core and charge radius using the
         electric dipole moment}
\label{electronPARAMETERSdipol}
\begin{center}
\begin{tabular}{||l|l|l||}                                      \hline
  parameter           & Symbol        & value
                                                               \\ \hline
  velocity charge     & $ \beta  $    & $ < 1 $
                                                               \\ \hline
  angular velocity $ 2/3~e $
                      & $ \omega  $
                                      & $ = 5.1696321 \times 10^{20} $ 1/s
                                                               \\ \hline
  rest mass $ 2/3~e $ & $ m_{0} c^{2}  $
                      & $ =  0.425931083 $ MeV $ ( 83.3526386 \% ) $
                                                               \\ \hline
  rotational energy $ 2/3~e $
                      & $ E_{rot} $
                      & $ = 0.085067837 $ MeV $  ( 16.6473613  \% ) $
                                                               \\ \hline
  charge radius $ 2/3~e $
                      & $ r_{0} $ & $ =   5.7991062 \times 10^{-13} $ m
                                                               \\ \hline
  charge radius rest system
                      & $ r_{e} = r_{0} \sqrt{( 1- \beta^{2})} $
                      & $0.105 \times 10^{-28} $ m
                                                               \\ \hline
  model factor
                      & $ B $
                      & $0.998610351$
                                                               \\ \hline
\end{tabular}
\end{center}
\end{table}

The third line in Table \ref{electronPARAMETERSdipol} shows the
numerical value of the velocity of the external charges $ 2/3 \times e $ 
in the form of $ \beta $, calculated using Eq.\ref{BETAKfactor}.
This velocity is brought extremely close to $ \beta = 1 $ by the very small value
of the electric dipole moment. However, the numerical value is not one. 
When considering Eq.\ref{BETAKfactor}, it is obvious that $ \beta $ must 
be less than one, depending on the numerical value of the factor 
$ K $ in Eq.\ref{Kfactor}, which is always positive.
The factor $K$ is a function of the magnetic dipole moment $\mu_{e}$,
 the speed of light $c$, the charge $e$ and the contracted radius $r_{e}$, 
 which results from the electron's electric dipole moment. The numerically
  extremely small measured value of the electric dipole moment drives 
  $K$ to a very high value, as shown in the first line of Tab. \ref{PrecissionKbetaB}.
Consequently, the value of $\beta$ calculated from Eq. \ref{BETAKfactor} is
very close to one, as shown in the second line of Tab. \ref{PrecissionKbetaB}.

\newpage

\begin{table}
\caption{ High precession numerical values of the parameter $ K $, $ \beta $ and $ B $ }
\begin{center}
\begin{tabular}{||l|l||}                                      \hline
  parameter     & value
                                                               \\ \hline
  K                   & $ 3.0503068128782230372348481033722070616426133911863 \times 10^{33} $
                                                               \\ \hline
  $ \beta $           & 0.99999999999999999999999999999999983608206299476884
                                                               \\ \hline
  B                   & 0.99861035118674254715174743056685316037852026506779
                                                               \\ \hline
\end{tabular}
\end{center}
\label{PrecissionKbetaB}
\end{table}

The fourth line in Tab. \ref{electronPARAMETERSdipol} shows
the rest mass $m_{0} c^{2} $ of the electron, calculated
using Eq. \ref{restmassK}. This numerical value also matches the numbers in 
Tab.\ref{modelbeta1} for the case: $ m_{0} c^{2} ( 2/3 e ) $.

The fifth line in Tab. \ref{electronPARAMETERSdipol} shows
the rotational energy $E_{rot}$ of the electron, calculated
using Eq. \ref{Erotre}. The numerical value matches the numbers 
in Tab.\ref{modelbeta1} for the case: $ E_{rot} (2/3 e ) $.

The sixth line in Tab.\ref{electronPARAMETERSdipol} shows
the un-contracted radius $ r_{0} $ of the mass boundary of the
electron, which coincides with the radius at which the charges
$ 1/3 e $ are located in Fig.{\ref{Basic-ETAMFP},
calculated with Eq.\ref{r0fuctionK}.
This is the radius an observer would measure if they were 
located at the center of the electron rotating with $ \omega $.
The numerical value matches the Tab.
\ref{modelbeta1} for the case $ r_{0} (2/3 e ) $.

The seventh line in Tab.\ref{electronPARAMETERSdipol} shows
the contracted radius $ r_{e} $ of the mass boundary of the
electron, which coincides with the radius at which the charges
$ 1/3 e $ are located in Fig.{\ref{Basic-ETAMFP}.
This radius would be observed by an observer located outside 
the stationary electron in the laboratory frame. 
The numerical values are taken from Eq.\ref{ElectronDipolRadius} or
recalculated for test of the high numerical accuracy
function of $ r_{0} $ and $ \beta $ as shown in
Tab.\ref{electronPARAMETERSdipol}

It is also in our interest to investigate the influence of the measurement
the electric dipole moment of the electron on the mass core
of the electron as a Lorentz-contracted rigid sphere.
Therefore, we calculated using Eq. \ref{BfactorfunctionK} the numerical
value of the model factor $B$, which is given in the last line
of Tab. \ref{electronPARAMETERSdipol}.
Numerically, this factor in Tab. \ref{electronPARAMETERSdipol} is
in agreement with the factor used in Fig. \ref{BETAfunctionB} and
Fig. \ref{RexpfunctionB} for the upper limit of $B$ for $\beta = 1$ up to
very high accuracy, as shown in Tab.\ref{PrecissionKbetaB}
in the last line. The measurement of the electric dipole moment
of the electron localises the solution for $ \beta $, $ \omega $,
the energies $ m_{0} c^{2} $, $ E_{rot} $, the radius $ r_{0} $ and
$ r_{e} $ as a function of $ B $, shown in
Fig.\ref{BETAfunctionB} to Fig.\ref{RexpfunctionB} as a function of
$ B $ at the upper limit very close to $ \beta = 1 $, as shown in 
Tab.\ref{PrecissionKbetaB}.
It is interesting to note that in this extreme situation $ \beta $ is up to 33 digits 
close to $ \beta=1 $. We refrain from recalculating the influence of the high-precision factor $B$
on the parameters just discussed, since both factors in
Tab.\ref{PrecissionKbetaB} and Tab.\ref{electronPARAMETERSdipol}
 are very similar and the change in parameters will be negligible.

\subsubsection { Estimation of the size of the mass kernel of the electron }
\label{Kernel of the electron}
  
 After introducing a model factor B into the calculation of the rotating mass
of the electron, the calculation shows that the factor B taken from 
Tab.\ref{electronPARAMETERSdipol} is B = 0.998610351, which is less 
than one. From this fact, we conclude that the electron contains a non-rotating 
mass kernel, which is the heritage of the Planck scale, as discussed in 
Fig. \ref{Timing-FP}. The mass $m_{K}$ of the kernel is represented in 
Eq. \ref{MassKernel1} as a function of the mass of the electron $m_{0}$. 

\begin{equation}
       m_{K} = m_{0} ( 1-B )
\label{MassKernel1}
\end{equation}

Using Eq.\ref{MassKernel1}, the volume of the kernel $V_{K}$ can be calculated 
as a function of the kernel radius $r_{K}$, the density of the kernel $\varrho_{k}$, 
and the mass of $m_{0}$ as a function of $m_{0}=\varrho_{k}V_{K}+\varrho_{S}V_{S}$, 
where $\varrho_{S}$ is the density of the shell. And $V_{S}$ is the volume of the 
shell from Eq.\ref{Kernelvolume}.

\begin{equation}
V_{K}=\frac{4}{3}\pi r_{K}^{3}=\frac{m_{0}(1-B)}{\varrho _{k}}=\frac{(\varrho _{k}V_{K}+\varrho _{S}V_{S})(1-B)}{\varrho _{k}}
\label{Kernelvolume}
\end{equation}

We approximate the electron density by a mean value $ \varrho _{SM} $ according to the Eq.\ref{mass-middle}.

\begin{equation}
\varrho _{SM}=(\varrho _{K}+\varrho _{S})=m_{0}/V_{0}=m_{0}/((4/3)\pi r_{0}^{3})
\label{mass-middle}
\end{equation}

Based on the time evolution of the density of the mass kernel $ \varrho _{k} $ 
of the heritage of the ETAMP model described in Fig.{\ref{Timing-FP}, we estimate 
the numerical order of magnitude of $ \varrho _{PL}/\varrho _{k} $ to be 
$ \varrho _{GUT}/\varrho _{SM} $. The density of the mass nucleus 
$ \varrho _{k} $ is then given by Eq.\ref{KernelDensity1}.

\begin{equation}
 \varrho _{k}=\frac{\varrho _{PL}}{\varrho _{GUT}/\varrho _{SM}}
\label{KernelDensity1}
\end{equation}

The density of $ \rho _{GUT} $ depends on $ \Lambda _{GUT} $ as shown in Eq.\ref{GUTdensity}.

\begin{equation}
      \rho_{GUT} = \Lambda_{GUT}^{4}/(\hbar c )^{3}
\label{GUTdensity}
\end{equation}

Introduction of the volume of the electron $ V_{0} $ in Eq.\ref{evolium}

\begin{equation}
 V_{0}=\frac{4}{3}\pi r_{0}^{3}  
\label{evolium}
\end{equation}

enables the calculation of $V_{k}$ as a function of $V_{0}$ in Eq. \ref{volium11}.

\begin{equation}
V_{K}=V_{0}\left\{ \frac{1-B}{1+B(\varrho _{k}/\varrho _{SM}-1)}\right\} 
\label{volium11}
\end{equation}

By combining Eq.\ref{MassKernel1} with Eq.\ref{volium11}, the radius of the
kernel $ r_{K} $ in Eq.\ref{KernelRadius1} can be calculated.

\begin{equation}
      r_{K} = r_{0} \sqrt[3]{\frac{(1-B)}  {(1+B(\frac{\rho_{k}}{\rho_{SM}}-1))}}
\label{KernelRadius1}
\end{equation}

To estimate the size of the electron kernel radius $r_{K}$, we are interested in
calculating $r_{K}$ as a function of $\Lambda_{GUT}$ and $\varrho_{GUT}$.
Therefore, it is necessary to know $\rho_{GUT}$, which is represented in Eq. \ref{GUTdensity1}
and already in Eq.\ref{GUTdensity}.

\begin{equation}
      \varrho _{GUT}= \Lambda_{GUT}^{4}/(\hbar c )^{3}
\label{GUTdensity1}
\end{equation}

Only approximate numbers exist, for the size of the GUT scale $ \Lambda_{GUT} $.
The current values in the literature are varying in a big range as
shown in Eq.\ref{GUTscale} \cite{GUTscale}.

\begin{equation}
      10^{13}~GeV < \Lambda_{GUT} < 10^{17}~GeV
\label{GUTscale}
\end{equation}

The implementation of five different GUT scales $ \Lambda_{GUT} $ in the
range of Eq.\ref{GUTscale} in Eq.\ref{GUTdensity} enables
the calculation of the corresponding GUT densities $ \varrho _{GUT} $.
For $ \hbar $ and $ c $ we used the values from
Tab.\ref{Parameterelectron}. Inserting $  \varrho _{GUT} $,
$  \varrho _{PL} = 2,15 \times 10^{76}~GeV^{4} $ and
$  \varrho _{SM} = 4.7977 \times 10^{-15} ~ GeV^{4} $ in
Eq.\ref{KernelDensity1} allows to calculate the
density of the kernel of the electron $  \varrho _{K} $
using Eq.\ref{KernelDensity1}.  Assuming
the shell density of the electron is approximately
the middle density of the electron
$  \varrho _{SM} \sim \rho_{e} $ and taking
the numerical values for $ B $ and $ r_{0} $ from
Tab.\ref{electronPARAMETERSdipol} we calculate for
five different GUT scales the kernel radius of the
electron $ r_{K} $ with Eq.\ref{KernelRadius1}.
The numerical values are shown in Tab.\ref{ElectronKernelRadius}.

\newpage

\begin{table}
\caption{ The density $  \varrho _{K} $ of the electron kernel and its
          radius $ r_{K} $ as function of five possible GUT scales $ \Lambda_{GUT} $ }
\begin{center}

\begin{tabular}{||l|l|l|l|l|l||}                                \hline
 $ \Lambda_{GUT} $[GeV]& $ 10^{13} $
                       & $ 10^{14} $
                       & $ 10^{15} $
                       & $ 10^{16} $
                       & $ 10^{17} $
                                                               \\ \hline
 $  \varrho _{GUT}~[GeV^{4}] $ & $ 9,95 \times 10^{51} $
                       & $ 9,95 \times 10^{55} $
                       & $ 9,95 \times 10^{59} $
                       & $ 9,95 \times 10^{63} $
                       & $ 9,95 \times 10^{67} $
                                                              \\ \hline
 $ r_{K}~[m] $         & $ 5.02 \times 10^{-22} $
                       & $ 1.08 \times 10^{-20} $
                       & $ 2.33 \times 10^{-19} $
                       & $ 5.02 \times 10^{-18} $
                       & $ 1.08 \times 10^{-16} $
                                                              \\ \hline
\end{tabular}
\end{center}
\label{ElectronKernelRadius}
\end{table}

The first line shows five different GUT scales.
These scales include those used in the literature 
and allow the investigation of the possible 
size range of the electron kernel.
The second line shows the corresponding 
density of the mass kernel. The last line shows the radius of the 
electron kernel for the five different GUT scales.
It is interesting to note that with increasing
$ \Lambda_{GUT} $ also $  \varrho _{GUT} $ increases, but originated
from the also increasing ratio $  \varrho _{GUT} /  \varrho _{SM} $ is
the electron kernel density via Eq.\ref{KernelDensity1}
decreasing which causes via Eq.\ref{KernelRadius1}
an increase of the kernel radius with increasing GUT scale.

The investigation of the size of the electron kernel shows that, within 
the framework of our ETAMFP model, the largest radius is still located 
far inside the electron and the smallest value does not interfere with the Planck scale.

Taking into account the running coupling constants $\alpha_{1}$, $\alpha_{2}$, and
$\alpha_{3}$ in Fig. \ref{Running-coupling}, the mean value from the range of 
$\Lambda_{GUT}$ is approximately $\Lambda_{GUT}\sim 10^{15}$ [GeV].
This corresponds approximately to the intersection of the means 
of $\alpha_{1}$ - $\alpha_{2}$ and $\alpha_{2}$ - $\alpha_{3}$. The kernel size 
for this $\Lambda_{GUT}\sim 10^{15}$ [GeV] is $r_{K}=2.33\times 10^{-19}$ [m].


\subsubsection { Conclusion on the application of special relativity in a classical approach to the study of the electron as a GYROSCOPE. }

The classical approach to estimating the size of an electron based on its experimentally
 measured parameters is an important benchmark test for the ETAMFP model.
The model developed in the chapter \ref{Empirical toy ansatz}: ``Empirical Toy Ansats about a Microstructure of Fundamental
Particles (ETAMFP)''  uses as input parameters parts of the
SM theory and the timing data of the Big Bang model. In particular, the running 
coupling constants of the SM theory shown in Fig.{\ref{Running-coupling} and the very 
early timing data of the Big Bang model shown in Fig. \ref{Timing-BP} lead to the geometric approach
in Fig. \ref{Timing-FP}.
The numerical input data for the electron parameters in the ETAMFP model are 
measured within the framework of the the SM theory. The test shows that the 
geometric structure of the electron agrees with the discussed schemes of the 
electron from the chapters ``Scheme of the lightest left-handed fundamental 
particles'' \ref{Scheme lightest left-handed FB} to ``Scheme of the lightest left-handed anti fundamental particles''
 \ref{Scheme lightest anti left-handed FB}.
To test the ETAMFP model, we follow the philosophy of using the simplest possible 
solutions to the problems in question and refining them step by step.

In the first step, we assume that the electron, which carries spin, 
behaves like a classical gyroscope. Under these circumstances, 
the charge attached to the electron at a distance $r_{e}$ from the 
center has a certain velocity. Calculating the velocity of the 
electron's charges around the center shows that it exceeds the speed of light.

In the next step, we assume the velocity of light $ c $
as absolute limit for a group velocity must be respected.
As orbit of the charges around the electron center
we use the in the standard model common ansatz about 
the flashing vacuum chapter \ref{The flashing vacuum}. 
This ansatz allows, for example, an electron to orbit a 
proton in a hydrogen atom in a high energy state 
without radiating energy. We therefore assume that the 
charge orbit around the electron is free of radiation energy loss.
The path of the charge follows the circumference according to 
its distance from the electron's center. Since the speed of the 
charge is close to the speed of light, the length of the path 
corresponds to the size of the circumference contacted by
the Lorentz equations. Comparing the measurement of the magnetic moment of the
electron and the electric dipole moment, taking this contraction 
into account, shows that the experiment for measuring the 
magnetic moment is sensitive to the non-contracted radius $ r_{0} $ where the
charges $ 1/3 \times e $ are located, while the experiment for the
electric dipole moment is sensitive to the contracted radius $ r_{e} $. 
This explains why the magnetic moment is a precise numerical 
measurement, while measuring the electric dipole moment is so difficult.
Lorentz contraction pushes the measurable dipole moment in the laboratory 
frame to decrease to zero as the charge velocity approaches the speed of light. 
Assuming $\beta = 1$ for the charge velocity, the important parameters of the 
electron can be estimated: the angular velocity of the charge, the rest mass, the 
rotational energy, and the un-contracted radius $r_{0}$ at which the charge is located.

Including in the next step in the ETAMFP model an ansatz about the shape
of the mass core of the electron as a rigid sphere and taking
into account the possibility of a non rotating inner mass kernel
implements an important limit of the model. 
It is only possible to find a solution for the electron parameters if 
only $ 2/3 \times e $ of the charge generates the magnetic moment. The 
solution for $ 1 \times e $ is excluded. It also turns out that a non-rotating 
inner mass nucleus could exist.

In the last step of the model development so far, we include
the limiting value of the electric dipole moment in the ansatz. This confirms
a possible inner mass kernel and the parameters angular velocity,
rest mass, rotational energy, and the un-contracted charge radius of the
electron, which were already calculated for the case $ \beta = 1 $.
In particular it highlights the fact that the numerical value
of $ \beta $ get pushed from the measurement of the electric dipole
moment of the electron extreme close to $ \beta = 1 $.
We highlight this problem in Fig.\ref{PrecissonradiusElectron}.

\newpage

\begin{figure}
\begin{center}
 \epsfig{file=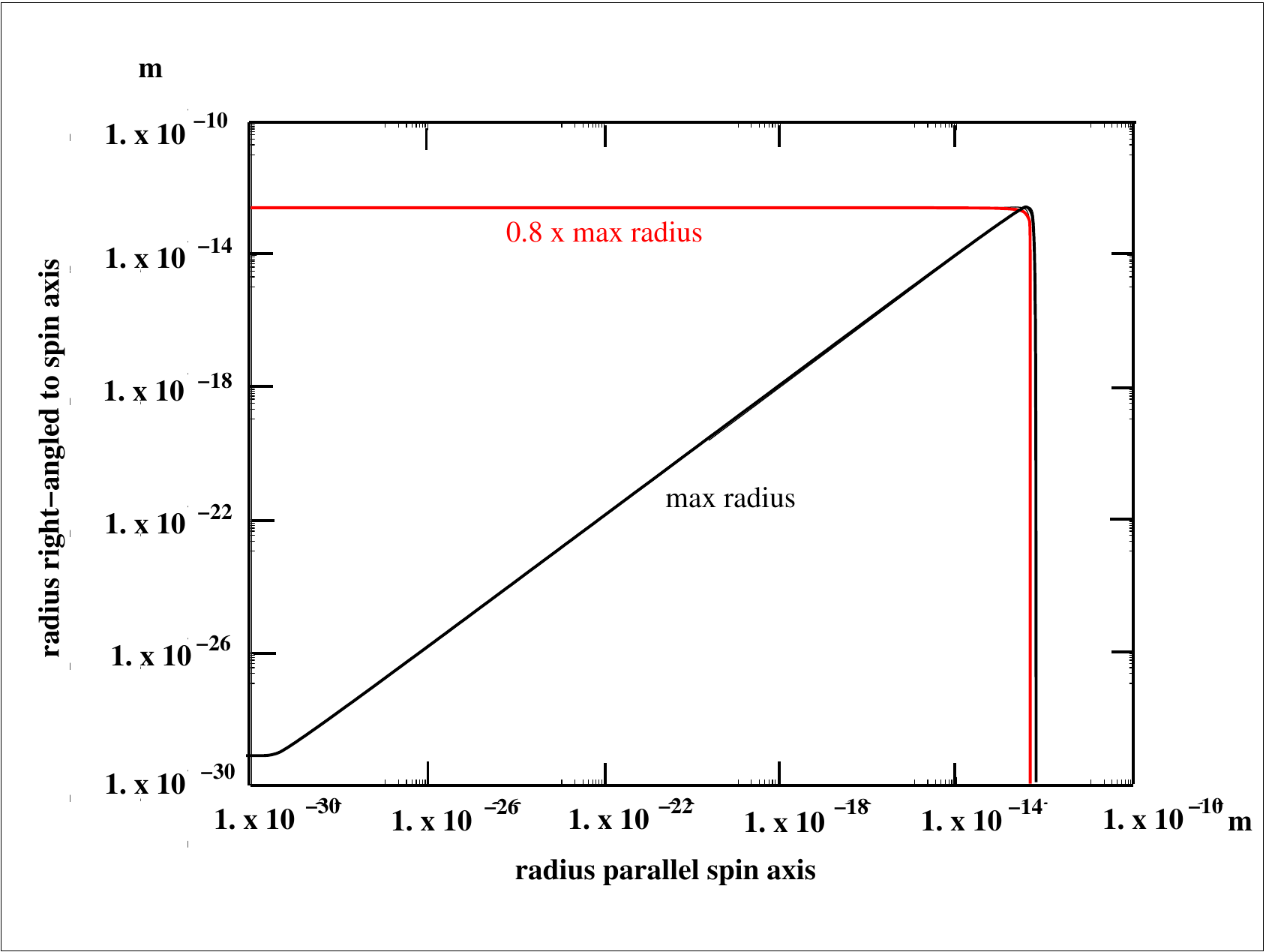,width=16.0cm,height=12.0cm}
\end{center}
\caption{x-axis un-contracted radius $ r_{0} $; y-axis: contracted electron radius $ r_{e} $. }
\label{PrecissonradiusElectron}
\end{figure}

The plot shows the contraction of the infinitely thin outer spherical shell, 
which rotates around the spin axis with $ \beta $ from Tab. \ref{PrecissionKbetaB}. 
The black solid line shows the contracted radius $ r_{e} $ perpendicular to the electron's spin axis
(y-axis in Fig. {\ref{PrecissonradiusElectron}} or x-axis in Fig. {\ref{Basic-ETAMFP} )
as a function of the non-contracted radius $ r_{0} $ parallel to the spin axis
(x-axis in Fig. {\ref{PrecissonradiusElectron}} or z-axis in Fig. {\ref{Basic-ETAMFP} )
of the electron The charges are located at their minimum at the coordinates
$ x = 0 $ (or z = 0 in Fig.{\ref{Basic-ETAMFP} ) and $ y = r_{e} = 0.105 \times 10^{-28} $ m 
(or $ x = r_{e} $ in Fig.{\ref{Basic-ETAMFP} ) in the lower left corner of the graph.
The contraction is concentrated exclusively in the outermost spherical 
shell of the electron nucleus. If the same behaviour is calculated for a 
shell located 20\% deeper in the nucleus, where the velocity $ \beta $ is 
correspondingly smaller, the contraction almost completely 
disappears, a s shown by the solid red line in 
Fig.\ref{PrecissonradiusElectron}.

To visualise the external shape of the electron from the perspective of an 
observer in the CM-system rotating with angular velocity $ \omega $, 
compared to a stationary observer in the LAB-system, we show 
Fig.\ref{SketchElectronContracted}. This figure is an artist's view 
of  Fig.\ref{PrecissonradiusElectron} in Fig.\ref{SketchElectronContracted}.

\newpage

\begin{figure}[htbp]
\begin{center}
 \includegraphics[width=16.0cm,height=12.0cm]{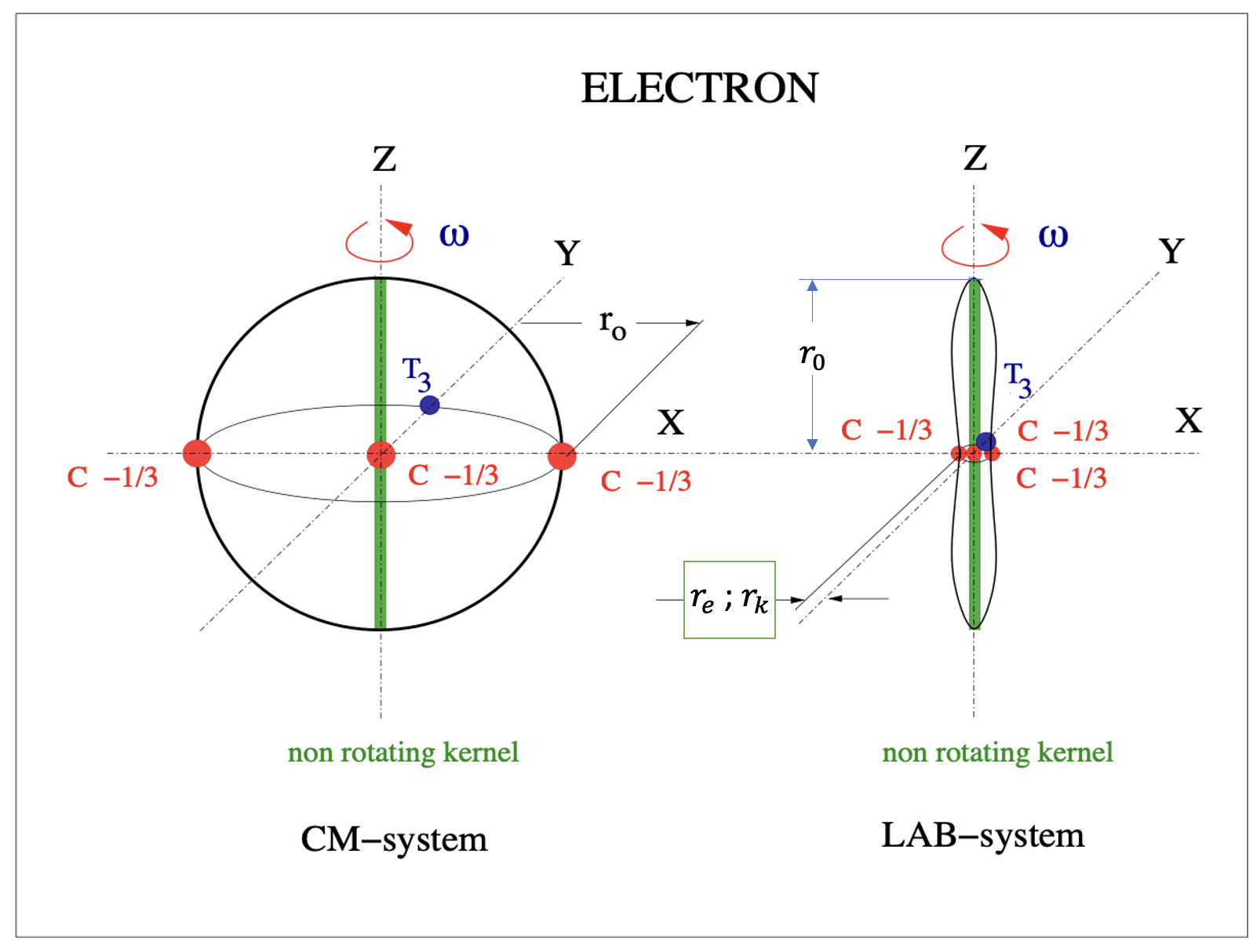}
\end{center}
\caption{Sketch of electron including the electric dipole moment.}
\label{SketchElectronContracted}
\end{figure}

The left side of Fig.\ref{SketchElectronContracted} shows the shape of the 
electron mass core (black circle) including the location of the three electric 
charges $ 1/3 \times e $ on the x-axis (red dots) in  the CM-system.
The maximum radius of the sphere is assigned as $ r_{0} $.
The weak charge $T_{3}$ is located on the y-axis (blue dot).
The spin axis is chosen to be parallel to the z-axis. We expect 
this kernel to be spherical, but for the sake of simplicity, we 
use a possible non-rotating mass kernel as a green solid strip on 
the z-axis in Fig.\ref{SketchElectronContracted}.

The right side of Fig.\ref{SketchElectronContracted} shows exactly 
the same electron, measured by a stationary observer in rest of the LAB-system. 
Only the contracted outer shell, including the electric and 
weak charges at radius $r_{e}$, is shown.

Comparing our initial assumption of a geometrically extended electron in 
Fig. {\ref{Basic-ETAMFP} with the actual shape of the electron after applying 
all the discussed conditions, a remarkable agreement is noticeable. This 
agreement becomes particularly visible when comparing Fig. {\ref{Basic-ETAMFP} 
with Fig. {\ref{PrecissonradiusElectron} and Fig. {\ref{SketchElectronContracted}.

An important benchmark test for our entire model discussion is
the comparison of the size of the electron radii $ r_{0} $, $ r_{e} $ and $ r_{K} $ 
with experimental data from completely different experiments. The parameters
of the electron in Tab.\ref{Parameterelectron} are measured
at low energies. The numerical values are experimental measured 
and calculated primarily with the standard field theory. The measurement of the magnetic
moment is a particular example for this experiments.  This
numerical values of Tab.\ref{Parameterelectron} are a part of the
input for our classical model of an geometrical extended FB. The
radius  $r_{0}^{\mu _{e}}$ calculated from the magnetic moment $ \mu _{e} $
is $r_{0}^{\mu _{e}}=8.39\times 10^{-13}$[m] very similar to the 
radius $r_{0}^{d_{e}}=5.79\times 10^{-13}$[m] calculated from the electric dipole 
moment $ d_{e} $. The velocity of $ \beta \sim 1$ of the charges
at the outside of position of the x - axis push $ r_{e} $ to $ r_{e}=0.105\times 10^{-28} $[m].

To estimate the radius of the kernel $ r_{K} $ , we used
a completely different input value derived from the time evolution of the
Big Bang theory and the running coupling constants of the
SM model. We compare the numerical results of these
conditions, which represent a mixture of field theory and complete
classical models, with the numerical values measured
in the direct contact term reaction at GeV energies.
The calculation of the radius of the electron in our ETAMFP model
uses as input the numerical values of the parameters of
in Tab. \ref{Parameterelectron} and all discussed properties from 
the SM and the Big Bang model, in particular the 
GUT - scale energy $\Lambda_{GUT}$, which is discussed in 
Tab.\ref{ElectronKernelRadius}, where the three interactions 
shown in Fig. \ref{Running-coupling} converge. An approximate 
value for the kernel size for this $\Lambda _{GUT}\sim 10^{15} $[GeV]
$ r_{K}=2.33\times 10^{-19}$ [m]. 
  
 It is interesting that the approximate size of the radii is about 
 $ r_{0}\gg r_{K}\gg r_{e} $. The radius $ r_{0} $ is not 
 Lorentz-contracted in the ETAMFP model. 
The radius $ r_{K} $ is approximated under the condition that
spin exists throughout the entire development of the ETAMFP model.
A Lorentz contraction is therefore implicitly contained in the radius $ r_{K} $.
The radius $ r_{e} $ is  pushed below $r_{K}$ because the charges $1/3\times e$ lie 
outside the radius $ r_{K} $ on the x-axis. Considering Fig.\ref{PrecissonradiusElectron}, 
$ r_{K} $ is less strongly contracted, since $\beta \leqslant 1$ represents a small 
depression inside the electron.

The entire estimation in Chapter \ref{GYROSCOPE}, which describes the electron as a classical GYROSCOPE:
``Application of special relativity in a classical approach to study the electron as a Lorentz contracted GYROSCOPE'',
leads to Fig. \ref{SketchElectronContracted}: ``Sketch of the electron including the electric dipole moment''.
Under these conditions, the electron is compared to Quantum Mechanics, NOT longer a POINT 
any more, but a geometrically extended object with two boundaries. In other words, 
if such a particle is scattered on a target, two different scattering results are possible depending 
on the experimental conditions, since the particle is NOT a POINT.
This fact opens up a possible discussion about the explanation of  ``Wave - particle duality''.
 
\subsection{Particle - like electro - vacuum soliton related to General Relativity.}
\label{Electro vacuum soliton}

In this paper so fare, we have summarised the results of our 
investigations into the non-point-like behaviour of fundamental particles.
With the introduction of special relativity, it became possible to 
define limits for the size of an electron using a GYROSCOPE,
from $r_{0}^{\mu _{e}}=8.39\times 10^{-13}$[m], as shown in 
Fig.\ref{SketchElectronContracted}.

So far, we have used special relativity to model a geometrically extended 
FB. However, it is very important to also apply general relativity for the same purpose.
Under these circumstances, it is very interesting to investigate the consequences 
of this fact for our basic scheme of the fundamental particle model discussed in 
Fig.{\ref{Basic-ETAMFP}. For this reason, in the following chapter we reiterate the main 
arguments for the particle-like structure in relation to the gravity of paper
\cite{SizePaper2003} and analyse the impact of the new data on our 
ETAMFP model using the self-gravitating particle-like structure 
with a de Sitter vacuum core.

The ansatz in this paper was to model an extended particle-like
object by a de Sitter-Schwarzschild geometry, which describes a
smooth transition from a de Sitter vacuum at the origin to a
Minkowski vacuum at infinity \cite{IRINA1}.

The issue is inspired by the attempt discussed in the actual 
article to enclose a gravitational object like that in Fig.{\ref{Forces1}, 
whose masses are comparable to the masses of FP's.
The size of a FP particle cannot be defined
by the Schwarzschild gravitational radius $r_g=2G m
c^{-2}$ ( Where $m$ is the gravitational mass, measured in rest by a distant observer).
The size is bounded below by the Planck length
$l_{Pl} \sim 10^{-33}$ cm, and for every elementary particle, its
Schwarzschild radius $r_g$ is many orders of magnitude smaller than $l_{Pl}$.
The Schwarzschild radius of gravity is derived from the Schwarzschild 
solution, which implies a point-like mass and is singular at $r=0$. 
The Schwarzschild metric can be modified by replacing the singularity 
with a regular de Sitter kernel. \cite{IRINA1,DeSitter44,lambda,lambda11}.
This modified solution, the de Sitter-Schwarzschild geometry, depends 
on the limiting vacuum density $\rho_{vac}$ at $r=0$, which 
satisfies the equation of state $p=-\rho_{vac}$ there.
The idea dates back to the mid-60s papers by
Sakharov, who suggested that $p=-\rho$ can arise at superhigh
densities \cite{sakharov}, by Gliner who interpreted it as the
vacuum equation of state and suggested that it can be achieved as
a result of a gravitational collapse \cite{gliner}, and by
Zeldovich who connected $\rho_{vac}$ with gravitational
interaction of virtual particles \cite{zeldovich}.

In the context of spontaneous symmetry breaking $\rho_{vac}$ is
related to the potential of a scalar field  in its symmetric
(false vacuum) phase. In this context, as well as in the context
of Einstein-Yang-Mills-Higgs (EYMH) self-gravitating non-Abelian
structures including black holes, $\rho_{vac}$ is related to
symmetry restoration in the origin \cite{dynamics}. 
In a neutral branch of the EYMH black hole solutions, a non-Abelian structure can
be approximated as a sphere of a uniform vacuum density
$\rho_{vac}$ whose radius is the Compton wavelength of the massive
non-Abelian field ( See \cite{maeda} and references therein).
The basic fact of de Sitter-Schwarzschild geometry is that a mass of
an object is generically related to an interior de Sitter vacuum
and smooth breaking of space-time symmetry from the de Sitter
group in the origin to the Poincar\'e group at infinity
\cite{ID2002CQG}.

In de Sitter-Schwarzschild geometry ( The particular exact analytic
solution was found in Ref.\cite{IRINA1}), there exists a
certain critical value of  the mass $m_{cr}$ which selects two
types of objects: a neutral non-singular black hole 
for $m\geq m_{cr}$, and for $m <m_{cr}$ a neutral self-gravitating
particle-like structure with de Sitter vacuum core related to its
gravitational mass \cite{DeSitter44}}. This fact is generic 
for de Sitter-Schwarzschild geometry. In the case when stress energy
tensor satisfies the dominant energy condition ( Density
non-negative for any observer, speed of sounds does not exceed
speed of light), the geometry has two characteristic surfaces. The
surface of zero scalar curvature $r=r_s$ defines the gravitational
size $r_s$ of a particle-like structure. 
The solution \cite{IRINA1} belongs to this class, and for an object described
by it most of the mass is within $r_s$. The surface of zero
gravity, $r=r_c<r_s$, there exists also in the case when only weak
energy condition is satisfied (density no-negative for any
observer) and defines a size of an inner vacuum core. 
Beyond $r=r_c$, gravitational attraction transforms into gravitational repulsion.
Both surfaces lie on the characteristic scale
$\sim{(m/\rho_{vac})^{1/3}}$.

\subsubsection{Self-gravitating, particle-like structure with a de Sitter vacuum core \\ for particles with $spin = 0$.}
\label{Electro vacuum soliton spin ZERO}

De Sitter-Schwarzschild geometry has appeared as describing a
black hole whose singularity is replaced with de Sitter core of
some fundamental scale \cite{Irina66,IRINA1,particle11}. Several
solutions have been obtained by direct matching de Sitter metric
inside to Schwarzschild metric outside of a junction surface
\cite{IRINA1}. Typical for matched solutions is that they have a
jump at the junction surface since the O'Brien-Synge junction
condition $T^{\mu\nu}n_{\nu}=0$ is violated there \cite{werner}.
Poisson and Israel analysed de Sitter-Schwarzschild transition and
came to the conclusion that a layer of ``non-inflationary'' material
should be introduced at the interface \cite{werner}. 
In the Ref. \cite{lambda} this material was specified as a spherically
symmetric anisotropic vacuum ( Inflationary in the radial
direction, $p_r=-\rho$), with the continuous density and pressure,
responsible for a class of regular metrics asymptotically de
Sitter at the center \cite{ID2002CQG}. The exact analytical
solution for a neutral spherically symmetric black hole with a
regular de Sitter interior was found in \cite{lambda}.

The main steps to find this solution are to insert the spherically
symmetric metric Eq.\ref{eq.7a}.

\begin{equation}
ds^{2}=e^{\nu }c^{2}dt^{2}-e^{\mu }dr^{2}-r^{2}
(d\theta^{2}+sin^{2}\theta d\phi ^{2})
\label{eq.7a}
\end{equation}

into the Einstein equations
\( R _{\mu \nu }-\frac{1}{2}Rg_{\mu \nu }=\frac{8\pi G}{c^{4}}T_{\mu \nu }
\)
which then take the form Eq.\ref{eq.7b}, Eq.\ref{eq.7c} and Eq.\ref{eq.7d}.

\begin{equation}
\frac{-e^{\mu }}{r^{2}}+\frac{\mu ^{\prime }
e^{-\mu}}{r}+\frac{1}{r^{2}}=\frac{8\pi G}{c^{4}}T_{t}^{t}
\label{eq.7b}
\end{equation}
\begin{equation}
\frac{-e^{\mu }}{r^{2}}-\frac{\nu ^{\prime }
e^{-\mu}}{r}+\frac{1}{r^{2}}=\frac{8\pi G}{c^{4}}T_{r}^{r}
\label{eq.7c}
\end{equation}
\begin{equation}
\frac{1}{2}e^{-\mu }(\nu ^{\prime \prime }+\frac{\nu ^{\prime 2}}{2}+\frac
{\nu ^{\prime }-\mu ^{\prime }}{r}-\frac{\nu ^{\prime }\mu ^
{\prime}}{2})=\frac{8
\pi G}{c^{4}}T_{\theta}^{\theta}=\frac{8\pi G}{c^{4}}T_{\phi}^{\phi}
\label{eq.7d}
\end{equation}

The requirement of regularity and weak energy condition leads to
the existence of a family of spherically symmetric solutions
\cite{ID2002CQG}, which includes the class of solutions which
connect smoothly the de Sitter metric inside to the Schwarzschild
metric outside. In this class asymptotical behavior of a
stress-energy tensor is  \( T_{\mu \nu }\rightarrow 0 \) as \(
r\rightarrow \infty \) and \( T_{\mu \nu }\rightarrow
\rho_{vac}g_{\mu\nu} \) as
 \( r\rightarrow 0  \), with \( \rho _{vac} \) as de Sitter vacuum density
at \( r = 0 \). The algebraic structure  of the stress-energy
tensor \( T_{\mu \nu } \) is Eq.\ref{eq.7e} \cite{lambda}.

\begin{equation}
T_{t}^{t}=T_{r}^{r}\,\, \textrm{and} \,\, T_{\theta }^{\theta }=T_{\phi }^{\phi }
\label{eq.7e}
\end{equation}

The stress-energy tensor of this structure describes a spherically
symmetric (anisotropic) vacuum, invariant under the boosts in the
radial direction (Lorentz rotations in $(r,t)$ plane)
\cite{lambda}. It can be associated with a time-dependent
spatially inhomogeneous cosmological term \cite{lambda11}.
It smoothly connects the de Sitter vacuum at the origin with the Minkowski
vacuum at infinity, and satisfies the equation of state Eq.\ref{eq.8} \cite{lambda,werner}

\label{alleqs8}
\begin{equation}
p_r=-\rho;~~p_{\perp}=p_r+{r\over 2} {dp_r\over{dr}}
\label{eq.8}
\end{equation}

where $p_r=-T_r^r$ is the radial pressure and
$p_{\perp}=-T_{\theta}^{\theta} = - T_{\phi}^{\phi}$
is the tangential pressure. In this class of solutions
the metric Eq.(\ref{eq.7a}) takes the  form Eq.\ref{eq.9}

\label{alleqs9}
\begin{equation}
ds^2=\biggl(1-\frac{R_g(r)}{r}\biggr) dt^2 -
\biggl(1-\frac{R_g(r)}{r}\biggr)^{-1} dr^2 - r^2 d{\Omega}^2,
\label{eq.9}
\end{equation}

where $d{\Omega}^2$ is the metric on the unit two-sphere and $ R_{g}(r) $ is Eq.\ref{eq.10}

\label{alleqs10}
\begin{equation}
R_g(r)=\frac{2GM(r)}{c^2};~ ~ M(r)= \frac{4\pi}{c^2}\int_0^r{\rho(r)r^2 dr}
\label{eq.10}
\end{equation}

In the model of Ref. \cite{lambda}, the density profile $T^t_t(r)= \rho(r) c^2$ 
has been chosen as Eq.\ref{eq.12}

\label{alleqs12}
\begin{equation}
\rho = \rho_{vac} e^{ -4\pi\rho_{vac} r^3/3 m }
\label{eq.12}
\end{equation}

which in the semiclassical limit describes the vacuum polarisation in the gravitational field
\cite{particle333}. Substituting Eq.(\ref{eq.12}) into Eq.(\ref{eq.10}) shows that \( R_g(r) \) takes
the form Eq.\ref{eq.13}

\label{alleqs13}
\begin{equation}
 R_{g}(r)=r_{g}(1-e^{-4\pi \rho_{vac}r^{3}/3m})
=r_g(1-e^{-r^3/r_0^2r_g})
\label{eq.13}
\end{equation}

where the radius $ r_{0}^{2} $ Eq.\ref{eq.14}

\label{alleqs14}
\begin{equation}
r_0^2 = \frac{3 c^2}{ 8 \pi G \rho_{vac}}
\label{eq.14}
\end{equation}

is the de Sitter radius, $ r_g = 2 G m / c^2 $ is the Schwarzschild radius, and \( m \) is
the gravitational mass of an object. In the limit of \( r <\!\!< (r_{0}^{2}r_{g})^{1/3} \)
Eq.(\ref{eq.13}) shows that \( R_g \rightarrow r^3 / r^2_0 \), and the metric
of Eq.(\ref{eq.9}) takes the de Sitter form with Eq.\ref{eq.15}

\label{alleqs15}
\begin{equation}
g_{tt}=1- \frac{R_g(r)}{r}=1 - \frac{r^2}{r_0^2}
\label{eq.15}
\end{equation}

In the de Sitter geometry the horizon $r_0$ bounds the causally connected region.
An observer at $r=0$ cannot get information from the region beyond the surface $r=r_0$.

For $r\gg (r_0^2 r_g)^{1/3}$ the metric takes the Schwarzschild form Eq.\ref{eq.16}

\label{alleqs16}
\begin{equation}
g_{tt}= 1 - \frac{R_g(r)}{r} = 1 - \frac{r_g}{r}
\label{eq.16}
\end{equation}

in agreement with boundary conditions. \\

The metric \( g_{tt}\left( r\right) \) Eq.\ref{eq.15} is shown in Fig.{\ref{fig.5}.

\newpage

\begin{figure}[htbp]
\vspace{-8.0mm}
\begin{center}
 \includegraphics[width=16.0cm,height=12.0cm]{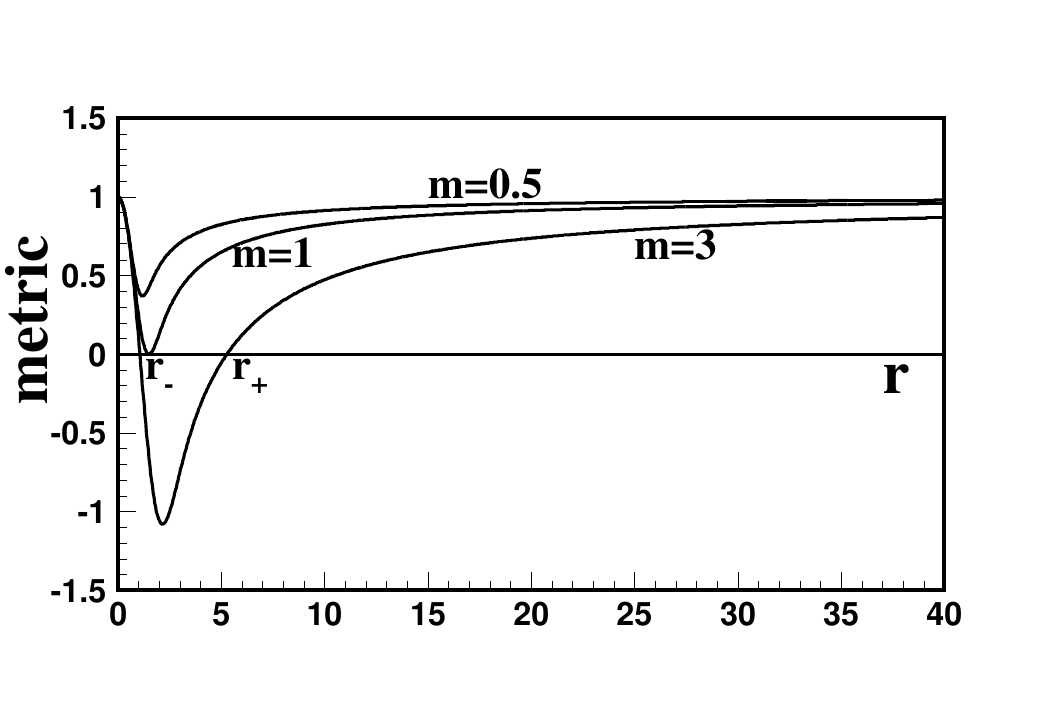}
\end{center}
\caption{De Sitter-Schwarzschild metric $g_{tt}(r)=1-R_g(r)/r$, 
             (Eq.\ref{eq.9},Eq.\ref{eq.15}). The mass  $m$
             is  normalized to $m_{cr}$. The radius is  normalized to $r_{o}$.
             For $m>1$ we have a black hole, $m=1$ corresponds to the extreme
             black hole, and configuration with $m<1$ represents
             a vacuum self-gravitating particle-like structure without horizons.}
\label{fig.5}
\end{figure}

The fundamental difference from the Schwarzschild case is that the de 
Sitter-Schwarzschild black hole has two horizons: the black hole 
horizon $r_{+}$ and the inner Cauchy horizon $r_{-}$.

The object is a black hole for $m\geq m_{cr}\simeq{0.3
m_{Pl}\sqrt{\rho_{Pl}/\rho_{vac}}}$, which loses its mass via
Hawking radiation until a critical mass \( m_{cr} \) is reached
where the Hawking temperature drops to zero \cite{particle333}.
At this point, the horizons meet. The critical value $m_{cr}$ sets the
 lower limit for the mass of a black hole. Below $m_{cr}$, 
 the Sitter-Schwarzschild geometry Eq.\ref{eq.9} describes a 
 neutral, self-gravitating, particle-like structure made of a 
 vacuum-like material Eq.(\ref{eq.7e}), where $T_{\mu\nu}\rightarrow
\rho_{vac}g_{\mu\nu}$ holds at the origin \cite{particle333}.
This fact does not depend on particular form of a density profile
\cite{particle333,dynamics}, which must only satisfy requirement of
regularity at the origin and guarantee the finiteness of the mass $ m $ 
as measured by a distant observer Eq.\ref{eq.17}

\label{alleqs17}
\begin{equation}
m = 4\pi \int_0^{\infty} {\rho(r) r^2 dr}
\label{eq.17}
\end{equation}

The interest of this paper is focused on the particle-like structure.
The case of $ m \geq m_{cr} $ is discussed in
\cite{particle333,dynamics,Irina10,Irina11}.

The De Sitter-Schwarzschild geometry with the density profile
Eq.\ref{eq.12} has two characteristic surfaces
at the characteristic scale $ r \sim(r_{0}^2 r_{g})^{1/3} $\cite{DeSitter}. 

The first is the surface of zero scalar curvature.
The scalar curvature \( R = 8 \pi GT \) changes sign at the surface
which contains the most of the mass $m$. Characteristic
gravitational size of a self-gravitating particle-like structure
can be defined by the radius $r_s$ Eq.\ref{eq.18}.

\label{alleqs18}
\begin{equation}
r = r_s =
 \biggl ( \frac{m}{\pi \rho_{vac}}\biggr)^{1/3} =
 \frac{1}{\pi^{1/3}}   \biggl(\frac{m}{m_{Pl}}\biggr)^{1/3}
 \biggl(\frac{\rho_{Pl}}{\rho_{vac}}\biggr)^{1/3}l_{Pl}
\label{eq.18}
\end{equation}

The second is related to the strong energy condition 
of the singularity theorems. It reads $
(T_{\mu\nu} - g_{\mu\nu}T/2)u^{\mu}u^{\nu}\geq 0$, where $ u^{\nu}
$ is any time-like vector. The strong energy condition is
violated, i.e., gravitational acceleration changes sign, at the
surface of zero gravity $r_c$ Eq.\ref{eq.19}.

\label{alleqs19}
\begin{equation}
 r = r_c =
 \biggl ( \frac{m}{2\pi \rho_{vac}}\biggr)^{1/3} = \frac{1}
{(2\pi)^{1/3}} \biggl(\frac{m}{m_{Pl}}\biggr)^{1/3}
  \biggl(\frac{\rho_{Pl}}{\rho_{vac}}\biggr)^{1/3}l_{Pl}
\label{eq.19}
\end{equation}

The globally regular configuration with de Sitter
core instead of a singularity arises as a result
of the balance between attractive gravity
outside and repulsion inside of the
surface $r = r_c$. This surface defines the characteristic
size of an inner vacuum core.
For a particle-like structure with $ m <\!\!< m_{Pl} $,
both these sizes are much bigger than
the Schwarzschild radius $r_g$. The ratio of a size of a  vacuum core
to the Schwarzschild radius $r_g$ is given by Eq.\ref{eq.20}.

\label{alleqs20}
\begin{equation}
\frac{r_c}{r_g}=\frac{1}{2}\frac{1}{(2\pi)^{1/3}}\biggl(\frac{m_{Pl}}{m}\biggr)^{2/3}
\biggl(\frac{\rho_{Pl}}{\rho_{vac}}\biggr)^{1/3}
\label{eq.20}
\end{equation}

The horizons and characteristic surfaces of de Sitter-Schwarzschild 
geometry are shown in the Fig.{\ref{fig.6} where they are normalised to $r_0$. 

\newpage

\begin{figure}[htbp]
\vspace{-10.0mm}
\begin{center}
 \includegraphics[width=15.0cm,height=12.0cm]{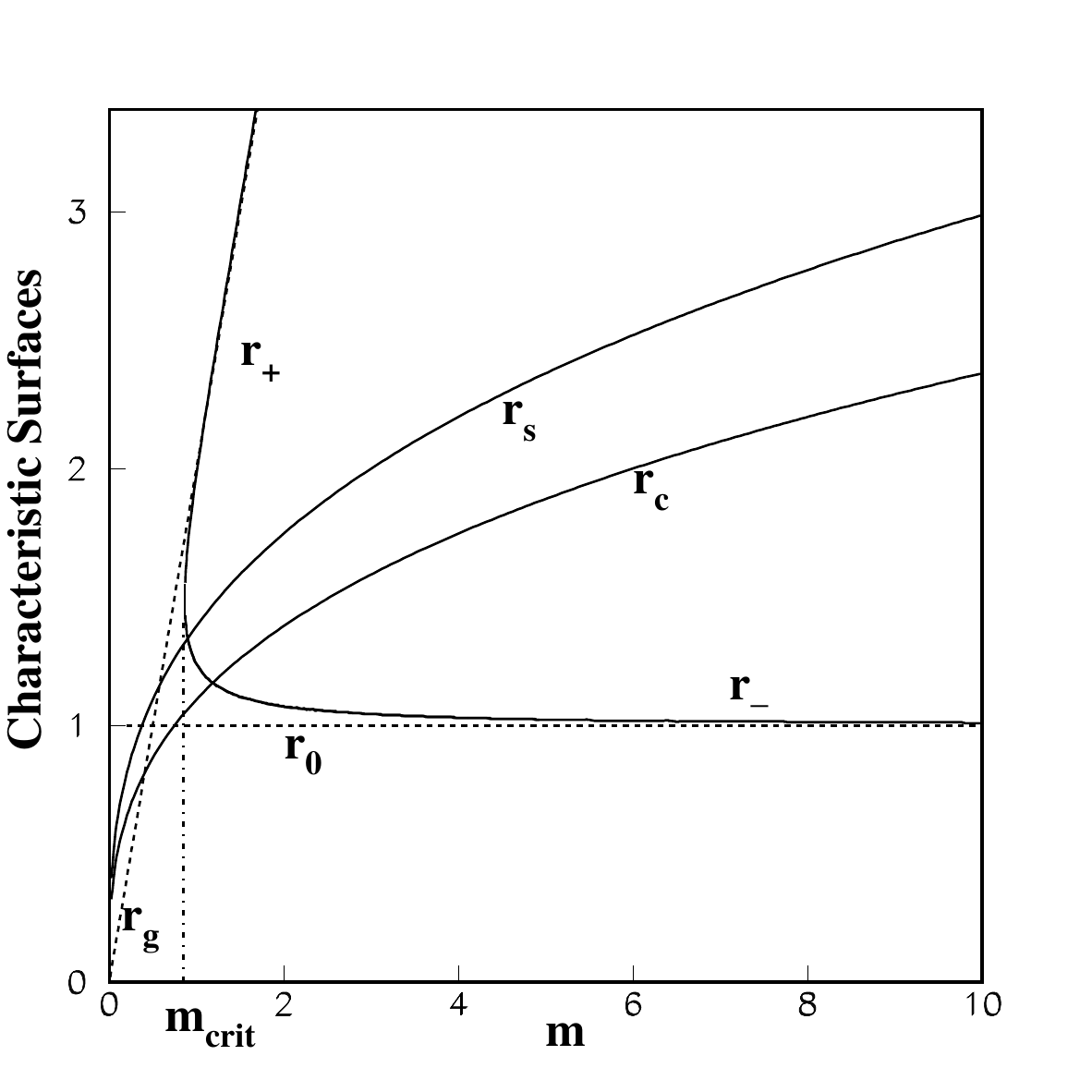}
\end{center}
\caption{Horizons $r_{\pm}$ and surfaces of zero curvature $r_s$ and zero
              gravity $r_c$ of de Sitter-Schwarzschild geometry. Schwarzschild
             radius $r_g$ and de Sitter radius $r_0$ are also shown.}
\label{fig.6}
\end{figure}

As we see from above, de Sitter-Schwarzschild
geometry gives rise to a self-gravitating particle-like structure, 
kind of a gravitational vacuum soliton \cite{Irina10}, which is stable 
for a wide range of density profiles including the profile Eq.\ref{eq.12}
\cite{Irina22}. The size of such a vacuum soliton is defined by its 
mass m and the vacuum density $  \varrho _{vac} $ 
in the vicinity of $ r = 0 $.

\subsubsection{Spinning superconducting electrovacuum soliton \cite{IrinaSOLITON}.}
\label{Electro vacuum soliton with spin}

To describe the key developmental steps of the ``Spinning superconducting electrovacuum soliton'' is very similar to 
describtion of the ``Self-gravitating, particle-like structure with a de Sitter vacuum core for particles with $spin = 0$'' 
in chapter \ref{Electro vacuum soliton spin ZERO}. It is first necessary to introduce a new metric for the spinning core. 
The Boyer-Lindquist coordinates $ ds^{2} $ are represented in Eq. \ref{Bo-Lin} \cite{IrinaSOLITON,Newman11}.

\begin{equation}
ds^{2}=\frac{2f-\sum }{\sum }dt^{2}+\frac{\sum }{\Delta }dr^{2}+\sum d\theta ^{2}-\frac{4afsin^{2}\theta }{\sum }dtd\phi +\left ( r^{2} +a^{2}+\frac{2fa^{2}sin^{2}\theta }{\sum }\right )sin^{2}\theta d\phi ^{2}
\label{Bo-Lin}
\end{equation}

The spherical coordinates are $ r,\theta $ and $ \phi $. The electric charge is $ e $, the mass of
 the object $ m $, the angular momentum is $ J = ma $ and the magnetic moment 
 $ \mu =ea $ . For an electron with $ I=0 $ is $ a=(I+s)/m=s/m $ a direct function of the spin $ s $.
 The parameters $ f $, $ \sum $  and $ \Delta $ are summarised in Eq.(\ref{Bo-Lin-detail}).

\begin{equation}
f=mr-e ^{2}/2  \qquad  \sum =r^{2}+a^{2}cos^{2} \theta \qquad  \Delta =r^{2}+a^{2}-2f
\label{Bo-Lin-detail}
\end{equation}

The spherical coordinates $ r,\theta $ and $ \phi $ from the Boyer-Lindquist coordinates 
are linked to the Kerr-Newman coordinates $ x, y, z$ via the equation Eq.\ref{Bo-Lin-detail-1}.

\begin{equation}
x^{2}+y^{2}=(r^{2}+a^{2})sin^{2}\theta \qquad z=rcos\theta
\label{Bo-Lin-detail-1}
\end{equation}

This forms in the Kerr-Newman geometry is an oblate ellipsoid Eq.\ref{Bo-Lin-detail-2}
shown in Fig.{\ref{Elips11}

\begin{equation}
\frac{x^{2}}{a^{2}}+\frac{y^{2}}{b^{2}}+\frac{z^{2}}{c^{2}}=1
\label{Bo-Lin-detail-2}
\end{equation}


\begin{figure}[htbp]
\begin{center}
 \includegraphics[width=16.0cm,height=9.0cm]{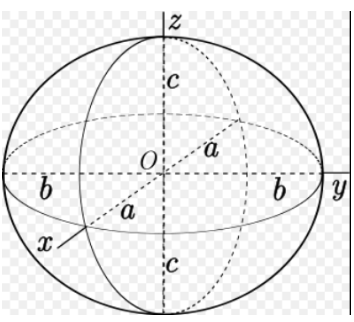}
\end{center}
\caption{Ellipsoid with $ a=b> c\to $ oblate ellipsoid.}
\label{Elips11}
\end{figure}

linked to the condition of the equation from Kerr-Newman Eq.\ref{Bo-Lin-detail-3}.

\begin{equation}
r^{4}-(x^{2}+y^{2}+z^{2})r^{2}-a^{2}z^{2}=0
\label{Bo-Lin-detail-3}
\end{equation}

If the spin is ZERO, then $ a = b = 0 $  and the ellipsoid collapses to ZERO.
In Elementary Superconductivity is the basic equation in nonlinear 
electrodynamics coupled to gravity obtained from the Lagrangian $ S $.
Eq.\ref{Bo-Lin-detail-4}.

\begin{equation}
S=\frac{1}{16\pi G}\int d^{4}x\sqrt{-g}\left [ R-L(F) \right ]
\label{Bo-Lin-detail-4}
\end{equation}

The electromagnetic tensor R is  $ R\equiv F=F_{\mu \nu }F^{\mu \nu } $ and $ L(F) $ is 
an arbitrary function in the Maxwell weak field limit $ L(F)\to F $ for large $ r $.

The dynamical equation is Eq.\ref{Bo-Lin-detail-5}

\begin{equation}
\triangledown _{\mu }=(L_{F}F^{\mu \nu })=0
\label{Bo-Lin-detail-5}
\end{equation}

with $ L_{F}=dL/dF $ and the Bianchi identities given by ( Asterisk Hodge dual )
Eq.\ref{Bo-Lin-detail-6}

\begin{equation}
\triangledown _{\mu }^{\ast }F^{\mu \nu }=0
\label{Bo-Lin-detail-6}
\end{equation}

The non-zero field components with axial symmetry $ F_{01},F_{02},F_{13},F_{23} $ are
Eq.\ref{Bo-Lin-detail-7}.

\begin{equation}
aF_{23}=(r^{2}+a^{2})F_{02}\qquad F_{31}=asin^{2}\theta F_{10}
\label{Bo-Lin-detail-7}
\end{equation}

The field invariant $ F=F_{\mu \nu }F^{\mu \nu } $reduces under these conditions to
Eq.\ref{Bo-Lin-detail-8}.

\begin{equation}
F=2\left ( \frac{F_{20}^{2}}{a^{2}sin^{2}\theta } -F_{10}^{2}\right )
\label{Bo-Lin-detail-8}
\end{equation}

The vector field is then defined as. Eq.\ref{Bo-Lin-detail-9}.

\begin{equation}
E=\left\{ F_{\beta 0}\right\}\qquad D=\left\{ L_{F}F^{0\beta }\right\}\qquad B=\left\{ ^{\ast }F^{\beta 0}\right\}\qquad H=\left\{ L_{f}^{\ast }F_{0\beta }\right\}
\label{Bo-Lin-detail-9}
\end{equation}

\subsubsection{Superconductive conditions on the DISC.}
\label{Superconductive DISC1}

The ``Superconductive conditions on the DISC'' chapter \ref{Superconductive  DISC1} are discussed ion details in reference 
\cite{IrinaSOLITON}. To introduce these conditions, we follow in the mean steps reference \cite{IrinaSOLITON}.

In the discussed de Sitter region it is $ \varepsilon _{r}=\varepsilon _{\theta }=L_{F};\mu _{r}=\mu _{\theta }=L_{F}^{-1} $ what leads to the
equations Eq.\ref{Disc1}.

\begin{equation}
L_{F}\sum {_{}}^{2}F_{10}=e(r^{2}-a^{2}cos^{2}\theta )\qquad L_{F}\sum {_{}}^{2}F_{20}=-era^{2}sin(2\theta )
\label{Disc1}
\end{equation}

The stress-energy tensor $ T_{\nu }^{\mu } $ of a nonlinear electromagnetic field from the Lagrangian $ L_{F} $ 
shown in Eq.\ref{Stress1}

\begin{equation}
\kappa T_{\nu }^{\mu }=2L_{F}F_{\nu \alpha }F^{\nu \alpha }-\frac{1}{2}\delta _{\nu }^{\mu }L
\label{Stress1}
\end{equation}

and the equation of state in the co-rotating reference frame ($ \kappa=8\pi G $ , $ p $ pressure, $ \rho $ density) 
represented in Eq.\ref{Stess2}

\begin{equation}
\kappa (p_{\perp}+\rho )=2\left ( L_{F}F_{10}^{2}+L_{F} \frac{F_{20}^{2}}{a^{2}sin^{2}\theta }\right )
\label{Stess2}
\end{equation}

allows to investigate the behaviour of the fields on the de Sitter vacuum disc \cite{IrinaSOLITON}}.

For spherically symmetric solutions, regularity, together with the equation of state on the disk, 
requires that the pressure $p_{\perp}$ acting perpendicular to the disk surface is equal 
to the density $-\rho$ on the disk, as shown in Eq. \ref{Density1}.

\begin{equation}
p_{\perp}=-\rho 
\label{Density1}
\end{equation}

This condition requests that the components of the field tensors $ F_{10} $ and $ F_{20} $  in
equation Eq.\ref{kfaktor1}.

\begin{equation}
 \kappa (p_{\perp}+\rho )=2(L_{F}F_{10}^{2}+L_{F}\frac{F_{20}^{2}}{a^{2}sin^{2}\theta })
\label{kfaktor1}
\end{equation}

must vanish on the disk independently of $ L_{F} $ to zero in equation Eq.\ref{Lff1}.

\begin{equation}
 L_{F}\frac{F_{20}^{2}}{a^{2}sin^{2}\theta }=0 \qquad L_{F}F_{10}^{2}=0
\label{Lff1}
\end{equation}

The magnetic induction B on the disk is zero, independent of the magnetic permeability, as shown below in Eq.\ref{B01}.

\begin{equation}
 \frac{2e^{2}(B^{r})^{2}}{\kappa(p_{\perp}+\rho )(r^{2}+a^{2})^{2}}=0  \qquad \frac{2e^{2}(B^{\theta})^{2}}{\kappa(p_{\perp}+\rho )(r^{2}+a^{2})^{2}}=0 
\label{B01}
\end{equation}

This is only possible B vanishes faster than ( $ p_{\perp}+\rho $ ). The Meissner effect takes place for a 
single spinning soliton and occurs at its de Sitter vacuum disk. This is requests a 
superconductor behaviour on the disk shown in Fig.{\ref{Meissner}, for temperature $  T>T_{C} $ above 
superconductivity and $  T< T_{C} $ below superconductivity  temperatures.

\newpage

\begin{figure}[htbp]
\begin{center}
 \includegraphics[width=16.0cm,height=16.0cm]{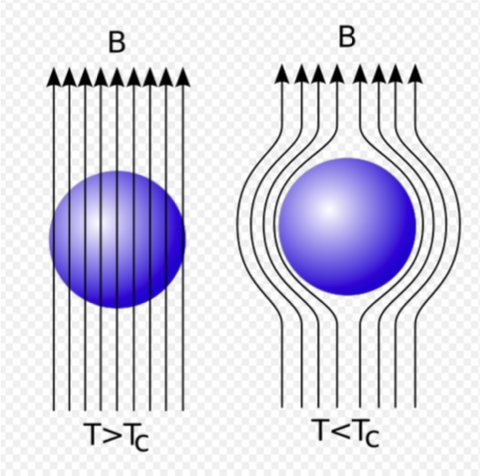}
\end{center}
\caption{Meissner effect of the displacement of an external magnetic field by a superconductor \cite{Meissner}.}
\label{Meissner}
\end{figure}

On the equatorial plane is the Lagrangian $ L_{F} $ shown in equation Eq.\ref{Lff11}.

\begin{equation}
L_{F}=\frac{2e^{2}}{\sum {_{}}^{2}\kappa(p_{\perp}+\rho )}
\label{Lff11}
\end{equation}

Using Eq.\ref{Lff11} leads to equation Eq.\ref{prho1}.

\begin{equation}
p_{\perp}=-\rho 
\label{prho1}
\end{equation}

The conditions on the disk are under these circumstances limiting $ \varepsilon $,
$ \mu $ and $ L_{F} $ shown in Eq.\ref{eee1} and Eq.\ref{uuu1}.

\begin{equation}
\varepsilon _{r}=\varepsilon _{\theta }=L_{F}\to \infty 
\label{eee1}
\end{equation}

\begin{equation}
\mu _{r}=\mu _{\theta }=L_{F}^{-1}\to 0
\label{uuu1}
\end{equation}

\subsubsection{Transformation of the E and B - Field from Boyer-Lindquist into cgs units.}
\label{Boyer-Lindquist in cgs}

The calculations of E and B - Fields in reference \cite{IrinaSOLITON} were performed in Boyer-Lindquist coordinates (BLC).
To compare these calculations with experiments, it is useful to convert them to $ ``cgs'' $ units.
In the first following first step, we summarise the most important equation from \cite{IrinaSOLITON,IrinaSOLITON1} for the
E  and B - field in the important De Sitter Region in Boyer-Lindquist coordinates.
In the second step, we focus on the ELECTRON data for the electric and magnetic fields in ``cgs'' units. \\

\textbf {1.5.4.1 E and B - Field in BLC - coordinates.}\\

We introduce from \cite{IrinaSOLITON} equation 2, 34, 36 and 41 the $ \overline{E} $  and $ \overline{B} $ - field
displayed in Eq.\ref{E-B1} and Eq.\ref{E-B11}.

\begin{equation}
\overline{E}=\left\{ F_{\beta 0}\right\}=\left ( \begin{matrix}
 F_{10}\\F_{20}
 \\F_{30}
\end{matrix} \right )=\left ( \begin{matrix}
 E_{r}\\
 E_{\theta }\\
 E\phi \\
\end{matrix} \right ) \qquad
\overline{B}=\left\{ ^{\ast }F^{\beta 0}\right\}=\left ( \begin{matrix}
 ^{\ast }F^{10}\\^{\ast }F^{20}
 \\^{\ast }F^{30}
\end{matrix} \right )=\left ( \begin{matrix}
 B_{r}\\
 B_{\theta }\\
 B\varphi  \\
\end{matrix} \right )
\label{E-B1}
\end{equation}

\begin{equation}
 \begin{matrix}
 E_{r}=F_{10}&  B^{r}=\frac{F_{23}}{\sum \cdot sin\theta }\\
 E_{\theta }=\frac{F_{20}}{\sum }&  B^{\theta }=\frac{F^{31}}{\sum \cdot sin\theta }\\
\end{matrix}
\label{E-B11}
\end{equation}
\\

\textbf {1.5.4.2. $ E_{r} $ and $ E_{\theta } $ - in De Sitter Region using BLC - coordinates. }\\

The field $  E_{r} $ and $  E_{\theta } $ is in Eq.\ref{Desitter11} a function of the parameters 
$ e, J, r, \mu, a, \theta, \sum $ and $ L_{F} $ shown below and taken from Eq. 47 reference \cite{IrinaSOLITON} .

\begin{equation}
E_{r}=F_{10}=\frac{e(r^{2}-a^{2}\cdot cos^{2}\theta )}{L_{F}\cdot  \sum {_{}}^{2} } \qquad
E_{\theta }=\frac{F_{20}}{\sum }=\frac{-e \cdot r \cdot a^{2} \cdot sin2\theta }{L_{F} \cdot \sum {_{}}^{3} } 
\label{Desitter11}
\end{equation}

\begin{itemize}
\item
 $e=charge \ e  \qquad   J=a\cdot m \qquad  J=AngularMomentum $
\item
$ r=radius \qquad \mu =e\cdot a \qquad \mu =GyromagneticRatio $
\item
$ a=parameter \qquad \theta =PolarAngle \qquad \sum =r^{2}+a^{2} \cdot cos^{2}\theta $
\item
$L_{F}=arbitrary-function=\varepsilon _{r}=\varepsilon _{\theta }$
\end{itemize}

Inserting the discussed parameters ( e - $ L_{F} $ ) in Eq.(\ref{Desitter11}) and  using Eq. 47 from
\cite{IrinaSOLITON}  gives Eq.\ref{DesitterX} and Eq.\ref{DesitterXX}.

\begin{equation}
E_{r}=f(r,\theta ,\epsilon _{r})=\frac{e\cdot (r^{2}-a^{2}\cdot cos^{2}\theta )}{\epsilon _{r}(r^{^{2}}+a^{2}\cdot cos^{2}\theta )^{2}} \qquad 
E_{\theta }=f(r,\theta ,\epsilon _{r})=-\frac{e\cdot r \cdot a^{2} \cdot sin2\theta }{\varepsilon _{r} \cdot (r^{2}+a^{2} \cdot cos^{2}\theta )^{3}}
\label{DesitterX}
\end{equation}

\begin{equation}
E_{r}=f(r,\theta ,\epsilon _{\theta })=\frac{e\cdot (r^{2}-a^{2}\cdot cos^{2}\theta )}{\epsilon _{\theta }(r^{^{2}}+a^{2}\cdot cos^{2}\theta )^{2}} \qquad 
E_{\theta }=f(r,\theta ,\epsilon _{\theta })=-\frac{e\cdot r \cdot a^{2} \cdot sin2\theta }{\varepsilon _{\theta } \cdot (r^{2}+a^{2} \cdot cos^{2}\theta )^{3}}
\label{DesitterXX}
\end{equation}

It is interesting to note that the equations Eq.\ref{DesitterX} and Eq.\ref{DesitterXX} are very similar,
only the parameters $\varepsilon_r$ and $ \varepsilon _{\theta }$ are significant.\\

\textbf {1.5.4.3. $  B_{r} $ and $  B_{\theta } $ - in De De Sitter Region using BLC - coordinates.}\\

The Eq.48 in the reference \cite{IrinaSOLITON} allows us to calculate $ B_{r} $ and $ B_{\theta } $, shown in Eq.\ref{BrrrBtheta}.

\begin{equation}
B^{r}=\frac{F_{23}}{\sum \cdot sin\theta }=(r^{2}+a^{2})\cdot\frac{e \cdot r \cdot a \cdot sin2\theta }{L_{F}\cdot \sum {_{}}^{3}sin\theta }  \qquad 
B^{\theta }=\frac{F_{31}}{\sum \cdot sin\theta }=a\cdot sin\theta \cdot\frac{e\cdot (r^{2}-a^{2}cos^{2}\theta )}{L_{F}\cdot \sum {_{}}^{3}}
\label{BrrrBtheta}
\end{equation}
\\

The paramers $ \mu _{r}, \mu_{\theta } $ and $ L_{F}^{-1} $ are defined Eq.\ref{Lffff} and Eq.\ref{mur-mut}.

\begin{equation}
\mu _{r}=\mu _{\theta }=L_{F}^{-1}
\label{Lffff}
\end{equation}

\begin{equation}
\mu _{r}=magnetic-permaeability-r \qquad \mu_{\theta }=magnetic-permeabilty-\theta 
\label{mur-mut}
\end{equation}
\\

By inserting Eq.\ref{Lffff} and the discussed parameters $e$ to $L_{F}$ into Eq.\ref{BrrrBtheta}, 
together with the behavior of the Meissner effect Fig.{\ref{Meissner} and Eq.48 from the 
reference \cite{IrinaSOLITON}, the magnetic field Eq.\ref{Br-Btheta} and Eq.\ref{Br-Btheta11} can be calculated.

\begin{equation}
B^{r}=f(r,\theta ,\mu _{r})=e \cdot \mu _{r}\frac{r \cdot a \cdot (r^{2}+a^{2})}{(r^{2}+a^{2} \cdot cos^{2}\theta )^{3}}\cdot \frac{sin2\theta }{sin\theta }
 \qquad B^{\theta }=f(r,\theta ,\mu _{r})=e \cdot \mu _{r}\frac{a \cdot (r^{2}-a^{2}cos^{2}\theta )}{(r^{2}+a^{2}cos^{2}\theta )^{3}}\cdot sin\theta 
\label{Br-Btheta}
\end{equation}

\begin{equation}
B^{r}=f(r,\theta ,\mu _{\theta })=e \cdot \mu _{\theta }\frac{r \cdot a \cdot (r^{2}+a^{2})}{(r^{2}+a^{2} \cdot cos^{2}\theta )^{3}}\cdot \frac{sin2\theta }{sin\theta }
 \qquad B^{\theta }=f(r,\theta ,\mu _{\theta })=e \cdot \mu _{\theta }\frac{a \cdot (r^{2}-a^{2}cos^{2}\theta )}{(r^{2}+a^{2}cos^{2}\theta )^{3}}\cdot sin\theta 
\label{Br-Btheta11}
\end{equation}
\\

Similar to the E - field, it is interesting to note that the equations Eq.\ref{Br-Btheta} and 
Eq.\ref{Br-Btheta11} are very similar. Only the parameters $\mu_{r}$ and $\mu_{\theta}$ are significant.
\\

\subsubsection {ELECTRON data for the E and B fields in cgs units.}
\label{Boyer-Lindquist E B cgs}

The main focus of our investigation is to study of the electron. After calculating the E and B fields in BLC units, 
it is possible to use all ``cgs'' units of the electron to calculate the E and B fields of the electron in ``cgs'' units.
\\

\textbf {1.5.5.1. ELECTRON parameters.}\\

In Fig.{\ref{E-parameter} a summary of all parameters for the calculation of the E and B fields of the 
ELECTRON in ``cgs'' units is shown.

\newpage

\begin{figure}[htbp]
\begin{center}
 \includegraphics[width=18.0cm,height=8.0cm]{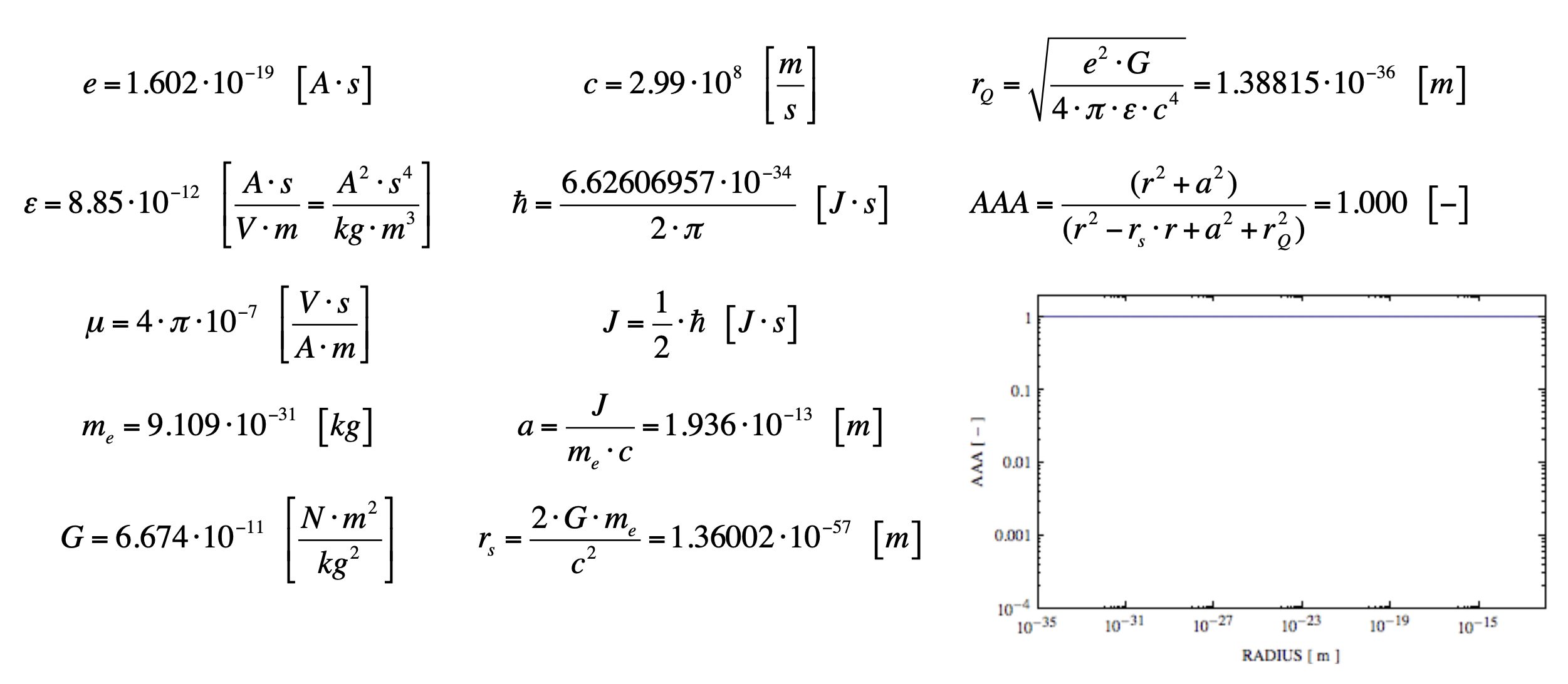}
\end{center}
\caption{The parameters of the ELECTRON in ``cgs'' units .}
\label{E-parameter}
\end{figure}

The parameter $ AAA = 1 $ in Fig.{\ref{E-parameter} simplifies $  \epsilon $ and $ \mu $ shown in Eq.\ref{epsilon1}
and Eq.\ref{mu1}.

\begin{equation}
\epsilon _{r}=\epsilon _{\theta }=\epsilon =L_{F}\qquad (De Sitter-CommonRegion) 
\label{epsilon1}
\end{equation}

\begin{equation}
\mu _{r}=\mu _{\theta }=\mu =L_{F}^{-1}\qquad (DeSitter-CommonRegion)
\label{mu1}
\end{equation}
\\

\textbf {1.5.5.2. Electron E - field B - field for AAA = 1}\\

The parameter AAA = 1 for the Electron in the range $  10^{-35}(m)<r<10^{-11}(m) $
reduces the equations of the E - field Eq.\ref{DesitterX}, Eq.\ref{DesitterXX} to Eq.\ref{cgsEr1}, Eq.\ref{cgsEtheta11} and 
the equation of the B - field Eq.\ref{Br-Btheta}, Eq.\ref{Br-Btheta11} to Eq.\ref{cgsEtheta1}, Eq.\ref{cgsBtheta1}.

\begin{equation}
E_{r}=f(r,\theta ,\epsilon )=\frac{e\cdot(r^{2}-a^{2}\cdot cos^{2}\theta )}{\varepsilon 
(r^{2}+a^{2}\cdot cos^{2}\theta )^{2}}\cdot \frac{1}{4\cdot \pi }\left [ \frac{V}{m} \right ]
\label{cgsEr1}
\end{equation}

\begin{equation}
E_{\theta}=f(r,\theta ,\epsilon )= - \frac{e\cdot r\cdot a^{2}\cdot sin2\theta }{\varepsilon 
(r^{2}+a^{2}\cdot cos^{2}\theta )^{3}}\cdot \frac{1}{4\cdot\pi }\cdot\sqrt{\frac{G\hbar}{c^{3}}}\left [ \frac{V}{m} \right ]
\label{cgsEtheta11}
\end{equation}

\begin{equation}
B^{r}=f(r,\theta ,\mu )= e\cdot \mu \cdot \frac{r \cdot a \cdot (r^{2}+a^{2})}{(r^{2}+a^{2}
\cdot cos^{2}\theta )^{3}} \cdot \frac{sin2\theta }{sin\theta }\cdot \frac{c}{4\cdot\pi}\left [ T \right ]
\label{cgsEtheta1}
\end{equation}

\begin{equation}
B^{\theta }=f(r,\theta ,\mu )= e\cdot \mu \cdot \frac{a\cdot (r^{2}-a^{2}\cdot cos^{2}\theta )}
{(r^{2}+a^{2}\cdot cos^{2}\theta )^{3}} \cdot sin\theta \cdot\frac{1}{4\cdot\pi } \cdot \sqrt{\frac{G\cdot \hbar }{c}}\left [ T \right ]
\label{cgsBtheta1}
\end{equation}
\\

The parameters of the Electron do not distinguish between the DeSitter and Common regions in the
E - field and B - field.

The comparison of the classical E - Field  of the Electron Eq.\ref{Eclassiqu11}

\begin{equation}
 E_{C}=\frac{e}{4\cdot\pi \cdot r^{2}}\left [ \frac{V}{m} \right ]
\label{Eclassiqu11}
\end{equation}

with the Soliton - Electron Field $ E_{tot} $ Eq.\ref{E-tot1}, as a function off 
$ E_{r} $  Eq.\ref{cgsEr1}  and $ E_{\theta } $ Eq.\ref{cgsEtheta11}

\begin{equation}
E_{tot}=\sqrt{E_{r}^{2}+E_{\theta }^{2}}\left [ \frac{V}{m} \right ]
\label{E-tot1}
\end{equation}

is shown in Fig.{\ref{E-field33} for $  \theta =5^{0}, 45^{0}, 85^{0} $ and $ 88^{0} $. 

\begin{figure}[htbp]
\begin{center}
 \includegraphics[width=18.0cm,height=8.0cm]{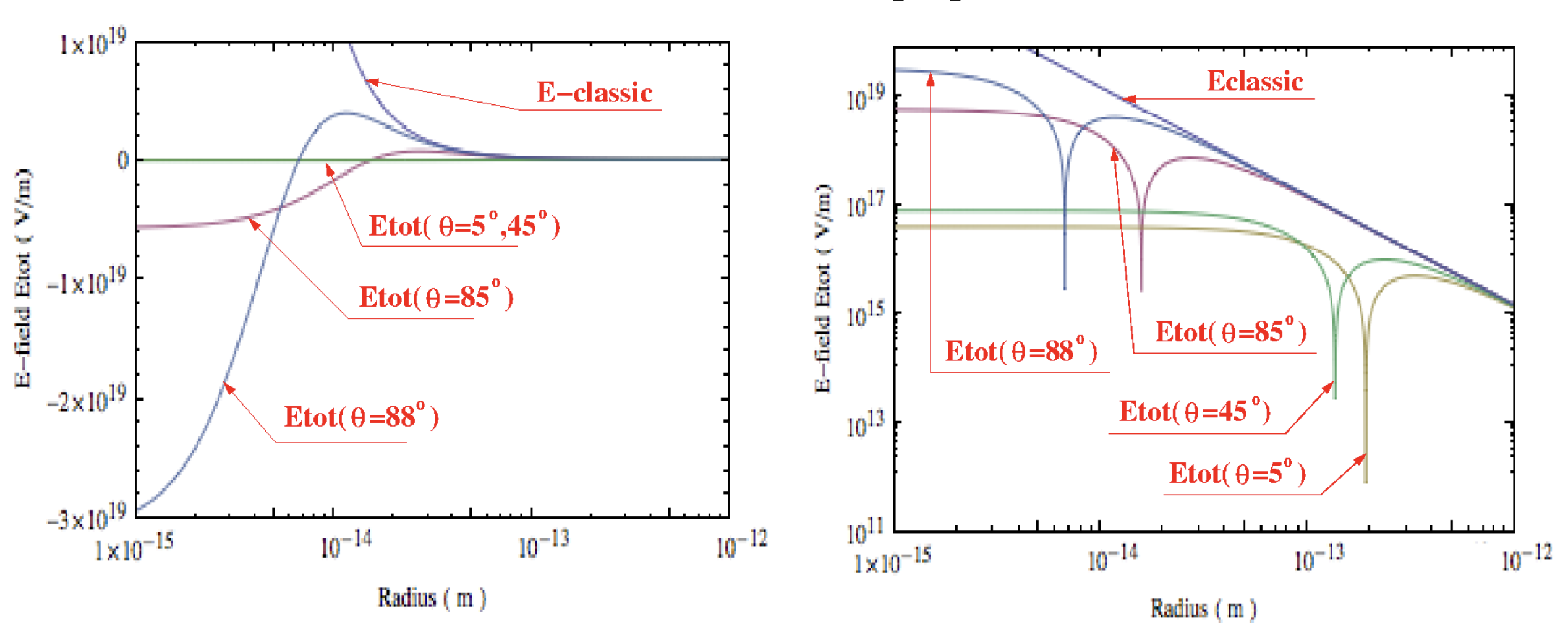}
\end{center}
\caption{Comparison of the classical E - Field with Soliton Electron Etot-Field ( left loglin-scale right loglog-scale ) .}
\label{E-field33}
\end{figure}

For obvious mathematical reasons, it is not possible to represent the field
linearly, as the variation of the field is too large to allow a linear representation.
For this reason is in Fig.{\ref{E-field33} $ E_{tot} $ after Eq.\ref{E-tot1} shown in loglin - scale ( left ) and in 
a loglog - scale ( right ). In the loglog - scale are of course only positive values of 
$ E_{tot} $ possible. The tip in the functions displays the zero transition.

It is interesting in Fig.{\ref{E-field33} that above approximately a Radius $ =10^{-14}\left [ m \right ] $ the classical field 
of Eq.\ref{Eclassiqu11} is similar to Soliton Electron  Etot - Field  Eq.\ref{E-tot1} for $ 5^{0}<\theta< 88^{0} $. The impact
of the the ``Soliton Electron'' is visible below the Radius $ =10^{-14}\left [ m \right ] $. 
This behavior will be more clearly recognisable in a VECTOR plot of the E - field, 
discussed in the next chapter. \\

\subsubsection {The VECTOR Plot of the Soliton Electron E - field }
\label{Vector E - field}

The VECTOR plot of Fig.{\ref{E-Vector-plot} summarises the information from Eq.\ref{cgsEr1}, Eq.\ref{cgsEtheta11}, Eq.\ref{E-tot1}
and Fig.{\ref{E-field33}. 

\begin{figure}[htbp]
\begin{center}
 \includegraphics[width=17.0cm,height=15.0cm]{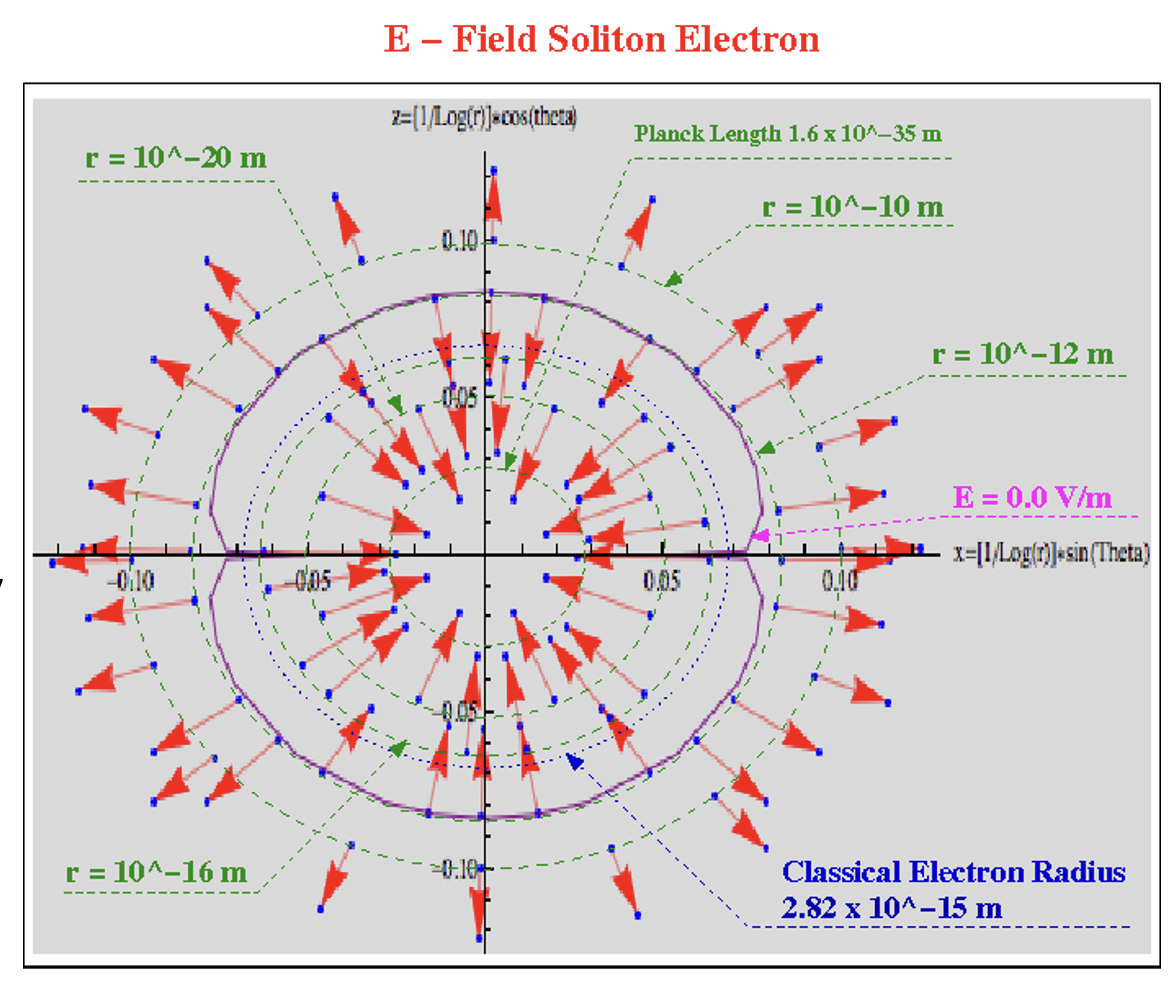}
\end{center}
\caption{Vector Plot of Soliton E - field .}
\label{E-Vector-plot}
\end{figure}

The size of the plot $ r $ ranges from $  10^{-20}\left [ m \right ]< r< 10^{-10}\left [ m \right ] $.

The E - field changes its sign at approximately at $ r\sim 10^{-12}\left [ m \right ] $ 
at the line $ E=0.0 \left [ V/m \right ] $ for the same sign of the charge
in Fig.{\ref{E-Vector-plot} according Eq.\ref{cgsEr1}. 

The term $ r=a\cdot cos\theta $ sets the E-field in Eq.\ref{cgsEr1} to zero. This radius is the 
most important radius for distinguishing between the classical description of the Electron 
and the de Sitter description.

The field increases from the common region at $ E\approx 10^{15}\left [ V/m \right ] $ to the disk for $ r \to 0 $ to
$ E\approx 10^{30}\left [ V/m \right ] $ and higher values. The field is close to the
Planck length $ L_{Planck}=1.6\times 10^{-35}\left [ m \right ] $ and increases sharply for $ \theta \to 90^{0} $.
This is an indication of a charge kernel.

The classical Electron radius is inside within the radius of $ E=0.0 \left [V/m \right ] $. 

It is also important to note that the size of the de Sitter structure 
($ E=0.0 \left [ V/m \right ] $ radius) is far outside the Planck length. In general,
the Planck length represents the limit for the quantisation of QED.

To account for the drastic changes in the E - field strength, the vector representation is 
calculated on a log - scale and the length of the arrows is adjusted accordingly. \\

\subsubsection {SKETCH to compare the E - field of the ETAMFP with SOLITON Model. }
\label{SKETCH E - field}

The sketch in Fig.{\ref{SKETCH1} compares the ETAMFP model with the SOLITON model,
from the perspective of the E-field.

\newpage

\begin{figure}[htbp]
\begin{center}
 \includegraphics[width=17.0cm,height=13.0cm]{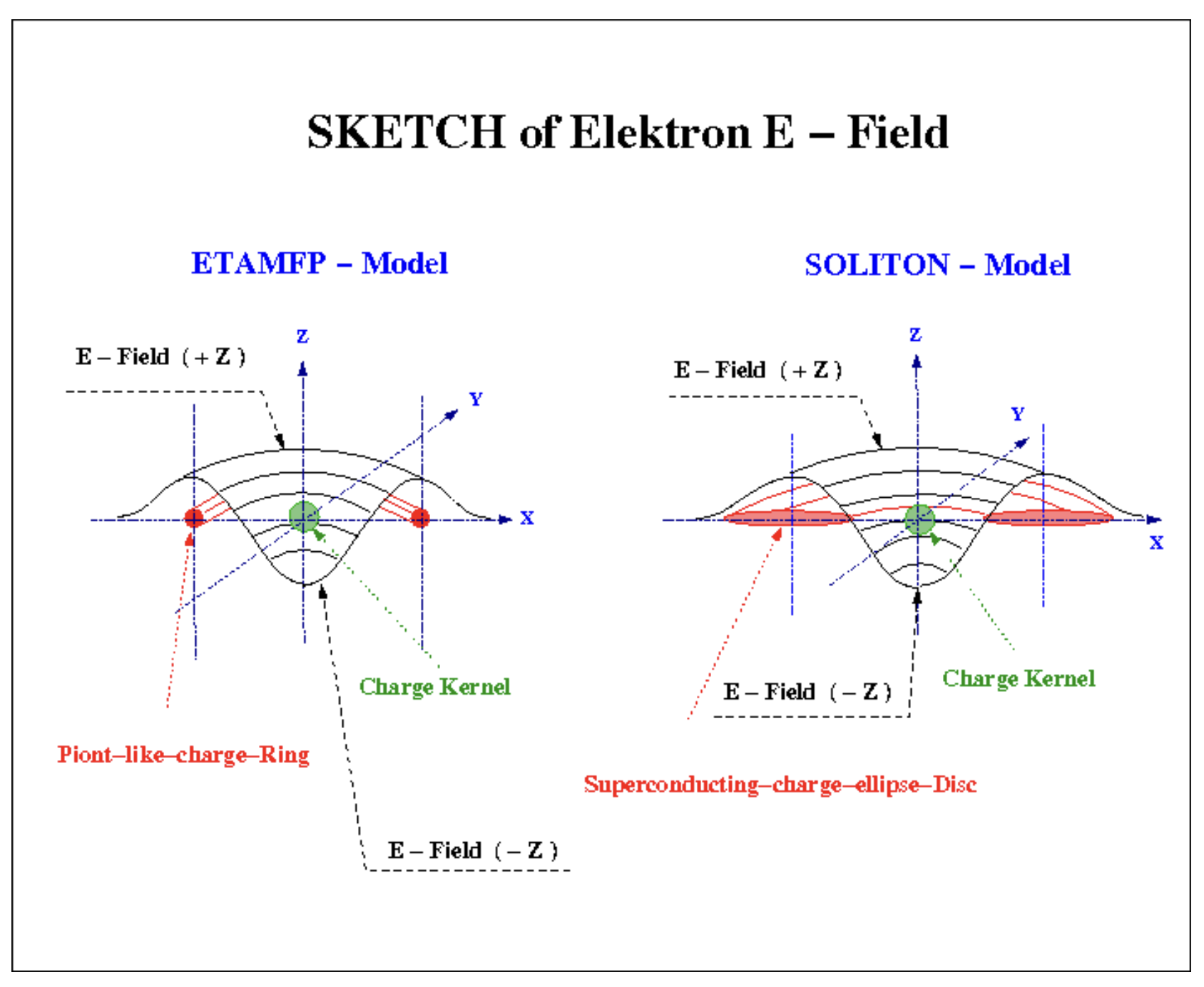}
\end{center}
\caption{Comparison of ETAMFP and SOLITON model via E - field .}
\label{SKETCH1}
\end{figure}

The ETAMFP model suggests an E - field shape that changes the direction of 
the field force from the outside to the center of the electron. 
However, the sign of the charge remains unchanged.
Similar to the behavior of the De Sitter gravitational field. Such an assumption would 
lead to a stable state of a geometrically extended Electron. The model also 
requires a non-rotating kernel at the center.
The SOLITON Model confirms the assumption that the field direction changes 
with the same sign of the charge, but within the framework of General 
Relativity. The point-like charge is replaced by a superconducting disk.
The E - field  has a kernel at $r \to 0$ and $\theta \to 90^{0}$. 

\subsubsection{ The VECTOR Plot of the common B - field region. } 
\label{The VECTOR B }

The VECTOR plot of Fig.{\ref{B-field-common} summarises the information 
from Eq.\ref{cgsEtheta1} and Eq.\ref{cgsBtheta1}.

\newpage

\begin{figure}[htbp]
\begin{center}
 \includegraphics[width=17.0cm,height=15.0cm]{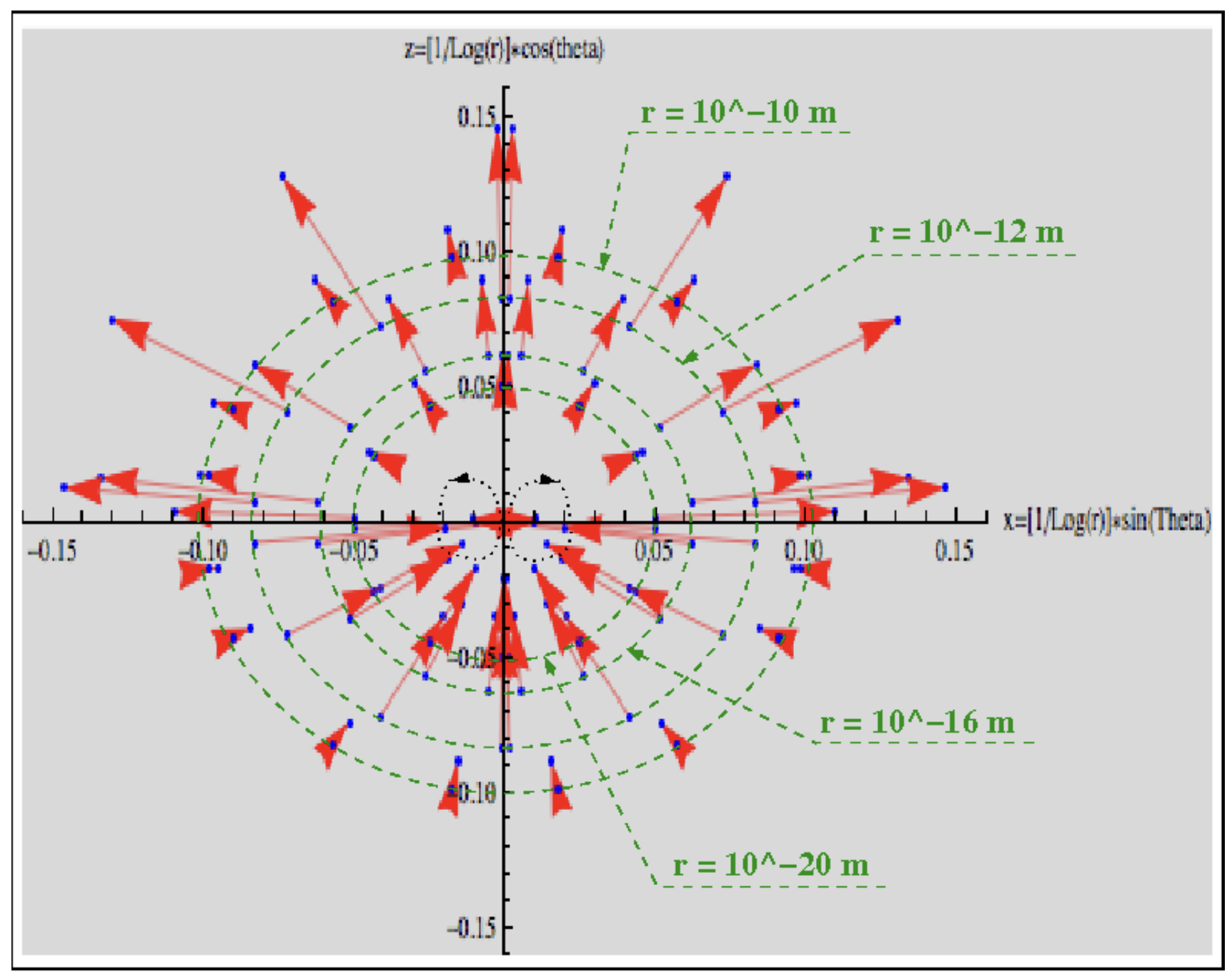}
\end{center}
\caption{Vector plot of B - field in the common region.}
\label{B-field-common}
\end{figure}

The size of the plot $ r $ is in the range $ 10^{-20}\left[m\right] < r < 10^{-10}\left[m\right] $.
The B-field changes direction the angle $\theta = 90^{0}$ for the same sign of the charge.
The transition of the B - field at $ \theta =90^{0} $ is performed with a high gradient. The field is in
the common region $ B\approx 10^{8}\left [ T \right ] $. Close to the disc for $ r \to 0 $ to
and $ \theta \to 90^{0} $ the field is up to $ B\approx 10^{40}\left [ T \right ]$. The field circles right 
hand at $ \theta =90^{0} $ and left hand at $ \theta =270^{0} $, forming an elliptic tube about the angle $ \phi $,
indicated in the inner part of the PLOT. To account for the drastic changes in B-field strength, the vector plot is 
calculated on a logarithmic scale and the length of the arrows is adjusted accordingly.

\subsubsection{ The VECTOR Plot of the inner region of the B -  field. }
\label{The VECTOR B inner }

The behavior of the E-field in the center of the electron becomes clearer in plot in Fig.{\ref{B-field-inner}
B-field inside: A change in the direction of the B-field near the Planck scale.

\begin{figure}[htbp]
\begin{center}
 \includegraphics[width=18.0cm,height=15.0cm]{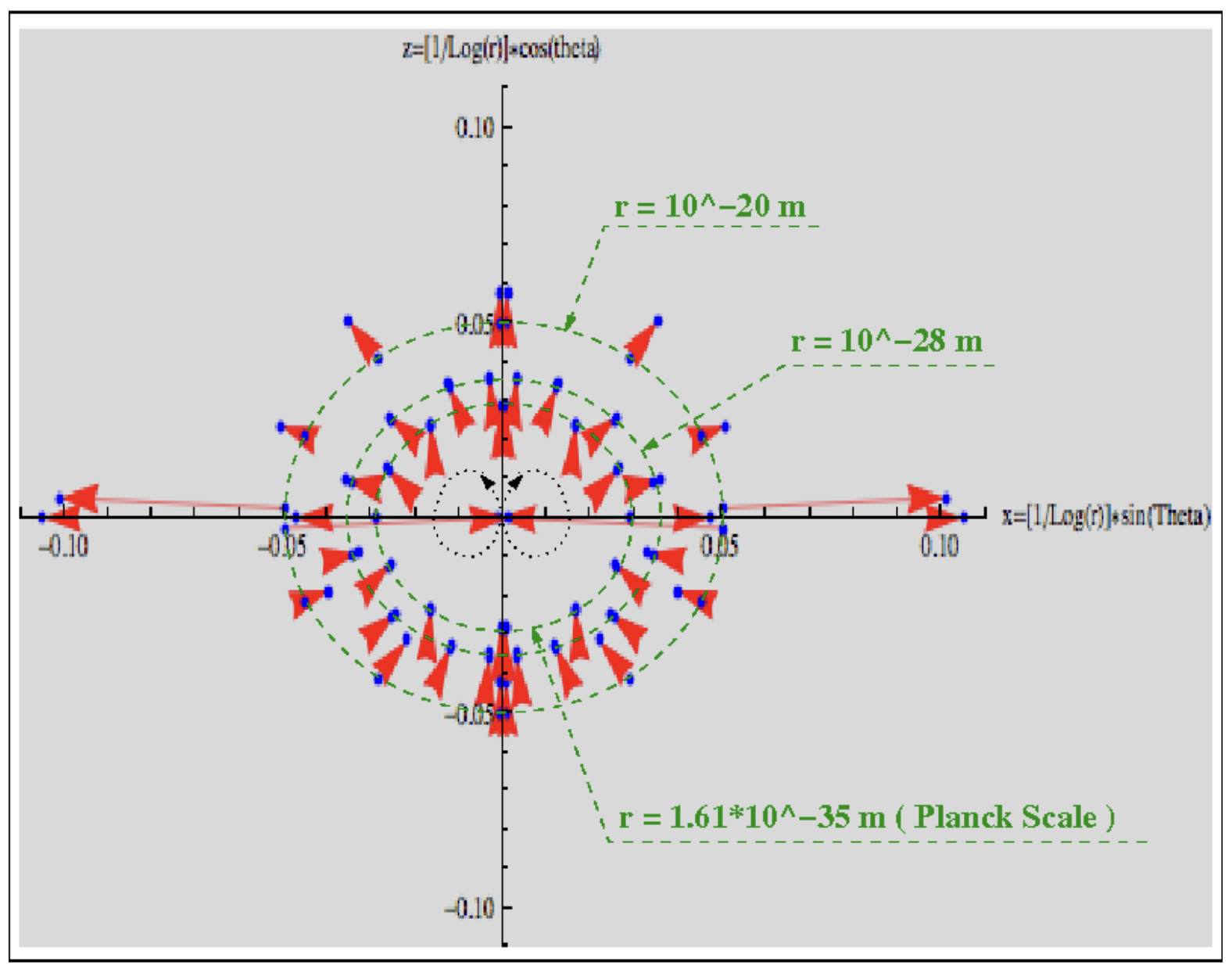}
\end{center}
\caption{The VECTOR Plot of the inner region of the B -  field.}
\label{B-field-inner}
\end{figure}

The right-handed circular behavior from the common area is superimposed by a 
left-handed behavior, which is visible at $ r=10^{-28}\left [ m \right ] $ and $ r=10^{-35}\left [ m \right ]$,
indicated in the centre of Fig.{\ref{B-field-inner}. This suggests that an inner part of the 
superconducting disc generates an B -  field in opposite rolling direction to the common 
behaviour in the out site region. The reason for this effect is the increase of the 
$ B_{\theta } $ - field compared to the $ B_{r} $ - field close to the centre of the electron.
Interestingly, the change in the direction of the circle is starts still outside the Planck scale.

\subsubsection{ SKETCH comparing the B-field of the ETAMFP with the SOLITON Model.}
\label{SKETCH  ETAMPF SOLITON}

The SKETCH in Fig. {\ref{SKETCH-B1} compares the ETAMFP model with the SOLITON model,
from the perspective of the B-field.

\begin{figure}[htbp]
\begin{center}
 \includegraphics[width=17.0cm,height=13.0cm]{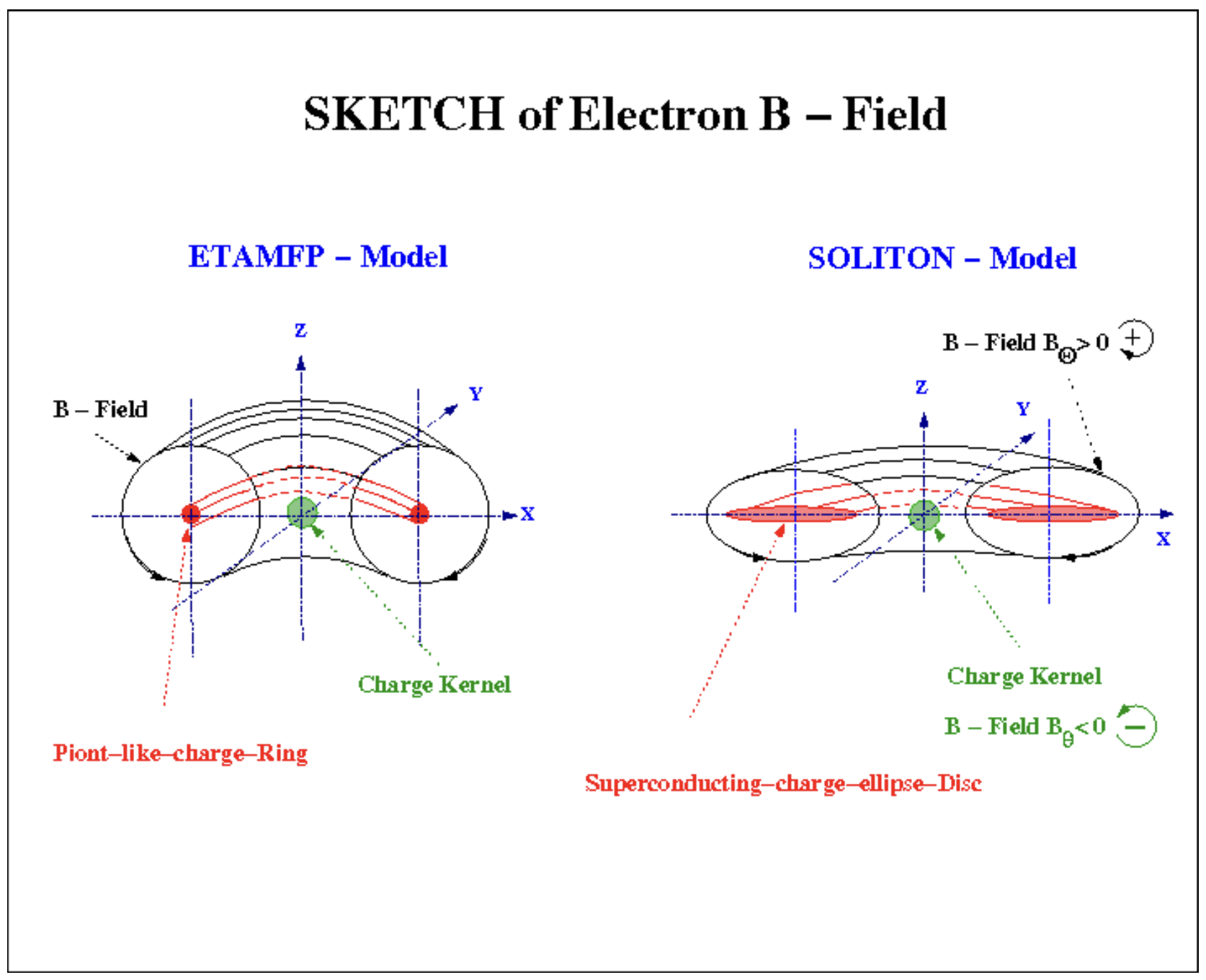}
\end{center}
\caption{Comparison of ETAMFP and SOLITON model via B - field .}
\label{SKETCH-B1}
\end{figure}

The ETAMFP Model suggest an B - field circling the point-like charge at the outer region 
of the electron. Forming a tube shape of a B -  field around the centre of the electron. 
It also suggests a non-rotating center of the electron.

The SOLITON model replaces the point - like charge with a superconducting disk.
The B - field is an elliptical tube surrounding the center of the electron. In 
Fig.{\ref{B-field-inner} appear the possibility of two boundaries of a rotating  outer area and
inner region of the Soliton Election in two opposite rolling directions.
This is shown in Fig.{\ref{SKETCH-B1} (top/bottom right side), the outer region of the 
Soliton Electron rolls in the opposite direction of the Charge Kernel of the Soliton Electron. 
This behavior is a consequence of General Relativity, a completely different ansatz than in the ETAMFP model.

\subsection{Conclusion of ETAMFP and SOLITON - Model }

In the ETAMFP model, the electron behaves like a rotating gyroscope. 
It is absolute obvious that this classical ansatz is only an overall
approximation for the concept of an geometrical extended electron.
Since the size of the object is larger than the Planck scale, where 
the quantisation of the fields becomes absolutely important, the 
model discussed seem to describe, to some extent, the real 
possibility of an extended Electron beyond the Standard Model.

In the SOLITON model, the electron would be an extended, rotating gyroscope 
with a superconducting center. This center would be created by a SOLITON. 
This would be ideal to explain the extremely long lifetime of the 
Electron: $ \tau _{e}> 4.6 \cdot 10^{26}\left [ y \right ] $.

The comparison of the ETAMFP model with the SOLITON model shows
a general agreement in behavior. The agreement of the absolutely simple
classical ETAMFP model with the SOLITON model of General Relativity 
is important, that two absolute independent models support the idea of
 a non pointe-like electron.

\subsection{Gross-Pitaevskii equation and ELECTRON wave function.}
\label{Gross-Pitaevskii}

To investigate the non-point-like character of FUNDAMENTAL PARTICLES, 
we first introduced the ETAMPF model in a theoretical chain to demonstrate 
the possibility of sorting all FB's in a common scheme.

Secondly, in a classical ansatz that describes FB's as
gyroscopes incorporating Special Relativity, it was possible to set limits on the size
of FP's, especially the Electron.

Third, we extended this ansatz to a spinning superconducting electro-vacuum soliton 
using General Relativity. 

Fourth, we introduce the character of a superconducting electro-vacuum soliton, the E-field of an Electron, the soliton character
of the Electron into the Gross-Pitaevskii equation to extract a wave function of the Electron.
This step opens a window between Quanten Mechanics an General Relativity shown in Fig.{\ref{Gross-11}.
This concludes the theoretical chain in Fig.{\ref{Invest-timing} with the very important calculation 
of a wave function using the substructure of FP's elements over the E - field of a non-point-like Electron.

\newpage

\begin{figure}[htbp]
\begin{center}
 \includegraphics[width=17.0cm,height=9.0cm]{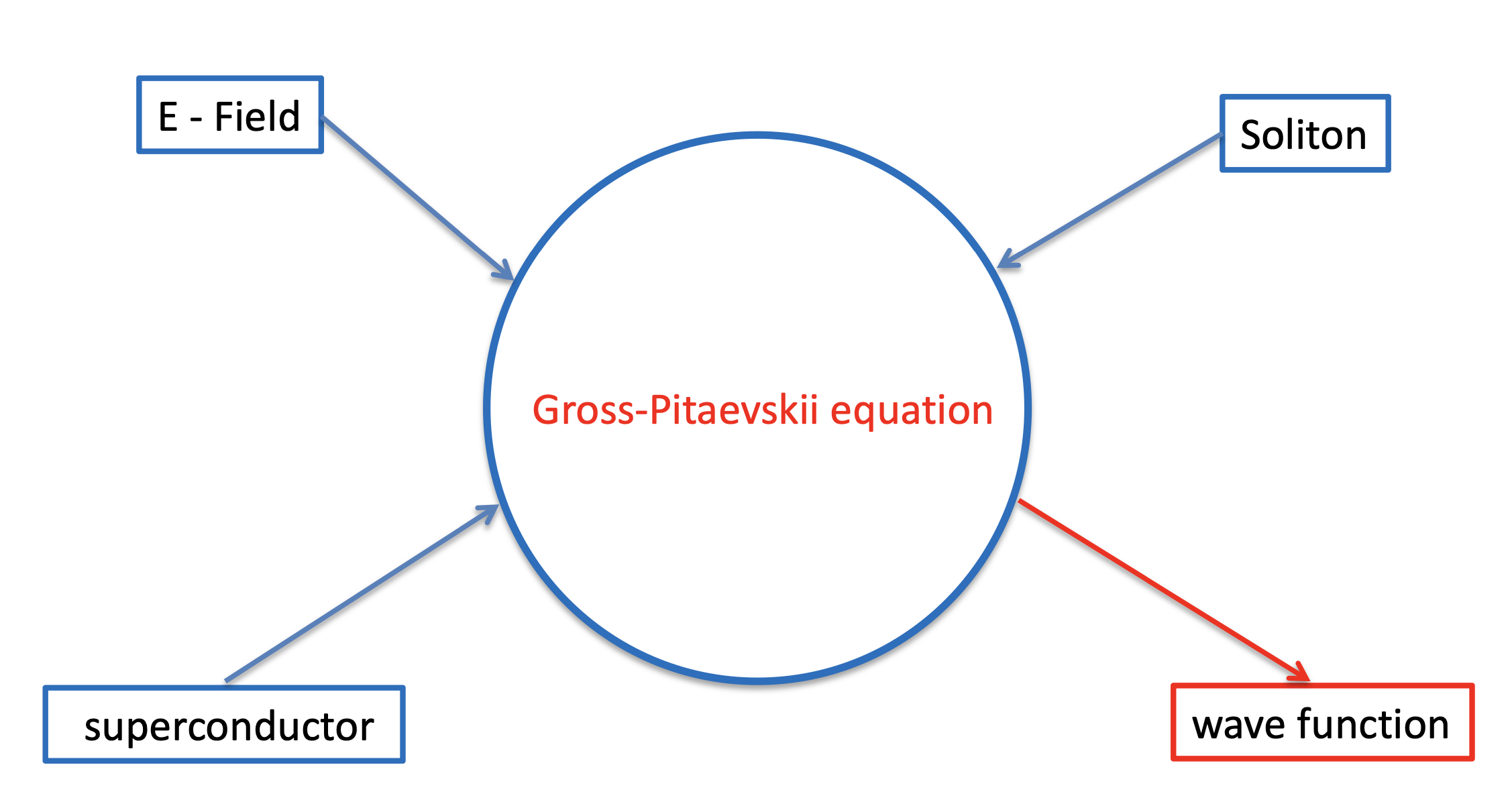}
\end{center}
\caption{ The Gross-Pitaevskii equation at  the source for the wave function of an Electron.}
\label{Gross-11}
\end{figure}

\subsubsection{The Gross-Pitaevskii equation.}
\label{Gross-Pitaevskii-1}
 
The Gross-Pitaevskii equation is a model equation for the single-particle wave function 
in a Bose-Einstein condensate, introduced in 1949 \cite{Gross11}, presented in Eq.\ref{GP-1}. 
The equation is ideally suited to investigate the knowledge discussed so far 
about our ETAMPF model, especially for the well known electron.  
The electron has a lifetime of almost infinity. For this reason, the solution to the 
Gross-Pitaevskii equation should not depend on time. Therefore, it is important
to use the time-independent Gross-Pitaevskii equation.
 
 \begin{equation}
 \mu \Psi (r)=\left ( -\frac{\hbar^{^{2}}}{2m}\bigtriangledown ^{2} +V(r)+g\left| \Psi (r)\right|^{2}\right )\Psi (r)
\label{GP-1}
\end{equation}

In Eq.\ref{GP-1} , $\mu$ is the chemical potential, $\Psi$ is the wave function, $\hbar$ is Planck's constant,
$m$ is the mass of the object, $V(r)$ is the external potential, and $g$ is the coupling constant.
The electron, as a soliton, has no external potential $V(r) = 0$. The chemical potential $\mu \approx 0$.
This simplifies the Gross-Pitaevskii equation to Eq.\ref{GP-2}.

\begin{equation}
0=\left ( -\frac{\hbar^{2}}{2m}\bigtriangledown ^{2}+g\left| \Psi (r)\right|^{2} \right )\Psi (r)
\label{GP-2}
\end{equation}

\subsubsection{Wave function solution for a SOLITON.}
\label{Solution SOLITON}

In the chapter \ref{Forces and stability} we already discussed the possibility of soliton formation,
via a combination of an attractive/repulsive force. The Gross-Pitaevskii equation 
offers precisely this possibility. A one-dimensional soliton can 
form a Bose-Einstein condensate (BEC). If the interaction is attractive or repulsive, either a 
light or a dark soliton is formed \cite{Gross11}.

If the BEC is repulsive such that $g > 0$, then one possible solution of the Gross-Pitaevskii equation is Eq.\ref{GP-3}.

\begin{equation}
\Psi (x)=\psi _{0}\cdot tanh\left ( \frac{x}{\sqrt{2\zeta }} \right )
\label{GP-3}
\end{equation}

$ \Psi _{0} $ is the weave function of the condensate at $ x\to \infty $ and $ \zeta =\hbar/\sqrt{2mn_{0}g} $ is the coherence length.

For an attractive interaction $ g < 0 $ the solution is Eq.\ref{GP-4}.

\begin{equation}
 \Psi (x,t)=\Psi (0)e^{-i\mu t/\hbar}\frac{1}{cosh\left [ \sqrt{2m\left| \mu \right|/\hbar} \cdot x\right ]}
\label{GP-4}
\end{equation}

The chemical potential for the bright soliton is $ \mu =g\left| \Psi (0)^{2}\right|/2 $.

\subsubsection{Mathematical Engineering of a toy model for an ELECTRON as SOLITON.}
\label{Engineering SOLITON}

Currently, no theoretical method exists for calculating an ELECTRON - SOLITON
using the Gross-Pitaevskii equation, since the details of equations Eq.\ref{GP-3} and Eq.\ref{GP-4}
are unknown. Therefore, we pursue the possible approach of using the main characteristics of the solutions
of Eq.\ref{GP-3} and Eq.\ref{GP-4} to develop a wave function for the ELECTRON.
The general idea is to use the electric field of the SOLITON solution Eq.\ref{cgsEr1}, 
since the structure of this field already contains the Bose-Einstein condensate
(BEC) for $g > 0$ and $g < 0$.

The extremely long lifetime of the ELECTRON could be explained by a stable SOLITON. 
As already discussed, it would be necessary to form an object with a repulsive and 
an attractive interior. If these two interactions compensate each other out, an infinite
 lifetime would be possible.

The behavior of the Bose-Einstein condensate (BEC) for $g > 0$ and $g < 0$ would satisfy this 
condition exactly. For this reason, we use equation Eq. \ref{Wplus1} for the simplified approach 
to the wave function for the repulsive component $g > 0$.

\begin{equation}
 \Psi (x)=C_{1} \cdot tangh(C_{2}\cdot x)
\label{Wplus1}
\end{equation}

For the attractive part of the function $ g < 0 $ the equation: Eq.\ref{Wminus1}.

\begin{equation}
 \Psi (x)=C_{3}\frac{1}{cosh\left ( C_{4}\cdot x \right )} 
\label{Wminus1}
\end{equation}

The constants $ C_{1} $, $C_{2}$, $ C_{3} $ and $ C_{4} $ must be adjusted the parameters of the Electron
electric field Eq.\ref{cgsEr1}. For the discussion we show the electric field of the ELECTRON 
SOLION in Eq.\ref{cgsEr11} of  Eq.\ref{cgsEr1} again.

\begin{equation}
E_{r}=f(r,\theta ,\epsilon )=\frac{e\cdot(r^{2}-a^{2}\cdot cos^{2}\theta )}{\varepsilon \cdot
(r^{2}+a^{2}\cdot cos^{2}\theta )^{2}}\cdot \frac{1}{4\cdot \pi }\left [ \frac{V}{m} \right ]
\label{cgsEr11}
\end{equation}
 
 In Fig.{\ref{E-parameter} all parameters of the electron are shown: charge e, $ \varepsilon $, $ \hbar $, mass m,
speed of light c, angular momentum J and a. Substituting this number in the $ E_{r} $ equation Eq.(\ref{cgsEr11})
it is possible to plot this function in 3D in Fig.{\ref{Er-3D11}.
 
 \begin{figure}[htbp]
\begin{center}
 \includegraphics[width=17.0cm,height=8.0cm]{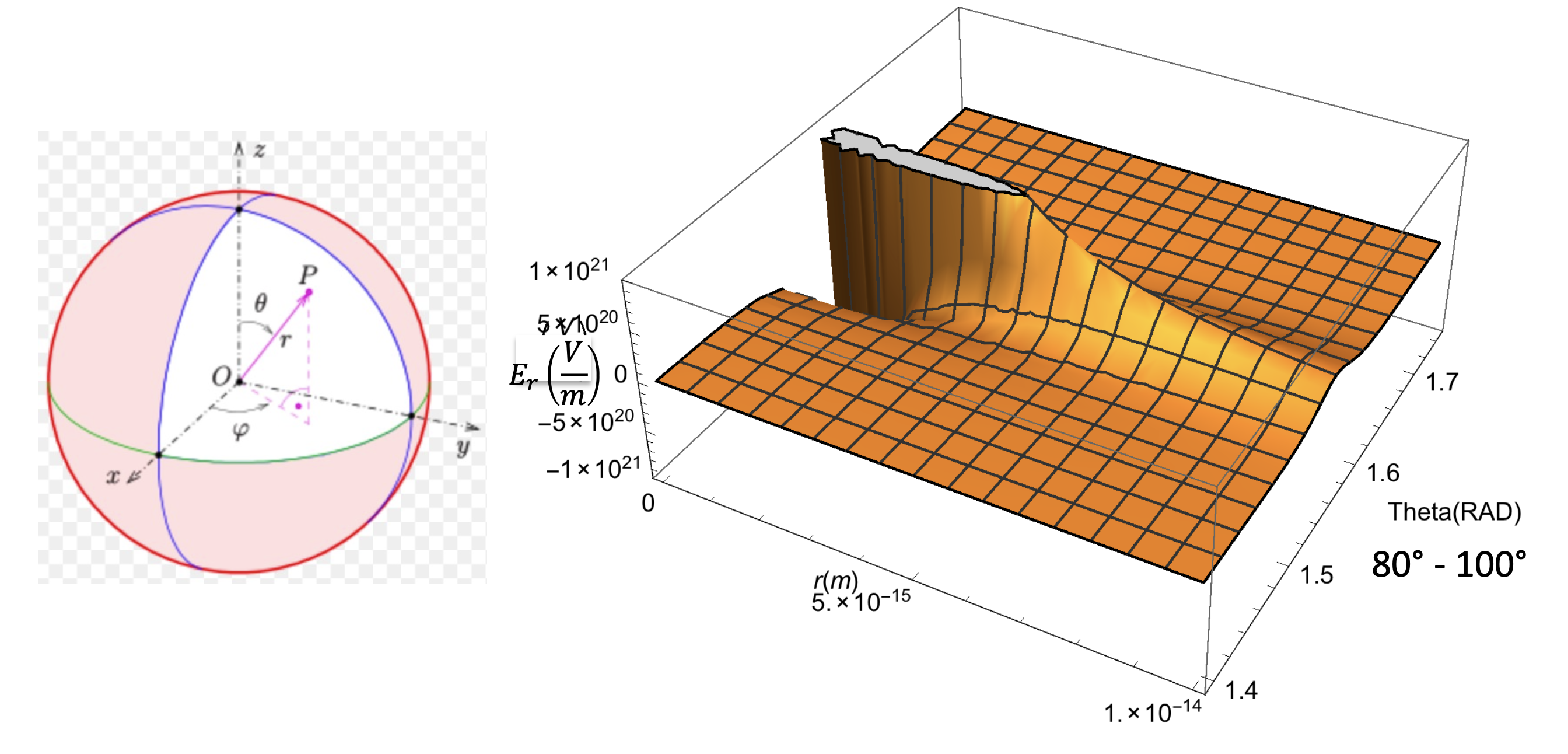}
\end{center}
\caption{ The Er - field of the Electron in 3D plot.}
\label{Er-3D11}
\end{figure}

Fig.{\ref{Er-3D11}  clearly shows that the field is both attractive and repulsive, exhibiting a very high maximum 
of approximately $E_{r}=10^{21}\left ( V/m \right ) $ when $r < 10^{-15}(m) $ and $Theta=\theta =90^{0} $.
The figure also shows that the field is not zero at $Theta=\theta =90^{0} $ because a constant $\epsilon $ is used.

In the preceding discussion chapter \ref{Superconductive  DISC1} it was demonstrated that at $ \theta =90^{0} $ the values of 
$ \epsilon _{r}=\epsilon _{\theta }=L_{F}\to \infty $ set the E-field at $ \theta =90^{0} $ to zero. 
In the equation of $ E_{r} $  Eq.(\ref{cgsEr11}) is  $ \epsilon $ not a function of $ r $ and the angle $ \theta $.
This function is currently unknown. For this reason, it is necessary to introduce such a function $ \epsilon = f(r,\theta ) $ which sets the field 
$ E_{r}=0 $ at $ \theta = 90^{0} $. We used the following test function Eq. (\ref{Sec11}).

\begin{equation}
\epsilon (r,\theta )=1-Sech\left [ m(\theta -\pi /2) \right ]
\label{Sec11}
\end{equation}

In this function $r$ runs in the range $0 < r < \infty$. The function Sech is equal to ONE for
 $\theta = 90^{0}$, which sets $\epsilon$ to ZERO as requested. The parameter $m$ is a 
 constant that defines the width of the delta function  $ ( delta \ E )=f(E_{r}\times \epsilon (r,\theta )) $
 according to Eq. (\ref{cgsEr11}) and Eq. (\ref{Sec11}). The value of $\epsilon$ remains constant in Eq. \ref{cgsEr11}.
Important is the condition for $ r $ in the range  $ 0 < r < \infty $. This condition sets the electric field $E_{r}$ to 
ZERO even when $r = \infty$. The constant $m$ will be adjusted in detail later. 
The effect of this test function for m = 1000 is shown in Fig.{\ref{DEelta11}.

 \begin{figure}[htbp]
\begin{center}
 \includegraphics[width=17.0cm,height=13.0cm]{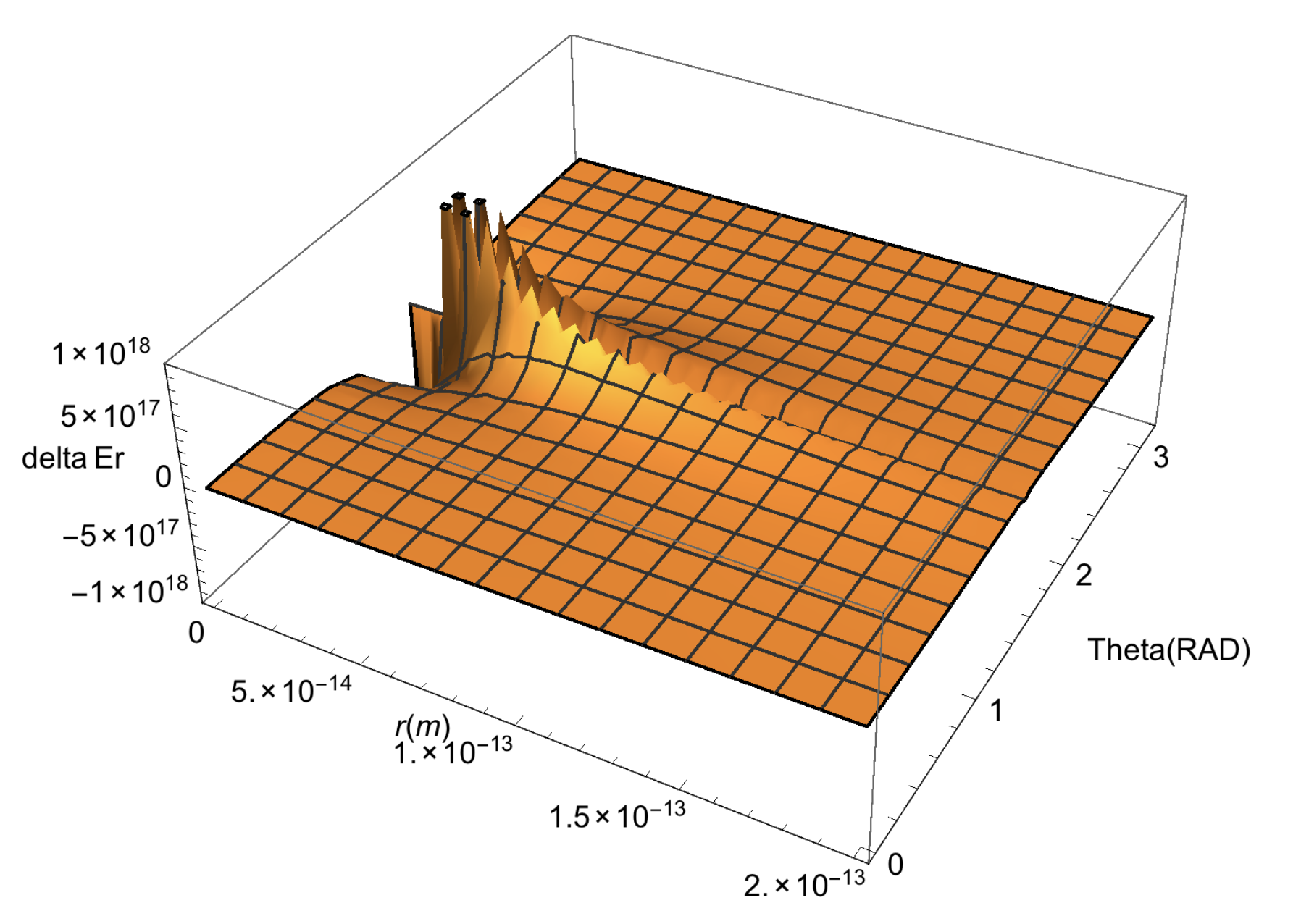}
\end{center}
\caption{ The Er - field including the Sech function Eq.(\ref{Sec11}) of the electron in 3D - plot.}
\label{DEelta11}
\end{figure}

In Fig.{\ref{DEelta11} it is clear visible that for $ \theta =90^{0} $ $ (\pi /2) $ the field $ (delta \ E_{r})=0 $ 
from  $ 0 < r < \infty $ is as requested.

After introducing an E - field function that satisfies the condition of the Gross-Pitaevskii equation,
i.e., repulsive for $g > 0$, attractive for $g < 0$, and superconducting at $\theta = 90^{0}$, it is possible to introduce
a test function for the wave function $ \Psi ^{2} $ of the Gross-Pitaevskii equation.

We first modified the electric field $ E_{r} $ Eq.(\ref{cgsEr11}) in simple form Eq.\ref{Test-E}.
We inserted the field $ E_{r} $ Eq.\ref{Test-E} in simple form into the solutions of
the wave function for a one-dimensional Soliton-Bose-Einstein condensate previously 
discussed in chapter \ref{Solution SOLITON}.

\begin{equation}
E(r)_{Test}=K_{01}\cdot \epsilon (r,\theta )\cdot K_{02}\cdot\frac{(r^{2}-a^{2}\cdot cos^{2}\theta )}{(r^{2}+a^{2}\cdot cos^{2}\theta)^{2}}
\label{Test-E}
\end{equation}

The function $ E(r)_{Test} $ Eq.(\ref{Test-E}) contains the two constants $ K_{01} $, $ K_{02} $, the constant $ ( a ) $
and the function $ \epsilon (r,\theta ) $ Eq.(\ref{Sec11}). 

After extensive testing of possible wave functions $ \Psi $ for a SOLITON ELECTRON, the wave function 
for $ \Psi ^{2} $  Eq. \ref{Wave11} is the most adequate solution for the parameters $ g > 0 $ and 
$ g < 0 $ of the Gross-Pitaevskii equation.
 
\begin{equation}
\Psi ^{2}=(\frac{1}{cosh\left [ E(r)_{Test} \right ]})^{2}\cdot (tanh\left [ E(r)_{Test} \right ])^{2}
\label{Wave11}
\end{equation}

The Eq.\ref{Wave11} contains the Eq.\ref{Wplus1} and Eq.\ref{Wminus1}. It is possible to plot the 
function $\Psi^2$ taking into account the following constants and conditions set be mathematics 
and parameters of the ELECTRON.

\begin{itemize}
\item 
The wave function is normalised to the maximum $ \Psi^{2}_{max}=1 $.
 \item 
 The constant $ m=1.0 \cdot 10^{7} $
 \item 
 The constant $ K_{02}=e/(4\cdot \pi \cdot \varepsilon )=1.44049\cdot 10^{-9} $
 \item 
 The constant $ K_{01}=9.0\cdot 10^{-20} $
 \item 
 The constant $ a=1.936 \cdot 10^{-13} $
 \end{itemize} 

The free parameter m adjusts the width of the zero of the $E_{r}$ field at $\theta = 90^{0}$. The parameter
$K_{01}$ is necessary to weaken the very strong $E_{r}$ field so that it can be used mathematically in the
tanh and cosh functions. The constants $K_{02}$ and $(a)$ are determined from the electron. Under these conditions, 
the wave function $\Psi^{2}$ exhibits two structures A and B, which are shown in the 3D 
representation of Fig.{\ref{Wave-33}.

\newpage

\begin{figure}[htbp]
\begin{center}
 \includegraphics[width=18.0cm,height=12.0cm]{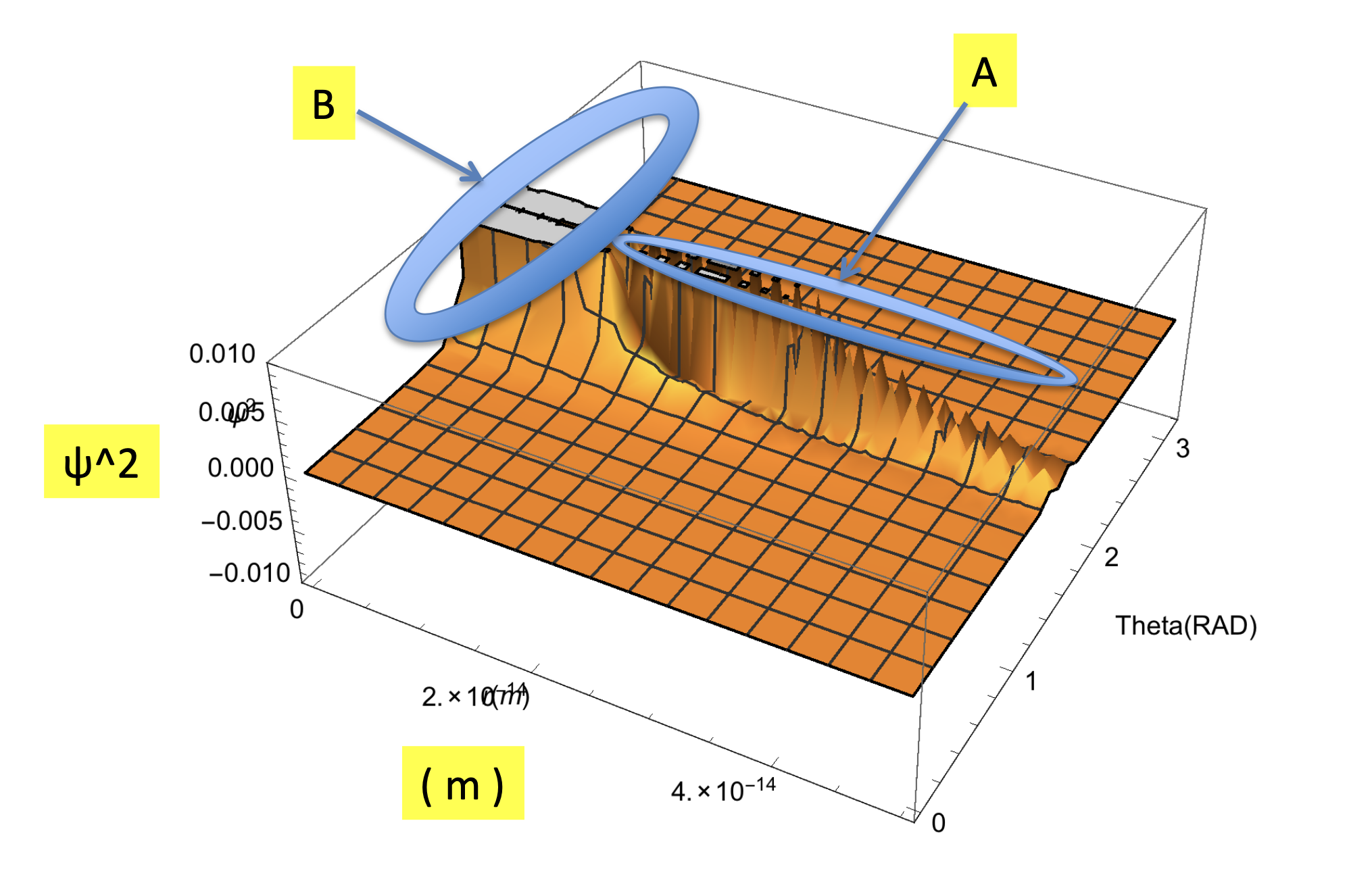}
\end{center}
\caption{ 3D plot of wave function $ \Psi^{2} $ including attractive (A) $ g<0 $ and a repulsive (B) superconducting core with $ g>0 $.  }
\label{Wave-33}
\end{figure}

From the findings of the preceding discussion, the existence of a superconducting disk at 
$ \theta =90^{0} $ with an attracting area ( A ) $ g<0 $ and a repulsing area ( B ) 
superconducting core with $ g>0 $ is evident. If we imagine this structure of a 
superconducting ELECTRON, the ELECTRON would look like the one shown in Fig.{\ref{E-soliton}.

\newpage

\begin{figure}[htbp]
\begin{center}
 \includegraphics[width=16.0cm,height=5.0cm]{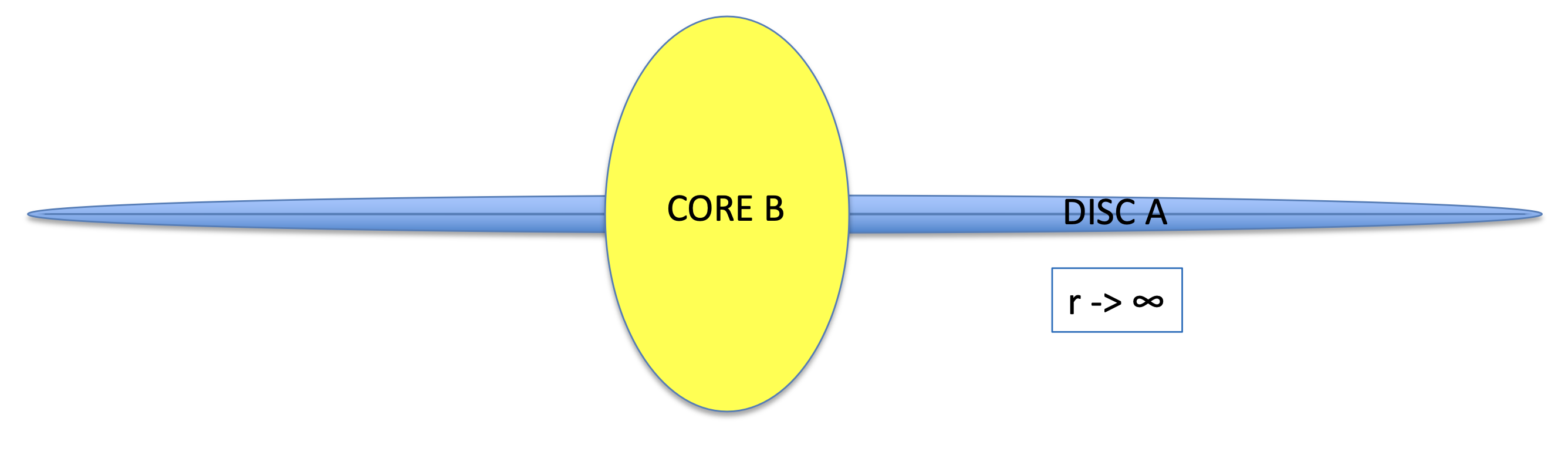}
\end{center}
\caption{ The sketch plot of a ELECTRON - soliton.}
\label{E-soliton}
\end{figure}

The structure A of the wave function $ \Psi^{2} > 0 $ close to the  $ \theta =90^{0} $ disc 
would be originated from the superconducting attractive $ g < 0 $ disc A.
The structure B of the wave function $ \Psi^{2} > 0 $ close to $ r > 0 $ would be 
originated from the superconducting repulsive $ g > 0 $ core.

\subsubsection{g - function SOLUTION for a SOLITON.}
\label{g - function SOLUTION}

After introducing a wave function for the Gross-Pitaevskii equation, it is 
important to investigate whether the g-function Eq. \ref{g-function} is 
finite and exhibits $g < 0$ attractive and $g > 0$ repulsive behavior.

\begin{equation}
g=\frac{\hbar ^{2}}{2m}\cdot \frac{\Delta \Psi }{\left| \Psi \right|^{2}\cdot\Psi }
\label{g-function}
\end{equation}

We investigate the g-function on two different scales and as a function of $\theta$ and $r$. \\

First, a 2D representation of the g-function using $\theta$ and a 3D 
representation of the g-function in the interval
$ 1.0 \cdot 10^{-14}<r<3.\cdot 3.0 \cdot 10^{-13}\left[m\right]$.
Using the constant $\hbar^{2}/2m = 6.10426 \cdot 10^{-39}$ and
the previously used parameters for the wave function m and $K_{01}$,
as well as the constants $(a)$ and $K_{02}$, it is possible to represent 
the 2D and 3D g-function in Fig.{\ref{g-function-1}}.

\newpage

\begin{figure}[htbp]
\begin{center} 
 \includegraphics[width=16.0cm,height=10.0cm]{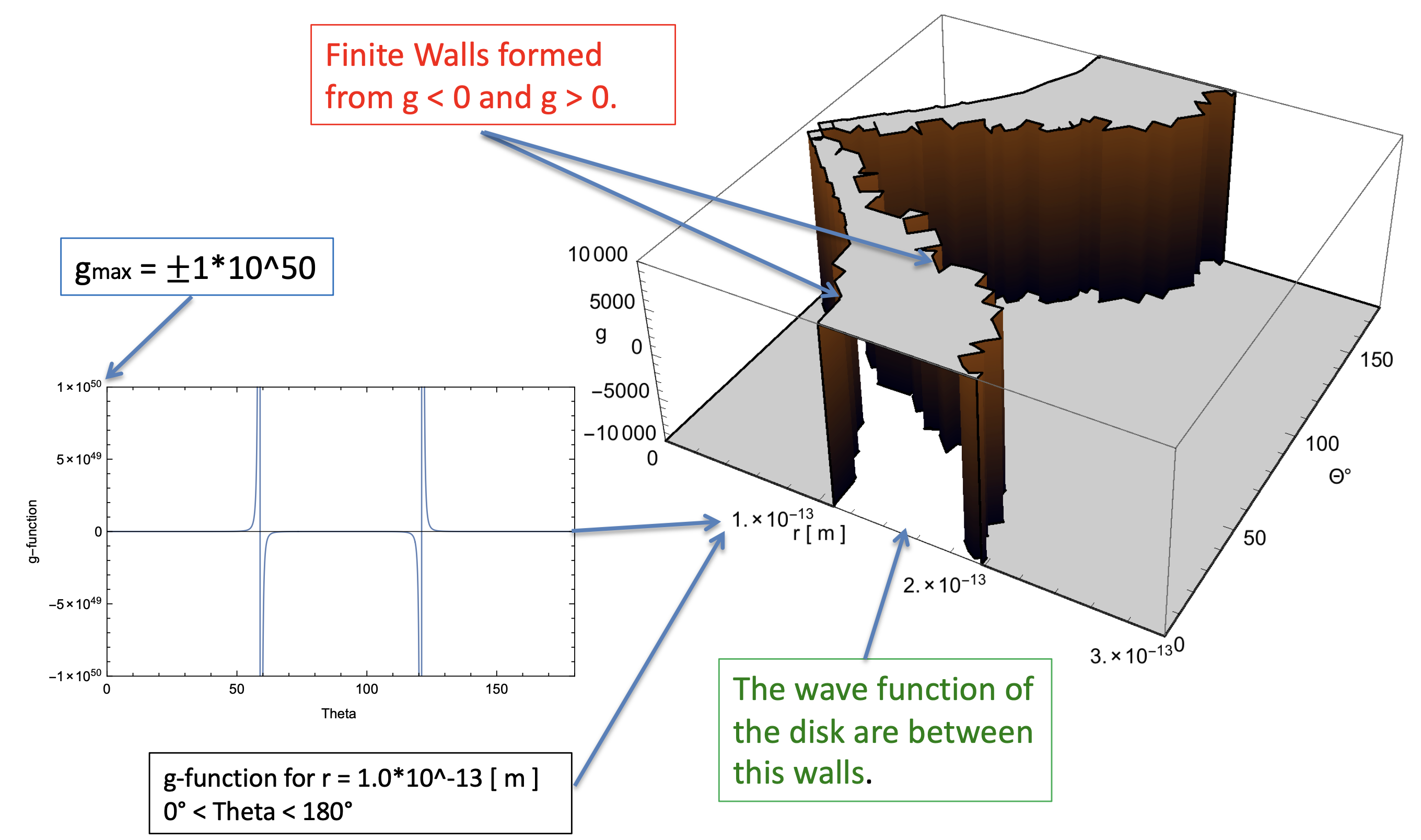}
\end{center}
\caption{ 2D plot via $ \theta $ of g - function plus 3D - plot of g - function.}
\label{g-function-1}
\end{figure}

The g - function forms very high walls up to 
$ g_{max}=\pm 1 \cdot 10^{50} $ for $ g < 0 $ and $ g > 0 $. Between these walls 
the wave function of the disc is located.

\newpage

Secondly, it is interesting that a wall appears at $r < 1.0 \cdot 10^{-15}\left [ m \right ] $, 
which closes off the disk structure. This points to the kernel discussed in Fig.{\ref{E-soliton}. 
The 3D representation as a function of $r$ and $\theta$ is shown in Fig.{\ref{g-function-2} in 
the region $r < 1.0 \cdot 10^{-14}\left [ m \right ] $ and $80^{0} < \theta < 100^{0}$.

\begin{figure}[htbp]
\begin{center}
 \includegraphics[width=16.0cm,height=10.0cm]{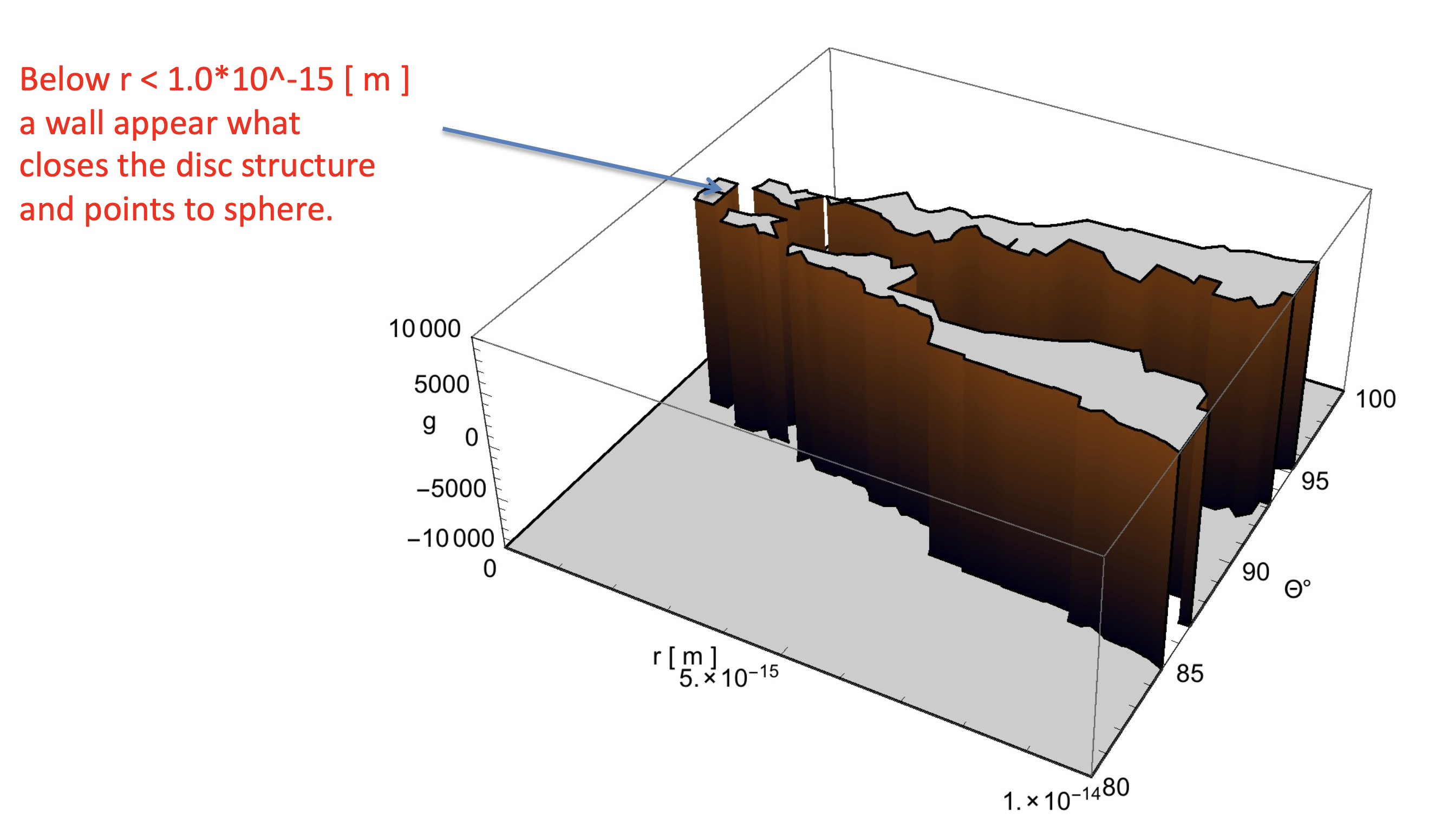}
\end{center}
\caption{  3D - plot of g - function via $ r $ and $ \theta $ including a wall below $ r < 1.0 \cdot 10^{-15}\left [ m \right ] $.}
\label{g-function-2}  
\end{figure}

In conclusion, it should be emphasised based on these diagrams Fig.{\ref{g-function-1}} and Fig.{\ref{g-function-2},
 that it is possible to calculate the function $ g=f(r,\theta ) $ using common mathematical methods and 
to show that g is positive and negative in the range $ -1\cdot 10^{50}<g<1\cdot 10^{50} $.

\subsubsection{ Application of the ELECTRON ETAMPF to the  SOLITON - model.}
\label{Application of the ELECTRON ETAMPF}

In the discussion of the extended fundamental particle scheme of the ETAMPF model, 
the ELECTRON  had a charge distribution of $  (1/3)\times e $ in the KERNEL and $  (2/3)\times e $ OUTSIDE. 
This is illustrated in Fig.{\ref{ETAMPF-charge} as reminder.

\newpage

\begin{figure}[htbp]
\begin{center}
 \includegraphics[width=16.0cm,height=12.0cm]{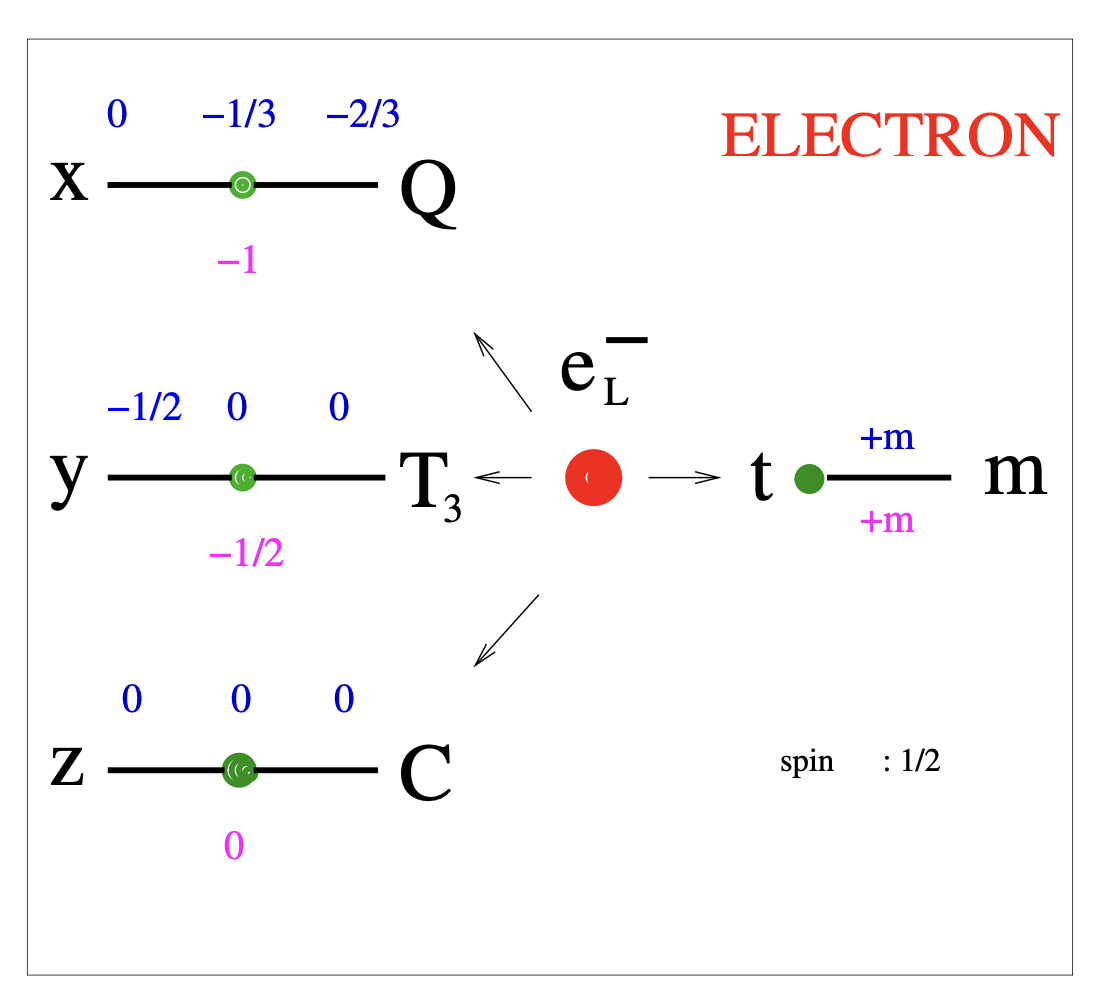}
\end{center}
\caption{ ETAMPF ELECTRON charge $ Q $  IN $  (-1/3)\times e $ OUT $  (-2/3)\times e $.}
\label{ETAMPF-charge}
\end{figure}

The question is whether the ratio of the integral of the wave function $\Psi^{2}$ Eq. \ref{Wave11} 
of the INTEGRAL KERNEL /  INTEGRAL TOTAL WAVE FUNCTION, follows this $1/3<->2/3$ behavior.
And how does the geometric extension of this wave function at radius $r$ compare to the experiment?

To investigate this question, we compare the ETAMPF model Fig.{\ref{ETAMPF-charge} to the 
3D representation of $ \Psi ^{2} $ for $ 1.0\cdot 10^{-14}<r<2.0\cdot 10^{-13}\left
 [ m \right ]$ and $ 0^{0}<\theta <70^{0} $ in Fig.{\ref{wave333} of the SOLITON model.

\newpage

\begin{figure}[htbp]
\begin{center}
 \includegraphics[width=16.0cm,height=12.0cm]{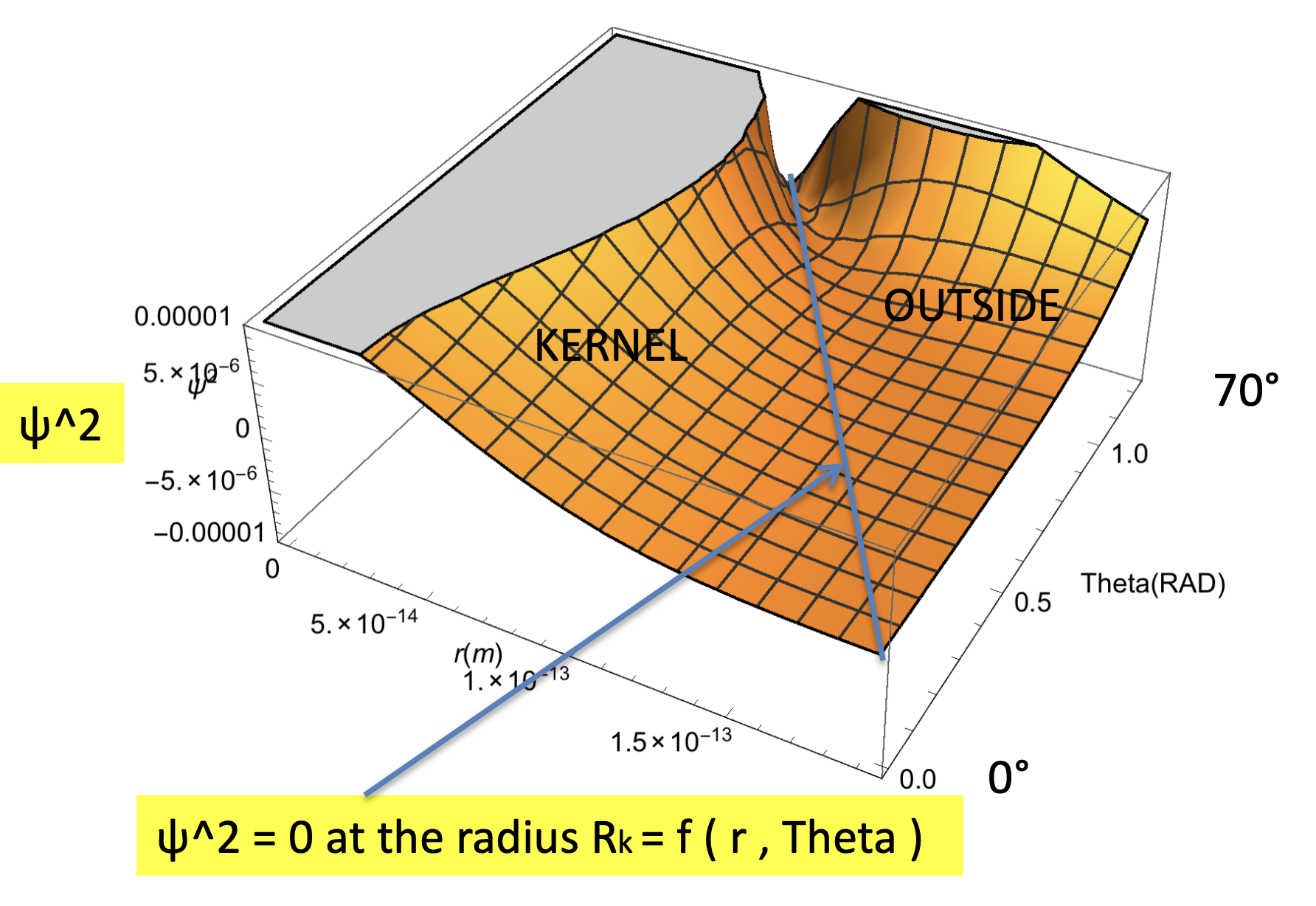}
\end{center}
\caption{ SOLITON model 3D plot of the well separated region of  ELECTRON \\ KERNEL/OUTSIDE.}
\label{wave333} 
\end{figure}

Two clearly separated regions of the wave function are evident between KERNEL and OUTSIDE.

To test the radio range from $1/3$ to $2/3$, we numerically integrated $\Psi^{2}$ in the KERNEL region.

\begin{itemize}
\item 
$0<r<R_{k}=f(r,\theta )\Rightarrow $The radius of the repulsive KERNEL area.
 \item 
$ 0<\theta <180^{0} $
\end{itemize}
  
 and in total structure
 
 \begin{itemize}
 \item 
 $ 0<r<\infty $ the total size of the structure.
 \item 
 $ 0<\theta <180^{0} $
 \end{itemize}

 In Eq. \ref{Rk11} the integral over $ \Psi ^{2} $ in the kernel area $ dr, d\theta $ and $d\varphi $ is shown.
 
\begin{equation}
INT(R_{k})=\int_{0<R_{k}}\int _{0<\theta <180^{0}}\int _{0<\varphi <180^{0}}\Psi ^{2}drd\theta d\varphi 
\label{Rk11}
\end{equation}

In Eq.\ref{Rk22} the integral over  $ \Psi ^{2} $ in the total area $ dr, d\theta $ and $d\varphi $ is displayed.

\begin{equation}
INT(tot)=\int_{0<r<\infty }\int _{0<\theta <180^{0}}\int _{0<\varphi <180^{0}}\Psi ^{2}drd\theta d\varphi 
\label{Rk22}
\end{equation}

Substituting Eq. \ref{Rk11} , Eq..\ref{Rk22} into Eq.\ref{Rk33} and using the limits discussed 
above, the ratio Eq.\ref{Rk33} $ RATIO(Kernel) \approx 46 \% $ is obtained.

\begin{equation}
RATIO(Kernel)=\frac{INT(R_{k})}{INT(tot)}\approx 46\%
\label{Rk33}
\end{equation}

This is not very much, but is approximately 13\% above the expected value of 33\%.
In our discussion of the electron, we concluded that the electron's wave-function 
originates from the electron's superconducting disk and a superconducting 
KERNEL. According to the rules of SUPERCONDUCTIVITY, the wave-function on 
the disk and in the INNER-KERNEL $R_{k0}$ should be zero.

It is important to examine the microstructure of the electron more closely. So far, the 
wave-function on the disk has been ZERO at $\theta = 90^{0}$. The following figure 
shows a 3D representation of $\Psi^{2}$ for $0.0 < r < 3.5 \cdot 10^{-15} \left [ m \right ]$ 
and $89^{0} < \theta < 90^{0}$. But, very close to $\theta = 90^{0}$ and
$r = 0 [ m ]$, the wave-function is not ZERO, which is clearly visible in the following 
figure Fig.{\ref{wave444}}.

\newpage

\begin{figure}[htbp]
\begin{center}
 \includegraphics[width=17.0cm,height=10.0cm]{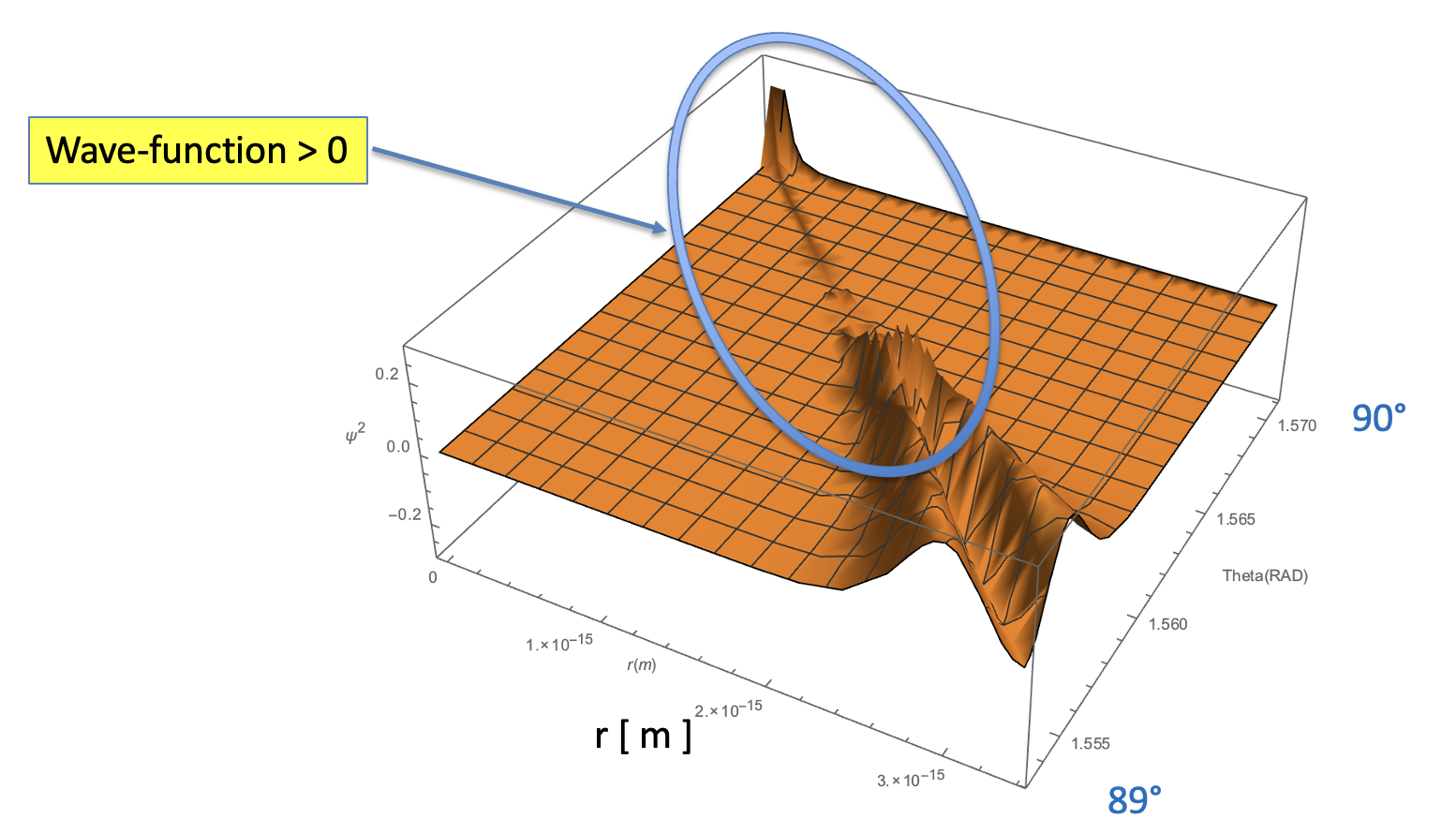}
\end{center}
\caption{SOLITON model 3D plot of $ \Psi ^{2} $ close to $ \theta =90^{0} $.}
\label{wave444}
\end{figure}

To follow the ansatz of SUPERCONDUCTIVITY, it is necessary to introduce an INNER-KERNEL 
wave-function $\Psi^{2}$ with radius $R_{k0}$ to set the wave-function to ZERO.
For this purpose, we explicitly introduce a function $E(r)_{TestA}$, where the 
REPULSIVE kernel $B$ encompasses a radius range $0<r<R_{k0}\left [ m \right ]$ 
with a ZERO wave-function. For simplification, a Unit-Step function was used Eq. \ref{Step11}.

\begin{equation}
STEP(R_{k0})=UnitStep\left [ r-R_{k0} \right ]
\label{Step11}
\end{equation}

To introduce this $ STEP(R_{k0}) $ Eq.\ref{Step11} function into Eq..\ref{Test-E}, a 
modification of electric field E(r)test Eq.\ref{Test-E} to E(r)TestA Eq.\ref{E-test11}} is performed.

\begin{equation}
E(r)_{TestA}=STEP(R_{k0})\cdot K_{01}\cdot \varepsilon (r,\theta )
\cdot K_{02}\cdot\frac{(r^{2}-a^{2}(cos\left [ \theta  \right ])^{2})}{(r^{2}+a^{2}(cos\left [ \theta  \right ])^{2})}
\label{E-test11}
\end{equation}

This also change the wave function $ \Psi ^{2} $ Eq. \ref{Wave11} to the equation below Eq.\ref{Wave555}.

\begin{equation}
\Psi ^{2}=(\frac{1}{cosh\left [ E(r)_{TestA} \right ]})^{2}\cdot(tangh\left [ E(r)_{TestA} \right ])^{2}
\label{Wave555}
\end{equation}

After introducing this additional kernel $R_{k0}$ from Eq. (\ref{Step11}) into Eq. (\ref{E-test11}}),
RATIO(Kernel) Eq. \ref{Rk33} can be recalculated in  Eq. (\ref{Rk55}). 
The following SOLITON parameters and constants are used
to recalculate RATIO(Kernel) in Eq. (\ref{Rk55}).

\begin{equation}
RATIO(Kernel)=\frac{INT(R_{k})}{INT(tot)}=33.42 \%
\label{Rk55}
\end{equation}

Open SOLITON paramaters are:
\begin{itemize}
\item 
$R_{k0}<r<R_{k}=f(r,\theta )\Rightarrow $The radius of the repulsiv KERNEL area.
\item 
$  R_{k0}=3.3\cdot 10^{-15}\left [ m \right ] $ Superconducting INNER-Kernel RADIUS with $ \Psi ^{2}=0 $.
\item 
$ m=1.0\cdot10^{7} $ Superconducting disc width.
\item 
$K_{01}=9.0\cdot10^{-20}$ Mathematic reason ( CUT infinity )
\end{itemize}

Constant factors from the ELEKTRON:
\begin{itemize}
\item 
$ K_{02}=e/(4\pi \varepsilon )=1.44049\cdot10^{-9} $
\item 
$ a=1.936\cdot10^{-13} $ Spin mass of the ELECTRON-
\end{itemize}

In the discussion of the scheme of extended fundamental particles, the ETAMPF model,
Fig.{\ref{ETAMPF-charge} , the ELECTRON had a charge distribution of  KERNEL [ $ ( 1 / 3 )\cdot e $ ] 
and [$( 2 / 3 )\cdot e $] OUTSIDE, as shown in RATIO(Kernel,ETAMPF) Eq.\ref{RkETAMPF}.

\begin{equation}
RATIO(Kernel,ETAMPF)=1/3=33.3333\%
\label{RkETAMPF}
\end{equation}

This charge ratio distribution must be compared with the ratio of the SOLITON model Eq.\ref{RkSOLITON}.
This ratio includes two important radius cuts: $R_{k0}<r<R_{k}=f(r,\theta )\Rightarrow $Radius of the repulsive KERNEL region,
and the cut $R_{k0}=3.3\cdot 10^{-15}\left [ m \right ] $ Radius of the superconducting inner nucleus with $\Psi ^{2}=0$.

\begin{equation}
RATIO(Kernel,SOLITON)=33.42 \%
\label{RkSOLITON}
\end{equation}

Both RATIOS agree very well numerically. The result is not sensitive to $m$ and $K_{01}$.
As already explained for $R_{k0}=0$, the RATIO(Kernel,SOLITON) would increase by approximately 10\%. \\

\subsubsection{Shape of ELECTRON according to ETAMPF and SOLITON model.}

In accordance with the current findings, the shape of the electron decomposes into TWO geometric AREAS. \\

A) A superconducting DISC with a width defined by the parameter m plus a KERNEL 
in the range of the radius $ 0 < R_{k0} $ in which the wave function is ZERO shown in Fig.{\ref{Shape-11}.

\newpage

\begin{figure}[htbp]
\begin{center}
 \includegraphics[width=16.0cm,height=3.5cm]{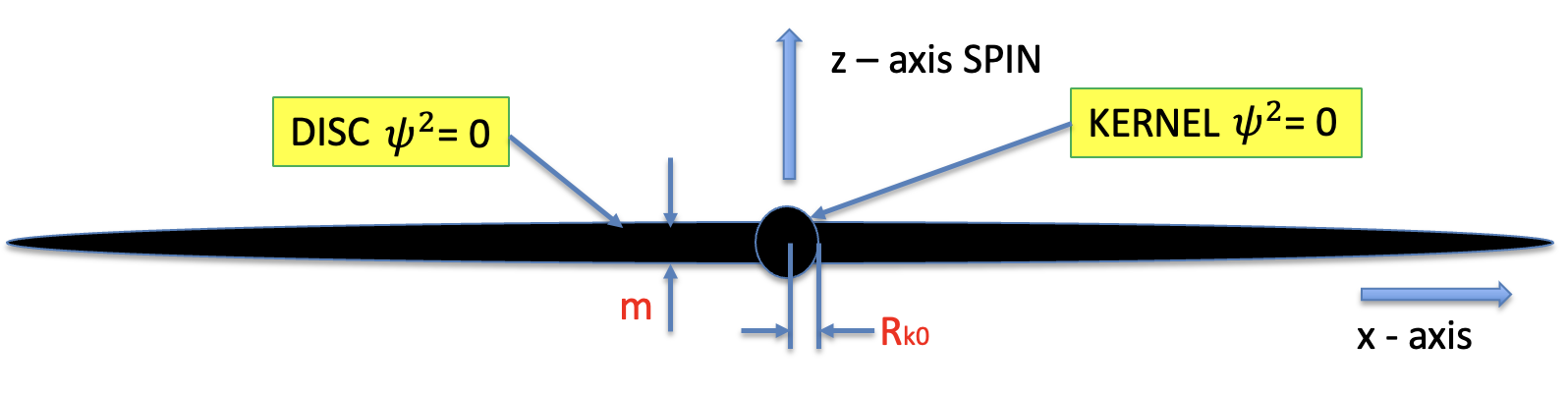}
\end{center}
\caption{Soliton superconductive DISC including a KERNEL .} 
\label{Shape-11}
\end{figure}

B) The superconducting DISK and a region of a wave function outside with attractive 
$ \Psi ^{2}> 0 $ and repulsive $ \Psi ^{2}> 0 $ between the disk and the KERNEL, 
which is shown in Fig.{\ref{Shape-22}.

\begin{figure}[htbp]
\begin{center}
 \includegraphics[width=16.0cm,height=6.0cm]{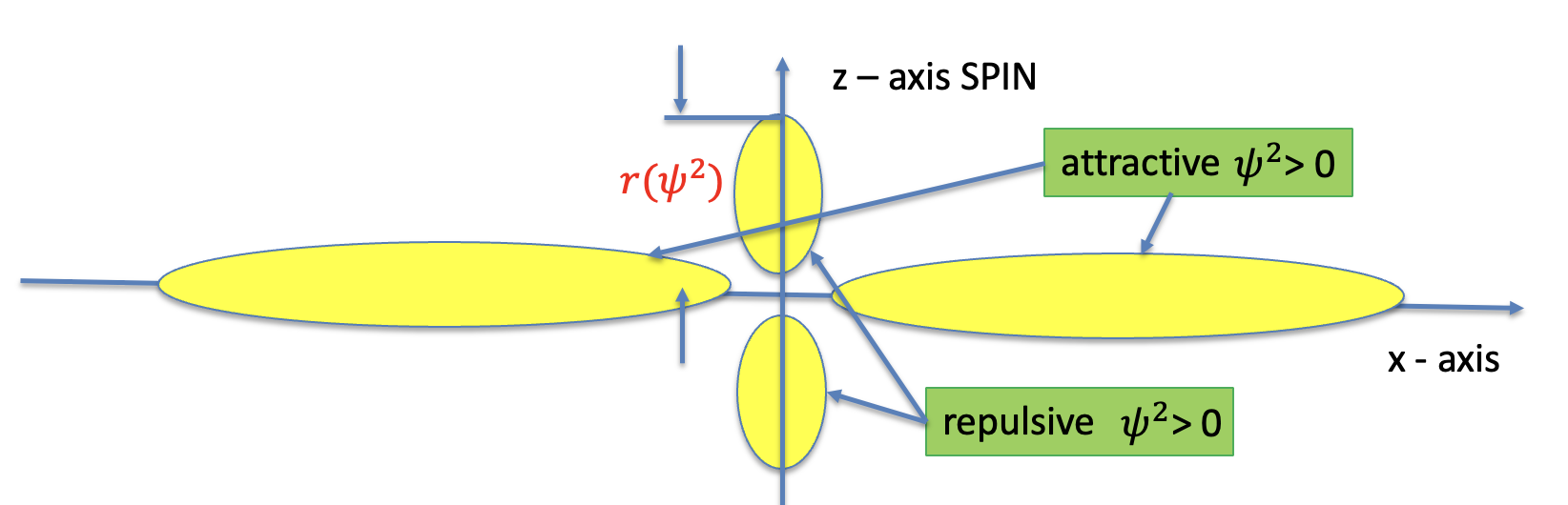}
\end{center}
\caption{Soliton superconductor with attractive $ \Psi ^{2}> 0 $ and repulsive $ \Psi ^{2}> 0 $.}
\label{Shape-22}
\end{figure}

By superimposing Fig.{\ref{Shape-11} and Fig.{\ref{Shape-22}, the overall view of the SOLITON-ELECTRON 
solution can be represented in Fig.{\ref{Shape-33}. In particular, the important input of the 
ETAMPH model for defining the size of the inner CERNEL region $ R_{k0} $
via the repulsive integral $1/3 $ and the attractive integral $2/3$ is shown.

\begin{figure}[htbp]
\begin{center}
 \includegraphics[width=16.0cm,height=8.0cm]{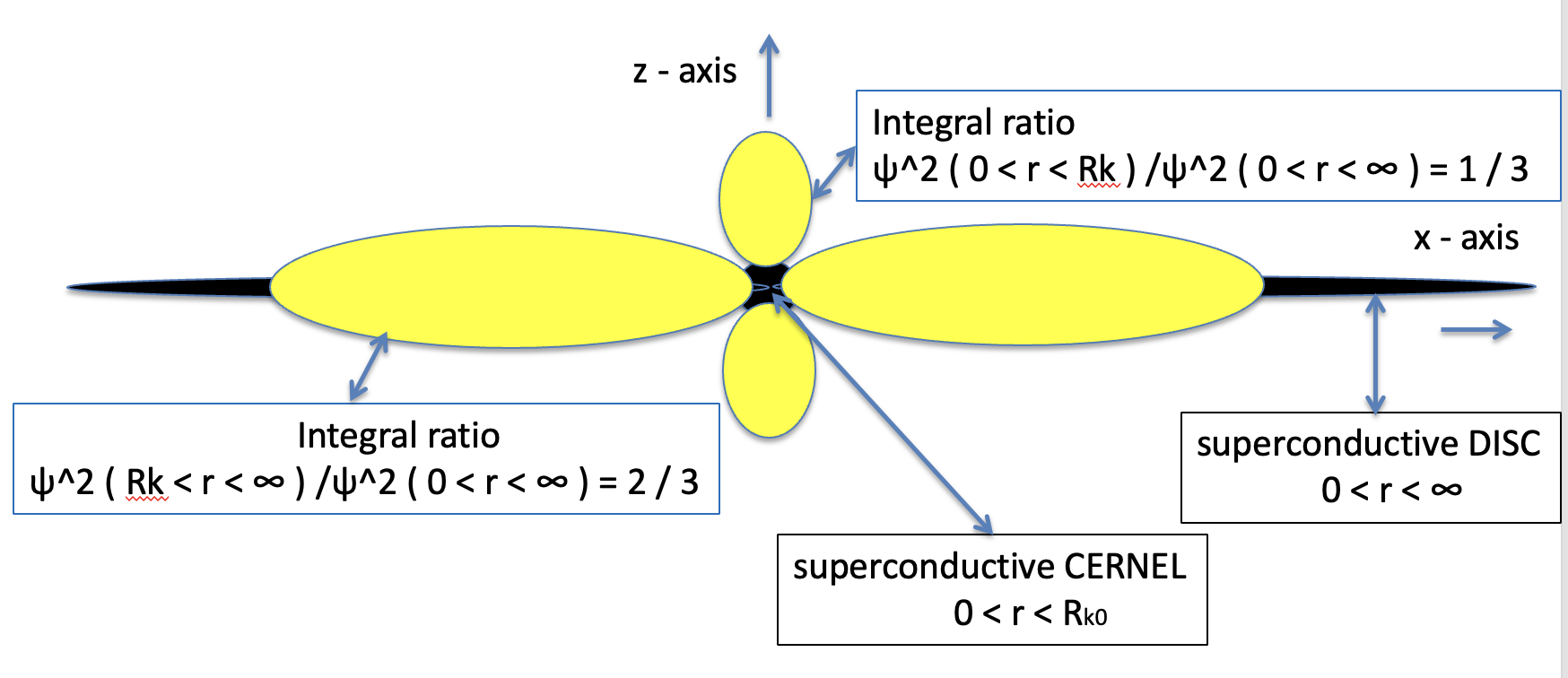}
\end{center}
\caption{Total view of Soliton ELECTRON superconductive DISC.}
\label{Shape-33}
\end{figure}

\newpage

\subsection{Conclusion of the theoretical consideration of a geometrical extended ELECTRON.}
\label{Conclusion of the theoretical consideration}

As discussed in Fig.{\ref{Invest-timing}  ``Timing of the investigation of the finite size of FP's'', performed by ETHZ and USTC, 
we approached the ``Substructure of FUNDAMENTAL PARTICLES (FP's)'' in a theoretical chain in four steps.
 
In a FIRST step we introduced the idea to describe all fundamental particles in common scheme of a 
``Empirical Toy Ansats about a Microstructure of Fundamental Particles (ETAMFP)'' 
chapter \ref{Empirical toy ansatz} the ETAMFP-model.
If one combines the temporal development of the Standard Theory from the Planck Era to the 
present day with the temporal development of the Big Bang Theory in the same time frame 
shown in Fig. {\ref{Timing-FP}, a ``Geometrical Extension of Fundamental Particle'' can 
be derived from this history. The charge distribution and size of the FB's are, in this sense, a 
legacy of the history of the Standard and Big Bang theory. 
These considerations enable the development of a ``Basic scheme of extended 
fundamental particles in the ETAMFP model'', discussed in chapter 
\ref{Empirical toy ansatz} and chapter \ref{Scheme of geometrical extended FB's } and 
shown in Fig.{\ref{Basic-ETAMFP} and Fig.{\ref{Electron11},
The lifetime of the electron $\tau >4.6\times 10^{26} $ yr is longer stable than the 
lifetime of the universe. If only ONE attractive gravitational force existed that could generate 
such a lifetime, this would be extremely complicated, more or less impossible, 
since the ansatz would end in a black hole.
One solution to this problem is to introduce a force that acts attractively and repulsively 
from a certain distance, as shown in Fig.{\ref{Forces1} and Fig.{\ref{Forces22} 
chapter \ref{Forces and stability}. A mass core subjected to  an attractive Schwarzschild 
force would transform into a repulsive force, similar to
the De Sitter force \cite{Desiter1}, and thus enable a stable lifetime.
Finally, in the first step, it was explicitly shown that it is possible to sort ALL FB's into a 
common scheme, which is shown in Fig.{\ref{FM-left-handed} to Fig.{\ref{Higgs-boson},
 and to calculate an approach for excited states of fermions, which is shown in 
 Fig.{\ref{Excited-states2} chapter \ref{Scheme of geometrical extended FB's }

In the SECOND step, we use special relativity in a classical approach to investigate the electron as a GYROSCOPE.
By inserting the magnetic moment $ \mu_{e} $ and the electric dipole moment $ d_{e} $ of the electron in a 
classical GYROSCOPE, the geometric shape of an ELECTRON in a CM - system and LAB - system 
can be approximated shown in Fig.\ref{SketchElectronContracted}. 
It is important to note that in this approach, the electron exists in the LAB system. The two charges $ 1/3 $
outside rotate at nearly the speed of light. This leads to an extreme Lorentz contraction
at every radius $ r $ perpendicular to the spin axis at $ \theta = 90^{0} $, but no contraction 
parallel to the spin axis at $ \theta = 0^{0} $.
The model allows the calculation of the TWO geometric parameters of the GYROSCOPE, 
in shown Fig.\ref{SketchElectronContracted}. The GYROSCOPE dimensions $ r_{0} $ in CM - system
and $ r_{L} = r_{0} $ in LAB - system run parallel to the axis of rotation. As input in the GYROSCOPE
can be used to calculate the radius of the ELECTRON from the magnetic moment $ \mu _{e} $  of the electron or the electric 
dipole moment $  d_{e} $ of the electron. The value of $  r_{L} = r_{0}^{\mu }=8.39035\times 10^{-13}\left [ m \right ]$ is calculated
from the magnetic moment $ \mu _{e} $ Eq.\ref{RadiusL}. 
The radius calculated from $  d_{e} $ is very similar divided in two parts $  r_{0}=r_{o}^{de} $
is calculated after Tab. \ref{modelbeta1} direct for $ \beta =1 $ $ r_{0}(2/3 \ e) $ and $  r_{L}=r_{o}^{de}=r_{o} $ 
in Tab. \ref{electronPARAMETERSdipol} takes into account that $ \beta $ a little pit smaller as ONE, shown 
in Tab.\ref{PrecissionKbetaB}. It is interesting to notice that the value of  $  r_{L} = r_{0}^{\mu } $ and 
$  r_{0}=r_{o}^{de} $ are not the same but very similar.

It is important to note that in the GYROSCOPE ansatz, the two charges $ 1/3 $ outside the kernel generate 
a well-defined magnetic moment $ \mu _{e} $. This agrees with the experimentally measured magnetic moment,
which is presented with very high statistical significance in Tab. ~\ref{Parameterelectron} refernce \cite{Electron-Parameter}. 
In contrast, the electron in the GYROSCOPE ansatz should not generate an electric dipole moment $ d_{e} $,
since, after the Fig.{\ref{ETAMPF-charge} NO charge anti-charge combination exists outside the electron kernel. 
Only the introduction of the de Sitter ansatz of an attractive - repulsive force for the same charge, which is 
discussed in the chapter \ref{Electro vacuum soliton with spin}, could generate a small electric dipole moment.
The measurement of the electron's electric dipole moment shows a value, of very low significance, 
still in agreement with ZERO shown Table ~\ref{Parameterelectron} refernce , \cite{Electron-Parameter}
and \cite{Dipol11}. The similarity of the different electron radii, calculated from $\mu_{e}$ and $d_{e}$, discussed 
above, supports the GYROSCOPE ansatz.

The estimation of the size of the electron kernel chapter \ref{Kernel of the electron} 
including the nucleus radius $ r_{k} $ is calculated from the history of the universe. The important parameter
for calculating the kernel size is the intersection of the running coupling constants 
$ \alpha _{1},\alpha _{2},\alpha _{3} $ as a function of the GUT scale $ \Lambda _{GUT} $. 
The predictions of the running coupling constant in SM and SUSY models for $\Lambda_{GUT}$ 
in Fig. {\ref{Running-coupling} and Fig. {\ref{Timing-BP} show that the physical conditions at 
these energy scales have changed dramatically. Currently, experimental knowledge about the crossing 
point energy at $  \Lambda _{GUT} $ in the SM or SUSY model is very incomplete. 
For example, according Tab.\ref{ElectronKernelRadius} we took for $\Lambda _{GUT}=10^{15} \left [ GeV \right ]$
$r_{K}=2.33\times 10^{-19}\left [ m \right ]$, however depending about the type of coupling constant $ \alpha $ values
up too $\Lambda _{GUT}=10^{17} \left [ GeV \right ]$ $r_{K}=1.08\times 10^{-16}\left [ m \right ]$ are possible.
Size of different parameters are given in Tab.~\ref{Parametrs-Electron}.
The strongly contracted radius $r_{e}$ is taken directly from $d_{e}$ Tab. \ref{electronPARAMETERSdipol}.
According to the figure Fig.\ref{PrecissonradiusElectron}, the contraction is only relevant in the outermost kernel region.

In the THIRT step a ``Spinning superconducting electrovacuum solition'' chapter \ref{Electro vacuum 
soliton with spin} is introduced to describe a geometrical extended electron 
``Particle-like electro vacuum soliton related to General Relativity'' chapter \ref{Electro vacuum soliton}. The
important difference to step TWO is general relativity is used in this step. It is first necessary to
study a ``Self gravitating particle-like structure with de Sitter vacuum core for $spin = 0$ particles''
chapter \ref{Electro vacuum soliton spin ZERO}. A metric and the associated equations of a de 
Sitter-Schwarzschild geometry describe a black hole 
whose singularity is replaced by a de Sitter core of a fundamental scale. This ansatz is 
extended from particles with spin = 0 chapter \ref{Electro vacuum soliton spin ZERO}  to 
particles with spin ``Spinning supraconducting electrovacuum soliton''  
chapter \ref{Electro vacuum soliton with spin} and \ref{Superconductive DISC1}. The introduction of Boyer-Lindquist coordinates, 
together with the corresponding equations, leads to an object with a superconducting disk
shown in Fig.{\ref{Meissner}.  The electric and magnetic fields of this 
object can then be calculated in the more common cgs system,
``Transformation of the E and B - Field from Boyer-Lindquist into cgs units'' chapter \ref{Boyer-Lindquist in cgs}.
Using the equations for the E - field of a soliton, a vector diagram of the E - field is shown in Fig.{\ref{E-Vector-plot}.
Based on this information, a sketch comparing the electron E - field of the ETAMFP model and the soliton model can 
be created in Fig.{\ref{SKETCH1}. The attractive and repulsive nature of the E - field, resulting from the de Sitter structure of the soliton, 
is clearly visible in Fig.{\ref{E-Vector-plot}. Importantly, the geometric structure of the Electron E - field 
in the ETAMFP - model is very similar to that of the SOLITON - model, as shown in Fig.{\ref{SKETCH1}. 
The point-like charge in the ETAMFP -  model is replaced by  a ``Superconducting - charge - ellipse - Disc'' 
in the SOLITON - model. See Chapter \ref{Boyer-Lindquist E B cgs}, \ref{Vector E - field} and \ref{SKETCH E - field}.
Similarly, by including the necessary equations, the B - field of the SOLITON - model can 
be calculated in Fig. {\ref{B-field-common} and Fig. {\ref{B-field-inner}, as well as the geometric structure of 
``Electron B - field in the ETAMFP - model and SOLITON - model'' in Fig.{\ref{SKETCH-B1}..
Two diagrams of the B-field are calculated because the B-field for radii $r < 10^{-20}\left [ m \right ]$ 
points in the direction of a kernel structure. Again Fig. $\ref{SKETCH-B1}$ also shows that the ETAMFP - 
model and the SOLITON - model are very similar. ``The - Point - like - charge - Ring'' is replaced by a 
``Superconducting - charge - ellipse - disc''. See chapter \ref{The VECTOR B }, \ref{The VECTOR B inner } 
and \ref{SKETCH  ETAMPF SOLITON}.

Of particular importance is the fact that the simple classical 
ETAMFP- model is very similar to the mathematically sophisticated SOLITON -  model.

In the last FOURTH step is a ``Calculation of a wave function using the substructure of FP, via the E - field of a non 
point like Electron withe the Gross - Pitaevskii equation'' is introduced see Fig. {\ref{Invest-timing}. 
The entirely new approach to calculating this wave function is explained in the chapter ``Gross-Pitaevskii equation 
and ELECTRON wave function'' \ref{Gross-Pitaevskii} . The details necessary for the calculation; 
superconductors, electric field, and solitons introduced in the Gross-Pitaevskii equation, 
are summarised in Fig.{\ref{Gross-11}. It is necessary to examine ``The Gross-Pitaevskii equation'' 
chapter \ref{Gross-Pitaevskii-1} and find the solution for the wave function of a soliton in the chapter
``Wave function solution for a SOLITON'' chapter \ref{Solution SOLITON}. Subsequently, a toy model 
for an electron as a soliton is introduced in the chapter ``Mathematical Engineering of a toy model for an 
ELECTRON as SOLITON'' chapter \ref{Engineering SOLITON}. 
The known equations of the Electron soliton's E - field must be related to a superconducting disk 
see Fig.{\ref{DEelta11}. The 3D representation of the E - field shows the attractive and repulsive 
structure of the E - field according to the De - Sitter structure and the structure of the superconducting 
disk with the E - field $E_r=0$ at $\theta=90^0$. Substituting this E - field into the solutions of 
the Gross-Pitaevskii equation leads to a solution for the wave function $\Psi^2$ as a function of the radius 
$r$ and the angle $\theta$ Fig.{\ref{Wave-33}. 
From the preceding explanations, it follows that at $\theta = 90^{0}$ a superconducting disk 
exists with an attractive (A) $g < 0$ and a repulsive (B) superconducting core with $g > 0$. 
This $\Psi^{2}$ function must be tested in the Gross-Pitaevskii equation ``g - function 
SOLUTION for a SOLITON'' see chapter \ref{g - function SOLUTION}. A finite solution for 
a g-function is shown in Fig. \ref{g-function-2}. Importantly, that below 
$ r< 1.0\times 10^{-15}\left [ m \right ] $ a wall appears assigned as $ R_{ko}^{g-function} $.
Next the chapter ``Application of the ELECTRON ETAMPF to the  SOLITON - model'' 
chapter \ref{Application of the ELECTRON ETAMPF} is then introduced.
In the discussion of the extended fundamental particle scheme of the ETAMPF model, the ELECTRON has a 
charge distribution with a KERNEL  charge of $  (1/3)\times e $ and an OUTSIDE charge of $  (2/3)\times e $.
This is illustrated in Fig.{\ref{ETAMPF-charge}. According to the ETAMPF - model, the integral of the 
$ \Psi ^{2} $ function shown in Fig.{\ref{wave333} over the KERNEL and OUTSIDE charge must be 
RATIO = KERNEL/OUTSIDE = 33.33\%.
To satisfy this condition of the ETAMPF - model, it was necessary to set $\Psi^{2}$ to ZERO below a 
radius $ R_{ko}=3.3\times 10^{-15}\left [ m \right ] $ . The geometric shape of the SOLITON  Electron is shown in 
Fig.{\ref{Shape-11}, Fig.{\ref{Shape-22} and Fig.{\ref{Shape-33}.
The important parameter in Fig.{\ref{Shape-11} is $R_{ko}$, and in Fig.{\ref{Shape-22}, the distance radius $r(\Psi^{2})$.
This distance radius $r(\Psi^{2})$ indicates the radius beyond which the repulsive force $\Psi^{2}$ is reduced to ZERO.
In comparison to the ETAMPF - model, this is the radius $r_{0}$ in Fig.\ref{SketchElectronContracted}.
Depending on the details, the ETAMPF model calculates two very similar radii, $r_{0}^{\mu}\left[m\right]$
 and $r_{0}^{de}\left[m\right]$. The radius $r_{0}^{\mu}\left[m\right]$ is calculated after taking into 
 account in Tab.\ref{PrecissionKbetaB} that the velocity of the charges outside the molecule does 
 not exactly correspond to the speed of light, $\beta \leq 1$. For the radius $r_{0}^{de}\left[m\right]$, 
 however, it is assumed that $\beta = 1$. 

To assess the quality of the ETAMPF model and the SOLITON model, it is helpful to
compare the radius of the ETAMPF model $ r_{0}^{\mu }$ with the radius of the SOLITON model
radius of $ r(\Psi ^{2}) $, where the wave function $ \Psi ^{2}=f(\theta =0,r_{S})=0 $.
For a Gyroscope, the non-contracted radius, measured parallel to the spin axis $  r_{0}^{\mu } $, must 
be equal to the radius of the repulsive nucleus of the soliton's wave function, 
parallel to the spin axis, $ r_{S}$, where $ \Psi ^{2}=0$.
Calculating the position of $r_{S}$ is mathematically complex because the wave 
function $\Psi^{2}$ gradually decreases to zero. In a 3D - plot  of 
$\Psi^{2}=f(\theta =0,r)$ in Fig.{\ref{PSI-theta-11}, it is clearly visible that the 
wave function $\Psi^{2}$ decreases from $r = 0$ through $r > \infty$ to $\Psi^{2} = 0$.

\begin{figure}[htbp]
\begin{center}
 \includegraphics[width=16.0cm,height=11.0cm]{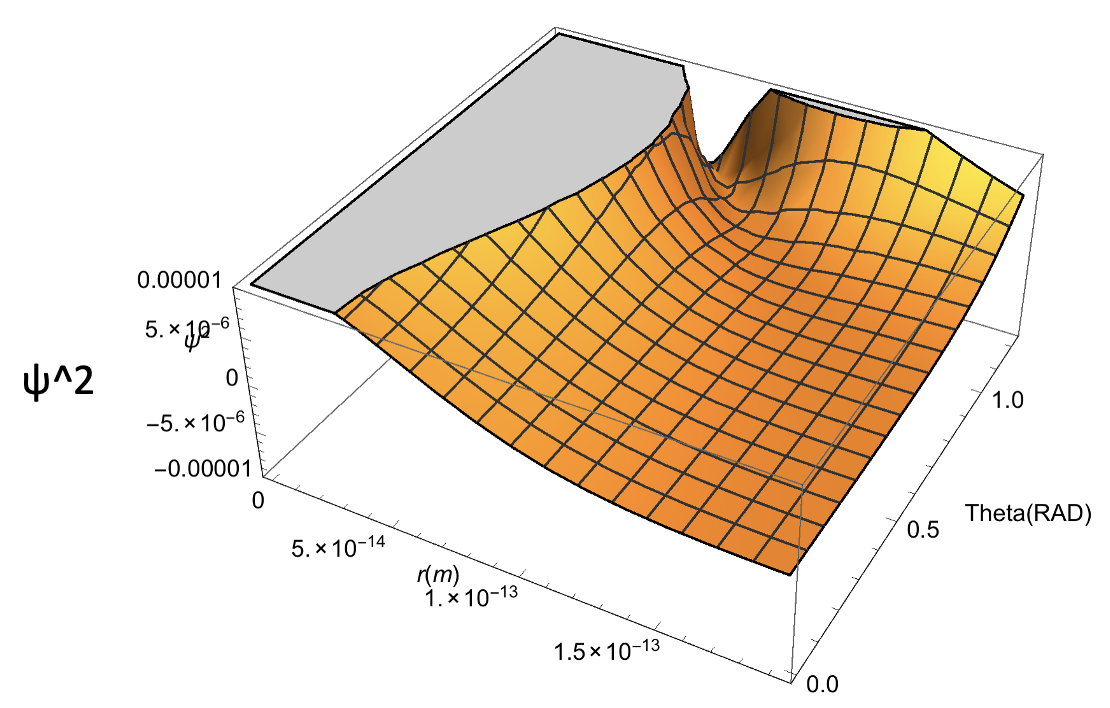}
\end{center}
\caption{The wave function $ \Psi ^{2}=f(\theta =0,r) $ parallel to the spin axis.}
\label{PSI-theta-11}
\end{figure}

A detailed calculation shows that the wave function at a radius of $ r_{S}= 8.4\times 10^{-13}\left [ m \right ] $
$ \Psi ^{2}=2.4\times 10^{-8}\simeq 0 $. This value is very close to zero and is a good approximation for 
comparing $ r_{0}^{\mu } $ with $ r_{S} $.

A summary of all important parameters of the ETAMPF- model and the SOLITON - model is 
presented in the Tab.~\ref{Parametrs-Electron}.

\newpage

\begin{table}
\caption{Parameters of the size of ELECTRON Gyroscope - Soliton.}
\label{Parametrs-Electron}
\begin{center}
\begin{tabular}{||l|l|l|l|l|l|l|l|l|l||}                         \hline
$Size$   &$ r_{0}^{\mu }\left [ m \right ]$&$r_{0}^{de}\left [ m \right ]$&$r_{k}\left [ m \right ]$&$r_{e}\left [ m \right ]$
                                                           \\  \hline
 $Gyroscope$   &$8.39035\times 10^{-13}$&$5.79910\times 10^{-13}$&$1.08\times 10^{-16}$&$0.105\times 10^{-28}$                                                        
                                                           \\  \hline
 $Size$   &$r_{S}(\psi ^{2})\left [ m \right ]$&$R_{k0}\left [ m \right ]$&$ R_{k0}^{g-function}\left [ m \right ]$&$----$                                                          
                                                           \\  \hline
 $Soliton$   &$ 8.4\times 10^-13$&$3.3\times 10^{-15}$&$ 1.0\times 10^{-15}$&$----$                                                                                                                    
                                                            \\  \hline
\end{tabular}
\end{center}
\end{table}

After discussing the conclusion chapter \ref{Conclusion of the theoretical consideration}, it is important to note that
the electron has an outer radius $r_{Outside}$ and an inner - kernel radius $r_{Inner}$. 
The radius $r_{Outside}$ is similar to $r_{0}^{\mu}$, $r_{0}^{de}$, and $r_{S}$, which were derived 
from very different approaches. The inner radius $r_{Inner}$, in turn, is similar to 
$R_{k0}\left[m\right]$, $R_{k0}^{g-function}\left[m\right]$ and $r_{k}\left[m\right]$.
The radius of the SOLITON $ R_{k0} $ of the wave function $ \Psi ^{2} $ must be set to 
$ \Psi ^{2} =0 $ to satisfy the charge distribution condition of the ETAMPF model 
$ 1/3 e $ in the center and $ 2/3 e $ outside, and the KERNAL 
radius $ R_{k0}^{g-function} $, which appears after examining 
the g-function of the solution of the Gross - Pitaevskii equation.
The radius $r_{k}\left [ m \right ] = 1.08\times 10^{-16}$ is derived from the heritage
of the Big Bang model, which is shown in Fig.{\ref{Running-coupling}.
Considering Tab.~\ref{Parametrs-Electron}, $r_{k}\left [ m \right ]$ is a function of $\Lambda_{GUT} [GeV]$
in the range $\Lambda_{GUT} [GeV] = 10^{13} - 10^{17}$, which is not known experimentally with high precision.
For comparison with the nuclear radius $R_{k0}\left [ m \right ]$ and $R_{k0}^{g-function}\left [ m \right ]$,
we have chosen the extreme value of $\Lambda_{GUT}$ in Tab.\ref{ElectronKernelRadius}, 
since this value limits the possibilities for $R_{k0}\left [ m \right ]$ and $R_{k0}^{g-function}\left [ m \right ]$.
The radius $r_{e} $ is the strongly Lorentz-contracted quantity of the Gyroscope, which runs perpendicular to 
the spin axis of rotation.

In the chronological sequence of the discussion in Fig.{\ref{Invest-timing} , the theoretical part of the investigation 
of the ELECTRON as a geometrically extended object is treated. We begin and end with the 
``Experiments for testing the size of FP's using the direct contact term''.

\section{Experiments to test the finite size of FP's.}
\label{Experiment11}

Three groups of experiments are used to test the size of an Electron via direct 
contact term. ``Is the Non - Pointness of the Electron Observable in $ {\EEG} $ Annihilation at 
Center-of-Mass Energies 55 - 207 GeV?'' \cite{Diffcross11}, ``Hint for a minimal interaction length in 
$ {\EEG} $ annihilation in total cross section of center-of-mass energies 55 - 207 GeV'' \cite{Tot-cross11} 
and ``Hint for axial-vector contact interactions in the data on $ {\EEEEG} $ at centre-of-mass 
energies 192 - 208 GeV'' \cite{Bhabha-cross11}. 
In the experimental section, we discuss in detail the first two groups of studies conducted by the 
USTC-ETHZ-GAMMA group. We only briefly describe the BHABHA work, as it was not 
carried out by the USTC-ETHZ group.

\subsection{Finite Size of Electron in $ {\EEG} $ Annihilation in differential cross section of centre-of-mass energies 55 - 207 GeV.}
\label{Diff-Cross}

\subsubsection{Introduction.}
\label{Diff-Cross-Introduction}

The electron is one of the most fundamental building blocks of matter,
and its discovery over a century ago revolutionised our understanding of
the physical world. Since then, it has been the subject of countless investigations,
revealing properties that continue to challenge our understanding of nature at the deepest level.
A historical account of the electron's discovery and subsequent study can
provide valuable insights into the foundations of modern particle~physics.
 
The discovery of Coulomb's law in 1785 and magnetism in 1820 paved the way for 
the investigation of charged particle beams using cathode - ray tubes.
This possibility was realised in 1869.

Charles-Augustin de Coulomb's law states that the magnitude of the electrostatic
force between two point charges is directly proportional to the product of the
charges and inversely proportional to the square of the distance between them~\cite{COULOMB1,COULOMB2}.

Three discoveries made in 1820 laid the groundwork for magnetism. Firstly, Hans Christian $\ddot{\rm O}$rsted
demonstrated that a current-carrying wire produces a circular magnetic field around itself~\cite{OERSTED,OERSTED1}.
Secondly, Andr\'e-Marie Amp\`ere showed that parallel wires with currents attract each other
if the currents flow in the same direction and~repel if they flow in opposite directions~\cite{AMPERE,AMPERE1,AMPERE2}.
Thirdly, Jean-Baptiste Biot and F\'elix Savart experimentally determined the forces that a current-carrying,
long, straight wire exerts on a small magnet. They found that the forces were inversely
proportional to the perpendicular distance from the wire to the magnet~\cite{BIOT}.

Cathode rays, also known as electron beams or $e$-beams, are streams of electrons observed
in discharge tubes. They are generated by applying a voltage across two electrodes
in an evacuated glass tube, causing electrons to be emitted from the cathode (the negative electrode).  
The~phenomenon was first observed in 1869 by Julius Pl\"{u}cker and Johann Wilhelm 
Hittorf~\cite{CATHODE,CATHODE2} and later named ``cathode rays'' (Kathodenstrahlen)~\cite{CATHODE2,CATHODEa}
by Eugen Goldstein in 1876. In~1897, Joseph John Thomson discovered that cathode rays were
 composed of negatively charged particles, which were later named electrons.
 Cathode-ray tubes (CRTs) use a focused beam of electrons deflected by
 electric or magnetic fields to create images on a screen~\cite{CATHODEb}.
 
Sir Joseph John Thomson is credited  with the discovery of the electron in 1897,
the first subatomic particle  discovered.He showed that cathode rays were 
composed of previously unknown negatively charged particles, which
must have bodies much smaller than atoms and a very large charge-to-mass ratio~\cite{Thomson}.
Finally, Robert A. Millikan and Harvey Fletcher measured in 1909 
the mass and charge separately in the oil drop experiment~\cite{Millikan,Millikan1,Millikan2}.

After the discovery of the electron,  Max Abraham~\cite{Abraham} and Hendric 
Lorentz~\cite{Lorentz1,Lorentz2} in 1903 proposed the first models of the electron, as an
extended spherical electrical charged object with its total energy concentrated in the 
electric field. However, these models were based on the assumption of a homogeneous 
distribution of charge density. Although the model offered a way to explain the electromagnetic origin of the 
electron mass, it also raised the problem of how to prevent the electron from flying apart 
under the influence of Coulomb repulsion. Abraham proposed a solution to this 
inconsistency by suggesting that non-electromagnetic forces (such as, for~example, 
Henri Poincar\'e stress)  were necessary to prevent the electron from exploding.
It can be said that modelling the electron within the framework of electromagnetism 
was deemed impossible at that time. Later, Paul Dirac \cite{Dirac} proposed a point-like model 
of the electron and recognised the appealing aspect of the Lorentz model \cite{Lorentz1} regarding 
the electromagnetic origin of the electron mass. Nonetheless, at~that time,
this idea was found to be inconsistent with the existence of the neutron.
Dirac highlighted~\cite{Dirac} that although the electron can be treated as a point
charge to avoid difficulties with  infinite Coulomb energy in equations, its finite size reappears
in a new sense in the physical interpretation. Specifically, the~interior of the electron
can be viewed as a region of space through which signals can be transmitted faster than~light.

Arthur Compton firstly introduced the idea of electron spin in 1921. In a  paper on investigations into
ferromagnetic substances with X-rays~\cite{Compton,Compton1}, he writes: ``Perhaps the most natural,
and certainly the most generally accepted view of the nature of the elementary magnet,
is that the revolution of electrons in orbits within the
atom give to the atom as a whole the properties of a tiny permanent magnet ''.
The electron's spin magnetic moment, $\mu_s$, is related to 
electron's spin amgular momentum, $S$, through
$\mu_s=-g_s\mu_{\rm B} S/\hbar$, where the spin $g$-factor, $g_s\approx2$,
 $\mu_{\rm B}$ is the Bohr magneton, and $\hbar$ is the reduced
Planck constant. The Stern--Gerlach experiment, first proposed by Otto Stern in 1921 and
conducted by Walther Gerlach in 1922~\cite{Stern}, inferred the existence of
quantised electron spin angular momentum. In the experiment, spatially varying
magnetic fields deflected silver atoms with non-zero magnetic moments on 
their way to  a glass slide detector screen,
providing evidence for the existence of electron spin.
The existence of electron spin can also be inferred theoretically from the
spin-statistics theorem and the Pauli exclusion principle.
Conversely, given the electron's spin, one can derive the Pauli exclusion principle~\cite{Stern,Pauli}.

The existence of quantised particle spin enables the investigation of spin-dependent interactions 
by scattering polarised particle beams at  different targets. In nuclear physics, polarised beam 
sources \cite{POLOVERVIEW,CROSSBEAM} are used for scattering experiments. The polarised beam
get accelerated in an electrostatic accelerators, such as Tandem accelerators, which can achieve 
center-of-mass energies of 1.2 MeV \cite{Tandem1,Tandem11} to 20 MeV \cite{Tandem2,Tandem21}.
Three types of polarised beams were developed: the atomic beam source, 
which utilises the technique of the Stern--Gerlach experiment~\cite{ATOM,ATOM1}, 
the Lamb-shift source  \cite{LAMB,LAMB1,LAMB2} developed after
the discovery of the Lamb shift in 1947~\cite{LAMB-SHIFT}  
and the crossed-beam source~\cite{CROSSBEAM,CROSSBEAM1}.

Since 1926, various classical models of spinning point particles have been developed.
However, these models face the challenge of constructing a stable point-like particle
that includes only a single repulsive Coulomb force over a range from zero to infinity.

One model of point-like particles related to electron spin is the Schr$\ddot{\rm o}$dinger
suggestion~\cite{Schroedinger1} that connects electron spin with its Zitterbewegung motion, ``A trembling
motion due to the rapid oscillation of a spinning particle about its classical world line''.
The Zitterbewegung concept was motivated by attempts to understand the intrinsic
nature of electron spin and involved fundamental studies in quantum mechanics
\cite{Schroedinger2,Schroedinger21,Schroedinger22,SchroedingerBefore23,Schroedinger23}.

Other types of classical models of point-like spinning particles have been developed. The~Yang--Mills model
 is a class of gauge theories that describe the strong and electroweak interactions in
the Standard Model of particle physics. The~Weyl model is a spinor field theory
that describes massless spin-1/2 particles that do not follow the Dirac equation.
The Thirring model is a $1+1$ dimensional field theory that describes a system of Dirac
fermions coupled to a massless bosonic field. It is exactly solvable and has been
used as a toy model for studying many-body problems in condensed matter physics.
The Gross--Neveu model is a $2+1$-dimensional field theory that describes a system of
fermions with an interaction term that is quadratic in the fermion field.
It is also exactly solvable and has been used to study critical phenomena in
condensed matter physics. The~Proca model is a relativistic quantum mechanical
model that describes a massive vector boson. It is used to describe the massive
vector bosons in the electroweak interaction and has been used extensively in the development of the Standard Model.
Many reviews and textbooks  consider various classical models of point-like spinning particles;
some examples include~Refs. \cite{pointTextbook1,pointTextbook2,pointTextbook3,pointTextbook4,pointTextbook5}.
In general, these kinds of models encounter the problem of divergent self-energy for a point charge.
The approach this problem in the frame of various generalisations of the classical Lagrangian terms with higher derivatives
or extra variables~\cite{CLMODEL1,CLMODEL11,CLMODEL12,CLMODEL13,
CLMODEL14,CLMODEL15,CLMODEL16,CLMODEL17,CLMODEL18,CLMODEL19} and then restrict
undesirable effects by applying geometrical~\cite{CLMODEL2,CLMODEL21}
or symmetry~\cite{CLMODEL3,CLMODEL31} constraints.

The discovery of the Kerr--Newman solution~\cite{Kerr} to the Einstein--Maxwell equations in 1965
led to new possibilities for investigating the electron's structure. Recently, this solution has been 
used in Ref.~\cite{Irina11DeSitter} to propose a model that considers the interplay between 
electrodynamics and gravity in the electron's structure. Coupling electrodynamics with gravity 
introduces the geometry of De Sitter spaces~\cite{DeSitter,DeSitter1},
which provides attractive/repulsive forces dependent on distance from the origin and can distinguish
between Schwarzschild and De Sitter black holes, ensuring the electron's stability from the Coulomb repulsion.
The theoretical aspects of the question whether the electron is point-like or not are discussed in Ref.~\cite{Irina1}.

The modelling of electron structure is driven by the desire to understand
fundamental issues, such as the number of fermion families, fermion mass hierarchy
and mixing properties, that the Standard Model cannot explain.
For example, a~natural consequence of the so-called composite models
approach~\cite{compMod1,compMod2,compMod3,compMod4} to
addressing the aforementioned questions is the assumption that quarks and leptons possess substructure.
According to this approach, a~quark or lepton might be a
bound state of three fermions~\cite{compMod5} or a fermion and a boson~\cite{compMod6}. 
In~many models along this type, quarks and leptons are composed of a scalar and a spin-1/2 preon.Composite models~\cite{compMod1,compMod2,compMod3,compMod4} predict a rich spectrum of
excited states~\cite{compMod1,compMod2,compMod3,compMod4,compMod7} of known particles.
Discovering the excited states of quarks and leptons would be the most convincing proof of their substructure.
Assuming that ordinary quarks and leptons represent the ground states, it is natural to assign the excited fermions with the
same electroweak, color and~spin quantum numbers as their low-lying partners.
Excited states are transferred to ground states through generalised magnetic-type transitions,
where photons (for leptons) or gluons (for quarks) are emitted, as~described in Ref.~\cite{HeavyElMu1}.
When excited states have small masses, radiative transition is the main decay mode,
but when their masses approach that of the $W$ boson,
a large fraction of three-particle final states appear in the decay of excited~states.

As discussed above, there is currently no fully predictive model
capable of describing the substructure of quarks and leptons.
Therefore, one must rely on phenomenological investigations of substructure effects,
which can manifest in various reactions (see~\cite{HeavyElMu333} for a review).
The search for excited charged and neutral fermions has been ongoing 
for over 30 years, but so far without success. Quantum electrodynamics (QED) 
provides an ideal framework for investigating potential lepton substructures.
Any deviation from QED's predictions in differential or total cross-section
of $e^+ e^-$ scattering can be interpreted as non-point-like
behavior of the electron or the presence of new physics.
Note that the case of excited quarks~\cite{excQuark555} is a direct generalisation of the lepton case.
Theoretical predictions suggest that the transition mechanism of excited quarks, $q^*$, is
through gluon, $g$, emission, $q^*\rightarrow qg$. However, distinguishing this effect from the
standard background of three-jet events poses a challenge.
As a result, the~lepton sector remains the most favorable field for searching for substructure
effects from an experimental standpoint. Among~the various channels in the $e^+ e^-$ scattering experiment,
the process of photon pair production $\EEG$
 stands out due to its negligible contribution
from weak interactions.  Additionally, another process of fermion pair production with $e^+ e^-$ final state,
which is used as the luminosity meter at low angle,
is also highly suitable for QED testing in the search for the electron's substructure. Both processes
are presented in Figure~\ref{Feynman1}.

\newpage

\begin{figure}[htbp]
\begin{center}
 \includegraphics[width=14.0cm,height=10.0cm]{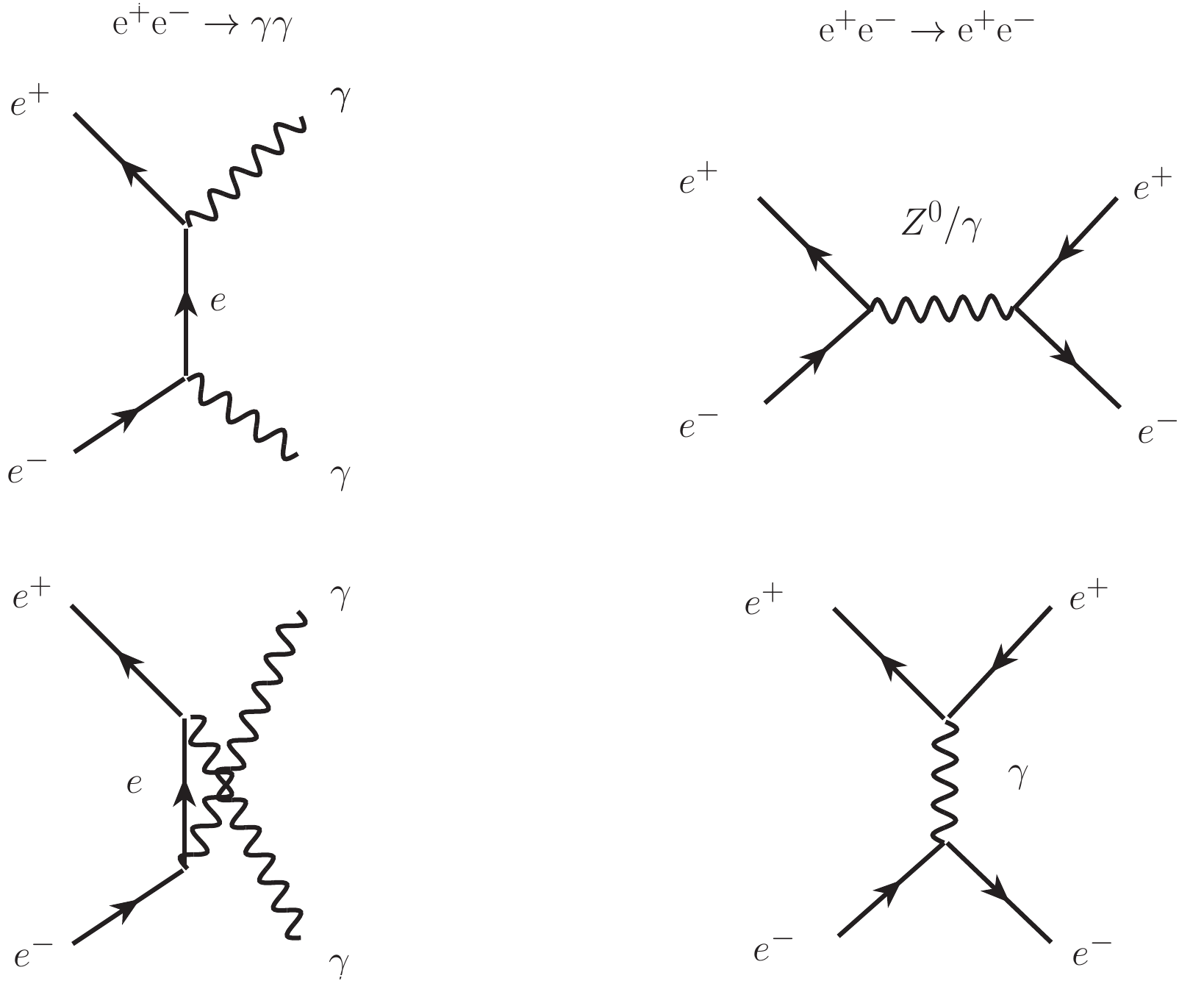}
\end{center}
\caption{The lowest-order Feynman diagrams for the $ \EEG $ and $ \EEEE $ reactions.}
\label{Feynman1}
\end{figure}

In the $ \EEG $ process, the~reaction proceeds through the $t$- and $u$-channels. The~Bhabha scattering $ \EEEE $
proceeds through the $s$- and $t$-channels. 

The two real photons in the final state of the $ \EEG $ reaction are indistinguishable,
therefore the reaction proceeds via the $t$ - and $u$ - channels, while the
$s$ - channel is forbidden due to the conservation of angular momentum.
The $s$-, $t$-, and $u$-channels (Mandelstam variables) represent different 
types of particle interaction and energy and momentum exchange.
The reaction is highly sensitive to long-range QED interactions,
because the two photons in the final state have left-handed and right-handed
polarisation which results in total spin of ZERO, forbidding the
$s$-channel with spin ONE for $ \gamma $ and $ Z^0 $.
As a pure annihilation reaction, the~$ e^+ $ and $ e^- $ in the initial state
completely annihilate into two photons in the final state,
making it straightforward to subtract the background~signal. 

The Bhabha reaction ${\EEEE}$ is a mixed reaction that occurs through scattering in 
both the $s$-channel and $t$-channel. At energies around the $Z^0$ pole, 
the~$Z^0$ contribution dominates.The elastic scattering and annihilation channel 
are superimposed since the $e^+$ and $e^-$ in the initial and final states are identical.
Therefore, the~${\EEEE}$ reaction serves as a test for the superimposition of 
short-range weak interaction and long-range QED~interaction.

In this paper, we analyse deviations from QED by combining data on the differential
cross-section of the $\EEG$ reaction measured via various $e^+ e^-$ storage ring experiments.
Specifically, the~VENUS Collaboration investigated this reaction in 1989~\cite{VENUS333} at the
center-of-mass energies, $\sqrt{s} $ = 55 GeV--57 GeV, while the OPAL Collaboration~\cite{OPAL444}
studied it in 1991 at the $Z^0$ pole at $\sqrt{s} $ = 91 GeV. The~TOPAS Collaboration also investigated
this reaction in 1992 at $\sqrt{s} $ = 57.6 GeV~\cite{TOPAS555}, while the ALEPH Collaboration studied it in
1992 at the $Z^0$ pole at $\sqrt{s} $ = 91.0 GeV~\cite{ALEPH666}. 
Furthermore, the~DELPHI Collaboration
investigated the reaction from 1994 to 2000 at energies ranging from $\sqrt{s} $ = 91.0 GeV 
to 202 GeV~\cite{DELPHI,DELPHI1,DELPHI666}, while the L3 Collaboration studied it 
from 1991 to 1993 at the $Z^0$ pole, with $\sqrt{s} $ = 88.5 GeV to 93.7 GeV~\cite{L3A666}. The~L3 Collaboration
also studied the reaction in 2002 at $\sqrt{s} $ ranging from 183 GeV to 207 GeV~\cite{L3B888},
and the OPAL Collaboration investigated it in 2003 at \mbox{$\sqrt{s} $ ranging from 183 GeV} to 207 GeV~\cite{OPAL999}}.
Deviations from QED were investigated through the study of contact interactions and excited electron exchange displaced as in Figure~\ref{Feynman1}. Colleagues of some of the authors of this paper have reviewed experimental
studies and models of deviations from QED in their Theses~\cite{ETHZ-USTC,ETHZ-USTC1,ETHZ-USTC555}. 
 For earlier review, see \cite{Models}.

The effective Lagrangian governing a contact interaction exhibits a dependence on the inverse cutoff scale,
$1/\Lambda$, where the power of $1/\Lambda$ is determined by the dimensionality of the fields 
involved. Additionally, this Lagrangian conserves the helicity of fermion currents. This ensures known 
particle masses are much less than $\Lambda$. Different helicity choices for the fields used in 
the Lagrangian result in different predictions for angular distributions and polarisation
observables in reactions where the contact interaction is present. Figure~\ref{Feynman2}b depicts 
the QED direct contact term, characterised by scale parameters $ \Lambda_{+} $ and $ \Lambda_{-} $, 
 which represent  for the cutoff parameter in case of positive and negative interferences, as described in Section~\ref{Helectron}. 
These parameters are subsequently interpreted as being indicative of an extended annihilation radius in the $\EEG$ reaction.
Figure~\ref{Feynman2}c depicts Feynman diagrams that are sensitive to the mass
of the excited electron, $m_{e^{*}}$, with the cutoff parameter, $\Lambda_{e^*}$, being a function of $m_{e^{*}}$.

In 1989, the~VENUS Collaboration~\cite{VENUS} established initial limits
of $ \Lambda_{+}> $ 81 GeV and $ \Lambda_{-}> $ 82 GeV. Table~11 
 in Ref.~\cite{VENUS} provides an overview of other Collaborations that have been studying the same subject.
The significance of all analyses was below $1\sigma$ (one standard deviation)
and~the fitted values of the parameters $1/\Lambda_{-}^4$ and $1/\Lambda_{+}^4$ were negative.
In 2002, the~L3 Collaboration~\cite{L3BBB44} established limits on $ \Lambda_{+} > $ 400 GeV,
\mbox{$ \Lambda_{-} >$ 300 GeV} and \mbox{$ m_{e^{*}} >$ 310 GeV},
including negative fit parameters with a significance below $ 1\sigma$.
In their latest publication on the subject, in~2013 the LEP (Large Electron-Positron collider) 
Electroweak Working Group~\cite{LEP2} conducted an analysis of data from
the differential cross-section of all LEP detectors in the energy range
of $ \sqrt{s} $ = 133 GeV to 207 GeV. The~group established limits of
$ \Lambda_{+} > $~431~GeV, $ \Lambda_{-} >$ 339 GeV and~$ m_{e^{*}} >$ 366 GeV,
which included negative fit parameters with a significance of nearly 2$\sigma $.

Thus, comparing the results obtained by combining the data from all LEP2
 Collaborations with those from L3 alone, it can be observed that the increased statistics
have led to higher confident results. Based on this observation, we conduct a global fit using data from all
six research projects mentioned above to investigate $ \Lambda_{+} $, $ \Lambda_{-} $,
and $ m_{e^{*}} $ for $ \sqrt{s} $ ranging from 55 GeV to 207 GeV, including the corresponding luminosities.  
An initial attempt to perform the global fit, which involved some of the authors of this paper, has been previously
published in detail in Refs.~\cite{5SIGMA,5SIGMA1,5SIGMA2}.  It is noteworthy that the global fit
revealed a significant deviation of the differential cross-section of the annihilation reaction $\EEG$ 
from QED predictions, with a statistical significance of 5$\sigma$.

In the current paper, we scrutinise the global fitting procedure
by examining all technical details used in the $\chi^2$-analysis.
Section~\ref{Experiment11} provides a detailed description of the theoretical
and experimental framework, used for calculating the differential and total cross-sections
of the $ \EEG $ reaction in QED, including radiative corrections and
modifications due to contact interactions and models with excited electrons
shown in Figure~\ref{Feynman2}. 

\begin{figure}[htbp]
\begin{center}
 \includegraphics[width=13.0cm,height=8.0cm]{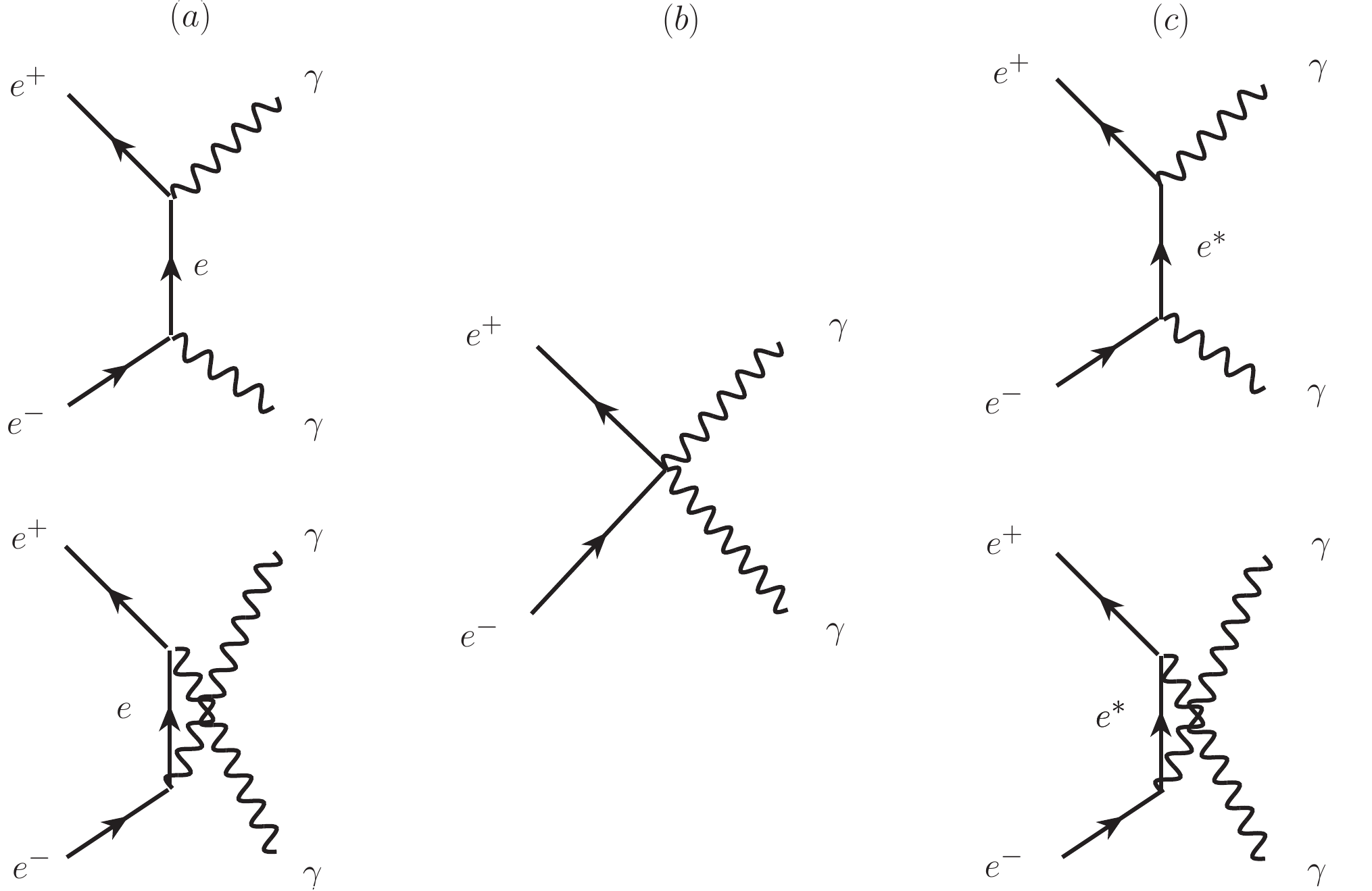}
\end{center}
\caption{ The lowest-order Feynman diagrams of the $\EEG$ reaction are shown, with~(\textbf{a}) representing QED, (\textbf{b}) contact interaction and~(\textbf{c}) excited electron~exchange.}
\label{Feynman2}
\end{figure}

Section Section~\ref{Experiment11} is in detail divided in 5 subsections :
Section~\ref{Diff-Cross} presents all the data of ``Finite Size of Electron in $ {\EEG} $ Annihilation in differential
 cross section of centre-of-mass energies 55 - 207 GeV''. 
Section~\ref{TotalCross} presents ``Hint for a minimal interaction length in $ \EEG $ annihilation in total cross 
section of center-of-mass energies 55 - 207 GeV''.
Section~\ref{BABHA1}} presents ``Hint for axial-vector contact interactions in the data on $ \EEEEG $ at centre-of-mass 
energies 192 - 208 GeV''.
Section~\ref{Experiment-Theory} presents ``Comparison of theoretical CALCULATIONS of the size of the 
ELECTRON with the EXPERIMENT''.
Section~\ref{OVERALLconclusion} presents ``Conclusion of the investigation of finit size of FP's, performed by ETHZ and USTC''.

\subsubsection{Theoretical Frameworks}
\label{Theory-Diff-cross}

Physical interactions in nature are governed
by the principles of local gauge invariance,
which are connected to conserved physical quantities of a
local region of space. Lagrangian formalism helps
to establish the connection between symmetries and conservation~laws.

The Dirac-Lagrange density describes a free particle with spin 1/2 as follows Eq. \ref{DIRAC}:

\begin{equation}
\label{DIRAC}
\mathcal{L}_{\rm Dirac}=\overline{\Psi }(i\gamma^{\mu}\partial _{ \mu }-m)\Psi ,
\end{equation}

In  \ref{DIRAC} is $\Psi$ is the fermion field, $\bar{\Psi} = \Psi^\dagger\gamma^0$
is its adjoint spinor with $\Psi^\dagger$ the Hermitian conjugate 
of  $\Psi$, $\gamma^\mu$ are the Dirac  $4\times 4$ $\gamma$-matrices
$\partial^\mu\equiv\partial/\partial x^\mu$ is the derivative (with $x^\mu$  being a four-vector with
dimensions of length, where  $\mu=0$ denotes time component, and $\mu= 1$, 2,  and 3 the space components),
and $m$ is the mass of the particle. The~requirement of local gauge invariance leads to the QED Lagrangian 
Eq. \ref{QED1}.

\begin{equation}
\label{QED1}
{\Lagr}_{\rm QED}=\overline{\Psi }(i\gamma^\mu \partial _\mu -m)\Psi 
+e\overline{\Psi }\gamma ^\mu A_\mu \Psi -\frac{1}{4}F_{\mu \nu }F^{\mu \nu }\ ,
\end{equation}

where $A_{\mu}$ is the gauge field, $m_{A} = m_{\gamma} = 0$, $e$
is the electron charge, $e\overline{\Psi }\gamma ^\mu A_\mu \Psi$
is the interaction term, and~$F_{\mu\nu}=\partial_{\mu}A_{\nu}-\partial_{\nu}A_{\mu}$.\\

\textbf {The Lowest Order Cross-Section of $ {\EEG} $.}\\ 

The Born-level cross-section~\cite{pointTextbook1,Berends,Berends1,Berends12,Berends13,Manat1} 
of the $ {\EEG} $ reaction, also known as the leading-order cross-section, is defined 
by the $ \it {M} $-matrix given by Eq.  \ref{QED2}.

\begin{equation}
\label{QED2}
M_{fi}=-e\int \overline{\Psi }\gamma ^{\mu }\Psi A_{\mu }d^{4}x\ ,
\end{equation}

where $f$ and $i$ denote the final and initial states, respectively.
At high energies centre of mass energies $ s $ ($s \gg m_{e}^2$), 
the~mass of the electron can be neglected.
Therefore, the~differential cross-section of the reaction depicted in
Figure~\ref{Feynman2}a can be expressed as follows after averaging over
the spin states of the initial particles Eq. \ref{BORN} :

\begin{equation}
\label{BORN}
\frac{d\sigma _0}{d\Omega}=\frac{S}{64 \pi ^2 s}\frac{p_f}{p_i}\left | \it {M}\right |^2=\frac{\alpha ^2}{s}\frac{1+\cos^2\theta}{
k^2 - \cos^2\theta}\ ,
\end{equation}

where $\Omega\equiv\cos\theta$, $\left | \it {M}\right |^2$ is the matrix element, $ S = 1/2 $ is the
statistical factor, the~momentum $ p_{f} = p_{i} $,
 $ k=E_{e^+}/\left | \vec{p}_{e^+}\right |\simeq 1 $ for high energies $ E_{e^+} $, where $e^+$ denotes the positron, and
$ \alpha = e^{2} / 4 \pi $. The~angle $ \theta $ is the photon-scattering
angle with respect to the $ e^+ e^- $ beam axis.
The Born-level total cross-section is expressed as Eq. \ref{BORN1} 

\begin{align}
\label{BORN1}
\sigma ^0&=\frac{1}{2}\frac{\alpha ^2}{s}\int_{0}^{2\pi }d\phi \int_{-1}^{+1}\frac{1+\cos^2\theta }{k-\cos^2\theta }d(\cos\theta ) & \vspace{0.5cm} \\
 &=\frac{2\pi \alpha ^2}{s}\left(\ln\frac{s}{m^2_{e}}-1\right). \nonumber
 \end{align}

As the statistics of the measurements of differential and total cross-sections increases,
it becomes essential to account for the radiative corrections discussed~below.

\subsubsection{Radiative~Corrections}
\label{radcor1}

In our analysis, we consider radiative corrections using a Monte Carlo
method~\cite{Berends,Berends1,Berends12,Berends13}, which incorporates a complete third-order
calculation that accounts for electron-mass effects.

The calculations involve six particles: the positron $e^+$ with momentum $p_+$,
the electron $e^-$ with momentum $p_-$, virtual photons and~soft
initial photons $\gamma(k_3)$ with momentum $k_3$, as~well as hard radiation photons
$\gamma(k_1)$ and $\gamma(k_2)$ with momentum $k_1$ and $k_2$, respectively.
The set of eight corrections for virtual photons are shown in the Feynman graphs in Figure~\ref{CORR1}.
Figure~\ref{CORR2} shows the lowest-order Feynman graphs  for two-photon annihilation and the 
associated corrections, consisting of six corrections for soft real photons in the initial 
state and eight corrections for hard photons \cite{Berends,Berends1,Berends12,Berends13}.

\begin{figure}[htbp]
\begin{center}
 \includegraphics[width=14.0cm,height=7.5cm]{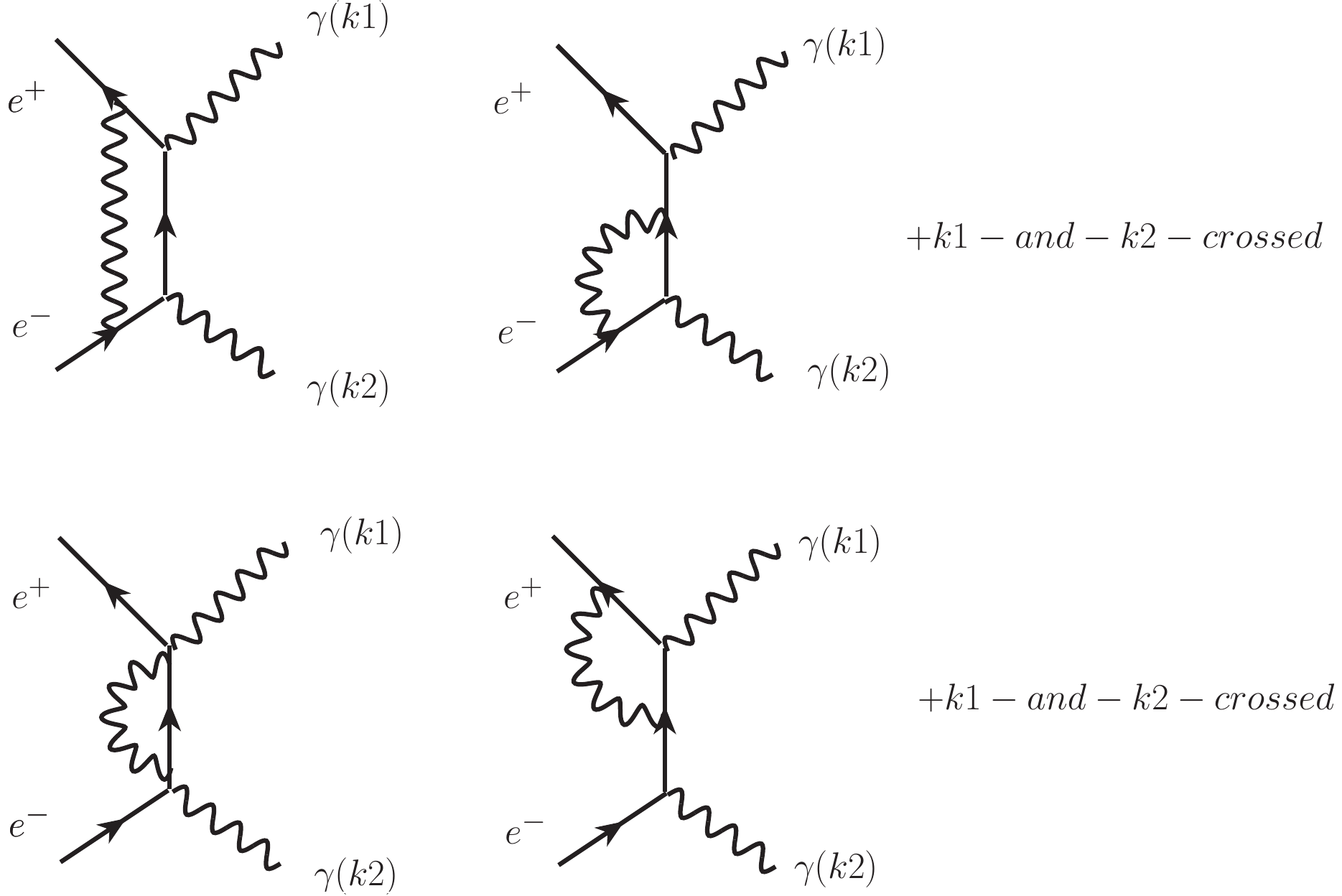}
\end{center}
\caption{The eight virtual photon corrections of third order Feynman graphs
for the $\EEG$ reaction. $k_i$ represents the momentum of the $i$-th photon.}
\label{CORR1}
\end{figure}

\newpage

\begin{figure}[htbp]
\begin{center}
 \includegraphics[width=17.5cm,height=13.0cm]{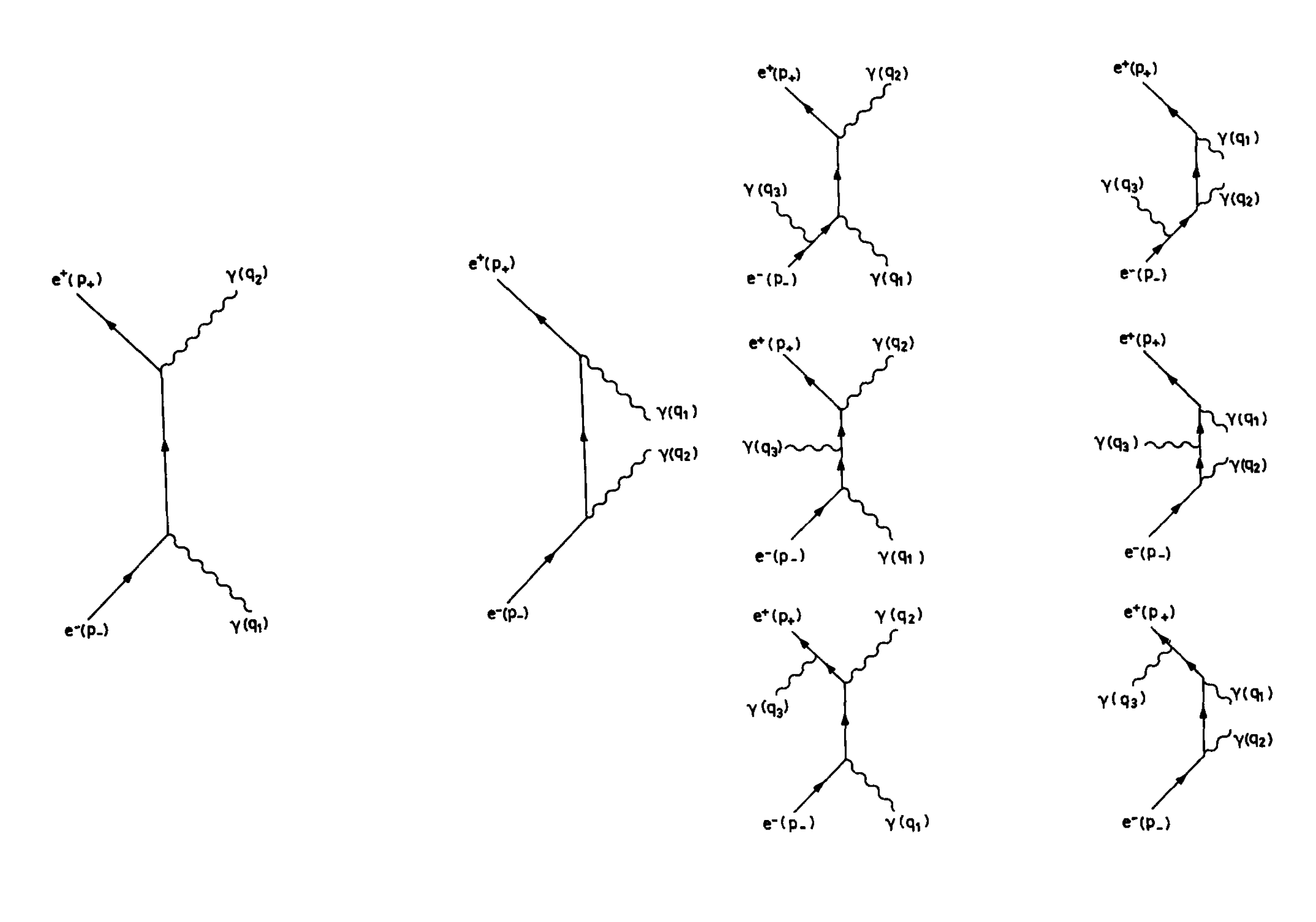}
\end{center}
\caption{Feynman diagrams for $\EEGG$ annihilation and radiative corrections.}
\label{CORR2}
\end{figure}

The left side in Figure~\ref{CORR2} shows the two lowest-order Feynman graphs for two-photon annihilation, 
which are also illustrated in Figure~\ref{Feynman1}. The momentum labelling used in this 
representation is based on the conventions described in Ref.~\cite{Berends12}.
The right side shows the set of corrections, comprising of six soft
initial state corrections for real photons, along with eight hard photon
corrections. The diagrams are adapted from Ref.~\cite{Berends12}. Here, 
$q_{i}=k_i$ is the momentum of the $i$th photon.

Since an exact analytical expression is not available, numerical
simulations are used to calculate the corrections of the Feynman
diagrams shown in Figures~\ref{CORR1} and \ref{CORR2}.
An event generator~\cite{Berends,Berends1,Berends12,Berends13} was used
to simulate the reaction $ \EEGG $. The~differential cross-section, including
radiative corrections up to $O(\alpha^3)$, can be expressed as in Eq. \ref{CORR-33}.

\begin{equation}
\label{CORR-33}
\left ( \frac{d\sigma }{d\Omega } \right )_{\alpha ^3}=\left
( \frac{d\sigma }{d\Omega } \right )_{\rm Born}(1+\delta _{\rm virtual}+\delta _{\rm soft}+\delta _{\rm hard})\ ,
\end{equation}

In Eq. \ref{CORR-33} is $\left(\frac{d\sigma}{d\Omega}\right)_{\rm Born}$ represents the lowest-order
cross-section. While $\delta_{\rm virtual}$, $\delta_{\rm soft}$
and $\delta_{\rm hard}$ correspond to the virtual, soft-Bremsstrahlung
and hard-Bremsstrahlung corrections, respectively.
To differentiate between soft  and hard Bremsstrahlung,
we introduce a dimensionless discriminator, $k_0 \ll 1$, into the generator.
If the momentum $k_3$ of the photon from initial state radiation
(Soft Bremsstrahlung) satisfies $k_3/|p_{e^+}|<k_0$, the reaction is considered as
two-photon final state; otherwise, it is treated as three-photon final state
(Hard Bremsstrahlung). To align with the notation used in Equation (\ref{BORN}), we use $p_{+}$ as a
shorthand for $p_{e^+}$. The~two cases are further elaborated on in 
Sections~\ref{virtSoft} and \ref{virtHard}.

\subsubsection{ Virtual and Soft Radiative~Corrections}
\label{virtSoft}

If the energies of the photons from initial state radiation
(Soft Bremsstrahlung) are too small to be detected, i.e.,~$k_3/|p_{+}| < k_0$,
the reaction is treated as a two-photon final state process Eq.\ref{Virtsoft1}:

\begin{equation}
\label{Virtsoft1}
e^+(p_{+})+e^-(p_{-}) \rightarrow \gamma (k_{1})+\gamma (k_{2})
\end{equation}

where $\delta_{\rm virtual}+\delta_{\rm soft} $ are expressed as follows Eq.\ref{BORNCORR1}.

\begin{equation}
\label{BORNCORR1}
\begin{matrix}
  \delta_{\rm soft}+\delta_{\rm virtual}= -\frac{\alpha }{\pi } \{2(1-2v)(\ln k_{0}+v)+\frac{3}{2}  \vspace{0.2cm} \\
  -\frac{1}{3}\pi^2 +\frac{1}{2(1+\cos^2\theta )} \vspace{0.2cm} \\
\times [-4v^2(3-\cos^2\theta) -8v\cos^2\theta  \\
+4uv(5+2\cos\theta +\cos^2\theta ) \\
+4wv(5-2\cos\theta ) + \cos^2\theta    \\
-u(5-6\cos\theta +\cos^2\theta ) \\
-w(5+6\cos\theta +\cos^2)  \\
-2u^2(5+2\cos\theta +\cos^2\theta ) \\
-2w^2(5-2\cos\theta +\cos^2\theta ) ] \}\ ,
\end{matrix}
\end{equation}

where the parameters $ v $, $ u $, $ w $ and $ m $ are shown in Eq.\ref{BORNCORR2}, Eq.\ref{BORNCORR3},
Eq.\ref{BORNCORR4} and Eq.\ref{BORNCORR5}.

\begin{align}
\label{BORNCORR2}
v&=\frac{1}{2}\ln\frac{s}{m^2_{e}}\ ,\\
\label{BORNCORR3}
u&=\frac{1}{2}\ln\frac{2(k +\cos\theta )}{m^2}\ ,\\
\label{BORNCORR4}
w&=\frac{1}{2} \ln\frac{2(k -\cos\theta )}{m^2}\ ,  \\
\label{BORNCORR5}
m&=\frac{m_{e}}{\left | p_{+} \right |}.
\end{align}

Note, that at a low energy regime the expression above includes  the mass  $m_{e}$
of the electron. \\

The total cross-section with two $\gamma$'s in the final state reads Eq.\ref{TWOgamma}:

\begin{equation}
\label{TWOgamma}
\sigma ^{2\gamma }=\sigma _{0}+\frac{2\alpha ^3}{s}\left[2(2v-1)^2\ln k_{0}+\frac{4}{3}v^3+3v^2\vspace{0.2cm} \\
+\left(\frac{2}{3}\pi ^2-6\right)v-\frac{1}{12}\pi ^2
\right].
\end{equation}

\subsubsection{ Hard Radiative~Corrections}
\label{virtHard}

If the energies of the photons from initial state radiation satisfy the condition
$k_3/|p_{+}| > k_0$, then the reaction is treated as a three-photon 
final state process Eq.\ref{twophoton22}:

\begin{equation}
\label{twophoton22}
e^+(p_{+})+e^-(p_{-}) \rightarrow \gamma (k_{1})+\gamma (k_{2})+\gamma (k_{3})\ .
\end{equation}


To obtain the differential cross-section of $\EEGG$, one needs to introduce two additional
parameters in the phase space. The computation (see~\cite{Mandl} for details) is performed in 
the ultra-relativistic regime, which is as follows Eq.\ref{HARTCORR1}:

\begin{equation}
\label{HARTCORR1}
\frac{d\sigma }{d\Gamma _{ijk}}=\frac{d\sigma }{d\Omega_{i} d\Omega _{k} dx_{k}}=\frac{\alpha ^3}{8\pi ^2s}w_{ijk}F(i,j,k)\ ,
\end{equation}
\\
where $ w_{ijk} $ is Eq.\ref{HARTCORR2}, $ y(z_{j}) $ is Eq.\ref{HARTCORR3} and $ z_{j} $ is Eq.\ref{HARTCORR4}
 
\begin{align}
\label{HARTCORR2}
w_{ijk}&=\frac{x_{i}x_{k}}{y(z_{j})},x_{i}=\frac{k_{i0}}{\left | \vec{p_{+}} \right |}\ ,  \\
\label{HARTCORR3}
y(z_{j}) &=2e -x_{k}+x_{k}z_{j}\ ,\\
\label{HARTCORR4}
z_{j} &=\cos(\alpha _{ik})\ ,
\end{align}

the parameter $ F(i,j,k) $ is Eq.\ref{HARTCORR6}

\begin{align}
\label{HARTCORR6}
F(i,j,k)&=\sum_{P} \left [ -2m^2 \frac{{k_{j}}'}{k^2_{k}{k_{i}}'}-2m^2\frac{k_{j}}{{k}'^2_{k}k_{i}}+\frac{2}{k_{k}{k}'_{k}} \left (  \frac{k^2_{j}+{k}'^2_{j}}{k_{i}{k}'_{i}}\right )\right ] \nonumber \\
&=\sum_{P}M(i,j,k)
\end{align}

and $ \alpha_{ik} $ is the angle between $ k_{i} $ and $ k_{j} $. The quantities $ k_{i} $  and $ k'_{i} $ 
are given by Eq.\ref{HARTCORR77} and Eq.\ref{HARTCORR7_1} .
$ P $ binds all permutations of (1, 2, 3).

\begin{align}
 \label{HARTCORR77}
 k_{i}&=x_{i}(k -\cos \theta _{i})
 \end{align}
 
 and
 
\begin{align}
  \label{HARTCORR7_1}
{k}'_{i}&=x_{i}(k +\cos \theta _{i}),
\end{align}

where $\theta _{i} $ is the angle between the momentum of the $ initial $  photon and $ | \vec{p}_{+}| $.
In its turn, the total cross-section with three $\gamma$'s in the final state reads as Eq.\ref{HARTCORR888}:

\begin{align}
\label{HARTCORR888}
\sigma ^{3\gamma }=\frac{1}{3!}\int d\Gamma _{ijk},i,j,k\in {\{1,2,3\}},
\end{align}

In Eq.\ref{HARTCORR888} the integral is taken over the phase spaces defined by $ k_{0} < x_{i} < 1 $.
In practice, Eq.\ref{HARTCORR888}) can be approximated by an analytical expression,
in which the photons are sorted by their energies, such that $E_{\gamma 1} \geq E_{\gamma 2} \geq E_{\gamma 3}$.
Here, $\gamma_{1}$ and $\gamma_{2}$ are treated as annihilation photons, and $\gamma_{3}$ is 
treated as a hard-Bremsstrahlung photon. Integrating Eq. \ref{HARTCORR888} (performed in 
Refs.~\cite{Berends,Berends1,Berends12,Berends13}), one arrives at Eq.\ref{3gamma555}:

\begin{align}
 \label{3gamma555}
 \sigma ^{3 \gamma } =\frac{2\alpha ^3}{s}[3-(\ln\frac{4}{m^2}-1)^2(2\ln k_{0}+1)].
 \end{align}
 \\
 \textbf {The Total Cross-Section in $\EEGG$}\\ 
 
 The total cross-section for $\EEGG$ is obtained by summing Eq. \ref{totcross888}) results in Eq.\ref{totcross999}.
 
\begin{align}
\label{totcross888}
\sigma _{\rm tot}&=\sigma^{2\gamma }+\sigma^{3\gamma }\\
\label{totcross999}
&=\sigma ^{0}+\frac{2\alpha ^3}{s}\left[\frac{4}{3}v^3-v^2+\left(\frac{2}{3}\pi ^2-2\right)v+2-\frac{1}{12}\pi ^2\right].
\end{align}

\subsubsection{ The Numerical Calculation of the $ \EEGG $ Differential Cross-Section}
\label{QEDnumeric}

The third-order differential cross-section is calculated using a
Monte Carlo generator~\cite{Berends,Berends1,Berends12,Berends13}. 
The generator produces events with three photons sorted in descending 
order of their energies ($E_{\gamma 1} \geq E_{\gamma 2} \geq E_{\gamma 3}$).
The photons are located at an angle $\alpha$ between the energies $E_{\gamma 1}
$ and $E_{\gamma 2}$, with the correct mix of soft ($k_{3}/\vert
p_{+}\vert < k_{0}$) and hard QED corrections ($k_{3}/\vert p_{+}
\vert > k_{0}$), as shown in Figures~\ref{CORR1} and~\ref{CORR2}.
The acceptance range for $\gamma$ - events is defined by $\alpha_{\rm min} <
\alpha < \alpha_{\rm max} $. The angle $\alpha$ is analytically related to the 
scattering angle $\theta$, thus connecting the limits with $|\cos \theta |$.

The binned differential cross-section is calculated as Eq.\ref{DIFF1},

\begin{align}
\label{DIFF1}
{\left(\frac{d\sigma }{d\Omega }\right)}^{\rm bin}_{i}=\frac{1}{2\pi \Delta (|
\cos\theta |)}\sigma_{\rm tot}\frac{N_{i}}{N}\ ,
\end{align}
\\
where $|\cos\theta| = (|\cos\theta_1| + |\cos\theta_2|)/2$ is the
scattering angle, with~$\theta_1$ and $\theta_2$ being the
scattering angles of photons with energies $E_{\gamma 1}$ and $E_{\gamma 2}$,
respectively. $N_i$ is the number of events in an angular bin width
$\Delta(|\cos\theta|)$ and~$N$ is the total number of generated events.

To search for potential deviations from QED, the~generated cross-section~(\ref{DIFF1}) 
is fitted as a function of $| \cos \theta |$ using a six-parameter $\chi^2$-fitting for each 
 analysed  $\sqrt{s}$.

As an illustration, we generated one million events at a
center-of-mass energy of 189 GeV, within an acceptance range of
$14^{\circ}<\alpha<166^{\circ}$, which corresponds to
$| \cos \theta | < 0.97$.  We used a soft/hard
discriminator of $k_0=0.01$ and distributed the events over 50
$\Delta (| \cos \theta |)$ bins. The~resulting $\chi^2$-fit
of the differential cross-section exhibits the following polynomial
behavior Eq.\ref{QEDfit222}.

\begin{align}
\label{QEDfit222}
&\left ( \frac{d\sigma }{d\Omega } \right )_{\rm QED}=\left ( \frac{d\sigma }{d\Omega } \right )_{\rm Born} \times \\
&\left ( 1+p_{1}+p_{2}\exp{\left({-\frac{x^{1.2}}{2p^2_{3}}}\right)}+p_{4}x+p_{5}x^2+p_{6}x^3 \right )\ , \nonumber
\end{align}

where $x=|\cos\theta |$ and parameters $ p_{1} $ to $ p_{6} $ are shown in Eq.\ref{QEDparameter}.

\begin{align}
\label{QEDparameter}
p_{1}&=0.2869\ ,\\ \nonumber
p_{2}&=-0.51851\ ,\\ \nonumber
p_{3}&=0.19946\ , \\ \nonumber
p_{4}&=-0.39652\ , \\ \nonumber
p_{5}&=-0.41213\ , \\ \nonumber
p_{6}&=0.70428.  \nonumber
\end{align}

We note that $ \EEGG $ channel lacks a comprehensive analysis of
theoretical uncertainty, specifically the uncertainty associated
with the third-order Monte Carlo prediction. In~a QED process,
higher-order effects can be approximated to be 10\% ($\simeq
\sqrt{\alpha}$) of the correction caused by the highest-order
corrections. The theory uncertainty can be estimated to be 10\% of
the radiative correction for each experiment,
with a minimum of 0.5\%.

\subsubsection{Deviations from~QED}
\label{QEDdev1}

If QED is a fundamental theory, it should be 
able to describe the experimental parameters of the $\EEGG$ reaction 
up to the Grand Unification scale. Currently, however, QED has only been
tested up to a center-of-mass energy of $\lesssim 100$ GeV.
Therefore, new, non-QED phenomena could become observable at higher 
energy scales. If a cutoff scale parameter is found, it can serve
as a threshold point for the breakdown of QED and the emergence of
new underlying physics.  This paper focuses on the mass of the
excited electron and the cutoff scale of the contact interaction, which can
be interpreted in terms of the size of the electron.

\subsubsection{Heavy Electron~Mass}
\label{Helectron}

This model assumes the existence of an excited state of the electron, and the reaction
$\EEG$ occurs through the exchange of a virtual excited electron $e^*$ in the $t$- and $u$-channels,
as shown in the Feynman graph in Figure~\ref{Feynman2}c. The interaction is characterised by a
coupling between the excited electron and the ordinary electron, as~well as between the excited
electron and the photon. A magnetic interaction term is introduced~\cite{HeavyElMu111,Litke} to account
for this interaction in the form Eq.\ref{Litke1111},

\begin{align}
\label{Litke1111}
{\Lagr}_{e^{*}}=\frac{e\lambda }{2 m_{e^*}}\overline{\Psi }_{e^*}\sigma _{\mu \nu }\Psi _{e}F^{\mu \nu }\ ,
\end{align}

where $\lambda$ is the relative magnetic coupling strength to the QED magnetic coupling,
and $\sigma ^{\mu \nu }=\frac{1}{2}\left [ \gamma ^{\mu },\gamma ^{\nu } \right ]$.
The differential cross section of the QED modified by this interaction is given by Eq. \ref{Litke2222}:

\begin{align}
\label{Litke2222}
\frac{d\sigma }{d\Omega }=\left ( \frac{d\sigma }{d\Omega } \right )_{\rm QED}\left
[ 1+\frac{s^2}{2}\frac{\lambda ^2}{m^4_{e^*}}\left ( 1-\cos^2\theta  \right )F(\cos\theta ) \right ]\ ,
\end{align}

where $F(\cos\theta )$ is given by Eq.\ref{Litke3333}

\begin{align}
\label{Litke3333}
F(\cos\theta )=&\left ( 1+\frac{s}{2m^2_{e^*}}\frac{1-\cos^2\theta }{1+\cos^2\theta } \right ) \times \\
&\left [ \left ( 1+\frac{s}{2m^2_{e^*}} \right )^2-\left ( \frac{s}{2m^2_{e^*}} \right )^2\cos^2\theta  \right ]^{-1}. \nonumber
\end{align}

At condition $s / m^2_{e^2} << 1$, the expression Eq. (\ref{Litke2222} is reduced to Eq. (\ref{Litke4444},

\begin{align}
\label{Litke4444}
\frac{d\sigma }{d\Omega }&=\left ( \frac{d\sigma }{d\Omega } \right )_{\rm QED}(1\pm \delta _{\rm new}) \\
&=\left ( \frac{d\sigma }{d\Omega } \right )_{\rm QED}\left [ 1\pm \frac{s^2}{2}\frac{1}{\Lambda ^4_{\pm }}(1-\cos^2\theta ) \right ]\ ,
\nonumber
\end{align}

where the scale $\Lambda _{+} =  \Lambda_{e^{*}} $ is related to 
$m_{e^*}$ by $ \Lambda ^2_{+} =m^2_{e^*} / \lambda $
and negative contribution $ \Lambda _{-} $ is added for symmetry.

\subsubsection{Minimal Interaction Length and Non-Pointness of the~Electron }

The effective Lagrangian for a contact interaction Figure~\ref{Feynman2} b uses current fields of known particles
and is proportional to the lowest power of $1/\Lambda$, which depends on the dimensionality of
the fields used. When constructing this Lagrangian, it is important to ensure that
fermion fluxes conserve helicity, which is necessary for composite models.
This condition guarantees that the known particle masses are significantly 
smaller than the energy scale of $\Lambda$.
Different choices of helicity for the fields used in the Lagrangian result
in different predictions for the angular distributions and polarisation observables
in reactions where the contribution of the contact interaction is considered.

In the references \cite{eeggGeneral1,Models1,Models11,Models12,Models13,Models14,Models1111}, 
the contact interaction between two fermions and two bosons in the general case was investigated.
In the following discussion, we focus on the simplest operator of dimension 6, which is described 
by the effective Lagrangian Eq. \ref{DIR222},

\begin{align}
\label{DIR222}
{\Lagr}_{6}=i\overline{\Psi }\gamma _{\mu }({\vec{D}}_{\nu }\Psi )(g_{6}F^{\mu \nu }+\tilde{g}_{6}\tilde{F}^{\mu \nu })\ ,
\end{align}

where the coupling constant $g_{n}$ with $n=6$ is related to the energy scale of the $\Lambda$ in Eq.:
$g_{n}=\sqrt{4\pi }/\Lambda ^{(n-4)}$. The QED-covariant derivative is represented 
by: $D_{\mu }=\partial_{\mu }-ieA_{\mu}$, where $A_{\mu}$  denotes the covariant 
four-potential of the electromagnetic field. The dual electromagnetic tensor
is $\tilde{F}^{\mu \nu }$, given by $\tilde{F}^{\alpha \beta }=\frac{1}{2}\varepsilon ^{\alpha \beta \mu \nu }F_{\mu \nu }$.
The modified differential cross section is shown in Eq. \ref{DIR333}.

\begin{align}
\label{DIR333}
\left ( \frac{d\sigma }{d\Omega } \right )_{T}&=\left ( \frac{d\sigma }{d\Omega } \right )_{\rm QED}\left [ 1 \pm \delta _{\rm new} \right ] \\
&=\left ( \frac{d\sigma }{d\Omega } \right )_{\rm QED}\left [ 1 \pm \frac{s^2}{2\alpha }\left ( \frac{1}{\Lambda ^4} +
\frac{1}{\tilde{\Lambda ^4}}\right )(1-\cos^2\theta ) \right ] \nonumber\ ,
\end{align}

In the following, we use $\Lambda = \tilde{\Lambda} = \Lambda_6$ and omit higher-order terms such as $\Lambda_7$ or
$\Lambda_8$ in $\delta_{\rm new}$. \\

A common method for searching for a deviations from QED is to use a $\chi^2$ test to compare experimentally 
measured cross-sections with predicted QED cross-sections. To incorporate a non-QED direct contact term 
into the QED cross section, an energy scale $\Lambda$ is introduced according to Eq. \ref{DIR333}.
$\Lambda$ can be interpreted as defining of the size of the object at which
annihilation occurs. This can be calculated either using the generalised uncertainty principle
~\cite{RES:RES11, RES:RES22, RES:RES33} or the electromagnetic energy $E$ 
and the wavelength $\lambda_{\gamma}$~\cite{EMenergy1} of the light emitted by the object.
The~wavelength  $\lambda_{\gamma}$ must be smaller or equal to the size of the interaction area. If  the $\chi^2$-test exhibits a
minimum for a certain $\Lambda$, this defines the region in which $e^{+}e^{-}$
annihilation must occur via $\Lambda=E=\hbar \nu_{\gamma}$, where $\nu_{\gamma}=c/\lambda_{\gamma}$ 
with $c$ the speed of light. We assume that $\lambda_{\gamma}=r_{e}$ regulates the size of the 
electron according to Eq. \ref{DIR111}

\begin{align}
\label{DIR111}
r_{e}= {\hbar c}/{\Lambda }\ .
\end{align}
 
Eq. (\ref{DIR111}) provides a generic formula for calculating the size of an object,
which can be obtained using the generalised uncertainty principle~\cite{MLI,QG:Qgrav4, QG:Qgrav5}.
It is worth noting that as $\Lambda$ approaches infinity in Equation~(\ref{DIR111}),
the size of the object tends to zero:~$r_{e}\rightarrow 0$.

\subsubsection{The Measurement of the Total and Differential Cross-Section}
\label{data}

The $ \EEGG $ reaction initiates in  storage $ e^+ e^- $ ring accelerators a 
background-free signal in a detector. For example, Figure~\ref{GAMMA11} shows a typical event
display recorded in the central detector's cross-section of the L3 detector at  LEP. 
For event selection, the $\EEGG$ type exhibits a characteristic signature in the detector.
Almost all of its energy is deposited in the electromagnetic calorimeter (BGO), and there 
is no trace in the charge-sensitive Time Expansion Chamber (TEC) within the BGO.
To select an event, at least two photon candidates with polar angles must be present.

\newpage

\begin{figure}[htbp]
\begin{center}
\includegraphics[width=17.0cm,height=9.0cm]{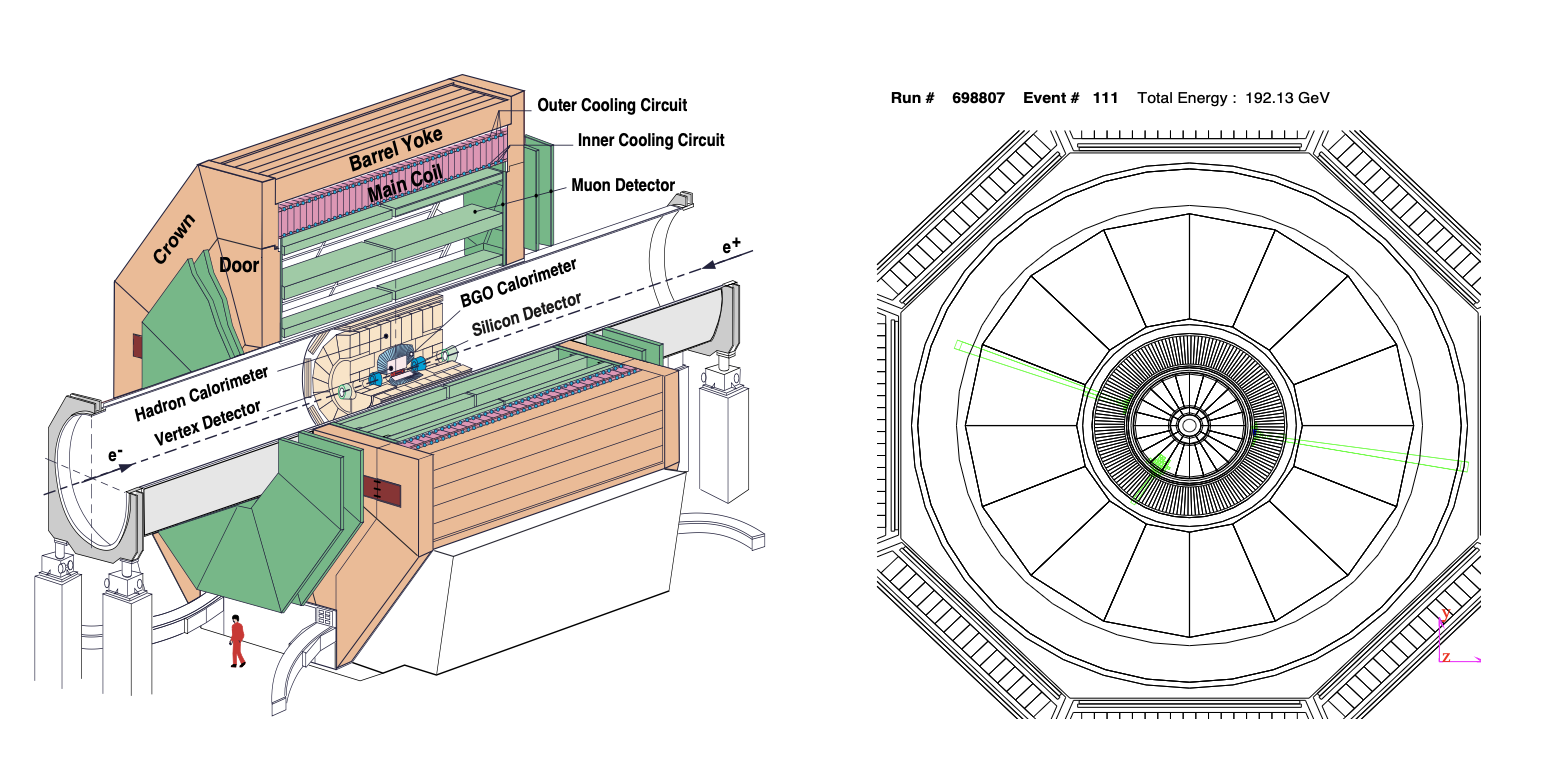}
\end{center}
\caption{The L3 detector (left side) and a typical  $\EEGG$ event ( right side ). The columns show the energy deposit in the BGO crystals \cite{gamma_L3}.}
\label{GAMMA11}
\end{figure}

Similar signals were observed in all LEP detectors, as well as in the VENUS and TOPAS detectors.
The channel topology is unambiguous, and event selection is based on the presence of two 
high-energy clusters in the electromagnetic calorimeter (ECAL).
The two highest-energy clusters must meet a minimum energy requirement.
The cuts on acollinearity or~missing longitudinal momentum,
as well as the allowed range in polar angle, of~the observed clusters,
have been applied. Charged tracks are generally not allowed,
except when they can be associated with a photon conversion in one hemisphere,
in order to remove background, particularly from Bhabha events $ \EEEEG $.

The limited coverage of the ECAL, along with selection cuts to
reject events with charged tracks, reduces the signal efficiency.
The impact of the above-mentioned cuts varies significantly depending on the detector geometry,
resulting in uncorrelated systematic errors across LEP experiments, VENUS and TOPAS.

The total cross-section is calculated as the ratio of the number $N$ of detected events
within the full angular coverage of the ECAL to the efficiency, $\varepsilon$, and the 
integrated luminosity, $L$, in Eq. \ref{Stot111},

\begin{align}
\label{Stot111}
\sigma _{\rm tot}=\frac{N}{\varepsilon L}.
\end{align}

The binned differential cross-section is calculated as Eq.\ref{SIGMAdiff1},

\begin{align}
\label{SIGMAdiff1}
{\left ( \frac{d\sigma }{d\Omega } \right )}^{\rm bin}_{i}=
\frac{1}{2\pi \Delta (|\cos\theta |)_{i}}\cdot \frac{N_{i}}{L\cdot \epsilon_{i}},
\end{align}

where $i$ defines the angular bin interval. To compare the measured differential-
cross section according to equation (\ref{SIGMAdiff1}) with QED predictions
in the $i$-th interval, the average value of $\cos\theta$ is calculated in Eq. \ref{COS1}.

\begin{align}
\label{COS1}
\left | \cos \theta \right |_{i}=\frac{\int_{|\cos\theta |\in i} |\cos\theta |
\left ( \frac{d \sigma }{d\Omega }(|\cos\theta |) \right )_{e^+e^-
\rightarrow \gamma \gamma }^{\rm Born}d|\cos\theta |}
{\int_{|\cos\theta |\in  i}\left ( \frac{d\sigma }{d\Omega }(|\cos\theta |)
\right )_{e^+e^-\rightarrow \gamma \gamma }^{\rm Born} d|\cos\theta |}.
\end{align}

In Eq.\ref{COS1}, $\cos\theta$ is defined as $0.5(|\cos\theta_1|+|\cos\theta_2|)$,
where $\cos\theta_1$ and $\cos\theta_2$ are the scattering angles of the first and
second photon, respectively. The average calculation is based on the QED Born-level prediction.
The Collaborations present their data in bins of $\left|\cos\theta\right|_i$ where
the cross-section is calculated based on the number of events in each bin.
Therefore, each value of $\left|\cos\theta\right|_i$ corresponds to the low edge of the~bin.

\subsubsection{ Differential Cross-Section Datasets }
\label{difCS}

This Section describes the data on the differential cross-section of the
reaction $\EEG$ that are included in the global fit analysis.
The data are reweighted for a single center-of-mass energy, $E_{\rm scale}$, and
presented by plotting the reweighted results along with the QED-$\alpha^3$
cross-section Eq.\ref{CORR-33}. Since the measurements of the 
cross-section are obtained with varying event numbers $N_i$,
and at different center-of-mass energies $E_i$, a reweighting is
necessary to enable the visual comparison of different datasets.
This is illustrated in Figures~\ref{VENUS} to \ref{cross-tot}.

The reweighting is performed using the equation Eq.\ref{SCAL1}.

\begin{align}
\label{SCAL1}
\left ( \frac{d\sigma }{d\Omega } \right )_{|\cos\theta|_{j} }^{\rm scale}
=\frac{\sum_{i}^{r}N_{i}\left ( \frac{d\sigma }{d\Omega } \right )
^{i}_{|\cos\theta |_{j}}\cdot \left ( \frac{E_{i}}{E_{\rm scale}} \right )^{2}}{\sum_{i}^{r}N_{i}}\ .
\end{align}

In Eq.~(\ref{SCAL1}), $N_i$ is used to derive the differential cross-section at 
$\sqrt{s}=E_i$, where $i=1$ to $r$,which represents the number of values
 $ \left ( \frac{d\sigma }{d\Omega } \right )^{i}{|\cos\theta|{j}} $ to be scaled.
Here, $j$ is the bin number of $\cos\theta$, and $\cos\theta_j$ is calculated 
using Eq. \ref{COS1}. The diagrams in Figures \ref{VENUS} to \ref{cross-tot} 
were created with $E_{\rm scale} = 91.2$ GeV. A deviation from QED in the 
differential cross section of the $\EEG$ response would manifest as an observed 
difference between the QED and experimental differential cross sections.

The VENUS Collaboration presented the luminosity, $ {\EEG} $ candidates,
angular distribution and~differential cross-section at four
center-of-mass energies $ \sqrt{s} $  = 55.0 GeV, 56.0 GeV, 56.5 GeV 
and 57.6 GeV; see Tables~2-4, and 8 in Ref.~\cite{VENUS}.
The TOPAS Collaboration presented the luminosity and differential
cross-section, with bin width, at $ \sqrt{s} $  = 57.0 GeV; see 
Tables~1 and 2 in Ref.~\cite{TOPAS}. Figure~\ref{VENUS} displays the
reweighted data from the VENUS and TOPAS experiments, obtained using Eq. \ref{SCAL1}
with $ E_{\rm scale} = 91.2 $ GeV, with the QED-$\alpha^{3}$ Eq.\ref{CORR-33}
differential cross-section at $ \sqrt{s} $ = 91.2 GeV (black line).

\begin{figure}[htbp]
\begin{center}
 \includegraphics[width=17.0cm,height=13.0cm]{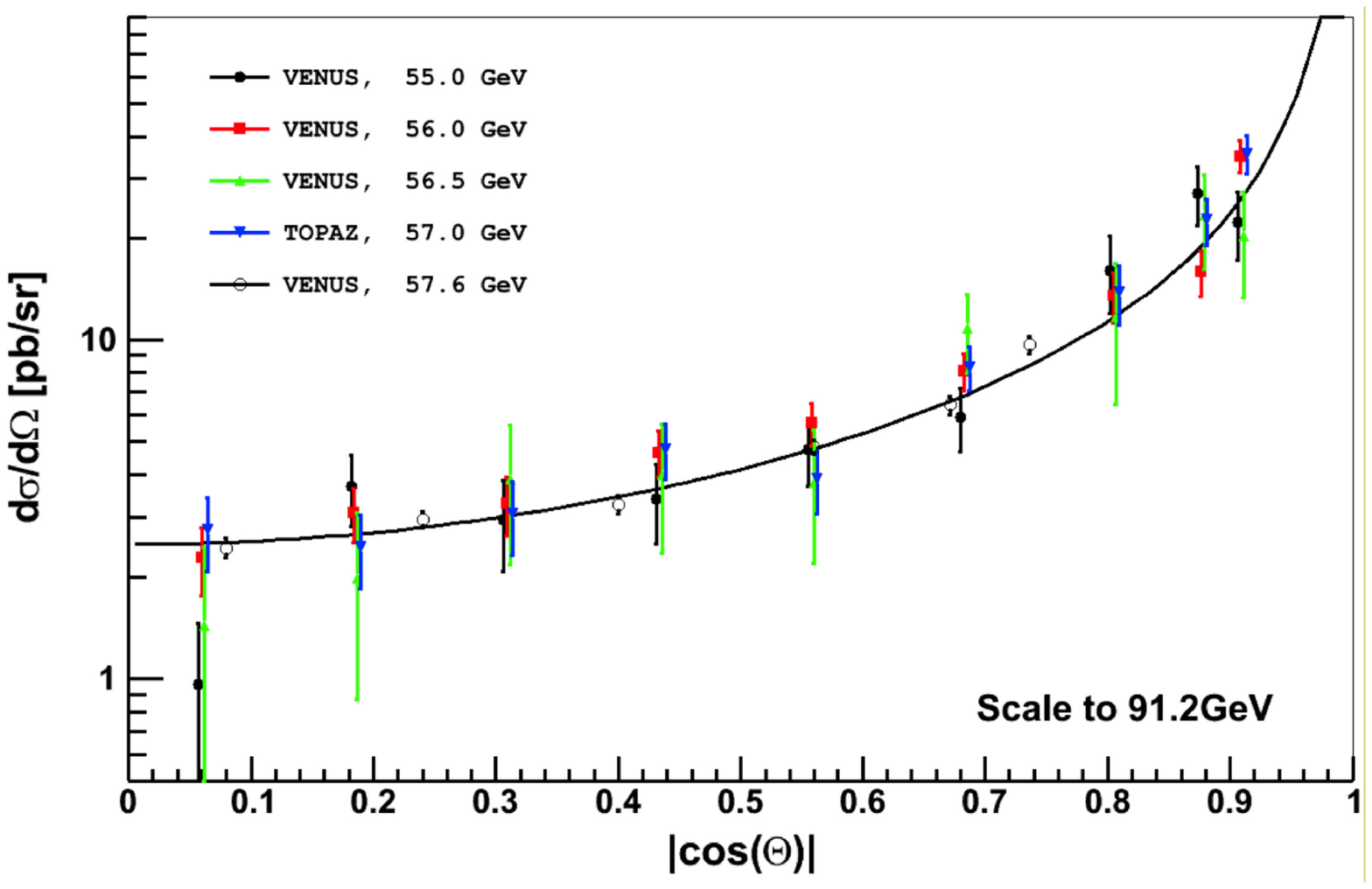}
\end{center}
\caption{The differential cross-section of the $\EEG$ reaction from VENUS \cite{VENUS} and TOPAS \cite{TOPAS} experiments.}
\label{VENUS}
\end{figure}

In Figure~\ref{VENUS}, the bars represent statistical uncertainties.The black line is the 
QED-$ \alpha ^3 $ cross-section Eq.\ref{CORR-33}. The graphis obtained with 
$E_{\rm scale}=91.2$~GeV. A deviation from QED in the differential cross-section of the $\EEG$
reaction would manifest as an observation of a difference between the QED 
calculations and experimental measurements.}

The ALEPH Collaboration provided measurements at
$\sqrt{s}=91.3$ GeV, as well as the bin width; see~Table~8.2 in Ref. 
\cite{ALEPH}. Figure~\ref{ALEPH} shows the ALEPH
data scaled to $\sqrt{s}=91.2$ GeV using Equation~(\ref{SCAL1}).
Only statistical uncertainties are displayed. The~black line corresponds
to the QED-$\alpha^3$ differential cross-section at $\sqrt{s}=91.2$ GeV 
Eq.\ref{CORR-33}.

\begin{figure}[htbp]
\begin{center}
 \includegraphics[width=16.0cm,height=12.0cm]{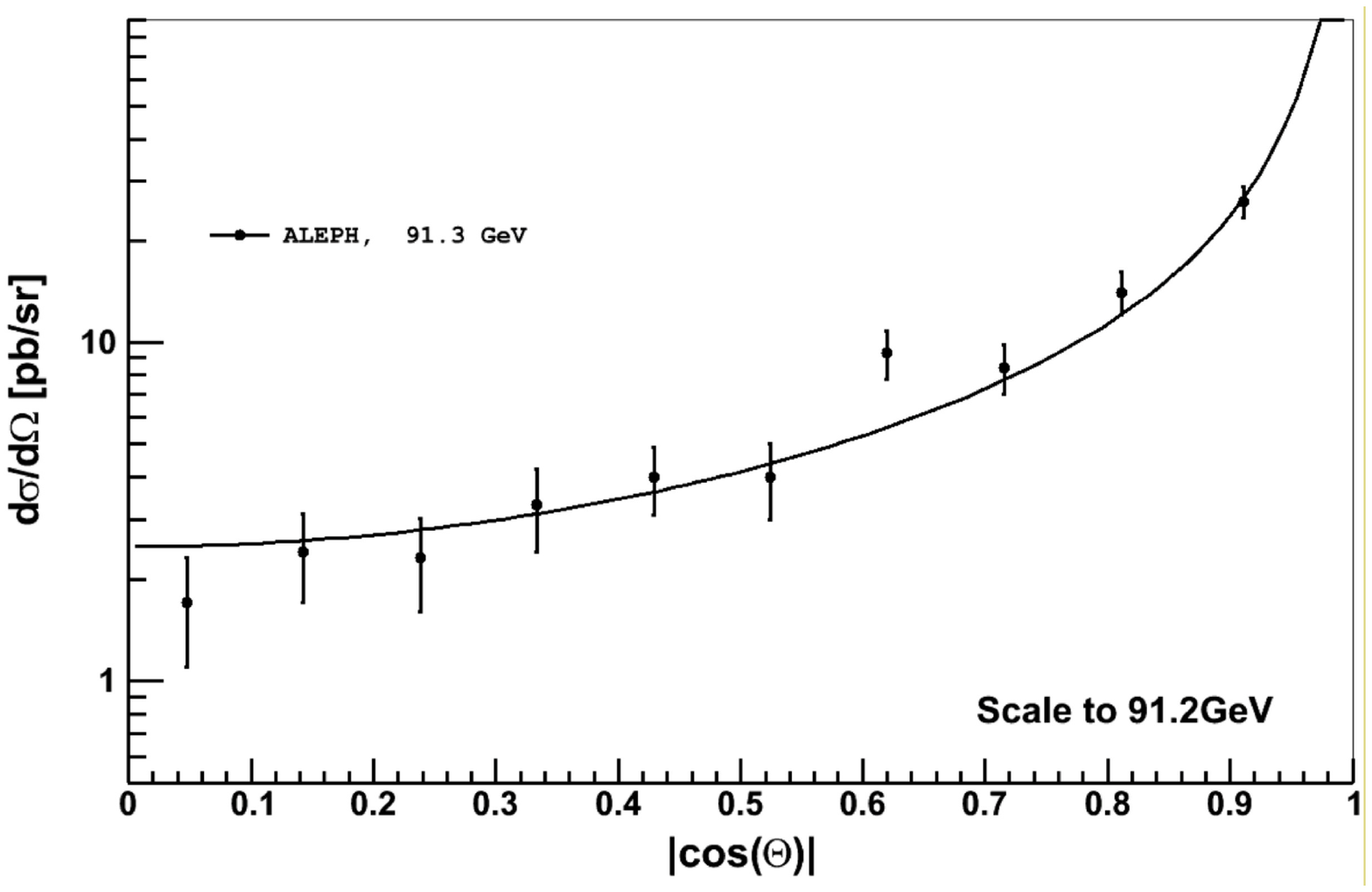}
\end{center}
\caption{Same as Figure~\ref{VENUS}, but with the data by ALEPH \cite{ALEPH}.}
\label{ALEPH}
\end{figure}

The DELPHI collaboration published data from 1994, 1998, and 2000 
in references \cite{DELPHI,DELPHI1,DELPHI2}.
The results from 1994 in reference \cite{DELPHI} show the luminosity at
$ \sqrt{s} $ = 91.25 GeV and the differential cross section $ \EEG $,
as well as the bin width and the number of events per bin.
The results from 1998 in Ref.~\cite{DELPHI1} show the same 
measurements at \mbox{$ \sqrt{s} $ = 91.25~GeV}, 130.4~GeV,
136.3~GeV, 161.5~GeV 172.4~GeV and 182.7~GeV. 
The measurements at $\sqrt{s}$ = 188.63~GeV,
191.6~GeV, 195.5~GeV, 199.5~GeV and 201.6~GeV are 
presented in Ref.~\cite{DELPHI2}  describing the year 2000 results. 
Figure~\ref{DELPHI} displays the DELPHI data scaled 
with Equation (\ref{SCAL1}) to $ \sqrt{s} $ = 91.2 GeV. 
Again, the black line represents the QED-$\alpha^{3}$
differential cross-section at $ \sqrt{s} $ = 91.2 GeV Eq.\ref{CORR-33}
and only statistical uncertainties are~shown.

\newpage

\begin{figure}[htbp]
\begin{center}
 \includegraphics[width=17.0cm,height=13.0cm]{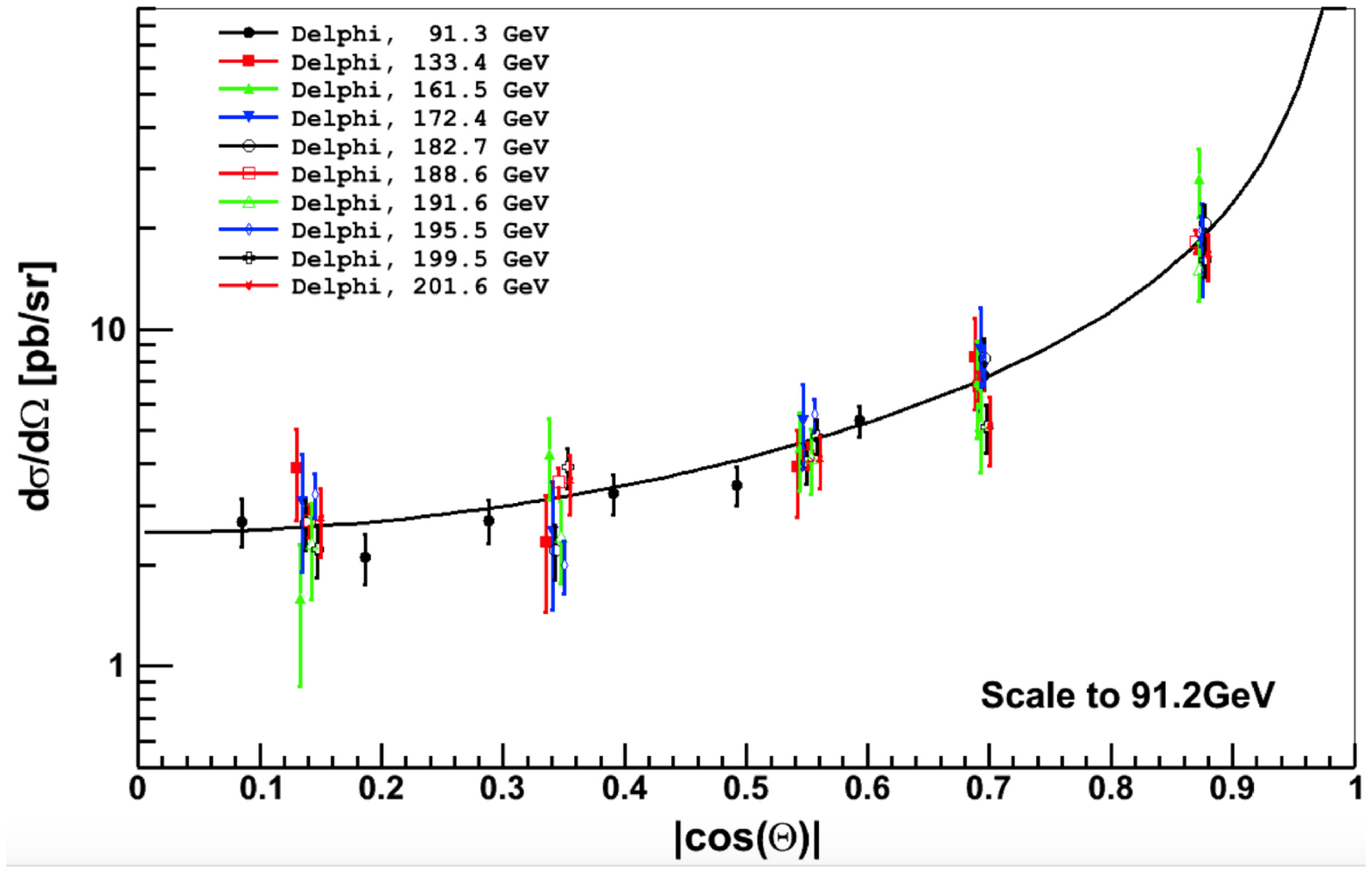}
\end{center}
\caption{Same as Figure~\ref{VENUS}, but with the data by DELPHI~\cite{DELPHI,DELPHI1,DELPHI2}.}
\label{DELPHI}
\end{figure}

In Ref. \cite{L3A} L3 Collaboration published measurements of the differential cross-section,
the data on bin size and event counts obtained in 1995 at $\sqrt{s}$ = 91.2 GeV.
The results from 2000 in Refs.~\cite{L33,L331,L332,L3333} show the same measurements
at $\sqrt{s} $ = 183~GeV and 189~GeV. The measurements at $\sqrt{s}$ = 192~GeV, 196~GeV, 200~GeV,
202~GeV, 205~GeV and~207~GeV are presented in Ref.~\cite{L3B}
describing the year 2002 results. 
Figure~\ref{L3} displays all L3 differential cross-section data at 
$ \sqrt{s} $ = 91~GeV to 207 GeV, scaled with Eq. \ref{SCAL1}
to 91.2 GeV; only statistical uncertainties are shown. As above, the black line represents the
QED-$\alpha^{3} $ differential cross-section at $ \sqrt{s} $ = 91.2 GeV Eq.\ref{CORR-33}.     

\newpage

\begin{figure}[htbp]
\vspace{-20.0mm}
\begin{center}
 \includegraphics[width=15.0cm,height=12.0cm]{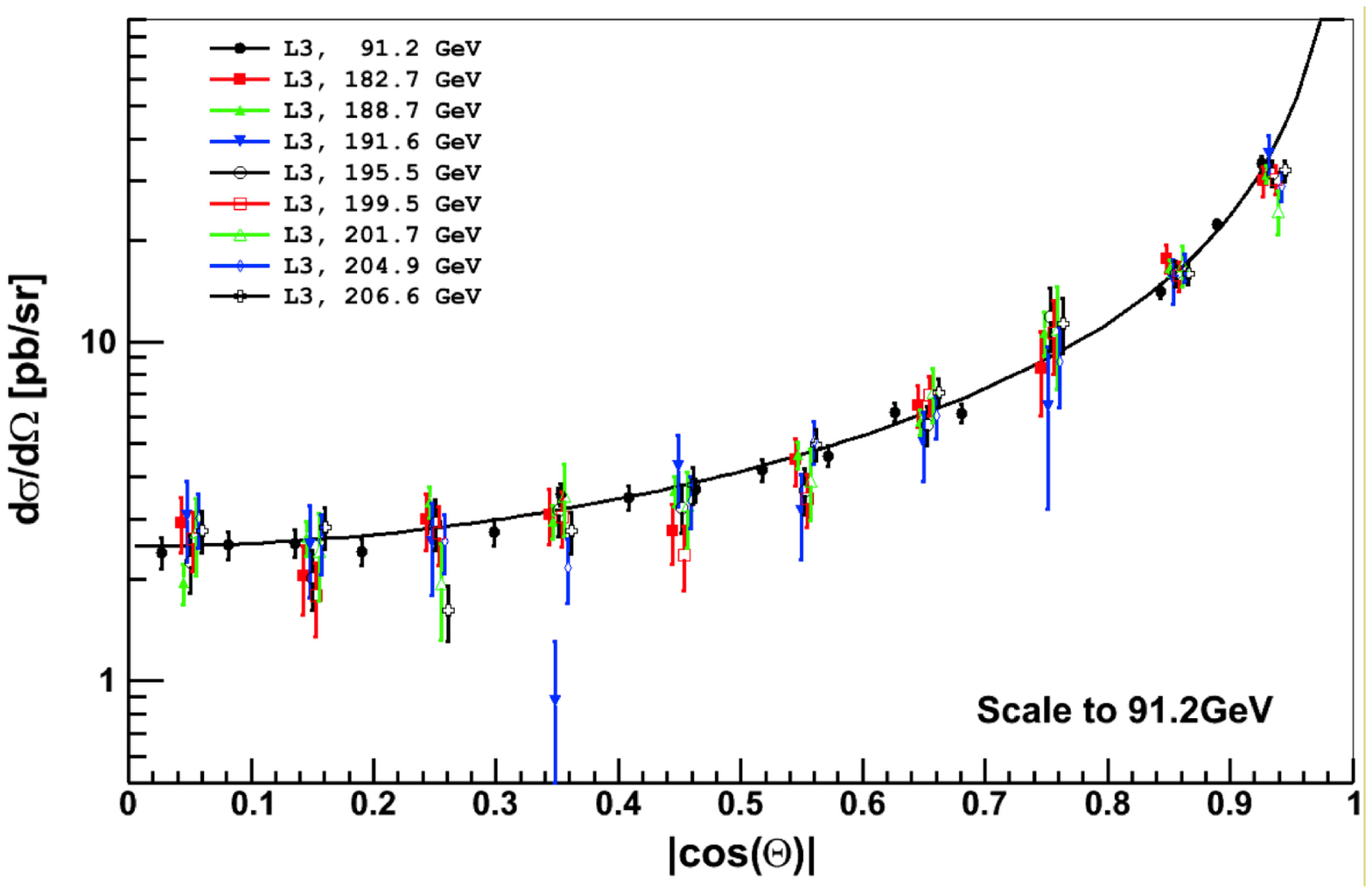}
\end{center}
\caption{Same as Figure~\ref{VENUS}, but with the data by L3~\cite{L3A,L3B,L33,L331,L332,L3333}.}
\label{L3}
\end{figure}

In Ref.~\cite{OPAL1}, OPAL Collaboration published measurements of the differential cross-section,
the data on bin size and event counts obtained in 1991 at $\sqrt{s}$ = 91.0~GeV. The year 2003 results in Ref.~\cite{OPAL2}
show the same measurements at $ \sqrt{s} $ = 183~GeV, 189~GeV, 192~GeV, 196~GeV, 200 GeV,
202~GeV, 205~GeV and 207~GeV. 
Figure~\ref{OPAL} displays all the OPAL differential cross-section
data from \mbox{$ \sqrt{s} $ = 91~GeV} to 207~GeV, scaled with (\ref{SCAL1}) to 91.2~GeV.
Only statistical uncertainties are shown. 
Again, the~black line is the QED-$\alpha^{3}$
differential cross-section at $ \sqrt{s} $ = 91.2~GeV Eq.\ref{CORR-33}.

\newpage

\begin{figure}[htbp]
\vspace{-20.0mm}
\begin{center}
 \includegraphics[width=16.0cm,height=14.0cm]{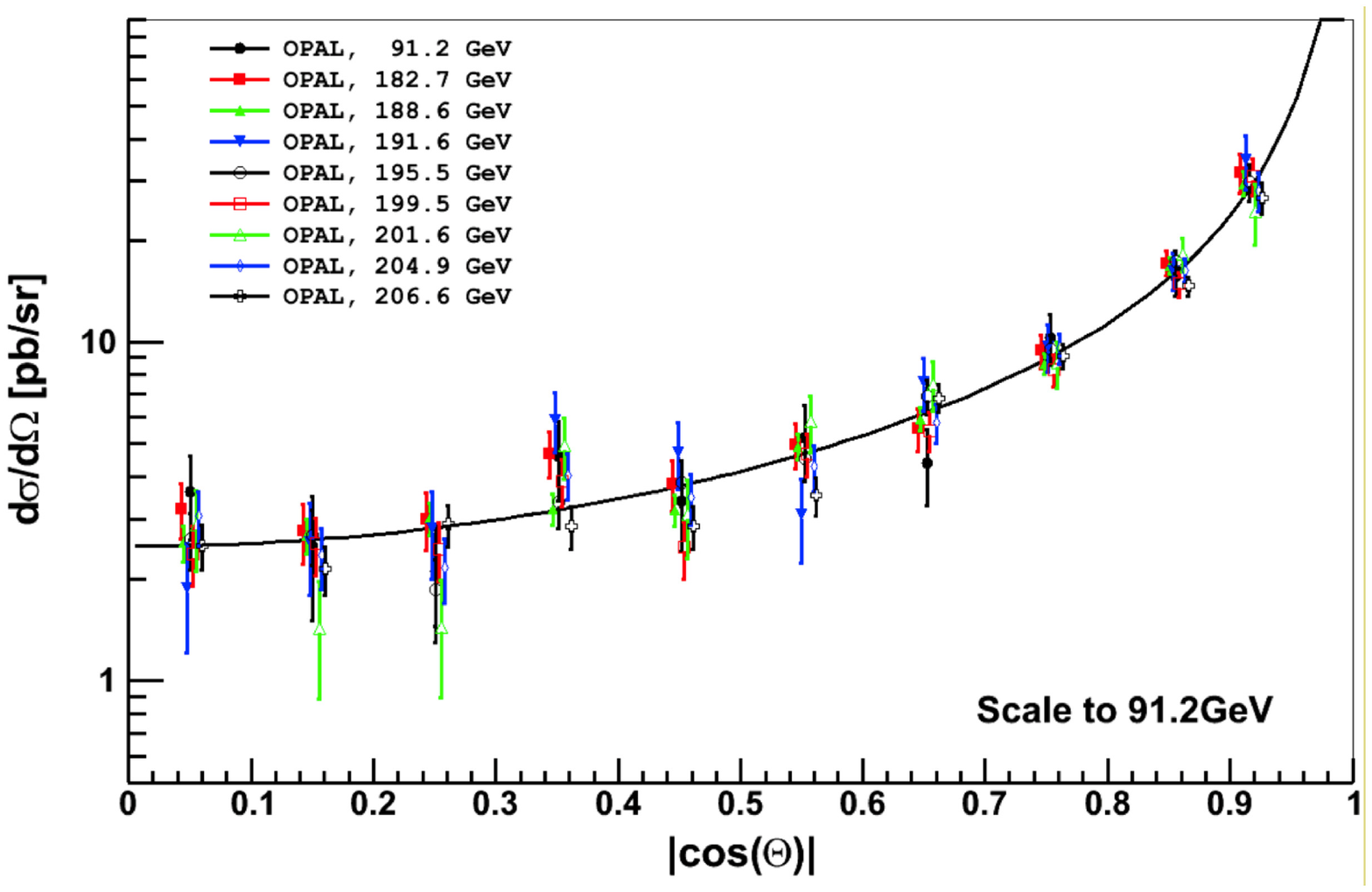}
\end{center}
\caption{Same as Figure~\ref{VENUS}, but with the data from OPAL \cite{OPAL1,OPAL2}.}
\label{OPAL}
\end{figure}

Figure~\ref{cross-tot} combines the differential cross-sections measured by VENUS, TOPAS, ALEPH, L3 and~OPAL at
the range $ \sqrt{s} $ = 55~GeV to 207~GeV. All data points are scaled with Equation~(\ref{SCAL1}) to 91.2~GeV 
and only statistical uncertainties are displayed. The black line represents the QED-$\alpha^{3}$ differential 
cross-section at $\sqrt{s}$ = 91.2~GeV Eq.\ref{CORR-33}.

\newpage

\begin{figure}[htbp]
\vspace{-20.0mm}
\begin{center}
 \includegraphics[width=17.0cm,height=14.0cm]{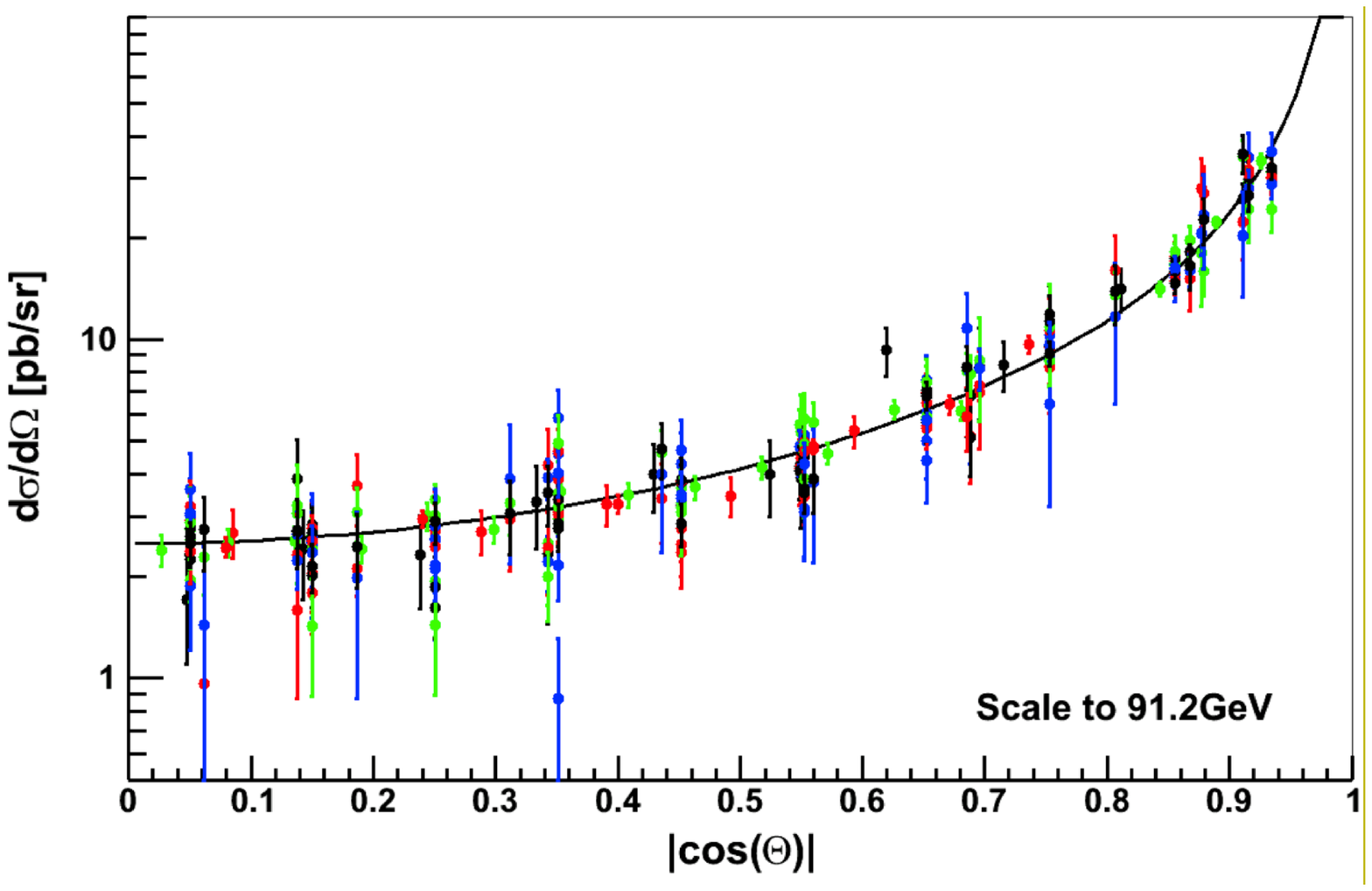}
\end{center}
\caption{Differential cross-section of the $ \EEG $ reaction from VENUS, TOPAS, ALEPH, L3 and OPAL.
The black line is the QED-$ \alpha ^3 $ cross-section~(\ref{CORR-33}). The~colours and symbols are 
same as in Figures~\ref{VENUS}--\ref{OPAL}.}
\label{cross-tot}
\end{figure}

No significant deviations from the QED predictions are apparent in Figures~\ref{VENUS} to \ref{cross-tot}.
It is well known from physics that deviations from the Standard Model become more significant when 
using a $\chi^2$-fit, which compares experiment and theory. For this reason, we perform a global $\chi^2$-fit to 
the combined dataset in in Section \ref{gFit} below.

\subsubsection{Global \boldmath{$ \chi ^2 $}-Test of the Differential Cross-Section }
\label{gFit}

The non-QED model parameters, discussed in Section~\ref{QEDdev1}, are determined by applying a
$\chi ^2$-test to the combined differential cross-section data measured by the VENUS, TOPAS, 
OPAL, DELPHI, ALEPH and~L3 Collaborations, just discussed in Section \ref{difCS}. The $\chi^2$ test 
searches for a minimum of the $\chi^2$ function that indicates a maximum of $\Lambda$.
This $\Lambda$ value, is after Eq. \ref{Litke2222}, Eq. (\ref{Litke4444} and Eq. \ref{DIR111}
is a direct indicator of the mass of a heavy electron $ m_{e^{\ast }} $ and the size of the 
electron $  \Lambda =f(m_{e^{\ast }},r_{e}) $. The following equation Eq. \ref{chi-222}, implemented in the 
MINUIT code \cite{MINUIT,MINUIT1}, is used to find this minimum.

\begin{align}
\label{chi-222}
\chi ^{2}=\sum_{i,j}\left \{ \frac{\frac{d\sigma }{d\Omega }^{\rm meas}
(|\cos\theta|_{i},E_{j} )-\frac{d\sigma }{d\Omega }^{\rm new}
(|\cos\theta |_{i},E_{j},\Lambda )}{\Delta \left [ \frac{d\sigma }
{d\Omega }^{\rm meas}(|\cos\theta |_{i},E_{j}) \right ]} \right \}^{2}.
\end{align}

In Eq. \ref{chi-222}  $\frac{d\sigma}{d\Omega}^{\rm meas}(|\cos\theta|_i, E_j)$ is the measured differential 
cross-section at an angular bin ($i$) and a center-of-mass energy bin ($j$),
while and $\frac{d\sigma}{d\Omega}^{\rm new}(|\cos\theta|i, E_j, \Lambda)$ is
the QED-$\alpha^3$ differential cross-section in the same bins and a test parameter $\Lambda$, as~defined in
Equations~(\ref{Litke4444}) and (\ref{DIR333}). The  ``$\pm$'' sign in front of
$\delta_{\rm new}$ in Eq.~\ref{Litke4444} and \ref{DIR333} allows
the $\chi^2$-test to search for positive and negative interference.
The term $\Delta\left[\frac{d\sigma}{d\Omega}^{\rm meas}(|\cos\theta|_i, E_j)\right]$
is the uncertainty of the mean value of the measurements, which is represented
by sum in quadratures of the statistical and systematic uncertainty,
to be discussed in Section  \ref{sys}. The $\chi^2$-test requires 
details of the differential cross-section and the luminosity
 at the different center-of-mass energies at which the data were taken.
The Section~\ref{difCS} provided a description of the dataset utilized in the
analysis, which included individual sub-sets published by the Collaborations.
Table~\ref{LUMI111} provides a summary of the luminosities
for all sub-sets from VENUS, TOPAS, ALEPH, DELPHI, L3,
and OPAL used in the $\chi^2$-test.

\newpage

\begin{table}
\caption{ The integrated luminosity used from the VENUS, TOPAS, ALEPH, DELPHI, L3 and OPAL~{experiments} at various
center-of-mass energies, $\sqrt{s}$. }

\begin{center}
\resizebox{1.0\textwidth}{!}{
\begin{tabular}{||l|l|l|l|l|l|l||}                               \hline
 GeV       & VENUS \cite{VENUS}           & TOPAZ \cite{TOPAS}            & ALEPH \cite{ALEPH}
           & DELPHI \cite{DELPHI}          & L3               & OPAL
                                                              \\ \hline
 55        & \small $ 2.34 pb^{-1} $  
           & & & & &
                                                             \\ \hline
 56        & \small $ 5.18 pb^{-1} $ 
           & & & & &
                                                             \\ \hline
 56.5      & \small $ 0.86 pb^{-1} $  
           & & & & &
                                                             \\ \hline
 57        & \small $ 3.70 pb^{-1} $  
           & & & & &
                                                             \\ \hline
 57.6      & & \small $ 52.26 pb^{-1} $ 
           & & & &
                                                             \\ \hline
 91        & &
           & \small $ 8.5   pb^{-1} $  
           & \small $ 36.9  pb^{-1} $ 
           & \small $ 64.6   pb^{-1} $  \cite{L3A}
           & \small $ 7.2   pb^{-1} $  \cite{OPAL1}
                                                              \\ \hline
 133       & & &
           & \small $ 5.92   pb^{-1} $  
           & &
                                                              \\ \hline
 162       & & &
           & \small $ 9.58   pb^{-1} $  
           & &
                                                              \\ \hline
 172       & & &
           & \small $ 9.80   pb^{-1} $  
           & &
                                                              \\ \hline
 183       & & &
           & \small $ 52.9 pb^{-1} $  
           & \small $ 54.8  pb^{-1} $  \cite{L3B}
           & \small $ 55.6  pb^{-1} $  \cite{OPAL2}
                                                              \\ \hline
 189       & & &
           & \small $ 151.9 pb^{-1} $  
           & \small $ 175.3 pb^{-1} $  \cite{L3B}
           & \small $ 181.1 pb^{-1} $  \cite{OPAL2}
                                                              \\ \hline
 192       & & &
           & \small $ 25.1  pb^{-1} $  
           & \small $ 28.8  pb^{-1} $  \cite{L3B}
           & \small $ 29.0  pb^{-1} $  \cite{OPAL2}
                                                              \\ \hline
 196       & & &
           & \small $ 76.1  pb^{-1} $ 
           & \small $ 82.4  pb^{-1} $  \cite{L3B}
           & \small $ 75.9  pb^{-1} $  \cite{OPAL2}
                                                              \\ \hline
 200       & & &
           & \small $ 82.6  pb^{-1} $  
           & \small $ 67.5  pb^{-1} $  \cite{L3B}
           & \small $ 87.2  pb^{-1} $  \cite{OPAL2}
                                                              \\ \hline
 202       & & &
           & \small $ 40.1  pb^{-1} $  
           & \small $ 35.9  pb^{-1} $  \cite{L3B}
           & \small $ 36.8  pb^{-1} $  \cite{OPAL2}
                                                              \\ \hline
 205       & & & &
           & \small $ 74.3  pb^{-1} $  \cite{L3B}
           & \small $ 79.2  pb^{-1} $  \cite{OPAL2}
                                                              \\ \hline
 207       & & & &
           & \small $ 138.1 pb^{-1} $  \cite{L3B}
           & \small $ 136.5 pb^{-1} $  \cite{OPAL2}
                                                              \\ \hline
\end{tabular}
} 
\end{center}
\label{LUMI111}
\end{table}

\subsubsection{ Global  {$ \chi ^2$}-Test for Heavy Electron $ m_{e*} $}
\label{gFitme}

In order to perform the $\chi^2$-fit in Eq.\ref{chi-222}, it is necessary to use the
differential QED cross-section from Section \ref{difCS} and the deviation from QED
Eq. \ref{QEDfit}) for a given center-of-mass energy, as well as the
differential cross-section calculated for the heavy electron, see Eq. \ref{Litke2222}) and \ref{Litke3333}.
The $\chi^2$-fit can be performed separately for every data sub-set
at respective $\sqrt{s}$ by either utilising the experimentally measured
differential cross-section, if available,
or by calculating it using the luminosity, N umber of events per angular bin width and efficiency,
as described in Eq. \ref{SIGMAdiff1} and \ref{COS1}.
The theoretical differential cross section for QED-$\alpha^3$ is computed using the numerical 
calculations of the $\EEGG$ reaction discussed in the Section \ref{QEDnumeric}. 
The parameter $1/\Lambda_{+}^4$ is used in the cross section \ref{Litke4444} for the test.
Table~\ref{SINGLE-FIT} displays the resulting fit
parameters ($1/\Lambda_{+}^4$ in $1/\mathrm{GeV}^4$)
and fit quality parameter $\chi^2/$dof (degrees of freedom)
of the $\chi^2$-test for every data subset.


\begin{table}
\caption{ The fit parameter $1/\Lambda_{+}^{4}$ for each data sub-set, along with the 
corresponding fit quality parameter $\chi^{2}/{\rm dof}$ (degrees of freedom).}

\begin{center}
\begin{tabular}{||l|l|l|l|l|l|l||}                               \hline
           & VENUS            & TOPAZ            & ALEPH
           & DELPHI           & L3               & OPAL          \\
 GeV,$ \sqrt{s}$       & $(1/\Lambda^{4})$& $(1/\Lambda^{4})$&$(1/\Lambda^{4})$
           & $(1/\Lambda^{4})$& $(1/\Lambda^{4})$&$(1/\Lambda^{4})$
                                                                 \\ \hline
\scriptsize 55        & \tiny  \color{red} $-(4.26\pm2.52) 10^{-8}$
                      & & & & &                                  \\
                      & \tiny  \color{red}$ \chi^{2}/dof = 12.9/8$
                      & & & & &
                                                            \\ \hline
\scriptsize 56        & \tiny  $ (3.24\pm1.88) 10^{-8}$
                      & & & & &                             \\
                      & \tiny  $ \chi^{2}/dof = 9.48/8$
                      & & & & &
                                                            \\ \hline
\scriptsize 56.5      & \tiny  \color{red}$ -(2.11\pm3.96) 10^{-8}$
                      & & & & &                             \\
                      & \tiny  \color{red}$ \chi^{2}/dof = 4.93/8$
                      & & & & &
                                                             \\ \hline

\scriptsize 57        & \tiny  \color{red}$ -(1.49\pm2.02) 10^{-8}$
                      & & & & &                             \\
                      & \tiny  \color{red}$ \chi^{2}/dof = 8.82/8$
                      & & & & &
                                                             \\ \hline
\scriptsize 57.6      & & \tiny  \color{red}$ -(1.59\pm5.61) 10^{-9}$
                      & & & &                              \\
                      & & \tiny  \color{red}$ \chi^{2}/dof = 7.32/5$
                      & & & &
                                                             \\ \hline
\scriptsize 91        & &
                      &\tiny $ (0.07\pm2.98) 10^{-9}$
                      &\tiny  \color{red}$ -(2.29\pm1.70) 10^{-9}$
                      &\tiny  \color{red}$ -(6.88\pm8.00) 10^{-10}$
                      &\tiny  \color{red}$ -(0.93\pm3.59) 10^{-9}$
                                                                 \\
                      & &
                      & \tiny  $ \chi^{2}/dof = 9.96/9$
                      & \tiny  \color{red}$ \chi^{2}/dof = 3.54/6$
                      & \tiny  \color{red}$ \chi^{2}/dof = 11.1/15$
                      & \tiny  \color{red}$ \chi^{2}/dof = 6.92/8$
                                                              \\ \hline
\scriptsize 133       & & & & \tiny  \color{red}$ -(0.48\pm1.26) 10^{-9}$
                      & &                                     \\
                      & & & & \tiny  \color{red}$ \chi^{2}/dof = 2.60/3$
                      & &
                                                             \\ \hline
\scriptsize 162       & & & & \tiny  \color{red}$ -(2.35\pm5.40) 10^{-10}$
                      & &                                    \\
                      & & & & \tiny  \color{red}$ \chi^{2}/dof = 4.59/4$
                      & &
                                                             \\ \hline
\scriptsize 172       & & & & \tiny  $ (0.74\pm5.19) 10^{-10}$
                      & &                                    \\
                      & & & & \tiny  $ \chi^{2}/dof = 1.09/4$
                      & &
                                                             \\ \hline
\scriptsize 183       & & &
                      &\tiny  \color{red}$ -(2.54\pm1.60) 10^{-10}$
                      &\tiny  \color{red}$ -(1.48\pm1.37) 10^{-10}$
                      &\tiny   $ (2.05\pm1.43) 10^{-10}$
                                                                 \\
                      & & &
                      & \tiny  \color{red}$ \chi^{2}/dof = 5.27/4$
                      & \tiny  \color{red}$ \chi^{2}/dof = 11.0/9$
                      & \tiny  $ \chi^{2}/dof = 5.86/9$
                                                              \\ \hline

\scriptsize 189       & & &
                      &\tiny  $ (0.14\pm1.01) 10^{-10}$
                      &\tiny  \color{red}$ -(8.58\pm7.16) 10^{-11}$
                      &\tiny  \color{red}$ -(2.05\pm6.89) 10^{-11}$
                                                                 \\
                      & & &
                      & \tiny  $ \chi^{2}/dof = 2.67/4$
                      & \tiny  \color{red}$ \chi^{2}/dof = 17.2/9$
                      & \tiny  \color{red}$ \chi^{2}/dof = 5.13/9$
                                                              \\ \hline
\scriptsize 192       & & &
                      &\tiny  \color{red}$ -(3.59\pm2.07) 10^{-10}$
                      &\tiny  \color{red}$ -(5.79\pm1.41) 10^{-10}$
                      &\tiny   $ (0.13\pm1.63) 10^{-10}$
                                                                 \\
                      & & &
                      & \tiny  \color{red}$ \chi^{2}/dof = 1.03/4$
                      & \tiny  \color{red}$ \chi^{2}/dof = 16.9/9$
                      & \tiny  $ \chi^{2}/dof = 12.6/9$
                                                              \\ \hline
\scriptsize 196       & & &
                      &\tiny  \color{red}$ -(0.43\pm1.19) 10^{-10}$
                      &\tiny  \color{red}$ -(1.93\pm0.89) 10^{-10}$
                      &\tiny  \color{red}$ -(1.62\pm9.37) 10^{-10}$
                                                                 \\
                      & & &
                      & \tiny  \color{red}$ \chi^{2}/dof = 16.4/4$
                      & \tiny  \color{red}$ \chi^{2}/dof = 7.84/9$
                      & \tiny  \color{red}$ \chi^{2}/dof = 7.48/9$
                                                              \\ \hline
\scriptsize 200       & & &
                      &\tiny  \color{red}$ -(0.88\pm1.12) 10^{-10}$
                      &\tiny  \color{red}$ -(2.58\pm0.90) 10^{-10}$
                      &\tiny  \color{red}$ -(1.65\pm0.84) 10^{-10}$
                                                                 \\
                      & & &
                      & \tiny  \color{red}$ \chi^{2}/dof = 8.07/4$
                      & \tiny  \color{red}$ \chi^{2}/dof = 13.8/9$
                      & \tiny  \color{red}$ \chi^{2}/dof = 8.63/9$
                                                              \\ \hline
\scriptsize 202       & & &
                      &\tiny  \color{red}$ -(1.11\pm1.51) 10^{-10}$
                      &\tiny  \color{red}$ -(1.49\pm1.24) 10^{-10}$
                      &\tiny  \color{red}$ -(1.47\pm1.16) 10^{-10}$
                                                                 \\
                      & & &
                      & \tiny  \color{red}$ \chi^{2}/dof = 2.94/4$
                      & \tiny  \color{red}$ \chi^{2}/dof = 15.2/9$
                      & \tiny  \color{red}$ \chi^{2}/dof = 17.8/9$
                                                              \\ \hline
\scriptsize 205       & & & &
                      &\tiny  \color{red}$ -(1.07\pm0.84) 10^{-10}$
                      &\tiny  \color{red}$ -(3.81\pm7.99) 10^{-11}$
                                                                 \\
                      & & & &
                      & \tiny  \color{red}$ \chi^{2}/dof = 12.9/9$
                      & \tiny  \color{red}$ \chi^{2}/dof = 6.26/9$
                                                              \\ \hline
\scriptsize 207       & & & &
                      &\tiny  \color{red}$ -(9.14\pm5.99) 10^{-11}$
                      &\tiny  \color{red}$ -(1.52\pm0.57) 10^{-10}$
                                                                 \\
                      & & & &
                      & \tiny  \color{red}$ \chi^{2}/dof = 23.6/9$
                      & \tiny  \color{red}$ \chi^{2}/dof = 10.7/9$
                                                              \\ \hline
\end{tabular}

\end{center}
\label{SINGLE-FIT}
\end{table}

Approximately 80\% of the data sub-sets in Table~\ref{SINGLE-FIT} exhibit a preference
for negative values of $1/\Lambda_{+}^4$.
This trend is also evident in Figure~\ref{FITgroups} and Table~\ref{fit-num}.
Figure~\ref{FITgroups} displays the results of the $\chi^2$ tests for the data sub-sets,
grouped according to their Collaborations (ALEPH, DELPHI, L3, OPAL) and~combinations
of Collaborations such as TRISTAN (TOPAS and VENUS), LEP1 and~LEP2.
The trend towards negative values of $1/\Lambda_{+}^{4}$ becomes more
pronounced as the statistics of the grouped combinations~increases.

Table~\ref{SINGLE-FIT} shows the values of $1/\Lambda^4$, obtained from the fits to the
data combined in groups of Collaborations. For~TRISTAN, the~values are positive with 
the large error bar for energies ranging from $\sqrt{s}=55$~GeV
to 57.6~GeV. At~LEP1, where the data were taken at the $Z^0$ pole with lower luminosity,
the values are already negative with a statistical significance of approximately one standard deviation.
This is reflected in the size of the error bars in Table~\ref{fit-num}.
The LEP2 data, covering energies from 133~GeV to 207~GeV, have much higher
luminosity and dominate the global fit result. The~fitted parameter is negative
with a significance of approximately five standard deviations in this energy range.
Note that the $\chi^2$-distribution exhibits a good parabolic shape, as~shown in Figure~\ref{CHItotal}.
This shape remains almost unaffected when applying the non-parabolic $\chi^2$ in the
 MINOS routine of the MINUIT program~\cite{MINUIT,MINUIT1}.
Therefore, the~significance deduced from the parabolic shape remains the~same.


\begin{figure}[htbp]
\begin{center}
 \includegraphics[width=14.0cm,height=18.0cm]{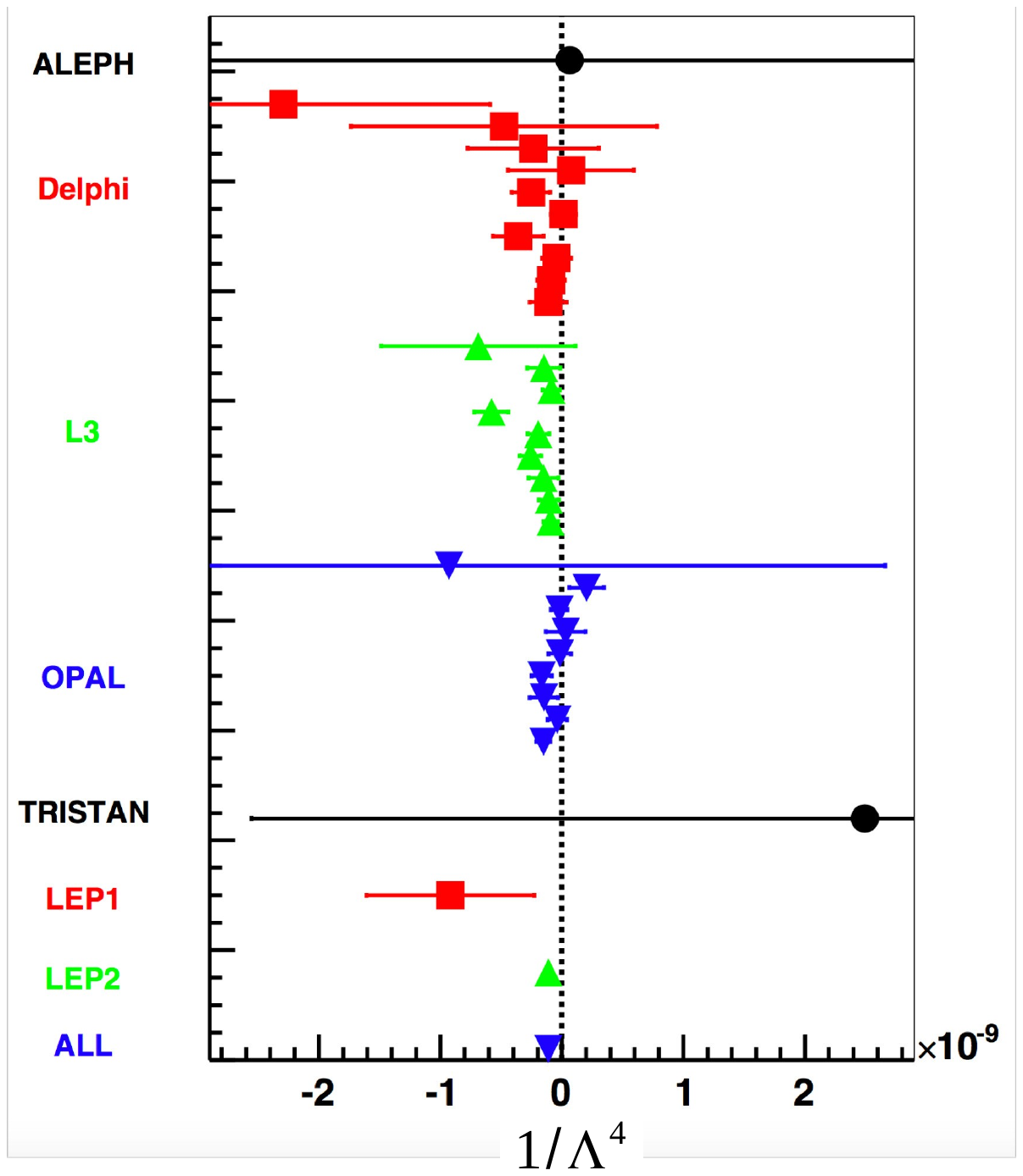}
\end{center}
\caption{The fit parameter $1/\Lambda^{4}$  (in $1/{\rm GeV}^{4}$), 
displayed for the data sub-sets, grouped according to the 
Collaborations (ALEPH, DELPHI, L3, OPAL) and~combinations
of the Collaborations such as TRISTAN (TOPAS and VENUS), LEP1 and~LEP2.
The global combination is labeled  ``ALL''.}
\label{FITgroups}
\end{figure}

\newpage

\begin{table}
\caption{The values of $1/\Lambda^4$ from the fits to data, combined in groups of Collaborations.}
\begin{center}
\begin{tabular}{||l|l||}                                       \hline
  TRISTAN       & $ ( 2.49\pm5.05)\times 10^{-9} $           \\
                & $ \chi^{2}/dof = 50.0/41 $
                                                              \\ \hline
  LEP I         & $ -( 9.20\pm6.90)\times 10^{-10} $           \\
                & $ \chi^{2}/dof = 32.3/41 $
                                                              \\ \hline
  LEP II        & $ -( 1.10\pm0.20)\times 10^{-10} $           \\
                & $ \chi^{2}/dof = 267/203 $
                                                              \\ \hline
  All Data      & $ -( 1.11\pm0.20)\times 10^{-10} $           \\
                & $ \chi^{2}/dof = 351/287 $
                                                              \\ \hline
\end{tabular}
\end{center}
\label{fit-num}
\end{table}

The best fit value of the parameter $1/\Lambda^{4}$ obtained from the $\chi^2$-fit is shown in
Figure~\ref{CHItotal} and given in Table~\ref{electron1}. It has a significance $  \Omega \approx 5.5 $
of about FIVE standard deviations discussed in  Section \ref {Sig444}, which implies the existence of an excited electron
with a mass of $m_{e^*} = 308 \pm 14$~GeV, as~interpreted through $\Lambda^2_{+} =m^2_{e^*}/\lambda$,
where $\lambda = 1.0 $.

\begin{figure}[htbp]
\begin{center}
 \includegraphics[width=16.0cm,height=8.0cm]{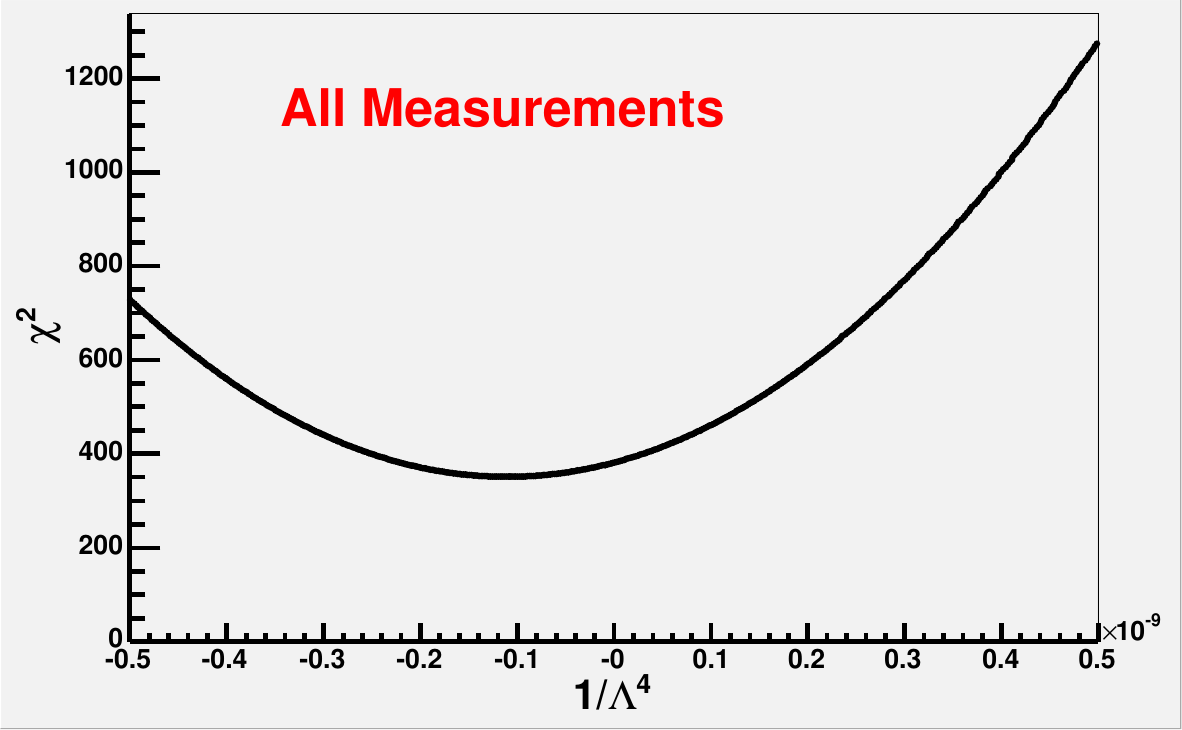}
\end{center}
\caption{$ \chi^{2} $ as function of $ 1/\Lambda^{4} \times 10^{-9} $ for all data~sets.}
\label{CHItotal}
\end{figure}

\newpage

\begin{table}
\begin{center}
\caption{ Summary of excited electron $\chi ^2$-tests $1/\Lambda_{+}^{4}$. See text for details}
\label{electron1}
\begin{tabular}{ | l | l | l | }
  \hline
  \multicolumn{3}{ | c | }{ Heavy electron mass $ m_{e^{*}} $  } \\
  \hline
   $ -(1.11\pm 0.20) \times 10^{-10 }$ & $ \Lambda ^2_{+} =m^2_{e^*} / \lambda $ & $ m_{e^{*}} = 308 \pm 14 $ GeV  \\
   \hline
 \end{tabular}
\end{center}
\end{table}

\subsubsection{Global $ \chi ^2$-Test for a Non-Pointness of the~Electron}
\label{fitContact11}

The dependence of the differential cross section on $(1-\cos^2\theta)$ is the same for
both the excited electron model (\ref{Litke4444}) and the contact interaction (\ref{DIR3}).
Therefore, the same MINUIT formalism can be used to perform the $\chi^2$ test by 
substituting $\Lambda =\tilde{\Lambda }=\Lambda _{6}$ into equation (\ref{chi-222}).
The difference between Equation~(\ref{DIR3}) and Equation~(\ref{Litke4444}) lies only in the
presence of the constant $\alpha$, but~it does not affect the significance of the fit,
as shown in Table~\ref{electron2}. The~result indicates the existence of
a contact interaction with a significance $ \Omega \approx 5.5 $  of about five standard 
deviations, discussed in  Section \ref {Sig444} and a cutoff scale of $\Lambda = 1253.53 \pm 226$~GeV.


\begin{table}[H]
\begin{center}
\caption{ $ \chi ^2 $ tests $ ( 1/\Lambda )^{4} [ {\rm GeV} ^{-4} ] $  finite size and heavy~electron.  }
\label{electron2}
\begin{tabular}{ | l | l | l | }
  \hline
  \multicolumn{3}{ | c | }{   Summary $ r_{electron} $ and  $ m_{e^{*}} $ }\\
  \hline
   & Test  $ r_{electron} $ &  \\
    \hline
    $ -(4.05\pm 0.73) \times 10^{-13 }$ & $ r= ( \hbar \times c )/ \Lambda $ & $ r=(1.57\pm 0.07) \times 10 ^{-17} $ cm  \\
   \hline
   & Test  $ m_{e^{*}} $ & \\
   \hline
  $ -(1.11\pm 0.20) \times 10^{-10 }$ & $ \Lambda ^2_{+} =m^2_{e^*} / \lambda $ &  $ m_{e^{*}}  = 308 \pm 14 $ GeV  \\
  \hline
\end{tabular}
\end{center}
\end{table}

Note that the p-value discussed in chapter \ref{Sig444} gives a significance result, that is similar to
the $\chi^2$-test, as~demonstrated in detail in the total cross-section analysis~\cite{Totcross}.

\subsubsection{ Indicati on of a Signal in the Total Cross-Section}
\label{totalXsection}

If a signal of finite electron size exists in the differential cross section of the $\EEGG$ reaction,
it is of great importance to prove that such a signal also exists in the total cross section of this reaction.
This analysis is carried out in detail in the Total Cross-Section chapter \ref{TotalCross} and chapter \ref{totalXsectionratio}..

 \subsubsection{Systematic Uncertainties in differential section analysis.}
 \label{sys}

Various sources of systematic uncertainties affect  the measurement, including uncertainties from the
luminosity evaluation, selection efficiency, background estimation, choice of QED-$\alpha^3$ theoretical
cross-section, fit procedure, fit parameter and~the choice of scattering angle in $|\cos\theta|$
for comparison between data and theory. The maximum estimated uncertainties from luminosity,
selection efficiency and~background evaluations contribute approximately $\delta\Lambda/\Lambda=0.01$
to the total systematic uncertainty in the estimated fit parameter. The choice of theoretical QED cross-section 
was validated using about 2000 $\EEGG$ events generated and processed with the geometry and selection cuts of the L3 detector~\cite{Manat1,Berends,Berends1,Berends12,Berends13}.
For scattering angles close to 90$^\circ$ where $|\cos(\theta)|_{\mathrm{experiment}} \sim 0.05$,
the systematic uncertainty contribution $(\delta\Lambda/\Lambda)_{\delta|\cos\theta|}$ is approximately 0.01.
The combined effect of these two systematic uncertainties yields the uncertainty of
$\delta\Lambda/\Lambda\approx0.015$. In a small sample of $\EEGG$ events, fit values were compared using $\chi^2$, maximum likelihood,
Smirnov--Cramer von Mises and~Kolmogorov tests with and without binning~\cite{Isiksal1}.
An additional uncertainty of $\delta\Lambda/\Lambda=0.005$ was
inferred from the fit procedure variation study.
In summary, while multiple sources of systematic uncertainties
have been identified, they are all smaller than the statistical uncertainty of the experimental~data.

\subsubsection{Concluding~Remarks}
\label{conclusions}

The VENUS, TOPAS, OPAL, DELPHI, ALEPH, and L3 collaborations measured the differential cross section
of the $\EEGG$ reaction to test quantum electrodynamics (QED). With the exception of ALEPH, all collaborations observed a negative
deviation from QED, albeit with low significance. The test of the total cross section of LEP2 and
comparison with the measurements in Figure~\ref{Ratio-1} and Figure~\ref{Ratio-2}  confirm these negative trends.
We performed a comprehensive $\chi^{2}$ analysis using all available data to search for evidence for an excited 
electron, $e^{*}$, and a finite annihilation length. We used a direct contact term approach.
By conducting a global analysis of the combined datasets revealed, for the first time, a significance of 
approximately 5$\sigma$ for the mass of an excited electron, which is $m_{e^{*}} = 308\pm 56$~GeV.
A similar 5$\sigma$ significance effect was detected for a charge distribution
radius of the electron, $r_e=(1.57\pm0.07)\times10^{-17}$ cm.
Therefore, combining the full statistical power of all available LEP and non-LEP high-efficiency experiments
on measurements of the cross-section of the reaction of annihilation in $e^+e^-$ collisions
allowed us to identify the signal of existence of excited electron and contact interaction
 at a high level of significance. Previous analyses have restricted themselves to the data collected only with the 
LEP detectors and at not all and limited range LEP energies as well as with non-LEP detectors.
Therefore, combining the full statistical power of all available LEP and non-LEP
high-efficiency experiments on measurements of the cross-section of the reaction
of annihilation in $e^+e^-$ collisions, we were able to recognise the signal
of the existence of an excited electron and contact interaction at a high level of significance.

\subsection{Hint for a minimal interaction length in $ \EEG $ annihilation in total cross section of center-of-mass energies 55 - 207 GeV.}
\label{TotalCross}

As discussed in section \ref{Diff-Cross}, ``Finite Size of Electron in $ {\EEG} $ 
Annihilation in differential cross section of centre-of-mass energies 55 - 207 GeV''
a search for a signal in the differential cross section of a finit size of the electron 
in the $ \EEG $ annihilation was performed.  As already mentioned in 
chapter \ref{totalXsection}, it is of fundamental importance that also a 
signal in the annihilation of the total cross section exist. In detail this analysis was 
performed in reference \cite{Totcross}. The analysis of the differential cross section 
of the $ \EEG $ annihilation performed in section \ref{Diff-Cross} is very 
similar to the analysis in section \ref{TotalCross}. Therefore, we will focus 
in the following exclusively on the technical differences between the two analyses.

\subsubsection{ Introduction }
\label{Introduction-TotalCross}

The introduction in the section \ref{Diff-Cross-Introduction} already describes
in great detail the analysis of ${\EEG} $ annihilation, in particular with a focus 
on the electron. In this introduction \ref{Introduction-TotalCross},
we concentrate on extending the analysis of the differential cross 
section to the total cross section.

The VENUS, TOPAS, OPAL, DELPHI, L3 and ALEPH  collaboration 
used the differential cross section of the $ \EEG $ reaction to search 
for a deviation from QED. This was performed for certain energy ranges and luminosities. 
We perform a global $ \chi ^2 $ FIT, using the data from these six 
research projects to investigate $ \Lambda_{+} $, $ \Lambda_{-} $ 
and $ m_{e^{*}} $ for energies from $ \sqrt{s} $ = 55 GeV to 207 GeV 
including the associated luminosities \cite{QED-1,Diffcross11}.
This analysis allowed to set an approximately 5$\times \sigma$ limit 
on the finite size of the electron $ r=(1.57\pm 0.28) \times 10 ^{-17} [cm] $
and  on the mass of an exited electron of $ m_{e^{*}}  = 308 \pm 56 [ GeV ] $. 
The deviation in the differential experimental cross section from the QED 
values of approximately 4 $ \% $ was only visible in the fit results but 
not direct in the comparison of the experimental and theoretical QED cross section.

The aim of this investigation is to prove, using a $ \chi ^2 $ fit, that using the total 
cross section of all data leads to a similar result. First, the theoretical 
framework for calculating the total QED cross section must be discussed, 
in particular the fact that an analytically precise QED cross section must 
be calculated using a Monte Carlo program. Secondly, since no 
total cross section exists for the QED response $\EEG$ from 
$\sqrt{s}$ = 55 GeV to 207 GeV, we introduced a model 
for a total cross section using data from the six collaborations. 
Incorporating this information, it is possible to perform a total 
$\chi^2$ fit with all the data. Finally, the results can be discussed.

\subsubsection{Theoretical Frameworks}
\label{Theoretical-Fram-Sigma-tot}

The theoretical frameworks from section \ref{Theory-Diff-cross} is the same as 
for the actual framework \ref{Theoretical-Fram-Sigma-tot} 
from the Dirac equation \ref{DIRAC} to equation of the total cross section 
of the $ \EEG $ reaction equation \ref{totcross888}. After the equation  \ref{totcross888} 
both theoretical framework from chapter \ref{Theory-Diff-cross} and 
chapter  \ref{Theoretical-Fram-Sigma-tot} are similar but not the same any more.
Two different Mode Carlo generators are used for the numerical calculation of the
total QED cross section $ \sigma _{tot} $. For this reason the analysis focus on this part next.

\subsubsection{ The numerical calculation of the $ \EEGG $ cross sections.} 
\label{Numerical calculation} 

For practical reasons, a numerical calculation is used to calculate the differential and total cross section.
This cross sections are inserted in the $ \chi ^2 $ fit.

The 6 collaborations under discussion, used an event generator \cite{Berends111} 
for the reaction $ \EEGG $ Eq. \ref{141} to calculate the $ \EEGG $ differential cross 
sections and the total cross section.

\begin{equation}
\label{141}
e^+(p_{+})+e^-(p_{-}) \rightarrow \gamma (k_{1})+\gamma (k_{2})+\gamma (k_{3})
\end{equation}

The lowest order differential cross section $ \left ( \frac{d\sigma }{d\Omega } \right )_{Born} $
is numerically corrected  by adding the higher order correction of Figure~\ref{CORR1} and Figure~\ref{CORR2} 
to $ O(\alpha ^3) $ in Eq. \ref{BORNCORR-11} .

\begin{equation}
\label{BORNCORR-11}
\left ( \frac{d\sigma }{d\Omega } \right )_{\alpha ^3}=\left ( \frac{d\sigma }{d\Omega } \right )_{Born}(1+\delta _{virtual}+\delta _{soft}+\delta _{hard})
\end{equation}

Where in Eq.\ref{BORNCORR-11}, $ \delta _{virtual} $ is the virtual correction and 
$ \delta_{soft} $ and $ \delta_{hard} $ are the soft- and hard-Bremsstrahlung 
corrections, respectively.

If the energy of the photons from initial state radiation (soft Bremsstrahlung) are 
too small to be detected, the reaction can be treated as 2-photon final state. 
This is valid if the parameter $ k_{3}/|p_{+}|=k_{0}<<1 $ of the  $ e^+ e^- $ - reaction is 
fulfilled Eq. \ref{142}.

\begin{equation}
\label{142}
e^+(p_{+})+e^-(p_{-}) \rightarrow \gamma (k_{1})+\gamma (k_{2}))
\end{equation}

For the third order differential cross section, the program generates  three $ \gamma $'s events sorted after the energies 
$ E_{\gamma 1 } \geq E_{\gamma 2 } \geq E_{\gamma 3 } $ , with the correct mixture for soft 
$ k_{3}/\vert p_{1}\vert = k_{0} \ll 1 $ and hard QED corrections shown in Figure~\ref{CORR1} and Figure~\ref{CORR2}.
The angle $ \beta $ between the $ E_{\gamma 1 } $ and $ E_{\gamma 2} $ event
 $ \beta_{min} < \beta < \beta_{max} $ is connected to the scattering angle $\theta $ by $ | cos \theta | $.

The differential cross section  $ \overline{(\frac{d\sigma }{d\Omega })}_{i} $ at 
an angle $ \theta $, an energy $ E_{tot}  $ and  for an angle bin width $ \Delta (|cos\theta |) $
can be expressed with Eq. \ref{DIFF11} .
 
\begin{align}
\label{DIFF11}
\overline{(\frac{d\sigma }{d\Omega })}_{i}=\frac{1}{2\pi \Delta (|cos\theta |)}\sigma _{tot}\frac{N_{i}}{N}
\end{align}

The scattering angle is $ | cos \theta | = ( | cos \theta_{1} | + | \cos \theta_{2} | ) /2 $ , where 
$ \theta_{1} $ is the scattering angle of $ E_{\gamma 1 } $ and  $\theta_{2} $ is the scattering angle of $ E_{\gamma 2 } $,
$ N_{i} $ is the number of events in an angle bin width $ \Delta (|cos\theta |) $ and $ N $ is the total amount of
events used in the generator. The Monte Carlo generator together with Eq. \ref{DIFF1} is used to calculate distributions 
of differential cross sections as a function of $ | cos \theta | $ including six discussed parameters. 

In the past, the Monte Carlo generator \cite{Berends111} was used from all 6 collaborations, to generate the
QED cross section of the $ \EEGG $ reaction. Our interested is focused on a 
new Monte Carlo generator BabaYaga@nlo \cite{BABAYAGA}. To use the BabaYaga QED generator
 requests a comparison of QED cross section data from both generators.
The new generator implements the same radiative corrections of Figure~\ref{CORR1} and Figure~\ref{CORR2}. 
To generate BabaYaga@nlo $ \EEGG $ events, the following seven parameters are used Eq. \ref{BabaYagaPar}.

\begin{align} 
\label{BabaYagaPar}
p_{11}&=final-\gamma-state=gg\\ \nonumber
p_{22}&=\sqrt{S}-ecms-ENERGY \left [ GeV \right ]\\ \nonumber
p_{33}&=minimum-angle-\theta_{min} - thmin \left [ deg \right ]\\ \nonumber
p_{44}&=maximum-angle-\theta_{max} - thmax \left [ deg \right ]\\ \nonumber
p_{55}&=Accollinear-angle-acoll.-\beta \left [ deg \right ] \\ \nonumber
p_{66}&=Minimum-energy-emin\left [ GeV \right ]  \\ \nonumber
p_{77}&=Number-Generating-Events=N  \nonumber
\end{align}

To perform a $\chi^{2}$ fitting for an electron of finite size, an analytical expression 
for the differential cross section Eq. \ref{DIFF1} ) is required. The angular distribution 
of the differential cross section Eq. \ref{DIFF1}) is fitted using a $\chi^{2}$ fitting 
with a polynomial with 6 parameters $p_{1}$ to $p_{6}$, as shown in Eq.\ref{QEDfit}.

\begin{align}
\label{QEDfit}
&\left ( \frac{d\sigma }{d\Omega } \right )_{QED}=\left ( \frac{d\sigma }{d\Omega } \right )_{Born} \times \\
&\left ( 1+p_{1}+p_{2}e^{-\frac{x^{1.2}}{2p^2_{3}}}+p_{4}x+p_{5}x^2+p_{6}x^3 \right )|_{x=|cos\theta |} \nonumber
\end{align}

To calculate cross sections for different detectors, it is necessary for the calculation of the QED cross sections to
regard the various parameters how every detector is able to measure events. We normalise all cross section
measurements and calculation of the QED cross section to the L3 parameters Eq. \ref{BabaYagaL3}  \cite{L3B}. 

\begin{align} 
\label{BabaYagaL3}
p_{11}&=final-\gamma-state=gg\\ \nonumber
p_{22}&=\sqrt{S}-ecms-ENERGY \left [ GeV \right ]\\ \nonumber
p_{33}&=minimum-angle-\theta_{min} - thmin=16.0 \left [ deg \right ]\\ \nonumber
p_{44}&=maximum-angle-\theta_{max} - thmax=164.0\left [ deg \right ]\\ \nonumber
p_{55}&=Accollinear-angle-acoll.-\beta =90.0\left [ deg \right ] \\ \nonumber
p_{66}&=Minimum-energy-emin=0.48 \left [ GeV \right ]  \\ \nonumber
p_{77}&=Number-Generating-Events=N=100000  \nonumber
\end{align}

Considering the experimental data of the ratio of the total measured cross section divided by the total QED cross
section $ R=\sigma (exp)/\sigma (QED) $ of the L3 collaboration shown in Fig.2 of \cite{L3B} , a small deviation from the 
QED - ratio R from 90 GeV to 207 GeV is visible. To test the QED deviation as a function of $ \sqrt{s} $,  
we performed two sets of $ \chi^{2} $ tests. One set from 55 GeV to 207 GeV with 17 energy points and the other set 
from 91.2 GeV to 200 GeV with 7 energy points.

Including the parameters of Eq.\ref{BabaYagaL3}, it is possible to calculate and fit all 17 angular distributions of
the energies under request from $ \sqrt{s} $ = 55 GeV to 207 GeV. The 6 fit parameters  $ p_{1} $  to $ p_{6} $  Eq. \ref{QEDfit} 
are summarised in Table~\ref{diffpar} (upper part). The same parameters for 7 $ \sqrt{s} $ energies from $ \sqrt{s} $ = 91.2 GeV 
to 200 GeV of the generator \cite{Berends} are summarised in Table~\ref{diffpar} (lower part).

For an example at $ \sqrt{s}=91.2$ GeV, the differential QED cross section fitted to Eq. \ref{QEDfit}  is shown in 
Figure \ref{QEDdiff}. 

\newpage

\begin{figure}[htbp]
\vspace{0.0mm}
\begin{center}
 \includegraphics[width=15.0cm,height=15.0cm]{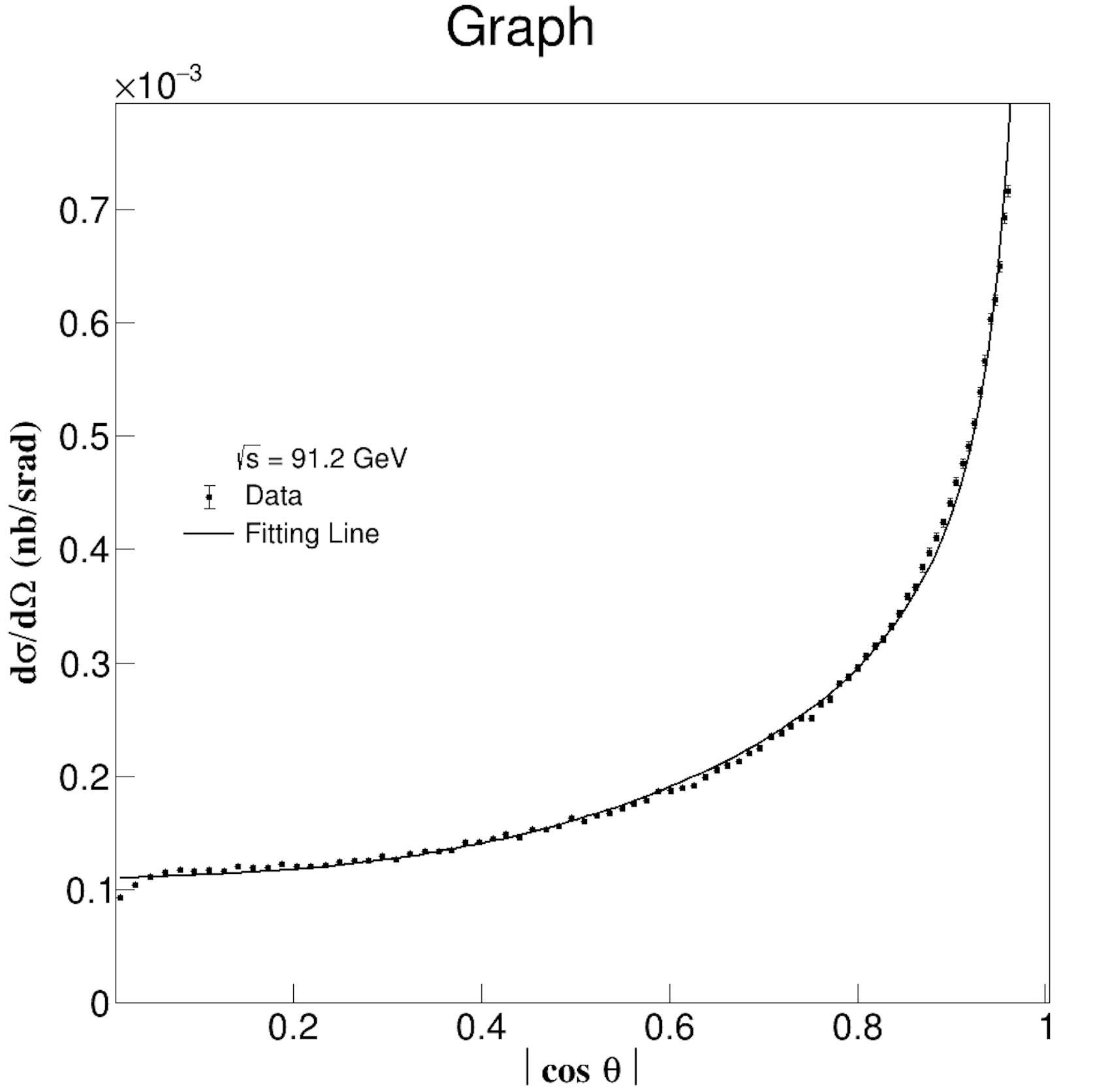}
\end{center}
\vspace{-1.0mm}
\caption{ Differential QED BabaYaga cross section of the $ \EEGG $ reactions at $ \sqrt{s} = 91.2 $ GeV.  
Normalised to the parameters of L3. Points represent the cross section and black line is the fit. }
\label{QEDdiff}
\end{figure}

By including the fitting parameters from Table \ref{diffpar} together with Eq.\ref{144}, the entire
QED-BabaYaga cross section from $ \sqrt{s}=55 $ GeV to 207 GeV can be calculated 
shown in Figure \ref{BHABAtot}. 

\begin{equation}
\label{144}
\sigma _{tot}(QED)=\int_{\theta =16.0^{\circ}}^{\theta =164.0^{\circ}}\left ( \frac{d\sigma }{d\Omega } \right )_{QED}d\left | cos\theta  \right |
\end{equation}

\newpage

\begin{figure}[htbp]
\begin{center}
 \includegraphics[width=14.0cm,height=12.0cm]{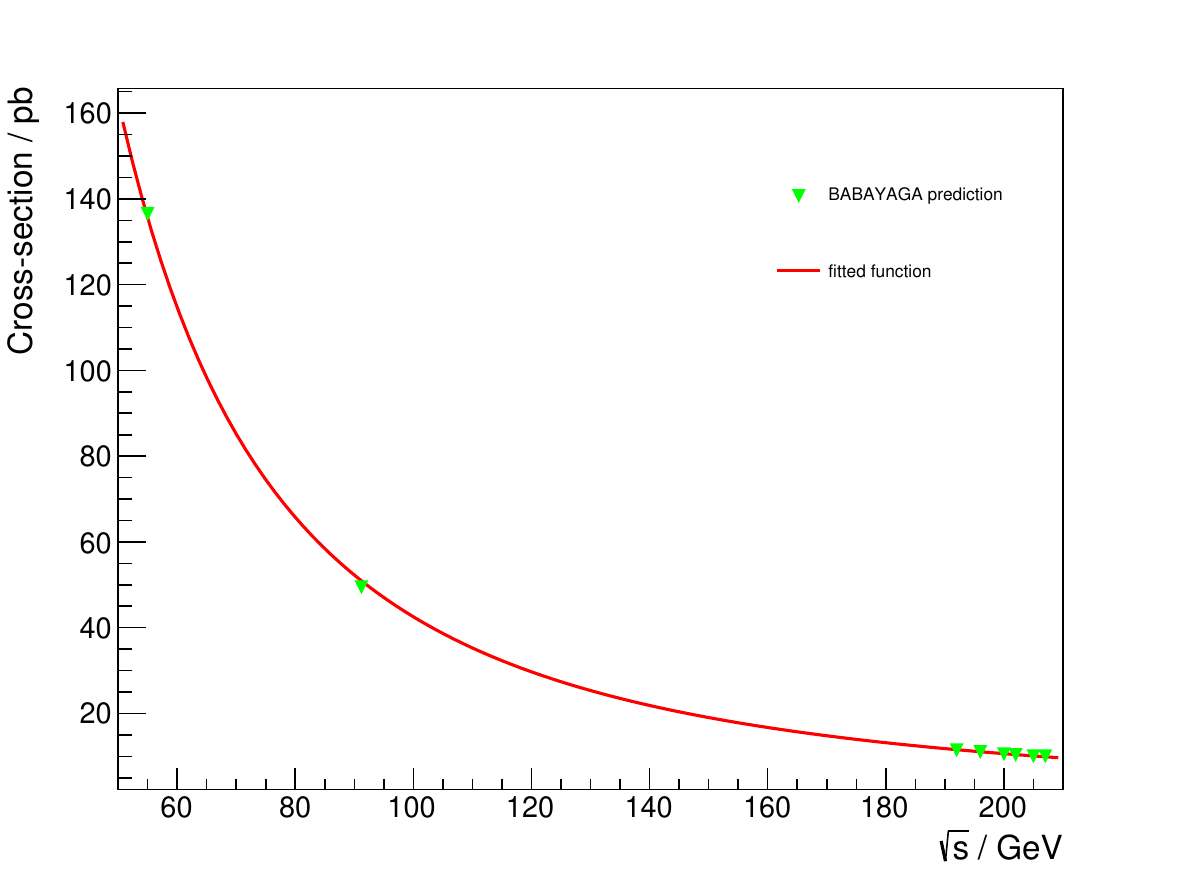}
\end{center}
\caption{ Total QED BabaYaga cross section of the $ \EEGG $ reactions. The points are the cross section and the red line is a fit to this points. }
\label{BHABAtot}
\end{figure}
 
The points represent the cross section. The red line is a fit using Eq. \ref{SIGMAQED11} 
and Eq. \ref{SIGMAQED22} to obtain an analytical expression for the total QED cross section.
The two fitting parameters are $ a = - 3.4 \times10^{4} $ and $ b = 4.8\times10^{7} $. 
 
\begin{align}
\label{SIGMAQED11}
\sigma (QED)_{tot} ^{L3}=\sigma(2\gamma )+korr\cdot  \sigma(3\gamma )\\
\label{SIGMAQED22}
korr=a\cdot \sqrt{s}+b
\end{align}

To compare all experimental data with the QED data, the sixth collaboration used the Monte Carlo QED generator \cite{Berends111}.
In this analysis, we used the generator BabaYaga@nlo \cite{BABAYAGA}. Therefore, it was necessary to look for 
discrepancies between the two QED generators.

The total cross section of the Monte Carlo QED generator \cite{Berends111} and the \cite{BABAYAGA} is equal 
within the statistical margin of error. For energies between $ \sqrt{s} $ = 91.0 GeV and 207.0 GeV, the mean 
value of \cite{Berends111} is $ \sigma _{tot}(QED) $ = 16.2 [pb] and of \cite{BABAYAGA} is 
$ \sigma _{tot}(QED) $ = 16.1 [pb], which corresponds to a total deviation of $ \Delta \sigma _{tot}(QED)$ = - 0.1 [pb]. 
This is approximately the statistical error of the QED generator \cite{Berends} of $ \Delta \sigma _{tot}(QED) $ = $\pm$ 0.1 [pb]. 
The total cross section $ \sigma _{tot}(QED) $ of \cite{BABAYAGA} above $ \sqrt{s} $ = 91.0 GeV is approximately 0.9 $ \% $ 
smaller than in \cite{Berends111}. In particular the $ \chi^{2} $ test was performed for both QED generators in 
chapter \ref{Baba Yaga 111}  and \ref{Berents 111}.

\subsubsection{Deviations from QED} 

If the QED is a fundamental theory, the experimental parameters of the $\EEGG$ reaction should be correctly 
calculated using the Monde-Carlo generator \cite{Berends111} or \cite{BABAYAGA} up to the Grand Unification 
Scale. All collaborations \cite{VENUS} to \cite{L33} investigated deviations from the QED reaction $\EEGG$ 
in the energy range from $\sqrt{s}$ = 55 GeV to $\sqrt{s}$ = 207 GeV.
Novel, non-QED phenomena could become visible at high energy scales, possibly below the Grand Unification scale.
An energy scale characterised by a truncation parameter $\Lambda$ GeV can serve as a threshold for QED breakdown
and for underlying new physics. Various models are discussed in \cite{Models1}.
Deviations from quantum electrodynamics (QED) with the reaction $\EEGG$ are investigated in a 
program initiated between 1991 and 2020 by ETH Zurich and USTC Hefei (University of Science 
and Technology of China). In this part of the investigation, we focus on a finite interaction length model in the 
reaction $\EEGG$.

\subsubsection{In the direct contact term annihilation a finite interaction length is a measure for the size of the electron} 
\label{Direct contact11}

The QED differential and total cross section of the $ \EEGG $ reaction are calculated under the condition that
the electron is point-like, without limited interaction length and coupled to the vacuum as shown
in Figure~\ref{CORR1} and Figure~\ref{CORR2}. To date, no experiments exist to test, in particular, the unlimited interaction length
of the electron up to the Planck scale. If, at a certain energy scale $ \Lambda $ between  $ ZERO <  \Lambda $  $ <  M_{P} $
a finite interaction length occurs in the $\EEGG$ reaction, this $\Lambda$ defines the size of an object 
in which the annihilation takes place. The size of the object can be calculated using the generalized 
uncertainty principle \cite{RES:RES1, RES:RES2, RES:RES3} or via the electromagnetic energy E 
and the wavelength $\lambda$ \cite{EMenergy} of the light emitted by the object.

It is well known that the wavelength $ \lambda $ of the gammas must be smaller or equal to the size of 
the interaction area. If the energy  $ \Lambda $ of the size of the interaction area is known, the 
frequency $\nu $ of the gammas is known via $ \Lambda=E=\hbar \times \nu$. This frequency $\nu $  is
 connected to the size of the object via the wave length $ \lambda $ 
to the equation $ \nu \times \lambda =c $. The energy scale $ \Lambda $ defines under these circumstances the size of the 
interaction area and in consequence the size of the electron involved in the annihilation area.

It is possible to construct several effective Lagrangians containing non-standard $\gamma e^+ e^-$ or
$\gamma \gamma e^+ e^-$ couplings, which $U(1)_{em}$ are gauge-invariant and differ only in their dimensions \cite{Models1}.
In the lowest-order effective Lagrangian, this reaction contains operators of dimensions 6, 7, and 8 \cite{Models2}.
In the further discussion, we concentrate on the simplest operator of dimension 6, with the effective Lagrangian of Eq.  \ref{DIR22}.

\begin{align}
\label{DIR22}
{\Lagr}_{6}=i\bar{\Psi }\gamma _{\mu }({\vec{D}}_{\nu }\Psi )(g_{6}F^{\mu \nu }+\tilde{g}_{6}\tilde{F}^{\mu \nu })
\end{align} 

The coupling constant $ g_{n}=\sqrt{4\pi }/\Lambda ^{(n-4)} $  $  ( n = 6 ) $, is related to the mass scale $ \Lambda $,
$ D_{\mu }=\partial _{\mu }-ieA_{\mu }$ is the QED covariant derivative, and $ \tilde{F}^{\mu \nu } $ is the dual of the electromagnetic tensor
$ \tilde{F}^{\alpha \beta }=\frac{1}{2}\varepsilon ^{\alpha \beta \mu \nu }F_{\mu \nu } $. The differential cross section 
corresponds to $ {\Lagr}_{6} $ is shown in Eq. \ref{DIR3}.

\begin{align}
\label{DIR3}
\left ( \frac{d\sigma }{d\Omega } \right )_{T}&=\left ( \frac{d\sigma }{d\Omega } \right )_{QED}\left [ 1+\delta _{new} \right ] \\
&=\left ( \frac{d\sigma }{d\Omega } \right )_{QED}\left [ 1+\frac{s^2}{2\alpha }\left ( \frac{1}{\Lambda ^4} +
\frac{1}{\tilde{\Lambda ^4}}\right )(1-cos^2\theta ) \right ] \nonumber
\end{align}

We use $ \Lambda =\tilde{\Lambda }=\Lambda _{6} $, higher order terms like $ \Lambda _{7} $ or $ \Lambda _{8} $ of 
$ \delta_{new} $ are omitted.  To search for a deviation of QED, it is common to use $ \chi^{2} $ tests. 
This test compares the QED cross section to the experimental measured cross section. In the test, 
the QED cross section is modified by a non QED direct contact term threshold energy scale $ \Lambda $ after Eq. \ref{DIR3}. 

If the $ \chi^{2} $ test indicates a minimum of a finite threshold energy scale $ \Lambda $, the energy of the 
truncation parameter, it defines, via  the two equations $ \Lambda=E=\hbar \times \nu $ and $ \nu \times \lambda =c $ 
a finite size of the area where the $ e ^ {+} $ $ e ^ {-} $ annihilation must occur. 
For $ \lambda=r_{e} $ is this a measure for size of the electron shown in Eq. \ref{DIR11}  where $ \hbar $ is 
the Planck constant and c is the speed of light.

\begin{align}
\label{DIR11}
r_{e}=\frac{\hbar c}{\Lambda }
\end{align}

Eq. \ref{DIR11}  is generic, the calculation using the generalised uncertainty principle generates the same equation.
It is interesting to notice that in Eq. \ref{DIR11}  for $ \Lambda \rightarrow \infty $ the size of the object will be $ r_{e}\rightarrow 0 $.
In consequence the point-like QED would be correct to infinite energies.

\subsubsection{The measurement of the total cross section $ \sigma _{tot} $ }

The $ \EEGG $ reaction initiates in a storage of $ e^+ e^- $ ring accelerators, a very clean background 
free signal of $  \gamma $'s in the detectors.  For example, Figure~\ref{GAMMA1} shows an event from 
LEP in the L3 detector. The position and energy storage of an $ \EEGG $ event perpendicular to the $ e^+ e^- $ beam axis in 
the electromagnetic BGO calorimeter is shown. The charge-sensitive detectors of the inner tracker 
detectors, the outer hadron calorimeter, and the muon chambers are free of any signal \cite{gamma_L3}.


\begin{figure}[htbp]
\vspace{-5.0mm}
\begin{center}
 \includegraphics[width=10.0cm,height=10.0cm]{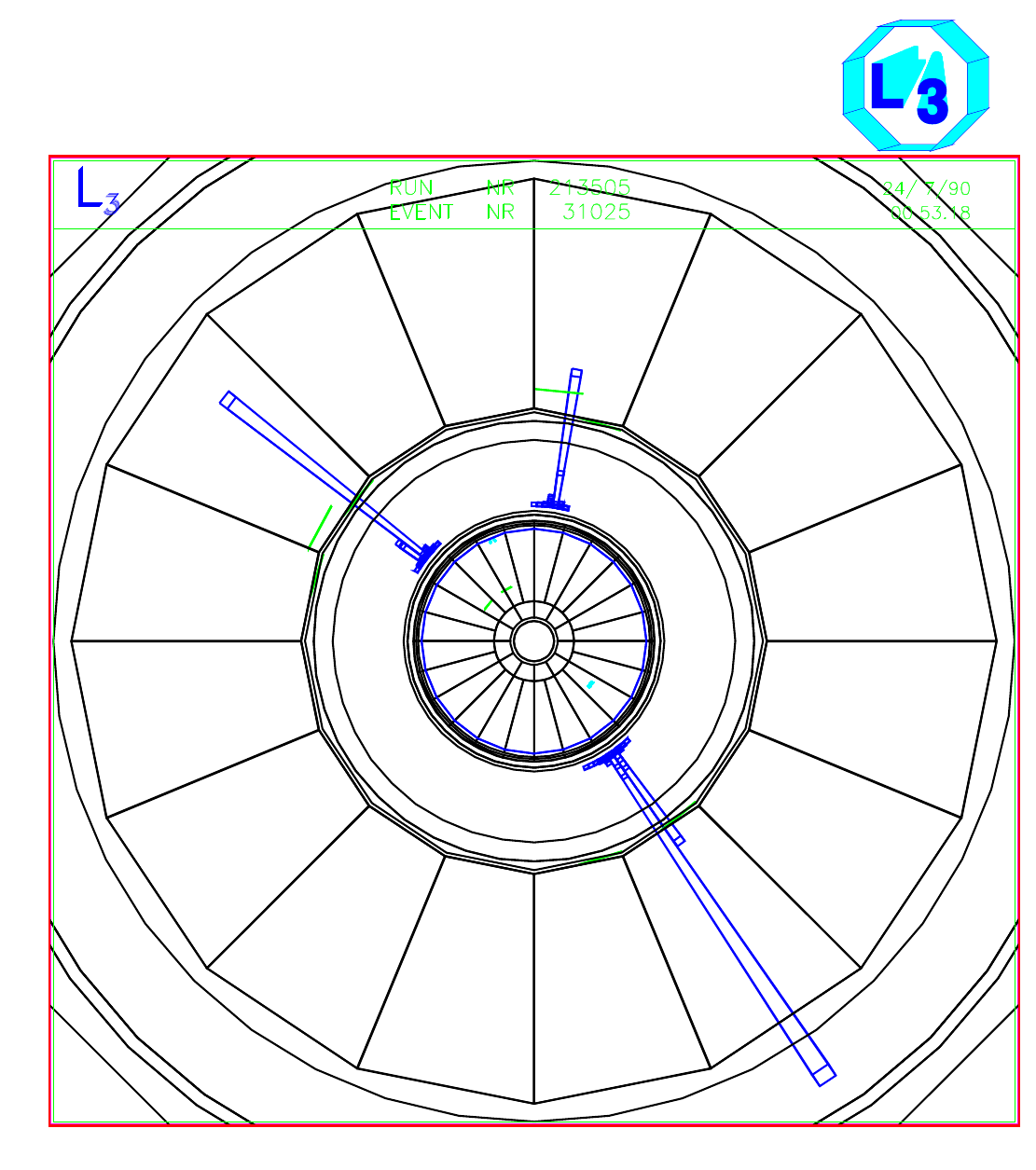}
\end{center}
\caption{ A typical $ \EEGG $ event ( Blue columns ) in the L3 detector at CERN \cite{gamma_L3}}
\label{GAMMA1}
\end{figure}

The total cross section can be described as a function of the number of events $ N $ in an angular range of 
the scattering angle  $ 0 < | cos \theta | < | cos \theta |_{max} $, the luminosity $ L $ and the 
efficiency $ \varepsilon $ of the detector for $ \EEG $ events Eq. \ref{SIGMAtota}.

\begin{align} 
\label{SIGMAtota}
\sigma _{tot}=\frac{N}{\varepsilon L}
\end{align} 

All six collaborations measured the total cross section $ \sigma _{tot} $ at different energies $ \sqrt{s} $.
To perform a total $ \chi ^2 $ fit of all experimental data, a model is needed to sum over all the information
together in one total cross section at one $ \sqrt{s} $. According to Table~\ref{LUMI}, seventeen data sets exist 
from $ \sqrt{s} $ = 55 GeV to 207 GeV. Table~\ref{LUMI} shows the luminosities and the references that the total 
cross section is published. At LEP nine data sets exist measuring $ \sigma _{tot} $ from more 
as ONE detector. For eight data sets only ONE detector has measured $ \sigma _{tot} $.  A  total $ \chi ^2 $ test 
would have in total 17 degrees of freedom, and if more as one detector is involved in one energy $ \sqrt{s} $,  
the statistical error $ \Delta\sigma(stat) $ would decrease.

All the detectors of the six collaborations use different scattering angles  $ 0 < | cos \theta | < | cos \theta |_{max} $, 
luminosities $ L $ and the efficiencies $ \varepsilon $ for the $ \EEG $ events. In common all the collaborations used for 
 $ \chi ^2 $ tests the CERNLIB program MINUIT  \cite{MINUIT} or Monte Carlo programs using the same radiative 
correction as discussed in Figure~\ref{CORR1} and Figure~\ref{CORR2}. The Monte Carlo program generates event
number including different scattering angles  range $ 0 < | cos \theta | < | cos \theta |_{max} $  and efficiency $ \varepsilon $.
This fact allows to normalise the experimental data of the six collaborations to an imaginary L3 detector using the 
QED cross section of the different collaborations.

 \subsubsection{Calculation of $ \sigma (tot) $, $ \Delta \sigma (stat) $, ratio R(exp) and $ \Delta R(stat) $ 
                           for more than one detector normalised to L3.}  
 \label{Sigma-more-as-one}

At LEP, there are nine data sets exist measuring $ \sigma _{tot} $ from more as ONE detector. To calculate 
in the model normalised to L3, the total experimental cross section $ \sigma(exp)_{tot}^{(sum)} $ 
Eq. \ref{SIGMAtot4}  , the statistical error  $ \Delta\sigma(exp)_{tot}^{(sum)} Eq. $ \ref{SIGMAtot5} , 
the ratio  R(exp) = $ R^{sum } $  Eq. \ref{SIGMAtot6}  and the statistical error of this ratio 
$ \Delta R(stat) $ = $ \Delta R(stat)^{(sum)} $ Eq. \ref{SIGMAtot8}, the following information 
is required in the next three steps, in addition to the corresponding input parameters:

First Step: To calculate the equations in the normalised L3 model it is first necessary to introduce,
the total experimental cross section $ \sigma(exp)_{tot}^{(det_i)} $ [pb] from VENUS, TOPAS, ALEPH, DELPHI, OPAL 
and total L3-N(exp) event rate shown Table~\ref{EXPtot},  the total QED cross section $ \sigma(QED)_{tot}^{(det_i)} $ [pb] 
from VENUS, TOPAS, ALEPH, DELPHI, OPAL and total L3-$ N(QED)^{L3} $ event rate from Table~\ref{QEDtot} and
the luminosity used from the VENUS, TOPAS, ALEPH, DELPHI, L3 and OPAL experiment shown in Table~\ref{LUMI} .
Including the information from Table~\ref{EXPtot}, Table~\ref{QEDtot} and Table~\ref{LUMI} the model first 
normalises the experimental total cross section of a detector$_{i} $ $ \sigma(exp)_{tot}^{(det_i)} $  
at one energy $ \sqrt{s} $ to the L3-normalized experimental total cross section 
$ \sigma(exp)_{tot}^{(L3-det_i)} $ Eq. \ref{SIGMAtot1}  via the total QED cross section of 
L3 $ \sigma(QED)_{tot}^{(L3)} $ to the total QED cross section of the 
detector$_{i} $ $ \sigma(QED)_{tot}^{(det_i)} $ at $ \sqrt{s} $ under investigation.
For the detailed calculation,  $ \sigma(exp)_{tot}^{(det_i)} $ is used for the numerical values summarised in table ~\ref{EXPtot},
including the statistical error $\Delta\sigma (exp)_{tot}^{det}$.
The values of $ \sigma(QED)_{tot}^{(det_i)} $ are summarised in Table~\ref{QEDtot}. The
value of $ \sigma(QED)_{tot}^{(L3)} $ is calculated using Eq. \ref{SIGMAQED11}  , Eq. \ref{SIGMAQED22} , and it is 
also listed in Table~\ref{QEDtot} for simplicity. \\
 
In the second step, the model introduces an L3-QED normalised efficiency $\varepsilon^{L3}$, which includes the number of L3-QED events
$N(QED)^{L3}$ of the total cross section at the investigated $\sqrt{s}$ energy, the total QED cross section
$\sigma(QED)_{tot}^{(L3)}$, and the L3 luminosity $L^{(L3)}$ at $\sqrt{s}$ Eq. \ref{SIGMAtot2}.
The numerical values of $N(QED)^{L3}$ are taken from \cite{L3B}. From this reference, 
the numbers from page 33, Table 2, were taken: ``The expected 2$\gamma$ events'' that 
agree with the QED Monte Carlo generators. The 3$\gamma$ configuration was
implicitly integrated into the generator with the parameter $p_{11}=final-\gamma-state=gg$ and
$p_{55}=Accollinear-angle-acoll.-\beta =90.0\left [deg \right ]$. The values for L3 luminosity $ L^{ (L3) }$ 
 at the $ \sqrt{s} $ energy under investigation are taken from Table~\ref{LUMI}.\\
 
In the third step, the to-L3 normalised total QED cross section of the detector detector$_{i} $ $ \sigma(QED)_{tot}^{(det_i)} $ at $ \sqrt{s} $ 
under investigation Eq. \ref{SIGMAtot1}, the L3 QED efficiency $ \varepsilon^{L3} $ Eq. \ref{SIGMAtot2} 
and the luminosity of the different detectors $ L^{(det)}_{i} $ open the possibility to calculate the experimental 
counting rate of every detector normalised to L3 of this detector $ N(exp)_{i}^{L3} $ Eq. \ref{SIGMAtot3} .
The numerical values for $ L^{(det)}_{i} $ are summarized in Table~\ref{LUMI}. In this Table included
are the important references for Table~\ref{EXPtot} and Table~\ref{QEDtot}.\\

By including the detailed numerical numbers from Eq. \ref{SIGMAtot1}, Eq. \ref{SIGMAtot2} and Eq. \ref{SIGMAtot3} , 
it is possible to sum over $ N(exp)_{i}^{L3} $ and $ L^{(det)}_{i} $ to calculate for every $ \sqrt{s} $ under investigation
the total summed cross section $ \sigma(exp)_{tot}^{(sum)} $ Eq. \ref{SIGMAtot4}  
and the statistical error $ \Delta\sigma(exp)_{tot}^{(sum)} $ Eq. \ref{SIGMAtot5}.  
To calculate the statistical error $ \Delta R(stat)^{(sum)} $ Eq. \ref{SIGMAtot8} , the most conservative approach is used
via the maximum possible error $ \Delta Max R $ Eq. \ref{SIGMAtot7} . 
  
\begin{align}
\label{SIGMAtot1}
\sigma(exp)_{tot}^{(L3-det_i)} =\sigma(exp)_{tot}^{(det_i)}\frac{\sigma(QED)_{tot}^{(L3)}}{\sigma(QED)_{tot}^{(det_i)}}\\
\label{SIGMAtot2}
\varepsilon^{L3}=\frac{N(QED)^{L3}}{(\sigma(QED)_{tot}^{(L3)}L^{(L3)})}\\ 
\label{SIGMAtot3}
N(exp)_{i}^{L3}= \sigma(exp)_{tot}^{(L3-det_{i})}\varepsilon^{L3}L^{(det)}_{i} \\
\label{SIGMAtot4}
\sigma(exp)_{tot}^{(sum)} =\frac{\sum_{i}N(exp)_{i}^{L3}}{[\varepsilon^{L3}\sum_{i}L_{i}^{det}]}\\
\label{SIGMAtot5}
\Delta\sigma(exp)_{tot}^{(sum)} =\frac{\sqrt{\sum_{i}N(exp)_{i}^{L3}}}{[\varepsilon^{L3}\sum_{i}L_{i}^{det}]}\\
\label{SIGMAtot6}
R^{sum}=\frac{\sigma(exp)_{tot}^{(sum)}}{\sigma(QED)_{tot}^{(L3)}}\\
\label{SIGMAtot7}
\Delta Max R = \frac{\sigma (exp)_{_{tot}}^{sum}+\Delta \sigma(exp)_{tot}^{sum}}{\sigma(QED)_{tot}^{L3} }\\
\label{SIGMAtot8}
\Delta R(stat)^{(sum)}=\Delta Max R -R^{sum}
\end{align}
  
Not all of the various collaborations mark for a systematic error for $\sigma(QED)_{tot}^{(det_i)}$. 
Therefore, equations $\ref{SIGMAtot5}$ and $\ref{SIGMAtot8}$ only contain the statistical error, which 
dominates in this analysis. All collaborations use a Monte Carlo generator to produce many millions of 
events to compute the differential and total QED cross sections $e^{+}e^{-}\to\gamma\gamma$, respectively. 
This allows us to neglect the systematic error resulting from the Monte Carlo simulation 
compared to the statistical error of the data. In the actual analysis, we investigated 
one more possible potentially significant systematic error. We examined two different Monte 
Carlo generators and used different energy ranges for the $\chi^2$ test to investigate this systematic error. In the 
following sections, we will see that the interaction radii of all four different tests lie within the range of the 
statistical error ( Table~\ref{Comparison} ). Interestingly, even the completely independent analyses of different 
cross sections agree with respect to this interaction radius (Table~\ref{Comparison}). This proves that the 
systematic error does not affect the result of this analysis.

 \subsubsection{ Calculation of $ \sigma (tot) $, $ \Delta \sigma (stat) $, ratio R(exp) and $ \Delta R(stat) $ for one detector normalised to L3.} 
\label{Sigma-one-detector}
    
If only ONE detector contributes to the calculation of equations \ref{SIGMAtot9} to \ref{SIGMAtot13},
the input data and detailed information are again contained in Table \ref{EXPtot}, Table \ref{QEDtot}, 
and Table \ref{LUMI}. In this case, there are eight data sets for $\sigma_{tot}$. 
  
Under these conditions, the total experimental cross section normalized to L3 $ \sigma (exp)_{tot}^{(single)} $ Eq. \ref{SIGMAtot9}  
is like  Eq. \ref{SIGMAtot1} a function of the total experimental cross section of one detector $ \sigma (exp)_{tot}^{det} $ 
and the ratio of the total QED L3 cross section $ \sigma (QED)_{tot}^{L3} $ to the total QED detector cross 
section $ \sigma (QED)_{tot}^{det} $. It is similarly  possible to calculate from the experimental 
error of detector $ \Delta\sigma (exp)_{tot}^{det} $ via the ratio $ \sigma (QED)_{tot}^{L3} $ to  
$ \sigma (QED)_{tot}^{det} $ the to-L3 normalised error  $ \Delta\sigma (exp)_{tot}^{(single)} $ Eq. \ref{SIGMAtot10}. 
The ratio R $^{single} $ Eq. \ref{SIGMAtot11}  is a function of Eq. \ref{SIGMAtot9} and the total 
QED L3 cross section $ \sigma (QED)_{tot}^{L3} $. The $ \sigma (QED)_{tot}^{L3} $ 
cancels in Eq. \ref{SIGMAtot11} , which replaces $ \sigma (exp)_{tot}^{(single)} $ by $ \sigma (exp)_{tot}^{det} $ and
$ \sigma (QED)_{tot}^{L3} $  by $ \sigma (QED)_{tot}^{det} $.
To calculate the statistical error $ \Delta R(stat)^{(single)} $ Eq. \ref{SIGMAtot13} for simplicity, 
the maximum value of $ \Delta Max R^{(single)} $ Eq. \ref{SIGMAtot12} is used to form the 
difference between $ \Delta Max R^{(single)} $ and $ R^{single} $. The sum of Eq. \ref{SIGMAtot9} 
and Eq. \ref{SIGMAtot10}  divided  by $ \sigma (QED)_{tot}^{L3} $ is shown in Eq. \ref{SIGMAtot12}.
    
\begin{align}
\label{SIGMAtot9}
\sigma (exp)_{tot}^{(single)}=\sigma (exp)_{tot}^{det}\frac{\sigma (QED)_{tot}^{L3}}{\sigma (QED)_{tot}^{det}}\\
\label{SIGMAtot10}
\Delta\sigma (exp)_{tot}^{(single)}=\Delta\sigma (exp)_{tot}^{det}\frac{\sigma (QED)_{tot}^{L3}}{\sigma (QED)_{tot}^{det}}\\
\label{SIGMAtot11}
R^{single}=\frac{\sigma (exp)_{tot}^{single}}{\sigma (QED)_{tot}^{L3}} =\frac{\sigma (exp)_{tot}^{det}}{\sigma (QED)_{tot}^{det}}\\
\label{SIGMAtot12}
\Delta Max R^{(single)} = \frac{\sigma (exp)_{_{tot}}^{single}+\Delta \sigma(exp)_{tot}^{single}}{\sigma(QED)_{tot}^{L3} }\\
\label{SIGMAtot13}
\Delta R(stat)^{(single)}=\Delta Max R^{single} -R^{single}
\end{align}

As discussed in chapter \ref{Sigma-more-as-one}, not all different collaborations mark for $ \sigma(QED)_{tot}^{(det)} $ 
a systematic error, for this reason Eq. \ref{SIGMAtot10} and Eq. \ref{SIGMAtot12}  include only the statistical error.

 \subsubsection{Numerical calculation of $ \sigma (tot) $, $ \Delta \sigma (stat) $, ratio R(exp) and $ \Delta R(stat) $.} 

Inserting the numerical values from Table~\ref{EXPtot}, Table~\ref{QEDtot} and Table~\ref{LUMI} in 
Eq. \ref{SIGMAtot1}  until  Eq. \ref{SIGMAtot13}  allows to calculate $ \sigma (tot) $, $ \Delta \sigma (stat) $, 
ratio R(exp) and $ \Delta R(stat) $ shown in  Table~\ref{table6}.

\begin{table}[H]    
\begin{center}
\caption{ Summary $ \sigma (tot) $, $ \Delta \sigma (stat) $ and ratio R(exp) and $ \Delta R(stat) $} 
\label{table6}
\begin{tabular}{ | l | l | l | }
  \hline
  $ \sqrt{s} $ [ GeV ] &\multicolumn{1}{ | c | }{ $ \sigma (tot) $ $ \Delta \sigma (stat) $ [pb] }&\multicolumn{1}{ | c | }{ R(exp) $ \Delta R(stat) $  }\\
  \hline
   55   & 124.7   $ \pm  $ 13.2    & 0.92  $ \pm $ 0.10 \\ \hline
   56   & 150.6   $ \pm  $ 9.7     & 1.15  $ \pm $ 0.07 \\ \hline
   56.5 & 141.6   $ \pm  $ 22.9    & 1.10  $ \pm $ 0.18 \\ \hline
   57   & 135.5   $ \pm  $ 10.8    & 1.07  $ \pm $ 0.09 \\ \hline
   57.6 & 125.3   $ \pm  $ 2.0     & 1.01  $ \pm $ 0.02 \\ \hline
   91   & 50.3    $ \pm  $ 0.9     & 0.99  $ \pm $ 0.02 \\ \hline
   133  & 26.5    $ \pm  $ 5.8     & 1.10  $ \pm $ 0.24 \\ \hline
   162  & 16.1    $ \pm  $ 2.4     & 0.98  $ \pm $ 0.15 \\ \hline
   172  & 15.6    $ \pm  $ 2.6     & 1.08  $ \pm $ 0.18 \\ \hline
   183  & 12.6    $ \pm  $ 0.3     & 0.99  $ \pm $ 0.03 \\ \hline
   189  & 11.8    $ \pm  $ 0.2     & 0.99  $ \pm $ 0.02 \\ \hline
   192  & 11.0    $ \pm  $ 0.5     & 0.95  $ \pm $ 0.04 \\ \hline
   196  & 11.3    $ \pm  $ 0.3     & 1.02  $ \pm $ 0.03 \\ \hline
   200  & 10.1    $ \pm  $ 0.3     & 0.95  $ \pm $ 0.03 \\ \hline
   202  & 10.1    $ \pm  $ 0.4     & 0.97  $ \pm $ 0.04 \\ \hline
   205  & 10.0    $ \pm  $ 0.3     & 0.99  $ \pm $ 0.03 \\ \hline
   207  & 9.7     $ \pm  $ 0.2     & 0.98  $ \pm $ 0.02 \\ \hline
 \end{tabular}
\end{center}
\end{table}
     
The  $ \sigma (tot) $ values from $ \sqrt{s} $ = 55 GeV to 207 GeV together with  $ \Delta \sigma (stat) $ of Table~\ref{table6} 
compared to the total QED cross section $  \sigma(QED)_{tot}^{(L3)} $ Eq. \ref{SIGMAQED11}  is displayed 
in Figure~\ref{SIGMAtot}. 

\newpage


\begin{figure}[htbp]
\vspace{0.0mm}
\begin{center}
 \includegraphics[width=16cm,height=13.0cm]{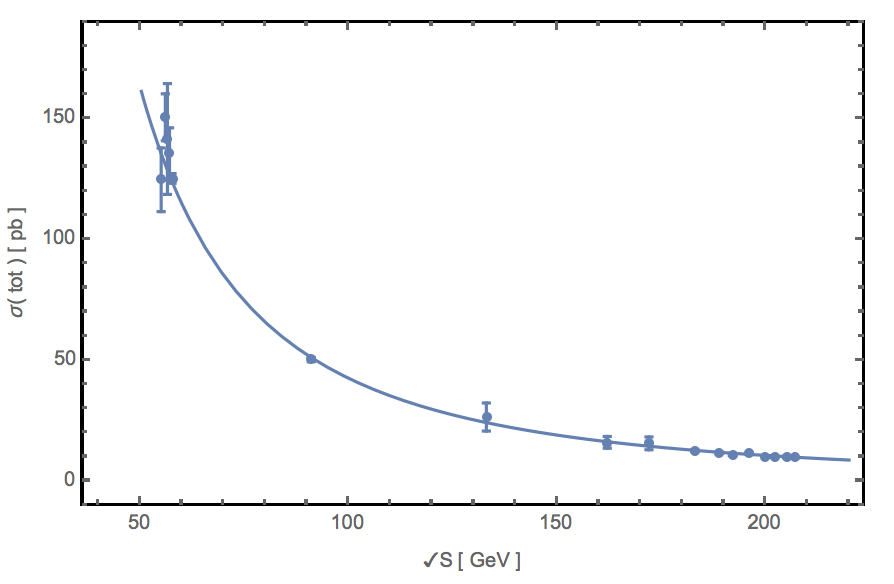}
\end{center}
\caption{  The $ \sigma $(tot)  of the $ \EEGG $ reaction of all detectors as a function of center-of-mass energy $ \sqrt{s} $. The data (points) are compared to QED prediction (solid line).}
\label{SIGMAtot}
\end{figure}

 Figure~\ref{SIGMAtot} shows a good agreement between the experimental measured values $ \sigma(tot) $ including the
 statistical error $ \Delta \sigma (stat) $ and the total QED cross section $ \sigma(QED)_{tot}^{(L3)} $ in the range of sensitivity. 
 
To search for deviation between measured values $ \sigma(tot) $ and the total QED cross section 
$ \sigma(QED)_{tot}^{(L3)}$, the graphic of Figure~\ref{SIGMAtot} is not sensitive enough 
because the deviation is on the $ \% $ level. A more sensitive graphic Figure~\ref{Ratio11} is plotted to display the 
ratio  $ \sigma $ (tot,measured.)/ $ \sigma$(tot,QED)  of the $ \EEGG $ reaction of all detectors as a function of  
center-of-mass energy $\sqrt{s}$. At $ \sqrt{s}>$ 90 GeV, a negative interference between 
measured values and QED of some percent is visible.   
  
 
 \newpage
 
\begin{figure}[htbp]
\vspace{0.0mm}
\begin{center}
 \includegraphics[width=16.0cm,height=11.0cm]{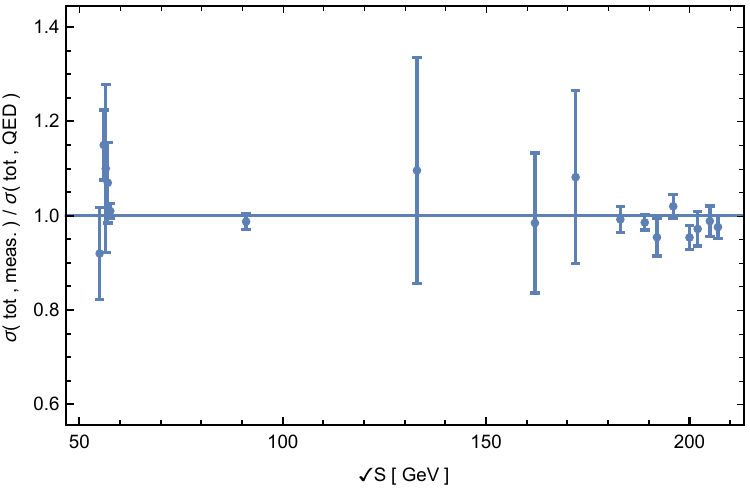}
\end{center}
\caption{  Ratio $ \sigma $(tot,measured)/$ \sigma$(tot,QED)  of the $ \EEGG $ reaction of all 
detectors as a function of  center-of-mass energy $ \sqrt{s} $ .The experimental data (points) and QED prediction (solid blue line).}
\label{Ratio11}
\end{figure} 

Furthermore, it is evident that the statistical errors for the 8 data points are significantly larger than 
those for the 9 data points. The reduction in statistical errors is due to the fact that at LEP, the nine 
data sets measure $\sigma_{tot}$ from more than one detector Eq. \ref{SIGMAtot5} ), while the eight 
data sets measure $\sigma_{tot}$ from only one detector. For example, in the model used, 
at $\sqrt{s}$ = 207 GeV, the statistical error $\Delta\sigma(stat)$ = 0.2 pb, compared to the statistical  
error of L3 $\Delta\sigma(stat)$ = 0.34 pb at the same energy $\sqrt{s}$ = 207 GeV \cite{L3B}.
This corresponds to a reduction of approximately 42\%.

\subsubsection{ Search for finite size of electron using a $ \chi^{2} $ test  of the  ratio \\ $ \sigma $ (tot,meas.) /  $ \sigma$(tot,QED) . }

The signal for the finite size of the electron is weak, not visible in the graphic of the total experimental cross section compared with the QED
total cross section in Figure~\ref{SIGMAtot}. A deviation of approximately some $ \% $ is discernible in the graphic of the ratio 
$ \sigma $ (tot,meas.) /  $ \sigma$(tot,QED)  of the $ \EEGG $ reaction in Figure~\ref{Ratio-1}. 
Based on the sensitivity difference between experiment and QED, a $ \chi^{2} $  test is performed 
on this ratio Eq. \ref{CHI-1} to find a minimum in $ \chi^{2} $.

\begin{align}
\label{CHI-1}
\chi ^{2}&=\sum_{i}\left \{ \frac{R(E_{i};exp)-R(E_{i};QED;\Lambda )}{\Delta R(E_{i};exp)} \right \}^{2}
\end{align} 

The ratio of the experimental data $ R(E_{i};exp) $ is in accordance with Table~\ref{table6}, the 
ratio of $ \sigma $ (tot,meas.) /  $ \sigma$(tot,QED) at a $ \sqrt{s} $ energy $ E_{i} $. The statistical 
error $ \Delta R(E_{i};exp) $ is the error at $ \sqrt{s} $ energy $ E_{i} $, in accordance with 
Table~\ref{table6}. The term for finding a deviation from the finite size of the electron is:
$R(E_{i};QED;\Lambda)$ Eq. \ref{Ratio-11} . The fitting parameter $(1/\Lambda_{6})^{4}$ is 
taken into account and acts as a function of the interaction quantity of the electron $r_{e}$ Eq. \ref{DIR11}.

\begin{align}
\label{Ratio-11}
R(E_{i};QED;\Lambda )=\frac{\int_{\Omega _{min}}^{\Omega _{max}}\left ( \frac{d\sigma }{d\Omega } 
\right )_{(E_{i};QED;\Lambda )}d\Omega }{\int_{\Omega _{min}}^{\Omega _{max}}\left ( \frac{d\sigma }{d\Omega } \right )_{(E_{i};QED)}d\Omega }
\end{align}

For the entire program of the $ \chi^{2} $ test, it is essential to use the fitting parameter 
$ \left ( 1/\Lambda _{6} \right )^{4} $, which is sensitive to the theoretical calculation of 
a deviation from the differential cross section of QED for positive and negative interference.
A parameter $( \pm \Lambda _{6} )$ would test ONLY the positive interference because the term  $ (\Lambda _{6})^{4} $ 
cut out the negative part. This problem is visible in Eq. \ref{Ratio-11}  and in particular Eq. \ref{NomRatio-11} . The parameter $ \Lambda _{6} $ would 
keep the sign in front of $ (1-cos^{2}\theta ) $ always positive independent of the $ \pm $ sign from $ \Lambda _{6} $. 
As a consequence, it always happens that 
$(d\sigma /d\Omega )_{(E_{i};QED;\Lambda )}> (d\sigma /d\Omega)_{(E_{i;QED})} $ and
$ R(E_{i};QED;\Lambda )>1 $ . The $ \chi^{2} $ test would NOT be able to find values $ R(E_{i};QED;\Lambda )<1 $.
For this reason a negative interference could NOT be detected. It is necessary for positive - negativ interference
to allow in Eq. \ref{NomRatio-11} in front of the $(1-cos^{2}\theta )$ in the $ \chi^{2} $ test a $ \pm $ sign 
independently of the constant $  \alpha $ like Eq. \ref{DIR333}.

\begin{align}
\label{NomRatio-11}
&\int_{\Omega _{min}}^{\Omega _{max}}\left ( \frac{d\sigma }{d\Omega } \right )_{(E_{i};QED;\Lambda) }d\Omega =\\
&\int_{\Omega _{min}}^{\Omega _{max}} \left [ \left ( \frac{d\sigma }{d\Omega }\right )_{(E_{i};QED)}(1+\frac{s^{2}}
{\alpha \Lambda _{6}^{4}}(1-cos^{2}\theta )) \right ]d\Omega \nonumber
\end{align}

The integrals Eq. \ref{Ratio-11} and  Eq. \ref{NomRatio-11}  included the differential QED 
cross section $ \left ( d\sigma /d\Omega  \right )_{(E_{i};QED)} $. 
To test the angular contribution part of $ R(E_{i};QED;\Lambda ) $ function, it is possible to
 integrate Eq. \ref{Ratio-11}  over the  $ \sqrt{s} $ and use only the contribution of the angular 
 distribution of the direct contact term Eq. \ref{DIR3} analogous Eq. \ref{Ratio-k} .

\begin{align}
\label{Ratio-k}
R(E_{i};QED;\Lambda )=\\ \nonumber
\frac{\int_{\Omega _{min}}^{\Omega _{max}}\left ( \frac{d\sigma }{d\Omega } \right )_{(E_{i};QED)}\left [ 1+\frac{s^{2}}{\alpha \Lambda _{6}^{4}}
(1-cos^{2}\theta ) \right ]d\Omega}{\int_{\Omega _{min}}^{\Omega ^{max}}\left ( \frac{d\sigma }{d\Omega } \right )_{(E_{i;QED})}d\Omega }=\\
\frac{k_{1}\int_{\Omega _{min}}^{\Omega ^{max}}\left [ 1+\frac{s^{2}}{\alpha \Lambda _{6}^{4}}(1-cos^{2}\theta ) \right ]d\Omega }
{k_{2}\int_{\Omega _{min}}^{\Omega ^{max}}d\Omega } \nonumber
\end{align}

The energy contribution to $ R(E_{i};QED;\Lambda ) $ is in the range of $ \% $ level, which opens the possibility 
to investigate the $ \chi^{2} $ test under the assumption that the constant $ k_{1} $ equals approximately $ k_{2} $.

\subsubsection{ Numerical calculation of the  $ \chi^{2} $ tests }

To investigate the sensitivity of the $\chi^{2} $-test to the Monte Carlo generator 
BabaYaga@nlo \cite{BABAYAGA} and the generator \cite{Berends}, a separate 
numerical calculation was performed for both tests. Additionally, an approximation 
was used to examine solely the impact of the direct contact term 
Eq. \ref{DIR3} in the $\chi^{2} $-test. The mathematical details of the various 
possible calculations of the significance and error of the 
interaction radius $ r $ are examined in the Appendix \ref{Appendices11}.

\subsubsection{ Calculation of the $ \chi^{2} $ test with QED BabaYaga }
\label{Baba Yaga 111}

The generators under discussion generate numbers of events respecting to L3 parameters Eq. \ref{BabaYagaL3}).
The events per angular range are used to fit a differential cross section as a function of $ d\sigma /d\Omega \left [ nb/srad \right ]$ 
and $ \left | cos \theta  \right | $ . An example of such a differential cross section is shown in Figure (\ref{QEDdiff}) for 
$ \sqrt{s} $ = 90.2 GeV. The numerical parameters of this fit $ p_{1} $  to $ p_{6} $ defined in Eq. \ref{QEDfit}  
for the 17 $ \sqrt{s} $ energies from 55 GeV to 207 GeV are summarised in Table~\ref{diffpar} ( Upper part ). 
The $ \chi^{2} $ test on the ratio Eq. \ref{CHI-1} as a function of  $ 1/\Lambda ^{4} $ for the $ \EEGG $ reaction 
is displayed in Figure \ref{CHIBhabha}. The relevant center-of-mass energy ranges from $ \sqrt{s} $ = 55 GeV to 207 GeV, 
and the used QED predictions are from Monte Carlo generator BabaYaga@nlo.

\newpage

\begin{figure}[htbp]
\vspace{-3.0mm}
\begin{center}
 \includegraphics[width=15.0cm,height=12.0cm]{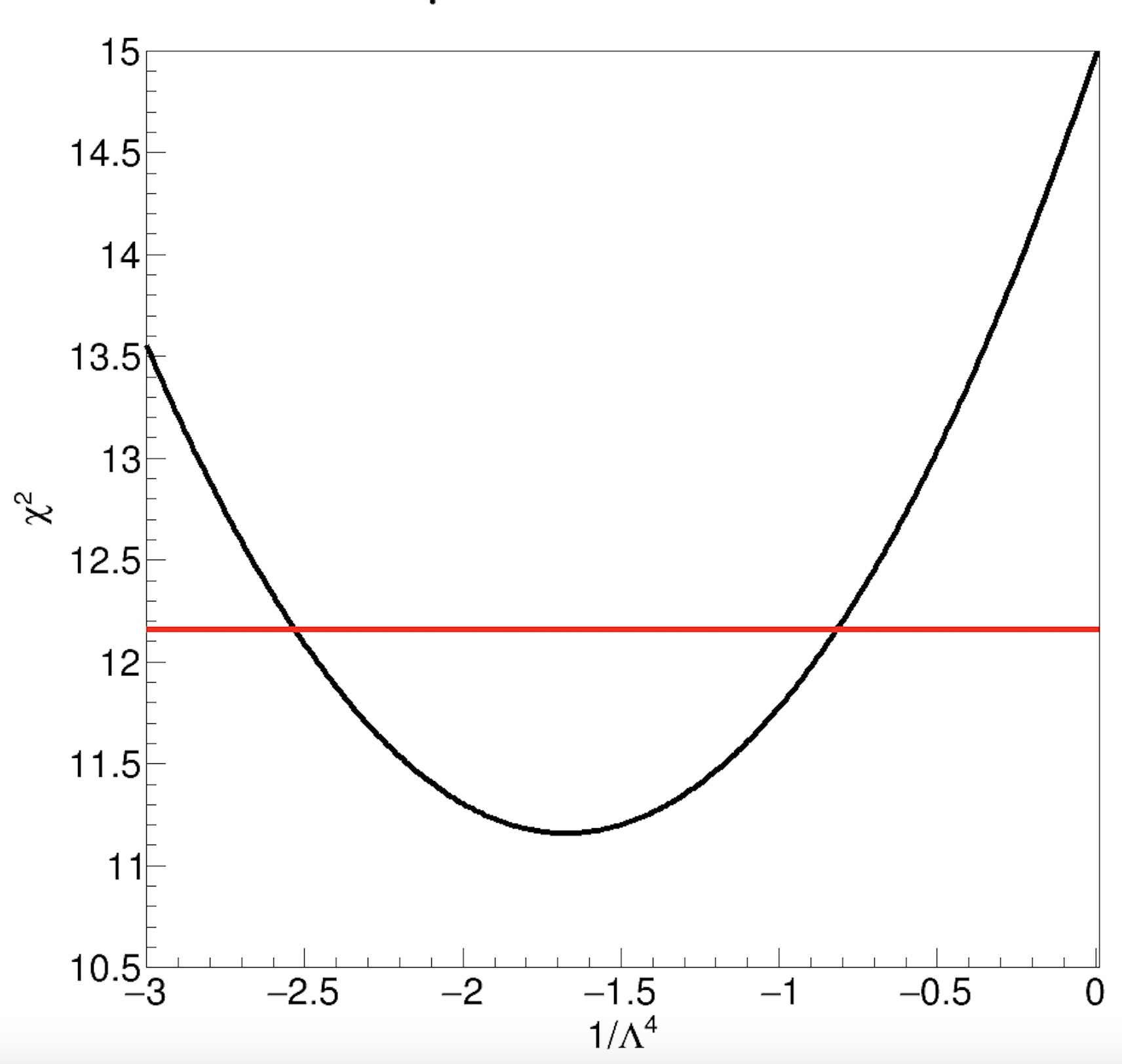}
\end{center}
\vspace{-2.0mm}
\caption{  $ \chi^{2} $ test on the ratio Eq. \ref{CHI-1}  of the $ \EEGG $ reaction from center-of-mass energy 
              $ \sqrt{s} $ = 55 GeV to 207 GeV using QED data from Monte Carlo generator BabaYaga@nlo. 
              The red line shows  $ \sigma $ limit. }
\label{CHIBhabha}
\end{figure} 

The minimum of the $ \chi^{2} $ test, including the error in Figure (\ref{CHIBhabha}) is 
$ 1/\Lambda ^{4} =-1.62_{-0.83}^{+0.84}\times 10^{-13}GeV^{-4} $.
The $ \chi^{2} $ value at the minimum is $ \chi^{2} $ = 11.21. The fit uses 17 degrees of freedom 
according to Table~\ref{table6}. The most important result of the $\chi^{2} $-fitting 
is that the fitting parameter $1/\Lambda^{4} $ has a negative sign. 
According to Figure \ref{Ratio-1}, the fitting is sensitive to the fact that the total QED cross section above 
approximately $\sqrt{s} $ = 92 GeV to 207 GeV is larger than the experimental total cross section. The 
discussed direct contact interaction term $ \delta _{new} $ Eq. \ref{DIR3}  has a negative sign. This indicates a 
negative interference of the direct contact interaction of the  $ {\EEG} $ reaction.

The significance of the fit $\chi^{2}$ is crucial in the results of the $\chi^{2}$ fitting. According to international 
rules, the physics community accepts $\sigma > 5$ as a discovery of new physical phenomena and 
$\sigma < 5$ as an indication of new physical phenomena.

A detailed mathematical calculation of significance is discussed in 
Appendix \ref{Sig444} ``Equations for the calculation of the significance of a $ \chi^{2} $ test''.
A first approximation for the significance of the $\chi^{2}$ fitting can be directly estimated from the 
error bars of the fitting: $\sigma = A/\Delta A$ = 1.62 / 0.83 = 1.95. Alternatively, the significance 
$\sigma$ can be calculated using the statistical probability function $p$ Eq. \ref{pvalue1}. The $p$ 
value of the $\chi^{2}$ test under consideration with 17 degrees of freedom (with a minimum
 $\chi^{2}$ value of 11.2) is $p$ = 0.15. According to the figure (\ref{pvaluesmall}), the 
 significance is approximately $\sigma$ = 1.5.
Similar to Appendix \ref{Sig444} , a detailed mathematical calculation of the error $ \Delta r_{e} $ of 
the interaction quantity $ r_{e} $ of the electron in the $ \chi^{2} $-test is discussed in the 
appendix \ref{deltar} ``The radius $ r_{e} $ and the error $ \Delta r_{e} $ of the interaction size of the electron in the  $ \chi^{2} $ test ''.

According to Eq. \ref{DIR11} ), the size of the interaction term is $ r_{e} =1.25 \times 10^{-17} $ [ cm ], and 
the error $ \Delta r_{e} = 0.16 \times 10^{-17}$  [ cm ]. The summary of all these results is given in Table~\ref{Baba}.

\begin{table}[H]  
 \begin{center}
 \caption{ Summary of $ \chi^{2} $ test with BabaYaga@nlo generator [Test 4.1.1].}
 \label{Baba}
 \begin{tabular}{ | l | c | r | r| }
 \hline
                    & $ (1/\Lambda )^{4} [GeV^{-4}] $     & $\sigma$ & $ ( r \pm \Delta r ) \times 10^{-17} [cm] $  \\  \hline
direct           & $ (-1.62 \pm 0.83)\times 10^{-13} $ & 1.95 &  \\  \hline
p - value      & $ (-1.62 \pm 0.83)\times 10^{-13} $ & 1.50 &  \\  \hline
$ r \pm \Delta r$    & $ (-1.62 \pm 0.83)\times 10^{-13} $ &  & $ (1.25\pm 0.16)$ \\  \hline
\end{tabular}
\end{center}
\end{table}
 
\subsubsection{ Calculation of the $ \chi^{2} $ test with QED generator \cite{Berends111}  }
\label{Berents 111}

The VENUS, TOPAS, OPAL, DELPHI, ALEPH and L3 collaborations used for the QED cross section
of the $ \EEGG $ reaction the generators \cite{Berends111}. As discussed in section \ref{Numerical calculation}, the deviation
between BabaYaga and \cite{Berends111} is approximately $ 0.9 \% $ under the condition that both generators used
the same L3 parameters Eq. \ref{BabaYagaL3}. Similar to the QED BabaYaga version, the events per angular 
range are used to fit a differential cross section as a function of  $ d\sigma /d\Omega \left [ pb/srad \right ]$ 
and $ \left | cos \Theta  \right | $ . The numerical parameters of this fit $ p_{1} $  to $ p_{6} $ are defined in Eq. \ref{QEDfit} 
for the 7 $ \sqrt{s} $ energies from 91.2 GeV to 200 GeV and summarised in the lower section of Table~\ref{diffpar}. 
Using QED predictions from Monte Carlo generator \cite{Berends111}, the $ \chi^{2} $ test on the ratio 
Eq. \ref{CHI-1} as a function of $ 1/\Lambda ^{4} $ for the $ \EEGG $ reaction is displayed in Figure (\ref{CHIZhao}).
The relevant center-of-mass energy ranges from $ \sqrt{s} $ = 91.2 GeV to 200 GeV.

\newpage
\begin{figure}[htbp]
\vspace{0.0mm}
\begin{center}
 \includegraphics[width=15.0cm,height=12.0cm]{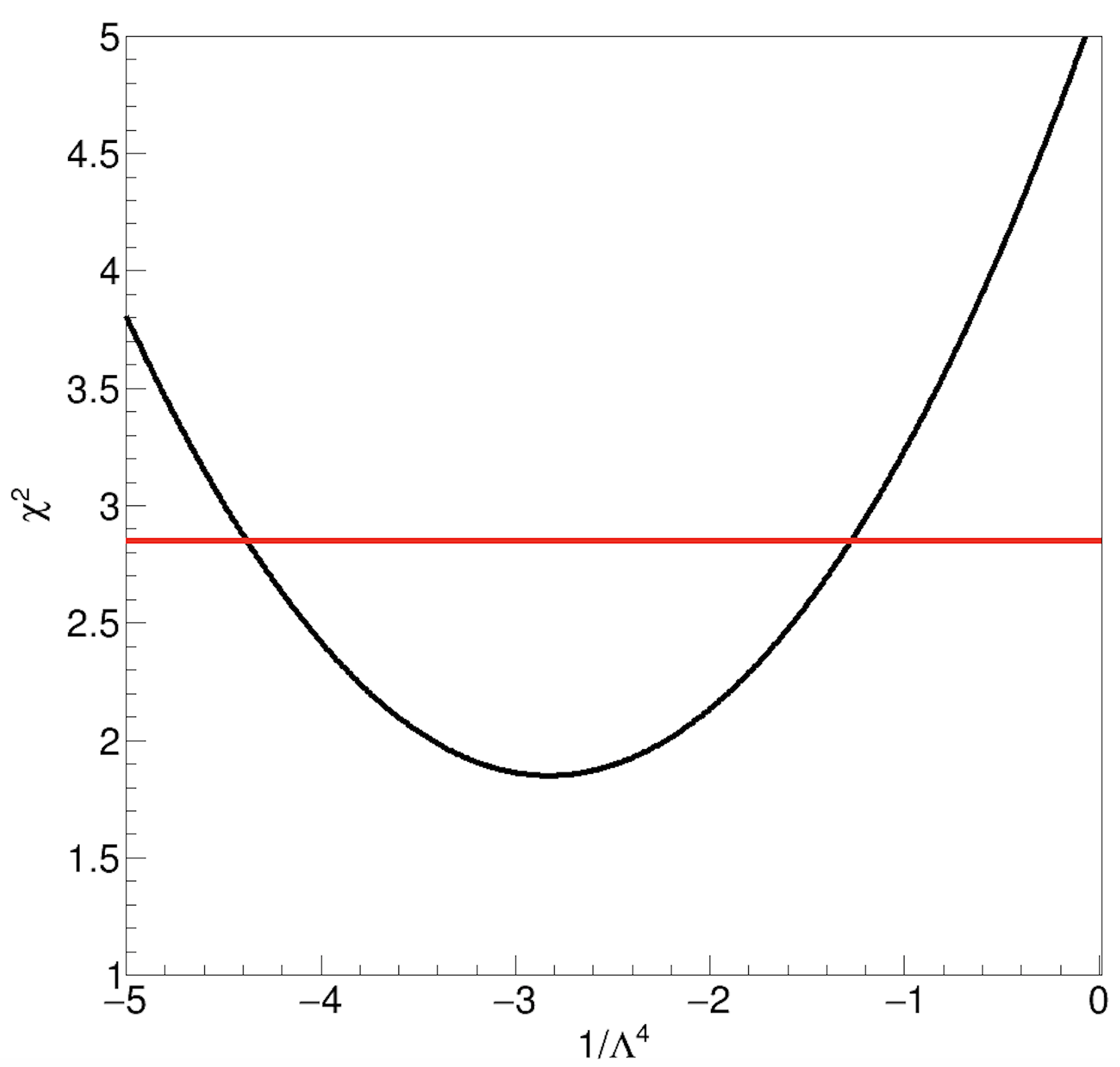}
\end{center}
\vspace{-3.0mm}
\caption{$ \chi^{2} $ test on the ratio Eq. \ref{CHI-1} of the $ \EEGG $ reaction from center-of-mass energy 
             $ \sqrt{s} $ = 91 GeV to 207 GeV using QED data from Monte Carlo generator \cite{Berends111}.
             The red line shows  one $ \sigma $ limit.  }
\label{CHIZhao}
\end{figure} 

The minimum of the $\chi^{2}$ test, including the error in Figure (\ref{CHIZhao}), is:
$ 1/\Lambda ^{4}=-2.83_{-1.55}^{+1.56}\times 10^{-13}GeV^{-4} $. The $\chi^{2}$ value at the minimum is:
$ \chi^{2} $ = 1.85. The fit uses 7 degrees of freedom.

Similar to the QED test with BabaYaga, $1/\Lambda^{4}$ is negative in the QED test \cite{Berends111}.
A first direct approximation for the significance is $\sigma = A/\Delta A$ = 2.83 / 1.55 = 1.83.
In a second approximation for calculating the significance $\sigma$, the formula for the statistical
probability function $p$ Eq. \ref{pvalue1} is used. The $p$ value of the discussed $\chi^{2}$ test 
with 7 degrees of freedom (with a minimum $\chi^{2}$ = 1.85) is: $p$ = 0.032. According to 
Figure \ref{pvaluesmall}, this corresponds to a significance of approximately $\sigma$ = 1.9.

According to Eq. \ref{DIR11} ), the size $ r $ of the interaction term is $ (1.4)\times 10^{-17} $ [ cm ].
The error $ \Delta r $ is $ \Delta r = 0.20 \times 10^{-17} $ [ cm ] from equation Eq. \ref{DIR18}. 
The summary of all these results is given in Table~\ref{Zhao1}. 

\begin{table}[H] 
 \begin{center}
 \caption{ Summary of $ \chi^{2} $ test with  generator \cite{Berends} [Test 4.1.2].} 
 \label{Zhao1}
 \begin{tabular}{ | l | c | r | r | }
 \hline
                    & $ (1/\Lambda )^{4} [GeV^{-4}] $     & $ \sigma $ & $ ( r \pm \Delta r ) \times 10^{-17} [cm] $  \\  \hline
direct           & $ (-2.83 \pm 1.55)\times 10^{-13} $ & 1.83 &  \\  \hline
p - value      & $ (-2.83 \pm 1.55)\times 10^{-13} $ & 1.90 &  \\  \hline
 $ r \pm \Delta r$   & $ (-2.83 \pm 1.55)\times 10^{-13} $ &  & $ (1.44\pm 0.20) $ \\  \hline
\end{tabular}
\end{center}
\end{table}

The  $ \chi^{2} $ tests performed with the QED BabaYaga and \cite{Berends111} generator request
the calculation of the differential QED cross section and fit this data with Eq. \ref{QEDfit} ).
Equation Eq. \ref{Ratio-k}  opens the possibility to test straight the direct contact term Eq. \ref{DIR3} .
The deviation between the measured R(exp) and R(QED) is on the $ \% $ level. This allows us  
to assume that the constant factors $ k_{1} $ and $ k_{2} $ are approximately the same, $ k_{1} \approx k_{2} $ . 

According to Table \ref{diffpar}, there are two R(QED) datasets: one with 7 degrees of freedom 
from $\sqrt{s}$ = 91.2 GeV to 200 GeV and one with 17 degrees of freedom from 
$\sqrt{s}$ = 55 GeV to 207 GeV. To check the significance $\sigma$ of the $\chi^{2}$ test, 
the full $\chi^{2}$ test is performed next for both datasets using the direct contact term.

\subsubsection{ Numerical calculation of the $ \chi^{2} $ test with 7 degrees of freedom in $ k_{1} \approx k_{2} $ approximation.}
\label{With 7 degrees}

The $ \chi^{2} $ test on the ratio Eq. \ref{CHI-1} ) as a function of $ 1/\Lambda ^{4} $ 
for the $ \EEGG $ reaction from center-of-mass energy $ \sqrt{s} $ = 91.2 GeV to 200 GeV is 
displayed in Figure (\ref{CHIk7d}). 

\newpage

\begin{figure}[htbp]
\begin{center}
 \includegraphics[width=15.0cm,height=12.0cm]{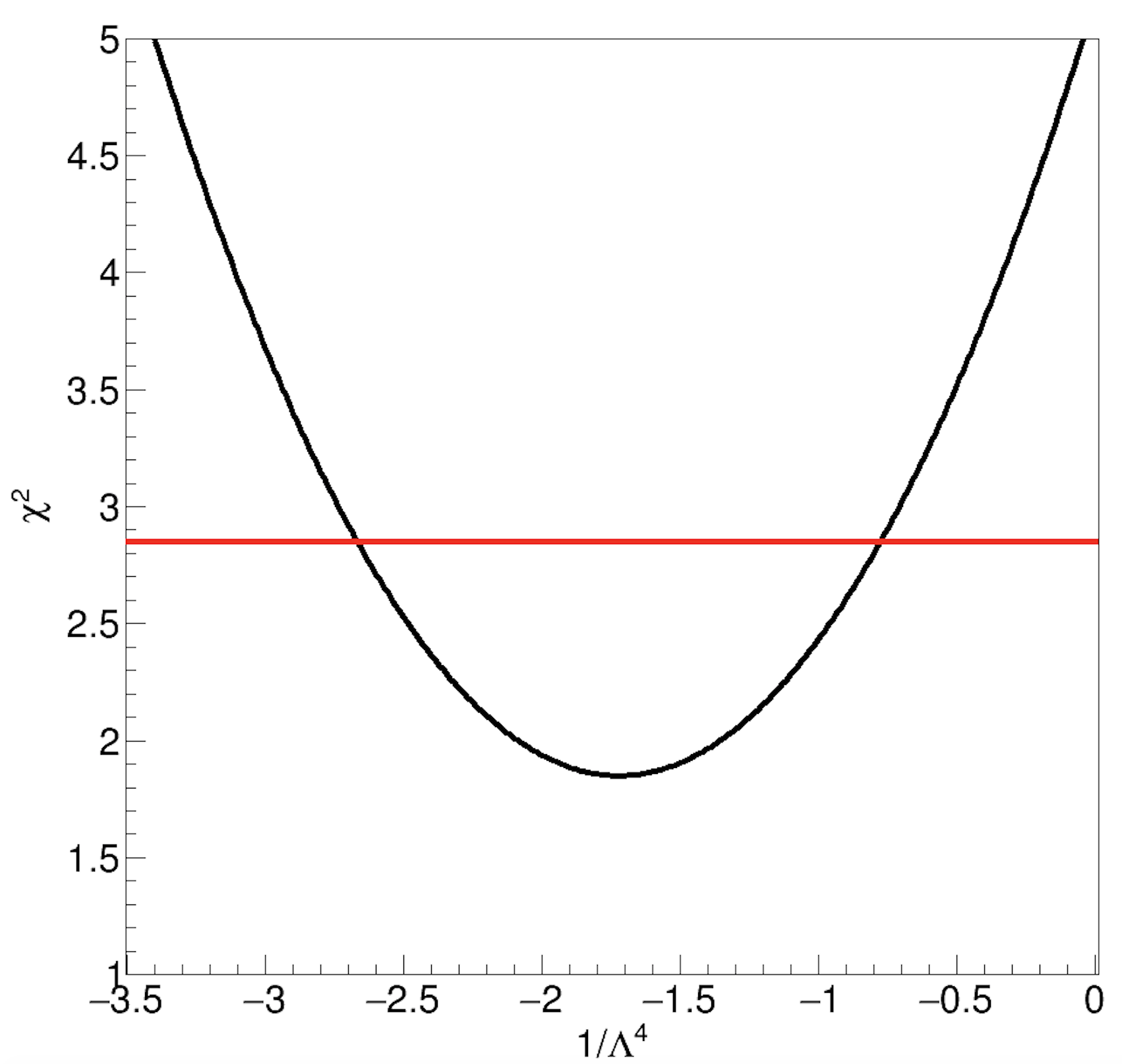}
\end{center}
\vspace{-4.0mm}
\caption{ $ \chi^{2} $ test on the ratio Eq. \ref{CHI-1} of the $ \EEGG $ reaction from center-of-mass energy $ \sqrt{s} $ = 91.2 GeV to 200 GeV using
               7 degrees of freedom in the $ k_{1} \approx k_{2} $ approximation. The red line shows  one $ \sigma $ limit.}
\label{CHIk7d}
\end{figure}

Inserted in the $ \chi^{2} $ test are the 7 $ \sqrt{s} $ energies from Table~\ref{diffpar} ( Lower part ) and the relevant 
R(exp) parameters in Table~\ref{table6}, including also  equation Eq. \ref{CHI-1},
Eq. \ref{Ratio-11} and Eq. \ref{Ratio-k} ) under the assumption $ k_{1} \approx k_{2} $.
The minimum of the $\chi^{2}$ test, including the error in Figure (\ref{CHIk7d}), is:
$ 1/\Lambda ^{4}=-1.72_{-0.95}^{+0.94}\times 10^{-13}GeV^{-4}$. The $\chi^{2}$ value at the minimum is:
$ \chi^{2}$ = 1.85. The fitting uses 7 degrees of freedom.

Similar to the QED test BabaYaga and \cite{Berends111}, $1/\Lambda^4$ is also negative in this approximation.
A first direct approximation for significance is $\sigma = A/\Delta A$ = 2.83 / 1.55 = 1.81.
In a second approximation for calculating the significance $\sigma$, the statistical probability 
function $p$ Eq. \ref{pvalue1} is used. The $p$ value of the discussed $\chi^2$ for 7 degrees 
of freedom (with a minimum $\chi^2$ = 1.85) is $p$ = 0.032. According to the figure 
\ref{pvaluesmall}, this corresponds to a significance of $ \sigma $ = 1.8.

According to Eq. \ref{DIR11} ), the size $ r $ of the interaction term is $ (1.27)\times 10^{-17} $ [ cm ].
The error $ \Delta r $ is after equation Eq. \ref{DIR14} and Eq. \ref{DIR18} $ \Delta r = 0.18 \times 10^{-17} $ [ cm ].
The summary of all these results is given in Table~\ref{Zhao}

\begin{table}[H] 
 \begin{center}
 \caption{ Summary of $ \chi^{2} $ test with 7 degrees of freedom in $ k_{1} \approx k_{2} $ approximation [ Test 4.1.4] } 
 \label{Zhao}
 \begin{tabular}{ | l | c | r | r | }
 \hline
                    & $ ((1/\Lambda )^{4} [GeV^{-4}] $)  & $ \sigma $ & $ ( r \pm \Delta r ) \times 10^{-17} [cm] $  \\  \hline
direct           & $ (-1.72 \pm 0.95)\times 10^{-13} $ & 1.81 &  \\  \hline
p - value      & $ (-1.72 \pm 0.95)\times 10^{-13} $ & 1.80 &  \\  \hline
$ r \pm \Delta r$  & $ (-1.72 \pm 0.95)\times 10^{-13} $ &  & $ (1.27\pm 0.18) $ \\  \hline
\end{tabular}
\end{center}
\end{table}

\subsubsection{ Numerical calculation of the $ \chi^{2} $ test with 17 degrees of freedom in $ k_{1} \approx k_{2} $ approximation.}
\label{Test with 17 degrees}

The $ \chi^{2} $ test on the ratio Eq. \ref{CHI-1}  as a function of $ 1/\Lambda ^{4} $ 
for the $ \EEGG $ reaction from center-of-mass energy $ \sqrt{s} $ = 55 GeV to 207 GeV is 
displayed in Figure \ref{CHIk17d}. 

\begin{figure}[htbp]
\vspace{0.0mm}
\begin{center}
 \includegraphics[width=15.0cm,height=12.0cm]{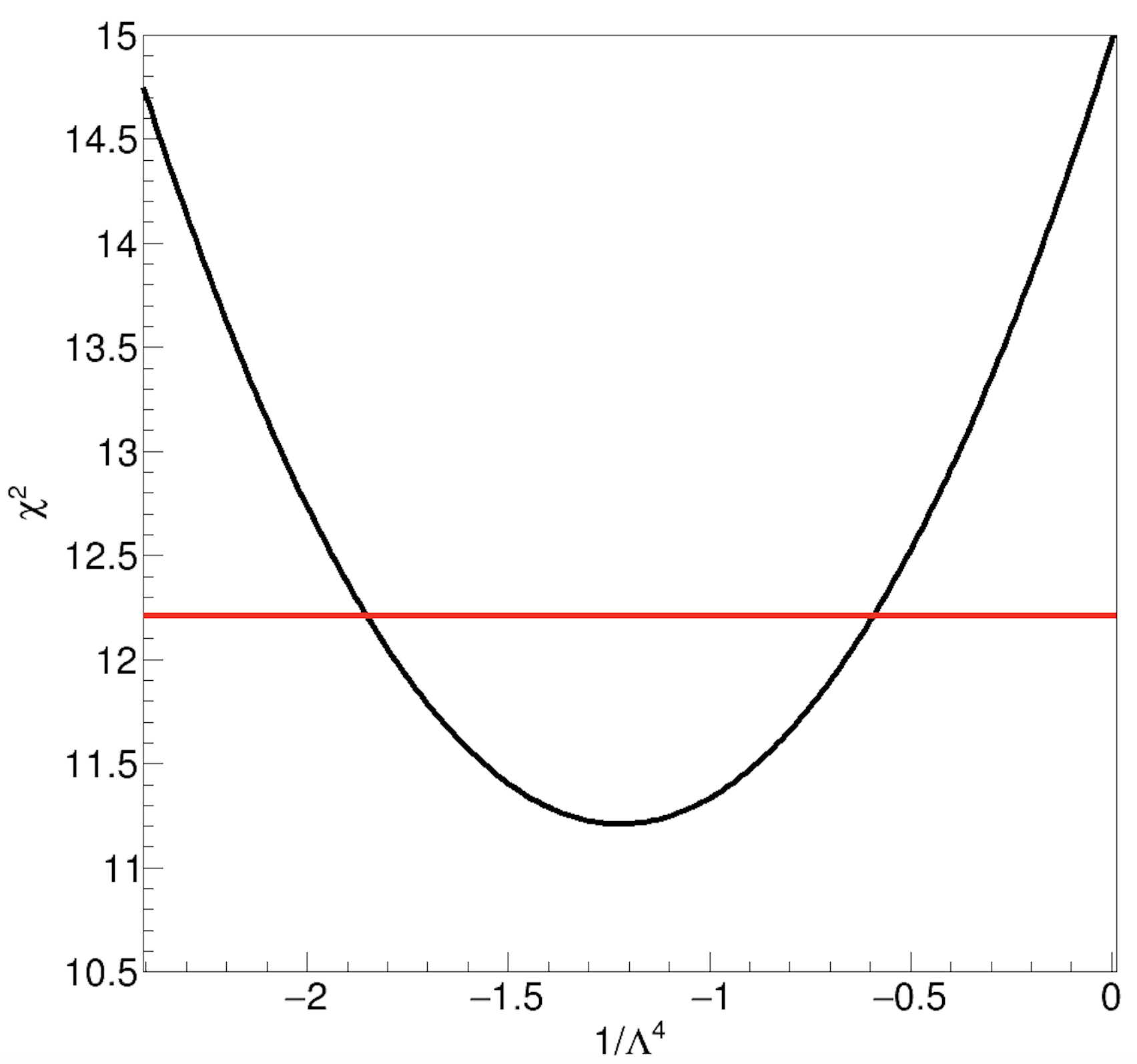}
\end{center}
\vspace{-3.0mm}
\caption{ $ \chi^{2} $ test on the ratio Eq. \ref{CHI-1} of the $ \EEGG $ reaction from center-of-mass energy $ \sqrt{s} $ = 52 GeV to 207 GeV using
               17 degrees of freedom in the $ k_{1} \approx k_{2} $ approximation. The red line shows one  $ \sigma $ limit.}
\label{CHIk17d}
\end{figure} 

In the $ \chi^{2} $-test,
17 $ \sqrt{s} $ energies R(exp) parameters from Table~\ref{table6}, equation Eq. \ref{CHI-1} ,
equation Eq. \ref{Ratio-11} and equation Eq. \ref{Ratio-k} are used under the assumption $ k_{1} \approx k_{2} $.

The minimum of the $\chi^{2}$ test, including the error in Figure (\ref{CHIk17d}), is:
$ 1/\Lambda ^{4}=-1.22_{-0.63}^{+0.63}\times 10^{-13}GeV^{-4} $. The $\chi^{2}$ value at the minimum is:
$ \chi^{2} $ = 11.2. The fit uses 17 degrees of freedom.

Similar to the QED - test BabaYaga and \cite{Berends111}, $ 1/ \Lambda ^{4} $ is also negative in this approximation. 
A first direct approximation of the significance is $ \sigma =A/\Delta A $  = 1.22 / 0.63 = 1.94. 
In a second approximation to calculate the significance $\sigma$, a statistical probability function $ p $ Eq. \ref{pvalue1}  
is used. The $ p $ value of the discussed $ \chi^{2} $ test for 17 degrees of freedom (with minimum $ \chi^{2} $ = 11.2) 
 equals  $ p $ = 0.15. Using Figure \ref{pvaluesmall}, this corresponds to a significance of approximate $ \sigma $ = 1.2.

According to Eq. \ref{DIR11} , the size $ r $ of the interaction term is $ (1.17)\times 10^{-17} $ [ cm ]. 
The error $ \Delta r $ is after equation Eq. \ref{DIR18} $ \Delta r = 0.15\times 10^{-17} $ [ cm ].
The summary of all these results is given in Table~\ref{CHI17d}.

\begin{table}[H] 
 \begin{center}
 \caption{ Summary of $ \chi^{2} $ test with 17 degrees of freedom in $ k_{1} \approx k_{2} $ approximation [Test 4.1.5]. } 
 \label{CHI17d}
 \begin{tabular}{ | l | c | r | r | }
 \hline
                    & $ ((1/\Lambda )^{4} [GeV^{-4}] $)  & $ \sigma $ & $ ( r \pm \Delta r ) \times 10^{-17} [cm] $  \\  \hline
direct           & $ (-1.22 \pm 0.63)\times 10^{-13} $ & 1.94 &  \\  \hline
p - value      & $ (-1.22 \pm 0.63)\times 10^{-13} $ & 1.20 &  \\  \hline
$ r \pm \Delta r$  & $ (-1.22 \pm 0.63)\times 10^{-13} $ &  & $ (1.17\pm 0.15) $ \\  \hline
\end{tabular}
\end{center}
\end{table}

\subsubsection{ Summary of the four different numerical calculations of the  $ \chi^{2} $ tests. }
\label{Summery of four calculations }

The four $ \chi^{2} $ tests on the ratio Eq. \ref{CHI-1}  as a function of $ 1/\Lambda ^{4} $ 
for the $ \EEGG $ reaction from center-of-mass energy $ \sqrt{s} $ = 55 GeV to 207 GeV are summarised in
Table~\ref{CHIsum}. 

\begin{table}[H] 
 \begin{center}
 \caption{ Summary of the numerical calculation of the  $ \chi^{2} $ tests . } 
 \label{CHIsum}
 \begin{tabular}{ | l | c | r | r | }
 \hline
Test        & $ \sigma $ &  interaction radius $  r \pm \Delta r )  $   \\  \hline
4.1.1      & $ 1.95 - 1.50 $ & $ 1.25\pm 0.16 \times 10^{-17} [cm] ) $   \\  \hline
4.1.2      & $ 1.90 - 1.83 $ & $ 1.44\pm 0.20 \times 10^{-17} [cm] ) $   \\  \hline
4.1.4      & $ 1.81 - 1.80 $ & $ 1.27\pm 0.18 \times 10^{-17} [cm] ) $   \\  \hline
4.1.5      & $ 1.94 - 1.20 $ & $ 1.17\pm 0.15 \times 10^{-17} [cm] ) $   \\  \hline
\end{tabular}
\end{center}
\end{table}

The Table contains Test 4.1.1 the ``Calculation of the $ \chi^{2} $ test with QED BabaYaga '', chapter \ref{Baba Yaga 111},
Test 4.1.2 ``Calculation of the $ \chi^{2} $ test with QED generator \cite{Berends111}'', chapter \ref{Berents 111}, 
Test 4.1.4 ``Numerical calculation of the $ \chi^{2} $ test with 7 degrees of freedom in $ k_{1} \approx k_{2} $ approximation'' , 
chapter \ref{With 7 degrees} and Test 4.1.5 ``Numerical calculation of the $ \chi^{2} $ test with 17 degrees of 
freedom in $ k_{1} \approx k_{2} $ approximation'', chapter \ref{Test with 17 degrees}.

All the four different $ \chi^{2} $ tests show a minimum, including a negative $ (1/\Lambda )^{4} [GeV^{-4}] $ value. 
This fact supports a negative interference effect of the direct contact term in Figure \ref{Feynman2}. 
In equation Eq. \ref{DIR3} ) has the term $\delta_{new}$ a negative sign. 

The sensitivities depend on the type of $ \chi^{2} $ test and the calculation methods. The directly calculated sensitivity
is approximately stable for all $ \chi^{2} $ tests and lies between $ \sigma = 1.95 $ to $ \sigma = 1.81 $.
Visible in Table~\ref{CHIsum} column $ \sigma $ left side. The sensitivity calculations via the statistics of
the p - values range from $ \sigma = 1.83 $ to $ \sigma = 1.20 $.Visible in Table~\ref{CHIsum} column $ \sigma $ right side.
It is important to notice that the approximation of the k1 - k2 method in $ \chi^{2} $ test is sensitive to 
the energy dependence of the k1 - k2 factors. This dependence is not included in the $ \chi^{2} $ test and 
lowers the p - values for large energy ranges, leading to a $ \sigma$ down to 1.20. Except the $ \sigma = 1.20 $ , 
the range of sensitivities between the direct calculation and the exact calculation of $ \sigma $ values  
is approximately the same. No significant differences among Monte Carlo generator of BabaYage, 
the generator \cite{Berends111} and the k1 - k2 approximation could be detected. 

The interaction radius $  r \pm \Delta r   $ is in the range of the statistical error $ \Delta r  $ the same 
for all $ \chi^{2} $ tests. Table~\ref{CHIsum} summarize all these information for the four $ \chi^{2} $ tests. 
In column 1 the method, column 2 the range of $ \sigma $ and in column 3 the interaction radius $  r \pm \Delta r $ is shown. 
 
It is well known that statistical tests depend on the amount of data and degrees of freedom  
available. This number defines ultimately the significance $ \sigma$. To test the differential cross 
section, every angular distribution bin of the differential cross section is one degree of freedom. 
For this reason, the differential cross-section test has significantly more degrees of freedom 
than the total cross-section test of the same data set, which increases the significance 
$ \sigma$ compared to the total cross-section tests $ \chi^{2} $.

Measurements of the differential cross section of the $\EEGG$ reaction by the VENUS, TOPAS, OPAL, 
DELPHI, ALEPH, and L3 collaborations, conducted between 1989 and 2003, are used for a $\chi^{2}$ 
test to search for a finite electron size. The USTC and ETHZ collaboration published a $\chi^{2}$ test 
of the differential cross section of the $\EEGG$ reaction in 2023, along with related  information from 
2006 to 2023 \cite{Diffcross11}. Data are available in the center-of-mass energy range from $\sqrt{s}$ = 55 GeV to 207 GeV at 17
 $\sqrt{s}$. In \cite{Diffcross11}, Table 2, and in this paper, Table \ref{SINGLE-FIT} shows the $\chi^{2}$ 
 test of the differential cross section of the $\EEGG$ reaction.
The 17 $\sqrt{s}$ energies comprise 254 degrees of freedom.

A comparison of the actual $ \chi^{2} $ test significance $ \sigma $ and the interaction radius $ r $ are shown in 
Table~\ref{Comparison}. To guide the eye, a copy of Table~\ref{CHIsum} is included.
In the last line, results of the ``diff-cross'' from \cite{Diffcross11} and chapter \ref{fitContact1} are also included.

\begin{table}[H] 
 \begin{center}
 \caption{ Comparison of $ \chi^{2} $ tests total cross section to differential cross section. } 
 \label{Comparison}
 \begin{tabular}{ | l | c | r | r | }
 \hline
Test        & $ \sigma $ &  interaction radius $  r \pm \Delta r )  $   \\  \hline
4.1.1      & $ 1.95 - 1.50 $ & $ 1.25\pm 0.16 \times 10^{-17} [cm] ) $   \\  \hline
4.1.2      & $ 1.90 - 1.83 $ & $ 1.44\pm 0.20 \times 10^{-17} [cm] ) $   \\  \hline
4.1.4      & $ 1.81 - 1.80 $ & $ 1.27\pm 0.18 \times 10^{-17} [cm] ) $   \\  \hline
4.1.5      & $ 1.94 - 1.20 $ & $ 1.17\pm 0.15 \times 10^{-17} [cm] ) $   \\  \hline
diff-cross      & $ 5.5 $ & $  1.57\pm 0.07 \times 10^{-17} [cm] ) $   \\  \hline
\end{tabular}
\end{center}
\end{table}
 
It is important to note that in \cite{Diffcross11}} and in Table \ref{electron2} of the finite size paper, 
the $ \chi^{2} $-test also yields a negative value of $ (1/\Lambda )^{4} [GeV^{-4}] = - ( 4.05 \pm 0.73 )\times 10^{-13} $ , 
which is similar to the $ \chi^{2} $-tests in tot cross section study, but with a significance of $ \sigma = 5.5 $ \cite{Diffcross11} ,
shown Table \ref{Comparison}.
 
The difference between the $\chi^{2}$ tests of the total and differential cross section
is visible in the significance $\sigma$. This confirms that the differential cross section test 
with 257 degrees of freedom yields a higher $\sigma$ than the total cross section test with 17 degrees of freedom.
It is important to note that the interaction radius $(r \pm \Delta r)$ is statistically the same for all tests.

 \subsubsection{ Ratio plot of the total cross section as a function of the finite size of the electron as indication of a signal in the Total Cross Section.}
 \label{totalXsectionratio}

It would be insightful to check whether a signal indicating the existence of the excited electron
and the contact interaction is also present in the overall cross-section of the ${\EEGG}$ reaction.

Investigating the sensitivity of the $\chi^2$ test to the total experimental cross section, $\sigma_{\mathrm{tot}}$,
of the $\EEG$ reaction, represented by combined data from various collaborations
in the energy range from 55 GeV to 207 GeV, presents a major challenge.

This is because different experiments have measured the total cross section
$\sigma_{\mathrm{tot}}$ at different angular ranges
$\theta$ and with different efficiencies $\varepsilon$ for the same or similar center-of-mass energy.
To solve this problem, we exploit the fact that all collaborations compare their measured total cross section
$\sigma_{{\rm tot}}$ with a Monte Carlo-simulated QED cross section, $\sigma_{\rm QED}$, which is either the same or very similar.
This opens up the possibility of using a reference detector to summarise  all the data.
Therefore, we use an L3 reference detector and normalise the total cross-sections 
measured by other detectors relative to those of L3 along with the corresponding 
number of events, as detailed in Ref. \cite{Totcross} and chapter \ref{QEDnumeric}.
This approach allows us to correctly combine the center-of-mass energy points 
where more than one detector has provided measurements of the 
total cross section. T his enables us to construct the ratios,
$R({\rm exp})=\sigma_{\rm tot}^{\rm comb}/\sigma_{\rm QED}$, by comparing
the combined measured total cross section, $\sigma_{\rm tot}^{\rm comb}$,
with the simulated cross section $\sigma_{\rm QED}$ at each available center-of-mass energy.
The uncertainties for $\sigma_{\rm tot}^{\rm comb}$ ($\Delta\sigma_{\rm tot}^{\rm comb}$)
and $R({\rm exp})$ ($\Delta R({\rm exp})$) are also calculated.
The processed numerical values, initially obtained in Ref.~\cite{Totcross},
are listed in Table~\ref{table6} and displayed in Figure~\ref{SIGMAtot}.

No significant discrepancy can be observed between the combined measured total cross-section,
$\sigma_{\rm tot}^{\rm comb}$, and the predicted value, $\sigma_{\rm QED}$,
of the $\EEG$ reaction in the center-of-mass energy range 55.0~GeV~$\leq$~$\sqrt{s}$~$\leq$~207~GeV
in figure~\ref{SIGMAtot}. The decrease in uncertainties is observed at higher center-of-mass energy bands
(LEP energies) due to the contribution of multiple detectors to each $\sqrt{s}$ point.

To illustrate the potential influence of the parameters derived at a significance of 5$\sigma$
from the $\chi^2$ test of the differential cross section, we plot the more sensitive plots for a deviation from QED,  
$R({\rm exp})$, $R({\Lambda_6})$ as a function of the center-of-mass energy 
$\sqrt{s}$ in Figure~\ref{Ratio-1}  and in a zoomed scale Figure~\ref{Ratio-2}.

\begin{figure}[htbp]
\begin{center}
 \includegraphics[width=16.0cm,height=10.0cm]{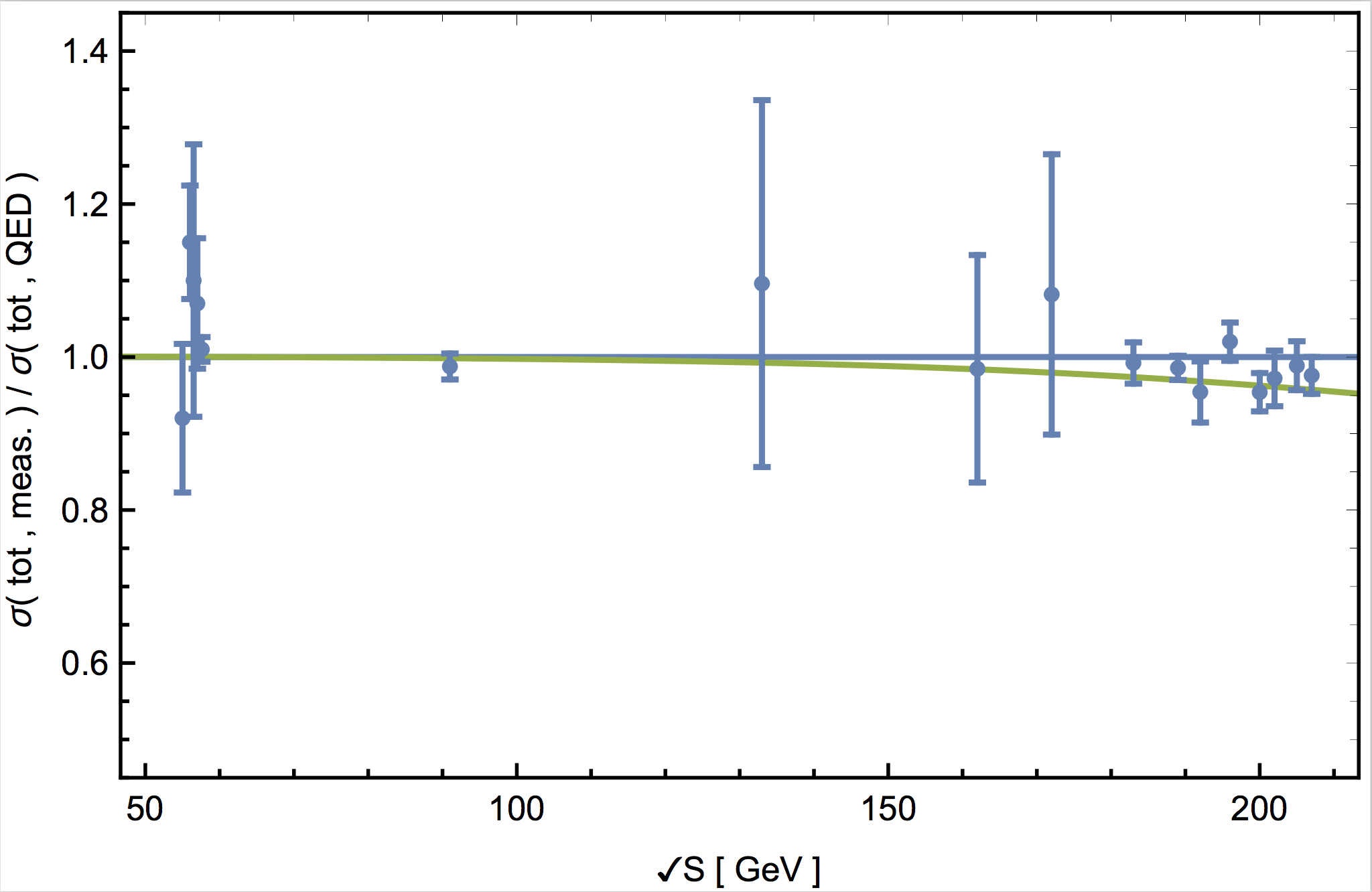}
\end{center}
\caption{ The ratio $R({\rm exp})$ (blue) along with ratio $R({\Lambda_6})=
\sigma_{\rm QED,\,  to t}^{\rm L3}/\sigma_{{\rm QED}+\Lambda_{\rm top},\, {\rm tot}}^{\rm L3}$ (green) }\label{fitContact1}
\label{Ratio-1}
\end{figure}

\newpage

\begin{figure}[htbp]
\begin{center}
 \includegraphics[width=16.0cm,height=10.0cm]{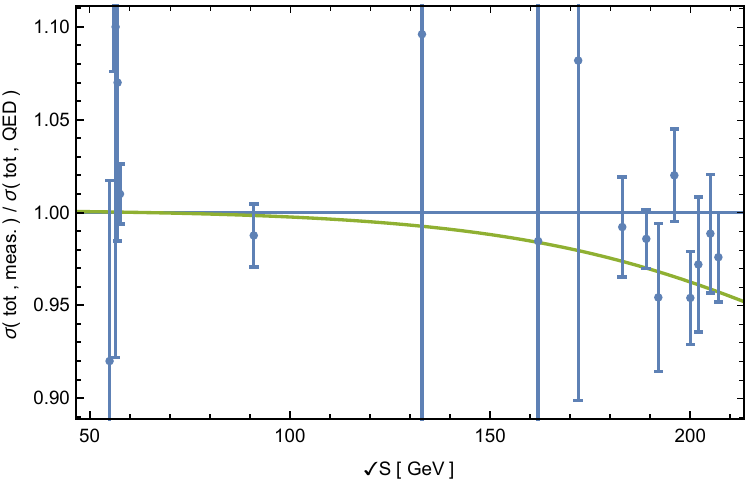}
\end{center}
\caption{ {Vertically} zoomed-in version of the Figure~\ref{Ratio-1} plot.}
\label{Ratio-2}
\end{figure}

In detail in Figure~\ref{Ratio-1} and Figure~\ref{Ratio-2} is the ratio $ R(exp)=\sigma (tot,meas)/\sigma (tot,QED) $ (blue) 
shown together with the  ratio 
$R({\Lambda_6})=\sigma_{\rm QED,\,  tot}^{\rm L3}/\sigma_{{\rm QED}+\Lambda_{\rm top},\, {\rm tot}}^{\rm L3}$ (green line). 
$ \sigma (tot,meas) $ is the total measured cross section shown in Table~\ref{table6}. $ \sigma (tot,QED) $ is the QED
total cross section calculated from the Monte Carlo generators discussed in chapter \ref{QEDnumeric}.
$\sigma_{\rm QED,\,  tot}^{\rm L3}$ and $\sigma_{{\rm QED}+\Lambda_{\rm top},\, {\rm tot}}^{\rm L3}$, 
are, the pure QED and modified total cross sections of the excited electron (contact interaction) at
$\Lambda = 1253.53$~GeV (see Chapter \ref{fitContact11}), normalised to the L3 detector.

The numerical results of $R({\Lambda_6})$ are calculated with the Monte Carlo generators for a CM energy
range 55.0~GeV~$\leq$~$\sqrt{s}$~$\leq$~207~GeV point by point of CM $\sqrt{s}$ energy. For the 
convenience of the reader we use the analytical approximation Eq. \ref{ToyR} including four parameters.
$C_1$ to $C_1$, what are fitted to the MC generator results.

\begin{align}
\label{ToyR}
R(\Lambda{_{6}} )=C_4-C_1\tanh (C_3\sqrt{s}+C_2)\ ,
\end{align}

The fit parameters of the constants are $C_1 = 0.0732964$, $C_2 = -3.06655$, 
$C_3 = 0.0127994$ and $C_4 = 0.928311$.

In summery, Figure~\ref{Ratio-1} and Figure~\ref{Ratio-2} indicates a deviation between the total cross-section of the
measured data and the QED prediction, in~contrast to the differential cross-section
test shown in Figures~\ref{VENUS}--\ref{cross-tot}. The data tend to lie below the 
horizontal line in the energy range $\sqrt{s} \gtrsim 180$~GeV,
with $R(\Lambda_{6})$ being approximately 4.0\% lower than the QED predicted~values.
The deviation $ \sigma $ (tot, meas.) / $ \sigma $ (tot, QED) $ < $ 1.0 above $\sqrt{s}$ = 180.0 GeV
indicates, as in all $\chi^{2}$ test of the actual paper, a negative interference. These results are consistent with the measurements of
L3 \cite{L3B} (Fig. 2) and DELPHI \cite{DELPHI2} (2000, Fig. 2), taking into account the large statistical error bars
of this papers.

\subsubsection{Appendices}
 \label{Appendices11}

The chapter \ref{TotalCross} uses a substantial amount of equations to calculate mathematical details. These
amount of equations disturb the clean view of the chapter \ref{TotalCross}. However, for the reader of these
articles it is important to write down once these details from chapter \ref{Appendices11} 
in chapter \ref{Appendices22} to chapter \ref{Appendices55} .

\subsubsection{ Virtual and soft radiative corrections of the $ \EEGG $ cross section. }
\label{Appendices22}

If the energy of the  photons from initial state radiation ( soft Bremsstrahlung ) 
is too small for detection $ k_{3}/|p_{+}|=k_{0}<<1 $, the reaction can be treated 
as 2-photon final state in (\ref{BORNCORR}).

\begin{equation}
\label{BORNCORR}
e^+(p_{+})+e^-(p_{-}) \rightarrow \gamma (k_{1})+\gamma (k_{2}))
\end{equation}

The equation for $ \delta _{virtual}+\delta _{soft} $ are in Eq. \ref{BORNCORR11}  to  Eq. \ref{BORNCORR44} .

\begin{equation}
\label{BORNCORR11}
\begin{matrix}
  \delta _{soft}+\delta _{virtual}= -\frac{\alpha }{\pi } \{2(1-2v)(lnk_{0}+v)+\frac{3}{2}  \vspace{0.2cm} \\  
  -\frac{1}{3}\pi ^2 +\frac{1}{2(1+cos^2\theta )} \vspace{0.2cm} \\
\times [-4v^2(3-cos^2\theta) -8vcos^2\theta  \\ 
+4uv(5+2cos\theta +cos^2\theta ) \\ 
+4wv(5-2cos\theta ) + cos^2\theta    \\ 
-u(5-6cos\theta +cos^2\theta ) \\
-w(5+6cos\theta +cos^2)  \\ 
-2u^2(5+2cos\theta +cos^2\theta ) \\
-2w^2(5-2cos\theta +cos^2\theta ) ] \}
\end{matrix}
\end{equation}

\begin{align}
\label{BORNCORR22}
v&=\frac{1}{2}ln\left ( \frac{s}{m^2_{e}} \right )\\ 
u&=\frac{1}{2}ln\left ( \frac{2(e+cos\theta )}{m^2} \right )\\
\label{BORNCORR33}
w&=\frac{1}{2} ln\left ( \frac{2(e-cos\theta )}{m^2} \right )  \\ 
\label{BORNCORR44}
m&=\frac{m_{e}}{\left | p_{+} \right |}
\end{align}
 
The mass of the electron $ m_{e} $ is still included in this equation. The total cross section
of the two $ \gamma $ final state is in Eq. \ref{Stot1} .

\begin{align}
\label{Stot1}
\sigma ^{2\gamma }&=\sigma _{0}+\frac{2\alpha ^3}{s}[2(2v-1)^2lnk_{0}+\frac{4}{3}v^3+3v^2\vspace{0.2cm} \\
&+(\frac{2}{3}\pi ^2-6)v-\frac{1}{12}\pi ^2] \nonumber 
\end{align}

\subsubsection{ Hard radiative corrections of the $ \EEGG $ cross section. }
\label{Appendices33}

The energy of the soft bremsstrahlung photons is limited by a value $k_{3}/\vert p_{1}\vert = k_{0} \ll 1 $.
If the energy of the photons from the initial-state radiation is greater than $k_{0} $, the process is 
treated as a 3-photon final-state reaction \ref{14}.

\begin{equation}
\label{14}
e^+(p_{+})+e^-(p_{-}) \rightarrow \gamma (k_{1})+\gamma (k_{2})+\gamma (k_{3})
\end{equation}

For the differential cross section of $\EEGG$ reaction, two additional parameters must be introduced in phase space.
The calculation in Eq. \ref{HARTCORR11} including the equation's up to Eq. \ref{Stot3} are performed in the 
extreme relativistic limit \cite{Mandl}.

\begin{equation}
\label{HARTCORR11}
\frac{d\sigma }{d\Gamma _{ijk}}=\frac{d\sigma }{d\Omega_{i} d\Omega _{k} dx_{k}}=\frac{\alpha ^3}{8\pi ^2s}w_{ijk}F(i,j,k)
\end{equation}

\begin{align}
\label{HARTCORR22}
w_{ijk}&=\frac{x_{i}x_{k}}{y(z_{j})},x_{i}=\frac{k_{i0}}{\left | \vec{p_{+}} \right |}  \\
\label{HARTCORR33}
y(z_{j}) &=2e -x_{k}+x_{k}z_{j}\\ 
\label{HARTCORR44}
z_{j} &=cos(\alpha _{ik})\\
\label{HARTCORR55}
x_{l}&=\frac{E_{l}}{\left | \vec{p_{+}} \right |} 
\end{align}

\begin{align}
\label{HARTCORR66}
F(i,j,k)&=\sum_{p} \left [ -2m^2 \frac{{k_{j}}'}{k^2_{k}{k_{i}}'}-2m^2\frac{k_{j}}{{k}'^2_{k}k_{i}}+\frac{2}{k_{k}{k}'_{k}} \left (  \frac{k^2_{j}+{k}'^2_{j}}{k_{i}{k}'_{i}}\right )\right ] \nonumber \\
&=\sum_{p}M(i,j,k)
\end{align}
 
 \noindent 
 $ \alpha_{ik} $ is the angle between $ k_{i} $ and $ k_{j} $. $ P $ binds all permutations of ( i, j, k $ \in $ ( 1, 2, 3 ).
 The quantities $ k_{i} $ and $ k'_{i} $ are Eq. \ref{HARTCORR7}.
  
 \begin{align}
 \label{HARTCORR7}
 k_{i}&=x_{i}(e-cos(\theta _{i})) \\
 \nonumber
{k}'_{i}&=x_{i}(e+cos(\theta _{i}))
\end{align}

\noindent
where $ \theta _{i} $ is the angle between the momentum of the i-th photon and $ | \vec{p}_{+}| $. The
total $ 3 \times \gamma $ cross section is Eq. \ref{HARTCORR8}.
 
\begin{align}
\label{HARTCORR8}
\sigma ^{3\gamma }=\frac{1}{3!}\int d\Gamma _{ijk},i,j,k\in {\{1,2,3\}}
\end{align}
 
 \noindent
 The integral runs over all phase space : $ k_{0} < x_{i} < 1 $ .
 The practical calculation Eq. \ref{HARTCORR8}  can be approximated by an analytical approach. The photons
 get sorted after energy, $ E_{\gamma 1 } \geq E_{\gamma 2 } \geq E_{\gamma 3 } $ where 
 $ \gamma_{1} $ and $ \gamma_{2} $ are treated as annihilation photons, and $ \gamma_{3} $ as 
 hard Bremsstrahlung photon. The total cross section after integrations is Eq. \ref{Stot3}  \cite{Berends} .
 
 \begin{align}
 \label{Stot3}
 \sigma ^{3 \gamma } =\frac{2\alpha ^3}{s}[3-(ln\frac{4}{m^2}-1)^2(2ln k_{0}+1)]
 \end{align}

\subsubsection{ Four tables needed for the numerical calculations of the $ \chi^{2} $ test. }
\label{Appendices44} 

To calculate the differential cross section Eq. \ref{QEDfit}  from the BabaYaga@nlo generator \cite {BABAYAGA} and \cite {Berends111},
the QED-Fit parameters $ p_{1} $ to $ p_{6} $ of the differential cross section of VENUS, TOPAS, 
ALEPH, DELPHI, OPAL, normalised to L3, are given in Table \ref{diffpar}.
The parameters are sorted after the 17 $ \sqrt {s} $ energies from $ \sqrt {s} $ = 55 GeV 
to 207 GeV for the BabaYaga generator Table~\ref{diffpar} ( Upper part ) and the 7 $ \sqrt {s} $ energies from $ \sqrt {s} $ = 91.2  GeV to 
200 GeV for the generator \cite {Berends} Table~\ref{diffpar} ( Lower part ).

\begin{table}[H] 
\caption{The QED fit parameters $ p_{1} $  to $ p_{6} $ of the differential cross section from VENUS, 
TOPAS, ALEPH, DELPHI, OPAL normalized to L3. Upper part BabaYaga, lower part \cite {Berends111} }
\label{diffpar}
\begin{center}
\begin{tabular*}{\textwidth}{@{\extracolsep{\fill}}|l|r|r|r|r|r|r|@{} }
\hline
GeV& \multicolumn{1}{c|}{$ p_{1} $} & \multicolumn{1}{c|}{$ p_{2} $} & \multicolumn{1}{c|}{$ p_{3} $} & \multicolumn{1}{c|}{$ p_{4} $} 
& \multicolumn{1}{c|}{$ p_{5} $}& \multicolumn{1}{c|}{$ p_{6} $} \\
\hline
$55$     & -5.31   & 8.97  & 0.65  &  9.41  &  0.42  & -7.30  \\ \hline
$56$     & -5.68   & 9.20  & 0.66  &  9.52  &  0.31  & -7.09  \\ \hline
$56.5$   & -0.39   & 2.29  & 0.52  &  4.53  & -1.28  & -3.59  \\ \hline
$57$     & -6.81   & 9.55  & 0.65  &  9.81  & -0.10  & -6.05  \\ \hline
$57.6$   & -7.92   & 7.34  & 0.62  &  7.44  & -0.98  & -1.46  \\ \hline
$91.2 $  & -9.28   & 9.08  & 0.61  &  9.90  & -1.68  & -2.06  \\ \hline
$133$    & -9.80   & 9.42  & 0.61  &  9.99  & -1.63  & -1.89  \\ \hline
$162$    & -4.02   & 6.29  & 0.58  &  8.95  & -1.87  & -4.74  \\ \hline
$172$    & -4.17   & 6.30  & 0.58  &  8.96  & -2.13  & -4.41  \\ \hline
$183$    & -4.42   & 6.57  & 0.59  &  9.11  & -1.91  & -4.60  \\ \hline
$189$    & -3.77   & 5.84  & 0.58  &  8.82  & -2.83  & -3.84  \\ \hline
$192$    & -4.89   & 6.81  & 0.59  &  9.27  & -2.10  & -4.27  \\ \hline
$196$    & -4.63   & 6.39  & 0.58  &  9.08  & -2.37  & -3.90  \\ \hline
$200$    & -4.29   & 6.44  & 0.59  &  9.07  & -2.02  & -4.56  \\ \hline
$202$    & -4.00   & 6.22  & 0.58  &  8.93  & -2.03  & -4.60  \\ \hline
$205$    & -4.11   & 6.27  & 0.58  &  8.98  & -2.04  & -4.55  \\ \hline
$207$    & -4.05   & 6.29  & 0.58  &  8.98  & -1.95  & -4.69  \\ \hline
$ 91.2 $ & 0.95    & -0.34 & 0.08  &  1.16  & -2.70  &  1.77  \\ \hline
$ 133 $  & 0.95    & -0.31 & 0.09  &  1.19  & -2.76  &  1.81  \\ \hline
$ 161 $  & 1.04    & -0.34 & 0.13  &  0.69  & -1.95  &  1.40  \\ \hline
$ 172$   & 1.22    & -0.47 & 0.18  & -0.17  & -0.62  &  0.75  \\ \hline
$ 183 $  & 1.27    & -0.51 & 0.19  & -0.37  & -0.38  &  0.66  \\ \hline
$ 189 $  & 1.29    & -0.52 & 0.20  & -0.40  & -0.41  &  0.70  \\ \hline
$ 200 $  & 1.29    & -0.53 & 0.20  & -0.40  & -0.42  &  0.72  \\ 
\hline
\end{tabular*}
\end{center}
\end{table}

The total experimental cross section $ \sigma(exp)_{tot}^{(det_i)} $ [pb] and the statistical 
error in [pb] from VENUS, TOPAS, ALEPH, DELPHI, OPAL and total L3 
event rate L3-N(exp) are shown in Table~\ref{EXPtot} sorted after 17 $ \sqrt {s} $ 
energies from $ \sqrt {s} $ = 55  GeV to 207 GeV. 

\begin{table}[H] 
\caption{The total experimental cross section $ \sigma(exp)_{tot}^{(det_i)} $ [pb] from VENUS, TOPAS, ALEPH, DELPHI, OPAL and total L3-N(exp) event rate.}
\label{EXPtot}
\begin{center}
\begin{tabular*}{\textwidth}{@{\extracolsep{\fill}}|l|r|r|r|r|r|r|@{} }
\hline
GeV& \multicolumn{1}{c|}{$VENUS  $} & \multicolumn{1}{c|}{$TOPAS$} & \multicolumn{1}{c|}{$ALEPH$} 
& \multicolumn{1}{c|}{$DELPHI$} & \multicolumn{1}{c|}{$L3-N(exp)$}& \multicolumn{1}{c|}{$OPAL$} \\
\hline
$55$ & 46.4 $\pm$ 4.9 & & &  & &\\ \hline
$56$ &    55.8  $\pm$ 3.6 & & &  & & \\ \hline
$56.5$ &   52.5  $\pm$ 8.5 & & &  & &\\ \hline
$57$ &  50.2  $\pm$ 4.0  & & &  & & \\ \hline
$57.6$ &   & 50.2  $\pm$ 0.8  & & &  &   \\ \hline
$91.2$ &    &  & 45.13 $\pm$ 2.6 & 17.4 $\pm$ 0.8 & 1882 & 32.4 $\pm$ 2.3\\ \hline
$133$ &    & & & 9.42 $\pm$ 2.06 & &  \\ \hline
$162$ &    & & &  5.76 $\pm$ 0.87  & &  \\ \hline
$172$ &   & & & 5.55 $\pm$ 0.94  & &  \\ \hline
$183$ &    &  & & 4.27 $\pm$ 0.35  & 439  & 10.05 $\pm$ 0.43  \\ \hline
$189$ &    &  & & 4.27 $\pm$ 0.20  & 1302  & 8.79 $\pm$ 0.23 \\ \hline
$192$ &    &  & & 3.43 $\pm$ 0.43  & 193  & 9.24 $\pm$ 0.58  \\ \hline
$196$ &    &  & & 4.22 $\pm$ 0.28  & 555 & 8.43 $\pm$ 0.34  \\ \hline
$200$ &    &  & & 3.73 $\pm$ 0.25  & 424  & 7.39 $\pm$ 0.31  \\ \hline
$202$ &    &  & & 3.50 $\pm$ 0.34  & 223  & 7.88 $\pm$ 0.47 \\ \hline
$205$ &    &  & &   & 459  & 7.40 $\pm$ 0.31 \\ \hline
$207$ &    &  & & & 863 & 6.78 $\pm$ 0.23 \\ 
\hline
\end{tabular*}
\end{center}
\end{table}

The total QED cross section $ \sigma(QED)_{tot}^{(det_i)} $ [pb] from 
VENUS, TOPAS, ALEPH, DELPHI, OPAL and L3 are shown in 
Table~\ref{QEDtot}. The Table~\ref{QEDtot} is sorted after the 17 $ \sqrt {s} $ 
energies from $ \sqrt {s} $ = 55  GeV to 207 GeV. As discussed in chapter 
\ref{Sigma-more-as-one} and \ref{Sigma-one-detector}, the 17 lines of the Table~\ref{QEDtot} include separate information if 
only one detector or more as one detector is involved in the calculation of the total QED cross section.

When only one detector is involved in the column $ VENUS - L3_{norm} $ from $ \sqrt {s} $ = 55 GeV to 57 GeV.
The left side is the VENUS QED total cross section in [pb] and the right side is the L3 normalised total cross section in [pb].
The same situation counts for $ TOPAS - L3_{norm} $ at $ \sqrt {s} $ = 57.6 GeV and $ DELPHI-(L3_{norm})$ from
$ \sqrt {s} $ = 133 GeV to 172 GeV. Recorded left side is the QED total cross section in  [pb] and the right side is the L3 
normalised total cross section in  [pb].

When more than one detector is involved at $ \sqrt {s} $ = 91.2 GeV and from 183 GeV to 207 GeV. In this case the QED total cross 
section from the detector in [pb] is recorded ( Not normalized to L3 ) with the exception of $ L3-N(QED)^{L3} $. Left side of 
the column shows the total QED cross section of L3 and right side shows the total event rate of the L3 at  that $ \sqrt {s} $ 
energy.

\begin{table}[H] 
\caption{The total QED cross section $ \sigma(QED)_{tot}^{(det_i)} $ [pb] from VENUS, TOPAS, ALEPH, 
               DELPHI, OPAL and total L3-$ N(QED)^{L3} $ event rate.}
\label{QEDtot} 
\begin{center}
\resizebox{1\textwidth}{!}{
\begin{tabular*}{1.185\textwidth}{@{}|l|r|r|r|r|r|r|@{} }
\hline
GeV& \multicolumn{1}{c|}{$VENUS-L3_{norm} $} & \multicolumn{1}{c|}{$TOPAS-L3_{norm} $} & \multicolumn{1}{c|}{$ALEPH$} 
& \multicolumn{1}{c|}{$DELPHI-L3_{norm}$} & \multicolumn{1}{c|}{ $ L3-N(QED)^{L3} $} & \multicolumn{1}{c|}{$OPAL$} \\
\hline
$55$ &   50.4 - 136  & & &  & &\\ \hline
$56$ &   48.5 - 131  & & &  & & \\ \hline
$56.5$ & 47.7 - 129  & & &  & &\\ \hline
$57$ &   46.9 - 127  & & &  & & \\ \hline
$57.6$ &    & 49.7 - 124 & &  &  &   \\ \hline
$91.2$ &    &  & 42.8 & 18.3  & 50.9 - 1890 & 32.0 \\ \hline
$133$ &     & & & 8.59 - 24.2 & &  \\ \hline
$162$ &     & & & 5.85 - 16.3 & &  \\ \hline
$172$ &     & & & 5.13 - 14.5 & &  \\ \hline
$183$ &    &  & & 4.57 & 12.7 - 457  & 9.32  \\ \hline
$189$ &    &  & & 4.28 & 11.9 - 1360 & 8.74 \\ \hline
$192$ &    &  & & 4.15 & 11.6 - 208  & 8.47  \\ \hline
$196$ &    &  & & 3.98 & 11.1 - 574  & 8.13  \\ \hline
$200$ &    &  & & 3.82 & 10.6 - 450  & 7.81  \\ \hline
$202$ &    &  & & 3.74 & 10.4 - 234  & 7.65  \\ \hline
$205$ &    &  & &   & 10.1 - 469     & 7.42 \\ \hline
$207$ &    &  & & & 9.90 - 845 & 7.29 \\ 
\hline

\end{tabular*}}
\end{center}
\end{table}

The luminosities used from the VENUS, TOPAS, ALEPH, DELPHI, L3 and OPAL experiments are shown in Table~\ref{LUMI}. 
Table~\ref{LUMI} show also the references where the values of luminosities are taken from.
The same references are also shown in Table~\ref{EXPtot} and Table~\ref{QEDtot} .

\begin{table}[H] 
\caption{The luminosity used from the VENUS, TOPAS, ALEPH, 
               DELPHI, L3 and OPAL experiment plus references the total cross section is published }
\label{LUMI}
\begin{center}
\resizebox{1\textwidth}{!}{
\begin{tabular*}{1.195\textwidth}{@{}|l|r|r|r|r|r|r|@{} }
\hline
GeV& \multicolumn{1}{c|}{$VENUS$} & \multicolumn{1}{c|}{$TOPAS$} & \multicolumn{1}{c|}{$ALEPH$} & \multicolumn{1}{c|}{$DELPHI$} & \multicolumn{1}{c|}{$L3$}& \multicolumn{1}{c|}{$OPAL$} \\
\hline
$55$ & 2.34 $pb^{-1}$ \cite{VENUS}& & &  & &\\ \hline
$56$ &    5.18 $pb^{-1}$ \cite{VENUS} & & &  & & \\ \hline
$56.5$ &   0.86 $pb^{-1}$ \cite{VENUS} & & &  & &\\ \hline
$57$ &    3.70 $pb^{-1}$ \cite{VENUS} & & &  & & \\ \hline
$57.6$ &   & 52.26 $pb^{-1}$ \cite{TOPAS} & & &  &   \\ \hline
$91$ &    &  & 8.5 $pb^{-1}$ \cite{ALEPH} & 36.9 $pb^{-1}$ \cite{DELPHI}& 64.6 $pb^{-1}$ \cite{L3A}& 7.2 $pb^{-1}$ \cite{OPAL1}\\ \hline
$133$ &    & & & 5.92 $pb^{-1}$ \cite{DELPHI} & &  \\ \hline
$162$ &    & & & 9.58 $pb^{-1}$ \cite{DELPHI} & &  \\ \hline
$172$ &   & & & 9.80 $pb^{-1}$ \cite{DELPHI} & &  \\ \hline
$183$ &    &  & & 52.9 $pb^{-1}$ \cite{DELPHI}& 54.8 $pb^{-1}$ \cite{L3B}& 55.6 $pb^{-1}$ \cite{OPAL2}\\ \hline
$189$ &    &  & & 151.9 $pb^{-1}$ \cite{DELPHI}& 175.3$pb^{-1}$ \cite{L3B}& 181.1 $pb^{-1}$ \cite{OPAL2}\\ \hline
$192$ &    &  & & 25.1$pb^{-1}$ \cite{DELPHI}& 28.8 $pb^{-1}$ \cite{L3B}& 29.0 $pb^{-1}$ \cite{OPAL2}\\ \hline
$196$ &    &  & & 76.1 $pb^{-1}$ \cite{DELPHI}& 82.4$pb^{-1}$ \cite{L3B}& 75.9 $pb^{-1}$ \cite{OPAL2}\\ \hline
$200$ &    &  & & 82.6 $pb^{-1}$ \cite{DELPHI}& 67.5 $pb^{-1}$ \cite{L3B}& 78.2 $pb^{-1}$ \cite{OPAL2}\\ \hline
$202$ &    &  & & 40.1 $pb^{-1}$ \cite{DELPHI}& 35.9 $pb^{-1}$ \cite{L3B}& 36.8 $pb^{-1}$ \cite{OPAL2}\\ \hline
$205$ &    &  & & & 74.3 $pb^{-1}$ \cite{L3B}& 79.2 $pb^{-1}$ \cite{OPAL2}\\ \hline
$207$ &    &  & & & 138.1 $pb^{-1}$ \cite{L3B}& 136.5 $pb^{-1}$ \cite{OPAL2}\\ 
\hline
\end{tabular*}}
\end{center}
\end{table}

\subsubsection{ Equations of significance and errors of the  $ \chi^{2} $ test. }
\label{SigCHI1}

The fit program Minuit calculates the minimum of the parameter $\Lambda$ and the error $\Delta\Lambda$.
This is illustrated in Figures \ref{CHIBhabha} to \ref{CHIk17d}. The program sets a limit (red line)
that allows the determination of $\Delta\Lambda$ values.
The crossing points of the $ \chi^{2} $ function with the red line 
give the size of $ \Delta \Lambda $ on the  $ 1/\Lambda ^{4} $ axis. It is common to calculate the signifcance
$ \sigma $ of a statistical test in a first approximation by dividing the value of a fit  
parameter $ A $ by the error bar of $ \Delta A $ like $ \sigma =A/\Delta A $. 

Furthermore, the significance $\sigma$ can be calculated using statistical theory  \cite{WIPI11}  and \cite{SIGNI} .
For the $\chi^{2}$ test, a statistical function f = ( NUMBER of degrees of freedom , Minimum of $\chi^{2}$ )
is used to calculate a $p$ value. The significance $\sigma$ can also be calculated using this $p$ value.

\subsubsection{ Equations for the calculation of the significance of a $ \chi^{2} $ test. }
\label{Sig444}

To calculate the statistical significance p of a  $ \chi^{2} $ test, a statistical probability function of 
f = ( NUMBER of degrees of freedom $ ( l ) $ , Minimum of the $ \chi^{2} $ $ T_{obs} $ ) is used.
The integral of this function f integrated from $ \chi^{2} $ = 0 to the minimum in $ \chi^{2} $ = $ T_{obs} $.
After equation 17 in \cite{SIGNI} is the p - value $ p = P(T<T_{obs}|l) $ Eq. \ref{pvalue1}.

\begin{align}
 \label{pvalue1}
 p=\int_{0}^{T_{obs}}d\chi ^{2}\frac{1}{2^{l/2}\Gamma (l/2)}e^{-\chi ^{2}/2}\left ( \chi ^{2} \right )^{-1+l/2}
 \end{align}

The gamma function $ \Gamma (\alpha ) $ Eq. \ref{pvalue2} contains the parameter $ \alpha $ = $ l / 2 $, 
where $ l $ are the degrees of freedom.

\begin{align}
\label{pvalue2}
\Gamma (\alpha )=\int_{0}^{\infty }y^{\alpha -1}e^{-y}dy
\end{align}

The $ p $ value is connected including the $ Erf $ function to the number of STANDART DEVIATIONS  $ \sigma $. 
The parameter $ sg $ in Eq. \ref{pvalue3} is $ sg = \sigma  $.

\begin{align}
\label{pvalue3}
p=1-Erf(sg/\sqrt{2})/2
\end{align}

In Figure \ref{pvaluebig} the function of STANDARD DEVIATIONS between $ sg=\sigma $ = 1 to 10 is represented via the p-value \cite{SIGNI}.

\begin{figure}[htbp]
\vspace{-10.0mm}
\begin{center}
 \includegraphics[width=15.0cm,height=13.0cm]{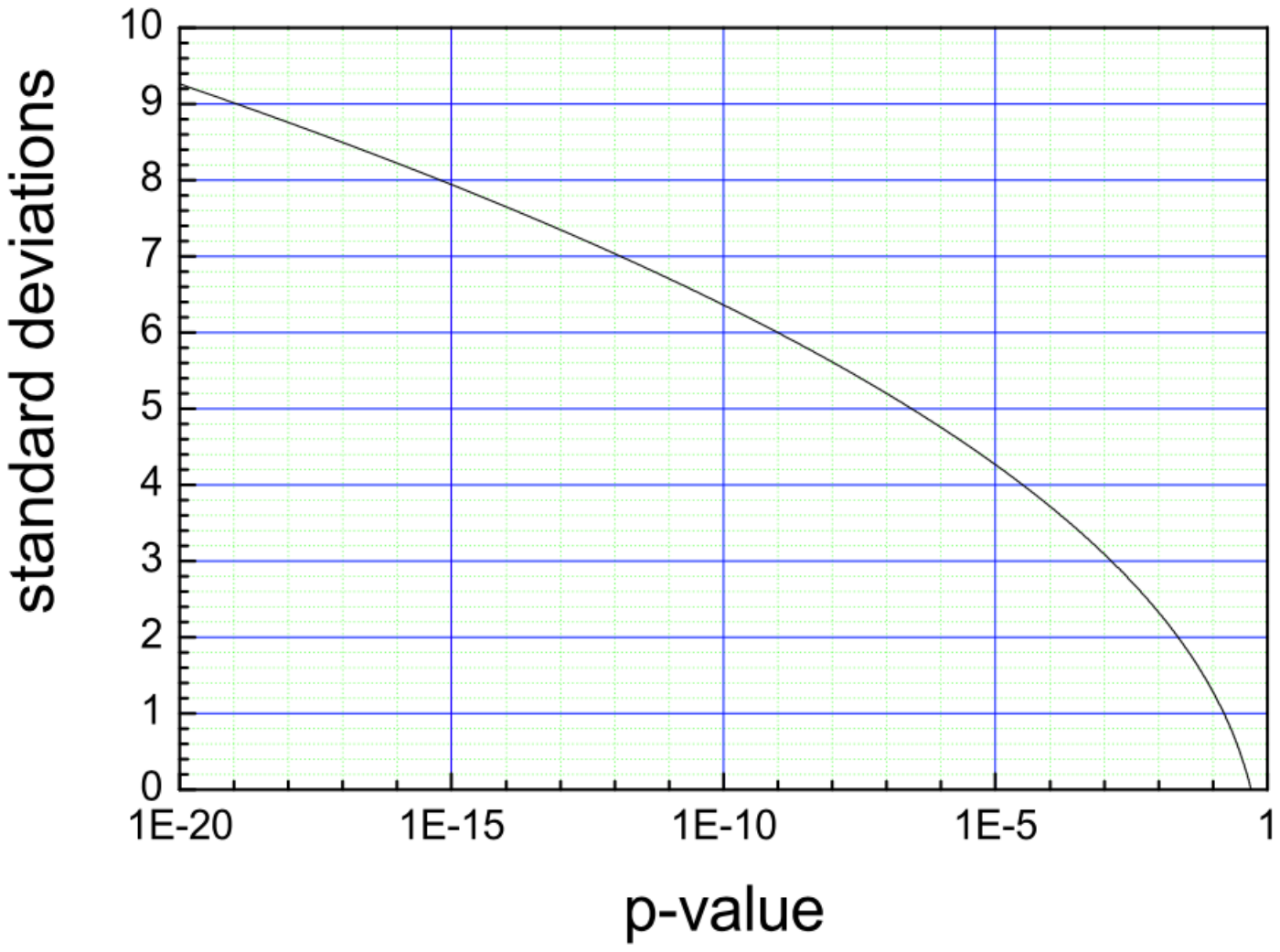}
\end{center}
\vspace{-5.0mm}
\caption{ The p - value Eq. \ref{pvalue1} as a function of STANDART DEVIATIONS $ sg=\sigma $ = 1 to 10 Eq. \ref{pvalue3}  }
\label{pvaluebig}
\end{figure} 

Figure \ref{pvaluesmall} shows the function of standard deviations between $ sg=\sigma $ = 1 and 4 using the p-value \cite{SIGNI}
on a lower scale.

\newpage

\begin{figure}[htbp]
\vspace{0.0mm}
\begin{center}
 \includegraphics[width=16.0cm,height=14.0cm]{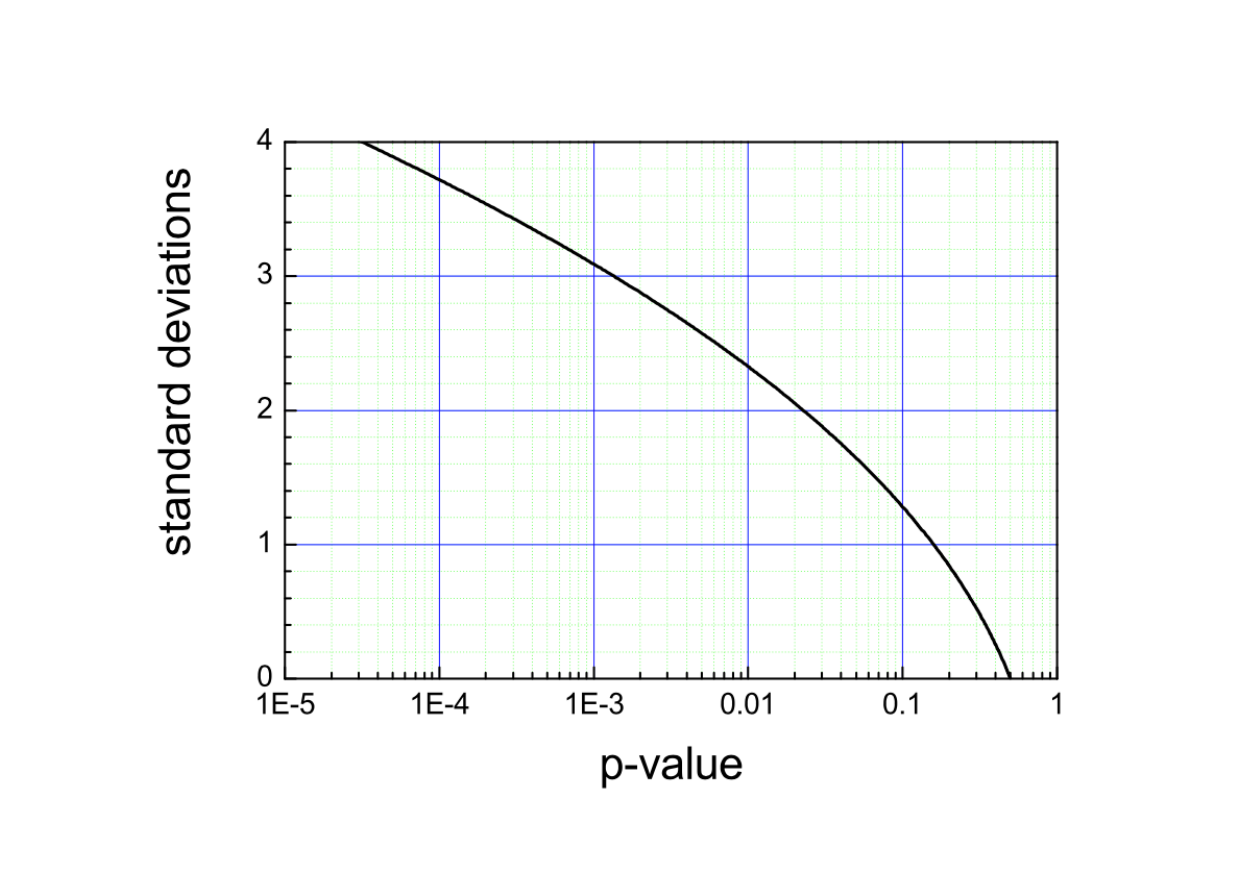}
\end{center}
\vspace{-10.0mm}
\caption{The p - value Eq. \ref{pvalue1} as a function of STANDART DEVIATIONS $ sg= \sigma $ = 1 to 4 Eq. \ref{pvalue3}  }
\label{pvaluesmall}
\end{figure} 

\subsubsection{ The radius $ r_{e} $ and the error $ \Delta r_{e} $ of the interaction size of the electron in the  $ \chi^{2} $ test. }
\label{deltar}

The minimum of the fit parameter $ 1/\Lambda ^{4} $ in the $ \chi^{2} $ test is a measure
of the interaction size $ r_{e} $ of the electron discussed in chapter \ref{Direct contact11} Eq.\ref{DIR11} and
repeated here for convenience of the reader Eq. \ref{DIR12}.

\begin{align}
\label{DIR12}
r_{e}=\frac{\hbar c}{\Lambda }
\end{align}

The minimum in the $ \chi^{2} $ test is given in units y = $ \frac{1}{\Lambda ^{4}}=y.yy_{-\Delta y.yy}^{+\Delta y.yy}\times GeV^{-4} $. 
For example,  according to Figure \ref{CHIk7d} y = $ \frac{1}{\Lambda ^{4}}=-1.72_{-0.95}^{+0.94}\times 10^{-13}GeV^{-4} $.
To develop the parameter $ \Lambda $ as a function of y in Eq. \ref{DIR13} 

\begin{align}
\label{DIR13}
\Lambda =\left ( \frac{1}{y} \right )^{1/4}
\end{align}

and use Eq. \ref{DIR12} allows to calculate the size $ r_{e} $ of the interaction radius.
In the discussed example, the interaction radius is $ r_{e} = (1.271)\times 10^{-17} $ [ cm ].

The error propagation of $ \Delta r $ is calculated in Eq.\ref{DIR14}.

\begin{align}
\label{DIR14}
\Delta r=\hbar \times c\times \frac{dr}{d\Lambda }\times \Delta \Lambda =-\frac{ \hbar \times c}{\Lambda ^{2}}\times \Delta\Lambda 
\end{align}

In Eq. \ref{DIR14}, $ \hbar $ is the Planck constant, $ c $ is the velocity of light, $ \Lambda $ is the cut off parameter from the
$ \chi^{2} $ test, and $ \Delta\Lambda $ is the error. The error calculation of the parameter $ \Lambda $ is 
not a linear function of the minimum in $ \chi^{2} $. 

Using y = $\frac{1}{\Lambda ^{4}}$, the error of y ($\Delta y$) in equation \ref{DIR15}, the error of 
$\Lambda$ ($\Delta\Lambda$) in equation \ref{DIR16}, and the first differential quotient $dy/d\Lambda$ 
in equation \ref{DIR17} can be calculated. Finally, the error of $\Delta\Lambda$ in equation \ref{DIR18} 
can also be calculated.

\begin{align}
\label{DIR15}
\Delta y=\frac{dy}{d\Lambda }\Delta \Lambda 
\end{align}

\begin{align}
\label{DIR16}
\Delta \Lambda =\frac{\Delta y}{(dy/d\Delta )}
\end{align}

\begin{align}
\label{DIR17}
dy/d\Delta =-4\Lambda ^{-5}
\end{align}

\begin{align}
\label{DIR18}
\Delta \Lambda =-\frac{1}{4}\Delta y\Lambda ^{5}
\end{align}

In the example discussed in Figure (\ref{CHIk7d}), the error is $ \Delta \Lambda = $ 214.41[ GeV ].
Substituting this value into Eq. \ref{DIR14} yields to $ \Delta r = 0.1755 \times 10^{-17} $ [cm].

\subsubsection{ Summary of equations for the calculation of significance and errors of the  $ \chi^{2} $ test. }
\label{Appendices55}

The equations Eq. \ref{pvalue1}  to Eq.\ref{pvalue3}  give an overview to calculate the significance $ \sigma $ of the
$ \chi^{2} $ - test. The significance in the chapter \ref{TotalCross} under discussion is in the range middle  $ 1.6 < \sigma < 1.9 $.
 It is common in the physics community, that such a  significance is a hint of new physics. The error $ \Delta r_{e} $ of the 
interaction size of the electron in the  $ \chi^{2} $ - test is described in the Eq. \ref{DIR14}  to Eq. \ref{DIR18}. 
It is interesting to note that, according to Table ~\ref{Comparison}, 
two independent $ \chi^{2} $ tests of the differential cross section and the total cross section 
yield the same interaction radius $ r_{e} $ within the statistical error limits.

\subsubsection{Systematic errors of $ \chi^{2} $ test on the total cross section. }
 \label{Appendices55}
 
 The default error in the ratio Eq. \ref{CHI-1}  $ \Delta R(E_{i};exp;sys) $  is usually the quadratic sum of the statistical 
 error and systematic error, provided both errors are independent. The data from the different groups show 
 in the total cross section the statistical error. The systematic error was not published in 
 detail for each group. Following the usual practice of  the various collaborations, only the 
 statistical error $ \Delta R(E_{i};exp) $ is used in Eq. \ref{CHI-1}.
  
As already discussed in chapter \ref{sys},  systematic errors arise from the 
evaluation of the luminosity, the selection efficiency, the evaluation of the background, the choice
of the theoretical cross section of the QED-$ \alpha^{3} $ as the reference cross section, the choice of the fitting method,
the type of fitting parameter and the scattering angle in $ | cos \theta | $ for the comparison between data and
theoretical calculations.
   
Concerning investigation in chapter \ref{TotalCross}, a slight deviation in $R({\rm exp})$ 
from the $\sigma({\rm QED})_{\rm tot}$ cross-section appears in
Table~\ref{table6} and Figure~\ref{Ratio-1} above $\sqrt{s}=91.2$~GeV. 
The systematic error of the measured total cross section of detector$_{i} $ 
$ \sigma(exp)_{tot}^{(det_i)} $ and total QED cross section of  $ \sigma(QED)_{tot}^{(det_i)} $  
are published, for L3 \cite{L3B} ( Table 3 ) above 
$  \sqrt{s} $ $>$ 91.2 GeV,  0.10$ <$ $ \Delta\sigma(meas.)_{sys} $ $<$ 0.13 [pb] and $ \Delta\sigma(QED)_{sys} $ = 0.1 [pb],
for DELPHI \cite{DELPHI} ( Table 4 ), year 2000 ) 0.09 $<$ $ \Delta\sigma(meas.)_{sys} $ $<$ 0.14 [pb] and
for OPAL \cite{OPAL2} ( Table 7 ) $ 0.05 <  \Delta\sigma(meas.)_{sys}  < 0.08 $ [pb]. 
According to these tables, the $ \Delta\sigma_{sys} $ values behave statistically 
and the systematic error $ \Delta\sigma(meas.)_{sys} $ is
much smaller than the statistical error $ \Delta\sigma(meas.)_{sys} < \Delta\sigma(meas.)_{stat} $.
No change as the function of the energy of the systematic errors above $ \sqrt{s} $ $ >$ 91.2 GeV has been observed. 
The statistical behaviour and the size of the $ \Delta\sigma(meas.)_{sys} $ exclude the possibility that the deviation 
of R(exp) from R(QED) could be originated from an energy $\sqrt{s}$ behavior of the systematic errors or the
size of the error $\Delta\sigma(meas.)_{sys}$.

Therefore, it is highly unlikely that the deviation of $R({\rm exp})$ from $R({\rm QED})$ is originate from systematic uncertainties. 
Further information can be found in Ref. \cite{Totcross}.

\subsubsection{Conclusion to chapter ``Hint for a minimal interaction length in $ \EEG $ annihilation in total cross 
                          section of center-of-mass energies 55 - 207 GeV''.}
\label{Conclusion total cross 11}                          

The total cross section of the $\EEGG$ reaction, measured by the VENUS, TOPAS, OPAL, DELPHI, ALEPH, and 
L3 collaborations, was used to test QED. The aim of chapter \ref{TotalCross} is to prove, using a global 
$\chi^2$ fit, that using the total cross section of all these measurements  implies a finite size of the electron.

First, the theoretical framework for calculating the total QED cross-section must be discussed,
in particular the fact that an analytically precise QED cross-section must be calculated using a Monte Carlo program.

Secondly, all the discussed collaborations measured the total cross-section for $ \EEG $ but for 
different parameters ( Like angular range and so on). Therefore, it is not possible to specify a 
total cross-section for the entire energy range from $ \sqrt{s} $= 55 GeV to 207 GeV. It is necessary 
to introduce a model for a total cross-section using all the data from the 6 collaborations. 
This model finally allows to perform a total $ \chi ^2 $ fit of all data.

Third, four different  $ \chi^{2} $  tests are used to verify the finite size of the electron.
All four  $ \chi^{2} $ tests show a minimum, including a negative value of $ (1/\Lambda )^{4} [GeV^{-4}] $.
This supports the assumption of a negative interference effect of the direct contact term. 
The maximum of the directly calculated sensitivity is approximately $ \sigma = 1.9 $. The 
sensitivity ranges of the four $ \chi^{2} $ tests are approximately equal. The 
interaction radius $  r \pm \Delta r $ is statistically identical. According to a common 
interpretation, this could be a  ``HINT'' of an effect of new physics.

The measurements of the differential cross section of the $ \EEGG $ reaction from the six collaborations, 
conducted between 1989 and 2003, are used to perform a global $ \chi^{2} $ test to search for a finite size of 
an electron \cite{QED-1}, \cite{SIZE}. The test detected also a negative 
$ (1/\Lambda )^{4} [GeV^{-4}]  = - ( 4.05 \pm 0.73 )\times 10^{-13} $, and a radius $ r $ of the electron  
of $ 1.57 \pm 0.28 \times 10^{-17} [cm] ) $. Considering the range of the statistical error, 
these values equal to the total cross section $ \chi^{2} $ test in the chapter \ref{TotalCross} under discussion.
The analysis of the differential cross section confirms the total cross section $ \chi^{2} $ 
statistic. An important difference between the $ \chi^{2} $  test of the 
differential cross section and the total cross section is the sensitivity. In the differential 
cross section test, $ \sigma $ = 5.5, and in the total cross section test 
maximum, $ \sigma $ = 1.9 only, see Table~\ref{Comparison}. This difference results from 
the different number of available data points (degrees of freedom) for the differential and the total cross section.

\subsection{ Hint for axial-vector contact interactions in the data on $ \EEEEG $ at centre-of-mass energies 192 - 208 GeV.}
\label{BABHA1}

In this chapter \ref{BABHA1} we give a brief overview of the BHABHA reaction $ \EEEEG $, 
which was investigated by D. Bourilkov, including the detailed information in the reference \cite{BOURILKOV}.

\subsubsection{ INTRODUCTION }
\label{BABHA11}

In the chapters \ref{Diff-Cross} and \ref{TotalCross} we investigated the $\EEGG$ reaction 
using the Lagrange equations Eq. \ref{DIR222} in the chapters \ref{Diff-Cross} and the same Eq. \ref{DIR22} 
in chapter \ref{TotalCross} for a direct contact term interaction. A deviation of the 
differential and total cross sections from the Standard Model (SM) was observed. 
In both investigations we detected significance of a 5 $\sigma$ and 2 $\sigma$ deviation.
This means that the size of the electron is finite and not ZERO, as predicted by the Standard Model ( SM ).

In Figure \ref{Feynman1} ``The lowest order Feynman diagrams for the reactions $ \EEG $ 
and $ \EEEE $'' is shown, the reaction in the $ \EEG $ process proceeds via the $t$ and $u$ channels. 
The Bhabha scattering $ \EEEEG $ proceeds via the $s$ and $t$ channels.
The two real photons in the final state of the $ \EEG $ reaction are indistinguishable,
therefore the reaction proceeds via the $t$ and $u$ channels, while the
$s$ channel is forbidden due to the conservation of angular momentum.
The $s$, $t$, and $u$ channels (Mandelstam variables) represent 
different types of particle interaction and energy and momentum exchange.
The reaction is highly sensitive to long-range QED interactions,
and the two photons in the final state possess left- and right-handed
polarisations, resulting in a total spin of zero and excluding the
$s$-channel with spin ONE for $\gamma$ and $Z^0$.
As a pure annihilation reaction, the~$ e^+ $ and $ e^- $ in the initial state
completely annihilate into two photons in the final state,
making it straightforward to subtract the background~signal. 

The Bhabha reaction $\EEEEG$ is a mixed reaction that occurs via scattering in both the $s$-channel and $t$-channel.
At energies around the $Z^0$ pole, the~$Z^0$ contribution dominates.
The elastic scattering and annihilation channel are superimposed since the $e^+$ and $e^-$
in the initial and final states are identical. Therefore, the~$\EEEEG$ reaction serves as a test 
for the superimposition of short-range weak interaction and long-range QED~interaction.

\subsubsection{ CONTACT INTERACTION }
\label{BABHA-CONTACT}

Contact interactions provide a general framework for describing a new interaction with a typical energy scale $\Lambda \gg \sqrt{s}$.
The presence of operators with canonical dimension $N>4$ in the Lagrangian leads to effects 
$\sim 1/\Lambda^{N-4}$. Such interactions can occur, for example, when Standard Model particles are 
composite or when new heavy particles are exchanged.

For fermion pair production, the lowest-order flavour diagonal and helicity-conserving operators have the dimension 
six \cite{compMod4} shown equation Eq. \ref{Bahba-direct}.

\begin{align}
\label{Bahba-direct}
\pounds _{\psi \psi }=(g^{2}/2\Lambda ^{2})\left [ \eta _{LL}\bar{\psi _{L}} \gamma _{\mu }\psi _{L}\bar{\Psi _{L}}\gamma ^{\mu }\psi _{L}+
\eta _{RR}\bar{\psi _{R}}\gamma _{\mu }\psi _{R}\bar{\psi _{R}}\gamma ^{\mu }\psi _{R}+
2\eta _{RL}\bar{\psi_{R}}\gamma _{\mu }\psi _{R}\bar{\psi _{L}}\gamma ^{\mu }\psi _{L}\right ]
\end{align}
\\
The Lagrange function of Eq. \ref{Bahba-direct} assumes the the standard $ SU(3)\otimes SU(2)\otimes U(1) $ gauge theory is correct,
and that $\Lambda \geq \Lambda _{EW}$  holds. Were $ \psi _{L} $ , $ \psi _{R} $ the left and right polarised wave functions of the 
distinct species. In equation Eq. \ref{Bahba-direct},  is the energy scale $ \Lambda $ very similar to 
Lagrange function  \ref{DIR222} of $ \EEG $ reaction for $ \Lambda $ - the size
of the particle would approach to ZERO equation Eq. \ref{DIR11}.
The Mandelstam variables are denoted by s, t, and u. The coupling is usually fixed, $g^{2}/4\pi =1$, 
and the interaction structure is parameterized by coefficients for the helicity 
amplitudes: $ \left| \eta _{i,j}\right|\leq 1 $ , where $ (i,j=L,R) $ labels the helicity of the incoming and outgoing fermions.
$ \gamma _{\mu } $ are the $ \gamma $ matrices.The differential cross section has the form of 
the equation Eq. $\ref{Bahba-diff-cross}$.

\begin{align}
\label{Bahba-diff-cross}
\frac{d\sigma }{d\Omega }=SM(s,t)+\varepsilon \cdot C_{int}(s,t)+\varepsilon ^{2}\cdot C_{CI}(s,t)
\end{align}

The first term in Eq. \ ref{Bahba-diff-cross} is the contribution of the Standard Model, the second results 
from the interference between the Standard Model SM and the contact interaction, and the third is the 
pure contact interaction effect. With the definition of Eq. \ref{Epsilon11}

\begin{align}
\label{Epsilon11}
\varepsilon =\frac{g^{2}}{4\pi }\frac{sgn(\eta )}{\Lambda ^{2}}
\end{align}

the sign of $\eta$ allows for the investigation of both positive and negative interference. 
The fit value of $ \varepsilon \left [ TeV^{-2} \right ] $
opens the possibility to define in a fit the significance $ \sigma $ of a fit-test.

\subsubsection{ EXPERIMENTAL DATA }
\label{BABHA-DATA}

Measurements of fermion pair production for the complete LEP2 collider dataset were published.
These include preliminary data in the center-of-mass energy range of 192 to 208 GeV, as well as previously 
published results at 183 and 189 GeV.  In the following the focus is on the Bhabha scattering measurements, for which 
large datasets were collected during the highly successful LEP acceleration runs from 1997 to 2000.

The ALEPH \cite{ALEPHBhabha,ALEPHBhabha1,ALEPHBhabha2,ALEPHBhabha3}, the 
DELPH \cite{DELPHIBhabha,DELPHIBhabha1,DELPHIBhabha2}, L3 \cite{L3Bhabha11,L3Bhabha22,L3Bhabha33}
and OPAL   \cite{OPALBhabha,OPALBhabha1,OPALBhabha2,OPALBhabha3} collaborations have presented results for the
total cross section $ \sigma _{tot} $ and the forward-backward asymmetry
$ A_{FB}=(\sigma_{F} -\sigma_{B} )/(\sigma _{F}+\sigma _{B}) $ \cite{FB-Asymmetry} of Bhabha scattering. 
In the case of DELPHI, L3 and OPAL the results are for all energy points and the scattering angle $ \theta $ is the angle between the incoming 
and the outgoing electrons in the laboratory frame. In the case ALEPH the forward-backward asymmetry is 
available only at 183 GeV and the scattering angle $ \theta ^{\ast } $  is defined in the outgoing $ e^{+}e^{-} $ rest frame. The acceptance 
is given by the angular range $ -0.9<\theta ^{\ast }<0.7 $ for ALEPH, by $ 44^{o}<\theta <136^{o} $ for DELPHI and L3, 
and by $ \left|cos\theta  \right|<0.7 $ for OPAL.
 
The experiments use different strategies to isolate the high energy sample, where the interactions
take place at energies close to the full available centre-of-mass energy.  
This sample is the main search field for new physics. DELPHI, L3 and OPAL apply an 
acollinearity cut of $ 20^{o} $ , $ 25^{o} $ and $ 10^{o} $ respectively. 
ALEPH defines the effective energy, $ \sqrt{s^{'}} $ , as the invariant 
mass of the outgoing fermion pair. It is determined from the angles of the outgoing fermions. 
 
For details on the selection procedures, as well as the statistical and 
systematic errors, we refer the reader to the publications of the LEP experiments.

\subsubsection{ ANALYSIS METHOD }
\label{BABHA-ANALYSIS}

The Standard Model predictions for Bhabha scattering are computed with the Monte Carlo generator BHWIDE \cite{MonteCarlo1}. 
We assign a theory uncertainty of 1.5 \% on the absolute scale of the predictions \cite{MonteCarlo2}. In all cases the individual 
experimental cuts of the selection procedures and the isolation of the high energy samples are taken into account. 
The results are cross-checked with the semi-analytic program 
TOPAZ0 \cite{MonteCarlo3} and the Monte Carlo generator LABSMC \cite{MonteCarlo4}.

The effects of new phenomena are computed as a function of the parameter $\epsilon$, which is defined in 
section \ref{BABHA-CONTACT}. Initial-state radiation (ISR) changes the effective center-of-mass 
energy in a large fraction of the observed events. These effects are taken into account by computing
the first order exponentiated cross section following \cite{Kleiss11 }. Other QED and electroweak
corrections give smaller effects and are neglected.

A total of 57 data points are available: 32 from the cross sections (eight energy points and 
four experiments) and 25 from the forward-backward asymmetries. A fitting procedure similar
 to that in \cite{Bourilkov24} is applied. A negative log-likelihood function is constructed by combining 
 all data points at the eight center-of-mass energies in the equation Eq. \ref{LOG-Like}.

\begin{align}
\label{LOG-Like}
-log\pounds =\sum_{r=1}^{n}\left ( \frac{\left [ prediction(SM,\varepsilon )-measurement \right ]^{2}}{2\cdot \Delta ^{2}} \right )_{r}
\end{align}

The prediction $ (SM,\epsilon )$ in Eq. \ref{LOG-Like} is the SM expectation for a given 
measurement (total cross section or forward-backward asymmetry) 
combined with the additional effect of contact interactions as a function 
of the scale $ \Lambda $,  where the measurement is the corresponding 
measured quantity and $ \Delta =error\left [ prediction(SM,\epsilon )-measurement \right ] $. 
The index $ r $ runs over all data points. The error on a 
deviation consists of three parts, which are combined in quadrature: a statistical error and a systematic error (as given by 
the experiments) and the theoretical error assigned above. 
The systematic errors account for small correlations between 
data points. The minimum of the negative log-likelihood function gives the central value of the fitted parameter $ \epsilon $ for each 
model. The interval containing 68\% of the total probability around the minimum is used to determine the values 
corresponding to one standard deviation in the positive and negative directions.
 
 \subsubsection{ RESULTS log-likelihood fit}
\label{BABHA-RESALTS}
 
The results of the fits for the different contact interaction models are summarised in Fig.{\ref{Bahaba-fit}. 
It is important to notice that Fig.{\ref{Bahaba-fit}, is a direct copy of Table I in the paper from 
D. Bourilkov \cite{BOURILKOV}.

\newpage

\begin{figure}[htbp]
\begin{center}
 \includegraphics[width=17.0cm,height=9.0cm]{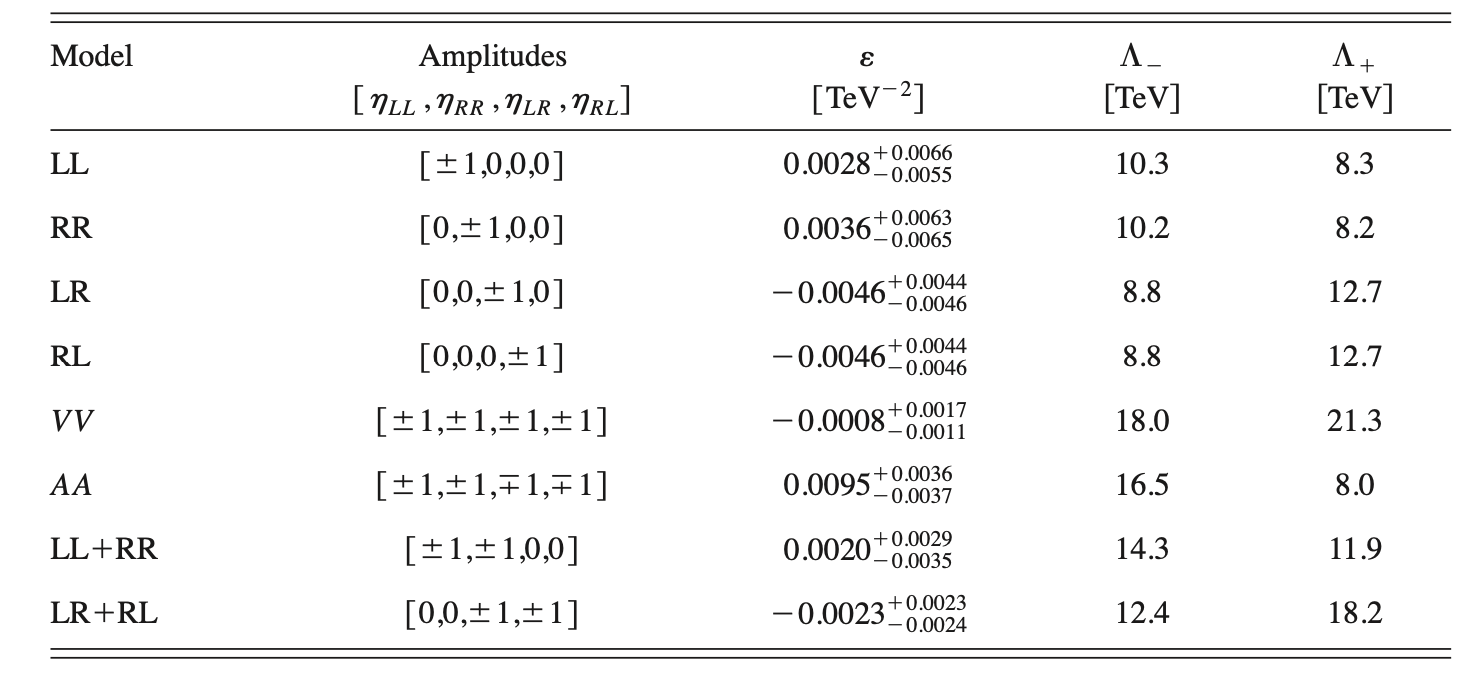}
\end{center}
\caption{TABLE I. Results of contact interaction fits to Bhabha scattering. The numbers in 
brackets are the values of   $\left [ \eta _{LL},\eta _{RR},\eta _{LR},\eta _{RL} \right ] $defining to which helicity amplitudes the contact 
interaction contributes. The models cover the interference of contact terms 
with single as well as with a combination of helicity amplitudes. The one-sided 95\% confidence 
level lower limits on the parameters $ \Lambda _{+} ( \Lambda _{-} ) $ given in $ TeV $ correspond to the
upper ( lower ) sign of the parameters $ \eta $, respectively.}
\label{Bahaba-fit}
\end{figure}

In Fig.{\ref{Bahaba-fit} the coefficients specifying the helicity structure of the investigated models are given. For all models, 
the central value of the parameter $ \varepsilon $ is no more than 1.05 standard deviations away from the Standard Model 
value $ \varepsilon =0 $, or, after equation Eq. \ref{Epsilon11}  $ \Lambda = \infty $ , with one exception. 

The fitted value for the AA model is in equation $ \epsilon $ for Eq. \ref{Fit-max}
 
\begin{equation}
\label{Fit-max}
\epsilon =0.0095_{-0.0037}^{+0.0036} \ TeV^{-2}
\end{equation}

 or using equation Eq. \ref{Epsilon11} for Eq. $ \Lambda $  for \ref{LAMDA11}

\begin{equation}
\label{LAMDA11}
 \Lambda =10.3_{-1.6}^{+2.8} \ TeV
\end{equation}

2.6 standard deviations from the SM expectation. 

The quality of all fits is good, with $ \chi ^{2} $ values $ \sim 52/56 $ degrees of freedom. 
For the AA model the $ \chi ^{2} $ value is $ 46.8/56 $ degrees of freedom. The differences 
between the experimental measurements and the standard model predictions are displayed in Fig.\ref{Bhabha-SIGMA}

\newpage

\begin{figure}[htbp]
\begin{center}
 \includegraphics[width=14.0cm,height=14.0cm]{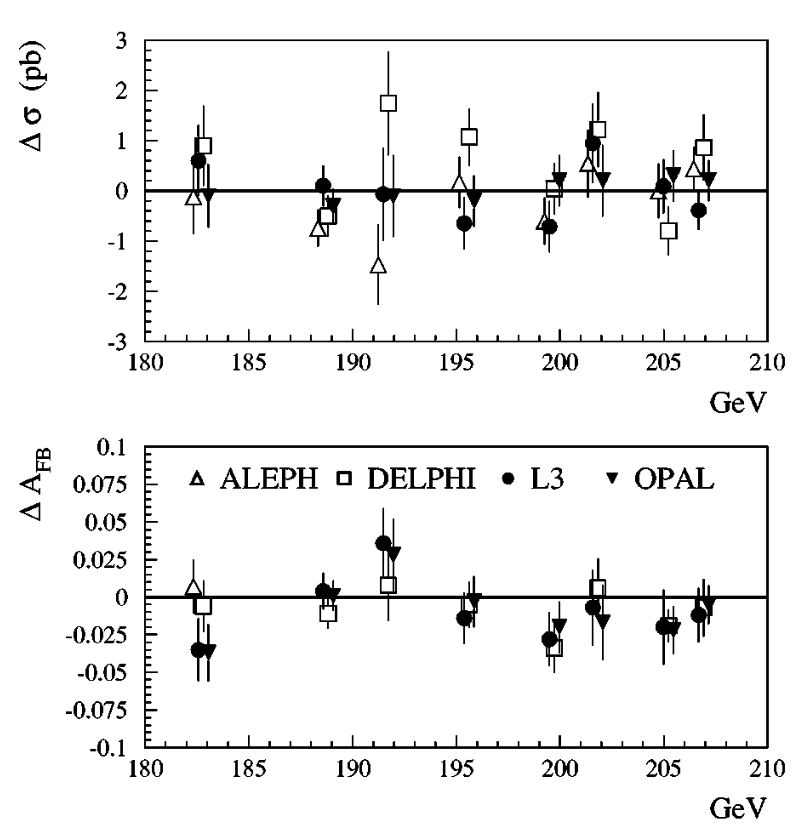}
\end{center}
\caption{Differences between the total cross sections and the forward-backward asymmetries 
for Bhabha scattering, presented by the LEP Collaborations, and the standard model predictions, 
computed with the program BHWIDE, as described in the text. The errors shown are statistical. 
For better visibility the points of the four experiments at the same center-of-mass energy are shifted 
slightly in the horizontal direction.}
\label{Bhabha-SIGMA}
\end{figure}

While good agreement is observed for the measurements of the total cross section, 
the measured forward-backward asymmetries above 192 GeV tend to be lower 
than predicted. The log-likelihood curves for the VV and AA models are shown in Fig.\ref{Bhabha-AA}.

\newpage

 \begin{figure}[htbp]
\begin{center}
 \includegraphics[width=14.0cm,height=14.0cm]{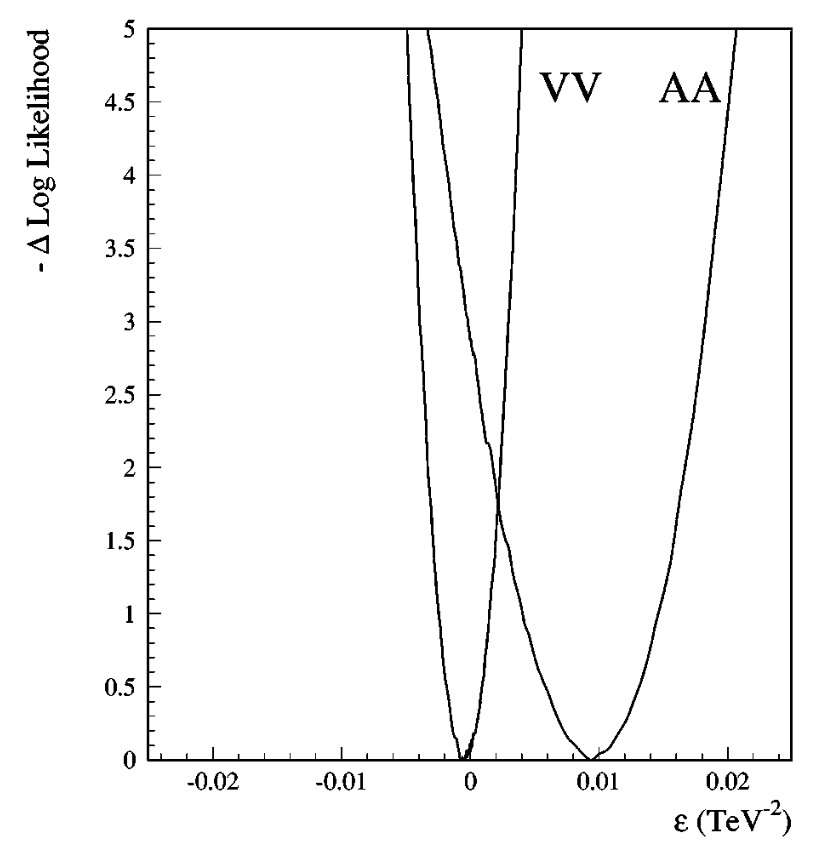}
\end{center}
\caption{Log-likelihood curves for the VV and AA contact interaction 
models of the combined fits to the data on Bhabha scattering from the four LEP experiments.}
\label{Bhabha-AA}
\end{figure}

In cases where the LEP collaboration data show no statistically significant 
deviations from the SM predictions, we can derive one-sided lower bounds 
for the $\Lambda$ scale of contact interactions with a confidence of 95\%.
This is done by integrating the log-likelihood functions in the physically allowed
range of parameters describing new physical phenomena, assuming a uniform
 prior distribution. The exact definition can be found in reference \cite{OPALBhabha1}. 
The limits for positive or negative interference are summarised in Table Fig.{\ref{Bahaba-fit}. 
The results presented here improve on the limits obtained by individual LEP 
experiments \cite{ALEPHBhabha,DELPHIBhabha,L3-25,L3-26,OPALBhabha1} 
or in a combined analysis \cite{BOURILKOV}.

 \subsubsection{ DISCUSSION }
 \label{DISCUSSION-RESALTS}

To discuss the $\EEEE$ reactions, the reader must be reminded of the helicity
of a particle beam. As explained in the chapter \ref{Diff-Cross-Introduction} on polarised 
ion sources, it is possible to correlate the vector of the spin axis to the vector of the movement of the beam
The technology allows to generate a beam the vector-axis of the spin is parallel of the vector-axis of the velocity of the beam 
( Longitudinal polarised )  or generate a beam both axis are perpendicular to each other ( Spin-UP or Spin-Down ).
The technology of the LEP accelerator ring is not equipped with a polarised ion source; instead, the magnetic system and
and accelerating systems in the ring generate an longitudinal polarised $  e^{+}-e^{-} $ beam. 
Regarding an experiment on a $\EEEEG$ reaction, a direct contact term test of the reaction is sensitive
to the size of the electron parallel to the spin axis. The $\EEEEG$ reaction is dominated by the short-range 
Weak Interaction $Z^{0}$, which is particularly evident in the size of the cross section of the $\EEGG$ 
reaction at the $Z^{0}$ polarization of $\sigma_{tot} ~ 50 [pb]$ (Figure \ref{SIGMAtot}) and 
the $\EEEEG$ reaction at the same energy of $\sigma_{tot} = 1505.2 [pb]$ \cite{TOT-cross-Bhabha}.

The AA solution fits perfect in the correlation to e+ e- SPIN - possibility of the LEP - beam to
 the  experiment shown in Fig. \ref{Bhaba-Spin}. The polarisation in the initial and final states
of the electrons is longitudinally polarised. As discussed in Chapter \ref{Theory111}, 
this situation opens up the optimal contact area between the electrons for the theory of the direct contact term.
 
\newpage  
  
 \begin{figure}[htbp]
\begin{center}
 \includegraphics[width=18.0cm,height=15.0cm]{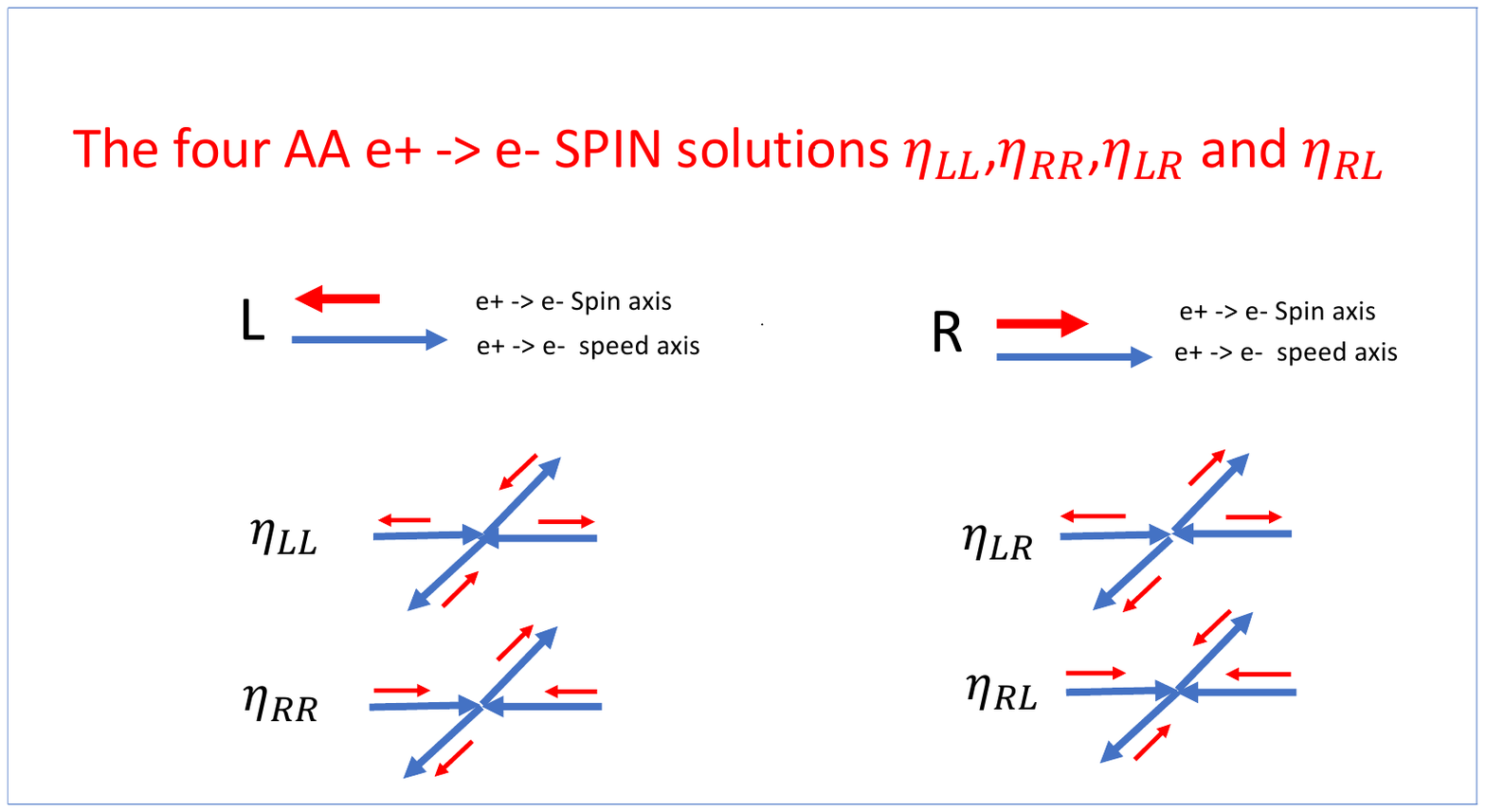}
\end{center}
\caption{The AA solution of the four possible spin correlations $ \eta _{LL},\eta _{RR},\eta _{LR},\eta _{RL} $ of the LEP-beam.}
\label{Bhaba-Spin}
\end{figure}
  
 The energy value $\Lambda$ for the direct contact term of the $\EEEEG$ reaction to the $\EEGG$ 
 reactions $\Lambda_{Bhabha} > \Lambda_{gamma}$ is greater than ONE. The data confirm that the 
 energy scale $\Lambda_{Bhabha} = 10.3 \ \left [ TeV \right ] > \Lambda_{gamma} = 1.3 \ \left [ TeV \right ]$ is.
 The size of the $\Lambda$ value finally defines the size of interaction area of the $\EEEEG$ and  $\EEGG$ reaction,
 since the size of the interaction area is inverse proportional to $\Lambda$.
 If $\Lambda_{Bhabha} > \Lambda_{gamma}$ is greater than ONE the channel of the $\EEEEG$ reaction records a smaller
 interaction size as the $\EEGG$ reaction.

This very interesting result requires a more detailed discussion of the AA model. The mean 
and the one-standard-deviation band of the equation Eq. \ref{Fit-max} correspond to a scale for 
axial-vector contact interactions Eq. \ref{LAMPDA111}.
  
 \begin{equation}
\label{LAMPDA111}
 \Lambda _{+}=10.3\left\{ _{8.7}^{13.1}\right\}_{(+1\sigma )}^{(-1\sigma )} \ TeV
\end{equation}
  
Let us begin the discussion with the central value of our investigation. It is possible to reverse the view point and ask:
What deviations from the predictions of the standard model are to be expected for the total 
cross section and the forward-backward asymmetry of the Bhabha scattering?
The result as a function of the center-of-mass energy is shown in Fig. \ref{Sigma-vvv}.
   
 \begin{figure}[htbp]
\begin{center}
 \includegraphics[width=16.0cm,height=12.0cm]{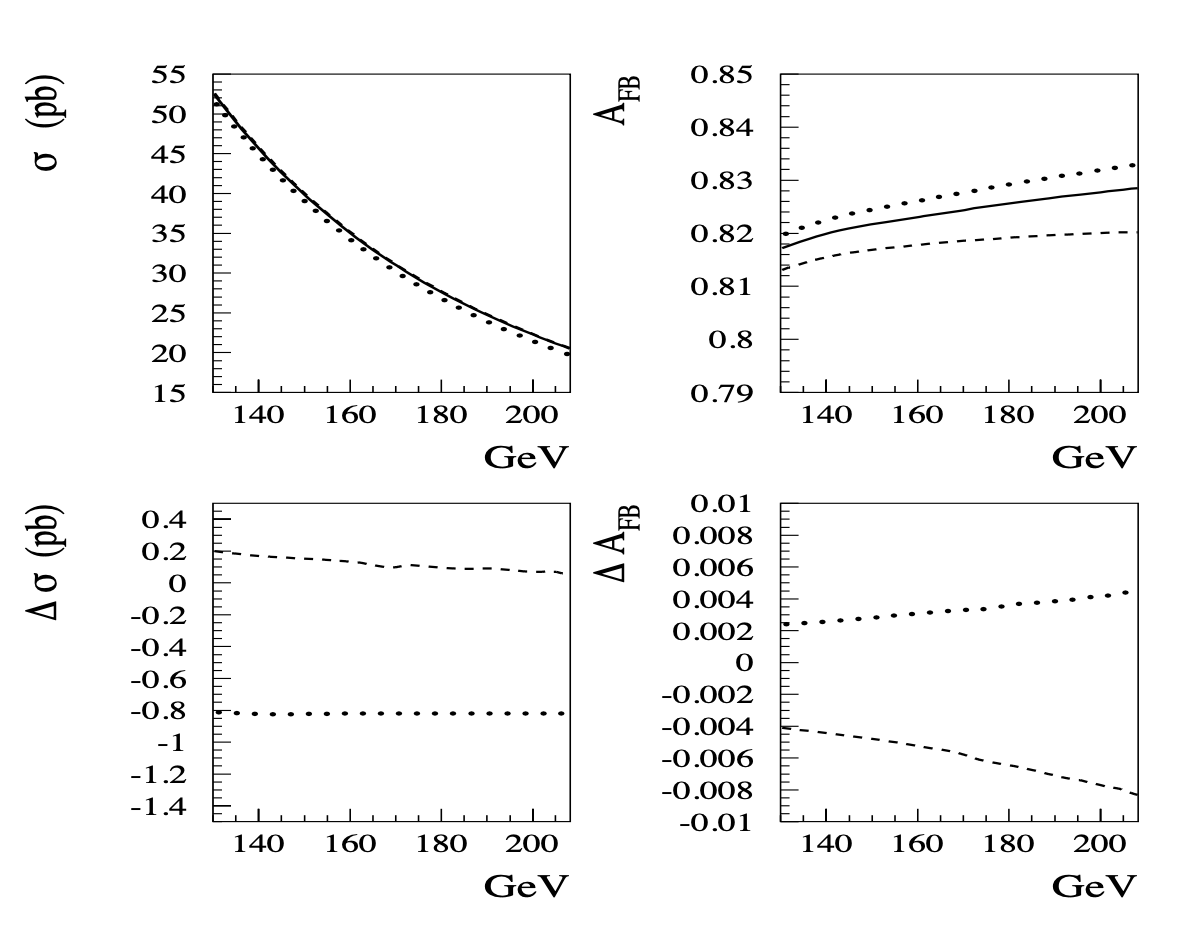}
\end{center}
\caption{Effects of contact interactions on the total cross section and the forward-backward 
asymmetry for Bhabha scattering as a function of the centre-of-mass energy (upper plots) and 
difference to the Standard Model predictions (lower plots): SM - solid, VV model - dotted and 
AA model - dashed. The energy scale is $ \Lambda _{+}=10.3 \ TeV $ in both cases. 
The angular range is $ 44^{0}<\theta <136^{0} $.}
\label{Sigma-vvv}
\end{figure}
      
As shown Fig. \ref{Sigma-vvv} in the AA model changes the total cross-section in this energy range by less than 0.5\%. The relative 
effect is practically constant with increasing energy. For comparison, the combined error 
of the total cross-section of the four experiments for the highest energy point (207 GeV) is 1.2\%.
 An additional factor is the theoretical uncertainty of 1.5\% on the absolute scale, discussed in Section \ref{BABHA-ANALYSIS}.
 On the contrary, the forward-backward asymmetry $ A_{FB}=(\sigma_{F} -\sigma_{B} )/(\sigma _{F}+\sigma _{B}) $ 
 is changed by (-0.0066) in absolute value (or -0.8\% relative) 
 at 183 GeV and by (-0.0083) or (-1.0\%) at 207 GeV. The effect is rising with energy. It is clear that the 
 AA model manifests itself mainly in the forward-backward asymmetry, which is not affected by the 
 uncertainty on the absolute scale.   Another favourable fact is that the statistical error of $A_{FB}$ 
 is given by $\sigma(A_{FB})=\sqrt{(1-A_{FB}^{2}/N}$.  For the same number of events N, this 
 results in a reduction factor of approximately 0.6 due to the large value of the asymmetry of 
 approximately 0.8. However, a disadvantage is this measurement requires recognition 
 of the electron and positron charges, and hence stricter event selection asking for good tracks.   
 
For comparison: The combined forward-backward asymmetry error for three experiments and the highest 
energy point at $ \sim $ 207 GeV is 1.1\%. Since the ALEPH collaboration did not present results 
for $ A_{FB}$ above 183 GeV, the result discussed here is based on the asymmetry 
measurements of the three other collaborations.    

The experiments apply charge-confusion corrections to their forward-backward asymmetry measurements,
equation Eq.\ref{charge-con}.

 \begin{equation}
\label{charge-con}
A_{FB}^{correction}=\frac{A_{FB}^{measured}}{1-2c}
\end{equation}

In Eq. \ref{charge-con}, c is the number of events with incorrect sign assignment. For example, 
if $ c = 0.05 $, an underestimation of charge confusion by 9\% leads to an underestimation of asymmetry by 1\%.
At $ c=0.02 $ one has to underestimate the charge confusion already by 24\% in order to 
underestimate the asymmetry by 1\%. The statistical error in the size of this correction is an 
important component of the systematic error, quoted by the experiments.
The forward-backward asymmetry can be influenced by the interference between initial and 
final state radiation, which is known to change the shape of the differential cross section.
This effect is estimated using the program TOPAZ0. When we change the interference, $A_{FB}$ increases
by 0.1\% with an acollinearity cut of $ 10^{\circ } $ and by 0.07\% with an acollinearity cut of $ 25^{\circ } $.
So the effects are very small, and as the experiments use differential efficiency in the scattering 
angle, the impact on the measured asymmetry values is much smaller.
We can conclude that the interference is under control.

Another interesting point are the preliminary results on contact interactions presented by DELPHI 
\cite{DELPHIBhabha2}, L3 \cite{L3-note-1111} and OPAL \cite{OPALBhabha3}. 
The deviations of $ \varepsilon $ from the SM value show a
positive attraction in all cases, but the contact interaction scale
under discussion here is at the sensitivity limit of individual
experiments. In \cite {BOURILKOV} a limit of $\Lambda _{+}>10.4 \ TeV $ at 95\% confidence level is obtained, 
which covers about 40\% of the favoured region from this analysis at higher energies.

The measurements of Bhabha scattering from 192 to 208 GeV are preliminary. The systematic 
errors and the central values may still change. In order to clarify the situation, the final results 
of the four LEP Collaborations are needed. The best option is to combine the measurements 
of the differential cross sections in the four experiments, as illustrated in Fig. \ref{Ratio-vvvv}..


 \begin{figure}[htbp]
\begin{center}
 \includegraphics[width=14.0cm,height=12.0cm]{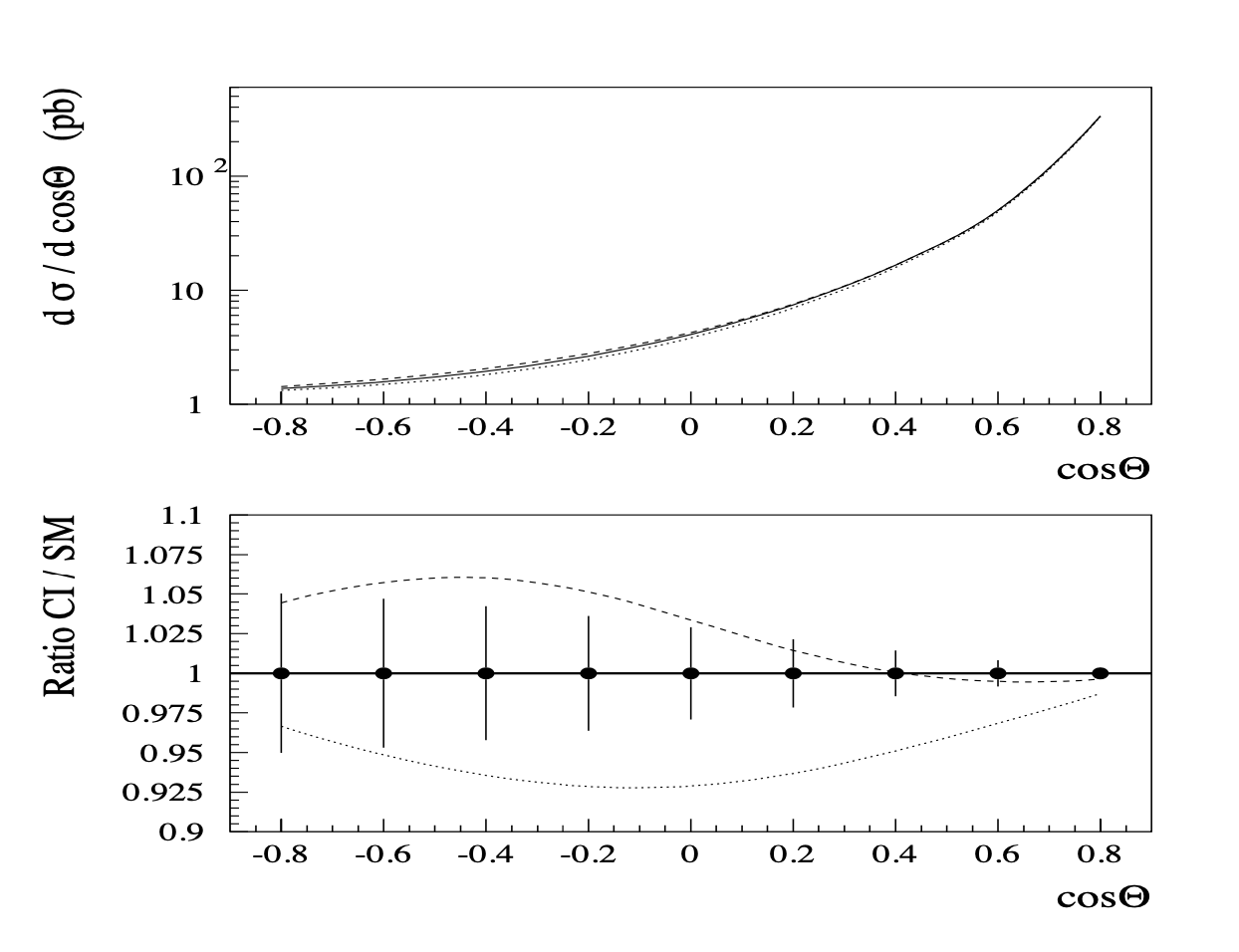}
\end{center}
\caption{Effects of contact interactions on the differential cross section for Bhabha scattering 
for centre-of-mass energy 206.8 GeV (upper plot) and the ratio to the Standard Model predictions 
(lower plot): SM - solid, VV model - dotted and AA model - dashed. The energy scale is 
$ \Lambda _{+}=10.3 \ TeV $ in both cases. The  $``data''$ points show the expected statistical error if we 
combine the measurements of the four LEP collaborations for the six energy points from 192 to 208 GeV.}
\label{Ratio-vvvv}
\end{figure}

The AA model gives a specific signature, enhancing the cross section in the 
central part and backwards, and only slightly reducing the forward peak.

When interpreting the physical meaning of contact interaction scales, we should remember 
that a strong coupling $  g^{2}/4\pi =1 $ for the novel interactions is postulated by convention. 
If we assume a coupling of different strength, $g^{2}/4\pi =\alpha ^{'}$, the limits can be translated as \\
$ \Lambda ^{'}=\sqrt{\alpha ^{'}}\cdot \Lambda $. For the extreme case of a coupling of electromagnetic 
strength the real scale can be around 1 TeV.
 
Measurements of Bhabha scattering above the Z resonance have reached a high level 
of precision. To make optimal use of the large datasets collected during the LEP project 
from 1997 to 2000, new or improved theoretical predictions are advantageous \cite{MonteCarlo2}. 
A joint effort by the four LEP collaborations is needed to answer the question: Does Bhabha 
scattering open a window to new physics in the TeV range? 
 
\subsection{Comparison of theoretical CALCULATIONS of the size of the ELECTRON with the EXPERIMENT.}
\label{Experiment-Theory}

To compare the theory to the experiment  of the electron as geometrical extended object the logic
chain is discussed in in Fig.{\ref{Invest-timing} ``Timing of the investigation of finit size of FP's, performed 
by ETHZ and USTC''. A summery of all these research is shown finally in the end of this chapter in Table \ref{ElectronSize}.

The research is divided into two parts. In the theoretical part:
First, the ETAMFP model is introduced to describe ALL FP's. Second, a 
classical theoretical approach is used to calculate the size of FP's. 
Third, a Spinning superconducting Electro-vacuum soliton is introduced. 
Finally, a wave function is calculated using the substructure of FP's,
via the electric field of a non-point-like electron with the 
Gross-Pitaevskii equation. The table \ref{ElectronSize}
in the upper theoretical part shows the various results of these investigations.

In the experimental  part: Experiments to test the size of FP's via direct contact term. Three
experiments are involved. First: Section \ref{Diff-Cross} ``Finite Size of Electron in $ {\EEG} $ Annihilation 
in differential cross section of centre-of-mass energies 55 - 207 GeV''. Second: Section \ref{TotalCross}
``Hint for a minimal interaction length in $ \EEG $ annihilation in total cross section of center-of-mass 
energies 55 - 207 GeV''.  Third:  section \ref{BABHA1} ``Hint for axial-vector contact interactions in 
the data on $ \EEEEG $ at centre-of-mass energies 192 - 208 GeV''. The Table \ref{ElectronSize} 
shows in the lower experimental part the different results of these investigations.

To compare the size of the electron in the theoretical part, it is particularly necessary, under the simple
assumption of a GYROSKOP, to respect  the Lorentz contraction, which depends on the direction
of the spin axis of the GYROSKOP to the plane of rotation of the GYROSKOP.
In the experimental part of the LEP collider, a different Lorentz contraction must be taken into account, 
because if the electron in the experiment is longitudinally polarised and moves at high speed, 
the size of the direct contact term is reduced by the Lorentz contraction.

The Lorentz contraction for the GYROSKOP is discussed in detail in the classical theoretical 
approach to calculating the size of FP's and is addressed in the explanation of the upper part 
of the table \ref{ElectronSize}. It is important to note that the electron is at rest in the laboratory system; 
the Lorentz contraction is an internal problem of the electron.

The Lorentz contraction of the experimental part of the LEP collider is an external 
problem for the electron, since the electron in the experiment moves at a specific velocity.
If the ELECTRON in the experiment is a geometrically extended object with kinetic energies 
between 55 GeV and 207 GeV, it will be longitudinally polarised along the e+ and e- beams axis.
This means the ELECTRONS will contact each other at a distance $ l $ at solid angle $ \theta =0^{\circ } $.
This Lorentz contraction is then calculated according to the well known equations
Eq.\ref{LC-1}, Eq. \ref{LC-2} and Eq. \ref{LC-3} .

\begin{equation}
\label{LC-1}
 l=l_{0}\sqrt{(1-(v /c^{2}))}
\end{equation}

\begin{equation}
\label{LC-2}
 \sqrt{(1-(v /c^{2}))}=m_{0}/m
\end{equation}

\begin{equation}
\label{LC-3}
l_{0}=l_{experiment}\ast \frac{m}{m_{0}}=l_{experiment}\ast \frac{E_{tot}}{E_{0}}
\end{equation}

In Eq.\ref{LC-1}, Eq. \ref{LC-2} and Eq. \ref{LC-3}, is $l$ is the length at velocity $v$, $l_{0}$ is the length at rest,
$m_{0}$ is the mass at rest, $m$ is the mass at velocity $v$, $E_{tot}$ is the center-of-mass energy
$\sqrt{s}$ of the e+ e- beam, and $E_{0}$ is the energy of the electron at rest. A graphical 
representation of these contractions can be found in Fig. \ref{Lorenz-11}.

\begin{figure}[htbp]
\begin{center}
 \includegraphics[width=16.0cm,height=6.0cm]{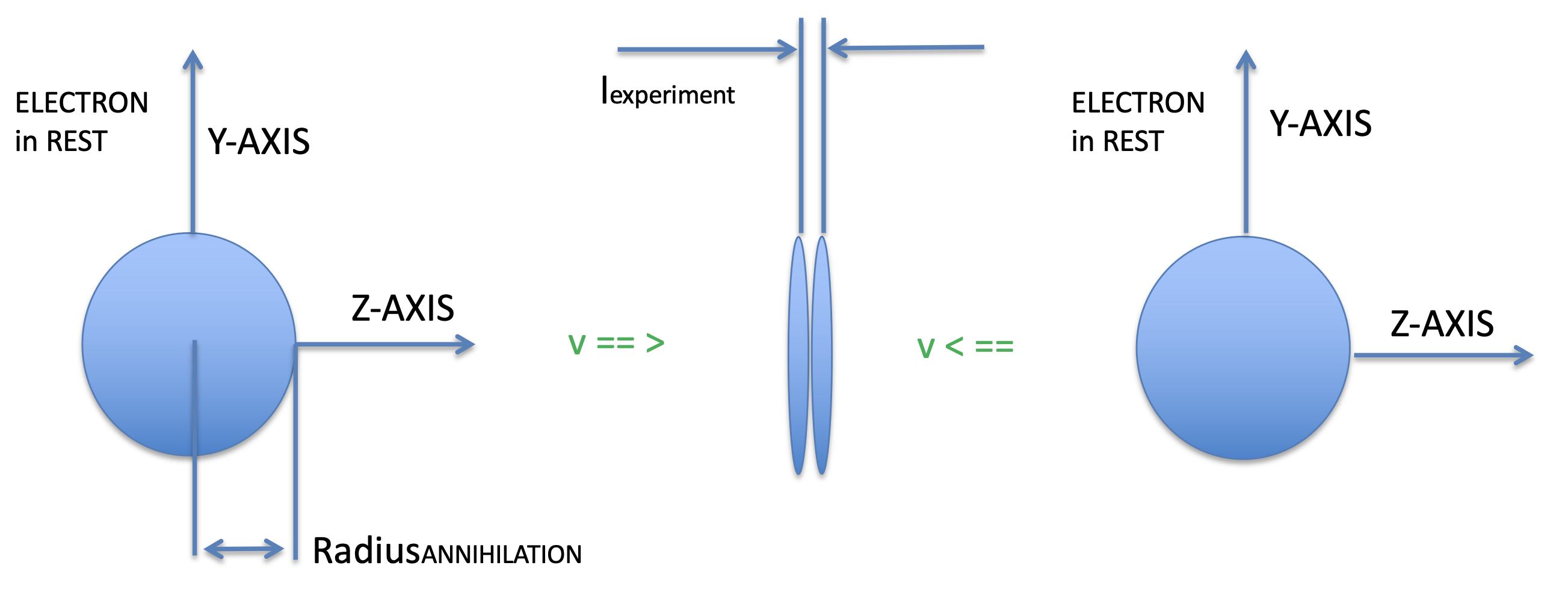}
\end{center}
\caption{Lorentz contraction of the experimental measure $ l_{experiment}$.}
\label{Lorenz-11}
\end{figure}

To compare the experiment with the theory, two different $l_{experiment}$ are important. \\

First the $ l_{experiment} $ were the annihilation of the $ \EEGG $ reaction starts $ Radius_{Anihilation} $
equation Eq. \ref{Lexp11} including the data taken from chapter \ref{Experiment11}.

\begin{equation}
\label{Lexp11}
Radius_{Anihilation}=\frac{1}{2}l_{experiment}\ast \frac{E_{tot}}{E_{0}}=\frac{1}{2}\ast 1.57\ast 10^{-19}\ast 
\frac{207000}{0.51099}=3.18\ast10^{-14}\left [  m \right ]
\end{equation}

It is important to check in the Soliton model at which size of the $ \psi ^{2} $ function the annihilation will start, because
the annihilation will not start at $ \psi ^{2} =0 $. A certain overlap of the wave function is certainly necessary. 
The situation is shown in Fig. \ref{PSI-theta-0}. The problem of the overlap range for $ \psi ^{2} =0 $ is 
discussed in more detail in chapter \ref{Discussion exp theory}.

\begin{figure}[htbp]
\begin{center}
 \includegraphics[width=16.0cm,height=8.0cm]{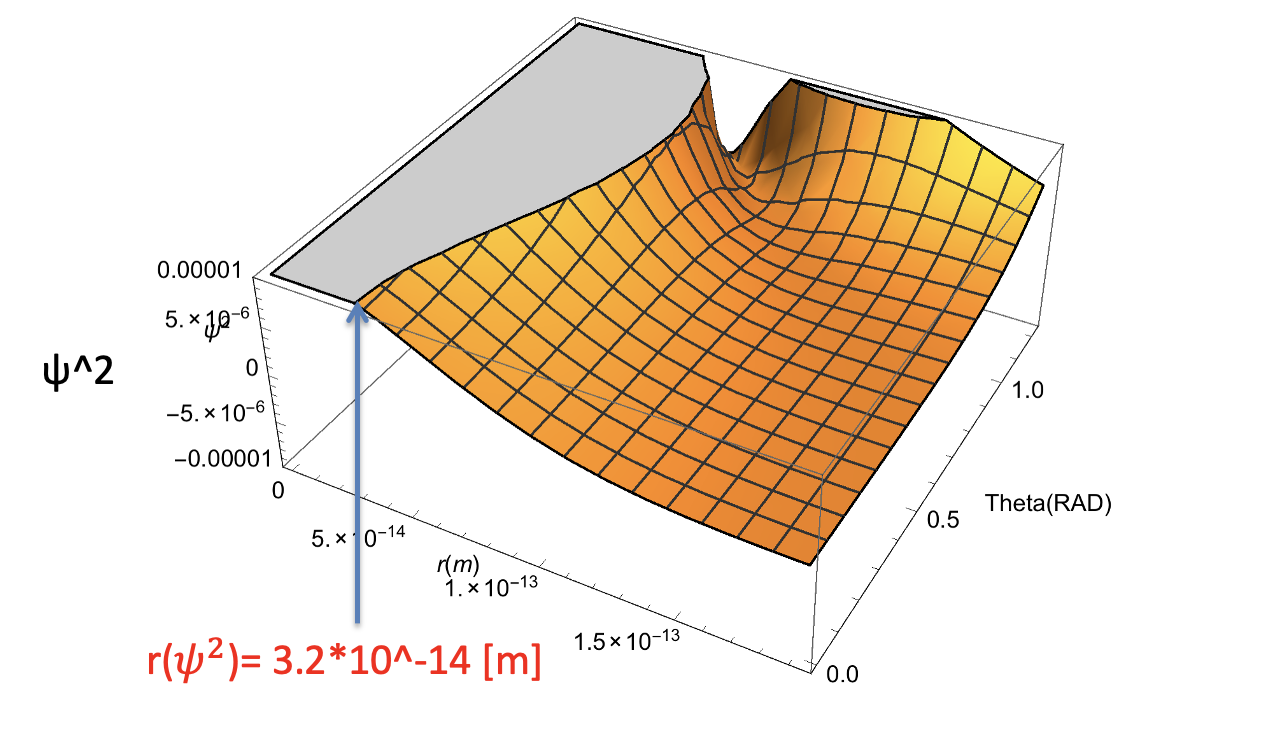}
\end{center}
\caption{The wave function $ \Psi ^{2}=f(\theta =0,r) $ parallel to the spin axis.}
\label{PSI-theta-0}
\end{figure}

The annihilation of the $ \EEGG $ reaction starts after Fig.\ref{PSI-theta-0} at approximately $ \psi ^{2}\approx 1.0\ast 10^{-5} $.\\

The second important parameter $l_{experiment}$ indicates where the annihilation stops, since in the soliton 
model the wave function $\psi^2$ is $\psi^2 = 0$ at this point. See chapter \ref{Gross-Pitaevskii}. This parameter is sensitive in the Bhabha 
reaction experiment $\EEEEG$ chapter \ref{BABHA1} with $l_{experiment} = 1.9^{-20}\left[m\right]$, calculated in equation Eq. \ref{Lexp22}.

\begin{equation}
\label{Lexp22}
Radius_{Anihilation-stop}=\frac{1}{2}l_{experiment}\ast \frac{E_{tot}}{E_{0}}=\frac{1}{2}\ast 1.9\ast 10^{-20}\ast 
\frac{207000}{0.51099}=3.9\ast10^{-15}\left [  m \right ]
\end{equation}

All the discussed parameters are shown in TABLE \ref{ElectronSize}.

\newpage

\begin{table*}
\caption{ Size of ELECTRON THEORY -  EXPERIMENT .}
\label{ElectronSize}
\begin{center}
\resizebox{1\textwidth}{!}{
\begin{tabular*}{1.060\textwidth}{@{}|l|r|r|r|r|r|r|@{} }

\hline

$I -  Theory$ &&&&
                                                            \\  \hline
$1 -  Size$   &$ r_{0}^{\mu }\left [ m \right ]$&$r_{0}^{de}\left [ m \right ]$&$r_{k}\left [ m \right ]$&$r_{e}\left [ m \right ]$
                                                           \\  \hline
$2 - Gyroscope$   &$8.39035\times 10^{-13}$&$5.79910\times 10^{-13}$&$2.33\times 10^{-19}$&$0.105\times 10^{-28}$                                                        
                                                           \\  \hline
$3 - Size - \psi _{1}^{2}$   &$r_{S}(\psi ^{2}=0)$&$R_{k0}$&$ R_{k0}^{g-function}$&$$                                                          
                                                           \\  \hline
$4 - Soliton$   &$ 8.4\times 10^{-13}$&$3.3\times 10^{-15}$&$ 1.0\times 10^{-15}$&$$                                                                                                                    
                                                            \\  \hline
$5 - Size - \psi _{2}^{2}$    &$ r_{S}(\psi ^{2}=1.0\times 10^{-5})$&&&
                                                             \\  \hline
$6 - Soliton$   &$3.2\times 10^{-14}$&&&
                                                             \\  \hline
$II - Experiment$ &&&&          
                                                            \\  \hline   
$7 - Size $   &r=($\EEGG$)&r=($\EEEEG$)&&
                                                           \\ \hline
 $8 - \Lambda \ \sqrt{s}=207 GeV$   &$ 1.254\ TeV $&$ 10.3 \ TeV $&&
                                                            \\ \hline
$9-r_{i}\ \sqrt{s}=207 GeV$   &$ 1.57\times 10^{-19}$&$ 1.9\times 10^{-20}$&&
                                                            \\ \hline  
 $10-r_{i}\ \sqrt{s}=0.0 GeV$   &$ 3.18\times 10^{-14}$&$ 3.9\times 10^{-15}$&&
                                                            \\ \hline 
                                                                                                        
\end{tabular*} 
}
\end{center}
\end{table*}

For clarity, the table \ref{ElectronSize} is sorted according to the logical structure of this paper shown in 
Figure \ref{Invest-timing}. First, the theoretical part is presented, divided into a Gyroscope  
and two Solution types for investigating fundamental particles, particularly the electron. 
This is followed by the experimental part, which covers the search for the vinite size 
of the electron in the reactions $\EEGG$ and $\EEEEG$.
In detail the TABLE \ref{ElectronSize} is sorted in line numbers 1 to 10, to connect
the writing of the different parameters of the whole investigation from Fig.{\ref{Invest-timing} .

NEW

\begin{itemize}
\item 
Line 1 - Size ( Gyroscope ): \\
$ r_{0}^{\mu }\left [ m \right ]$ - $r_{0}^{de}\left [ m \right ]$ - $r_{k}\left [ m \right ]$ - $r_{e}\left [ m \right ]$ are displyed in 
Fig.\ref{SketchElectronContracted}. \\
$ r_{0}^{\mu }\left [ m \right ]$ radius of the Gyroscope parallel to the spin axis calculate from the magnetic moment of the electron $ \mu $
Eq.(\ref{RadiusL}). \\
$r_{0}^{de}\left [ m \right ]$ radius of the Gyroscope parallel to the spin axis calculate from the electric moment of the electron $ d_{e} $
Eq.(\ref{r0fuctionK}) and Table \ref{electronPARAMETERSdipol}.\\
$r_{k}\left [ m \right ]$ Kernel radius from in Eq.(\ref{KernelRadius1}). \\
$r_{e}\left [ m \right ]$ Lorentz contracted radius of the electron perpendicular to spin axis of the electron Eq.(\ref{ElectronDipolRadius}).
\item 
Line 2 - Gyroscope:\\
Inserted are the numerical values from Eq.(\ref{RadiusL}) , Eq.(\ref{r0fuctionK}) and Tab.\ref{electronPARAMETERSdipol}, Eq.(\ref{KernelRadius1}) 
and Eq.(\ref{Comparisondeexpre}). The numerical value from $r_{k}=2.33 \times 10^{-19}\left [ m \right ]$ is the middle value from 
Table \ref{ElectronKernelRadius} for $ \Lambda _{GUT}=10^{15}\left [ GeV \right ] $. The energy at the tree interactions merge in on point
is not known very well. The whole possible so far known range of $ \Lambda _{GUT} $ is $ 10^{13}\left [ GeV \right ]<\Lambda _{GUT}<10^{17}\left [ GeV \right ] $
this in the range for $r_{k} $ from $5.02\times 10^{-22}\left [ m \right ]<r_{k}<1.08\times 10^{-16}\left [ m \right ]$.
\item 
Line 3 - Size - $\Psi _{1}^{2}$ ( Soliton ): \\
$r_{S}(\psi ^{2}=0)$ The radius of the wave function $\psi ^{2} $ at the angle 
$ \theta =0 $ parallel to the axis of the electron spin for $\psi ^{2}=0 $ displayed in Fig. \ref{PSI-theta-11} 
and table Tab.~\ref{Parametrs-Electron}.\\
$R_{k0}$: The radius in the Soliton model Eq.(\ref{Step11}) the wave function $\psi ^{2} $ must be set to $\psi ^{2}=0 $ to fulfil
the requirement of the ETAMFP model to distribute the outside charge of the electron $ e $ to inside $ 2/3-e$  to $ 1/3-e $. \\
$ R_{k0}^{g-function}$ the radius of a similar possibility to set $\psi ^{2}=0 $ shown in Fig.\ref{g-function-2}. 
\item 
Line 4 - Soliton : \\
All numerical values were displayed of $r_{S}(\psi ^{2}=0)$, $R_{k0}$ and $ R_{k0}^{g-function}$.
\item 
Line 5 - Size - $\Psi _{2}^{2}$ : \\
$ r_{S}(\psi ^{2}=1.0\times 10^{-5})$ is the radius of the $ \psi ^{2} $ function for the angle $ \theta =0 $ parallel to the spin
axis of the electron the experimental value shows the annihilation will start shown in Fig. \ref{PSI-theta-0}.
\item 
Line 6 - Soliton : \\
The numerical value for the radius in [ m ] taken from rom Fig. \ref{PSI-theta-0}.
\item 
Line 7 - Size : \\
The parameters $ \Lambda $ , $ r_{i} $ at $ \sqrt{s}=207 \left [ GeV \right ] $ and $ r_{i} $ at $ \sqrt{s}=0.0 \left [ GeV \right ]$
of the experiment of the direct contact term of the $ \EEGG $ and $ \EEEEG $ reaction.
\item 
Line 8 - $\Lambda \ \sqrt{s}=207 GeV$ :\\
Displayed are the numerical value of $\Lambda$ in TeV taken from the $ \EEGG $ reaction of the chapter \ref{fitContact11} and the $ \EEEEG $ reaction from Eq.\ref{LAMDA11}.
\item 
Line 9 - $ r_{i}\ \sqrt{s}=207 GeV$:\\
Displayed are the numerical value of $r_{i}$ calculated from Eq.\ref{DIR11} taken from the $ \EEGG $ reaction from Table \ref{electron2}
and the BHABHA reaction $ \EEEEG $ calculated via Eq.\ref{DIR11} for $\Lambda=10.3 TeV $.
\item 
Line 10 - $ r_{i}\ \sqrt{s}=0.0 GeV$:\\
Displayed are the numerical value of $r_{i}$ from Line - 9 but corrected to the electron in rest $ \sqrt{s}=0.0 GeV$. For the
$ \EEGG $ reaction Eq. \ref{Lexp11} and for the $ \EEEEG $ reaction Eq. \ref{Lexp22}.
\end{itemize}
 
\subsubsection{Discussion of comparison of the EXPERIMENT with the theoretical CALCULATIONS of the size of the ELECTRON.}
\label{Discussion exp theory}
 
To compare the theoretical and experimental development of this article, it is helpful to consider information from the theoretical section:
the electron as a gyroscope, the electron as a soliton, and finally the experimental investigation of the geometric extent of the electron
with the search for a finite size using the direct contact term. The various parameters are listed in Table \ref{ElectronSize};
these parameters are visualised for the convenience  of the reader in Fig. \ref{Gyro-Soliton}. 
 
\begin{figure}[htbp]
\begin{center}
 \includegraphics[width=17.0cm,height=6.5cm]{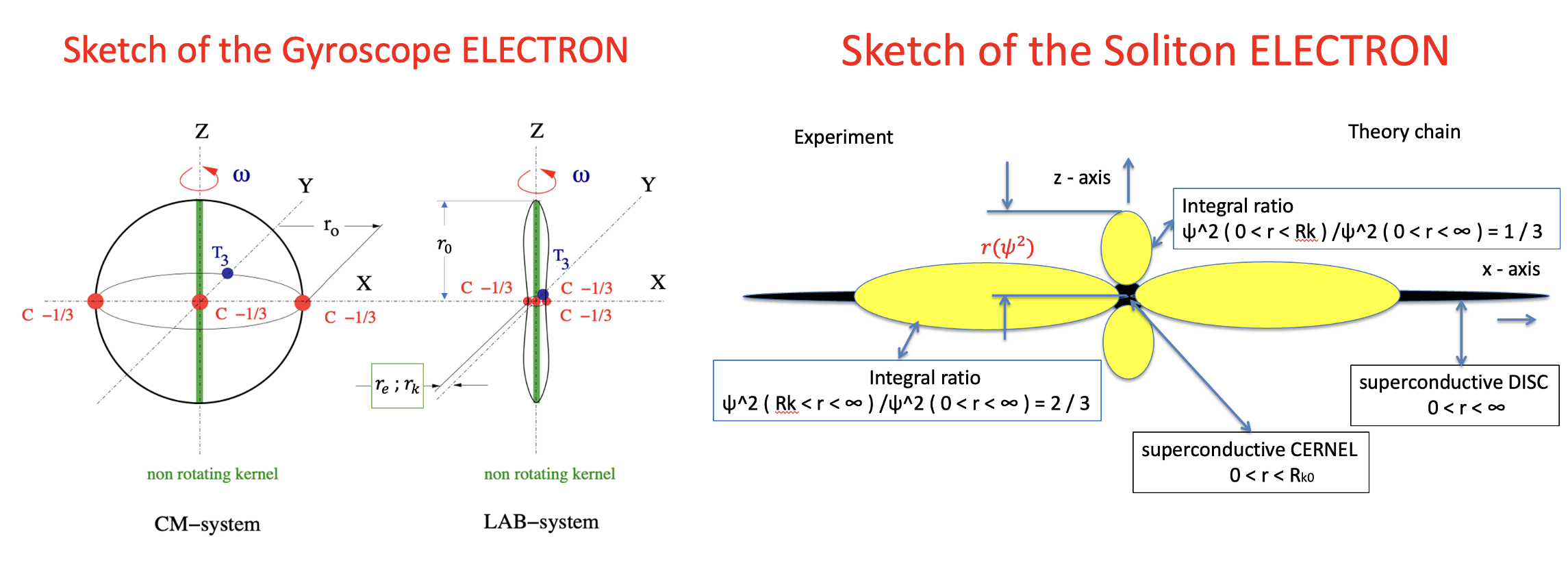}
\end{center}
\caption{Sketch of the Gyroscope and Soliton ELECTRON}
\label{Gyro-Soliton}
\end{figure}

In the Gyroscope model, the electron exists in the LAB system and is highly Lorenz-contracted.
The only non-contracted radii are $r_{0}^{\mu}$ and $r_{0}^{de}$. The radii, $r_{0}^{\mu}$,
and $r_{0}^{de}$, can be calculated from the magnetic moment $\mu_{e}$ and the 
electric dipole moment $d_{e}$ of the electron, displayed in table ~\ref{Parameterelectron}. 
Both radii agree within the margin of error. The kernel radius, $r_{k}$, is calculated from the 
history since the Big Bang. The radius is inside the electron with high probability NOT Lorentz corrected.
See for example Fig.\ref{PrecissonradiusElectron}. Table \ref{ElectronKernelRadius} 
shows an overview of the range $\Lambda_{GUT}$ as a 
function of the kernel radius $r_{K}$. Since the experiment deals with the weak and 
electromagnetic interactions, which are depicted in Fig. {\ref{Running-coupling}
the point at which both interactions converge is of interest for the search for the radius $r_{K}$.
This point lies in the upper range of Table \ref{ElectronKernelRadius}
$ \Lambda _{GUT}\approx 10^{17}\left [ GeV \right ] $ and $ r_{k}\approx 5.0\times 10^{-15}\left [ m \right ] $.
The radius $ r_{k} $ is in the range of the radius of the soliton model $R_{k0}=3.3\times 10^{-15}\left [ m \right ] $. 
This suggests that $ r_{k}\sim R_{k0}$.

When comparing the soliton model with the gyroscope model, it should be noted that distances such as $ r(\psi ^{2}) $
are not exactly defined at big distances, since the wave function of the soliton model is not described by sharp edges, but is a 
smooth function without sharp cuts. For this reason, it is necessary to verify the numerical value of $ 0.0<\psi ^{2}<1.0 $
as a function of the distance $ r $. Considering Fig. \ref{PSI-theta-0}, the wave function $ \psi ^{2} =0 $ is given by 
$ r_{S}=8.4\times 10^{-13}\left [ m \right ] $. This yields $r_{S}(\psi ^{2}=0)\approx r_{0}^{\mu }$, an interesting agreement
with two completely different models.

To compare the Gyroscope and Soliton models with the experiment, the Lorentz contraction
of the various parameters must be taken into account. It should also be noted that the spin 
axis of the $e+e-$ beam is parallel to the velocity axis of the particle.
Therefore, the $\EEGG$ reaction and the $\EEEEG$ reaction depend only on the radius
$r_{0}$ of the gyroscope model and $r(\psi^{2})$ of the soliton model, respectively.
In the $\EEGG$ reaction, the annihilation of the reaction starts 
at $r=(\EEGG)=3.18\times 10^{-14}\left [ m \right ]$. To verify the soliton model,
it is necessary to determine the value of $\psi^{2}$ at this distance. According to Fig. \ref{PSI-theta-0}, it is:
$\psi^{2}=1.0\times 10^{-5}$. This remains a very good agreement between experiment and the soliton model,
since annihilation requires an overlap of the wave functions. A precise calculation of 
this fact does not yet exist. In the Soliton model, there exists a kernel radius 
$R_{k0}=3.3\times 10^{-15}\left [ m \right ]$, which results from a step function to set
$\psi^{2}=0$. This radius is not affected by the problem $r(\psi^{2})$ for large distances, since
the step function calculates a well-defined number. This radius was tested in the $\EEEEG$ reaction and yields a value of
$r=(\EEEEG)=3.9\times 10^{-15}\left [ m \right ]$. Indeed, $R_{k0}\sim r(\EEEEG)$ represents 
a very important agreement between theory and experiment.

\subsection{Conclusion of the investigation of finit size of FP's, performed by ETHZ and USTC.}
\label{OVERALLconclusion}

The effort of the total investigation of finit size of FP's, performed by ETHZ and USTC is shown in Fig.{\ref{Invest-timing}.
We introduced two totally independent circles, a theoretical circle and an experimental circle to investigate the 
the ``Substructure of FUNDAMENTAL PARTICLE''. 

In the theoretical part four major steps are introduced: ``The ETAMFP model. TWO New particles describe ALL FP's'', 
``Classical theoretical approach to calculate the size of FP's'',
``Spinning superconducting Electro - vacuum soliton'' and ``Calculation of a wave function using the substructure of
FP's, via the E-field of a non point like electron with the Gross - Pitaevskii equation''.
In the experimental  circle we investigated, three experiments to test the size of FP's via direct contact term. In the 
$ \EEGG $ reaction we tested the differential and absolut cross section.  Finally the BHABHA reaction $ \EEEEG $
was testet. In the most important part of all these investigations we introduced a chapter \ref{Experiment-Theory}
``Comparison of theoretical data with experiment''.

To abandon the QED requirement that all FB's are points, we introduced the ETAMFP model. 
The model, we present in three chapters: Firstly, in the chapter \ref{History ETAMPF} ``The 
history of the ETAMFP model on a new substructure
of fundamental particles'', we discussed the history of the substructure
of fundamental particles from 1738 to the 20th century. 

Secondly, in the chapter \ref{Empirical toy ansatz} ``Empirical Toy Approach to the Microstructure of Fundamental Particles (ETAMFP)'',
we introduced a ``Schematic development of geometrical extended FP in the ETAMPF model''
in Fig. {\ref{Timing-FP} and the ``Basic scheme of extended fundamental particle in the ETAMFP model'' in 
Fig. {\ref{Basic-ETAMFP}. In the subchapter chapter \ref{Conclusion ETAMPF} ``Conclusion on the scheme of geometrically extended fundamental particles and antiparticles'', we concluded that the ETAMFP model is generic to sort all fundamental particles in on common scheme.

Thirdly, in the chapter in chapter  \ref{Scheme of geometrical extended FB's } 
``Scheme of geometrical extended fundamental particles and anti-particles'' 
we introduces all fundamental particles after Tab.~\ref{Hypercharge}
from the subchapter \ref{Scheme lightest left-handed FB} of the 
``Scheme of the lightest left-handed fundamental particles''
to the subchapter \ref{Scheme Gauge Bosons} ``Scheme of the Gauge Bosons''.
In the subchapter \ref{Exited-Fermions} ``Excited Fermions'', we investigate the possibility of the existence 
of fundamental vibrational states of fermions, as shown in Fig. \ref{Excited-states1}. 
Based on this, a simple model for excited fermions can be developed, as shown in Fig. \ref{Excited-states2}. 
In the subchapter \ref{Conclusion ETAMPF} ``Conclusion on the scheme of geometrically extended 
fundamental particles and antiparticles'', we summarise the development of the ETAMPF model, 
starting from the hypothesis that the temporal and spatial evolution of the history of the very young 
universe forms the geometric extension of the fundamental particles and antiparticles and makes 
it possible to classify all fundamental particles into a common scheme.

The ETAMPF scheme opens a new interpretation of the ``Wave-particle duality''. The
FP's are ONE single particle with a KERNEL pus an OUTSIDE HOMOGEN STRUCTURE.
The FP's are NOT points any more. In other words TWO distinct features are intergraded in
ONE particle. It is also important that the different charges of the FP's are measured ONLY for the entire
particle. The detailed distribution of these charges in the scheme has not yet been total clarified.

In chapter \ref{GYROSCOPE} ``Application of special relativity in a classical approach 
to study the electron as a Lorentz contracted GYROSCOPE'', we are able to calculate 
different sizes of the electron parallel to the SPIN axis and perpendicular to the SPIN axis. The
input quantities to this calculation was the magnetic moment $  \mu $ and electric 
dipole moment $ d_{e} $ of the electron respecting  ``Special Relativity''.

Chapter \ref{Electro vacuum soliton} ``Particle - like electro - vacuum soliton related to General Relativity'', is the theoretical
most demanding chapter,  based on the very important article from I. Dymnikove \cite{Irina111}. In detail the history of this calculation
is written in the beginning of Chapter \ref{Electro vacuum soliton}. In this paper a de Sitter vacuum core was used, forming 
a model for a super contacting disc SOLITON of the electron, out of an attractive and repulsive
force. It was possible to calculate electric and magnetic fields as $  E_{fiel},B_{field}=f(r,\theta) $ of this geometrical extended object. 

The equation $E_{fiel},B_{field}=f(r,\theta)$ from the chapter \ref{Electro vacuum soliton}, 
opens the possibility, in the chapter \ref{Gross-Pitaevskii} ``Gross-Pitaevskii Equation and 
Electron Wavefunction'', of developing a wave function for a geometrically extended electron.
This electron wave function differs fundamentally from the established QED electron wave function.
This type of function describes the internal structure of an electron such as $\psi_{electron} =f(r,\theta )$. 
Such a wave function is so far unknown and establishes a link between QED and general relativity.

In section \ref{Experiment11} ``Experiments to test the finite size of FP's'', 
three experiments are discussed in detail. These experiments serve to test the 
size of electrons using the direct contact term.

In section \ref{Diff-Cross} ``Finite Size of Electron in $ {\EEG} $ Annihilation in differential cross section 
of centre-of-mass energies 55 - 207 GeV'' the data of the differential cross section on the $ \EEGG $ reaction
were used to establish limits on the size of the electron. In section \ref{gFitme}  
``Global  {$ \chi ^2$}-Test for Heavy Electron $ m_{e*} $'' it was possible to determine a mass of an 
exited electron $ m_{e*} $.

In section \ref{TotalCross} ``Hint for a minimal interaction length in $ \EEG $ annihilation in total cross section 
of center-of-mass energies 55 - 207 GeV'', the data of the total cross section of the $
 \EEGG $ reaction were used to establish limits for the size of the electron.
 
In section \ref{BABHA1} ``Hint for axial-vector contact interactions in the data on $ \EEEEG $ at centre-of-mass energies 192 - 208 GeV''.
The BHABHA cross-section data for the $\EEEEG$ reaction were used to establish limits on the electron size.

The section \ref{Experiment-Theory} ``Comparison of theoretical CALCULATIONS of the size of 
the ELECTRON with the EXPERIMENT'' presents a comparison of theoretical calculations of electron size with experimental results.
This section is the important chapter to test are the theoretical models to calculate parameters of the size of the electron
in agreement with the three experiments unter discussion. The parameters of the theoretical and experiential part displayed in
TABLE \ref{ElectronSize}.

In the case of the ``Gyroscope'' model, the radius of the electron GYROSCOPE coincides parallel to the 
spin axis, which is calculated via the magnetic moment of the electron and the electric moment of 
the electron $ r_{0}^{\mu }\sim r_{0}^{de} $.
The kernel radius $r_{k}$ is a consequence of the Big Bang history and is not directly related to the Gyroscope.
The point of interaction between the weak and electromagnetic forces, which is important for the kernel radius, 
is shown in Fig. {\ref{Running-coupling} and is only known within a specific range.
However, let us consider the entire range $10^{13}\left[GeV\right] < \Lambda_{GUT} < 10^{17}\left[GeV\right]$
the meeting point of week and electromagnetic interaction is located at  the upper end of the Table.\ref{ElectronKernelRadius} at
$ \Lambda _{GUT}\approx 10^{17}\left [ GeV \right ] $ and $ r_{k}\approx 5.0\times 10^{-15}\left [ m \right ] $ ,
this is close to the range of the radius of the soliton model $R_{k0}=3.3\times 10^{-15}\left [ m \right ] $ this suggests $ r_{k}\sim R_{k0}$.
The radius $ r_{e} $ is shifted close to zero in the GYROSCOPE model, since the velocity of the 
outer charge is $ \beta \simeq 1.0 $ according to Tab.\ref{PrecissionKbetaB}.

In the SOLITON model, the radii parallel to the spin axis are similar to the radius of the GYROSCOPE.
$ r_{0}^{\mu }\simeq r_{0}^{de}\simeq r_{S}(\Psi^{2} =0) $.

To compare the radius $r_{s}$ parallel to the spin axis from theory and experiment, it is essential to consider Lorentz contraction.
We choose for comparison the CM enrgy $\sqrt{s}=207 \left [ GeV \right ]$, since the discrepancy between experiment and theory is maximal 
as shown in Figure \ref{Ratio-2}. Under these circumstances, the radius from the SOLITON model is very 
similar to the $\EEGG$ experiment $r_{s}(\Psi ^{2}=1.0\times 10^{-5})\simeq r \to (\EEGG)$ shown in Fig. \ref{PSI-theta-0}.
A precise calculation of the required overlap in $\Psi^{2}$ to trigger the annihilation of the $\EEGG$ 
reaction does not yet exist. However, a starting value of $ \Psi ^{2}=1.0\times 10^{-5} $ seems realistic. 
It is important to emphasise that this $\EEGG$ experiment is sensitive to long-range QED forces. 
The kernel radius $ R_{k0} $ of the soliton model is similar to the radius $ r $ of the 
$\EEEEG$ reaction $ R_{k0}\simeq r \to (\EEEEG) $.
The reaction $\EEEEG$ is sensitive to the short-range Weak interaction. This fact confirms the 
theoretical assumption of a second inner kernel radius of the electron. The same findings are visible 
in the comparison of the $ \Lambda $  and $ r_{i} $ values of the $ {\EEG} $ and $ \EEEEG $ reaction. 

In our research, we focused on a fined size of the electron, but as discussed in section
\ref{gFitme}, we also detected an excited electron mass $ m_{e^{*}} $ shown in Table \ref{electron1}. 
This result is incorporated into the following discussion of the investigation of the direct contact term.

Extensive measurements and analyses were conducted to search for quark and lepton compositeness in
contact interaction~\cite{BHABA}, specifically in $ \EEGG $ and the Bhabha channel as shown in Figure~\ref{Feynman1}.
At the Z$^{0}$ pole, the~$ \EEGG $ reaction exhibits a suppression of the $s$-channel, resulting in
$R(\Lambda_{6})= 0.999$, demonstrating very good agreement.

Alternatively, in a Bhabha-like reaction ($ e^{+} e^{-} \rightarrow e^{+} e^{-} ( \gamma ) $)
another QED test is used to search for $ m_{e^{*}} $. This utilizes pair production in the
s-channel through $\gamma$ and Z boson exchange, similar to Bhabha scattering 
\cite{BOURILKOV} as shown in Figure \ref{Feynman1}.
The mass values or limits for an excited electron depend on the test reaction used for its study
and the theoretical interpretation of $\Lambda$ values. For~example, in~the case of the
$ \EEGG $ reaction, these values can be obtained from the Lagrangian Eq.\ref{Litke1111} 
(or Equation~(4) in~Ref. \cite{L3B}).
The L3 Collaboration \cite{L3limit} set the lower limits on 95\% CL for pair production of neutral
heavy leptons in the mass range $m_{L}^{*} > 102.7$~GeV to $m_{L}^{*} > 80.5$~GeV, 
depending on the model used (Dirac or Majorana). 
L3 Collaboration also set the lower limits at 95\% CL for pair-produced
charged heavy leptons from $m_{L}^{*} > 102.6$~GeV to $m_{L}^{*} > 100.8$~GeV.
Similarly, the~OPAL Collaboration set the lower limits on 95\% CL for long-lived charged heavy
leptons and charginos by $ (m_{L^{*} }, m_{\rm chargino}) > 102.0$~GeV, as~well as lower
limits on the neutral ($ L^{0} $) and charged ($ L^{\pm}) $ heavy leptons~\cite{OPLALlimit,OPLALlimit1}.
The HERA H1 Collaboration searched for heavy leptons and obtained best-fit limits for  $e^{*}$
production in the HERA mass range in the $ \gamma $ final state, with~a composite scale parameter
$\Lambda$ excluding values below approximately 300~GeV \cite{HERA}.

The CMS Collaboration is searching for long-lived charged particles in pp
collisions~\cite{CMS1} using a modified Drell--Yan production process.
This involves the annihilation of a quark and an antiquark from two different hadrons,
producing a pair of leptons through the exchange of a virtual photon or
$Z^0$ in the s-channel. The~study excluded Drell--Yan signals with the charge,
$|Q|=1e$, below masses of 574~GeV/$c^2$.

The latest experimental data from hadronic machines ~\cite{H1E1,D0E1,ATLASE1,CMSE1} provide no evidence of excited leptons.
They establish an exclusion limit for the mass of the excited electron of $m_{e^{*}} = 3.0 $ TeV.
This applies to reactions with single production such as $ep\rightarrow e^*X\rightarrow\gamma X$ (where $X$ denotes all particles).
This reaction differs from the double $e^*$ production studied in the scattering reaction $e^{+} e^{-}$.
The LHC experiments rigorously investigate lepton--quark contact interactions by analyzing
high-mass oppositely charged lepton pairs produced
through the $q\bar q\rightarrow ll$ Drell--Yan process~\cite{dy1,dy2}.
However, it is important to note that the result obtained from the Drell--Yan
process analysis is not directly comparable to the result obtained from
the $e^+e^-$ annihilation analysis. 
These two methods involve different experimental techniques,
and are therefore sensitive to different types of cut-off parameters,
leading to different interpretations regarding the size of an electron.
Future particle accelerators with higher center-of-mass energy
and luminosity are expected to continue the search for excited leptons and contact interactions.
The production of two photons at large angles in $e^+e^-$ annihilation has been
suggested as a way to measure the luminosity of future circular and linear colliders.
These colliders, including the Future Circular Collider, FCC-ee~\cite{FCCee1},
the Circular Electron Positron Collider, CEPC~\cite{CEPC1}, the International Linear Collider, ILC~\cite{ILC1},
and the Compact Linear Collider,  CLIC~\cite{CLIC1}, 
 are designed to have polarised beams and can be used to test the accuracy
of the Standard Model and search for signals of new~physics.

The high-precision measurements of the electron's magnetic dipole moment,
$(g-2){e}$, provide a powerful tool for determining the electron radius
\cite{g_e1,g_e2, Dipol}. If non-standard contributions
to $(g-2){e}$ scale linearly with the electron mass, the estimated limit for 
the electron radius is on the order of $10^{-21}$ cm.
However, if~these contributions scale quadratically with the electron mass, 
as~predicted by chiral symmetry~\cite{g_e1}, the~bound becomes weaker and is at the
level of $10^{-16}$ cm. Importantly, this does not contradict the result obtained
from the measurements of the direct contact term in the annihilation reaction,
as obtained in our~analysis. We used this information, along with the electric dipole moment, to develop a 
completely different model of a GYROSKOP-like electron. This classical model 
agrees with the soliton model and ultimately with experimental data.

The exchange of the excited electron does not produce any non-zero polarisation effects
in the case of only one polarised beam \cite{polar1}, at least in the lowest order of perturbation theory.
This is because the reaction that produces the excited electron preserves spatial parity,
which can be derived from the expression for the Lagrangian Eq. \ref{Litke1111}.
On the other hand, the contact interaction influences the polarisation observables
in the reaction $\EEGG$ if the initial particles are polarised.
In the general case, the contact interaction violates spatial parity \cite{polar1}.
Therefore, non-zero observables only appear if
one of the beams is polarised. The pure QED mechanism of this reaction,
without considering radiation corrections, does not produce such polarisation effects.
However, electroweak corrections at the one-loop level, as shown in Ref. \cite{polar2})
can introduce an additional term to the amplitude of this process, which violates parity and
can therefore lead to non-zero polarisation magnitudes. The future colliders mentioned above,
which use polarised beams, offer a promising opportunity
for the experimental investigation of the polarisation effects of contact interactions.

Considering the question of whether the non-pointiness of the electron is observed in 
the annihilation reaction, an approach can be chosen using a model 
proposed in the references \cite{Irina,SIZE}, which is based on the superconducting 
cores  \cite{Super,Super1,Super2,Super3}.
The model proposes an electromagnetic, spinning soliton for the electron,
accompanied by a de Sitter vacuum disk that generates electric and magnetic fields.
Within this framework, a wave function for the electric field can be constructed.

In conjunction with the model from Ref. \cite{SIZE,SIZE01,SIZE1}, this wave 
function results in a outer radius of $ 3.2\times 10^{-14} $ [m] in the SOLITON model, compared to the 
experimental value of $ 3.18\times 10^{-14} $ [m]. Furthermore, using the 
SOLITON model in conjunction with the ETAMPF model, a Superconducting 
INNER-Kernel RADIUS for the electron wave function at $\psi^{2}=0$ of $R_{ko}=3.3\ast 10^{-15}$[m] 
was determined, compared to the experimental value of $R_{ko}=3.9\times 10^{-15}$[m]. The numerical 
results are given in line 4, 6, and 10 of the table \ref{ElectronSize}. All numbers refer to the LAB system.

The numerical agreement between the calculations in Ref. \cite{Irina,SIZE,SIZE01,SIZE1} and 
the experimental results suggests a possible manifestation of the non-point-like nature of the 
electron within the context of the studies in Ref. \cite{Irina,SIZE,SIZE01,SIZE1}.

Based on the discussion so far, we come to the conclusion that the electron exhibit two types of extended interiors, 
depending on the experimental investigations.  

The $\EEGG$ reaction investigates only the long-range QED interaction, while the Weak interaction is 
suppressed via $Z^0$ by the conservation of angular momentum. The Bhabha reaction, 
$e^+e^- \rightarrow e^+e^- (\gamma)$, includes short-range Weak interactions and QED interactions.
Due to the significantly larger differential cross section in the Bhabha channel compared to the pure 
QED channel, the Bhabha channel dominates.
The maximum deviation  from the Standard Theory to the Direct Contact term Ansatz was 
measured at a CM energy $\sqrt{s}$ = 207 GeV shown in Fig. \ref{Ratio-2}. 
At this CM energy $\sqrt{s}$ is the minimum value of $\Lambda_{top}$ for the 
$\EEGG$ reaction, of $\Lambda_{top}$  equal $1253.53 \pm 226$ GeV and
for the Bhabha reaction $e^+e^- \rightarrow e^+e^- (\gamma)$, $\Lambda_{top}$ 
is significantly higher at $10.3^{+2.8}_{-1.6}$ TeV.
This leads to an eightfold reduction in the geometric extension
described by $\Lambda_{top}$ of the electron, as shown in 
Table \ref{ElectronSize}, line 8. Considering the size of the electron at a center-of-mass 
energy of $\sqrt{s}$ = 0.0 GeV, the $\EEGG$ reaction of the long-term QED reaction 
results in a radius of $r=3.18\times 10^{-14} $[m], and the short-term Bhabha reaction, 
$e^+e^- \rightarrow e^+e^- (\gamma)$, results in a radius of $r$ for the electron, 
in a radius of $ r=3.9\times 10^{-15} $[m] , also an eightfold reduction 
shown line 10 in Table \ref{ElectronSize}). Based on the observed data and 
analyses, it is logical to assume that the electron could have not just one, 
but two different interiors: an outer shell and an inner core.
The electron combines these two properties, as theories for example \cite{Irina} suggest. 
The possibilities are truly very important and open up new window for further research and 
discoveries in the field of particle physics.
 
If the point-like character of the fundamental particles in Quantum Mechanics is abandoned,
the ETAMPF model opens up the possibility of sorting ALL fundamental particles in a common scheme,
and the soliton model developed within the framework of General Relativity,
opens a window for the merging of General Relativity and Quantum Mechanics.

\section{Acknowledgments.}

We express our gratitude to Andr\'{e} Rubbia, Claude Becker, Xiaolian Wang, Ziping Zhang,
Zizong Xu and Alexander and S. Sakharov for their years of unwavering support for this project. In particular, 
we express our gratitude to Irina Dymnicova for her direct contribution to the soliton model in this work. 
We also remember Hans Hofer and Hongfang Chen, whose dedication to this experiment was invaluable.


\begin{thebibliography}{99}


\bibitem{Irina111} I. Dymnikova, Spinning superconducting electrovacuum soliton. Physics Letters B 639 (2006) 368 - 372.


\bibitem{Michell} J. Michell, Phil. Trans. R. Soc. (London) 74 ( 1784 ) 35, http://www.jstor.org/pss/106576.


\bibitem{Brodbeck} O. Brodbeck, M. Heusler, N. Straumann and M. Volkov, Phys.Rev.Lett. 79 (1997) 4310,
                               arXiv:gr-qc/9707057; \\
                               M. S. Volkov and N. Straumann, Phys.Rev.Lett. 79 (1997) 1428, arXiv:hep-th/9704026 ; \\
                               M. S.Volkov, arXiv:hep-th/9609201 ; \\
                               Dmitri V. Gal'tsov and Jose'P. S. Lemos, Class.Quant.Grav. 18 (2001) 1715, arXiv:gr-qc/0008076
                               http://arxiv.org/abs/gr-qc/0008076 ; \\
                               G. Clement and D. Gal'tsov, Phys.Rev. D62 (2000) 124013, arXiv:hep-th/0007228 http://arxiv.org/abs/hep-th/0007228 ; \\
                               N. Okuyama and Kei-ichi Maeda, Phys.Rev. D67 (2003) 104012, arXiv:gr-qc/0212022 http://arxiv.org/abs/gr-qc/0212022 ; \\
                               Takashi Tamaki and Kei-ichi Maeda, Phys.Rev. D64 (2001) 084019, arXiv:gr-qc/0106008
                                http://arxiv.org/abs/gr-qc/0106008 ; \\
                                Piotr Bizo, Acta Cosmologica 22 (1996) 81,
                                arXiv:gr-qc/9606060
                                http://arxiv.org/abs/gr-qc/9606060 ; \\
                                Piotr Bizo, Acta Phys.Polon. B25 (1994) 877,
                                arXiv:gr-qc/9402016
                                http://arxiv.org/abs/gr-qc/9402016 ; \\
                                T. Torii, Kei-ichi Maeda and T. Tachizawa, Phys.Rev. D52 (1995) 4272, arXiv:gr-qc/9506018
                                http://arxiv.org/abs/gr-qc/9506018 ; \\
                                T. Tachizawa, K. Maeda and T. Torii, Phys.Rev. D51 (1995) 4054, arXiv:gr-qc/9410016
                                http://arxiv.org/abs/gr-qc/9410016 ; \\
                                T. Torii, Kei-ichi Maeda and T. Tachizawa, Phys.Rev. D51 (1995) 1510, arXiv:gr-qc/9406013
                                http://arxiv.org/abs/gr-qc/9406013 ; \\
                                K. Maeda, T. Tachizawa, T. Torii and T. Maki, Phys.Rev.Lett. 72 (1994) 450,
                                arXiv:gr-qc/9310015, http://arxiv.org/abs/gr-qc/9310015


\bibitem{Dirac1}  P. Dirac, Proc. Roy. Soc. (London) A133 (1931) 60.


\bibitem{Ali}  Ali H. Chamseddine and M. S. Volkov, Phys.Rev.Lett. 79 (1997) 3343, arXiv:hep-th/9707176 ; \\
                     V.V. Dyadichev and D.V. Gal'tsov, Phys.Rev. D65 (2002) 124026, arXiv:hep-th/0202177, 134
                      5http://arxiv.org/abs/hep-th/0202177 ; \\
                      D. V. Gal'tsov, arXiv:hep-th/0112038, http://arxiv.org/abs/hep-th/0112038 ;  \\
                      Dao-Jun Liu, Ying-Li Zhang and Xin-Zhou Li, 
                      Eur.Phys.J. C60 ( 2009 ) 495; \\
                      Ch. Hoelbling, C. Rebbi and V. A. Rubakov, Phys.Rev. D63 034506.
 
                      
   \bibitem{Jang1} C. N. Yang, R. Mills, Phys. Rev. 96 (1954) 191.
   
   
   \bibitem{Lipkin} H. Lipkin, A nuclear pseudo-potential , Proc. of the Rehovot conference on nuclear structure, ( 1957 ); \\
                             T. H. R. Skyrme, Skyrme A non linear theory of of strong interactons ,
                              Proc.Roy.Soc. A247 ( 1958 ) 260; \\
                              T. H. R. Skyrme, A unified model of K and Pi-Mesons , Proc.Roy.Soc. A252 ( 1959 ) 236; \\
                              T. H. R. Skyrme, A nonlinear field theory ,
                             Proc.Roy. Soc A260 ( 1961 ) 127; \\
                             T. H. R. Skyrme, Particle states in a quantized meson field , Proc.Roy.Soc. A262 ( 1961 ) 237; \\
                             Feldknoten als Teilchen , Spektrum der Wissenschaft April  2009.                   
   
                    
  \bibitem{Tokuno} A. Tokuno, Y. Mitamura, M. Oshikawa and I.F. Herbut, arXiv:0812.2736,
                              http://arxiv.org/abs/0812.2736; \\
                              S. Akiyama and M. Kawabata Phys.Rev. D76 ( 2007 ) 096002, arXiv:0709.2759,
                              http://arxiv.org/abs/0709.2759; \\
                              P. Bizo, T. Chmaj and A. Rostworowski, Phys.Rev. D75 ( 2007 ) 121702, arXiv:math-ph/0701037,
                              http://arxiv.org/abs/math-ph/0701037.
  
  
  \bibitem{Davies}  M. C. Davies and L. Marleau, Phys.Rev. D79 (2009) 074003, arXiv:0904.3337,
                              http://arxiv.org/abs/0904.3337 ; \\
                              W. T. Lin and B. Piette, Phys.Rev. D77 ( 2008 )125028,
                              arXiv:0804.4786, http://arxiv.org/abs/0804.4786; \\
                              B. M. A. G. Piette and R.S.Ward, Physica D201 (2005) 45,
                              arXiv:hep-th/0402210,
                              http://arxiv.org/abs/hep-th/0402210; \\
                               T. Portengen, J. R. Chapman, V. N. Nicopoulos and N. F. Johnson, Phys. Rev. B56 (1997) R10052,
                              arXiv:cond-mat/9707045,
                              http://arxiv.org/abs/cond-mat/9707045; \\
                               B. M. A. G. Piette, B. J. Schroers and W. J. Zakrzewski, Nucl.Phys. B439 (1995) 205, arXiv:hep-ph/9410256, http:// 
                                arxiv.org/abs/hep-ph/9410256.                
 
                    
 \bibitem{Byung} Byung-Yoon Park, Hee-Jung Lee and V. Vento, arXiv:0811.3731, 135
                            http://arxiv.org/abs/0811.3731; \\
                            J. Schliemann, Phys. Rev. B78 (2008) 195426 , arXiv:0809.0217, http://arxiv.org/abs/0809.0217; \\
                            M. Atiyah and P. Sutcliffe, Phys.Lett. B605 (2005) 106, arXiv:hep-th/0411052, http://arxiv.org/abs/hep-th/0411052; \\
                            Y. Brihaye, C. Hill and C. Zachos, Phys.Rev. D70 (2004) 111502, arXiv:hep-th/0409222,
                            http://arxiv.org/abs/hep-th/0409222; \\
                            J. G. Groshaus, V. Umansky, H. Shtrikman Y. Levinson and I. Bar-Joseph, Phys. Rev. Lett. 93 (2004) 096802,
                            arXiv:cond-mat/0406756, http://arxiv.org/abs/cond-mat/0406756; \\
                            Hee-Jung Lee, Byung-Yoon Park, M. Rho and V. Vento, Nucl.Phys. A726 (2003) 69, arXiv:hep-ph/0304066,
                             http://arxiv.org/abs/hep-ph/0304066; \\
                             Hee-Jung Lee, Byung-Yoon Park, Dong-Pil Min, M. Rho and V. Vento, Nucl.Phys. A723 (2003) 427,
                             arXiv:hep-ph/0302019, http://arxiv.org/abs/hep-ph/0302019.
 
 
 \bibitem{Manton}  N. S. Manton and St. W. Wood, arXiv:0809.3501,
                             http://arxiv.org/abs/0809.3501; \\
                             O. V. Manko, N. S. Manton and St. W. Wood, Phys.Rev. C76 ( 2007 ) 055203, arXiv:0707.0868,
                             http://arxiv.org/abs/0707.0868; \\
                             S. Bolognesi and M. Shifman, Phys.Rev. D75 ( 2007 ) 065020, arXiv:hep-th/0701065,
                             http://arxiv.org/abs/hep-th/0701065; \\
                             H. Sato, N. Sawado and N. Shiiki, Phys.Rev. D75 ( 2007 ) 014011, arXiv:hep-th/0609196,
                              http://arxiv.org/abs/hep-th/0609196; \\
                              M. Loewe, S. Mendizabal and J. C. Rojas, Phys.Lett. B638 (2006) 464, arXiv:hep-ph/0605305,
                              http://arxiv.org/abs/hep-ph/0605305; \\
                              R. Battye, N. Manton and P. Sutcliffe, Proc.Roy.Soc.Lond. A463 (2007) 261, arXiv:hep-th/0605284,
                              http://arxiv.org/abs/hep-th/0605284; \\
                             Jian Dai and V.P. Nair, Phys.Rev. D74 (2006) 085014, arXiv:hep-ph/0605090,
                              http://arxiv.org/abs/hep-ph/0605090; \\
                             R. Battye and P. Sutcliffe, Nucl.Phys. B705 (2005) 384, arXiv:hep-ph/0410157, http://arxiv.org/abs/hep-ph/0410157; \\
                              St. Krusch J.Phys. A36 ( 2003 ) 8141, arXiv:hep-th/0304264, http://arxiv.org/abs/hep-th/0304264; \\
                             M. Praszalowicz, A. Blotz and K. Goeke, Phys.Lett. B354 (1995) 415, arXiv:hep-ph/9505328,
                              http://arxiv.org/abs/hep-ph/9505328; \\
                             R. A. Leese, N. S. Manton and B. J. Schroers, Nucl.Phys. B442 (1995) 228, arXiv:hep-ph/9502405,
                             http://arxiv.org/abs/hep-ph/9502405; \\
                              Dong-Pil Min, Yongseok Oh, Byung-Yoon Park and Mannque Rho, Int.J.Mod.Phys. E4 (1995)47, arXiv:hep-ph/9412302,
                             http://arxiv.org/abs/hep-ph/9412302.
 
 
  \bibitem{Souza}  I. A. D'Souza and C. S. Kalman 1992 PREONS ( World Scientific Publishing Co. Pte. Ltd. ) ISBN 981-02-1019-1
 
 
 
 \bibitem{Ioannidou}  Th. Ioannidou, B. Kleihaus and J. Kunz, Phys.Lett. B643 ( 2006 ) 213, arXiv:gr-qc/0608110,
                                 http://arxiv.org/abs/gr-qc/0608110; \\
                                 I. F. Herbut and M. Oshikawa, Phys.Rev.Lett. 97 (2006) 080403, arXiv:cond-mat/0604557,
                                 http://arxiv.org/abs/cond-mat/0604557; \\
                                 E. Radu and D. H. Tchrakian, Phys.Lett. B632 (2006) 109, arXiv:hep-th/0509014,
                                 http://arxiv.org/abs/hep-th/0509014;\\ 
                                 R. A. Battye, St. Krusch and P. M. Sutcliffe, Phys.Lett. B626 (2005) 120, arXiv:hep-th/0507279,
                                  http://arxiv.org/abs/hep-th/0507279; \\
                                 G. Kaelbermann, J. M. Eisenberg and A. Schaefer, Phys.Lett. B339 (1994) 211, arXiv:hep-ph/9409299,
                                  http://arxiv.org/abs/hep-ph/9409299.                    


\bibitem{Brihaye} Y. Brihaye and T. Delsate, Mod.Phys.Lett. A21 ( 2006 ) 2043, arXiv:hep-th/0512339,
                                  http://arxiv.org/abs/hep-th/0512339;
                                  N. Shiiki, N. Sawado, T. Torii and Kei-ichi Maeda, Gen.Rel.Grav. 36 (2004) 1361, arXiv:gr-qc/0401020,
                                  http://arxiv.org/abs/gr-qc/0401020; \\
                                  I. G. Moss, N. Shiiki and E Winstanley, Class.Quant.Grav. 17 (2000) 4161.


\bibitem{Shiiki}  N. Shiiki, N. Sawado and S. Oryu, Phys.Rev. D70 (2004) 114023, arXiv:hep-ph/0409054,
                         http://arxiv.org/abs/hep-ph/0409054; \\
                         T. Ioannidou, B. Kleihaus and W. Zakrzewski, Phys.Lett. B600 (2004) 116, arXiv:gr-qc/0407035,
                          http://arxiv.org/abs/gr-qc/0407035.

\bibitem{Skyriom1} P, Jaikumar and R. Ouyed, Astrophys.J. 639 (2006) 354, arXiv:astro-ph/0504075,
                               http://arxiv.org/abs/astro-ph/0504075 ; \\
                               S.B. Popov and M.E. Prokhorov, Astron.Astrophys. 434 (2005) 649, arXiv:astro-ph/0412327,
                               http://arxiv.org/abs/astro-ph/0412327.


\bibitem{Skyriom2} B. Kleihaus, D. H. Tchrakian and F. Zimmerschied, J.Math.Phys. 41 (2000) 816, arXiv:hep-th/9907035,
                               http://arxiv.org/abs/hep-th/9907035; \\
                               C. Houghton, N. Manton and P. Sutcliffe, Nucl.Phys. B510 (1998) 507, arXiv:hep-th/9705151,
                               http://arxiv.org/abs/hep-th/9705151


\bibitem{Skyriom3} K. Nawa, A. Hosaka and H. Suganuma, arXiv:0901.3080,
                               http://arxiv.org/abs/0901.3080; \\
                                J. C. Collins and W.J. Zakrzewski, J.Phys. A42 ( 2009 ) 165102, arXiv:0809.0459,
                                http://arxiv.org/abs/0809.0459; \\
                                T. Ioannidou and P. Kevrekidis, Phys.Lett. A372 ( 2008 ) 6735,
                                arXiv:0807.0538, http://arxiv.org/abs/0807.0538; \\
                                T. D. Cohen, J. A. Ponciano and N. N. Scoccola, Phys.Rev. D78 ( 2008 ) 034040, arXiv:0804.4711,
                                 http://arxiv.org/abs/0804.4711; \\
                               R. Auzzi, S. Bolognesi and M. Shifman, Phys.Rev. D77 ( 2008 ) 125029, arXiv:0804.0229,
                                http://arxiv.org/abs/0804.0229; \\
                                J. A. Ponciano and N. N. Scoccola, Phys.Lett. B659 ( 2008 ) 551, arXiv:0706.0865,
                                http://arxiv.org/abs/0706.0865; \\
                                 B. M. A. G. Piette and G. I. Probert, Phys.Rev. D75 ( 2007 ) 125023, arXiv:0704.0527,
                                 http://arxiv.org/abs/0704.0527 ; \\
                                 M. Loewe, S. Mendizabal and J.C. Rojas, Mod.Phys.Lett. A22 ( 2007 ) 3003, arXiv:hep-ph/0608125,
                                 http://arxiv.org/abs/hep-ph/0608125 ; \\
                                 J. Yamashita and M. Hirayama, Phys.Lett. B642 (2006) 160, arXiv:hep-th/0605059,
                                 http://arxiv.org/abs/hep-th/0605059 ; \\
                                T. Ioannidou, B. Kleihaus and J. Kunz, Phys.Lett. B635 (2006) 161, arXiv:gr-qc/0601103,
                                 http://arxiv.org/abs/gr-qc/0601103 ; \\
                                Soon-Tae Hong, arXiv:hep-th/0502036,
                                http://arxiv.org/abs/hep-th/0502036 ; \\
                                M. Ohtani and K. Ohta, Phys.Rev. D70 (2004) 096014, arXiv:hep-ph/0406173,
                                http://arxiv.org/abs/hep-ph/0406173 ; \\
                                C. M. Savage and J. Ruostekoski, Phys.Rev.Lett. 91 (2003) 010403, arXiv:cond-mat/0306112,
                                http://arxiv.org/abs/cond-mat/0306112 ; \\
                                St. Krusch, J.Phys. A36 8141 ( 2003 ) 8141, arXiv:hep-th/0304264,
                                http://arxiv.org/abs/hep-th/0304264 ; \\
                                R. A. Battye, N. R. Cooper and P. M. Sutcliffe, Phys.Rev.Lett. 88 (2002) 080401, arXiv:cond-mat/0109448,
                                http://arxiv.org/abs/cond-mat/0109448 ; \\
                                J. A. Ponciano, L. N. Epele, H. Fanchiotti and C. A. Garcia Canal, Phys.Rev. C64 045205 (2001) 045205,
                                arXiv:hep-ph/0106150, http://arxiv.org/abs/hep-ph/0106150 ; \\
                                J. A. Neto, C. Neves, E. R. de Oliveira and W. Oliveira, J.Phys. A34 ( 2001 ) 5117, arXiv:hep-th/0105146,
                                http://arxiv.org/abs/hep-th/0105146 ; \\
                                J. Ruostekoski and J. R. Anglin, Phys.Rev.Lett. 86 ( 2001 ) 3934, arXiv:cond-mat/0103310,
                                http://arxiv.org/abs/cond-mat/0103310 ; \\
                                M. de Innocentis and R.S. Ward, arXiv:hep-th/0103046, http://arxiv.org/abs/hep-th/0103046 ; \\
                                V. Pasquier, Phys.Lett. B513 (2001) 241, arXiv:cond-mat/0012207, http://arxiv.org/abs/cond-mat/0012207 ; \\
                                N. S. Manton and B. M. A. G. Piette, arXiv:hep-th/0008110, http://arxiv.org/abs/hep-th/0008110 ; \\
                                Th. Ioannidou, Nonlinearity 13 (2000) 1217, arXiv:hep-th/0004174, http://arxiv.org/abs/hep-th/0004174 ; \\
                                Karl-Peter Marzlin, W. Zhang and B. Sanders, Phys. Rev. A62 (2000) 13602, arXiv:cond-mat/0003273,
                                http://arxiv.org/abs/cond-mat/0003273 ; \\
                                V. B. Kopeliovich, B. E. Stern and W. J. Zakrzewski, Phys.Lett. B492 (2000) 39, arXiv:hep-th/0002014,
                                http://arxiv.org/abs/hep-th/0002014 ; \\
                                G. Holzwarth, Nucl.Phys. A672 (2000) 167, arXiv:hep-ph/9905473, http://arxiv.org/abs/hep-ph/9905473 ; \\
                                Soon-Tae Hong, Yong-Wan Kim and Young-Jai Park, Phys.Rev. D59 (1999) 114026, arXiv:hep-th/9811066,
                                http://arxiv.org/abs/hep-th/9811066 ; \\
                               Y. Brihaye, B. Kleihaus and D. H. Tchrakian, J. Math.Phys. 40 (1999) 1136, arXiv:hep-th/9805059,
                                http://arxiv.org/abs/hep-th/9805059 ; \\
                                P. Bizo and T. Chmaj, Phys.Rev. D58 (1998) 041501, arXiv:gr-qc/9801012,
                                http://arxiv.org/abs/gr-qc/9801012 ; \\
                               Vladimir B. Kopeliovich, Nucl.Phys. A639 (1998) 75c,  arXiv:hep-ph/9712453,
                               http://arxiv.org/abs/hep-ph/9712453 ; \\
                                A. Acus, E. Norvaisas and D. O. Riska, Phys.Rev. C57 ( 1998 ) 2597 ; \\
                              F. Leblond and L. Marleau, Phys.Rev. D58 (1998) 05400, arXiv:hep-ph/9707397, 
                              http://arxiv.org/abs/hep-ph/9707397 ; \\
                              V. B. Kopeliovich, J.Exp.Theor.Phys. 85 (1997) 1060; Zh.Eksp.Teor.Fiz. 112 (1997) 1941, arXiv:hep-th/9707067,
                              http://arxiv.org/abs/hep-th/9707067; \\
                              Yuli V. Nazarov and A. V. Khaetskii, Phys. Rev. Lett. 80 (1998) 576, arXiv:cond-mat/9703159,
                              http://arxiv.org/abs/cond-mat/9703159 ; \\
                              R. Battye and P. Sutcliffe, Phys.Rev.Lett. 79 (1997) 363, arXiv:hep-th/9702089,
                              http://arxiv.org/abs/hep-th/9702089 ; \\
                             G. K\"albermann, Nucl.Phys. A612 (1997) 359, arXiv:hep-ph/9609254,
                             http://arxiv.org/abs/hep-ph/9609254 ; \\
                              N. Dorey and M. Mattis, Phys.Rev. D52 (1995) 2891, arXiv:hep-ph/9412373,
                               http://arxiv.org/abs/hep-ph/9412373 ; \\
                               B. Dion, L. Marleau and G. Simon, Phys.Rev. D53 (1996) 1542, arXiv:hep-ph/9408403,
                               http://arxiv.org/abs/hep-ph/9408403 ; \\
                               U. Z\"uckert, R. Alkofer, H. Reinhardt and H. Weigel, Mod.Phys.Lett. A10 ( 1995 ) 67, arXiv:hep-ph/9407305,
                               http://arxiv.org/abs/hep-ph/9407305 ; \\
                               T. Gisiger and M. B. Paranjape, Phys.Rev. D50 (1994) 1010, arXiv:hep-th/9401040,
                                http://arxiv.org/abs/hep-th/9401040 ; \\
                               J. Ananias Neto, C. Neves, E.R. de Oliveira and W. Oliveira, J.Phys. A34 ( 2001 ) 5117,
                               arXiv:hep-th/0105146, http://arxiv.org/abs/hep-th/0105146 ; \\
                               B. Dion, L. Marleau and G. Simon, Phys.Rev. D53 (1996) 1542, arXiv:hep-ph/9408403,
                               http://arxiv.org/abs/hep-ph/9408403 \\



\bibitem{Lasinio} Yoichiro Nambu and G. Jona-Lasinio, Phys.Rev. 122 ( 1961 ) 345, arXiv:hep-ph/0508263 \\



\bibitem{Lasinio1} M.K.Volkov and A.E. Radzhabor, arXiv:0812.3406v2 [ hep-ph ] 23 Dec . 2008 \\


\bibitem{Lasinio2}  V. A. Miransky, M. Tanabashi and K. Yamawaki Phys. Lett., B221 ( 1989 ) 177 \\
                                       W.A. Bardeen, C. T. Hill and M. Lindner, Phys.Rev. D41 ( 1990 ) 1647.


\bibitem{Lasinio3}  B. A. Arbuzow, M.K.Volkov and I.V.Zaitser, Int.J.Mod.Phys. A21 ( 2006 ) 5721


\bibitem{Lasinio4} A. E. Radshabov and M.K. Volkov, Eur.Phys.J. A19 ( 2004 ) 139


\bibitem{Rosen}  G. Rosen and J. Math. Phys. 9 (1968) 996.


\bibitem{Rosen1}  R. Friedberg, T.D. Lee and A. Sirlin, Phys. Rev. D13 (1976) 2739.


\bibitem{Rosen2}  S. Coleman, Nucl. Phys. B262 (1985) 263; erratum: B269 744 (1986).



\bibitem{Rosen3} M. S. Volkov and E. Woehnert, Phys.Rev. D66 ( 2002 ) 085003.



\bibitem{Sphaleron} F. R. Klinkhamer and N. S. Manton Phys. Rev. D bf 30 ( 1984 ) 2212, doi:10.1103/PhysRevD.30.2212


\bibitem{Sphaleron1}   K. Kainulainen, Nucl.Phys.Proc.Suppl. 43 (1995) 292, arXiv:hep-ph/9503335; \\
                                    K. S. Babu, R.N. Mohapatra and S. Nasri, Phys.Rev.Lett. 97 ( 2006 ) 131301, arXiv:hep-ph/0606144; \
                                    E. Nardi, Y. Nir, J. Racker and E. Roulet, JHEP 0601 (2006) 068, arXiv:hep-ph/0512052; \\
                                     A. Kovner, A. Krasnitz and R. Potting, Phys.Rev. D61 (2000) 025009, arXiv:hep-ph/9907381;
                                    D. Comelli, D. Grasso, M. Pietroni and A. Riotto, Phys.Lett. B458 (1999) 304, arXiv:hep-ph/9903227; \\
                                    G. D. Moore, arXiv:hep-ph/9902464; \\
                                    G. D. Moore, Nucl.Phys. B568 (2000) 367, arXiv:hep-ph/9810313; \\
                                    Bing-Lin Young, arXiv:hep-ph/9503298; \\
                                    G. D. Moore Phys.Rev. D62 (2000) 085011, arXiv:hep-ph/0001216; \\
                                    S. Mohan, Phys.Lett. B389 (1996) 551, arXiv:hep-ph/9605329.


\bibitem{Sphaleron2}  B. Kleihaus, J. Kunz and M. Leissner, arXiv:0810.1142; \\
                                  E. Radu and M. S. Volkov, arXiv:0810.0908; \\
                                  M. Laine and M. Shaposhnikov, Phys.Rev. D61 (2000) 117302, arXiv:hep-ph/9911473; \\
                                  A. Kovner, A. Krasnitz and R. Potting, Phys.Rev. D61 (2000) 025009, arXiv:hep-ph/9907381; \\
                                  P. M. Saffin and E. J. Copeland, Phys.Rev. D57 (1998) 5064, arXiv:hep-ph/9710343; \\
                                  M. N. Chernodub, F. V. Gubarev and E. M. Ilgenfritz, Phys.Lett. B424 (1998) 106


\bibitem{Sphaleron3}  G. Lavrelashvili, arXiv:hep-th/9410183; M. Hindmarsh, arXiv:hep-ph/9408241


\bibitem{Sphaleron4}   M. S.Volkov, D. D.Gal'tsov, Phys.Lett. B341 279; \\
                                   B. Kleihaus, J. Kunz and Abha Sood, Phys.Lett. B354 (1995) 240; \\
                                   B. Kleihaus, J. Kunz and A. Sood, Phys.Lett. B372 (1996) 204; \\
                                   M.S.Volkov, O.Brodbeck, G.Lavrelashvili and N.Straumann, Phys.Lett. B349 (1995) 438

\bibitem{Sphaleron5}  E. Radu and M. S. Volkov, Phys.Rept. 468 ( 2008 ) 101; \\
                                   B. Kleihaus, J. Kunz and M. Leissner Phys.Lett. B663 ( 2008 ) 438, arXiv:0802.3275

\bibitem{Sphaleron6} ] R. St. Millward and E. W. Hirschmann, Phys.Rev. D68 (2003) 024017; \\
                                   F. Bezrukov, C. Rebbi, V. Rubakov and P. Tinyakov, arXiv:hep-ph/0110109; \\
                                   Y. Brihaye and M. Desoil, Mod.Phys.Lett. A15 (2000) 889; \\
                                   B. Kleihaus, D. H. Tchrakian and F. Zimmerschied Phys.Lett. B461 (1999) 224; \\
                                    K. L. Frost and L. G. Yaffe, Phys.Rev. D60 (1999) 105021; \\
                                    J. Grant and M. Hindmarsh, Phys.Rev. D59 (1999) 116014; \\
                                   D. Diakonov, M. Polyakov, P. Sieber, J. Schaldach and K. Goeke, Phys.Rev. D53 (1996) 3366; \\
                                    J. Choi, Phys.Lett. B345 (1995) 253


\bibitem{Sphaleron7}  I. Zahed, Nucl.Phys. A715 (2003) 887; \\
                                   E. Shuryak and I. Zahed Phys.Rev. D67 (2003) 014006


\bibitem{Dilaton}  Y. Fujii, arXiv:gr-qc/0212030.


\bibitem{Dilaton1} M. Hayashi, T. Watanabe, I. Aizawa and K. Aketo, arXiv:hep-ph/0303029.


\bibitem{Dilaton2} F. Alvarenge, A. Batista and J. Fabris, arXiv:gr-qc/0404034.


\bibitem{Dilaton3} H. Lu, Z. Huang, W. Fang and K. Zhang, arXiv:hep-th/0409309.


\bibitem{Dilaton4}  M. S. Volkov and D. Maison, Nucl.Phys. B559 (1999) 591, arXiv:hep-th/9904174; \\
                             O. Sarbach, N. Straumann and M. S. Volkov, arXiv:gr-qc/9709081


\bibitem{Soliton}  J. Scott Russell, Report on waves ,
                           Fourteenth meeting of the British Association for the Advancement of Science, ( 1844 )


\bibitem{Soliton1}   M. S .Volkov, arXiv:hep-th/0401030; \\
                              M. S. Volkov and E. Woehnert, Phys.Rev. D67 ( 2003 ) 105066; \\
                              I. Dymnikova, Phys.Lett. B639 ( 2006 ) 368, hep-th/0607174; \\
                               Ali H. Chamseddine and M. S. Volkov, Phys.Rev. D57 (1998) 6242, \\ arXiv:hep-th/9711181 ; \\
                              M. S. Volkov and D. V. Gal?tsov, Phys.Rept. 319 (1999) 1, arXiv:hep-th/9810070; \\
                              M. S. Volkov, D. V. Gal?tsov, Phys.Rept. 319 (1999) 1, arXiv:hep-th/9810070 ; \\
                              O. Brodbeck, M. Heusler, N. Straumann and M. Volkov, Phys.Rev.Lett. 79 (1997) 4310, \\
                              arXiv:gr-qc/9707057 ; \\
                              D. V. Gal'tsov and E. A. Davydov, Phys.Rev. D75 ( 2007 ) 084016, arXiv:hep-th/0612273,
                              http://arxiv.org/abs/hep-th/0612273 ; \\
                              D. V. Gal'tsov and V. V. Dyadichev, arXiv:gr-qc/0101101, http://arxiv.org/abs/gr-qc/0101101 ; \\
                              P. Breitenlohner, D. Maison and G. Lavrelashvili, Class.Quant.Grav. 21 (2004) 1667, arXiv:gr-qc/0307029,
                              http://arxiv.org/abs/gr-qc/0307029 ; \\
                              J. Diaz-Alonso, D. Rubiera-Garcia and Rubiera-Garcia, arXiv:0712.1702,
                              http://arxiv.org/abs/0712.1702 ; \\
                              K. Goeke, J. Grabis, J. Ossmann, P. Schweitzer A. Silva and D. Urbano, Phys.Rev. C75 ( 2007 ) 055207,
                               arXiv:hep-ph/0702031, http://arxiv.org/abs/hep-ph/0702031 ; \\
                              H. Blas, JHEP 0703 ( 2007 ) 055, arXiv:hep-th/0702197,
                               http://arxiv.org/abs/hep-th/0702197 ; \\
                              M. Schmid and M. Shaposhnikov, Nucl.Phys. B775 ( 2007 ) 365, arXiv:hep-th/0702101,
                              http://arxiv.org/abs/hep-th/0702101 ; \\
                              K. Goeke, J. Grabis, J. Ossmann, M. V. Polyakov, P. Schweitzer, A. Silva and D. Urbano, 
                              Phys.Rev. D75 ( 2007 ) 094021,
                               arXiv:hep-ph/0702030, http://arxiv.org/abs/hep-ph/0702030 ; \\
                              D. Jurciukonis and E. Norvaisas, J.Math.Phys. 48 ( 2007 ) 052101, arXiv:hep-th/0702007,
                              http://arxiv.org/abs/hep-th/0702007


\bibitem{De Sitter}  M. S. Volkov, arXiv:hep-th/0204021; \\
                              M. S. Volkov and A. Wipf, Nucl.Phys. B582 (2000) 313, arXiv:hep-th/0003081
 

\bibitem{De Sitter1} W. De Sitter, Proc. Kon. Ned. Acad. Wet. 19 ( 1917 ) 1217; \\
                                 W. De Sitter, Proc. Kon. Ned. Acad. Wet. 20 229 ( 1917 ) 229


\bibitem{vacuum}  M. K. Volkov, E. S. Kokoulina and E. A. Kuraev, Phys.Part.Nucl.Lett. 1 (2004) 235 ,
                             arXiv:hep-ph/0402163.


\bibitem{vacuum1}  D. Walter and H. Gies,
                               Probing the quantum vacuum: perturbative effective action approach , Berlin: Springer ( 2000 ) ISBN 3540674284.


\bibitem{Yang-Mills}  D. V. Gal'tsov, E. A. Davydov and M. S. Volkov Phys.Lett. B648 ( 2007 ) 249, arXiv:hep-th/0610183; \\
                                 M. S. Volkov, Phys.Lett. B644 ( 2007 ) 203, arXiv:hep-th/0609112.


\bibitem{dipol}  B.C. Regan, E.D. Commins, C.J. Schmidt and D. DeMille,
                       Phys. Rev. Lett. 88 ( 2002 ) 071805; \\
                       E.D. Commins, St.B. Ross, D. DeMille, and B.C. Regan, Phys.Rev. A50 ( 1994 ) 2960


\bibitem{POINT-PAPER}  Yutao Chen, Chih - Hsun Lin, Minghui Liu, Alexander S. Sakharov, J\"{u}rgen Ulbricht and Jiawei Zhao
                                          Physics 2023, 5(3), 752-783; https://doi.org/10.3390/physics5030048 - 10 Jul 2023 
 
 
 \bibitem{PROTON-decay} F. Halzen A. D. Martin , 1984 Quarks and Leptons ( John Wiley and Sons, Inc. ) ISBN 0-471-88741-2\\
                                                  Bajc, Borut; Hisano, Junji; Kuwahara, Takomi; Omura, Yuji ( 2016 ) ``Threshold corrections
                                                  to dimension-six proton decay operators in non-minimal SUSY SU(5) GUTs'' 
                                                  Nucle. Phys. B 910  1 ; arXiv:1603.03568
                                                  
 
  \bibitem{Weeler}  J.A.Wheeler, Annals Physic 2 (1957 ) 604
  
  
  \bibitem{Weeler1}  L. Crane and L.Smolin, Gen.Ralativity and Gravitation 17 No.12 ( 1985 ); \\
                                L. A. Anchordoqui, Journal of Physics:Conf.Series. 60 ( 2007 ) 191, DOI:10.1088/1742-6596/60/1/039. 
 
 \bibitem{Instability}  L. Kofman, Physica Scripta. T36 ( 1991 ) 108; \\
                                T. Souradeep, PRAMANA Journal of Physics 67 No.4 ( (2006) 699, DOI: 10.1007/s12043-006-0063-4;
                                 T. Padmanabhan , arXiv:gr-qc/0503107v1; \\
                                 D. S. Chellone, J. Phys. A: Math. Gen. 14 ( 1981 ) 2339; \\
                                 M. Crawford and D. N. Schramm, Nature 298 ( 1982 ) 538
 
 
 \bibitem{Condenstate}  New Scientist 19 June ( 1999 ), http://ldolphin.org/qfoam.html
 
 
 \bibitem{Condenstate1}  B. L. Hu, Int.J.of Theoretical Phys. 44 ( 2005 ) No.10., DOI: 10.1007/s10773-005-8895-0
 
 
 \bibitem{Direct-Contact}  O. J. P. Eboli et. al. Phys.Lett. B271 ( 1991 ) 274; \\
                                        P. Mery, M. Perrottet and F.M. Renard Z.Phys. C38 ( 1988 ) 579; \\
                                        Stanley J. Brodsky and S.D. Drell Phys.Rev. D22 ( 1980 ) 2236.


\bibitem{Scharzschild} K. Schwarzschild, \"{U}ber das Gravitationsfeld eines Massenpunktes nach der Einsteinschen Theorie , \\
                                     Sitzungsberichte der Deutschen Akademie der Wissenschaften zu Berlin, Klasse fur Mathematik, Physik, und Technik           
                                     (1916) 189.


\bibitem{Desiter1}  E. Poisson and E. Israel, Class. Quant. Grav. 5, (1988) L201; \\
                             M. R. Bernstein, Bull. Amer. Phys. Soc. 16 (1984) 1016; \\
                              E. Fahri and A. Guth, A. Phys. Lett. B183 (1987) 149; \\
                              W. Shen and S. Zhu, Phys. Lett. A126, (1988) 229; \\
                              V. P. Frolov, M. A. Markov and V. F. Mukhanov, Phys. Rev. D41, (1990) 3831.
                                  

\bibitem{Desiter3} I. Dymnikova, Int.J.Mod.Phys. D12 (2003) 1015.


\bibitem{Bohr1}  N. Bohr, Philosophical Magazine 26 (1913) 1.


\bibitem{Vacuum1} D. Walter and H. Gies, Probing the quantum vacuum: perturbative 
                                         effective action approach , Berlin: Springer ( 2000 ) ISBN 3540674284


\bibitem{Hawking1} Stephen Hawking , A Brief History of Time Bantam Books, ( 1988 ).


\bibitem{Irina66} I. G. Dymnikova, Int. J. Mod. Phys. D5 (1996) 529


\bibitem{Dymnikova22} I. Dymnikova, Gen. Rel. Grav. 24, (1992) 235



\bibitem{Dymnikova33}  I. Dymnikova, Int.J.Mod.Phys. D12 (2003) 1015


\bibitem{Dirac22} P.A.M. Dirac, Nature, 168:906-7 ( 1951 ) ;\\
                                   P.A.M. Dirac, Nat. Rundschau, 6:441-6 ( 1953 );\\
                                    P.A.M. Dirac, The Scientific Monthly, 78:142-6, ( 1954 )

\bibitem{Gothe1} R. W. Gothe private communication, University of South Carolina, Columbia, SC 29208 (2006)


\bibitem{Millikan33}  Millikan R. A. 1911, Phys. Rev. 32 , ( 1911 ) 349


\bibitem{Particle1}  Particle Data Group, Phys. Lett. B592 (2004) 33 


\bibitem{Particle2} Particle Data Group, Phys. Lett. B592 (2004) 33; \\
                               J. Sapirstein, Phys.Rev.D20 (1979) 3246; \\
                                R. W. Robinett, Phys.Rev.D28 (1983) 1185; \\
                                J.Berabe?u et. al., hep-ph/9702222 v2; \\ 
                                M. C. Gonzalez-Garcia, hep-ph/9609393 v1; \\
                                J. Bernabe'u at. all., hep-ph/9411289 v2


\bibitem{Neutrino1} Particle Data Group, Phys. Lett. B592 (2004) 35



\bibitem{massUL}  Particle Data Group, Phys. Lett. B592 (2004) 37.


\bibitem{Dirac11}  B. Kayser, hep-ph/0504052 v1.


\bibitem{W-Z} Particle Data Group, Phys. Lett. B592 (2004) 31.


\bibitem{W-Z22}  Mark A. Samuel et.al., Phys. Rev. Lett. 67 (1991) 9; \\
                           Mark A. Samuel et.al., Phys. Lett. B 280 (1992) 124; \\
                           A. V. Strelchenko, Phys. Lett. B 542 (2002) 223.


\bibitem{Hasan1}  A. Hasan, J. Ulbricht and J.Wu Helv. Phys. Acta 70 (1997) 27



\bibitem{Ellis1} J. R. Ellis et.al., Phys. Lett. B213 (1988) 526; \\
                         The LEP Collaboration ALEPH, DELPHI, L3 and OPAL, Electroweak 
                          Parameters of the $ Z^{0} $  Resonance and the Standard Model, Phys. Lett. B276 (1992) 247; \\
                          The LEP Collaboration ALEPH, DELPHI, L3 and OPAL, the LEP Electroweak 
                          working group, the SLD Electroweak and Heavy Flavour Groups A Combination 
                          of Preliminary Electroweak Measurements and Constrains on the Standard Model, 
                          CERN-EP-2004- 069,page 160, hep-ex/0412015; \\
                          Particle Data Group, Phys. Lett. B592 (2004) 37

\bibitem{WAVE-PARICLE} https://en.wikipedia.org/wiki/Wave-particle$\_$duality.


\bibitem{PREON1} I. A. D'Souza $\&$ C. S. Kalman, PREONS, World Scientific Publishing Co. Pte.Ltd. ISBN 981-02-1091-1.,


\bibitem{CONTACT1}   D. Bourilkov, Phys.Rev. D64 (2001) R071701;\\
                                     Zeus collaboration, arXiv:hep-ex/0401009v2 2 Apr 2004;\\
                                     Zeus collaboration, arXiv:hep-ex/9905039v3 14 Jan 2000.



\bibitem{Electron-Parameter}  Particle Data Group, Phys. Lett. B592 (2004) 33.\\
                                               old; https://lss.fnal.gov/archive/openaccess/openaccess$\_$cpc$\_$40$\_$10$\_$100001.pdf \\
                                               new; https://journals.aps.org/prd/abstract/10.1103/PhysRevD.110.030001.\\
                                               X. Fan et.al. ``Measurement of the Electron Magnetic Moment'', PHYSICAL REVIEW LETTERS 130, 071801 (2023).\\
                                               B. C. Regan et. al. ``New Limit on the Electron Electric Dipole Moment'', Phys. Rev. Lett. 88, 071805.

\bibitem{Dipol11}  B.C. Regan, E.D. Commins, C.J. Schmidt and D. DeMille, \\
                        Phys. Rev. Lett. 88 ( 2002 ) 071805; \\
                        E.D. Commins, St.B. Ross, D. DeMille, and B.C. Regan, Phys.Rev. A50 ( 1994 ) 2960 \\
                        L. Caldwell, Elektrisches Dipolmoment des Elektrons, Spektrum der Wissenschaft 6.24 13.


\bibitem{GUTscale}  Steven A. Abel, Joerg Jaeckel and Valentin V. Khoze hep-ph/0703086 ( 2007 ); \\
                                  Alex G. Dias, Edison T. Franco and Vicente Pleitez hep-ph/0708.1009 ( 2007 )


\bibitem{SizePaper2003} Irina Dymnikova, Alexander Sakharov , J\"urgen Ulbricht and Jia-wei Zhao, hep-ph/0111302 ( 2003 ).



\bibitem{IRINA1}  E. Poisson and E. Israel, Class. Quant. Grav. 5, (1988) L201; 
                                    M. R. Bernstein, Bull. Amer. Phys. Soc. 16 (1984) 1016;
                                    E. Fahri and A. Guth, A. Phys. Lett. B183 (1987) 149;W. Shen and S. Zhu, Phys. Lett. A126, (1988) 229;
                                    V. P. Frolov, M. A. Markov and V. F. Mukhanov, Phys. Rev. D41, (1990) 3831


\bibitem{DeSitter44}  I. G. Dymnikova, Phys. Lett. B472, (2000) 33; gr-ge/9912116.


\bibitem{particle333} I. G. Dymnikova, Int. J. Mod. Phys. D5 (1996) 529.



\bibitem{lambda} I. Dymnikova, Gen. Rel. Grav. 24, (1992) 235.



\bibitem{lambda11}  I. G. Dymnikova, Phys. Lett. B472, (2000) 33; gr-ge/9912116


\bibitem{sakharov} E. B. Gliner, Sov. Phys. JETP 22, (1966) 378.


\bibitem{gliner} E. B. Gliner, Sov. Phys. JETP 22, (1966) 378.


\bibitem{zeldovich} Ya. B. Zeldovich, Sov. Phys. Lett. 6, (1967) 883.


\bibitem{dynamics} I. G. Dymnikova, in : Internal Structure of Black Holes and Spacetime Singularities ?, \\
                               eds. M. Bucko and A. Ori ( Bristol:   of Physics and the Isreal Physical Society )


\bibitem{maeda} K. Maeda, T. Tashizawa, T. Torii, M. Maki, Phys. Rev. Lett. 72 (1994) 450.


\bibitem{ID2002CQG}  I. G. Dymnikova, Class. Quant. Grav. 19 ( 2002 ) 725, gr-qc/0112052.



\bibitem{particle11}  E. Poisson and E. Israel, Class. Quant. Grav. 5 (1988) L201


\bibitem{werner} E. Poisson and E. Israel, Class. Quant. Grav. 5 (1988) L201.


\bibitem{Irina10} I. G. Dymnikova, Class. Quant. Grav. 19 ( 2002 ) 725, gr-qc/0112052


\bibitem{Irina11} I. G. Dymnikova, Int. J. Mod. Phys. 12 ( 2003 ) 1015; \\
                           gr-qc/0304110; gr-qc/0310031; hep-th/0310047 ;gr-qc/0112052
                           
        
\bibitem{Irina22}  I. G. Dymnikova, E. Galaktionov Phys. Lett. B645, ( 2007 ) 358    

 
 \bibitem{Irina33} I. Dymnikova, Int.J.Mod.Phys. D12 (2003) 1015    
                        

\bibitem{IrinaSOLITON} I. Dymnikova, Physics Letters B 639, 3-4 (2006) 368; \\
                                       Particle 2021, 4, 129-145. https : // doi.org/10.3390/particles4020013.



\bibitem{Newman11} E.T. Newman, E. Cough, K. Chinnapared, A. Exton, A. Prakash, \\
                                  and R. Torrence, J. Math. Phys. 6 (1965) 918
                          

 
\bibitem{Meissner} Meissner, W.; Ochsenfeld, R. (1933). Ein neuer Effekt bei Eintritt der \\
                               Supraleitf\"{a}higkeit. Naturwissenschaften. 21 (44): 787?788. \\
                               Bibcode:1933NW.....21..787M. doi:10.1007/BF01504252. S2CID 37842752


\bibitem{IrinaSOLITON1} I. Dymnikova, Physics Letters B 639, 3-4 (2006) 368.


\bibitem{Gross11} E. P. Gross (1961). ``Structure of a quantized vortex in boson systems''. 
                             Il Nuovo Cimento. 20 (3): 454 - 457. Bibcode:1961NCim...20..454G. 
                             doi:10.1007/BF02731494. S2CID 121538191. \\
                             L. P. Pitaevskii (1961). ``Vortex lines in an imperfect Bose gas''. Sov. Phys. JETP. 13 (2): 451 - 454. \\
                             Gross - Pitaevskii : https://en.wikipedia.org/wiki/Gross\%E2\%80\%93Pitaevskii$_{-}$equation \\
                              \url{https://uni-tuebingen.de/fileadmin/Uni_Tuebingen/Fakultaeten/MathePhysik/Institute/PIT/Festk}
                             

\bibitem{Diffcross11} Yutao Chen et. al. ``Is the Non - Pointness of the Electron Observable in $  e^{+}e^{-} $ Annihilation at Center-of-Mass 
                                  Energies 55 - 207 GeV?'' , Physics 2023, 1, 1 - 33. https://doi.org/. \\
                                  
                                  Yutao Chen et. al. ``Finite Size of Electron in $ {\EEG} $ Annihilation in differential
                                  cross section of centre-of-mass energies 55 - 207 GeV'' , arXiv:2210.11149v1 [hep-ex] 20 Oct 2022, \\
                                  
                                   Yutao Chen et. al. ``Possible Manifestation of a Non - Pointness of the Electron in $  e^{+}e^{-} $ Annihilation 
                                   Reaction at Centre of Mass Energies 55 - 207 GeV'', arXiv:2210.11149v2 [hep-ex] 30 Mar 2023 .\\
                                   
                                  I. Dymnikova, ``Spinning superconducting electrovacuum soliton'', Physics Letters B 639 (2006) 368 - 372. \\
                                  
                                   I. Dymnikova, ``Image of the Electron Suggested by Nonlinear Electrodynamics Coupled to Gravity''
                                   Particles 2021, 4, 129 - 145. https:// doi.org/10.3390/particles4020013. \\   
                                                                                                         
                                  I. Dymnikova, A. Sakharov, and J. Ulbricht,``Appearance of a Minimal Length in $ e^{+} e^{-} $ Annihilation'',
                                  Hindawi Publishing Corporation Advances in High Energy Physic \\ \textbf{vol. 2014,Article ID 707812} 9 pages \\  
                                                                
                                   Chih-Hsun Lin, Jurgen Ulbricht, Jian Wu, Jiawei Zhao
                                   \textit{``Experimental and Theoretical Evidence for Extended Particle Models''},
                                   arXiv:1001.5374 [hep-ph] ( 2010 ). \\
                                   
                                   Yutao Chen et. al. ``Possible Manifestation of a Non-Pointness of the Electron in $  e^{+}e^{-} $Annihilation 
                                   Reaction at Centre of Mass Energies 55 - 207 GeV'', arXiv:2210.11149v2 [hep-ex] 30 Mar 2023.



\bibitem{Tot-cross11} Y. Chen, M. Liu  and J. Ulbricht, ; arXiv:2112.04767v2 [hep-ex] 24 Mar 2022
                                  Yutao Chen, Minghui Liu, J\"{u}rgen Ulbricht, ``Hint for a minimal interaction length in $ \EEG $ annihilation in 
                                  total cross section of center-of-mass energies 55 - 207 GeV``, arXiv:2112.04767v2 [hep-ex]


 \bibitem{Bhabha-cross11} D. Bourilkov, arXiv:hep-ph/0104165v1 17 Apr 2001


\bibitem{COULOMB1}  {Coulomb, C.A.}
Premier m\'{e}moire sur l'\'{e}lectricit\'{e} et le magn\'{e}tisme. \emph{M\'{e}moire 
l'Acad. Sci. } {\bf 1785}, 569--577. Reprinted in {Coulomb, C.A.}
\emph{M\'{e}moires sur l'\'{e}lectricit\'{e} et le magn\'{e}tisme. Extraits des M\'emoires de l'Acad\'emie Royale des Sciences de
Paris, publi\'es dans le ann\'ees 1785 \`a 1789, avec planches at tableaux.} Bachelier, libraire: Paris, France, 1789.
 {https://doi.org/10.5479/sil.304245.39088000647479}


\bibitem{COULOMB2}  {Coulomb, C.A.} Second m\'{e}moire sur l'\'{e}lectricit\'{e} et le magn\'{e}tisme. \emph{M\'{e}moire 
 l'Acad. R. Sci. } {\bf 1785}, 578--611. Reprinted in {Coulomb, C.A.}
\emph{M\'{e}moires sur l'\'{e}lectricit\'{e} et le magn\'{e}tisme. Extraits des M\'emoires de l'Acad\'emie Royale des Sciences de
Paris, publi\'es dans le ann\'ees 1785 \`a 1789, avec planches at tableaux.} Bachelier, libraire: Paris, France, 1789.
 {https://doi.org/10.5479/sil.304245.39088000647479}


\bibitem{OERSTED}
Whittaker, E.T.  \emph{A History of the Theories of Aether and Electricity}; 
 {Longman, Green, and Co.: London, UK/Hodges, Figgis \& Co. Ltd.: Dublin, UK, }
 1910. 
 {Available online:} {https://archive.org/details/historyoftheorie00whitrich} (accessed on 5 June 2023).


\bibitem{OERSTED1} {Oersted,} {H.}C.  Experiments on the effect of a current of electricity on the magnetic needle.
                                 \textit{Ann. Philos.}  \textbf{1820},  \emph{16}, 273--276.\\
                                {Available online:} http://www.ampere.cnrs.fr/ice/ice-book-detail.phplang=fr;type=text;
                                bdd=kore-ampere;table=ampere-text;bookId=7;typeofbookId=2;num=0
                                 (accessed on 5 June 2023).


\bibitem{AMPERE}  Blondel, C. \emph{A.-M. {Amp\`ere} et la {Cr\'eation} de {l'\'Electrodynamique,} 1820 - 1827}; {Biblioth\`eque}
                                 Nationale: Paris, France, 1982. 


\bibitem{AMPERE1} {Blundell,} S.J. \textit{Magnetism:   A Very Short Introduction}; {Oxford University Press}: Oxford, UK, 2012;  p. 31.
                                  ISPN 9780191633720, {https://doi.org/10.1093/actrade/9780199601202.001.0001}


\bibitem{AMPERE2} {Tricker, }R.A.R.  \textit{Early Electrodynamics. {The First Law of Circulation}}; 
                                  {Pergamon Press Ltd.}: {Oxford, UK,} 1965; p. 23. {https://doi.org/10.1016/C2013-0-05377-7}


\bibitem{BIOT} {Biot,} J.-B.; Savart, F. Note sur le magn\'etisme de la pile de Volta. \emph{Ann. Chem. Phys.} \\
                         \textbf{1820}, \emph{15}, 222{--223}. {Available online:} \\
                         {http://www.ampere.cnrs.fr/ice/ice-book-detail.php?lang=fr\&type=text\&bdd=koyre-ampere\&table=ampere-text\&bookId=7\&typeofbookId=2\&num=0}
                         (accessed on 5 June 2023).


\bibitem{CATHODE}
Martin, A. Cathode ray tubes for industrial and military applications.  \emph{Adv. Electron. Electron Phys.}
\textbf{1986}, \emph{67}, 183--328.   


\bibitem{CATHODE2} Keithley, J.F. \textit{The Story of Electrical and Magnetic Measurements: From 500 B.C. to the 1940s};
                                    {IEEE Press/Instituteof Electrical and Electronics Engineers, Inc.: New York, NY, USA,} 1999; p. 205. 
 Available online:
https://www.scribd.com/document/424498727/The-Story-of-Electrical-and-Magnetic-Measurements-From-500-BC-to-the-1940s-pdf
(accessed on 5 June 2023).


\bibitem{CATHODEa} {Goldstein,} E.  {Vorl\"{a}ufige Mittheilungen \"{u}ber elektrische Entladungen in verd\"{u}nnten Gasen}
                                   (Preliminary communications on electric discharges in rarefied gases).
                                    \emph{Monatsber. der K\"{o}nigl. Preuss.
                                   Akad. Wissensch. Berlin (Month. Rep. R. Pruss. Acad. Sci. Berlin)}
                                    {\bf 1876},  {\it 4 May}, 279--295.
                                    Quoting p. 286:
                                   13. Das durch die Kathodenstrahlen in der Wand hervorgerufene Phosphorescenzlicht ist h\"{o}chst
                                   selten von gleichf\"{o}rmiger Intensit\"{a}t auf der von ihm bedeckten 
                                   Fl\"{a}che, und zeigt oft sehr barocke Muster.''  (13 ).
                                   The phosphorescent light that's produced in the wall by the cathode rays is very 
                                   rarely of uniform intensity on the surface that
                                   5it covers and [it] often shows very baroque patterns.) 
                                   Available online: https://books.google.ch/books?id=7-
                                   caAAAAYAAJ$\&$pg=PA279$\&$redir-esc=y\#v=onepage. (accessed on 5 June 2023).
                                  

\bibitem{CATHODEb} {Science} History Institute. Education. Scientific Biographies. {Joseph ``J.J.'' }  John Thomson.
                                    Available online: https://sciencehistory.org/education/scientific-biographies/joseph-john-j-j-thomson/
                                   (accessed on 6 June 2023).

 
\bibitem{Thomson} Davis, E.A.; Falconer, I.J.  \emph{J. J. Thompson and the Discovery of the Electron};
                               {CRC Press/Taylor \& Francis Group:} London, UK, 1997. 
                                https://doi.org/10.1201/9781482272994.

\bibitem{Thomson} {Thomson, J.J.} {XL. Cathode} rays. \emph{{London Edinburg Dublin} Philos. 
                               Magaz. {J. Sci.}}, {1897,} {\it 44}, 293--316.
                               https://doi.org/10.1080/14786449708621070.

\bibitem{Millikan}  The Nobel Prize in Physics 1923. Robert. A. Millikan. Nobel Lecture.
                             Available online: https://www.nobelprize.org/prizes/physics/1923/millikan/lecture. (accessed on 5 June 2023).
 

\bibitem{Millikan1} Fundamental Physics Constants: Elementary Charge. In ; The NIST Reference on Constants, Units and Uncertainty;
                               NIST: Gaithersburg, MD, USA. 2019.
                               Available online: https://physics.nist.gov/cgi-bin/cuu/Value?e (accessed on 5 June 2023).


\bibitem{Millikan2} Tessinga, E.; Mohr, P.J.; Taylor, B.N.; Newell, D.B.
                              CODATA recommended values of the fundamental physics constants: 2018.
                               \emph{J. Phys. Chem. Ref. Data} {\bf 2021}, {\it 50}, 033105.
                               https://doi.org/10.1063/5.0064853

\bibitem{Abraham} Abraham, M.  Prinzipien der Dynamik des Elektrons. \emph{Ann. d. Phys.} \textbf{1903},
\emph{315}, https://doi.org/10.1002/andp.19023150105.


\bibitem{Lorentz1} Lorentz, H.A. Electromagnetic phenomena in a system moving with any velocity smaller than that of light.
                              \emph{Proc. R. Netherl. Acad. Arts Sci.(KNAW) \textbf{1903--1904}, \emph{6}, 809--831.
                               Reprinted, in Lorentz, H.A. \emph{Collected Papers. Volume V}; 
                               Springer Science+Business Media, B.V.:  Dordrecht, The
                               Netherlands, 1937172--197. https://doi.org/10.1007/978-94-015-3445-1\_5}.
                              

\bibitem{Lorentz2} Lorentz, H.A. \emph{Theory of Electrons and its Applications to the Phenomena of Light and Radiant }
                              Heat Dover Publications, Inc.: New York, NY, USA, 1952. 


\bibitem{Dirac}  Dirac, P.A.M. Classical theory of radiating electrons. \emph{Proc. R. Soc. Lond. A Math. Phys. Engin.} 
                          \textbf{1938}, \emph{167}, 148--169.
                           https://doi.org/10.1098/rspa.1938.0124


\bibitem{Compton} Compton, A.H. The magnetic electron. \emph{J. Frankl. Inst.}  \textbf{1921}, \emph{192},
                              145--155. https://doi.org/10.1016/S0016-0032(21)90917-7.


\bibitem{Compton1} Enz, C.P.   {Heisenberg's applications of quantum mechanics (1926-33) or the settling of the new land};      
              \emph{Helv. Phys. Acta} {\bf 1983}, {\it 56}, 993--1001. https://doi.org/10.5169/seals-115429.


\bibitem{Stern} Gerlach, W.; Stern, O.   Der experimentelle Nachweis der Richtungsquantelung im Magnetfeld.  \emph{Z.
                        Phys.}. \textbf{1922}, \emph{9}, 349--352. https://doi.org/10.1007/BF01326983.


\bibitem{Pauli} Pauli, W.   \"{U}ber den Zusammenhang des Abschlusses der Elektronengruppen im 
                        Atom mit der Komplexstruktur der Spektren.
                        \emph{Z. Phys.} \textbf{1925}, \emph{31}, 765--783.  https://doi.org/10.1007/BF02980631. {220--236.}


\bibitem{POLOVERVIEW} Schmor, P.W.   A review of polarized ion sources. \emph{Proceedings} of the 1995 IEEE Conference on
                                           Particle Accelerator, 1-- 5 May 1995, Dallas, TX, USA; IEEE:   New York, NY, USA, 1996; pp. 853--857.
                                           10.1109/PAC.1995.504814.
 

\bibitem{CROSSBEAM} Haeberli, W.  Sources of polarized ions.  \emph{Annu. Rev. Nucl. Sci.}  \textbf{1967}, \emph{17}, 373--426.
                                           https://doi.org/10.1146/annurev.ns.17.120167.002105. 


\bibitem{Tandem1} Arnold, W.; Ulbricht, J.; Berg, H.; Keiner, P.; Krause, H.H.; Schmidt, R.; Clausnitzer, G.
                                The Giessen polarization : facility: II. 1.2 MeV tandem accelerator.
                                 \emph{Nucl. Instr. Meth.} \textbf{1977}, \emph{143}, 457,-- 465.
\bibitem{Tandem11} Krause, H.H.; Stock, R.; Arnold, W.; Berg, H.; Huttel, E.; Ulbricht, J.; Clausnitzer, G. The
                                 Giesssen polarization : facility: III. Multi-detector analyzing system.
                                 \emph{Nucl. Instr. Meth.} \textbf{1977}, \emph{143}, 467 -- 471.


\bibitem{Tandem2} According. to William Barletta, Director of USPAS, the US Particle Accelerator School;see Feder, 
                                T. Accelerator school travels university circuit; \emph{Phys. Today} {\bf 2010}, {\it 63}, 20--22.
                                https://doi.org/10.1063/1.3326981.

\bibitem{Tandem21}  Minehara, E.; Abe, S.; Yoshida, T.; Sato, Y.; Kanda, M.;  Kobayashi, C.; Hanashima, S.  On the
                                  production of the $  KrF^{-} $ and $ XeF^{-} $  ion beams for the tandem
                                  electrostatic accelerators. \emph{Nucl. Instrum. Meth. Phys. Res. B.
                                   1984, 5, 217 -- 220}.  https://doi.org/10.1016/0168-583X(84)90513-5.

\bibitem{ATOM} Szczerba, D.  Development of a Polarized Atomic Beam Source and Measurement of Spin Correlation Parameters.
                          Ph.D. Thesis,   Naturwissenschaften ETH Z\"{u}rich, Z\"urich, Switzerland, 2001. 
                          https://doi.org/10.3929/ethz-a-004176559. 

\bibitem{ATOM1} Nass, A.; Stancari, M.; Steffens, E.  Studies on beam formation in an atomic beam source.
                             \emph{AIP Conf. Proc.} \textbf{2009}, \emph{1149}, 863 -- 867. https://doi.org/10.1063/1.3215780.


\bibitem{LAMB}Clegg, T.B.  Lamb-shift polarized ion sources---After 15 years. \emph{AIP Conf. Proc.} \textbf{1982}, \emph{80}, 21--37.
                           https://doi.org/10.1063/1.33417. 

\bibitem{LAMB1} Arnold, W.; Berg, H.; Krause, H.H.; Ulbricht, J.; Clausnitzer, G. The Giessen poarization ,facility: I. Lambshift source.
                           \emph{Nucl. Instr. Meth.} \textbf{1977}, \emph{143}, 441--455.
                           
                           
\bibitem{LAMB2} Ulbricht, J.; Arnold, W.; Berg, H.; Huttel, E.; Krause, H.H.; Clausnitzer, G.  The polarised proton capture reaction
                           $^7$Li($\vec{{\rm p}}, \gamma$)$^8$Be in the energy range from 380 to 960 keV. \emph{Nucl. Phys. A} 
                           \textbf{1977}, \emph{287}, 220--236.

\bibitem{LAMB-SHIFT} Aruldhas, G. \emph{Quantum Mechanics}, 
                                     PHI Learning Private Limited: New. Delhi, India, 2009; Ch. 15.15. Available online: 
                                     https://www.scribd.com/document/   397840888/G-Aruldhas-Quantum-Mechanics. (accessed on 5 June 2023).

\bibitem{CROSSBEAM1}Raymond, R.S. An Intense Source of Negative Polarized Hydrogen Ions. Ph.D. Thesis,
                                         University of Wisconsin--Madison, Madison, WI, USA, 1979; pp. 123--126.


\bibitem{Schroedinger1} Schr\"{o}dinger, E.   \"{U}ber die kr\"{a}ftefreie Bewegung in der relativistischen Quantenmechanik.
                                        \emph{Sitzunber. Preuss. Akad. Wiss. Phys.-Math. Kl.} \textbf{1930}, XXIV, 418 -- 428.


\bibitem{Schroedinger2} Weyssenhoff, J.; Raabe,  A. Relativistic dynamics of spin-fluids and spin-particles. \emph{Acta Phys. Pol.} 
                                        \textbf{1947}, IX. Available online: Phttps://www.actaphys.uj.edu.pl/fulltext?series=T\&vol=9\&no=1\&page=19
                                         (accessed on 5 June 2023).

\bibitem{Schroedinger21} Pryce, M.H.L.  The mass-center in the restricted theory of relativity and its connexion with 
                                          the quantum theory of elementary
                                        particles. \emph{Proc. R. Soc. A Math. Phys. Engin. Sci.} \textbf{1948}, \emph{195}, 62 -- 81.
                                        https://doi.org/10.1098/rspa.1948.0103
 
 \bibitem{Schroedinger22}Fleming, C.N.   Covariant position operators, spin and locality. \emph{Phys. Rev. B} \textbf{1965},
                                           \emph{137}, 188--197. 
 
 
 \bibitem{SchroedingerBefore23} Riewe, F.   Generalized mechanics of a spinning particle. \emph{Lett. Nuovo Cim.} \textbf{1971},
                                                     \emph{1}, 807--808. https://doi.org/10.1007/BF02785231.
 
 \bibitem{Schroedinger23} Barut, A.O.; Zangh{i},  N. Classical model of the Dirac electron. \emph{Phys. Rev. Lett.} 
                                           \textbf{1984}, \emph{52}, 2009 -- 2012.

\bibitem{pointTextbook1} Srednicki, M. \emph{Quantum Field Theory}; Cambridge University Press: 
                                        Cambridge, UK, 2007; https://doi.org/10.1017/CBO9780511813917.
 

\bibitem{pointTextbook2} Peskin, M.E.; Schroeder, D.V. \emph{An Introduction to Quantum Field Theory};
                                        CRC Press/Taylor \& Francis Group LLC: Boca Raton, FL, USA, 1995.
                                        https://doi.org/10.1201/9780429503559.

\bibitem{pointTextbook3} Schwartz, M.D. \emph{Quantum Field Theory and the Standard Model};  
                                         Cambridge University Press: Cambridge, UK, 2014.

\bibitem{pointTextbook4} Thomson, M. \emph{Modern Particle Physics}; Cambridge University Press: 
                                        Cambridge, UK, 2013; https://doi.org/10.1017/CBO9781139525367.


\bibitem{pointTextbook5} Itzykson, C.; Zuber, J.-B. \emph{Quantum Field Theory};  
                                         McGraw-Hill, Inc.: New York, NY, USA, 1980. 


\bibitem{CLMODEL1} Frenkel, J.  Die Elektrodynamik des rotierenden Elektrons.   
                                   \emph{Z. Phys.} \textbf{1926}, \emph{37}, 243--262.
                                   https://doi.org/10.1007/BF01397099.

\bibitem{CLMODEL11} Mathisson, M. Neue Mechanik materieller Systeme. 
                                      \emph{Acta Phys. Pol.} \textbf{1937}, \emph{VI}, 163 -- 200.
                                      Available online: https://www.actaphys.uj.edu.pl/fulltext?series=T\&vol=6\&no=3\&page=163.
                                       (accessed on 5 June 2023).

\bibitem{CLMODEL12} Kramers, L.H.   \emph{Quantentheorie des Electron und der Strahlung}; 
                                      Akademische Verlagsgesellschaft: Leipzig, Germany, 1938. 

\bibitem{CLMODEL13} H\"{o}nl, H.; Papapetrou, A.   \"{U}ber die innere Bewegung des Elektrons. I. \emph{Z. Phys.}
                                      \textbf{1939}, \emph{112}, 512 -- 540.
                                        https://doi.org/10.1007/BF01341246.

\bibitem{CLMODEL14} Bhabha, H.J.; Corben, A.C.   General classical theory of spinning particles in a Maxwell field. 
                                     \emph{Proc.R. Soc. A Math. Phys. Engin.} \textbf{1941}, \emph{178}, 273 -- 314.
                                      https://doi.org/10.1098/rspa.1941.0056.

\bibitem{CLMODEL15} Bargman, V.; Michel, L.; Telegdi,  V.L. Precession of the polarization of particles moving in a 
                                       homogeneous electromagnetic field.
                                     \emph{Phys. Rev. Lett.} \textbf{1959}, \emph{2}, 435 -- 436. 
                                     
\bibitem{CLMODEL16} Nash, P.L.  A Lagrangian theory of the classical spinning electron. 
                                      \emph{J. Math. Phys.} \textbf{1984}, \emph{25}, 2104--2108.

\bibitem{CLMODEL17} Plyushchay, M.S.   Relativistic massive particle with higher 
                                      curvatures as a model for the description of bosons and fermions.
                                       \emph{Phys. Lett. B} \textbf{1990}, \emph{235}, 47 -- 51.  


\bibitem{CLMODEL18} Yee, K.; Bander,  M. Equations of motion for spinning particles in external 
                                      electromagnetic and gravitational fields. \emph{Phys. Rev. D}
                                   \textbf{1993}, \emph{48}, 2797 -- 2799.  


\bibitem{CLMODEL19} Bolte, J.; Keppeler, S.   Semiclassical form factor for chaotic systems with spin. \emph{J. Phys. 
                                     A Math Gen.} \textbf{1999}, \emph{32}, 8863--8880. https://doi.org/10.1088/0305-4470/32/50/307

\bibitem{CLMODEL2} Nesterenko, V.V.  Singular Lagrangians with higher derivatives. 
                                    \emph{J. Phys. A Math. Gen.} \textbf{1989}, \emph{22}, 1673 -- 1687. 

\bibitem{CLMODEL21} Rylov, Y.A.  Spin and wave function as attributes of ideal fluid. 
                                      \emph{J. Math. Phys.} \textbf{1999}, \emph{40}, 256--278. https://doi.org/10.1063/1.532771.
 

 \bibitem{CLMODEL3}  Rivas, M.K. \emph{Kinematic} Theory of Spinning Particles : Classical and 
                                      Quantum Mechanical Formalism of Elementare Particles;
                                     Kluwer, Academic Publishers: Dordrecht, The Netherlands, 2002.
                                     https://doi.org/10.1007/0-306-47133-7.

\bibitem{CLMODEL31} Rivas, M.   The dynamical equation of the spinning electron. 
                                      \emph{J. Phys. A Math. Gen.} \textbf{2003}, \emph{36}, 4703--4716.


\bibitem{Kerr}  Newman, E.T.; Cough, E.; Chinnapared, K.; Exton, A.; Prakash, A.; Torrence, R.   
                        Metric of a rotating, charged mass.  \emph{J. Math. Phys.}
                         \textbf{1965}, \emph{6}, 918--919.


\bibitem{Irina11DeSitter} Dymnikova, I.   Image of the electron suggested by nonlinear electrodynamics coupled to gravity.
                       \emph{Particles} \textbf{2021}, \emph{4}, 129 -- 145. https://doi.org/10.3390/particles4020013.
 
\bibitem{DeSitter} De Sitter, W.   On the relativity of inertia. Remarks concerning Einstein's latest hypothesis.
                              \emph{Proc. R. Netherl. Acad. Arts Sci. (KNAW)} \textbf{1917}, \emph{19}, 1217--1225.
                             Available online: https://dwc.knaw.nl/DL/publications/PU00012455.pdf (accessed on 5 June 2023).

\bibitem{DeSitter1} De Sitter, W.   On the curvature of space. \emph{Proc. R. Netherl. Acad. Arts Sci. (KNAW)} 
              \textbf{1918}, \emph{20}, 229--243.      
              Available online:https://dwc.knaw.nl/DL/publications/PU00012216.pdf (accessed on 5 June 2023).


\bibitem{Irina1} Dymnikova, I.   De Sitter--Schwarzschild black hole: Its particlelike core and thermodynamical properties. 
                         \emph{Int. J. Mod. Phys. D} \textbf{1996}, \emph{5}, 529--540.


\bibitem{compMod1} Terazawa, H.; Yasue, M.; Akama, K.; Hayashi, M. Observable effects of the possible substructure of leptons and quarks.
                                   \emph{Phys. Lett. B} \textbf{1982}, \emph{112}, 387--392. https://doi.org/10.1016/0370-2693(82)91075-9.


\bibitem{compMod2} Renard, F.M. Excited quarks and new hadronic states. \emph{Nuovo Cim. A} \textbf{1983}, \emph{77}, 1 -- 20.
                                  https://doi.org/10.1007/BF02768908.


\bibitem{compMod3} De Rujula, A.; Maiani, L.; Petronzio, R. Search for excited quarks. \emph{Phys. Lett. B} \textbf{1984}, \emph{140},  253--258.
                                  https://doi.org/10.1016/0370-2693(84)90930-4.


\bibitem{compMod4} Eichten, E.; Lane, K.D.; Peskin, M.E. New tests for quark and lepton substructure. \emph{Phys. Rev. Lett.} \textbf{1983}, \emph{50}, 811--814.
                                  https://doi.org/10.1103/PhysRevLett.50.811.


\bibitem{compMod5} Terazawa, H.; Chikashige, Y.; Akama, K. Unified model of the 
                                  Nambu-Jona-Lasinio type for all elementary particle forces.
                                  \emph{Phys. Rev. D} \textbf{1977},  \emph{15}, 480 -- 487.
                                   https://doi.org/10.1103/PhysRevD.15.480.


\bibitem{compMod6} Ne'eman, Y. Primitive particle model. \emph{Phys. Lett. B} \textbf{1979}, \emph{82},  69 -- 70.
                                  https://doi.org/10.1016/0370-2693(79)90427-1.


\bibitem{compMod7} Baur, U.; Spira, M.; Zerwas, P.M. Excited quark and lepton production at hadron colliders.
                                  \emph{Phys. Rev. D} \textbf{1990}, \emph{42},  815 -- 824.
                                  https://doi.org/10.1103/PhysRevD.42.815.

\bibitem{HeavyElMu1} Low, F.E. Heavy electrons and muons. \emph{Phys. Rev. Lett.} \textbf{1965}, \emph{14}, 238--239.
                                     https://doi.org/10.1103/PhysRevLett.14.238.


\bibitem{HeavyElMu2} Boudjema, F. Substructure effects at LEP100. \emph{Int. J. Mod. Phys. A} \textbf{1991}, \emph{6}, 1 -- 20.
                                     https://doi.org/10.1142/S0217751X91000022.


\bibitem{excQuark1} Harari, H. Colored leptons. \emph{Phys. Lett. B} \textbf{1985}, \emph{156}, 250--254.
                                   https://doi.org/10.1016/0370-2693(85)91518-7.

  
\bibitem{VENUS} Abe, K.; et al. [VENUS Collaboration]. Measurements of the differential cross-sections
                            of $ \EEG $ and $ \EEGGG $ at  $ \sqrt{s} $ = 55, 56, 56.5 and 57 GeV.
                             \textit{Z. Phys. C} \textbf{1989}, \emph{45}, 175 -- 191.
 

\bibitem{OPAL1} Akrawy, M.Z.et al.  [OPAL Collaboration]. Measurements of the cross-sections of the reaction
                             $ \EEG $ and $ \EEGGG $ at LEP. \textit{Phys. Lett. B} \textbf{1991}, \emph{257}, 531 -- 540.

\bibitem{TOPAS} Shimozawa, K.; et al. [TOPAZ Collaboration].  Studies of $ \EEG $ and $ \EEGGG $ reaction.
                            \textit{Phys. Lett. B} \textbf{1992}, \emph{284}, 144 -- 150.


\bibitem{ALEPH} Decamp, D.; et al. The ALEPH Collaboration]. Search for new particles in $Z$ decays using the ALEPH detector.
                            \textit{Phys. Rep.} \textbf{1992}, \emph{216}, 253 -- 340.

\bibitem{DELPHI} Abreu, P.;et al. [DELPHI Collaboration]. Measurement of the $ \EEGG $ cross-section at LEP energies.
                            \textit{Phys. Lett. B} \textbf{1994}, \emph{327}, 386 -- 396.


\bibitem{DELPHI1} Abreu, P.; et al. [DELPHI Collaboration]. Measurement of the $ \EEGG $ cross-section at LEP energies.
                            \textit{Phys. Lett. B} \textbf{1998}, \emph{433}, 429 -- 440. 


\bibitem{DELPHI2}  Abreu, P.; et al. [DELPHI Collaboration]. Determination of the $ \EEGG $ cross-section at center-of-mass
                               energies ranging from 189 GeV to 202 GeV. 
                             \textit{Phys. Lett. B} \textbf{2000}, \emph{491}, 67 ---80.


\bibitem{L3A} Acciarri, M.; et al. [L3 Collaboration]. Test of QED at LEP energies using $ \EEGG $ and $ \EELLGG $.
                       \textit{Phys. Lett. B} \textbf{1995}, \emph{353}, 136 ---144.


\bibitem{L3B} Achard, P.; et al. [L3 Collaboration] et.al.  Study of multiphoton final states and tests of QED in $ e^+ e^- $ collisions at $ \sqrt{s} $
                        up to 209 GeV. \textit{Phys. Lett. B} \textbf{2002}, \emph{531}, 28 -- 38.


\bibitem{OPAL2}  Abbiendi, G.; et al. [OPAL Collaboration]. Multi-photon production in $ e^+ e^- $ collisions at $ \sqrt{s} $ = 181--209 GeV.
                        \textit{Eur. Phys. J. C} \textbf{2003}, \emph{26}, 331 -- 344.


\bibitem{ETHZ-USTC} Xe, J. Physics with $ \gamma $ Final States at the $ Z^0 $ Energy Scale Using Electron-Positron Collision.
                                     Master's  Thesis. Chinese University of Science and Technology, Hefei,   China, 1992. 


\bibitem{ETHZ-USTC1} Wu, J. Tests of Quantum Electrodynamics at the $Z^0$ Scale.
                                     Master's  Thesis.  Chinese University of Science and Technology, Hefei,   China,  1997.   

\bibitem{ETHZ-USTC2} Zhao,  J. Tests of QED using $ \EEGG $ Reactions at LEP200 and Study of Inclusive Semileptonic D Meson Decays at BES.
                                      Master's  Thesis. Chinese University of Science and Technology, Hefei,   China, 2001.
                                      
  
\bibitem{Models} Altarelli, G., Kleiss, R., Verzegnassi, C., Eds. \emph{Z Physics at LEP1. 
                            Volume 2:  Higgs Search and New Physics;}
                           CERN: Geneva, Switzerland, 1989. https://doi.org/10.5170/CERN-1989-008-V-2.


\bibitem{LEP2}  The ALEPH Collaboration; The DELPHI Collaboration; The L3 Collaboration; 
                          The OPAL Collaboration; The LEP Electroweak Working Group.
                          Electroweak measurements in electron-positron collisions at $W$-boson-pair energies at LEP.
                           \textit{Phys. Rep.} \textbf{2013}, \emph{532}, 119--244. 


 \bibitem{5SIGMA} Dymnikova, I.; Sakharov, A.; Ulbricht, J.  Appearance of a minimal length in $ e^{+}e^{-} $ annihilation.
                              \emph{Adv. High Energy Phys.}   \textbf{2014}, \emph{2014},  707812.
                              http://dx.doi.org/10.1155/2014/707812. 
 
 
 \bibitem{5SIGMA1} Bajo, A.; Dymnikova, I.; Sakharov, A.; Sanchez, E.; Ulbricht, J.;   Zhao, J. QED test at LEP200 energies in the reaction
                                 $ \EEGG $. \textit{AIP Conf. Proc.} \textbf{2001}, \emph{564}, 255 --- 562.
                                https://doi.org/10.1063/1.1374992


\bibitem{5SIGMA2} Dymnikova, I.G.; Hasan, A.; Ulbricht, J.; Zhao, J.  
                                Limits on the sizes of fundamental particles and gravitational mass of
                             the Higgs particle. \textit{Gravit. Cosmol.} \textbf{2001}, \emph{7}, 122 ---130.


\bibitem{Berends}  F.A. Berends and R.  Kleiss, 
                                ``DISTRIBUTIONS FOR ELECTRON-POSITRON ANNIHILATION INTO 
                              TWO AND THREE PHOTONS'', 
                              Nucl. Phys.  \textbf{B 186}, 22 (1981) \\                            
                              CALKUL Collaboration, F.A. Berends et al., 
                              `` MUL TIPLE BREMSSTRAHLUNG IN GAUGE THEORIES AT HIGH ENERGIES
                               (IV). The process $ \EEGGGG $ '',                               
                              Nucl. Phys. \textbf{B 239 }(1984) 395.        


\bibitem{Berends1} Berends~F.A.; et~al. [CALKUL Collaboration]. Multiple bremsstrahlung in gauge theories at 
                                high energies: (IV) The process 
                                $e^+e^- \to \gamma \gamma \gamma \gamma$.
                                Nucl. Phys. B 239 (1984) 395 --- 409.


\bibitem{Berends12}  Berends, F.A.; Gatsmans, R.  Hard production corrections for $ {\EEG} $
                                  \emph{Nucl. Phys. B}  {\bf 1973} {\it 61}  414--428.


\bibitem{Berends13} Carloni Calame, C.M.; Montagna, G.; Nicrosini, O.; Piccinini, F.
                                  \emph{EPJ Web Conf.} {\bf 2019}, {\it 218}, 07004.
                                   https://doi.org/10.1051/epjconf/201921807004


\bibitem{Manat1}   Maolinbay, M. Study of Reactions  $ e^+ e^-\rightarrow \gamma\gamma$~/~$\gamma\gamma\gamma $ at LEP Energies.
                              Master's Thesis. Eidgen\"{o}ssische Technische Hochschule, Z\"{u}rich, Switzerland,   {1995}.



\bibitem{Mandl}  Mandl, F.; Skyrme, T.H.R. The theory of the double Compton effect.
                           \textit{Proc. R. Soc. A Math. Phys. Engin.} \textbf{1952}, \emph{215}, 497 --- 507.



\bibitem{HeavyElMu111} Low, F.E. Heavy electrons and muons.
                                     \emph{Phys. Rev. Lett.} \textbf{1965}, \emph{14}, 238 --- 239.
                                     https://doi.org/10.1103/PhysRevLett.14.238.


\bibitem{HeavyElMu333} Boudjema, F. Substructure effects at LEP100.
                                     \emph{Int. J. Mod. Phys. A} \textbf{1991}, \emph{6}, 1--20.
                                    https://doi.org/10.1142/S0217751X91000022.


\bibitem{excQuark555} Harari, H. Colored leptons.
                                 \emph{Phys. Lett. B} \textbf{1985}, \emph{156}, 250--254.
                                  https://doi.org/10.1016/0370-2693(85)91518-7. 1970.


\bibitem{Litke} Litke, A.M.Experiments with Electron-Positron Colliding Beams.Master's 
                       Thesis. Harvard University, Cambridge, MA, USA,
 
                       
\bibitem{eeggGeneral1} Mery, P.; Perrottet, M.; Renard, F.M. Anomalous effects in e+e- annihilation into boson pairs. 
                                       The e+e- $  \to $ ZZ, $\gamma $Z, $\gamma \gamma $. Z. Phys. C 1988, 38, 579 -- 591. 
                                       https://doi.org/10.1007/BF01624363.


\bibitem{Models1}  King, S. F.; Sharpe S. R .Exotic CERN events from exotic color states. Nucl. Phys. B1985, 253,1--13.


\bibitem{Models11} Leung, C .N.; Love, S. T.; Rao, S. Low-energy manifestations of 
                               a new interactions scale: Operator analysis. Z. Phys. C1986, 31, 433 -- 437.


\bibitem{Models12}  Drell, S. D.; Parke, S. J. Constraints on radiative $ Z^{0} $ decays. Phys. Rev. Lett. 1984, 53, 1993 --1995.


\bibitem{Models13} Dicus, D. A.;Tata, X .Anomalous photon interactions. Phys. Lett. B1985, 155,103 --106.



\bibitem{Models14} Dicus, D. A.New interactions and neutrino counting. Phys. Rev. D1985, 31, 2999--3001.


\bibitem{Models2} Eboli, O.J.P.; Natale, A.A.; Novaes, S.F. Bounds on effective interactions from the reaction 
                              $ \EEG $ at LEP. Phys. Lett. B 1991, 271, 274 -- 276.                    
        

\bibitem{RES:RES1} Kempf,A.;Mangano,G.;Mann,R.B. Hilbert space representation of the minimal 
                                  length uncertainty relation. Phys. Rev. D1995, 52, 1108?1118.


\bibitem{RES:RES2} Kempf,A.;Mangano, G. Minimal length uncertainty relation and
                                  ultraviolet regularization. Phys. Rev. D1997, 55, 7909 --7920.


\bibitem{RES:RES3} Kempf, A. Mode generating mechanism in inpation with a cutoff. Phys. Rev. D2001, 63, 083514.  


\bibitem{EMenergy} Khan Academy. Light: Electromagnetic Waves, the Electromagnetic Spectrum Photons. 
                                 Available online: https://www. khanacademy.org/science/physics/light-waves/
                                 introduction-to-light-waves/a/light-and-the-electromagnetic-spectrum (accessed on 5 June 2023).


\bibitem{MLI} Hossenfelder, S. Minimal length scale scenarios for quantum gravity. Living Rev. Relativ. 2013, 16, 2.


\bibitem{QG:Qgrav4} Bosso, P.;Das, S.; Todorinov, V. Quantum field theory with the generalized uncertainty principle II: 
                                  Quantum electrodynamics. Ann. Phys. 2021, 424, 168350.


\bibitem{QG:Qgrav5} Luciano, G.G.;Petruzziello, L. Generalized uncertainty principle and its implications on 
                                  geometric phases in quantum mechanics. Eur. Phys. J. Plus 2021, 136, 179. 


\bibitem{L33} Adriani, O.; et al. [L3 Collaboration]. A test of quantum electrodynamics in the 
                      reaction $ \EEGG $. Phys. Lett. B 1992, 288, 404 -- 411. 


\bibitem{L331}  Acciarri, M.; et al. [L3 Collaboration]. Observation of multiple hard photon final states at
                          $\sqrt{s}$= 130 -- 140 GeV at LEP. Phys. Lett. B 1996, 384, 323 -- 332. 


\bibitem{L332}  Acciarri, M.; et al. [L3 Collaboration]. Hard-photon production at $\sqrt{s}$= 161 and 172 GeV 
                         at LEP. Phys. Lett. B 1997, 413, 159 -- 166


\bibitem{L333} . Acciarri,M.; et al.[L3Collaboration].Hard-photon production and 
                          tests of QED at LEP. Phys. Lett. B2000, 475, 198 --205.


\bibitem{MINUIT} James, F. MINUIT, Function Minimization and Error Analysis. Version 94.1;
                             CERN:Geneva, Switzerland, 1994. Availableonline:
                             https://root.cern.ch/download/minuit.pdf (accessed on 5 June 2023).


\bibitem{MINUIT1} James, F. ;Roos, M. MINUIT : A system for function minimization and analysis of 
                               the parameter errors and corrections. Comput. Phys. Commun. 1975, 10, 343 -- 367.   


\bibitem{Totcross} Chen, Y.; Liu, M.; Ulbricht, J. Hint for a minimal interaction length in $ \EEG $ 
                              annihilation in total cross-section of center-of-mass energies 55-207 GeV. 
                              arXiv 2022, arXiv:2112.04767. https: / / doi.org/ 10.48550/ arXiv. 2112.04767

\bibitem{Isiksal1} Isiksal, E. Test der Quantenelektrodynamik bei LEP -- Energien.
                             Master's Thesis. Eidgen\"{o}ssische Technische Hochschule, Z\"{u}rich, Switzerland, 1991.


\bibitem{BHABA} Workman, R. L. et al. [ Particle Data Group ].The Review of Particle Prog.
                             Theor. Exp. Phys. 2022, 2022, 083C01; see Chapter 91 ?Searches for Quark and Lepton Compositeness?.


\bibitem{BOURILKOV} Bourilkov, D. Hint for axial-vector contact interactions in the data 
                                                     on $ \EEEEG $ at center-of-mass energies 192 -- 208 GeV.
                                                     Phys. Rev. D 2001, 64, 071701(R).

 \bibitem{L3limit} Achard, P.; et al. [L3 Collaboration]. Search for heavy neutral and charged leptons in
                             e+e- annihilation a tLEP. Phys. Lett. B2001, 517, 75--85.


\bibitem{OPLALlimit} Abbiendi, G.; etal. [OPALCollaboration]. Search for stable and long lived 
                                  massive charged particles in e+e-  collisions at s = 130 GeV -- 209 GeV. 
                                  Phys. Lett. B 2003, 572, 8 -- 20.


\bibitem{OPLALlimit1} Abbiendi, G.; et al.[TheOPAL Collaboration]. Search for unstable heavy
                                     and excited leptons at LEP2. Eur. Phys. J. C. 2000, 14, 73 -- 84.


\bibitem{HERA} Ahmed, T.; et al. [H1Collaboration]. A search for heavy leptons at HERA. Phys. Lett. B 1994, 340, 205--216.


\bibitem{CMS1}  Chatrchyan , S.; et al. [CMSCollaboration].
                          Search for long - lived charged particle in pp collisions at $\sqrt{s}$ = 7 and 8 TeV 
                          J. High Energy Phys. 2013, 2013, 122.


\bibitem{H1E1} Aaron, F.D.; et al. [H1 Collaboration]. Search for excited electrons in 
                          ep collisions at HERA. Phys. Lett. B 2008, 666, 131?139.
                          https://doi.org/10.1016/j.physletb.2008.07.014.


\bibitem{D0E1} Abazov,  V. M.; et al. [D0Collaboration].
                         Search for excited electrons in $ p\bar{p} $ collisions at $\sqrt{s}$ = 1.96 TeV. Phys. Rev. D2008, 77, 091102.
                         https://doi.org/10.1103/PhysRevD.77.091102.


\bibitem{ATLASE1} Aad, G.; et al. [ATLAS Collaboration]. Search for excited electrons and muons in
                               $\sqrt{s}$ = 8 TeV proton-proton collisions with the
                                ATLAS detector. New J. Phys. 2013, 15, 093011. 
                                https://doi.org/10.1088/1367-2630/15/9/093011.


\bibitem{CMSE1}  Khachatryan, V.; et al. [The CMS Collaboration]. Search for excited leptons in proton-proton collisions at
                              $\sqrt{s}$ = 8 TeV. J. High Energy Phys. 2016, 125.
                              https://doi.org/10.1007/JHEP03(2016)125.


\bibitem{dy1} Aaboud, M.; et al. [The ATLAS Collaboration]. Search for new high-mass phenomena in the dilepton 
                      final state using $ 36\ fb^{-1} $ of proton-proton collision data at $\sqrt{s}$ = 13 TeV 
                       with the ATLAS detector. J. High Energy Phys. 2017, 2017, 182.
                       https://doi.org/10.1007/JHEP10(2017)182.


\bibitem{dy2} Sirunyan,A.M.; et al. [TheCMSCollaboration].
                      Search for contact interactions and large extra dimensions in the dilepton mass
                       spectra from proton-proton collisions at $\sqrt{s}$ = 13 TeV. J. High Energy Phys. 2019, 2019, 114. 
                       https://doi.org/10.1007/JHEP04(2019)114.


\bibitem{FCCee1} Carloni Calame, C.M.; Chiesa, M.; Montagna, G.; Nicrosini, O.; Piccinini, F. Electroweak corrections to 
                              $ {\EEG} $  as a luminosity
                              process at FCC-ee. Phys. Lett. B 2019, 798, 134976. 
                              https://doi.org/10.1016/j.physletb.2019.134976.


\bibitem{CEPC1} The CEPC Study Group. CEPC Conceptual Design Report. August 2018. Volume1: Accelerator.arXiv2018,arXiv:1809.00285.
                             https://doi.org/10.48550/arXiv.1809.00285


\bibitem{ILC1} Behnke,T.; Brau,J.E.; Foster, B.; Fuster,J.; Harrison,M.; Paterson,J.M.; Peskin,M.;
                        Stanitzki,M.; Walker,N.; Yamamoto,H., Eds. The International Linear Collider Technical Design 
                        Report 2013. Volume 1: Executive Summary. arXiv 2013, arXiv:1306.6327. https://doi.org/10.48550/arXiv.1306.6327
 
 
\bibitem{CLIC1} Blondel, A.;  Janot, P.;  Circular and linear $  e^{+}e^{-} $ colliders: Another story of complementarity. 
                          arXiv 2019, arXiv:1912.11871. https://doi.org/10.48550/arXiv.1912.11871

\bibitem{g_e1} Pospelov,M.; Ritz,A. Electric dipole moments as probes of new physics. Ann. Phys.2005,318,119-169.


\bibitem{g_e2} Brodsky,S.J.; Drell,S.D. The anomalous magnetic moment and limits on fermion substructure. Phys. Rev. D1980,22,2236?2243.
                        https://doi.org/10.1103/PhysRevD.22.2236


\bibitem{Dipol} Kinoshita,T.; Yennie,D.R. High precision tests of quantum electrodynamics: An Overview. Adv. Ser. Direct. High Energy Phys.
                        1990, 7, 1-14. https://doi.org/10.1142/9789814503273$ _{-}$0001.


\bibitem{polar1} Gakh, G.I.; Konchatnij, M.I.; Merenkov, N.P.; Gakh, A.G. Effects of excited electron and contact 
                         $ee\gamma\gamma $ interaction in $ e^+e^-\rightarrow \gamma\gamma $ reaction. \emph{arXiv} \textbf{2022},
                         arXiv:2211.16306. https://doi.org/10.48550/arXiv.2211.16306


\bibitem{polar2} Bondarenko, S.; Dydyshka, Y.; Kalinovskaya, L.; Kampf, A.; Rumyantsev, L.; Sadykov, R.; Yermolchyk, V. 
                          One-loop radiative corrections to photon-pair production in polarized positron-electron annihilation.
                         Phys. Rev. D, \textbf{2023}, {\it 107}, 073003.
                         

\bibitem{Irina} Dymnikova, I.   Image of the electron suggested by nonlinear electrodynamics coupled to gravity.
                       \emph{Particles} \textbf{2021}, \emph{4}, 129--145. https://doi.org/10.3390/particles4020013.


\bibitem{SIZE}  Dymnikova, I. Spinning superconducting electro vacuum soliton.
                          \textit{Phys. Lett. B} \textbf{2006}, \emph{639}, 368-372. 


\bibitem{Super} Bardeen, J.; Cooper, L.N.; Schriffer, J.R.  Theory of superconductivity. \emph{Phys. Rev.} 
                          \textbf{1957},   \emph{108}, 1175-1204. 


\bibitem{Super1} Daintith, J.; Rennie, R.  \emph{The Facts on File Dictionary of Physics}; 
                            Infobase Publishing: New York, NY, USA, 2005;  p. 238.
                            

\bibitem{Super2} Gallop, J.C.   \emph{SQUIDs, the Josephson Effects and Superconducting Electronics};  CRC Press Taylor $\&$ Francis
                           Group, LLC: New York, NY, USA, 1991; pp. 1, 20. 
                           

\bibitem{Super3} Durrant, A., Ed.  \emph{Quantum Physics of Matter}; The Open University/Institute of Physics Publishing: Bristol, UK 2000;
                            pp. 102-103. 


\bibitem{SIZE01} Lin, C.-H.; Ulbricht, J.; Wu, J.; Zhao, J. Experimental and theoretical evidence for extended particle models. \emph{arXiv}
                            \textbf{2010}, arXiv:1001.5374. https://doi.org/10.48550/arXiv.1001.5374.
                            

\bibitem{SIZE1} Ulbricht, J.  Substructure of Fundamental Particles. Lecture at CICPI on 9 December 2019.
                           Available online: http://cicpi.ustc.edu.cn/indico/conferenceDisplay.py?confId=5724 (accessed on 5 June 2023).


\bibitem{QED-1} A. Bajo et. al. 
                           `` QED test at LEP200 energies in the reaction $ \EEGG $'', 
                            AIP conference proceedings,\textbf{Vol. 564}, 255 (2001) \\                         
                           I. Dymnikova et al. 
                           `` Limits on the sizes of fundamental particles and gravitational mass of the Higgs particle'',
                           Grav. Cosmology, \textbf{Vol. 7}, 122 (2001)\\                                                    
                           I. Dymnikova et al. 
                           `` Appearance of a Minimal Length in $ e^{+} e^{-} $ Annihilation'', 
                           Advances in High Energy Physics \textbf{Vol. 214}, (2014), Article ID 707812, 9 pages  


\bibitem{Berends111}  F.A. Berends and R.  Kleiss, 
                                ``DISTRIBUTIONS FOR ELECTRON-POSITRON ANNIHILATION INTO 
                              TWO AND THREE PHOTONS'', 
                              Nucl. Phys.  \textbf{B 186}, 22 (1981) \\                            
                              CALKUL Collaboration, F.A. Berends et al., 
                              `` MUL TIPLE BREMSSTRAHLUNG IN GAUGE THEORIES AT HIGH ENERGIES
                               (IV). The process $ \EEGGGG $ '',                               
                              Nucl. Phys. \textbf{B 239 }(1984) 395.                                                       

\bibitem{BABAYAGA}   BabaYaga@NLO; \\ https://inspirehep.net/literature/1740483 ; http://www.pv.infn.it/hepcomplex/babayaga.html;
                                     EPJ Web of Conferences 218, 07004 (2019) PhiPsi 2017


 \bibitem{L3BBB44}  B. Achard et al. [L3 Collaboration], 
                       `` Study of multiphoton final states and tests of QED in $ e^{+} e^{-} $ collisions at  $ \sqrt{s} $ up to 209 GeV'', 
                       Phys. Lett. \textbf{B 351}, 28 (2002)
                       

\bibitem{VENUS333} K. Abe et al. [VENUS Collaboration], 
                            `` Measurement of the differential cross sections of $ \EEG $ and $ \EEGGG $ at 
                             $ \sqrt{s} $ = 55, 56, 56.5 and 57 GeV and search for unstable photino pair production'', 
                            Z. Phys. \textbf{C 45}, 175 (1989)


\bibitem{OPAL444} M.Z. Akrawy et al. [OPAL Collaboration], 
                           `` Measurement of the cross sections of the reactions $ \EEG $ and $ \EEGGG $ at LEP'',
                           Phys. Lett. \textbf{B 257}, 531 (1991)
                           
 
\bibitem{TOPAS555} K. Shimozawa et al. [TOPAS Collaboration], 
                            `` Studies of $ \EEG $ and $ \EEGGG $ reactions'',
                           Phys. Lett. \textbf{B 284}, 144 (1992)
                           
 

\bibitem{ALEPH666} D. Decamp et al. [ALEPH Collaboration], 
                            `` SEARCH FOR NEW PARTICLES IN Z DECAYS USING THE ALEPH DETECTOR'',
                            Phys. Rept. \textbf{216}, 253 (1992)


\bibitem{DELPHI666} P. Abreu et al. [DELPHI Collaboration] 
                             `` Measurement of the $ \EEGG $ cross section at LEP energies'',
                             Phys. Lett. \textbf{B 327}, 386 (1994) \\
                             P. Abreu et al. [DELPHI Collaboration], 
                             `` Measurement of the $ \EEGG $ cross section at the LEP energies'', 
                             Phys. Lett. \textbf{B 433}, 429 (1998)\\
                             P. Abreu et al. [DELPHI Collaboration], 
                             `` Determination of the $ \EEGG $ cross-section at center-of-mass energies ranging from 189 GeV to 202 GeV'', 
                             Phys. Lett. \textbf{B 491}, 67 (2000)

 
\bibitem{L3A666}  M. Acciari et al. [L3 Collaboration], 
                        `` Tests of QED at LEP energies using $ \EEGG $ and $ {\EEllgg} $'', 
                        Phys. Lett. \textbf{B 353}, 136 (1995)   
                        
 

\bibitem{L3B888}  B. Achard et al. [L3 Collaboration], 
                       `` Study of multiphoton final states and tests of QED in $ e^{+} e^{-} $ collisions at  $ \sqrt{s} $ up to 209 GeV'', 
                       Phys. Lett. \textbf{B 351}, 28 (2002)

 


\bibitem{OPAL999}  G.  Abbiendi et al. [OPAL Collaboration], 
                             `` Multi-photon production in $ e^{+} e^{-} $ collisions at $ \sqrt{s} $ =181 - 209 GeV'', 
                             Eur. Phys. J. \textbf{C 26}, 331 (2003)



\bibitem{L3333} O. Adriani et al. [L3 Collaboration], 
                      `` A test of quantum electrodynamics in the reaction $ \EEGG $'', 
                       Phys. Lett. \textbf{B 288}, 404 (1992)\\
                      M. Acciari et al. [L3 Collaboration], 
                      ``  Observation of multiple hard photon final states at $ \sqrt{s} $ = 130- 140 GeV at LEP'', 
                      Phys. Lett. \textbf{B 384}, 323 (1996)\\
                      M. Acciari et al. [L3 Collaboration], 
                       `` Hard-photon production at $ \sqrt{s} $ = 161 and 172 GeV at LEP'', 
                       Phys. Lett. \textbf{B 413}, 159 (1997)\\                      
                      M. Acciari et al. [L3 Collaboration], 
                      `` Hard-photon production and tests of QED at LEP'', 
                      Phys. Lett. \textbf{B 475}, 198 (2000)                                          



\bibitem{Models1111} S.F. King and S.R. Sharpe, 
                             `` EXOTIC CERN EVENTS FROM EXOTIC COLOR STATES*'', 
                              Nucl. Phys. \textbf{B 253}, 1 (1985) \\
                             C.N. Leung, S.T. Love and S. Rao, 
                             `` Low-Energy Manifestations of a New Interactions Scale: Operator Analysis'',  
                             Z. Phys.\textbf{C 31}, 433 (1986) \\
                             S.D. Drell and S.J. Parke, `` Constrains on Radiative $ Z_{0} $ Decays '',                            
                             Phys. Rev. Lett. \textbf{53}, 1993 (1984) \\                            
                             D.A.  Dicus and X.Tata , 
                             `` ANOMALOUS PHOTON INTERACTION '', 
                             Phys. Lett. \textbf{B 155}, 103 (1985) \\ 
                             D.A.  Dicus,
                              `` New interactions and neutrino counting '', 
                              Phys. Rev. \textbf{D 31}, 2999 (1985)


\bibitem{ETHZ-USTC555}  E. Isiksal, No. 9479 Eidgen\"{o}ssische Technische Hochschule [ M. S. thesis ],
                                     ETH Z\"{u}rich, Z\"{u}rich, Switzerland, 1991.                                                                \\
                                     J. B. Xe, Chinese University of Science and Technology [ M. S. thesis ],
                                     USTC Hefei, Anhui 230 029, China 1992, unpublished.                                                           \\
                                     M. Maolinbay, No. 11028  Eidgen\"{o}ssische Technische Hochschule [ M. S. thesis ],
                                     ETH Z\"{u}rich, Z\"{u}rich, Switzerland, 1995.                                                               \\
                                     Jian Wu, No. JX1612-477, Chinese University of Science and Technology [ M. S. thesis ],
                                    USTC Hefei, Anhui 230 029, China 1997.                                                             \\
                                    Jiawei Zhao, No. JX1612-647 Chinese University of Science and Technology [M. S. thesis],
                                    USTC Hefei, Anhui 230 029, China 2001.  


\bibitem{RES:RES11} A. Kempf, G. Mangano, and R. B. Mann, ``Hilbert space representation of the minimal 
                         length uncertainty relation'', Physical Review \textbf{ D, vol. 52, no. 2,} 1108 - 1118 (1995)


\bibitem{RES:RES22} A. Kempf and G. Mangano, ``Minimal length uncertainty relation and 
                                  ultraviolet regularization'', 
                                  Physical Review \textbf{D, vol. 55, no. 12 }   7909 - 7920 (1997)



\bibitem{RES:RES33} A. Kempf, ``Mode generating mechanism in inpation with a cutoff'',  
                          Physical Review D, vol. 63, no. 8, Article ID 083514  5 pages, (2001)



 \bibitem{EMenergy1}  Khan Academy, `` Light: Electromagnetic waves, \\
                                   the electromagnetic spectrum photons '', \\
                                   https://www.khanacademy.org/science/physics \\
                                   /light-waves/introduction-to-light-waves \\
                                   /a/light-and-the-electromagnetic-spectrum

                  
 \bibitem{gamma_L3}  O.Adriani et. all. The L3 Collaboration, ``Results from the L3 Experiment at LEP'' , 
                                    CERN-PPE/93?31 February 22, 1993; \\
                                    Jiawei Zhao on the behalf of the L3 Collaboration , 
                                    ``QED Test at LEP200 Energies in the Reaction $ \EEGG $'',
                                    arXiv:hep-ph/0012095v4 6 Apr 2002;\\
                                    L3 Collaboration, Adeva, B., et al, ``The construction of the L3 experiment'' ,
                                    Nucl. Inst. Meth. A289, 35 (1990).\\
                                     l3.web.cern.ch/l3/l3pictures/events/gammas.html;

                                                                        
     
 
\bibitem{SIGNI} Frederik Beaujean , Allen Caldwell; ``A Test Statistic for Weighted Runs'',  \\ arXiv:1005.3233v2 [math.ST] 16 Jul 2010 



\bibitem{WIPI11}  Rouaud, M. (2013), `` Probability, Statistics and \\ 
                            Estimation: Propagation of Uncertainties in Experimental Measurement (PDF)'',  \\
                            $ http://www.incertitudes.fr/book.pdf $

\bibitem{ALEPHBhabha} R. Barate et al. ALEPH Collaboration, E. Phys. J. C 12, 183 (2000).


\bibitem{ALEPHBhabha1} ALEPH Collaboration, ALEPH 99-018 CONF 99 - 013, CERN, 1999


\bibitem{ALEPHBhabha2}ALEPH Collaboration, ALEPH 2000-047 CONF 2000- 030, CERN, 2000. 


\bibitem{ALEPHBhabha3} ALEPH Collaboration, ALEPH 2001-019 CONF 2001- 016, CERN, 2001.


\bibitem{DELPHIBhabha} P. Abreu et al. DELPHI Collaboration, Phys. Lett. B 485, 45 (2000).


\bibitem{DELPHIBhabha1}  DELPHI Collaboration, DELPHI 2000-128 OSAKA CONF 427, CERN, 2000.


\bibitem{DELPHIBhabha2} DELPHI Collaboration, DELPHI 2001-022 CONF 463, CERN, 2001.


\bibitem{L3Bhabha11}  M. Acciarri et al. L3 Collaboration, Phys. Lett. B 479, 101 (2000).


\bibitem{L3Bhabha22} L3 Collaboration, L3 Note 2563, CERN, 2000.


\bibitem{L3Bhabha33}  L3 Collaboration, L3 Note 2648, CERN, 2001.



\bibitem{OPALBhabha}  G. Abbiendi et al. OPAL Collaboration, Eur. Phys. J. C 6, 1 (1999).


\bibitem{OPALBhabha1}  G. Abbiendi et al. OPAL Collaboration, Eur. Phys. J. C 13, 553 (2000).


\bibitem{OPALBhabha2}  OPAL Collaboration, OPAL PN424, CERN, 2000.


\bibitem{OPALBhabha3}  OPAL Collaboration, OPAL PN469, CERN, 2001.


\bibitem{FB-Asymmetry} Journal of Physics: Conference Series 383 (2012) 012005 \\
                                        doi:10.1088/1742-6596/383/1/012005.
                                        
          
\bibitem{MonteCarlo1}  S. Jadach, W. Placzek, and B. Ward, Phys. Lett. B 390, 298 (1997).


\bibitem{MonteCarlo2}  Reports of the Working Groups on Precision Calculations for LEP2 physics; \\
                                     Two-fermion Production in Electron-positron
                                       Collisions, \\ CERN-YR-2000-009, hep-ph/0007180.


\bibitem{MonteCarlo3} G. Montagna et al., Nucl. Phys. B 401, 3 (1993).


\bibitem{MonteCarlo4}  A. B. Arbuzov, hep-ph/9907298.                                     
                                        


\bibitem{Kleiss11} R. Kleiss et al., Z Physics at LEP 1 (CERN, Geneva, 1989), Vol. 3.


\bibitem{Bourilkov24} D. Bourilkov, J. High Energy Phys. 08, 006 (1999).


\bibitem{L3-25} L3 Collaboration, M. Acciarri et al. Phys. Lett. B 433, 163 ( 1998 ).


\bibitem{L3-26} L3 Collaboration, M. Acciarri et al. Phys. Lett. B 489, 81 ( 2000 ).



\bibitem{TOT-cross-Bhabha} R. Boswell, Preliminary Wide-Angle Bhabha Results from 1994 Data, \\
                                               https://citeseerx.ist.psu.edu $ > $ document PDF.


\bibitem{L3-note-1111} L3 Collaboration, L3 Note 2563, CERN, 2000.


\bibitem{Firstpaper} D. Bourilkov, ``Hint for axial-vector contact interactions in the 
                                data on $  e^{+}e^{-}\to e^{+}e^{-}(\gamma ) $ at centre-of-mass energies 192-208 GeV'',
                                arXiv:hep-ph/0104165v1.


\bibitem{Runnig111}  U. Amaldi, Wim de Boer, Paul H. Frampton, H. Furstenau and J. T. Liu Phys.Lett.
                                       B281 ( 1992 ) 374


\bibitem{Timing555} Big history thresholds, https://esnarrative.weebly.com/threshold-1.html.



\bibitem{Timing222} Image Credit: Particle Data Group. \\
                                  https://blogs.unimelb.edu.au/nicole-bell/home/research/.


\bibitem{WMAP222} Wilkinson Microwave Anisotropy Probe, \\
                                 https://de.wikipedia.org/wiki/Wilkinson$_{-}$Microwave$_{-}$Anisotropy$_{-}$Probe


\bibitem{lambda1} K. Schwarzschild, On the gravitational field of a mass point according to Einstein's theory 
                             arXiv:physics/9905030

\end{thebibliography}
\end{document}